\documentclass[final,3p,times]{elsarticle}

\biboptions{sort&compress}

\graphicspath{{figs/}}

\usepackage{graphicx}
\usepackage{amssymb}
\usepackage{amsmath}
\usepackage{booktabs}
\usepackage{bm}
\usepackage{color}
\usepackage{multirow}
\usepackage{arydshln}
\usepackage{epsfig}
\usepackage{bm,bbm}
\usepackage{slashed}



\def\f(s){\left[(\alpha+\beta)m_c^2-\alpha\beta s\right]}
\def\FF(s){\left[(\alpha+\beta)m_c^2-\alpha\beta s\right]}
\def\HH(s){\left[m_c^2-\alpha(1-\alpha) s\right]}

\def\be{\begin{equation}}
\def\ee{\end{equation}}

\newcommand{\ver}{{\bm r}}

\newcommand{\vep}{{\bm p}}

\newcommand{\veS}{{\bm S}}
\newcommand{\veL}{{\bm L}}
\newcommand{\veJ}{{\bm J}}

\newcommand{\ven}{{\bm n}}

\newcommand{\veq}{{\bm q}}
\newcommand{\vek}{{\bm k}}
\newcommand{\xyz}{XYZ}
\newcommand{\xx}{X(3872)}
\newcommand{\y}{Y(4260)}

\newcommand{\zc}{Z_c(3900)}
\newcommand{\zcp}{Z_c(4020)}

\newcommand{\BR}{{\cal B}}

\newcommand{\pip}{\pi^+}
\newcommand{\pim}{\pi^-}
\newcommand{\piz}{\pi^0}

\newcommand{\hc}{h_c}
\newcommand{\pphc}{\pi^+\pi^- h_c}
\newcommand{\etac}{\eta_c}

\newcommand{\dstrbar}{\bar{D}^{*}}
\newcommand{\psp}{\psi(2S)}
\newcommand{\psip}{\psi(2S)}
\newcommand{\pspp}{\psi(3770)}
\newcommand{\jpsi}{J/\psi}

\newcommand{\EE}{e^+e^-}
\newcommand{\MM}{\mu^+\mu^-}
\newcommand{\LL}{\ell^+\ell^-}

\newcommand{\pp}{\pi^+\pi^-}
\newcommand{\ppp}{\pi^+\pi^-\pi^0}
\newcommand{\kk}{K^+K^-}
\newcommand{\ks}{K_{S}^{0}}

\newcommand{\ppjpsi}{\pi^+\pi^- J/\psi}

\newcommand{\ddpi}{D^0D^{*-}\pi^+ + c.c.}
\newcommand{\ch}{\textcolor{red}}

\usepackage[colorlinks, citecolor=blue,anchorcolor=red,menucolor=red, linkcolor=red,filecolor=red,runcolor=red,urlcolor=blue,frenchlinks=red]{hyperref}

\usepackage{lineno}

\journal{Physics Reports}

\begin{document}

\begin{frontmatter}


\title{The $XYZ$ states: experimental and theoretical status and perspectives}

\author[TUM,TUM2]{Nora Brambilla}
\ead{nora.brambilla@ph.tum.de}

\author[BINP,NSU,LPI]{Simon Eidelman}
\ead{eidelman@mail.cern.ch}

\author[FJ]{Christoph Hanhart\corref{cor1}}
\cortext[cor1]{Corresponding author}
\ead{c.hanhart@fz-juelich.de}

\author[LPI,MIPT,NRNU]{Alexey Nefediev}
\ead{nefediev@lebedev.ru}

\author[FD,BEIHANG]{Cheng-Ping Shen\corref{cor1}}
\ead{shencp@fudan.edu.cn}

\author[CU]{Christopher~E.~Thomas}
\ead{C.E.Thomas@damtp.cam.ac.uk}

\author[TUM]{Antonio Vairo}
\ead{antonio.vairo@ph.tum.de}

\author[IHEP,CASU]{Chang-Zheng Yuan}
\ead{yuancz@ihep.ac.cn}

\address[BEIHANG]{School of Physics and Nuclear Energy Engineering, Beihang University, Beijing 100191, China}
\address[BINP]{Budker Institute of Nuclear Physics, SB RAS, Novosibirsk 630090, Russia}
\address[CASU]{University of Chinese Academy of Sciences, Beijing 100049, China}
\address[CU]{DAMTP, University of Cambridge, Wilberforce Road, Cambridge CB3 0WA, UK}
\address[FD]{Key Laboratory of Nuclear Physics and Ion-beam Application and Institute of Modern Physics, Fudan University, Shanghai 200433, China}
\address[FJ]{Institute for Advanced Simulation, Institut f$\ddot{u}$r Kernphysik and J$\ddot{u}$lich Center for Hadron Physics,
Forschungszentrum J$\ddot{u}$lich, J$\ddot{u}$lich D-52425, Germany}
\address[IHEP]{Institute of High Energy Physics, Chinese Academy of Sciences, Beijing 100049, China}
\address[LPI]{P.N. Lebedev Physical Institute of the Russian Academy of Sciences, Leninskiy Prospect 53, Moscow 119991, Russia}
\address[MIPT]{Moscow Institute of Physics and Technology, Institutsky lane 9, Dolgoprudny, Moscow
Region 141700, Russia}
\address[NRNU]{National Research Nuclear University MEPhI, Kashirskoe highway 31, Moscow 115409, Russia}
\address[NSU]{Novosibirsk State University, Novosibirsk 630090, Russia}
\address[TUM]{Physik Department, Technische Universit$\ddot{a}$t M$\ddot{u}$nchen, Garching D-85748, Germany}
\address[TUM2]{Institute for Advanced Study, Universit$\ddot{a}$t M$\ddot{u}$nchen, Lichtenbergstrasse 2 a, Garching D-85748, Germany}

\begin{abstract}
  The quark model was formulated in 1964 to classify mesons as bound states made of a quark-antiquark pair, and baryons as bound states made of three quarks.
  For a long time all known mesons and baryons could be classified within this scheme.
  Quantum Chromodynamics (QCD), however, in principle also allows the existence of more complex structures, generically called exotic hadrons or simply exotics.
  These include four-quark hadrons (tetraquarks and hadronic molecules), five-quark hadrons (pentaquarks) and
  states with active gluonic degrees of freedom (hybrids), and even states of pure glue (glueballs).
  Exotic hadrons have been systematically searched for in numerous experiments for many years.
  Remarkably, in the past fifteen years, many new hadrons that do not exhibit the expected properties of ordinary (not exotic) hadrons have been discovered in the quarkonium spectrum.
  These hadrons are collectively known as $XYZ$ states. Some of them, like the charged states, are undoubtedly exotic.
  Parallel to the experimental progress, the last decades have also witnessed an enormous theoretical effort to reach a theoretical understanding of the $XYZ$ states.
  Theoretical approaches include not only phenomenological extensions of the quark model to exotics,
  but also modern non-relativistic effective field theories and lattice QCD calculations.
  The present work aims at reviewing the rapid progress in the field of exotic $XYZ$ hadrons over the past few years both in experiments and theory.
  It concludes with a summary on future prospects and challenges.
\end{abstract}


\end{frontmatter}

\tableofcontents

\section{Introduction}
\label{sect:1}
\subsection{Some generalities}

Hadrons are described in the Standard Model by Quantum Chromodynamics (QCD), the theory of the strong interactions.
QCD is based on the gauge symmetry group SU(3), whose quantum number is called `color'.
At the most fundamental level its degrees of freedom are quarks ($q$)\footnote{For a heavy quark a capital letter $Q$ ($\bar{Q}$) for the (anti)quark may be used in what follows to stress its large mass.}, antiquarks ($\bar q$) and gluons ($g$), the last ones being the SU(3) gauge bosons~\cite{Fritzsch:1973pi}.
The fact that SU(3) is a non-Abelian group makes QCD a very different theory from Quantum Electrodynamics (QED), the theory of the electromagnetic interactions,
and accordingly the phenomenology of hadrons is very different from the one of atoms.
Gluons, in contrast to photons, carry a (color) charge, self-interact, antiscreen the vacuum, and are at the origin of such highly nontrivial phenomena in QCD as asymptotic freedom and color confinement.
Asymptotic freedom guarantees that high-energy properties of QCD can be computed in weak-coupling perturbation theory~\cite{Gross:1973id,Politzer:1973fx}.
On the other hand, color confinement, which implies that all detected hadrons are singlets (neutral) under color SU(3), sets in at low energies, where weak-coupling perturbation theory fails.
Both energy scales are relevant for the formation of heavy hadrons.

The simplest color-singlet quark combinations that can be formed are $\bar qq$ and $qqq$. For a long time, all known hadrons could be explained in terms of the quark model where mesons are bound states of a constituent quark and antiquark, and baryons are bound states of three constituent quarks.
However, the requirement of color neutrality still allows
for more complicated structures, as was already recognised by Gell-Mann in one of the first publications on the quark model~\cite{GellMann:1964nj}, like $\bar q\bar q qq$, $\bar q\bar q\bar q qqq$, $\bar q qqqq$, and so on.
In QCD additional states are, in principle, possible
like hybrids whose quantum numbers are determined by their quark, antiquark and gluon content:
as gluons carry a color charge they can play a much more active role in the formation of a state than
photons in atoms. One even expects `glueballs' whose quantum numbers are determined exclusively by their gluonic content.
We will refer to any state which does not appear to fit with the expectations for an ordinary $\bar qq$ or $qqq$ hadron in the quark model as `exotic'.

Some of these exotics have quantum numbers that cannot be reproduced by ordinary hadrons.
In this case, the identification of these states as exotic is straightforward.
In the other cases, the distinction requires a careful analysis of experimental observations and theoretical
predictions as will be also discussed in some detail in this review.

There are various candidates for exotics in the light-quark sector, e.g., the light scalar mesons.
Nevertheless, at the beginning of the century, no candidates for states beyond ordinary hadrons containing heavy quarks had been found.
This was even more the case for mesons made of a heavy quark and a heavy antiquark, called quarkonia:
The quark model showed excellent agreement with the data both in the charmonium ($\bar cc$ quarkonia) and in the bottomonium ($\bar bb$ quarkonia) sectors.
The situation changed dramatically in 2003 when Belle observed a structure in the $\pi\pi J/\psi$ final state~\cite{Choi:2003ue}, the $X(3872)$.
The properties of the new state did not fit those of an ordinary quarkonium, although it definitely contained a charm-anticharm pair.
The assumption that the $X(3872)$ contains the charm-anticharm pair observed in the final state is motivated by the strong suppression of the heavy quark-antiquark pair creation within QCD.
After the discovery of the $X(3872)$, many more hadrons were found in processes with final states containing a heavy quark and antiquark,
but with properties at odds with those expected for ordinary quarkonia.
Among these new states, those that carry an electric charge stick out.
Since they must contain, in addition to a heavy quark-antiquark pair, at least a light quark-antiquark pair, they immediately qualify as exotics. For states that from their quantum numbers could be quark-antiquark mesons, but show properties
inconsistent with expectations from established quark models, sometimes the term crypto-exotic is used.

The candidates for exotic states were dubbed by the experimental collaborations either $X(\mbox{mass})$, or $Y(\mbox{mass})$, or $Z(\mbox{mass})$, and are usually referred to collectively as $XYZ$ states.
$Y(\mbox{mass})$ is usually used for exotics with vector quantum numbers, i.e., $J^{PC} = 1^{--}$ where $J$ is the spin, $P$ is the parity and $C$ is the charge-conjugation quantum number.
Meanwhile, the Particle Data Group (PDG) has proposed a new naming scheme~\cite{Tanabashi:2018oca} which is briefly reviewed in Sec.~\ref{sec:namingscheme}.

\begin{figure}[h!]
\begin{center}
\includegraphics[width=0.9\linewidth]{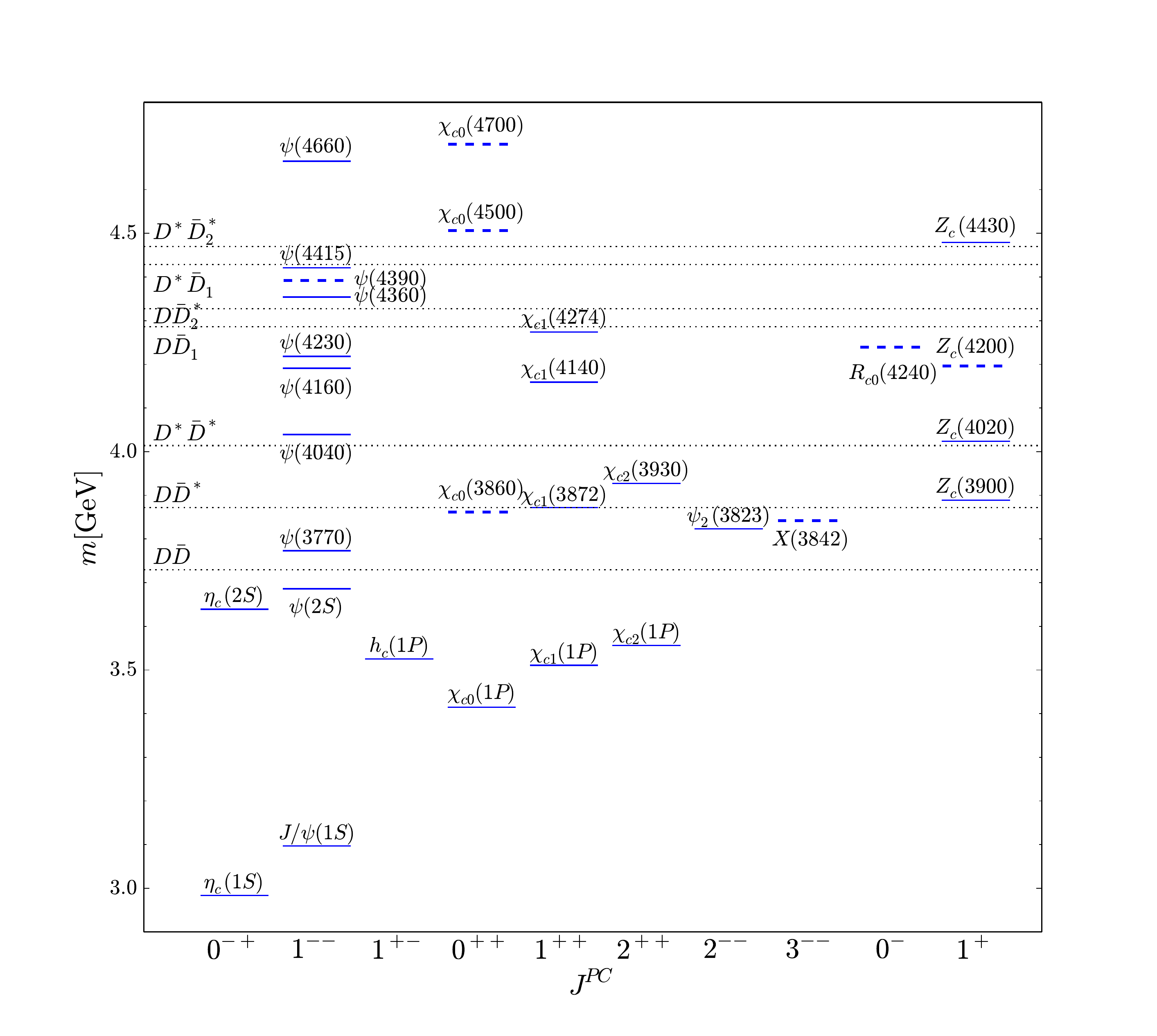}
\caption{The spectrum of states in the $\bar cc$ sector as of July 2019.
Thin solid lines represent the states established experimentally and dashed lines are for those that are claimed but not (yet) established
(following the approach used by PDG, we regard a state as established if it is seen in different modes).
States whose quantum numbers are undetermined are not shown.
States in the plot are labeled according to the PDG primary naming scheme --- see Sec.~\ref{sec:namingscheme} for further details and for the correspondence with the $XYZ$ naming scheme.
Dashed lines show some relevant thresholds that open in the considered mass range; here $D_1$ stands for $D_1(2420)$ and $D_2^*$ for $D_2^*(2460)$.
Thresholds with hidden strangeness or involving broad states are not shown.
The states shown in the two columns to the right are isovectors containing a $\bar c c$ pair; they are necessarily exotic. }
\label{fig:barccspec}
\end{center}
\end{figure}

\begin{figure}[h!]
\begin{center}
\includegraphics[width=0.9\linewidth]{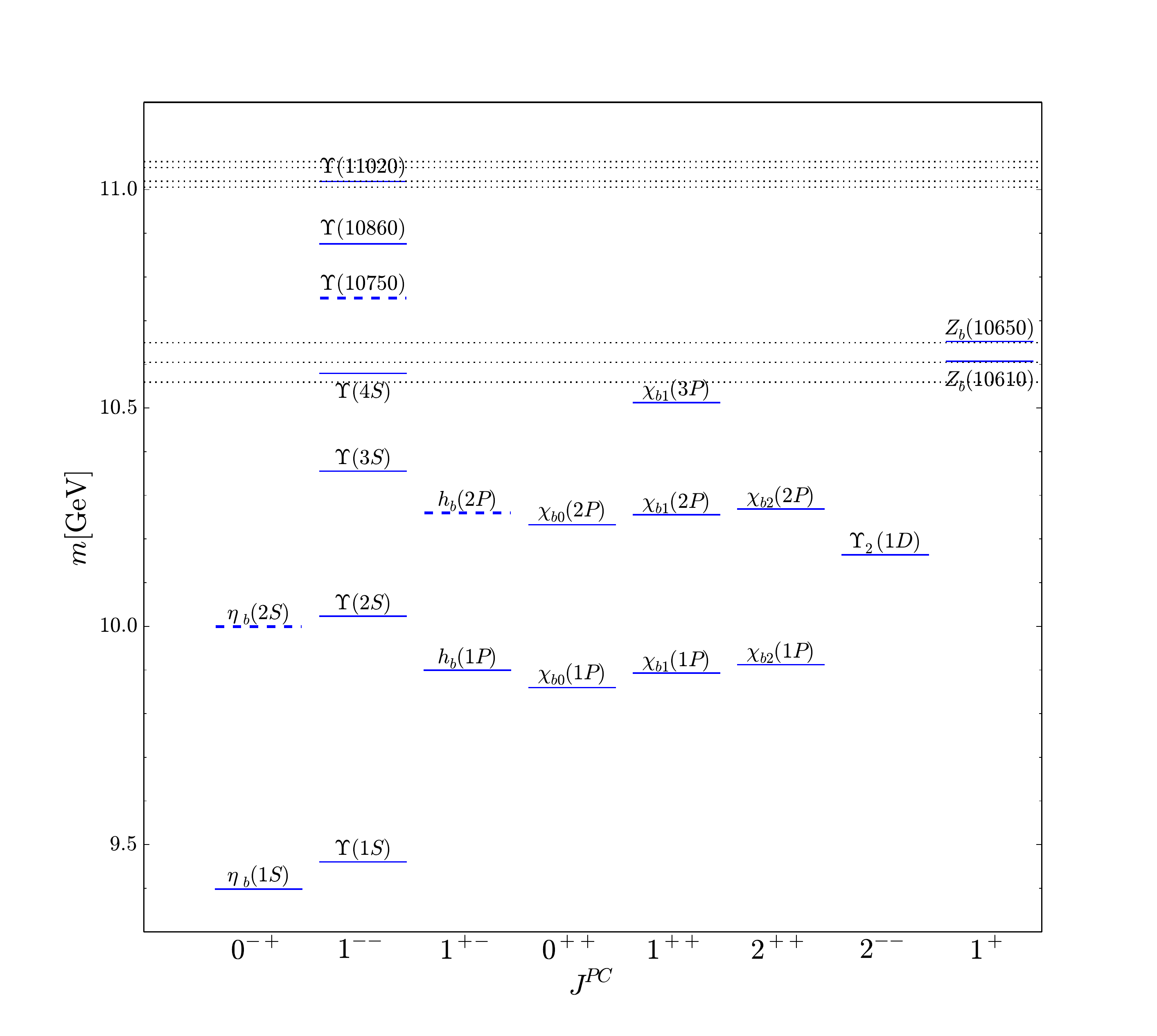}
\caption{The same as in Fig.~\ref{fig:barccspec} but for the states in the $\bar{b}b$ spectrum.
Dashed lines show the open-bottom thresholds (from the lowest upwards)
$\bar BB$, $\bar BB^*$, $\bar B^*B^*$, $\bar BB_1(5721)$, $\bar BB_2^*(5747)$, $\bar B^*B_1(5721)$, and $\bar B^*B(5747)$.
Given its large statistical significance (6.7 standard deviations) we have included the $\Upsilon(10750)$ signal recently seen by the Belle Collaboration~\cite{Abdesselam:2019gth},
although it is not in the current PDG list (see also Sec.~\ref{Sect:4.2.3bis}).}
\label{fig:barbbspec}
\end{center}
\end{figure}

The spectrum in the $\bar cc$ sector is shown in Fig.~\ref{fig:barccspec}.
Only states with known quantum numbers are included in the figure and, in all cases, the primary name according to the PDG is used.
In addition, we also show with dotted lines some thresholds for decays into open-flavour states.
To not overload the figure, thresholds with hidden strangeness are not shown.
States that show quantum numbers incompatible with the quark model are clearly exotic.
For the states in Fig.~\ref{fig:barccspec}, this applies to the isovector states displayed in the two rightmost columns.
In the other cases, there is no general rule when a state should carry the label exotic and when it is an ordinary $\bar c c$ state.
However, there is consensus that all states below the lowest open-flavour threshold ($\bar DD$) are ordinary states.
Moreover, also the properties of the vector states $\psi(3770)$, $\psi(4040)$ and $\psi(4160)$ and of the tensor state $\psi_2(3823)$ appear to agree with those of ordinary quarkonia.
All the other states may or may not be exotic and will be discussed in this report.
As of today, in the charmonium sector the number of experimentally established exotic candidates is similar to the number of ordinary states.
In the bottomonium sector, on the other hand, only two exotics are established and they are both charged. The states in the $\bar{b}b$ spectrum are shown in Fig.~\ref{fig:barbbspec}.

Some of the exotic candidates reside rather close to open-flavour thresholds, however, the impact these thresholds have on the states is not clear yet.
A possibility is that they induce kinematic enhancements, so that not all of the observed signals may correspond to new states in the QCD spectrum.
Nevertheless, as we will argue in this review, most of them certainly do correspond to new states.
From a comparison of Figs.~\ref{fig:barccspec} and~\ref{fig:barbbspec} one can see that
the distribution of states and thresholds in the charmonium sector and in the bottomonium sector is, at present, rather different.
If this hints to some physical differences in the two heavy-quark sectors or just reflects our limited knowledge of the bottomonium spectrum
is one of the challenges that future theoretical and experimental studies will have to face.

Exotic states may originate dynamically from different possible structures.
These can be grouped into two classes: (i) structures with active gluons and (ii) multiquark states.
The former class contains hybrids and glueballs.
The latter class contains exotics made of a heavy quark and its antiquark together with at least one light quark-antiquark pair.
\emph{A priori} the simplest system consisting of only two quarks and two antiquarks (generically called tetraquarks) is already a very complicated object
and it is unclear whether or not any kind of clustering occurs in it. However, to simplify the problem it is common nowadays to focus on
certain substructures and investigate their implications:
(i) In hadroquarkonia the heavy quark and antiquark form a compact core surrounded by a light-quark cloud.
(ii) In compact tetraquarks the relevant degrees of freedom are compact diquarks and antidiquarks. In this review we restrict ourselves to this definition of a compact tetraquark which is quite common in the contemporary literature. {It should be mentioned, however, that in the older literature  sometimes
the phrase ``diquonium'' was used to refer directly to the assumed diquark clustering~\cite{Chan:1978nk,Gavela:1978hq}.}
For a comprehensive review of diquarks we refer to Ref.~\cite{Anselmino:1992vg} while for
a critical discussion of the diquark clustering in heavy exotics we refer to Ref.~\cite{Richard:2018yrm}. (iii) In hadronic molecules the building blocks are color-neutral hadrons.
Obviously a complete picture may include some of or even all these configurations in the final description of the
exotic state; nevertheless, the question remains of what is the dominant configuration.
Finally, even an ordinary hadron, which has a dominant $\bar q q$ or $qqq$ configuration, will have subleading configurations with additional quark-antiquark pairs and active gluons. Depending on the state, they may turn out to be more or less relevant to describe its properties.

\medskip
\centerline{* \quad * \quad *}
\medskip

This report is devoted to exotic states in the charmonium and bottomonium spectrum, i.e. the $XYZ$ states, and also discusses related states in this energy region, such as double-charm/bottom systems, and pentaquarks containing heavy quarks.
Its aim is to review the experimental information we have gained on these states over the last fifteen years, and to summarize our theoretical understanding of them.

Data have been collected originally mostly at the $B$ factories (Belle, BaBar, CLEO) and also at the Tevatron experiments, and later at BES and LHCb.
They have led to the discovery of several new states (see Figs.~\ref{fig:barccspec} and~\ref{fig:barbbspec}) and
to the collection of an impressive amount of measurements in production and decay channels.
This clean and clear amount of data has no match with the few controversial exotic states in the light-hadron sector.

From the theoretical viewpoint, the study of quarkonium in the last few decades has witnessed two major developments:
(i) the establishment of non-relativistic effective field theories and (ii) progress in dynamical lattice QCD calculations of excited states and resonances, and calculations with light-quark masses at or close to the physical point.
Both allow for precise and systematically improvable computations that are (to a large extent) model-independent.
Although in many cases phenomenological quark models remain a useful resource,
it is the advancement in our understanding of quarkonium and quarkonium-like systems due to non-relativistic effective field theories and lattice QCD
that makes quarkonium exotica particularly valuable.
In fact, today we are not only confronted with a huge amount of high-quality data, which have provided for the first time
uncontroversial evidence for the existence of exotic hadrons, but also have
 modern theoretical tools that allow us to explore in a controlled way these new forms of matter
and get a unique insight into the low-energy dynamics of QCD.

The new quarkonium revolution that started in 2003 with the discovery of the $X(3872)$, with all its experimental and theoretical developments and challenges,
has been chronicled over the past years in several comprehensive reviews~\cite{Brambilla:2004wf,Brambilla:2010cs,Brambilla:2014jmp}.
The present report focuses on the $XYZ$ states and aims at portraiting, as precisely as possible, the status of the subject in the year 2019.
It is organized in the following main sections.
In Sec.~\ref{sect:2}, we review the main experimental facilities participating to the $XYZ$ searches.
The status of these searches and the relevant data that have been collected are summarized in Sec.~\ref{Sect:3}.
Theoretical methods, ranging from phenomenological quark models to effective field theories of QCD and lattice QCD,
and theoretical results and predictions are discussed in Sec.~\ref{sect:4}.
Finally, in Sec.~\ref{sect:5} some future prospects both for experiments and theory are highlighted.
We close with a short summary.

As was explained above, in recent years the field of exotic hadrons in general and especially those containing heavy quarks has attracted a lot of experimental and theoretical efforts. As an interesting fact it is worth mentioning that, despite a very rich physics programme at the Belle experiment and many interesting and exciting results obtained by this collaboration for the entire history of its operation, the paper on the discovery of the exotic $X(3872)$ charmonium-like state turns out to be its most cited publication. Thus it should not come as a surprise that the number of papers on the subject, both experimental and theoretical, including various reviews, grows fast. It is therefore important to make it clear from the very beginning what one will find here that has not been previously discussed. A unique feature of the present review is that it contains simultaneously the most recent experimental information and updates on the exotic states, and a discussion of various aspects of the theoretical interpretation of the existing and foreseen experimental data. In the theoretical part, although various models for exotic states presented in the literature are mentioned and discussed in some detail, the main emphasis is on model-independent approaches based on effective field theories and lattice QCD.

\subsection{Remarks on the naming scheme of the Particle Data Group}
\label{sec:namingscheme}

Exotic states that do not seem to fit with expectations of ordinary quarkonia have been dubbed $X$, $Y$, and $Z$ in their discovery publications, without any special naming criterion.
These names have been used so far in most of the literature on the subject.
With the number of $XYZ$ states growing, this way of naming the new states has become increasingly inadequate.
For this reason the PDG has recently developed a new naming scheme~\cite{Tanabashi:2018oca} with the intent to
extend the scheme used for ordinary quarkonia, and based on the quark model, to the newly discovered states.
The new names carry the information on the $J^{PC}$ quantum numbers of the states.
In addition, aside from the ground states, a mass label is added in brackets behind the name symbol.
This mass label may deviate from the actual mass
 whenever updated measurements give shifted mass values compared to the
earlier measurements.
They do not say anything about the nature of the state, as this is in many cases controversial.
For instance, even the $X(3872)$ is not unanimously accepted as an exotic quarkonium.
The only exception are the charged states, which clearly require at least a four-quark structure.

\begin{table}[ht]
\begin{center}
\begin{tabular}{|c|c|cccc|}
\hline
\multicolumn{2}{|r|}{$PC$}&${-+}$&${+-}$	&${--}$	&${++}$\\
\hline
Isospin & heavy quark content & & & & \\
\strut $I=0$ & with $c\overline c$&$\eta_c$&$h_c$&$\psi$&$\chi_c$\\
\strut $I=0$ & with $b\overline b$&$\eta_b$&$h_b$&$\Upsilon$&$\chi_b$\\
\strut $I=1$ & with $c\overline{c}$&$(\Pi_c)$&$Z_c$&$(R_c)$&$(W_c)$\\
\strut $I=1$ & with $b\overline{b}$&$(\Pi_b)$&$Z_b$&$(R_b)$&$(W_b)$\\
 \hline
\end{tabular}
\caption{The PDG naming scheme for quarkonium and quarkonium-like states. For $I=1$, $C$ refers to the charge-conjugation quantum number of the neutral state.
The quark model for ordinary quarkonia only allows even values of $J$ for the states in the first column, and odd values of $J$ for those in the second one.
Moreover, it does not allow a $0^{--}$ state -- see Sec.~\ref{Sect:4.1.1}.
Aside from the $\eta_Q$ states with $J=0$ and $h_Q$, $\psi$ and $\Upsilon$ states with $J=1$,
the value of $J$ is added to the name as a subscript.
E.g., a state containing a $\bar cc$ pair with the quantum numbers $1^{-+}$ is called $\eta_{c1}$.
Aside from the ground states a mass label is added in brackets to the name symbol.
In this table names in brackets indicate that there is no experimentally established state (yet) with the corresponding quantum numbers. \label{tab:PDGnames}}
\end{center}
\end{table}

The naming scheme adopted by the PDG is shown in Tab.~\ref{tab:PDGnames} and explained in the caption.
It serves the purpose of allowing an unambiguous identification of the quantum numbers of a given state from its name.
States whose quantum numbers are not yet fixed are named $X(\mbox{mass})$.
It is important to remark that, even if in this way a state may get the name of an ordinary quarkonium,
this is not meant to imply that the nature of the state is that of an ordinary quarkonium.
So far, only for the charged quarkonium-like states new names have been introduced --- here the names already used in the literature are used,
namely $Z_c(\mbox{mass})$ and $Z_b(\mbox{mass})$ for states containing a charm-anticharm and a bottom-antibottom pair, respectively, with the isospin 1 and $J^{PC}=1^{+-}$ (where $C$ refers to the charge-conjugation quantum number of the neutral member of the isotriplet).

To keep track of the names used so far in the literature, those are added in the listings with the remark `also known as', abbreviated as `aka'.
For instance, the states known as $X(3872)$ and $Y(4660)$ appear as
$$
\chi_{c1}(3872) \ \mbox{aka} \ X(3872) \quad \mbox{and} \quad \psi(4660) \ \mbox{aka} \ Y(4660).
$$
In this report, we use the $XYZ$-names for the states most of the time, but in various places we remind the reader of the PDG naming scheme, especially in the headings.

\section{Experiments}
\label{sect:2}

As a hot topic in experimental particle physics, exotic hadrons
containing heavy quarks
are studied in all experiments where their production is possible.
This includes not only experiments dedicated to hadron physics
such as BESIII, GlueX and CLEO-c, but also  experiments designed for
completely different purposes such as ATLAS and CMS for precision
eletroweak physics and beyond, and the $B$ factories, BaBar, Belle, and
LHCb, for $CP$ violation. Sometimes the study of  exotic states is
originally only a byproduct of some other measuremnts, but it becomes more
and more significant as more and more candidates for exotic hadrons
are observed.

Most of the exotic hadron studies are from $e^+e^-$ annihilation experiments
such as BESIII, Belle, BaBar, and CLEO, because of the very clean experimental
environment and various production mechanisms. Exotic hadrons
are produced directly in $\EE$ annihilation and in association with
another charmonium production (double charmonium production), two-photon
processes, initial-state radiation (ISR) processes, bottomonium
decays and $B$ decays.

More and more results are being reported from the LHCb experiment where
long-lived $b$ hadrons ($B$ and $B_s$ mesons as well as the $\Lambda_b$ baryon)
produced in $pp$ collisions are used as a source of exotic hadrons.
Essential information on production of exotic hadrons directly in parton
fusion and in $b$-hadron decays is also coming from the
ATLAS and CMS experiments at the LHC $pp$ collider,
and came from CDF and D0 at the Tevatron $p\bar p$ collider.

In this section, we give a short introduction to the experiments
which are involved in the studies of exotic hadrons discussed in this review,
including a description
of the detector components, data samples, and production mechanisms
of exotic hadrons. New experiments being built or planned
which can contribute to the study of exotic states in the future,
along with those which have recently started taking data such as GlueX and Belle II,
will be discussed in Sec.~\ref{sect:6.1}.

\subsection{Experiments at $e^+e^-$ colliders}
\label{Sect.2.0}
The first state that triggered studies of the new hadron spectroscopy is
the $\xx$ observed by the Belle experiment in 2003.
All the $\EE$ annihilation experiments operating at that time joined the
effort immediately,
if their energy could reach the production threshold of charmonium states.
The BaBar experiment with data samples comparable with Belle could do the
same study.
The CLEO experiment also had data in the bottomonium energy region but with
much smaller statistics, whereas somewhat later the CLEO-c detector could run
directly in the charmonium energy region allowing for new specific studies.
The BES experiment running at the BEPC had small data samples only, originally
taken for measurements of the $R$ value\ch{,} and a focused study of the $\xyz$
states became possible only after its upgrade to the BESIII experiment at
BEPCII with its first data sample taken in 2011.

\subsubsection{The Belle experiment}
\begin{figure*}
\centering
\includegraphics*[width=.65\textwidth]{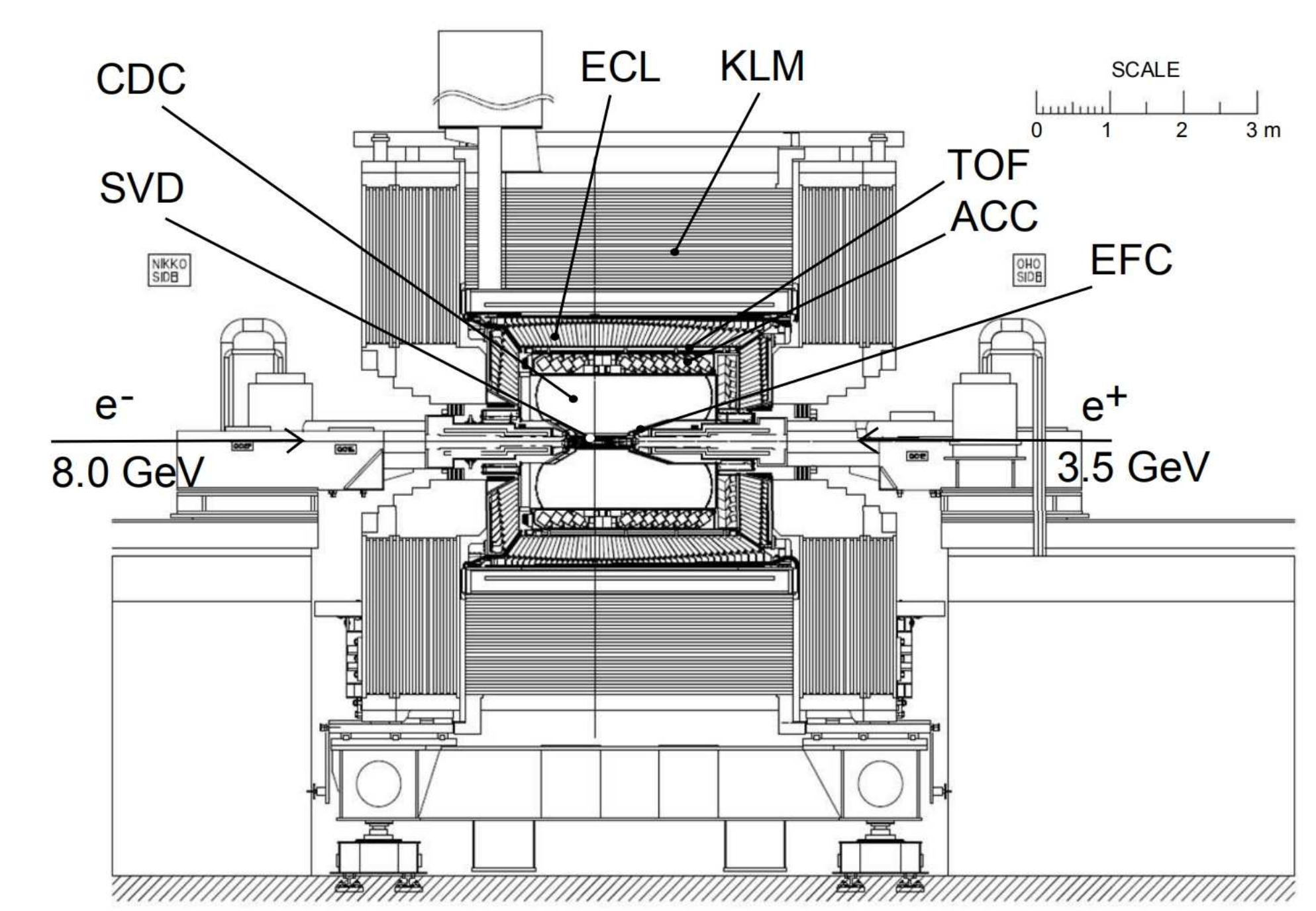}
\caption{Belle detector longitudinal section from Ref.~\cite{Abashian:2000cg}.} \label{belle_detector}
\end{figure*}

The Belle detector~\cite{Abashian:2000cg} operating at the KEKB
asymmetric-energy $\EE$ collider~\cite{Kurokawa:2001nw} is a large
solid-angle magnetic spectrometer that consists of a silicon
vertex detector (SVD), a 50-layer central drift chamber (CDC), an array of
aerogel threshold Cherenkov counters (ACC), a barrel-like arrangement of
time-of-flight (TOF) scintillation counters, and an electromagnetic
calorimeter (EMC) comprised of CsI(TI) crystals located inside a
superconducting solenoid coil that provides a 1.5~T magnetic
field. Information from  specific ionization in the
CDC, time measurements in the TOF and the response of the ACC
is combined to perform the charged particle identification (PID).
An iron flux-return yoke instrumented with resistive plate
chambers (RPC) located outside the coil is used to detect $K^{0}_{L}$
mesons and to identify muons. A detailed description of the Belle
detector can be found in Refs.~\cite{Abashian:2000cg,
Bevan:2014iga}. Figure~\ref{belle_detector} shows the structure of
the Belle detector.

The Belle detector as described above took data from 1999-2010
that are partially still under analysis. Since the beginning of 2019 the successor experiment
Belle II, which is described in Sec.~\ref{sect:6.1} of this review, has been in operation.
The analyses from Belle utilize one or more of the following data
samples: 5.74~fb$^{-1}$ of data collected at the $\Upsilon(1S)$ peak
[102 million $\Upsilon(1S)$ events], 24.91~fb$^{-1}$
collected at the $\Upsilon(2S)$ peak [158 million $\Upsilon(2S)$
events], 89.5~fb$^{-1}$ collected at $\sqrt{s} =10.52$~GeV,
702.6~fb$^{-1}$ data collected at $\sqrt{s} =10.58$~GeV
[$\Upsilon$(4S) peak], and 121.1~fb$^{-1}$ data collected at
$\sqrt{s} =10.867$~GeV [$\Upsilon$(5S) peak]. There are also some
data points between $\Upsilon(4S)$ and $\Upsilon(6S)$ with low
statistics for a measurement of the inclusive hadronic cross sections.

\subsubsection{The BaBar experiment}
\begin{figure*}
\centering
\includegraphics*[width=.65\textwidth]{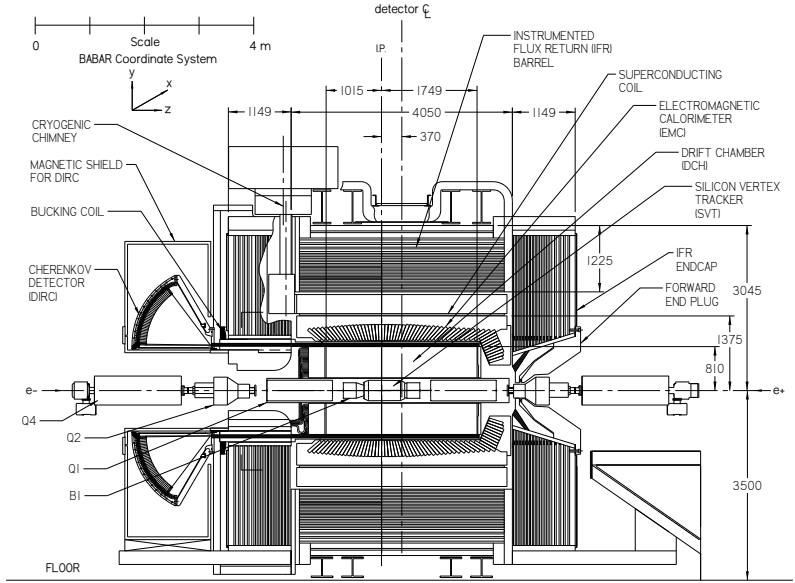}
\caption{BaBar detector longitudinal section from Refs.~\cite{Aubert:2001tu}.}
\label{babar_detector}
\end{figure*}

The PEP-II $B$ Factory was an asymmetric $\EE$ collider designed to
operate at a center-of-mass (c.m.) energy of 10.58~GeV, the mass of the
$\Upsilon(4S)$ resonance. The BaBar detector was the
experiment running at this collider. Figure~\ref{babar_detector}
shows a longitudinal section through the detector center~\cite{Aubert:2001tu,TheBABAR:2013jta}.

The charged particle tracking system was made of two components\ch{:}
the silicon vertex tracker (SVT) and the drift chamber (DCH).
Pulse height information from the SVT and DCH was also used to
measure ionization losses for charged PID.
The SVT was composed of five layers of double-sided silicon strip
detectors and the DCH of 40 layers of small, hexagonal cells. The
Detector of Internally Reflected Cherenkov light (DIRC)
provides separation of pions and kaons up to the kinematic limit of 4.5~GeV.
Cherenkov light was produced in 4.9-m long bars of
synthetic fused silica of rectangular cross section, and
transported to a large array of photomultiplier tubes (PMT). The
EMC was a finely segmented array of CsI(Tl)
crystals of projective geometry to detect electromagnetic
showers with excellent energy and angular resolution over the
energy range from 20~MeV to 4~GeV. The instrumented flux return was
designed to identify muons and to detect neutral hadrons.
The BaBar experiment took data from 1999-2008, however, some analyses are
still ongoing.

The analyses from BaBar utilize one or more of the following data
samples: 13.6~fb$^{-1}$ of data collected at the $\Upsilon(2S)$ peak,
27.96~fb$^{-1}$ collected at the $\Upsilon(3S)$ peak,
424.7~fb$^{-1}$ collected at the $\Upsilon(4S)$ peak,
and 43.9~fb$^{-1}$ collected at $\sqrt{s} =10.54$~GeV.

\subsubsection{The CLEO(-c) experiment}
The CLEO experiment ran successfully for nearly thirty years,
from 1979 to 2008, recording particles produced in
electron-positron collisions at the Cornell Electron-positron
Storage Ring (CESR). The experiment took data in the bottomonium
energy region before 2003 and CLEO-c, a modified version
of the CLEO III detector, accumulated
data in the charmonium energy region between 2004
and 2008. The main differences between CLEO III and CLEO-c
detectors are the SVD replaced with a wire
vertex chamber and the magnetic field reduced from 1.5~T to
1.0~T to maintain high efficiency for low momentum tracks and
reasonably high momentum resolution.

The CLEO-c detector was equipped to measure the momenta and
directions of charged particles, identify charged hadrons, detect
photons, and determine with good precision their directions and
energies. The muons above $p=1.1$~GeV could also be identified
with the muon detector. The detector was almost cylindrically
symmetric with everything but the muon detector inside a
superconducting magnet coil supplying 1.0~T field. The charged
tracks were reconstructed using the 47-layer drift chamber and the
coaxial 6-layer vertex drift chamber. For tracks that traverse all
layers of the drift chamber, the root-mean-square (rms) momentum
resolution was approximately 0.6\% at $p = 1$~GeV.
In the whole of this review we set $c=1$.
Photons were detected in an EMC containing about
7800 CsI(Tl) crystals, whose rms photon energy resolution was 2.2\%
at $E_\gamma = 1$~GeV, and 5\% at $E_\gamma = 100$~MeV. The solid
angle for detection of charged tracks and photons was 93\% of
$4\pi$. PID information to separate kaon
from pion was provided by measurements of ionization ($dE/dx$) in
the CDC and by a cylindrical ring-imaging
Cherenkov (RICH) detector.

The CLEO experiment had 22.2~fb$^{-1}$ of $e^+e^-$ data taken in
the $\Upsilon(1S-5S)$ region, which were used for $\xyz$ and
charmed meson studies\ch{.} The CLEO-c experiment accumulated 27
million $\psi(2S)$ events, 818~pb$^{-1}$ of data at the $\psi(3770)$
peak, 586~pb$^{-1}$ collected at 4.17~GeV, and 60~pb$^{-1}$ of data
between 3.97 and 4.26~GeV (including 13~pb$^{-1}$ of data at 4.26~GeV) --
these data were also used for $\xyz$ studies.

\subsubsection{The BESIII experiment}
The BESIII experiment~\cite{Ablikim:2009aa} at the BEPCII storage
ring started its first collisions in the tau-charm energy region
in 2008. The BESIII detector has an effective geometrical
acceptance of 93\% of $4\pi$. It contains a small cell
helium-based multilayered drift chamber (MDC) that provides
momentum measurements of charged particles; a TOF based on plastic
scintillator which helps to identify
charged particles; an EMC made of CsI(Tl) crystals which is used to
measure the energies of photons
and provide trigger signals; and a muon system (MUC) made of
RPCs. The momentum resolution of the
charged particles is $0.5$\% at 1~GeV in a 1~T magnetic field;
the energy loss  measurement provided by the MDC has a
resolution better than 6\% for electrons from Bhabha scattering;
the photon energy resolution can reach $2.5$\% ($5$\%) at 1~GeV in
the barrel (endcaps) of the EMC; and the time resolution of TOF is
$80$~ps in the barrel and $110$~ps in the endcaps. In 2015, the
endcap TOF was replaced with a Multigap Resistive Plate Chamber (MRPC),
and the time resolution improved to 60~ps.

\begin{figure*}
\centering
\includegraphics*[width=8cm]{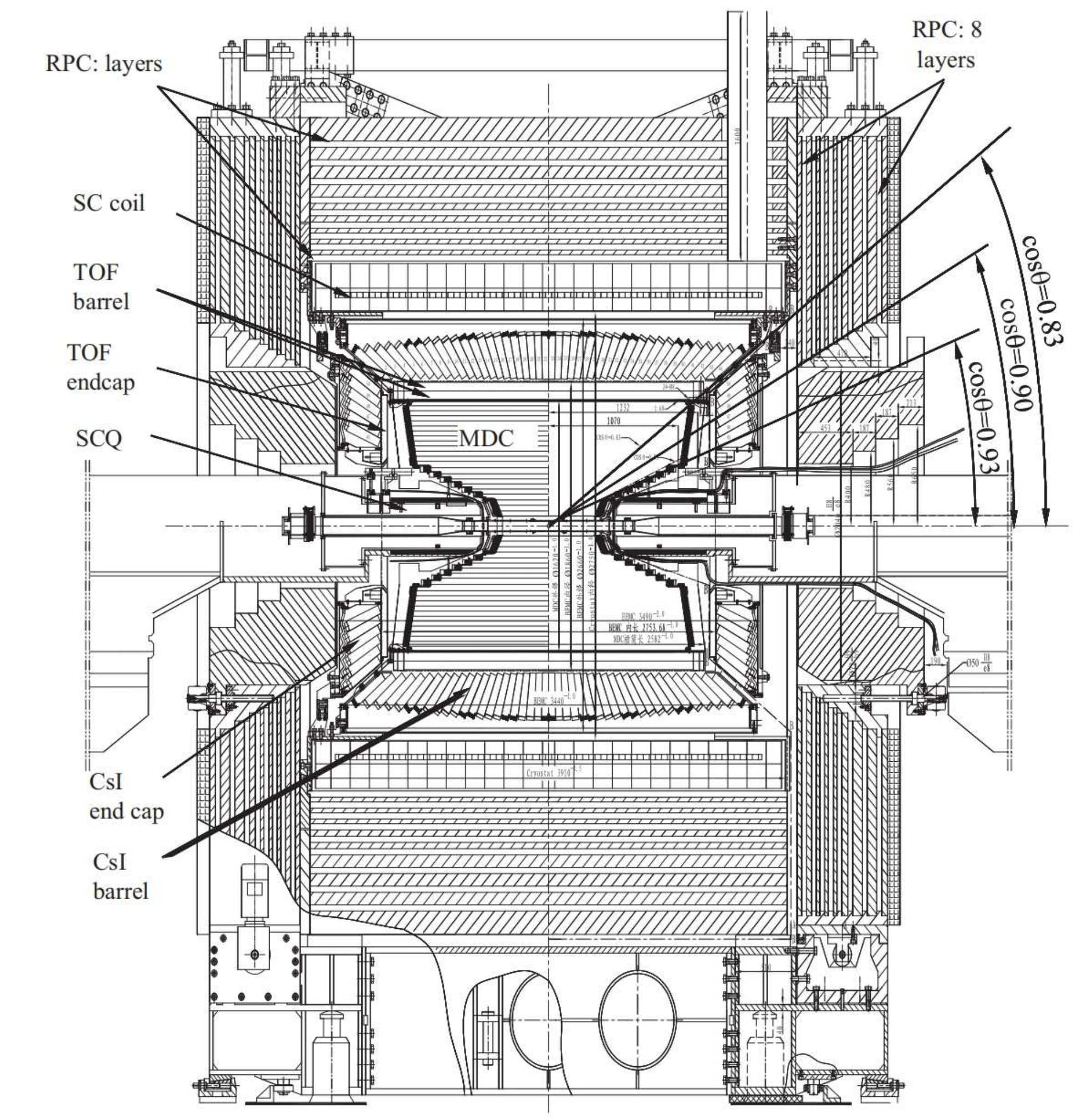}
\caption{Schematic drawing of the BESIII detector from Ref.~\cite{Ablikim:2009aa}.}
\label{bes3_detector}
\end{figure*}

After a few years of running at energies of the $\jpsi$, $\psp$, and $\pspp$
peaks~\cite{Asner:2008nq},
the BESIII experiment started collecting data for
the study of $\xyz$ particles. At the moment the maximum c.m.\ energy point
that BESIII can reach is 4.6~GeV.
In a data sample of 525~pb$^{-1}$ collected during one month from December
14, 2012 to January 14, 2013 at the c.m.\ energy of $4.26$~GeV, the charged
charmonium-like state $\zc$ was discovered~\cite{Ablikim:2013mio}.
This motivated BESIII to focus on c.m.\ energies above 4~GeV
resulting in the largest data sample above 4~GeV for $\xyz$ studies.

The data samples for the $\xyz$ study (``$\xyz$ data'' hereafter)
are presented in Table~\ref{ecm_lum_xyz}, which lists the nominal
c.m.\ energy, measured c.m. energy (when it is available), and
integrated luminosity at each energy point. These data were used
for all the analyses presented in this report.
BESIII also did a fine scan between 3.8 and 4.6~GeV at more than 100 energy
points (``$R$-scan data'' hereafter) in 2014, with a total integrated
luminosity of about 800~pb$^{-1}$. This data sample is also used for $\xyz$
studies although the statistics are low~\cite{Ablikim:2017wlt}.

\begin{table}[htbp]
 \centering
\begin{tabular}{ccr} \hline\hline
Data sample & c.m.\ energy~(MeV)  & ${\cal L}$ ($\rm pb^{-1}$) \\
\hline
3810       &   3807.65$\pm$0.10$\pm$0.58            &50.54$\pm$0.03     \\
3900       &   3896.24$\pm$0.11$\pm$0.72            &52.61$\pm$0.03     \\
4009       &   4007.62$\pm$0.05$\pm$0.66            &481.96$\pm$0.01    \\
4090       &   4085.45$\pm$0.14$\pm$0.66            &52.63$\pm$0.03     \\
4130       &         --                             &  $\sim 400$       \\
4160       &         --                             &  $\sim 400$       \\
4180       &             4178                       &  $\sim 3190$      \\
4190       &   4188.59$\pm$0.15$\pm$0.68            &43.09$\pm$0.03     \\
4190       &         --      &   $\sim 500$ \\
4200       &         --      &   $\sim 500$ \\
4210       &   4207.73$\pm$0.14$\pm$0.61            &54.55$\pm$0.03     \\
4210       &         --      &   $\sim 500$ \\
4220       &   4217.13$\pm$0.14$\pm$0.67            &54.13$\pm$0.03     \\
4220       &         --      &   $\sim 500$ \\
4230       &   4226.26$\pm$0.04$\pm$0.65            &1091.74$\pm$0.15  \\
4237       &         --      &   $\sim 500$ \\
4245       &   4241.66$\pm$0.12$\pm$0.73            &55.59$\pm$0.04    \\
4246       &         --      &   $\sim 500$ \\
4260       &   4257.97$\pm$0.04$\pm$0.66            &825.67$\pm$0.13   \\
4270       &         --      &   $\sim 500$ \\
4280       &         --      &   $\sim 200$ \\
4290       &         --                             &  $\sim 500$       \\
4310       &   4307.89$\pm$0.17$\pm$0.63            &44.90$\pm$0.03   \\
4315       &         --                             &  $\sim 500$     \\
4340       &         --                             &  $\sim 500$       \\
4360       &   4358.26$\pm$0.05$\pm$0.62            &539.84$\pm$0.10  \\
4380       &         --                             &  $\sim 500$       \\
4390       &   4387.40$\pm$0.17$\pm$0.65            &55.18$\pm$0.04   \\
4400       &         --                             &  $\sim 500$       \\
4420       &   4415.58$\pm$0.04$\pm$0.72            &1073.56$\pm$0.14  \\
4440       &         --                             &  $\sim 570$       \\
4470       &   4467.06$\pm$0.11$\pm$0.73            &109.94$\pm$0.04   \\
4530       &   4527.14$\pm$0.11$\pm$0.72            &109.98$\pm$0.04   \\
4575       &   4574.50$\pm$0.18$\pm$0.70            &47.67$\pm$0.03    \\
4600       &   4599.53$\pm$0.07$\pm$0.74            &566.93$\pm$0.11   \\
 \hline\hline
\end{tabular} \label{ecm_lum_xyz}
\caption{The measured c.m.\ energy~\cite{Ablikim:2017wlt} and integrated
luminosity ${\cal L}$~\cite{Ablikim:2015nan} of each data sample collected for the
BESIII study of $\xyz$ states.
The uncertainties on the c.m.\ energies are statistical and systematic.
The uncertainties on the integrated
luminosities are statistical only; a 1\% systematic uncertainty
common to all the data points is not listed. ``--" means not
available yet and numbers without error are rough estimates.}
\end{table}

Compared with the $B$ factories, BaBar and Belle, BESIII has some
advantages in the study of the $\xyz$ states, especially those $Y$ states
with vector quantum numbers. BESIII collects $\EE$ annihilation
data at c.m.\ energies that go directly into the production of $Y$ states,
while the $B$ factories use data produced via ISR, so BESIII has a much
higher detection efficiency and can take more data at any energy of
interest [for example, the efficiency is 46\% at
BESIII~\cite{Ablikim:2013mio} and about 10\% at
Belle~\cite{Yuan:2007sj} for selecting $Y(4260)\to \pp\jpsi\to \pp\LL$
($\ell=e$, or $\mu$) events]. This makes the study of the $Z_c$
states from the $Y$ decays also more efficient at BESIII than at
the $B$ factories. However, $B$ factories can measure the cross
sections in a wide energy range since all the events are produced
at the same time, while BESIII needs to tune the c.m.\ energy point
by point to collect data, thus can only cover a limited energy range.
Finally, the $B$ factories can study the $\xyz$ states
with $B$ decays, two-photon fusion, double-charmonium
production, and $\Upsilon(nS)$ ($n=1-6$) decays, while
BESIII is limited to $\EE$ annihilation.

\subsection{Experiments at proton-antiproton colliders}
\label{Sect.2.2}

\begin{figure}[ht]
\begin{center}
\includegraphics[width=0.4\textwidth]{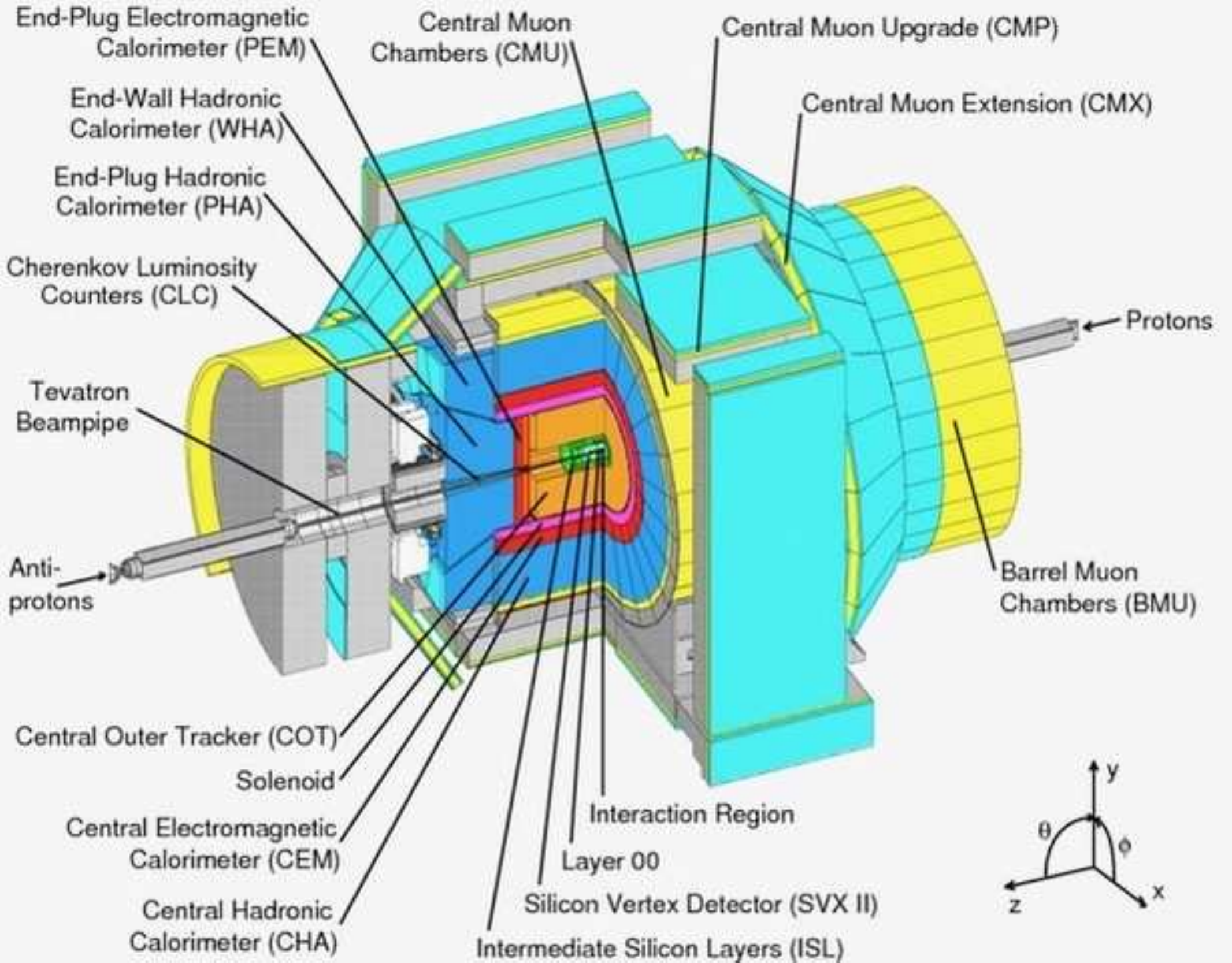}
\hspace*{9mm}
\includegraphics[width=0.4\textwidth]{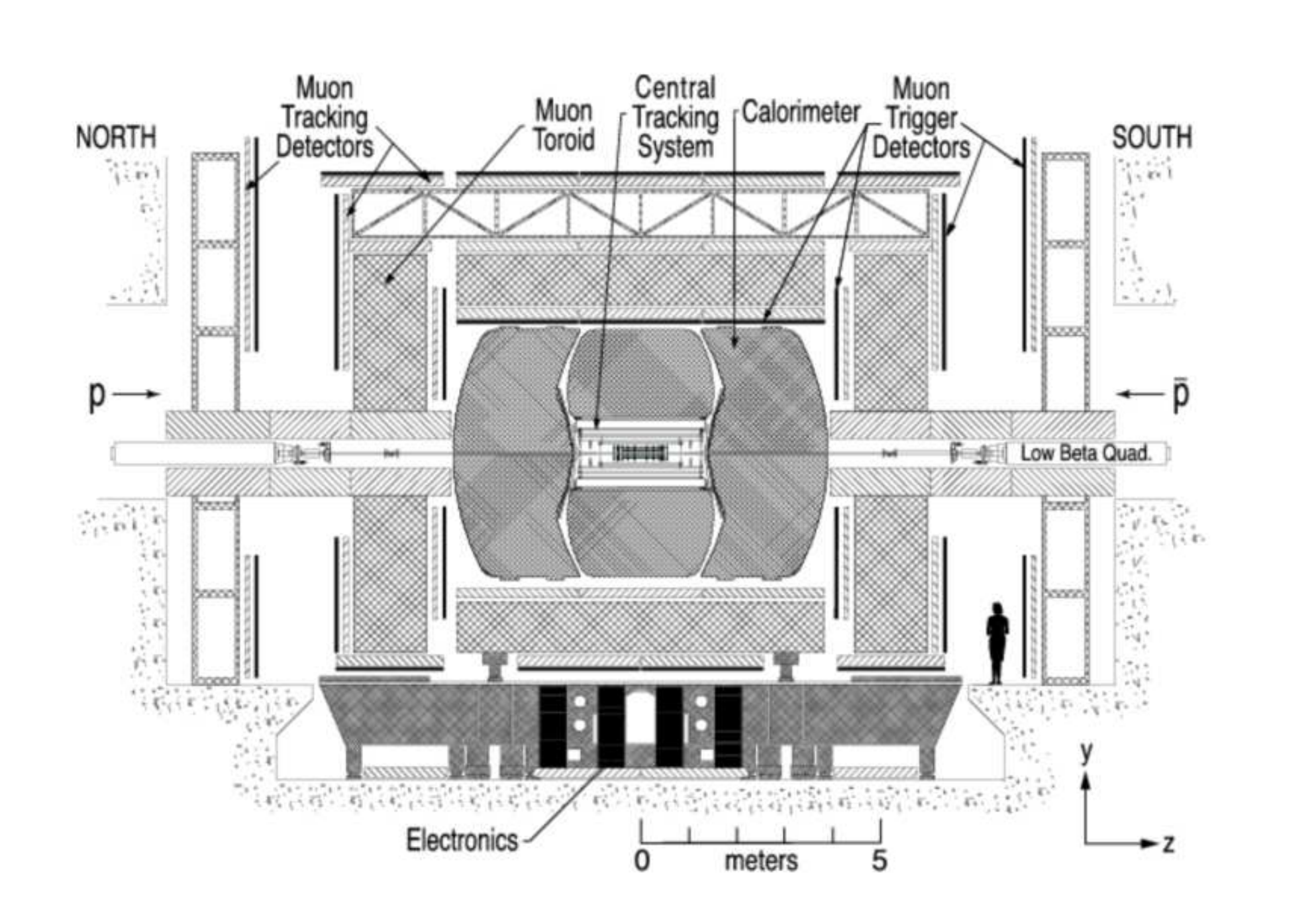}
\caption{General layout of the CDF (left) and D0 (right) detectors from Refs.~\cite{Abe:1988me,Abachi:1993em}.}
\label{fig:tev}
\end{center}
\end{figure}

The Tevatron was a synchrotron at Fermilab that accelerated protons and
antiprotons in a 6.28 km ring to energies of up to 1 TeV. First
collisions were in 1986 and it ran until 2011. The maximum luminosity
achieved was $4 \times 10^{32}$ cm$^{-2}$ s$^{-1}$. Two large general-purpose
detectors, CDF~\cite{Abe:1988me} and D0~\cite{Abachi:1993em}, shown in Fig.~\ref{fig:tev},
were taking data.
The CDF detector consists of\ch{:} a silicon detector used to track the paths
of charged particles
and composed of seven layers of silicon arranged in a barrel shape around
the beam pipe; the central outer tracker used to track the paths of charged
particles and located within a 1.5 T solenoidal magnetic field; the combined
electromagnetic and hadronic calorimeter that has approximately uniform
granularity in rapidity-azimuthal angle and extends down to $2^{\circ}$ from
the beam direction to measure the energy of light particles and hadrons;
the muon detector with
four layers of planar drift chambers, each with the capability of detecting
muons with a transverse
momentum $p_T>1.4$ GeV. The D0 detector consists of three major subsystems:
central tracking detectors,
uranium/liquid-argon calorimeters, and a muon spectrometer.
The central tracking system includes a silicon microstrip tracker and a
scintillating-fiber tracker
located within a 2 T solenoidal magnet to identify displaced vertices for
$b$-quark tagging.
The magnetic field enables measurement of the energy-to-momentum ratio for
electron identification and calorimeter calibration. The calorimeters were
designed to provide energy measurements for
electrons, photons, and jets in the absence of a central magnetic field,
as well as assist in identification of electrons, photons, jets, and muons and
measure the transverse energy balance in events. The outermost layer of the
D0 detector
is the muon spectrometer consisting of a central muon system proportional
drift tubes and toroidal magnets,
central scintillation counters, and a forward muon system.

In 1995, the CDF and D0 Collaborations announced the discovery of the
top quark~\cite{Abe:1995hr,D0:1995jca} and then systematically studied its
characteristics. By now the combination of their measurements of the top
quark mass gives $(172.9 \pm 0.4)$ GeV, a precision of about
0.23\%~\cite{TevatronElectroweakWorkingGroup:2016lid}.
In 2006, the CDF Collaboration reported the first measurement
of $B_s$ oscillations~\cite{Acosta:2004gt} and both detectors observed for the
first time various heavy baryons. In addition to that, CDF and D0 have also
contributed to studies of exotic states using data samples of 9.0~fb$^{-1}$
and 10.4 ~fb$^{-1}$ of proton-antiproton collisions  taken
at $\sqrt{s}=$ 1.96 TeV. For example, after the discovery of the
$\chi_{c1}(3872)$ aka $X(3872)$ both groups worked on determining
its characteristics and the most precise measurement of its mass still belongs to CDF~\cite{Aaltonen:2009vj}.
In 2009 CDF was the first to report an observation of the $\chi_{c1}(4140)$ aka
$X(4140)$~\cite{Aaltonen:2009tz} while in 2016 D0 announced an observation
of the $X(5568)$~\cite{D0:2016mwd}.

\subsection{Experiments at proton-proton colliders}
\label{Sect:2.3}
The Large Hadron Collider (LHC) is the world largest and most powerful
particle accelerator. It started operation on September 10, 2008. The LHC
consists of a 27-kilometer ring of superconducting magnets with a number
of accelerating structures to boost the energy of the particles up to 6.5 TeV.
The maximum luminosity achieved so far is $2.06 \times 10^{34}$ cm$^{-2}$ s$^{-1}$.
Four large detectors, ALICE, ATLAS, CMS, and LHCb, are taking data.

\subsubsection{The LHCb experiment}
\begin{figure}[ht]
\begin{center}
\includegraphics[width=0.5\textwidth]{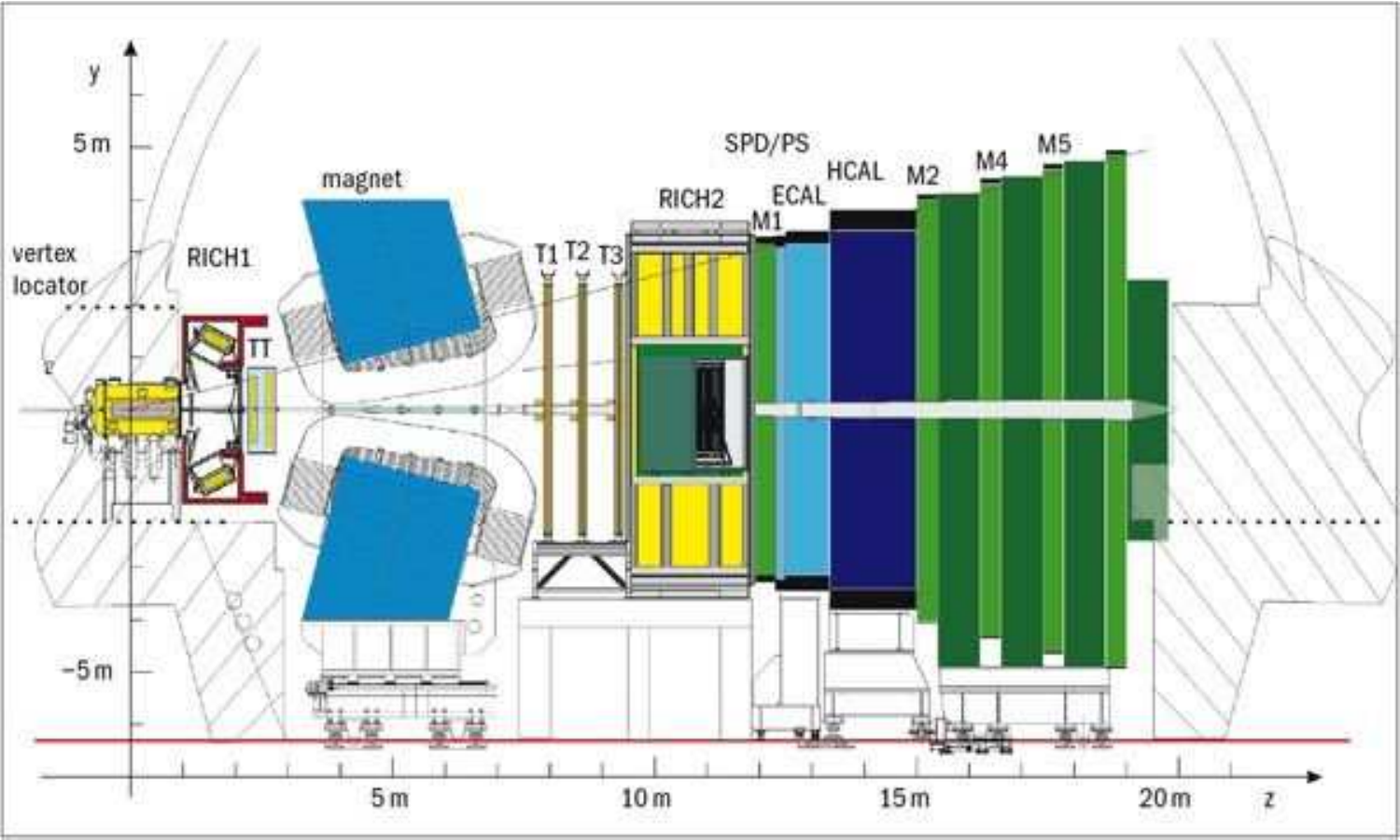}
\caption{General layout of the LHCb detector from Ref.~\cite{Aaij:2014jba}.}
\label{fig:lhc2}
\end{center}
\end{figure}

The LHCb detector shown in Fig.~\ref{fig:lhc2} investigates properties
of the charm and bottom quarks~\cite{Alves:2008zz}. It has contributed a lot to
various studies of mesons and baryons with open charm and bottom, and has
also succeeded in investigating exotics,
e.g. discovering pentaquarks~\cite{Aaij:2015tga}, determining the quantum
numbers of the $X(3872)$~\cite{Aaij:2013zoa} and
$Z(4430)$~\cite{Aaij:2014jqa}, and disentangling
the complicated structure of the $J/\psi\phi$ system around
4140 MeV~\cite{Aaij:2016iza}.

The LHCb detector is a single-arm spectrometer with a forward angular
coverage from approximately 15~mrad to 300 (250)~mrad in the
bending (non-bending) plane~\cite{Aaij:2014jba}.
The spectrometer magnet is a warm dipole magnet providing an
integrated field of about 4~Tm, which deflects charged particles
in the horizontal plane. The tracking system consists of the
vertex locator, and four planar tracking stations: the
Tracker Turicensis (TT) upstream of the dipole magnet, and
tracking stations T1--T3 downstream of the magnet. Charged
particles require a minimum momentum of $1.5$~GeV to reach the
tracking stations, T1--T3. Charged hadron identification in the
momentum range from 2 to 100~GeV is achieved by two RICHs (RICH1 and RICH2).
The calorimeter
system is composed of a scintillating pad detector, a
preshower, a shashlik type EMC
and a hadronic calorimeter. The muon detection system
provides muon identification and contributes to the L0 trigger of
the experiment. The minimum momentum that a muon must have to
traverse the five stations is approximately $6$~GeV.

The LHCb detector started its data taking in 2008.
The integrated luminosity recorded is
1.11~fb$^{-1}$ at 7~TeV in 2011, 2.08~fb$^{-1}$ at 8~TeV in 2012,
and 5.9~fb$^{-1}$ at 13~TeV from 2015 up to now.

Due to the large production cross sections, more $B$ mesons and
other particles are produced at LHCb than at other experiments
like BaBar and Belle. However, to avoid background from direct
$pp$ collision,
many of the studies looked for $\xyz$ states in the decays of
a mother particle which has a long decay length.
So far, most of the
results related to the $\xyz$ particles are from $B$ decays. The
large $B$ sample makes the determination of the quantum numbers
and the decay dynamics of the $\xyz$ states possible via partial
wave analysis (PWA) of the $B$ decays.

\subsubsection{The ATLAS and CMS experiments}
\begin{figure}[ht]
\begin{center}
\includegraphics[width=0.4\textwidth]{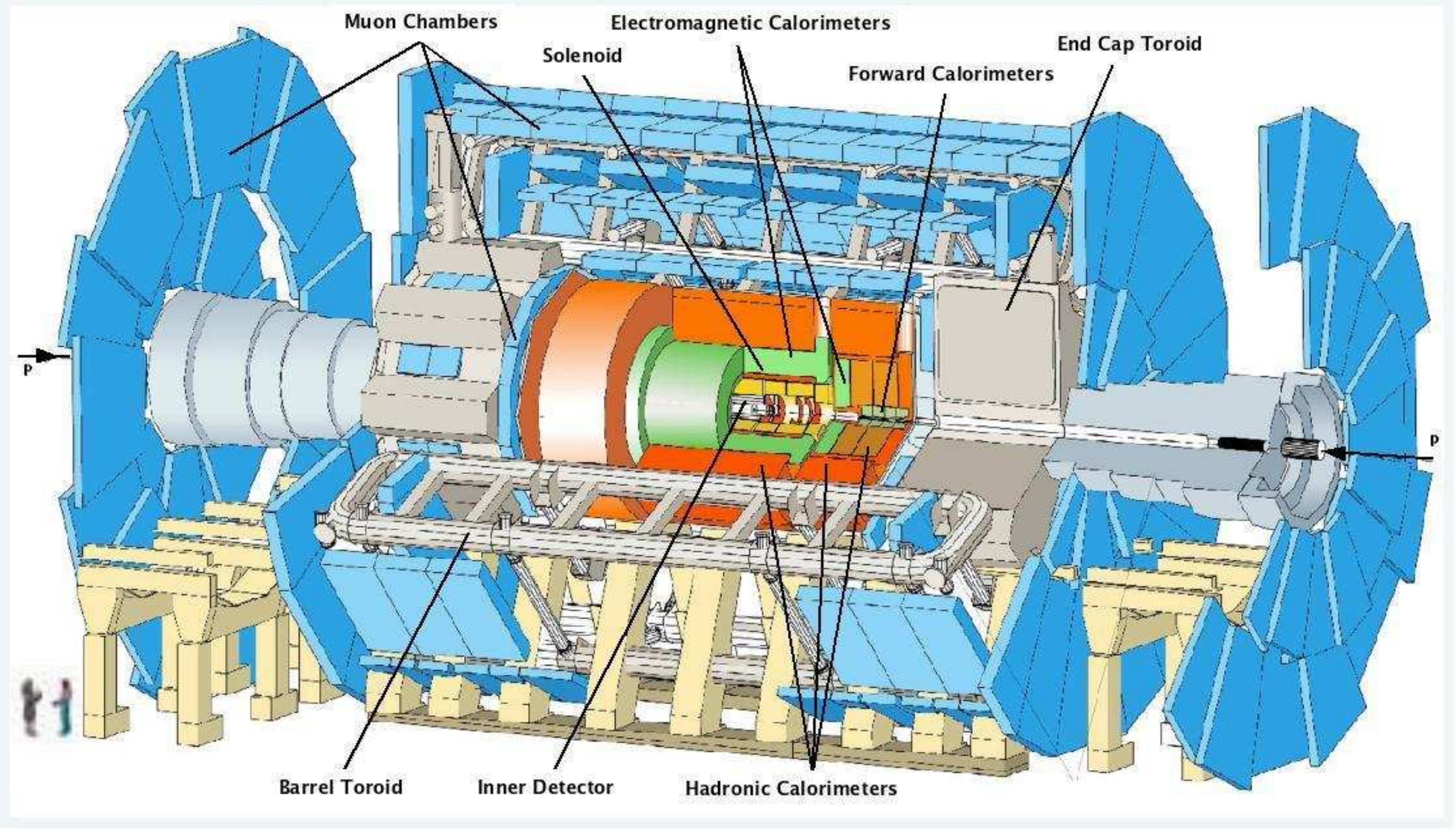}
\hspace*{9mm}
\includegraphics[width=0.4\textwidth]{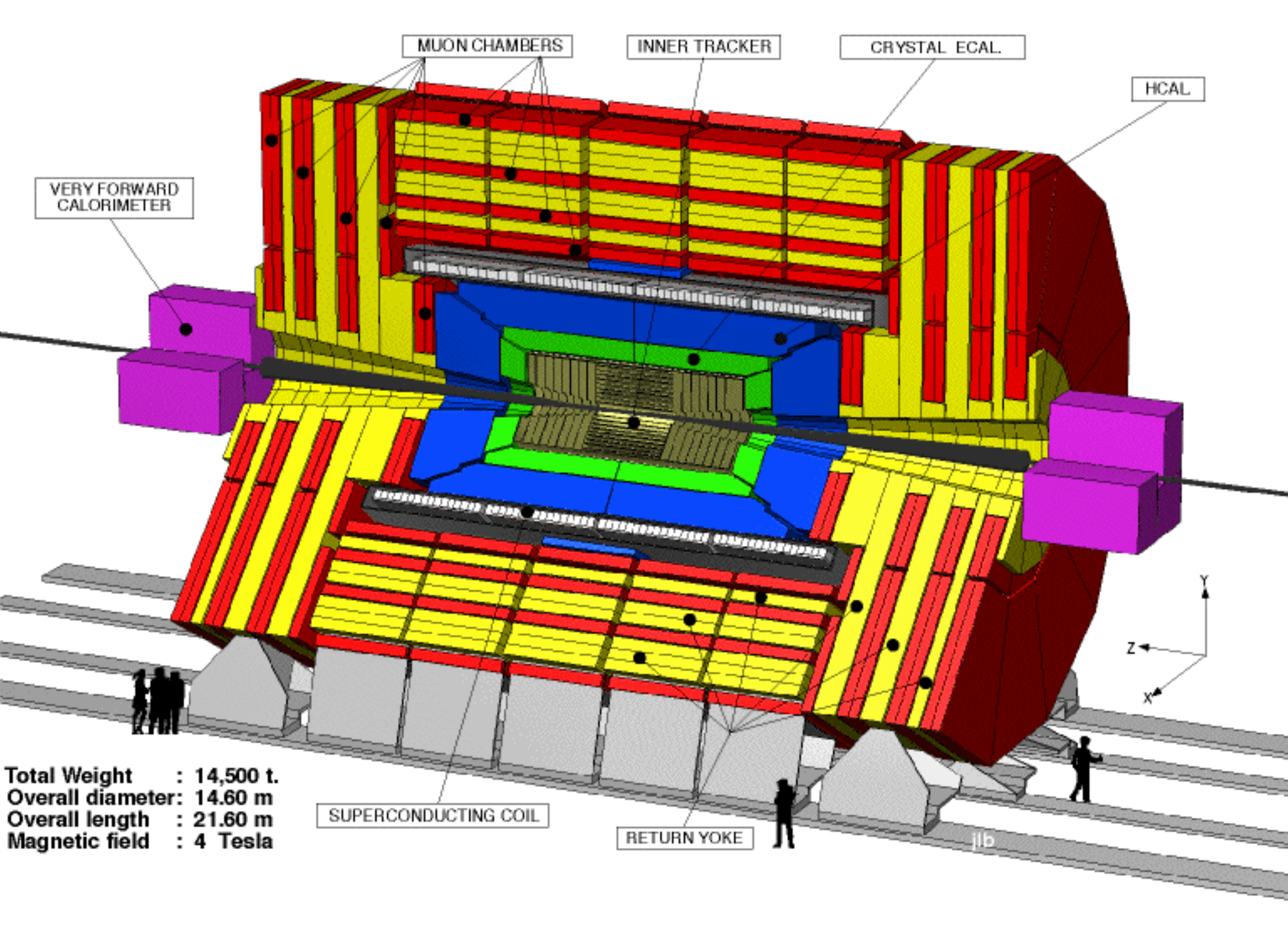}
\caption{General layout of the ATLAS (left) and CMS (right) detectors from Refs.~\cite{CMS:2012nga,ATLAS:2012oga}.}
\label{fig:lhc1}
\end{center}
\end{figure}

In 2012, the ATLAS and CMS Collaborations announced the discovery of the
Higgs boson -- a fundamental particle with a mass around 125 GeV predicted
by the Standard Model as manifestation of the mechanism generating the masses
of the elementary particles~\cite{CMS:2012nga,ATLAS:2012oga}.
These detectors\ch{,} shown in Fig.~\ref{fig:lhc1}\ch{,} extensively study its properties
and simultaneously search for various possible manifestations of new
physics beyond the Standard Model. They are also involved in looking for
new heavy states.

The CMS detector is built around a huge solenoid magnet. It is a cylindrical
coil of superconducting cables
that generates a field of 3.8 T. The complete detector is 21 meters long,
15 meters wide, and 15 meters high.
The subdetectors which constitute the CMS detector from inner to outer are
inner tracking system to measure the trajectories of charged particles and
reconstruct secondary vertices;
EMC to measure the energy of electrons and photons; hadronic calorimeter
to measure the energy of hadrons;
superconducting magnet providing a 3.8 T magnetic field parallel to the beam
axis to bend the tracks of charged particles; muon system to identify and
measure the trajectories of muons.
The ATLAS detector has cylindrical geometry with its central axis coinciding
with the beamlines.
It is 44 meters in length and 25 meters along the diameter of the transverse
round plane, and
provides almost full $4\pi$ solid angle coverage around the collision point.
The subdetectors which constitute the ATLAS detector from inner to outer are
inner tracking detector which consists of a silicon pixel detector,
silicon microstrip tracker and transition radiation tracker to provide good
charged-particle tracking reconstruction as well as vertex reconstruction;
EMC surrounded by the hadronic calorimeter;
solenoid superconducting magnetic system providing a 2 T axial magnetic field;
muon spectrometer with a whole coverage up to $|\eta|< 2.4$.

The integrated luminosity recorded by the CMS/ATLAS detectors is
45.0/45.0~pb$^{-1}$ at 7~TeV in 2010, 6.1/5.1~fb$^{-1}$ at 7~TeV in 2011,
23.3/21.3~fb$^{-1}$ at 8~TeV in 2012, and 162.9/147.0~fb$^{-1}$ at 13~TeV
from 2015 up to now.

\section{Exotic hadron candidates}
\label{Sect:3}
In this section, experimental evidence for the
candidates for exotic states is collected.
While some comments are already given about why those states might most probably not be regular quark-model
states, a detailed discussion of their nature is postponed to the subsequent sections --- for all states
the relevant theory sections are mentioned below.

\subsection{Isoscalar states}
\label{Sect:3.1}
Most of the states discovered recently in the quarkonium mass range
are isoscalar states and we will begin the discussion with those.
Isoscalar states could in principle be regular quarkonia ---
the arguments for why most authors still regard them as exotics are presented
in this article. The isovector states that call for going beyond
the most naive realisation of the quark model
are presented in section~\ref{Sect:3.2}.

\subsubsection{Neutral $X$ states}
\label{Sect:3.1.1}

\vspace{0.3cm}\noindent
$\bullet$ {\it The $\psi_2(3823)$ }
\vspace{0.3cm}

\begin{figure}
\centering
\includegraphics[height=5cm]{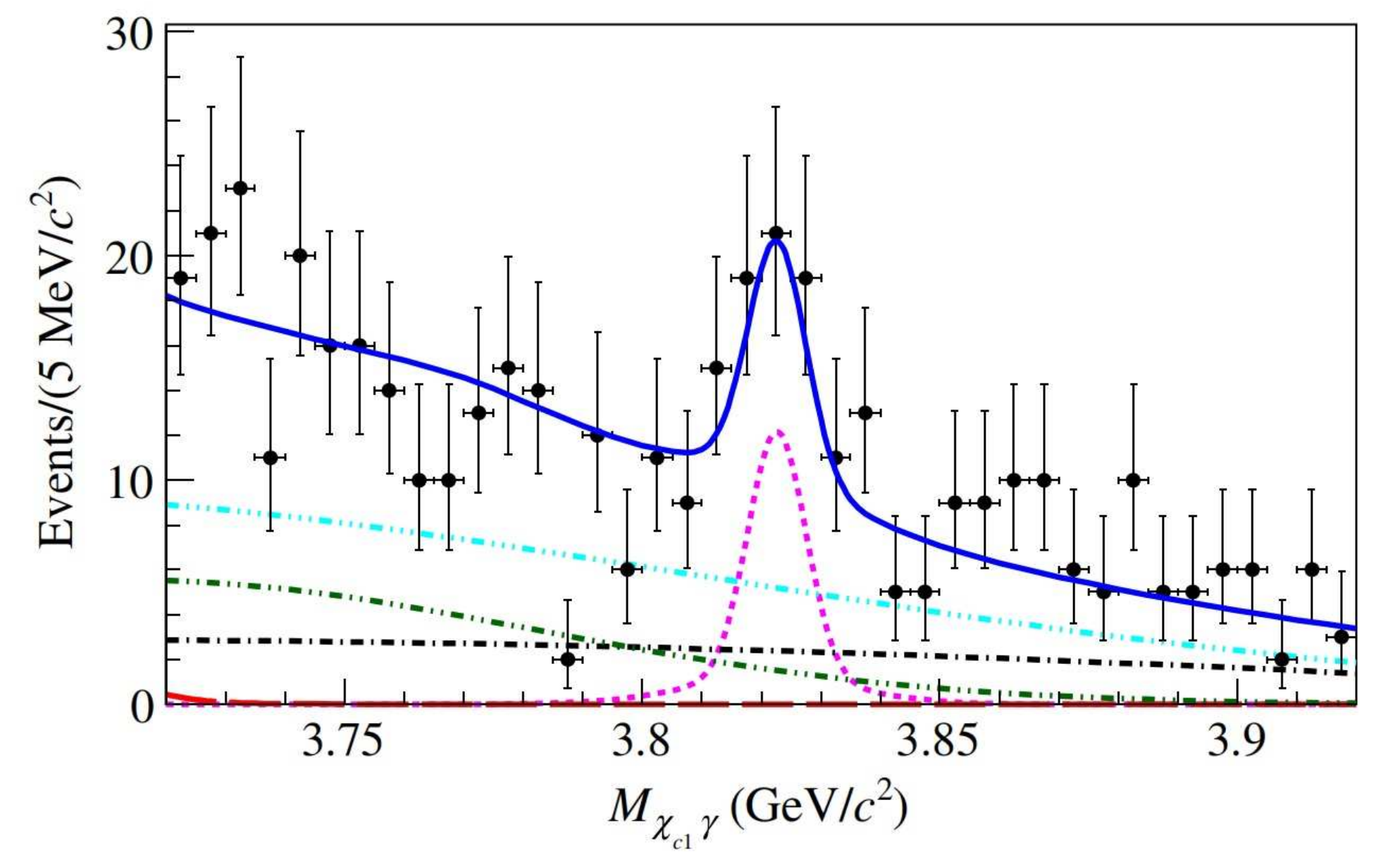}
\caption{The simultaneous fit projection on the $\gamma \chi_{c1}$
mass spectrum to the selected $B^{\pm}\to K^{\pm} \gamma
\chi_{c1}$ and $B^{0}\to K_S^0 \gamma \chi_{c1}$ signal
candidates~\cite{Bhardwaj:2013rmw}, where the solid blue line is
the best fit. The curves show the signal [large-dashed red line
for $\psi(2S)$, short-dashed magenta line for $\psi_2(3823)$, and dotted violet line for $\chi_{c1}(3872)$ aka
$X(3872)$] and the background components [dot-dashed black line for
combinatorial, two-dot-dashed dark green line for $B\to K
\psi(2S)(\nrightarrow \gamma \chi_{c1})$, and cyan
three-dot-dashed for peaking component].} \label{belle-x3823}
\end{figure}

In the charmonium spectrum, the $\psi_2(1^3D_2)$
is expected to dominantly decay into $\gamma \chi_{c1}$,
while the $\psi_3(1^3D_3)$ is expected to decay into $\gamma \chi_{c2}$
with a large branching fraction besides its dominant open charm decay
into $D\bar{D}$. To search for them, Belle measured the reactions
$B\to K \gamma \chi_{c1}\to K \gamma \gamma J/\psi$
and $B\to K \gamma \chi_{c2} \to K \gamma \gamma J/\psi$ using a data sample of
$772\times 10^6$ $B\bar{B}$ events~\cite{Bhardwaj:2013rmw}.
In this analysis, the most important technique is a correction of the
photon energy from $B$ decays by scaling the energy of the photon
so that the $\Delta E$ (the energy difference between
the beam energy and the reconstructed $B$ candidates) is equal to zero.
This improves the $\gamma \chi_{cJ}$ mass resolution a lot.
In the $\gamma \chi_{c1}$ mass spectrum for the selected
$B^{\pm}\to K^{\pm} \gamma \chi_{c1}$ signal candidates,
besides the clear $\psi(2S)$ signal, there is a significant narrow peak
at 3823~MeV, denoted as $\psi_2(3823)$ .
No signal of $X(3872)\to \gamma \chi_{c1}$ is seen.
After using a Breit-Wigner (BW) function convolved with a mass resolution to parameterize
the $\psi_2(3823)$ signal shape, the mass obtained from a
simultaneous fit to the selected $B^{\pm}\to K^{\pm} \gamma \chi_{c1}$ and
$B^{0}\to K_S^0 \gamma \chi_{c1}$ signal candidates is
$(3823.1\pm 1.8\pm 0.7)$~MeV with a significance of
4.0$\sigma$ with systematic uncertainties included, see Fig.~\ref{belle-x3823}.
The upper limit at 90\% confidence level (C.L.) on its width is estimated to be 24~MeV.
The $\psi_2(3823)$ mass agrees well with the potential model expectations for
$\psi_2(1^3D_2)$. Besides the mass, the measured product of branching
fractions $\BR[B\to K \psi_2(3823)]\BR[\psi_2(3823)\to \gamma\chi_{c1}]$
is approximately two orders of magnitude lower than for the $\psi(2S)$,
which also supports the interpretation of the $\psi_2(3823)$ as the
$\psi_2(1^3D_2)$ state~\cite{Suzuki:2002sq,Colangelo:2002mj}.
In the $\gamma \chi_{c2}$ mass spectrum,
no evidence is found for $\psi_2(3823) \to \gamma \chi_{c2}$.

\begin{figure}
\begin{center}
\includegraphics[height=5cm]{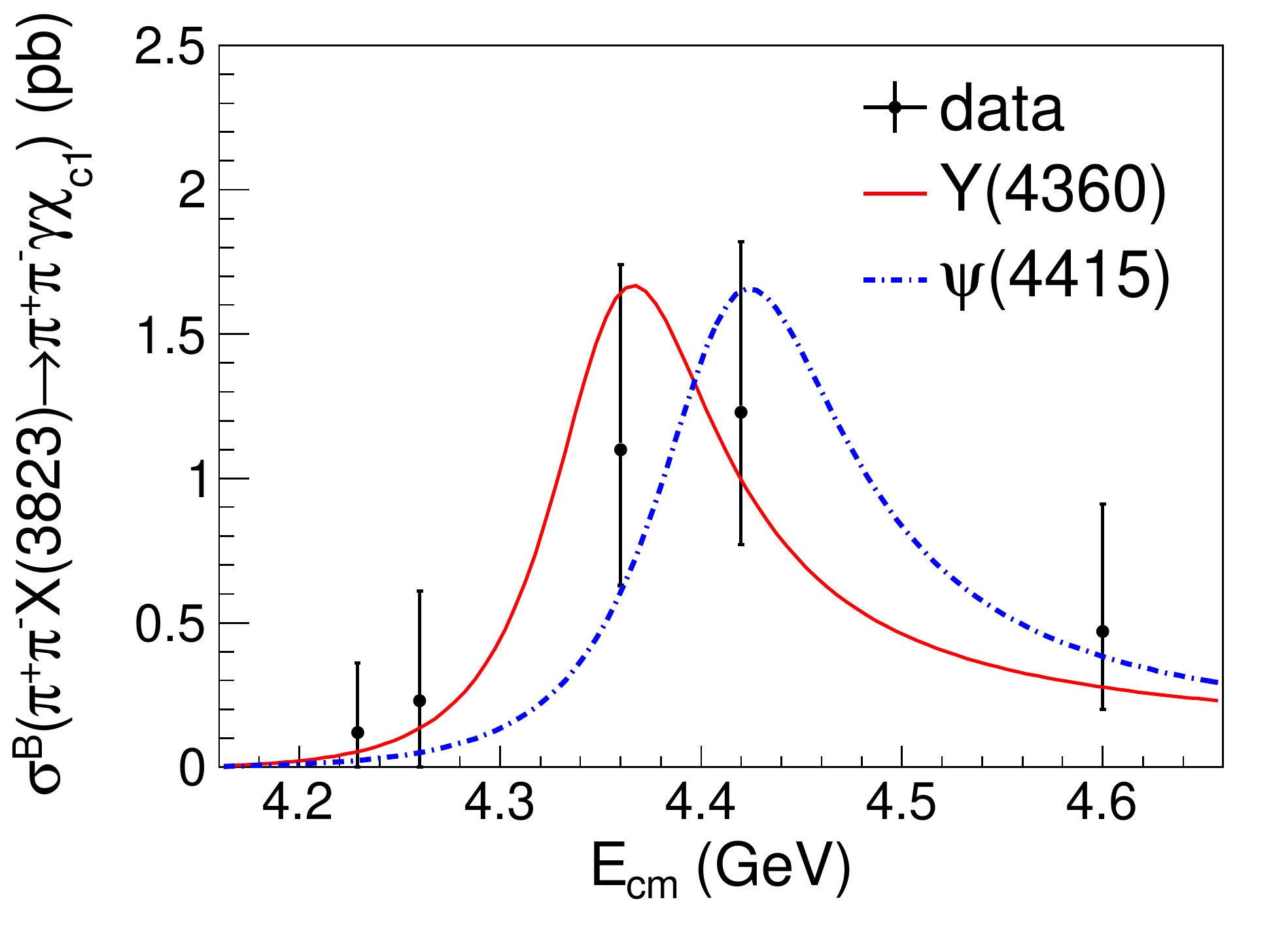}
\caption{The fits to the products of $\sigma[\EE\to\pp \psi_2(3823)]\BR[\psi_2(3823)\to\gamma\chi_{c1}]$
from the BESIII measurement~\cite{Ablikim:2015dlj} using a $\psi(4360)$ aka $Y(4360)$ or
$\psi(4415)$ line shape.
Dots with error bars are data, and the red solid/blue dot-dashed curve shows
the best fit with a $Y(4360)$/$\psi(4415)$ line shape.} \label{bes3-x3823-sec}
\end{center}
\end{figure}

Later BESIII searched for the $\psi_2(3823)$ in the
$\EE\to \pi^+\pi^-\psi_2(3823) \to \pi^+\pi^-\gamma\chi_{c1}$ process using
data samples at c.m.\ energies of 4.23, 4.26, 4.36, 4.42, and 4.60~GeV
corresponding to a total luminosity of 4.67 fb$^{-1}$~\cite{Ablikim:2015dlj}.
In the $\pp$ recoil mass distribution for events in the $\chi_{c1}$
signal region, the $\psi_2(3823)$ signals are observed clearly
with a statistical significance of $6.2\sigma$, while no evidence for
the $\psi_2(3823)$ signal can be seen for events in the
$\chi_{c2}$ signal region. The fit yields $19\pm 5$ $\psi_2(3823)$
 signal events in the $\gamma\chi_{c1}$ mode, with a measured mass of
$(3821.7\pm 1.3\pm 0.7)~{\rm MeV}$. The upper limit on its width is estimated
to be 16 MeV at 90\% C.L.
The measured parameters of $\psi_2(3823)$ are consistent
with those from the Belle measurement~\cite{Bhardwaj:2013rmw}.
The products of $\sigma[\EE\to\pp \psi_2(3823)] \BR[\psi_2(3823)\to \gamma\chi_{c1}]$
are measured at the above mentioned c.m.\ energies,
as shown in Fig.~\ref{bes3-x3823-sec} with dots with error bars.
Fitting this distribution using a $Y(4360)$ shape or
$\psi(4415)$ shape with their resonant parameters fixed to the PDG
values~\cite{Tanabashi:2018oca} can describe the data well in both cases.
The fit results are shown in Fig.~\ref{bes3-x3823-sec} with the solid and
dot-dashed lines. Larger data samples with more energy points are needed
to separate these two fits or more possibilities.

According to the measurements from BESIII and Belle,
the narrow $\psi_2(3823)$ resonance is a good candidate for the
$\psi_2(1\,^3D_2)$ charmonium state. According to potential
models~\cite{Godfrey:1985xj, Ebert:2002pp, Eichten:2004uh, Blank:2011ha, Kwong:1987mj},
the mass of $\psi_2(1\,^3D_2)$ is in the
$3.810\sim 3.840$~GeV range, which is above the $D\bar{D}$ mass threshold but
below the $D\bar{D}^*$ and $D\bar{D}\pi$ thresholds. Since the decay $\psi_2(1\,^3D_2)\to
D\bar{D}$ needs to proceed in a $D$--wave which would violate parity,
the $\psi_2(1\,^3D_2)$ state is expected to be
narrow, and $\psi_2(1\,^3D_2)\to \gamma\chi_{c1}$ should be the dominant decay
mode~\cite{Godfrey:1985xj, Ebert:2002pp, Eichten:2004uh, Blank:2011ha, Kwong:1987mj,Qiao:1996ve}.
All agree well with the experimental observations. From the cross section measurement,
BESIII obtained the ratio $$\frac{\BR[\psi_2(3823)\to \gamma\chi_{c2}]}
{\BR[\psi_2(3823)\to \gamma\chi_{c1}]}<0.42$$ at the 90\% C.L., which
also agrees with expectations for the $\psi_2(1\,^3D_2)$ state~\cite{Qiao:1996ve}.

It should be noted that the spin-parity of this state were not determined,
although the BESIII measurement of the angular distribution supports a
$J^P=2^-$ assignment. With much larger data samples at BESIII and Belle II in the
near future, the determination of the spin-parity and precise measurements of the
width and decay ratio between $\gamma \chi_{c1}$ and $\gamma \chi_{c2}$ are needed
to confirm the $\psi_2(3823)$ as the conventional $\psi_2(1\,^3D_2)$ charmonium state.

Very recently LHCb observed a new narrow charmonium state, the $X(3842)$ resonance,
in the decay modes $X(3842) \to D^0 \bar{D}^0$ and $X(3842) \to D^+D^-$
with its measured mass and width consistent with the unobserved spin-3 $\psi_3(1^3D_3)$
charmonium state using an integrated luminosity of 9 fb$^{-1}$ proton-proton collision data~\cite{Aaij:2019evc}.
This state can be checked in other experiments in the near future.

\vspace{0.3cm}\noindent
$\bullet$ {\it The $\chi_{c1}(3872)$ aka $X(3872)$ and its counterpart $X_b$}
\vspace{0.3cm}

The $X(3872)$ was first observed in $B^\pm\to K^\pm \ppjpsi$ decays
by Belle in 2003~\cite{Choi:2003ue}, and subsequently was confirmed
by several other experiments~\cite{Acosta:2003zx,Abazov:2004kp,Aubert:2004ns}.
The spin-parity quantum numbers $J^P$ of the state were restricted to
two possibilities, $1^{+}$ or $2^{-}$, by the CDF experiment
via an analysis of the angular correlations in the
$\pi^+\pi^-\jpsi \to \pi^+ \pi^-\MM$ final state~\cite{Abulencia:2006ma}. Using
1.0~fb$^{-1}$ of $pp$ collision data, LHCb rules out $J^{P}=2^{-}$ by
analyzing the angular correlations in the same decay chain,
with the $X(3872)$ state produced in $B^+\to K^+ X(3872)$ decays~\cite{Aaij:2013zoa}.
Twelve years after the $X(3872)$ was discovered,
its spin-parity quantum numbers
were finally determined in the LHCb experiment by performing
a five-dimensional angular correlation analysis in
$B^+\to K^+ X(3872)$, $X(3872)\to
\rho^0\jpsi$, $\rho^0\to \pi^+\pi^-$, $\jpsi\to \MM$ decays with
$1011\pm 38$ events selected from 3~fb$^{-1}$ data at $7$ and
$8$~TeV~\cite{Aaij:2015eva}. This analysis determined the $J^{PC}$
values to be $1^{++}$ for the $X(3872)$ which
thus could be a candidate for the not yet identified quark model state $\chi_{c1}(2^3P_1)$.

So far the $X(3872)$ is one of the best studied exotic
meson candidates with a $c\bar{c}$
content. Some properties including the enhancement of isospin-violating
$\rho^0 J/\psi$ decays, the mass right at the $D^0 \bar{D}^{*0}$ threshold,
and the spin-parity quantum numbers $J^{PC}$ are well established. Due to
its mass at the $D^0 \bar{D}^{*0}$ threshold
and large decay rate to $D^0 \bar{D}^{*0}$, a natural explanation of this
state is a loosely bound molecule.
However, its large prompt production rates at the
Tevatron~\cite{Abazov:2004kp} and the LHC~\cite{Aaij:2011sn,Chatrchyan:2013cld,Aaboud:2016vzw}
might point at a compact component. Its preferred decay mode of
$\gamma \psi(2S)$ over $\gamma J/\psi$ matches
the expectation for the $\chi_{c1}(2P)$ state. The various possible
interpretations of the $X(3872)$
are discussed in detail in various subsections of Sec.~\ref{sect:4}.

In order to better understand the nature of $X(3872)$ experimentally,
one might either look for more production mechanisms, or find more decay modes.

\vspace{0.3cm}\noindent
{\it (1) Production mechanisms of $\chi_{c1}(3872)$ aka $X(3872)$}
\vspace{0.3cm}

Since the quantum numbers of $X(3872)$ are $J^{PC}=1^{++}$, it can be produced
through the radiative transition of excited vector charmonium/bottomonium or
charmonium-like/bottomonium-like states. Belle reported the searches for
the $X(3872)$ as well as the $X(3915)$ and $X(4140)$
using $102\times10^6$ $\Upsilon(1S)$ and $158\times10^6$
$\Upsilon(2S)$ events~\cite{Shen:2010iu,Wang:2011qm}.
Belle also searched for the $X(3872)$ in
$\Upsilon(1S)$ inclusive decays~\cite{Shen:2016yzg}.
In all these attempts, no apparent $X(3872)$ signal was
observed and 90\% C.L.\ upper limits
of the production rates in $\Upsilon(1S, 2S)$ decays were set.

\begin{figure}
\begin{center}
\includegraphics[height=4cm]{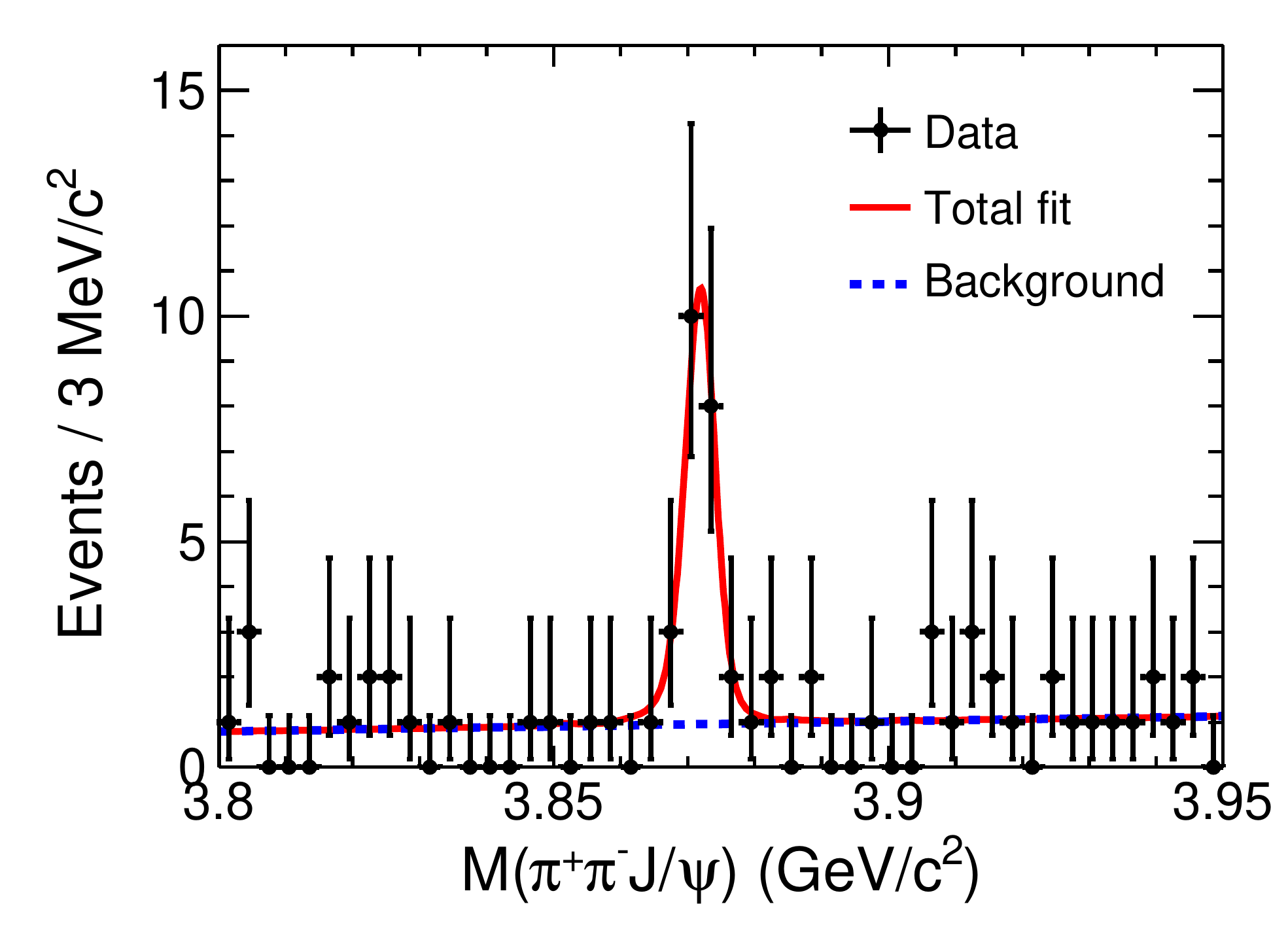}
\includegraphics[height=4cm]{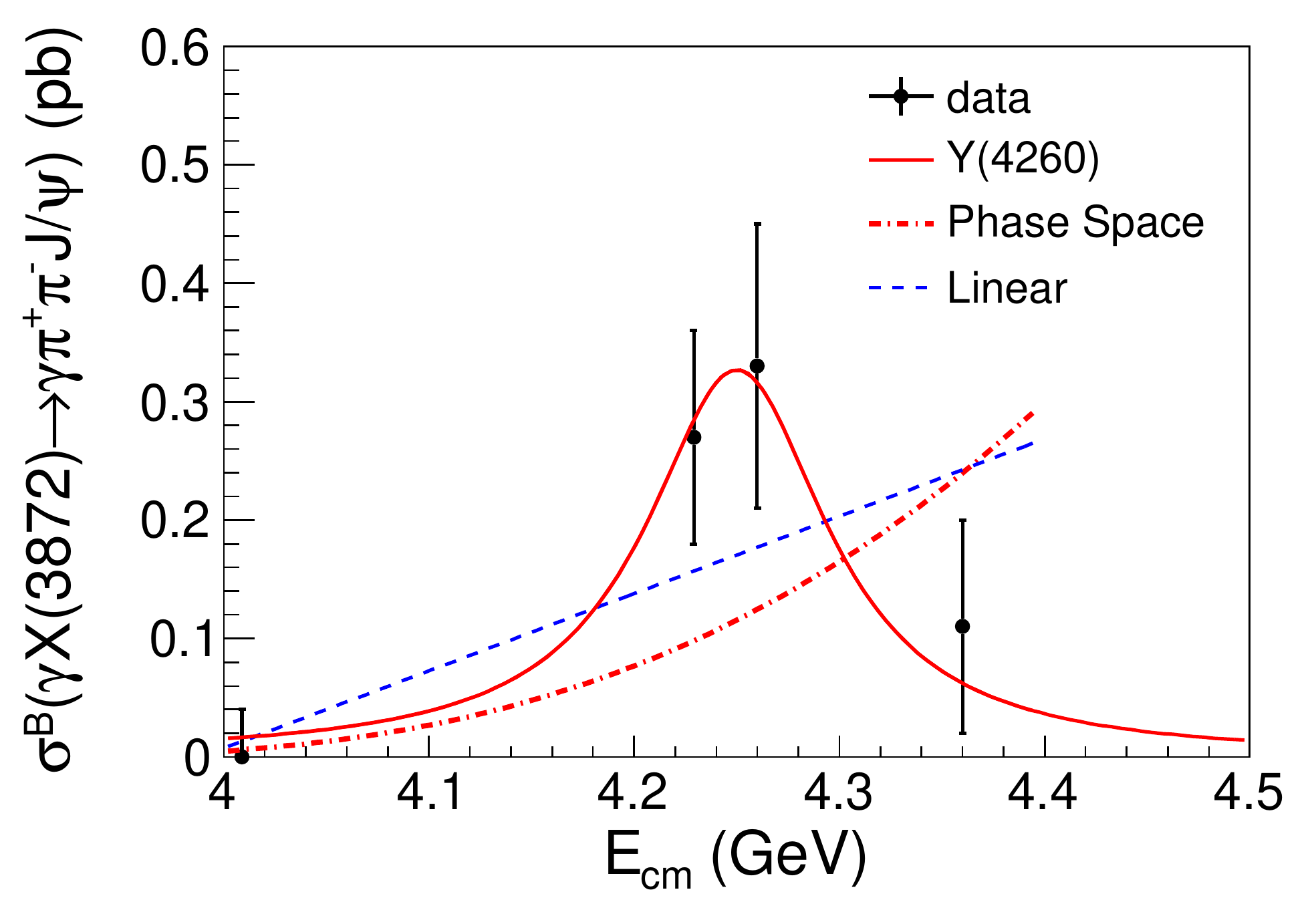}
\caption{Left panel: a fit to the $M(\ppjpsi)$ distribution observed at
BESIII~\cite{Ablikim:2013dyn}. Dots with error bars are data, and the red
solid curve is the best fit. Right panel: a fit to $\sigma[\EE\to \gamma X(3872)]\mathcal{B}[X(3872)\to \ppjpsi]$ measured by BESIII~\cite{Ablikim:2013dyn} (dots with error
bars) with a $\y$ resonance (red solid curve), a linear continuum
(blue dashed curve), or an $E1$-transition phase-space term (red
dotted-dashed curve). } \label{fit-mx}
\end{center}
\end{figure}

BESIII reported evidence for $X(3872)$ in $Y(4260)$ radiative
decay~\cite{Ablikim:2013dyn}. It searched for the process
$\EE\to \gamma X(3872)\to \gamma \ppjpsi$ with data collected at c.m.\ energies
of 4.01, 4.23, 4.26, and 4.36~GeV~\cite{Ablikim:2013dyn}.
The $M(\ppjpsi)$ distribution summed over all energy points, as
shown in Fig.~\ref{fit-mx}~(left plot), is fitted to extract the mass
and signal yield of $\xx$. The solid line shows the
best fit with the measured mass of $(3871.9\pm 0.7\pm 0.2)$~MeV and the width of less than 2.4~MeV at
the 90\% C.L.\ for the $\xx$. The statistical significance of $\xx$ is $6.3\sigma$.
The measured Born cross sections\footnote{
The Born cross section for $\EE\to \gamma X(3872)\to \gamma \ppjpsi$ at each c.m.\ energy point is given by
the formula
$\sigma[\EE\to \gamma X(3872)]\times \mathcal{B}[X(3872)\to \ppjpsi]=
\sigma^{\rm dressed}[\EE\to \gamma X(3872)]\times \mathcal{B}[X(3872)\to \ppjpsi]\times |1-\Pi|^2
= \sigma^{\rm vis} [\EE\to \gamma X(3872)]\times \mathcal{B}[X(3872)\to \ppjpsi] \times |1-\Pi|^2/(1+\delta)_{\rm ISR}=
N^{\rm obs}\times|1-\prod|^{2}/[\mathcal{L} \times
\mathcal{B}({J/\psi \to \ell^+ \ell^-}) \times \varepsilon \times (1+\delta)_{\rm ISR}]$,
where $\sigma^{\rm dressed}$ and $\sigma^{\rm vis}$ are called dressed cross section and visible
cross section, $N^{\rm obs}$ is the number of $X(3872)$ signal events obtained from the fit
to the $\pi^+ \pi^- J/\psi$ mass spectrum, $\mathcal{L}$ is the
integrated luminosity of the data sample, $\mathcal{B}({J/\psi \to \ell^+ \ell^-})$
is the branching fraction
of $J/\psi$ to lepton pair, $\varepsilon $ is the detection efficiency.
$(1+\delta)_{\rm ISR}$ is the radiative-correction factor, calculated using the formula given in Ref.~\cite{Kuraev:1985hb}, and
$|1-\prod|^{2}$ is the vacuum polarization factor, calculated according to Ref.~\cite{Actis:2010gg}.
For other processes, the calculation is similar.
}
are shown in Fig.~\ref{fit-mx}~(right plot)
with dots with error bars, together with the comparison with a $\y$
resonance (parameters fixed to PDG~\cite{Tanabashi:2018oca} values),
linear continuum, or $E1$-transition phase space ($\propto E^3_\gamma$) term.
The $\y$ resonance describes the data better than the other two options,
which supports the existence of the radiative transition $Y(4260)\to \gamma X(3872)$.
Together with the hadronic transition to the charged charmonium-like state
$\zc$~\cite{Ablikim:2013mio,Liu:2013dau,Xiao:2013iha}, this suggests that there
might be some common nature of $\xx$, $\y$, and $\zc$,
and so models developed to interpret any one of them should
also consider the other two --- in fact both the compact tetraquark model, discussed in Sec.~\ref{Sect:4.1.4},
as well as the molecular model, discussed in Sec.~\ref{Sect:4.1.5}, are
consistent with this request. The radiative decay
of $\y$ to the $\xx$ was even predicted within the latter model as a
necessary consequence of the observation of the reaction
$Y(4260)\to \pi Z_c(3900)$, if all the three mentioned states
are hadronic molecules~\cite{Guo:2013nza}.
 Recently BESIII updated the analysis of
$e^+e^- \to \gamma X(3872)\to \gamma \pi^+ \pi^- J/\psi$
using an about 9 fb$^{-1}$ data sample at c.m.\ energies above
4 GeV~\cite{Ryan:2019aaa}. Clear $X(3872)$ signals
are observed with a signal significance of 16.1$\sigma$, which confirms
the previous observation~\cite{Ablikim:2013dyn}. Besides the discovered
channel, Belle also observed clear $\xx$ signals in
$B^0 \to K^+ \pi^- X(3872)$ and $B^+ \to K^0_S \pi^+ X(3872)$ using
$772\times 10^{6}$ $B\bar{B}$ events~\cite{Bala:2015wep}, where the fraction
of the $K^{*}(892)^0$ signal component in the $K^+ \pi^-$ system is 34\%.

Combining the measured $\EE\to \ppjpsi$ cross sections
at c.m.\ energies of 4.23 and 4.26~GeV from BESIII~\cite{Ablikim:2016qzw},
we obtain $\sigma[\EE\to
\gamma X(3872)]\BR[X(3872)\to \ppjpsi]/\sigma(\EE\to \ppjpsi) =
(3.7\pm 0.9)\times 10^{-3}$, under the assumption that $\xx$ and
$\ppjpsi$ are only produced from $\y$ decays. If we take
$\BR[X(3872)\to \ppjpsi] = 5\%$ as an estimate, then we have
$$\mathcal{R} = \frac{\BR[Y(4260)\to \gamma X(3872)]}{\BR[Y(4260)\to \ppjpsi]}\sim
10 \% \ , $$
which is quite sizeable given the suppression from the small electromagnetic
coupling present in the numerator.

\begin{figure}
\begin{center}
\includegraphics[height=5cm]{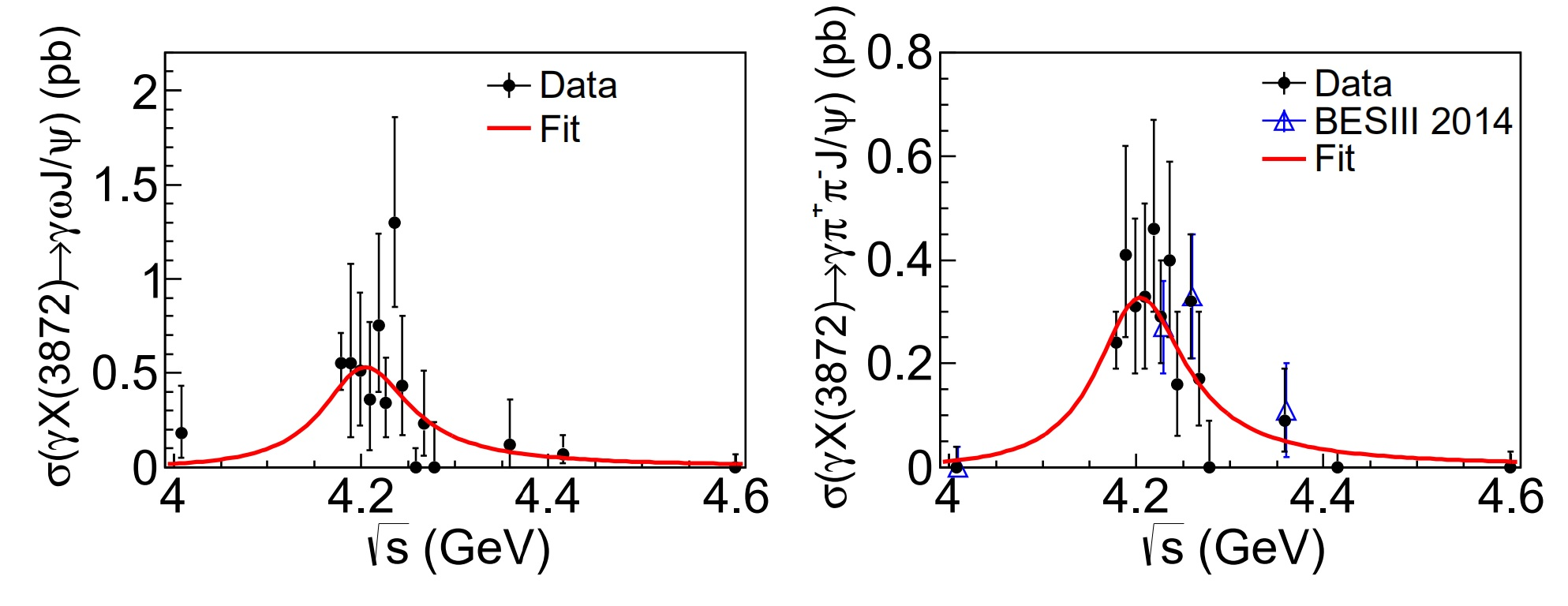}
\caption{The distributions of $\sigma[e^+e^-\to \gamma X(3872)]\BR[X(3872)\to \omega J/\psi]$ (left)
and $\sigma[e^+e^-\to \gamma X(3872)]\BR[X(3872)\to \pi^+ \pi^- J/\psi]$ (right)
from the BESIII measurement~\cite{Ablikim:2019zio}.
The solid lines are from a simultaneous fit to these two distributions
assuming a single BW resonance.} \label{bes3-goj}
\end{center}
\end{figure}

Since the $X(3872)$ can be produced in
$\EE\to \gamma X(3872)$, very recently BESIII
studied the $e^+e^- \to \gamma \omega J/\psi$ process using 11.6 fb$^{-1}$ of
data at c.m.\ energies from $\sqrt{s}=$4.008 to 4.600 GeV~\cite{Ablikim:2019zio},
where a signal peak consistent with the $X(3872)$
resonance is observed in the $\omega J/\psi$ mass spectrum.
The measured $\sigma[e^+e^-\to X(3872)]\BR[X(3872)\to \omega J/\psi]$ cross
section is shown in Fig.~\ref{bes3-goj} (left panel). Meanwhile, using the
same analysis method as
in Ref.~\cite{Ablikim:2013dyn}, BESIII updated the measurement of
$\sigma[e^+e^-\to \gamma X(3872)]\BR[X(3872)\to \pi^+ \pi^- J/\psi]$ as well,
as shown in the right panel of Fig.~\ref{bes3-goj}.
A simultaneous maximum-likelihood fit is performed to these two distributions
with a single BW resonance. The fit gives its mass
$M=(4200.6^{+7.9}_{-13.3}\pm 3.0)$ MeV
and width $\Gamma=(115^{+38}_{-26}\pm12)$ MeV, the values of which are consistent with
the $\psi(4160)$ or $Y(4230)$. The ratio
$\BR[X(3872)\to \omega J/\psi]$/$\BR[X(3872)\to \pi^+ \pi^- J/\psi]$
as a free parameter in the simultaneous fit is obtained to be
$1.6^{+0.4}_{-0.3}\pm0.2$.

Besides a few more $X(3872)$ production modes discovered by BESIII, the first observation of the
$X(3872)$ in the $\Lambda_b^0$ decay $\Lambda_b^0\to X(3872) p K^- \to J/\psi \pi^+ \pi^- p K^-$
with $J/\psi \to \mu^+ \mu^-$
was reported very recently by LHCb based on data collected in proton-proton collisions
corresponding to 1.0, 2.0 and 1.9 fb$^{-1}$ of integrated
luminosity at c.m.\ energies of 7, 8 and 13 TeV~\cite{Aaij:2019zkm}, respectively.

\begin{figure}
\begin{center}
\includegraphics[height=5cm]{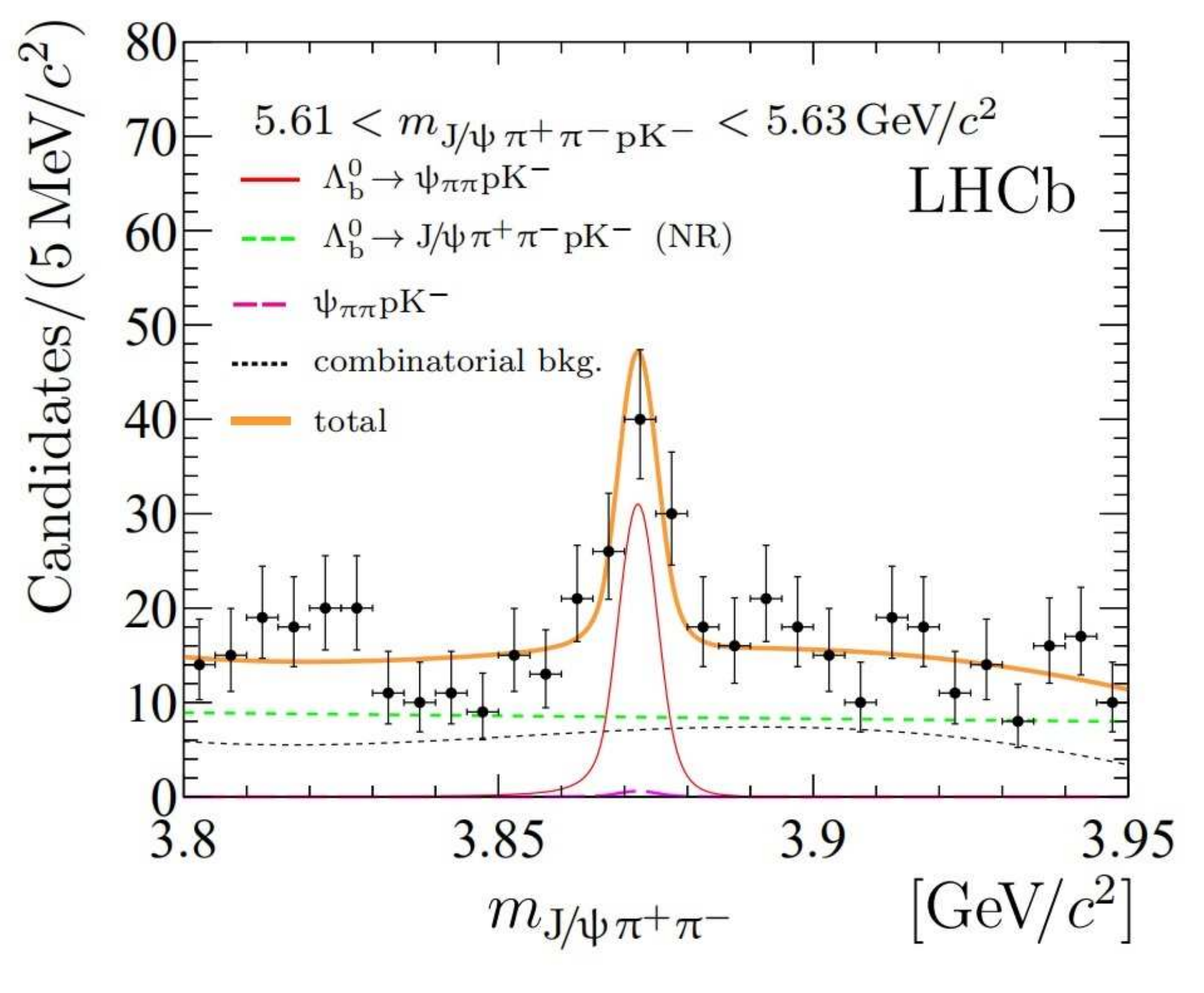}
\includegraphics[height=5cm]{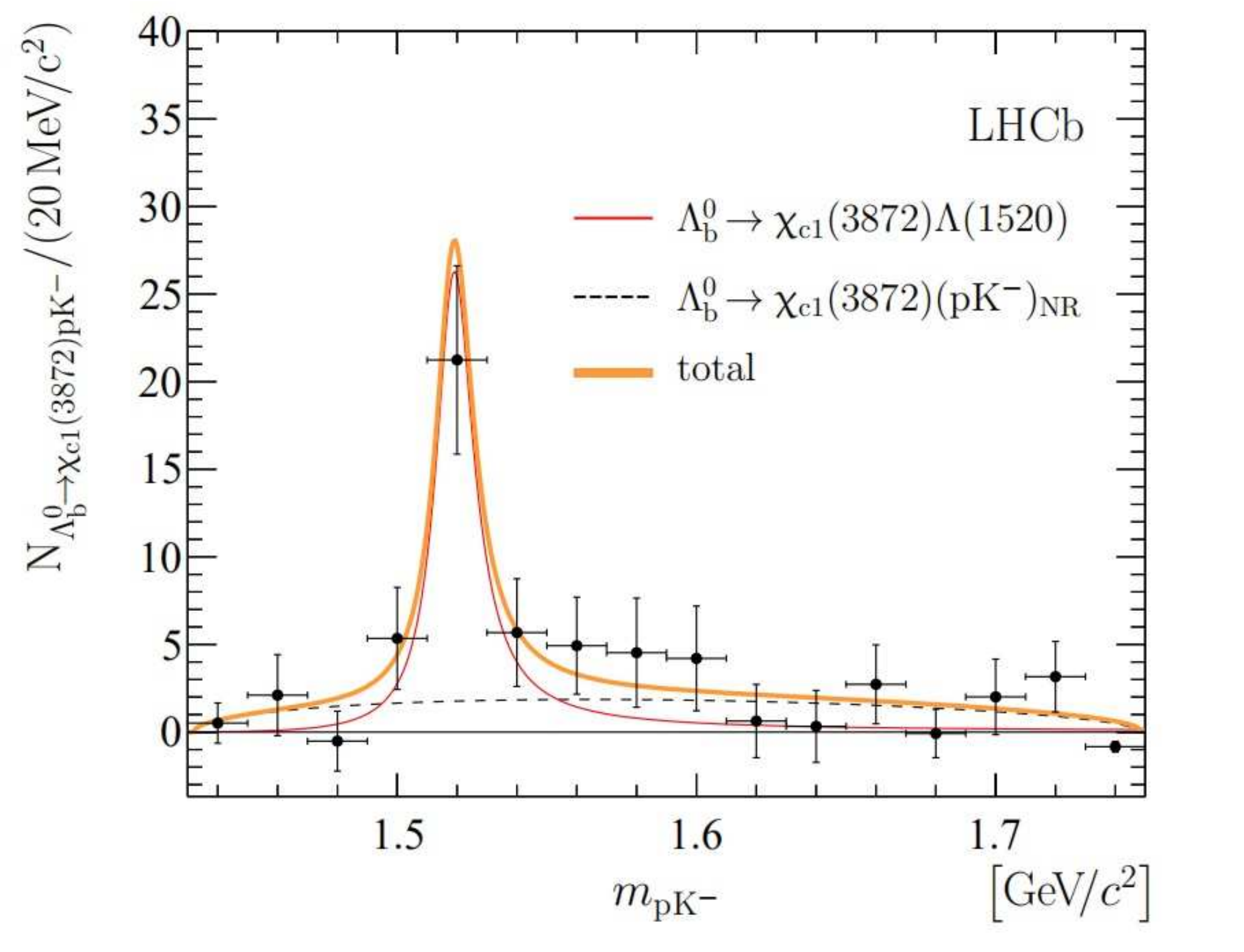}
\caption{The $J/\psi \pi^+ \pi^-$ and background-subtracted $pK^-$ mass spectra
within the $X(3872)$ signal region from $\Lambda_b^0 \to J/\psi \pi^+ \pi^- p K^-$ decay,
together with fitted results described
 in the text~\cite{Aaij:2019zkm}.} \label{lhcb-x3872}
\end{center}
\end{figure}

After event selection, clear $\Lambda_b^0$ signals in $J/\psi \pi^+ \pi^- p K^-$
mass spectrum, and $\psi(2S)$ and $X(3872)$ signals in $J/\psi \pi^+ \pi^-$ mass spectrum
are observed. To obtain the $\psi(2S)$ and $X(3872)$ signal yields from $\Lambda_b^0$ decay,
a 2D unbinned extended maximum-likelihood fir to the $J/\psi \pi^+ \pi^- p K^-$
and $J/\psi \pi^+ \pi^-$ mass spectra was performed with four components included:
$\psi(2S)$ or $X(3872)$ signals from $\Lambda_b^0$ decay ($\Lambda_b^0\to \psi_{\pi\pi} p K^-$);
a nonresonant (NR) component from $\Lambda_b^0$ decay with on $\psi(2S)$ or $X(3872)$
intermediate state; a component with $\psi(2S)$ or $X(3872)$ describing random combinations
that are not from $\Lambda_b^0$ decay ($\psi_{\pi \pi} p K^-$); and a combinatorial $J/\psi \pi^+ \pi^- p K^-$ component.
The fit yields $610\pm30$ and $55\pm11$ $\psi(2S)$ and $X(3872)$ signal events with a
statistical significance of 7.2$\sigma$ for the $X(3872)$. The projection to $J/\psi \pi^+ \pi^-$
mass distribution from the 2D fit in the $X(3872)$ signal region is shown in Fig.~\ref{lhcb-x3872} (left panel)
with fitted components indicated.
The background-subtracted $pK^-$ mass spectrum for the $X(3872)$ channel is shown
in Fig.~\ref{lhcb-x3872} (right panel), where a clear peak associated with the $\Lambda(1520)$ state
is seen. An unbinned maximum-likelihood fit with a $\Lambda(1520)$ signal and a nonresonant component
was performed. The fraction of the $\Lambda(1520)$ is $(58 \pm 15)\%$, where
the uncertainty is statistical only. The fitted results are shown in Fig.~\ref{lhcb-x3872} (right panel)
with fitted components indicated.
Using the $\Lambda_b^0\to \psi(2S) p K^-$ decay as a normalization channel,
the ratio of the branching fractions is measured to be
$$
R=\frac{\BR[\Lambda_b^0\to X(3872) p K^-]}{\BR[\Lambda_b^0\to \psi(2S) p K^-]}\times \frac{\BR[X(3872)\to J/\psi \pi^+\pi^-]}{\BR[\psi(2S)\to J/\psi \pi^+\pi^-]}=(5.4\pm1.1\pm0.2)\%,
$$
where the first uncertainty is statistical and the second is systematic.
This the first time that the $X(3872)$ was observed in the $\Lambda_b^0$ decay.

\begin{figure}[htbp]
\begin{center}
\includegraphics[height=5cm]{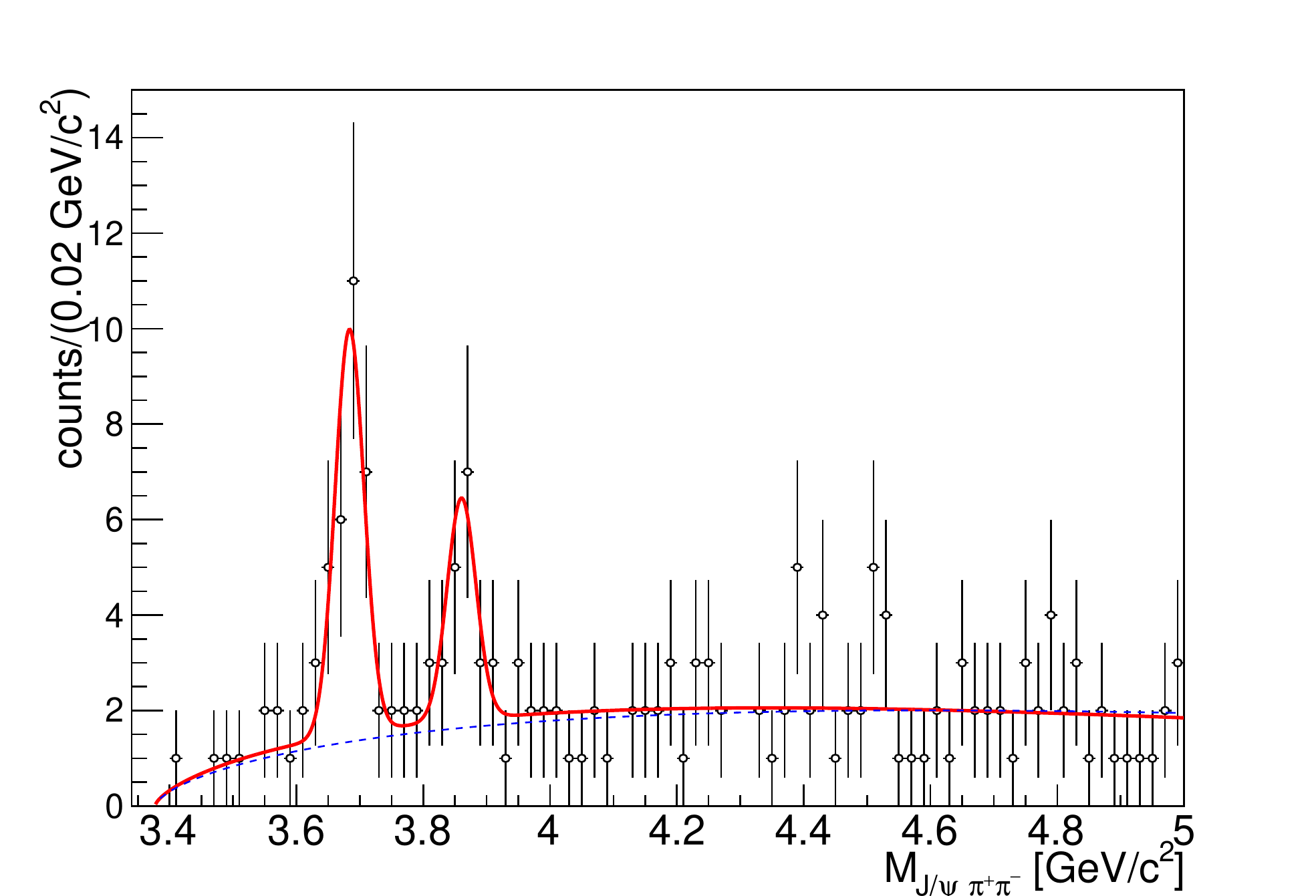}
\includegraphics[height=5cm]{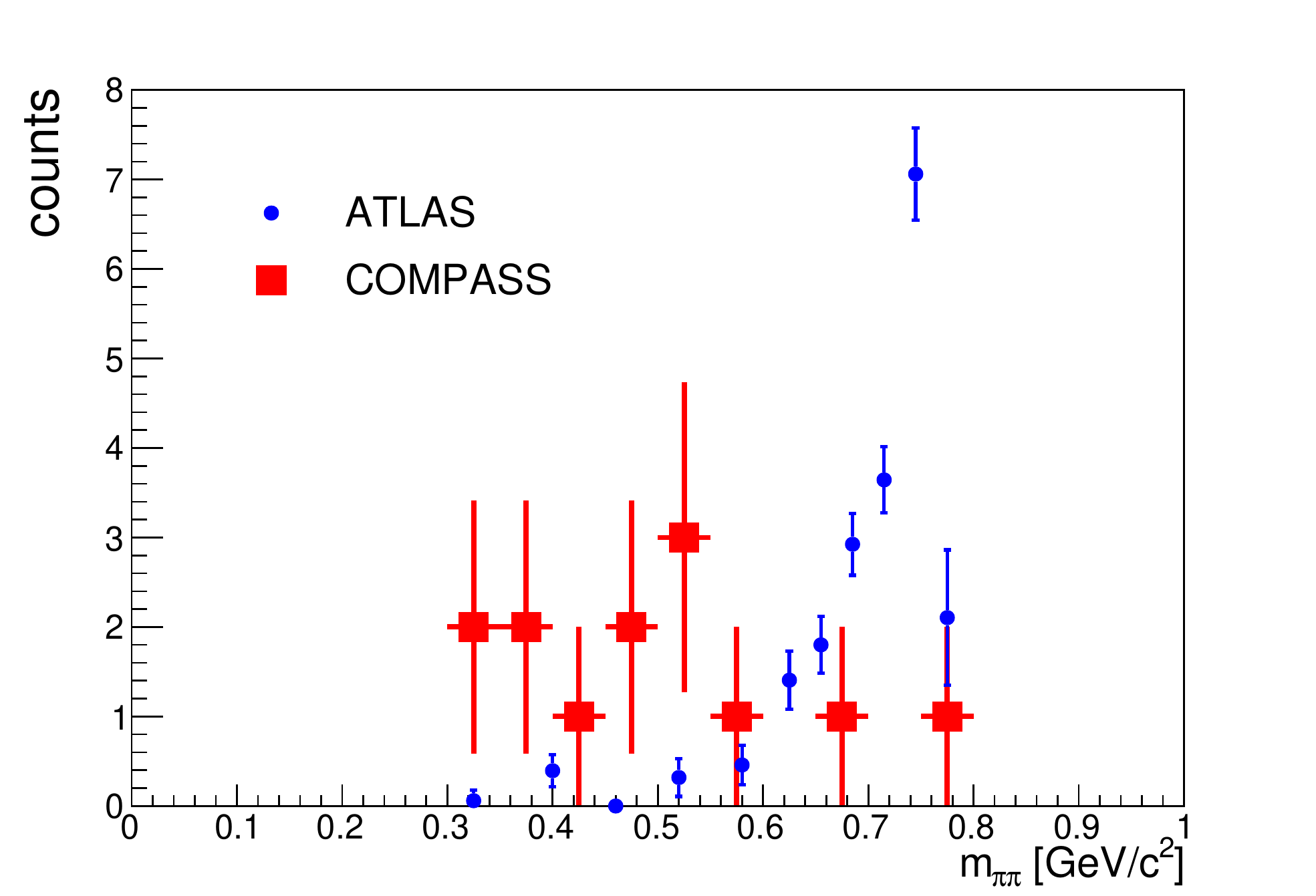}
\end{center}
\caption{The $M(\pi^+ \pi^- J/\psi)$ distribution (left) and
the corresponding $\pi^+ \pi^-$ mass spectrum with
$M(\pi^+ \pi^-J/\psi)$ within
the $X(3872)$ signal region
shown with the squares with error bars (right) from the selected
$\mu^+ N \to \mu^+ J/\psi \pi^+ \pi^- \pi^\pm N^{\prime}$
candidates~\cite{Aghasyan:2017utv}. The dots with error bars in the right
plot are from the decays of $X(3872) \to \pi^+ \pi^- J/\psi$ observed
by ATLAS for comparison~\cite{Aaboud:2016vzw}.
}\label{compass-x3872}
\end{figure}

A search for exotic charmonium-like states in exclusive photoproduction
reactions was proposed in Refs.~\cite{Li:2004sta, Liu:2008qx, He:2009yda}.
Recently COMPASS, a fixed-target experiment at CERN, analyzed
the full set of data collected with a muon beam between 2002 and 2011,
covering the range from 7 to 19 GeV in the c.m.\ energy of the virtual
photon-nucleon system, to search for the $X(3872)$
in photoproduction reactions.
The $X(3872)$ was searched for in the charge-exchange
reaction $\mu^+ N \to \mu^+ J/\psi \pi^+ \pi^- \pi^\pm N^{\prime}$, where $N$
denotes the target nucleon and $N'$ the unobserved recoil system.
The resulting $\pi^+ \pi^-J/\psi$ invariant mass distribution is shown
in Fig.~\ref{compass-x3872} (left plot),
where two peaks with positions and widths consistent with the $\psi(2S)$
and $X(3872)$ are evident. The significance of the
second peak is 4.1$\sigma$ with systematic effects included. However, the
shape of the $\pi^+\pi^-$ mass distribution
corresponding to the second peak shows disagreement with previous
observations for $X(3872)$ and is inconsistent with
quantum numbers $J^{PC}=1^{++}$, as shown in Fig.~\ref{compass-x3872}
(right plot) where the squares with error bars are from the COMPASS
measurement and the dots with error bars from the ATLAS
data set~\cite{Aaboud:2016vzw} for comparison.
Due to this critical difference, the COMPASS Collaboration concluded that
the observed signal is not the well-known $X(3872)$
giving possible evidence for a new charmonium-like state
denoted $\tilde{X}(3872)$.
The measured mass and
width of the $\tilde{X}(3872)$ are $M[\tilde{X}(3872)]=(3860.0\pm 10.4)$ MeV
and $\Gamma[\tilde{X}(3872)]<51$ MeV at 90\% C.L.,
and the product of the cross section and branching fraction of the
$\tilde{X}(3872)$ into $\pi^+ \pi^- J/\psi$
is determined to be $(71 \pm 28\pm 39)$ pb.
An independent confirmation of the observed $\tilde{X}(3872)$ signal from
high-precision experiments with high-energy virtual or real
photons is required.

Lepto(photo-)production of the $X(3872)$ was also
searched for in the neutral reaction
$\mu^+ N \to \mu^+ J/\psi \pi^+ \pi^- N^{\prime}$~\cite{Aghasyan:2017utv}.
Except for a peak from photodiffractive production of $\psi(2S)$ in the
$\pi^+ \pi^- J/\psi $ invariant mass spectrum, no statistically significant
signal at around 3872 MeV can be seen. The 90\% C.L.\ upper limit of the
$X(3872)$ production cross section in this reaction
multiplied by the branching fraction for the decay
$X(3872) \to J/\psi \pi^+ \pi^-$ was set to be 2.9 pb.

\vspace{0.3cm}\noindent
{\it (2) Decay patterns of $\chi_{c1}(3872)$ aka $X(3872)$}
\vspace{0.3cm}

The transition $X(3872) \to \gamma \jpsi$ was measured by
BaBar~\cite{Aubert:2008ae} with a statistical significance of $3.6\sigma$,
Belle~\cite{Bhardwaj:2011dj} with a statistical significance of
$5.5\sigma$, and LHCb clearly~\cite{Aaij:2014ala}. So the decay mode
$X(3872)\to \gamma \jpsi$ is well established,
while for $X(3872)\to \gamma \psip$, the results from some measurements
are not consistent. Evidence for it was first reported by
BaBar with a statistical significance of $3.5\sigma$ using
$(465\pm5)\times 10^6$ $B\bar{B}$ pairs~\cite{Aubert:2008ae}. The ratio
of the branching fractions is measured to be
$R = \BR[X(3872)\to \gamma\psp]/\BR[X(3872)\to \gamma\jpsi] = 3.4\pm 1.4$,
where statistical and systematic uncertainties are combined.
Using a 3~fb$^{-1}$ data sample at $\sqrt{s}=7$ and $8$~TeV, LHCb reported
strong evidence for the decay $X(3872)\to\gamma \psp$ with a
statistical significance of $4.4\sigma$~\cite{Aaij:2014ala}.
The measured value of the ratio $R$ is $2.46\pm 0.64\pm 0.29$,
which is in good agreement with the BaBar's measurement~\cite{Aubert:2008ae}.
In contrast, the negative result was reported by Belle using
$772\times10^6$ $B\bar{B}$ pairs and an
upper limit of $R < 2.1$ was set at the 90\% C.L.~\cite{Bhardwaj:2011dj}.
Using the information from Ref.~\cite{Bhardwaj:2011dj}, we obtain
$R=0.6\pm1.4$ as an estimation of the central value and uncertainty.
Although there is no serious disagreement between the measurements from
BaBar, LHCb, and Belle, there is some tension in the values for the decay
rates of $X(3827)\to \gamma\psp$.
Using the above mentioned results, a weighted average gives
$$\overline{R} = \frac{\BR[X(3872)\to \gamma\psp]}{\BR[X(3827)\to
\gamma\jpsi]} = 2.31\pm 0.57$$ without considering a possible
small correlation. Since BESIII has observed clear
$\EE\to \gamma X(3872)\to \gamma \ppjpsi$
signals~\cite{Ablikim:2013dyn,Ablikim:2019zio}, we also expect the results
for $\EE\to \gamma X(3872)\to \gamma \gamma \psp$ from this
experiment soon.

A relatively large branching fraction for $X(3872) \to \gamma \psi(2S)$
was claimed to be inconsistent with a purely $D^0 \bar{D}^{*0}$ molecular
interpretation of the $X(3872)$ based on a model calculation presented in Ref.~\cite{Swanson:2004pp}.
In the meantime, it is consistent with expectations if
the $\xx$ is a pure charmonium or a mixture of a molecule and a
charmonium~\cite{Chen:2016qju,Brambilla:2010cs,Aaij:2014ala}.
However, Ref.~\cite{Guo:2014taa} argues that the results of Ref.~\cite{Swanson:2004pp}
are model dependent and that an effective field theory approach allows
for the mentioned ratio even within a molecular approach --- we come back
to this issue in Secs.~\ref{Sect:4.1.5} and~\ref{Sect:4.2.4}.

\begin{figure}[htbp]
\begin{center}
\includegraphics[width=16cm]{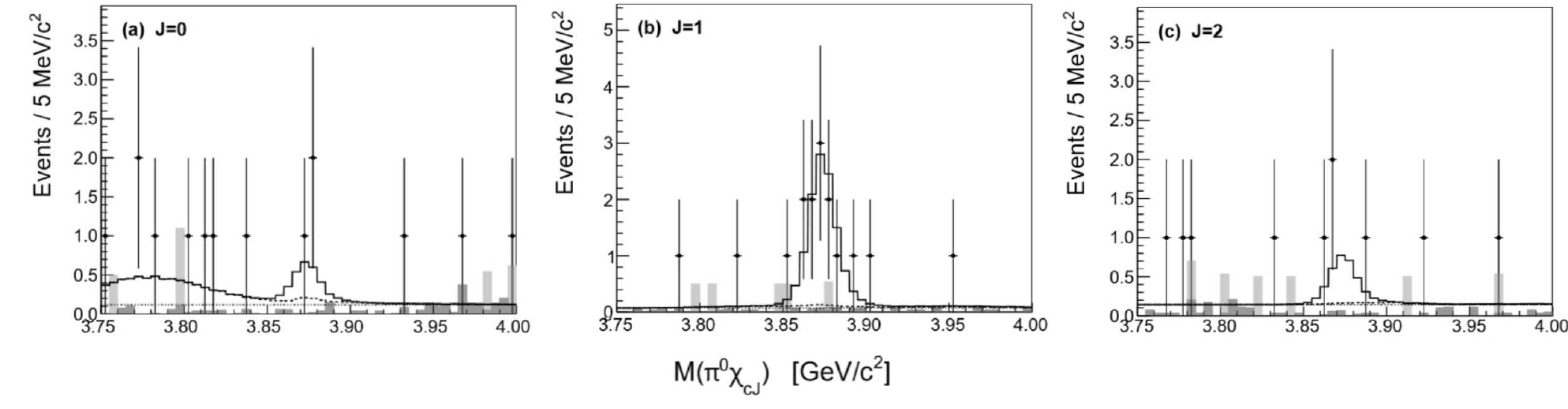}
\end{center}
\caption{The $\pi^0 \chi_{cJ}$ mass distributions, $M(\pi^0 \chi_{cJ})$, from
the process $e^+ e^- \to \gamma \pi^0 \chi_{cJ} $ for (a) $J = 0$, (b) $J = 1$, and (c) $J = 2$~\cite{Ryan:2019aaa}.
Points are data; lines are fits (solid is the total, the dotted is the
polynomial background, and the dashed line is the total background);
the darker histogram is a MC estimate of peaking $J/\psi$ backgrounds;
the lighter stacked histogram
is an estimate of non-peaking backgrounds using $J/\psi$ sidebands from data.}
\label{fig:bes3-x3872}
\end{figure}

A measurement of pionic transitions of the $X(3872)$ to
the $\chi_{cJ}$ ($J=0$, 1, 2) has been proposed to
be one way to distinguish between various interpretations.
If the $X(3872)$ is a conventional $c\bar{c}$ state,
transition to the $\pi^0 \chi_{c1}$
should be suppressed compared to $\pi^+ \pi^- \chi_{c1}$ due to isospin
breaking by the light quark
masses. If the $X(3872)$ is a compact tetraquark or molecular state, on the other
hand, these rates are expected to be enhanced~\cite{Dubynskiy:2007tj,Fleming:2008yn,Mehen:2015efa}.
In the search for $X(3872) \to \pi^+ \pi^- \chi_{c1}$ with negative results,
the Belle Collaboration determined the
branching fraction $\BR[B^+ \to K^+ X(3872)] \BR[X(3872)\to \pi^+ \pi^- \chi_{c1}]<1.5\times10^{-6}$
at 90\% C.L.~\cite{Bhardwaj:2015rju}.
Recently BESIII reported the first observation of the decay
$X(3872) \to \pi^0 \chi_{c1}$
in $e^+e^- \to \gamma X(3872)$ with a statistical significance of
5.2$\sigma$ using data at the c.m.\ energies above 4 GeV corresponding an
integrated luminosity of 9 fb$^{-1}$~\cite{Ryan:2019aaa}.
To calculate the ratio
$$
R^{X}_{\chi_{cJ}/\psi}\equiv \frac{\BR[X(3872)\to \pi^0 \chi_{cJ}]}{\BR[X(3872)\to \pi^+ \pi^- J/\psi]}=
\left\{6.6^{+6.5}_{-4.5}\pm1.1 (19), \ 0.88^{+0.33}_{-0.27}\pm0.10,
\ 0.40^{+0.37}_{-0.27}\pm0.04 (1.1)\right\}, $$
for $J=0,~1$, and 2, respectively, where the numbers in brackets are
90\% C.L.\ upper limits. The
normalization channel was $e^+e^- \to \gamma X(3872)$ with $X(3872) \to \pi^+ \pi^- J/\psi$
updated at the same time. In the BESIII measurement,
after initial event selection, a clear signal for the $X(3872)$ can be seen in the data of 4.15 to 4.30~GeV, while no evidence
for the $X(3872)$ is seen at other c.m.\
energy points. A fit to the invariant mass distribution of $\pi^0 \chi_{cJ}$
with a first-order polynomial background and a signal shape from the signal MC
simulation directly yields $16.9^{+5.2}_{-4.5}$ $X(3872)$
events with a statistical significance of 4.8$\sigma$.
After requiring the $\gamma J/\psi$ mass within the
$\chi_{cJ}$ $(J=0,~1,~2)$ mass region respectively,
the resulting distributions for $M(\pi^0 \chi_{cJ})$ with $J=0,~1,~2$
are shown in Fig.~\ref{fig:bes3-x3872}.
The fits to each $M(\pi^0 \chi_{cJ})$ distribution with a constant background
and a signal shape from MC simulation give $1.9^{+1.9}_{-1.3}$,
$10.8^{+3.8}_{-3.1}$, and $2.5^{+2.3}_{-1.7}$
$X(3872)$ signal events with signal significances of
1.6$\sigma$, 5.2$\sigma$, and 1.6$\sigma$
for $J=0,~1$, and 2, respectively. No significant $X(3872)$ signal is found in the $M(\pi^0 \chi_{c0,c2})$ distributions.
This is the first observation of a decay of the $X(3872)$ to a $P$-wave charmonium state and
it supports the non-$c \bar{c}$ interpretations of the
$X(3872)$~\cite{Dubynskiy:2007tj,Fleming:2008yn,Mehen:2015efa}.
This BESIII observation can be cross-checked by other experiments like Belle.
Previously, Belle measured $B^+ \to K^+ \pi^0 \chi_{c1}$
and provided the background-subtracted $\pi^0 \chi_{c1}$ mass distribution
without any structure at the $X(3872)$ mass~\cite{Bhardwaj:2015rju},
but the bin width of that mass spectrum is too large to claim a contradictory
result.
Considering this, very recently Belle updated the analysis of
$B^+ \to \chi_{c1} \pi^0 K^+$ decays to focus on the $X(3872)$ mass
region~\cite{Bhardwaj:2019spn} using $772 \times 10^{6}$ $B\overline{B}$ events.
No statistically significant $X(3872)$ signal
at around 3872 MeV can be seen in the $\chi_{c1} \pi^0$ mass spectrum even with a
much narrower bin width. An upper limit $R^{X}_{\chi_{c1}/\psi}$ $<$ 0.97 at 90\% C.L.
is set, which does not contradict the BESIII result~\cite{Ryan:2019aaa}.
In the future Belle II can utilize a similar study to provide
improved results for the ratio $R^{X}_{\chi_{cJ}/\psi}$.

\begin{figure}[htbp]
\begin{center}
\includegraphics[height=5cm]{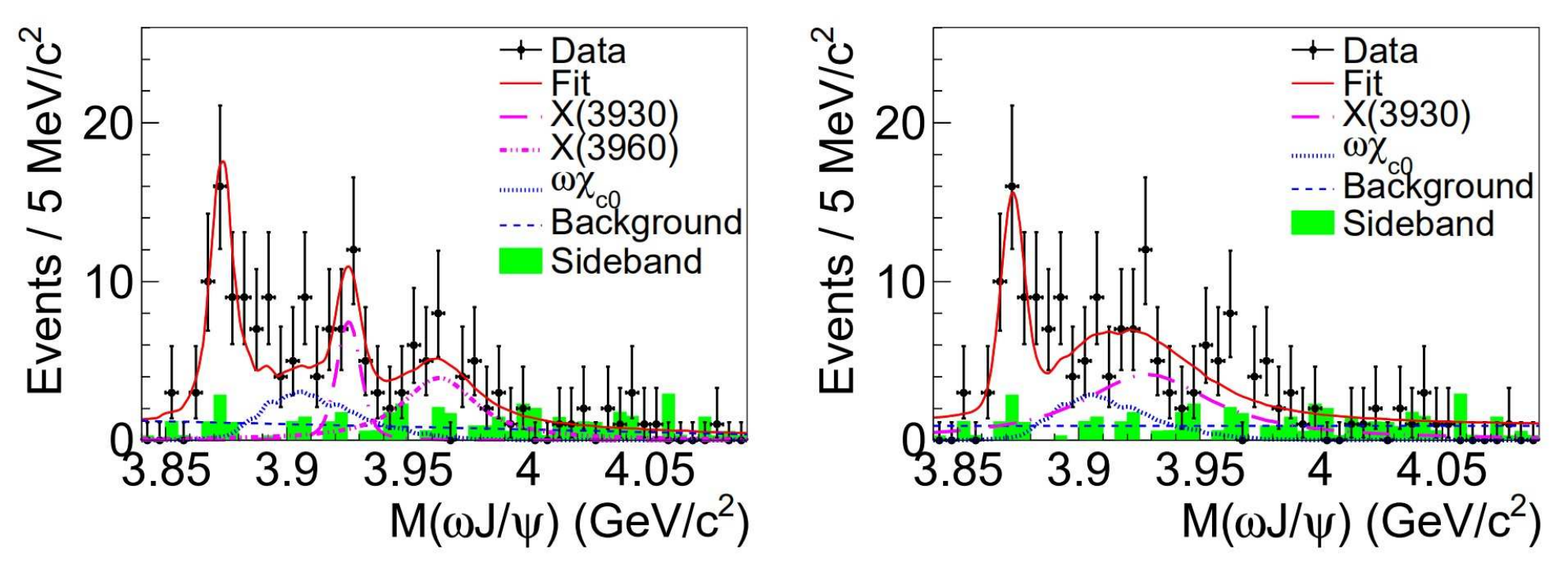}
\end{center}
\caption{The $\omega J/\psi$ mass distributions with
results of two fit scenarios described in the text.
Dots with error bars are data~\cite{Ablikim:2019zio},
the red solid curves show the total fit results,
the green shaded histograms
are the normalized contribution from the $J/\psi$- and $\omega$-mass
sidebands, and other components
included in the fits are indicated in the plots.}\label{bes3-xtooj}
\end{figure}

The hadronic decay $X(3872)\to \omega J/\psi$ was reported by the Belle
and BaBar Collaborations with less than $5\sigma$ evidence in $B$ decays
based on 275 million and 467 million $B\bar{B}$
pairs, respectively~\cite{delAmoSanchez:2010jr,Abe:2005ix}. Recently the
$X(3872)\to \omega J/\psi$ decay was firmly observed with
a more than 5$\sigma$ significance by the BESIII Collaboration in the process
$e^+e^- \to \gamma \omega J/\psi$ with a total integrated luminosity of about
11.6 fb$^{-1}$ at c.m.\ energies from $\sqrt{s}=$4.008 to 4.600 GeV~\cite{Ablikim:2019zio}.
The $\omega J/\psi$ invariant mass distribution from the BESIII measurement
is shown in Fig.~\ref{bes3-xtooj}
as the dots with error bars, where a signal peak consistent with the
$X(3872)$ resonance is observed together with
a broad structure from irreducible $e^+e^- \to \omega \chi_{c0}$ background
events. In addition, there are evident structures above 3.9~GeV.
Two fit scenarios were adopted for an approximation of the $\omega J/\psi$
mass spectrum:
(1) the incoherent sum of three BW resonances (denoted as $X(3872)$,
$X(3915)$, and $X(3960)$, respectively); (2) the incoherent sum of the
$X(3872)$ and $X(3915)$.
The corresponding fit results are shown in Fig.~\ref{bes3-xtooj} with
red solid lines. In both fits, the signal significance of the $X(3872)$ is larger than $5\sigma$ and the
$X(3872)$ mass was measured to be
$(3873.3\pm1.1\pm1.0)$~MeV.
The low statistics of the BESIII experiment prevent us from drawing a solid
conclusion about the states above 3.9~GeV.

With a data sample containing 447M $B\bar{B}$ events, Belle
observed a near-threshold $D^0\bar{D}^0\pi^0$ mass enhancement in
$B\to K D^0\bar{D}^0\pi^0$ decays that, when interpreted as
$X(3872)\to D^0\bar{D}^0\pi^0$, gave an $X(3872)$ mass of
$(3875.4\pm 0.7^{+1.2}_{-2.0})$~MeV~\cite{Gokhroo:2006bt}. BaBar
studied $B\to KD^{*0}\bar{D}^0$ with a sample of 383M $B\bar{B}$
pairs and found a similar near-threshold enhancement that, if
considered to be due to the $X(3872)\to D^{*0}\bar{D}^0$, gave a
mass of $(3875.1^{+0.7}_{-0.5}\pm
0.5)$~MeV~\cite{Aubert:2007rva}. Originally this state has been considered
to be a state different from the $X(3872)$ in literature.
However, a subsequent Belle study of $B\to KD^{*0}\bar{D}^0$ based
on 657M $B\bar{B}$ pairs found for the near-threshold peak a mass
of $(3872.9^{+0.6+0.4}_{-0.4-0.5})$~MeV~\cite{Adachi:2008sua} by
fitting the peak with a phase-space modulated BW
function, much closer to the value determined from the $\pi^+\pi^-
J/\psi$ decay channel. Now those data are taken as the open charm decay mode of the $X(3872)$.

LHCb searched for the $X(3872) \to p \bar{p}$
using a 3 fb$^{-1}$ $pp$ collision data sample~\cite{Aaij:2016kxn}.
No signals of the $X(3872)$ are seen, and a
95\% C.L.\ upper limit is obtained
$$\frac{\BR[B^+ \to K^+ X(3872)]\BR[X(3872) \to p \bar{p}]}{\BR(B^+\to K^+ J/\psi)\BR(J/\psi \to p \bar{p})}<0.25\% \ .$$
LHCb also searched for the $X(3872) \to \phi \phi$, but no signals of the
$X(3872)$ are seen either~\cite{Aaij:2017tzn}.
An upper limit for the inclusive production of
$X(3872)$ at 90\% C.L.\ is obtained on
$$\frac{\BR[b\to X(3872)X]\BR[X(3872)\to \phi \phi]}{\BR(b\to \chi_{c1}X)\BR(\chi_{c1}\to \phi \phi)}<0.34 \ .$$

\vspace{0.3cm}\noindent
{\it (3) Other studies related to $\chi_{c1}(3872)$ aka $X(3872)$}
\vspace{0.3cm}

So far, all the $X(3872)$ related measurements are product branching
fractions or relative branching ratios since the absolute
production rate of the $X(3872)$ is unknown in any of the
experiments. The only attempt to measure the production rate of
the $X(3872)$ is via inclusive $B$ decays into a $X(3872)$ and a kaon in
the BaBar and Belle experiments. But due to the high
multicombinational background level, no evidence for the
$X(3872)$ can be seen. BaBar set an upper
limit of the $X(3872)$ production rate in
the $B$-meson decays by measuring the momentum distribution of the
inclusive kaons from $B$-meson decays with 210~fb$^{-1}$
$\Upsilon(4S)$ data~\cite{Aubert:2005vi}: \( \BR[B^-\to K^-
X(3872)]<3.2\times 10^{-4} \) at the 90\% C.L.

A recent update comes from the Belle experiment with the full sample
of $772\times 10^6$ $B\bar{B}$ pairs~\cite{Kato:2017gfv}.
No significant $X(3872)$ signal is observed, and Belle sets a more stringent
upper limit of $\BR[B^-\to K^- X(3872)]<2.7\times 10^{-4}$
at the 90\% C.L., and the central value is $\BR[B^-\to K^-
X(3872)] = (1.2\pm 1.1\pm 0.1)\times 10^{-4}$.
Together with all the other measurements on the product branching
fractions $\BR[B^-\to K^- X(3872)]\BR[X(3872)\to {\rm
exclusive}]$ [${\rm exclusive}=\pp\jpsi$, $\pp\piz\jpsi$,
$\gamma\jpsi$, $\gamma\psp$, $D^0\bar{D}^{*0}+c.c.$]~\cite{Tanabashi:2018oca},
one obtains
\[
2.9\%<\BR[X(3872)\to \pi^+\pi^-J/\psi]<10\%,
\]
\[
0.9\times 10^{-4}<\BR[B^-\to K^- X(3872)]<2.7\times 10^{-4},
\]
at the 90\% C.L.~\cite{Yuan:2009iu}. We find that the decay width of the
$X(3872)$ to $\pi^+\pi^-J/\psi$ is larger and the
production rate of the $X(3872)$ is smaller than
for conventional charmonium states
such as $\eta_c$, $\psp$, and $\chi_{c1}$~\cite{Tanabashi:2018oca}.
In the forthcoming Belle II experiment, to measure the
absolute branching fraction of $X(3872)$,
the inclusive decays $B \to K X(3872)$ may be still the only way.
In order to observe the $X(3872)$ signals, we need to
improve the $B$ tagging efficiency and suppress the multicombinatorial
backgrounds.

An axial-vector $\chi_{c1}$ state should also be produced directly in
the $e^+e^-$ annihilation, however, the corresponding probability
is suppressed by an additional power of the fine structure constant
$\alpha$ because
positive-parity states cannot be produced in a single-photon annihilation
of an electron-positron pair. Theoretical predictions for
such a direct production of the generic $c\bar{c}$ charmonium $\chi_{c1}$
can be found in Refs.~\cite{Kuhn:1979bb,Kivel:2015iea,Czyz:2016xvc} while
a similar estimate for the $X(3872)$ was made in Ref.~\cite{Denig:2014fha}. The result obtained,
$\Gamma_{ee}[X(3872)]>0.03$~eV, is not in contradiction with the
upper limit on the probability of this production established
experimentally --- the most recent result is $\Gamma_{ee}[X(3872)] {\cal B}[X(3872)\to\pi^+\pi^-J/\psi]<0.13$~eV at the 90\% C.L.
\cite{Ablikim:2015ain}. It, therefore, remains to be seen whether or not
necessary statistics can be collected in future experiments to observe the
$X(3872)$ directly in $e^+e^-$ annihilation.

In a short summary, after 16 years of studies, the knowledge on the $\xx$
is still very limited: We know it is an isoscalar with $J^{PC}=1^{++}$, a very
precise mass close to the $D^0\bar{D}^{*0}$ threshold, and a very
small width. Significant decays
into $D^0\bar{D}^{*0}$, $\ppjpsi$, $\omega J/\psi$, $\pi^0 \chi_{c1}$,
and $\gamma\jpsi$ have been observed,
while the significances of other modes such as $\gamma\psp$ and light hadrons are
still less than $5\sigma$. The absolute decay rates to the above modes
are unknown. Although an isoscalar partner of $X(3872)$
was observed by COMPASS, further confirmation is needed.

Since more data have been accumulated at LHCb, some of the
measurements can be improved, such as the study of $X(3872)\to
\gamma\psp$ and possibly of $X(3872)\to \pp\piz\jpsi$, if the
$\piz$ background can be handled properly. The BESIII experiment accumulated
more data close to the $Y(4260)$ peak, which can be used to measure all
the final states since the background level is very low as has
been shown in the $\ppjpsi$ and $\omega\jpsi$ cases~\cite{Ablikim:2013dyn,Ablikim:2019zio}.
Since the production cross section of $\EE\to \gamma X(3872)$ is
at a few pb level~\cite{Ablikim:2013dyn,Ablikim:2019zio},
the $X(3872)$ sample has about $10^4$ events, which
allows measurements of final states with branching fractions
at the percent level only.

A very interesting proposal for a precise $X(3872)$ mass measurement was put forward
recently~\cite{Guo:2019qcn}. It exploits the interplay of a triangle singularity
and the pole of the $X(3872)$ on the resulting line shapes in this way enhancing
tremendously the sensitivity to the mass. For this method to work it is
necessary that the source of the reaction generates a large number of low-energy $D^*\bar D^*$
pairs which appears possible, e.g., at LHCb.

\vspace{0.3cm}\noindent
{\it (4) Search for the $X_b$ state}
\vspace{0.3cm}

Many theoretical works have been carried out in order to
understand the nature of $X(3872)$. It is also natural
to search for a similar state with $J^{PC} = 1^{++}$ (called $X_b$ hereafter)
in the bottomonium system~\cite{Ebert:2005nc,Hou:2006it}.
The search for $X_b$ supplies important information about the discrimination
of a compact multiquark configuration and a loosely bound hadronic molecule
configuration for the $X(3872)$.
The existence of the $X_b$ is predicted in both the
compact tetraquark model~\cite{Ali:2009pi} and those involving a molecular
interpretation~\cite{Guo:2013sya, Karliner:2013dqa}, although employing
heavy quark flavor symmetry within a common hadronic effective field theory
for states containing $b\bar b$ and $c\bar c$ appears not to be possible~\cite{Baru:2018qkb}.

The production of $X_b$ at LHC and Tevatron~\cite{Guo:2014sca}
has been extensively investigated. Since the mass of $X_b$ may be very large,
a search at LHC should be promising.
The $X_b$ state can also be searched for
at $B$-factories with radiative decays $\Upsilon(5S, 6S) \to \gamma X_b$, and
the production rates are predicted at the orders of $10^{-5}$ under the
assumption that the $X_b$ is a $B\bar{B}^*$ molecular state~\cite{Wu:2016dws}.
As for the $X_b$ decay modes, partial widths of the radiative decays of
$X_b \to \gamma \Upsilon(nS)$ ($n=1,~2,~3$) with $X_b$ being a candidate
for the $B\bar{B}^*$ molecular state
are found at about 1 keV level~\cite{Li:2014uia}, and the partial width of the
hadronic decay of $X_b \to \omega\Upsilon(1S)$ is about tens of keVs~\cite{Li:2015uwa}.

\begin{figure*}[htbp]
\begin{center}
\includegraphics[height=5cm]{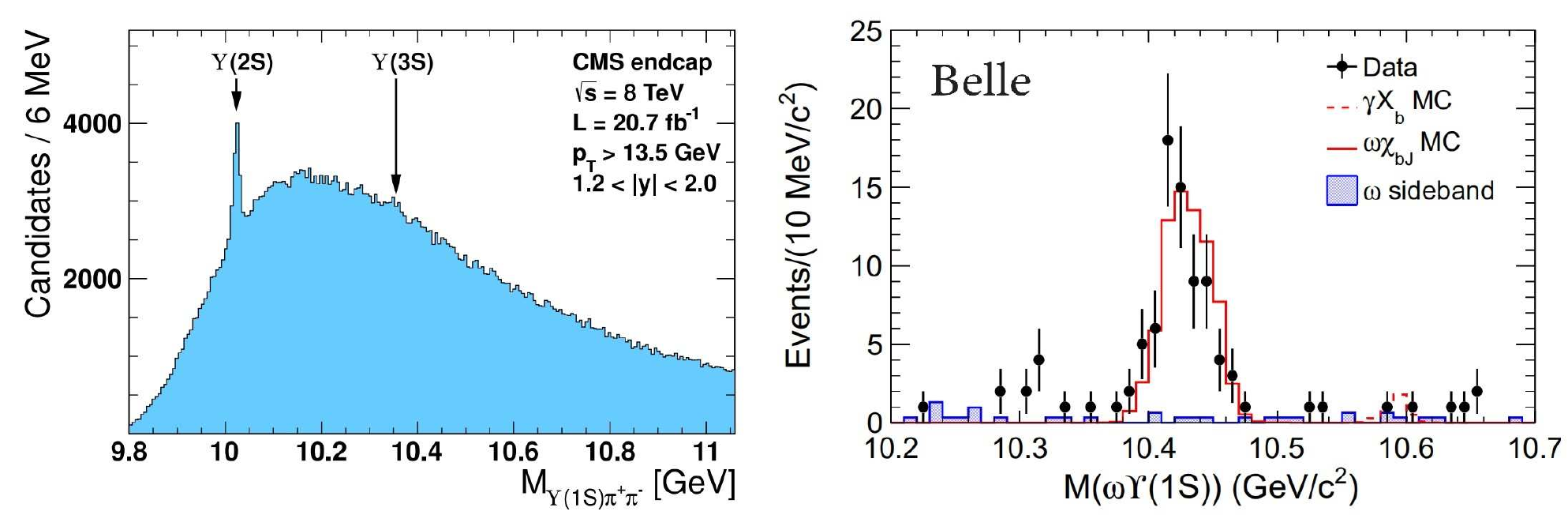}
\end{center}
\caption{Results on the search of the $X_b$ state from CMS in the $M[\pi^+ \pi^- \Upsilon(1S)]$ distribution~\cite{Chatrchyan:2013mea}
and Belle in the $M[\omega\Upsilon(1S)]$ distribution~\cite{He:2014sqj}.}\label{cms-belle-xb}
\end{figure*}

The ATLAS and CMS Collaborations searched for the $X_b$ decaying to
$\pi^+ \pi^-\Upsilon(1S)$ based on a sample of $pp$
collisions at $\sqrt{s} = 8$ TeV, corresponding to an integrated luminosity
of 16.2 fb$^{-1}$ and 20.7 fb$^{-1}$, respectively~\cite{Aad:2014ama,Chatrchyan:2013mea}. Figure~\ref{cms-belle-xb} (left plot) shows the
$\pi^+ \pi^-\Upsilon(1S)$ invariant mass distribution in CMS data~\cite{Chatrchyan:2013mea}.
Except the clear $\Upsilon(2S)$ signal, no evidence for an $X_b$ signal was
observed. However, unlike the $X(3872)$, whose decays
exhibit large isospin violation,
the $X_b$ should decay preferably into $\pi^+\pi^-\pi^0 \Upsilon(1S)$
rather than $\pi^+\pi^-\Upsilon(1S)$
if it exists~\cite{Guo:2013sya, Guo:2014sca,Karliner:2013dqa}. So Belle
performed a search for
an $X_b$ signal decaying to $\omega\Upsilon(1S)$
in $e^+e^- \to \gamma X_b$ at a c.m.\ energy of 10.867 GeV using a
118~fb$^{-1}$ data sample~\cite{He:2014sqj}.
Figure~\ref{cms-belle-xb} (right plot) shows the $\omega\Upsilon(1S)$
invariant mass distribution in a range from 10.55 to 10.65 GeV.
The dots with error bars are from data, the solid histogram is from the
normalized contribution of $e^+e^- \to \omega \chi_{bJ}$ ($J=0,~1,~2$).
No obvious $X_b$ signal is observed, and 90\% C.L.\ upper limits on the
product branching fraction
${\cal B}[\Upsilon(5S)\to\gamma X_b]{\cal B}[X_b\to\omega\Upsilon(1S)]$
vary smoothly from $2.6\times10^{-5}$ to $3.8\times10^{-5}$ between
$10.55$ and $10.65$ GeV.
We also note that the absence of the $X_b$ state in ATLAS, CMS, and Belle experiments
may be understood in a few theoretical models. For example, in Ref.~\cite{Zhou:2018hlv}
the $X_b$ is regarded as a virtual state about 10 MeV below the $BB^*$ threshold,
which means that it has no significant observable effects.

\vspace{0.3cm}\noindent
$\bullet$ {\it The $X(3915)$ and $X(3860)$}
\vspace{0.3cm}

\begin{figure}[htbp]
\centering
\includegraphics[height=4.55cm]{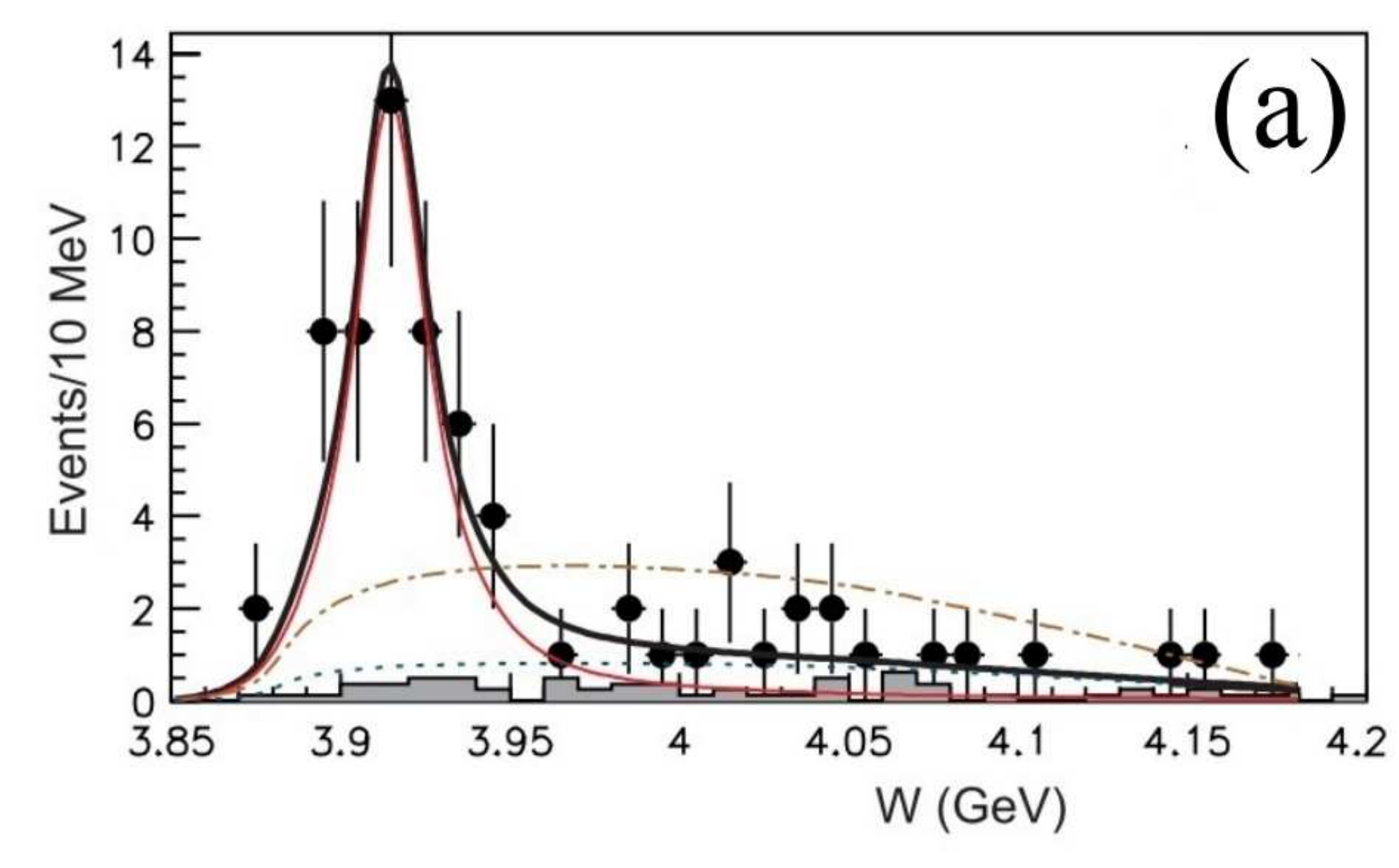}
\includegraphics[height=4.6cm]{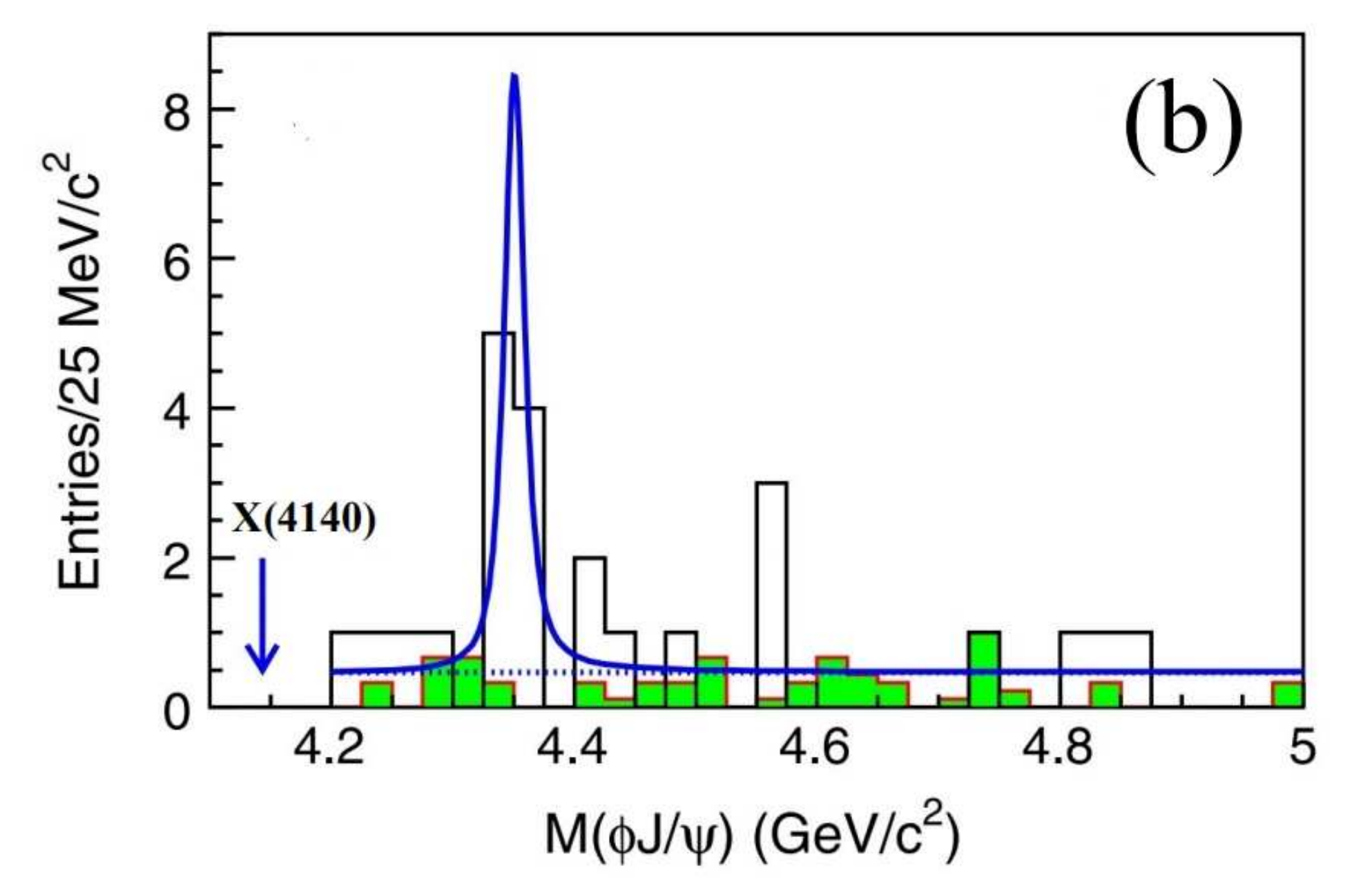}
\caption{Distributions of $\omega \jpsi$ and $\phi \jpsi$
invariant mass spectra from the selected (a) $\gamma \gamma \to \omega \jpsi$~\cite{Uehara:2009tx}
and (b) $\phi \jpsi$~\cite{Shen:2009vs} candidates at Belle.
The dots with error bars in (a)
and the open histogram in (b)
are from experimental data. The shaded histograms are from non-$\omega \jpsi$
and non-$\phi \jpsi$
backgrounds. The solid curves are the best fits with an (a) $X(3915)$ and (b)
$X(4350)$ signal component.
The dot-dashed curve in (a) is the fit without a resonance.
\label{belle-twophoton}}
\end{figure}

The Belle experiment studied the process $\gamma \gamma \to \omega \jpsi$ using an
integrated luminosity of 694~fb$^{-1}$~\cite{Uehara:2009tx}.
The obtained $\omega \jpsi$ mass spectrum is shown in Fig.~\ref{belle-twophoton}(a),
where a prominent resonance-like peak around 3.92~GeV is observed.
Using an $S$-wave BW function with a variable width for the resonant component,
the obtained resonance parameters are:
$M=(3915\pm3\pm2)$~MeV and $\Gamma=(17\pm10\pm3)$~MeV. This structure is
called $X(3915)$.
Subsequently, the $X(3915)$ was confirmed by BaBar using a data sample of
519.2~fb$^{-1}$
in the same process~\cite{Lees:2012xs}. Besides the confirmation of the
existence of $X(3915)$,
BaBar also did a spin-parity analysis, which supports the assignment
$J^{PC}=0^{++}$.
Therefore, BaBar identified the $X(3915)$ as the $\chi_{c0}(2P)$ resonance.

Actually, the $X(3915)$ was firstly observed by Belle
in $B \to J/\psi \omega K$ decays~\cite{Abe:2004zs} with an original name
of $Y(3940)$,
and then was confirmed by BaBar~\cite{Aubert:2007vj}.
However, in 2010 BaBar restudied the decays $B^{0,+} \to \jpsi \pi^+ \pi^- \pi^0 K^{0,+}$
using $467\times 10^6$ $B\bar{B}$ pairs~\cite{delAmoSanchez:2010jr}.
Considering the phase space limitation, with a less restrictive
$\pi^+ \pi^- \pi^0$ mass region requirement
a detailed study on the $\omega \jpsi$ mass spectrum revealed that the
original $Y(3940)$ was made up of two structures,
the $X(3915)$ and $X(3872)$. As mentioned above,
due to the assignment of $J^{PC}=0^{++}$ for the $X(3915)$,
it was identified as the $\chi_{c0}(2P)$ in the 2014 PDG tables.
However, this assignment has some problems~\cite{Guo:2012tv,Olsen:2014maa}.
Moreover, in Ref.~\cite{Zhou:2015uva} it was shown that the data for the
$X(3915)$ could even be consistent with the $\chi_{c2}(3930)$,
if the helicity-1-dominance constraint is relaxed in the analysis.
As a result of these considerations, the $X(3915)$ is no longer identified
as the $\chi_{c0}(2P)$ in the more recent PDG tables.

Some theoretical models explained the $X(3915)$ as a $D_s\bar{D}_s$ molecule,
a compact tetraquark state, a $c\bar{c}$-gluon hybrid, etc.
If $X(3915)$ were a compact $cs\bar{c}\bar{s}$ tetraquark, although the decay
$X(3915) \to \omega J/\psi$ is allowed,
the decay rate should be much lower than the mode of $\eta \eta_c$
considering the $\eta$ meson has a relative large
$s\bar{s}$ component. Experimental search for $X(3915) \to \eta \eta_c$ was
done by Belle in $B^{\pm}\to K^{\pm} \eta \eta_c$,
but a negative result was reported~\cite{Vinokurova:2015txd}.
This disfavors the interpretation of the $X(3915)$ as a compact $cs\bar{c}\bar{s}$ tetraquark.
If $X(3915)$ were a $c\bar{c}$-gluon hybrid, for a $0^{++}$ state,
the $X(3915)$ mass would however not be a good match for a light hybrid~\cite{Liu:2012ze,Cheung:2016bym}.
Therefore none of the above explanations is satisfactory and the nature of the $X(3915)$ is still unknown.

\begin{figure*}[htbp]
\begin{center}
\includegraphics[height=5.5cm]{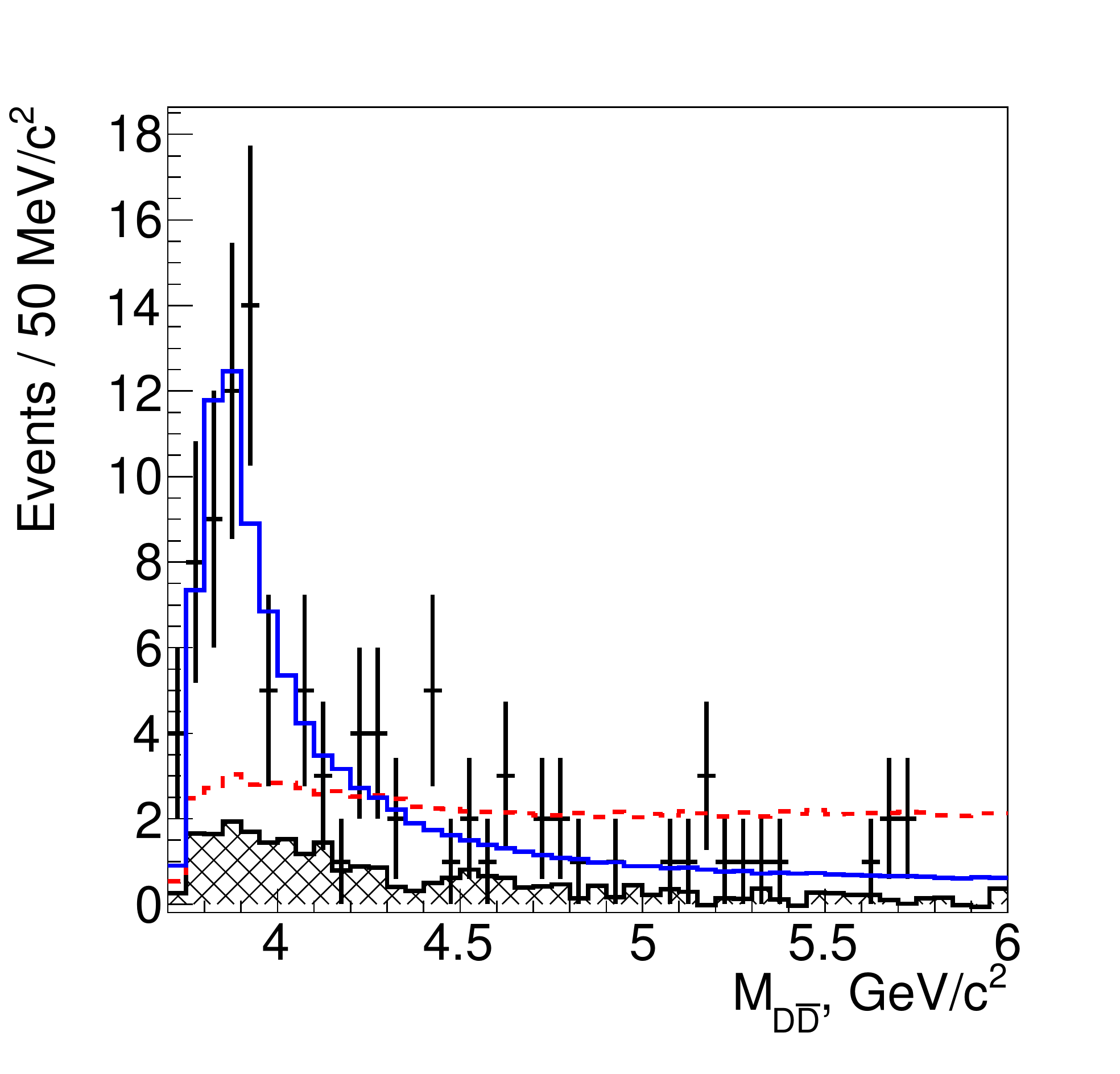}
\end{center}
\caption{Projection of the signal fit results onto $M(D\bar{D})$ in
the analysis of $e^+ e^- \to \jpsi D \bar{D}$ by Belle~\cite{Chilikin:2017evr}.
The points with error bars are the data, the hatched histogram is the
background, the blue solid line is the fit with a new $X^{*}$ resonance with
$J^{PC}=0^{++}$, and the red dashed line is the fit with a nonresonant
amplitude only.}\label{belle-x3860}
\end{figure*}

A promising process that is suitable for a search for the $\chi_{c0}(2P)$ and
other charmonium states with positive $C$-parity is double-charmonium
production in association with a $J/\psi$. The $X(3940)$ state was observed
by Belle in the inclusive $e^+ e^- \to J/\psi X$ spectrum and in the
process $e^+ e^- \to J/\psi D^* \bar{D}$~\cite{Abe:2007jna,Abe:2007sya},
and the $X(4160)$ was observed in the process
$e^+ e^- \to J/\psi D^* \bar{D}^*$~\cite{Abe:2007sya}.
Recently, Belle performed a full amplitude analysis of the process
$\jpsi D \bar{D}$ ($D$=$D^0$ or $D^+$)
based on a 980 fb$^{-1}$ data sample~\cite{Chilikin:2017evr}.
A new charmonium-like state $X({3860})$ that decays to $D \bar{D}$ is
observed with a significance of $6.5\sigma$.
Its mass is $(3862^{+26+40}_{-32-13})$~MeV and width is $(201^{+154+88}_{-67-82})$~MeV.
The $J^{PC}=0^{++}$ hypothesis is favored over the $2^{++}$ hypothesis at the
level of $2.5\sigma$ and the new state is now called $\chi_{c0}(3860)$ by
the PDG. Its mass is close to the
potential model expectations for the $\chi_{c0}(2P)$, so
it is a better candidate for the $\chi_{c0}(2P)$ charmonium state than the
$X(3915)$.
Figure~\ref{belle-x3860} shows the projection of the signal fit results
onto the $M(D \bar{D})$ distribution. The points with error bars are the data,
the hatched histogram is the background, the blue solid line is the fit with
the new resonance with $J^{PC}=0^{++}$, and the red dashed line is the fit
with a nonresonant amplitude only.

The resonant parameters of the $X(3860)$ are very close to the
phenomenological analyses~\cite{chao:2007it,Guo:2012tv}
of the Belle~\cite{Uehara:2005qd} and BaBar~\cite{Aubert:2010ab} data
on $\gamma\gamma\to D\bar{D}$ where the $\chi_{c2}(3930)$
was observed. Further experimental investigation is needed to
check the angular distribution of the two-photon process and
to confirm the resonant nature of the events. The search for
$\chi_{c0}(3860)$ and $\chi_{c2}(3930)$ in the radiative transitions
of the excited $\psi$ states is also very helpful to understand whether
they are charmonium $2P$ states.

\vspace{0.3cm}\noindent
$\bullet$ {\it The $\chi_{c1}(4140)$ aka $X(4140)$ and other related states}
\vspace{0.3cm}

In 2008 the CDF experiment claimed a $3.8\sigma$ evidence for a near-threshold
$X(4140)\to J/\psi \phi$ in $B^+ \to J/\psi \phi K^+$
decays using a data sample corresponding to an integrated luminosity
of 2.7 fb$^{-1}$~\cite{Aaltonen:2009tz}.
The mass and width of this structure are measured to be $(4143.0\pm2.9\pm1.2)$~MeV
and $(11.7^{+8.3}_{-5.0}\pm 3.7)$ MeV, respectively.
Much larger widths are expected for charmonium states in this mass range
because of open-flavor decay channels, which makes the observation of
the $X(4140)$ very interesting.
It has been suggested that the $X(4140)$ structure
could be a molecular state, a $\chi_{c1}(3P)$ state,
a compact tetraquark state, a hybrid state, a rescattering effect and so on.

Searches for the $X(4140)$ did not confirm
its presence in the analyses of the same $B$ decays performed by the
Belle~\cite{ChengPing:2009vu} (unpublished) and BaBar~\cite{Lees:2014lra}
experiments.
With a much larger data sample, LHCb did not find evidence for the narrow
$X(4140)$ peak in the same $B$ decays either~\cite{Aaij:2012pz}.
Later CDF updated the analysis of $B^+ \to J/\psi \phi K^+$ using a larger data
sample of 6.0 fb$^{-1}$~\cite{Aaltonen:2011at}.
This time, besides the confirmation of the $X(4140)$
with a mass of $(4143.4^{+2.9}_{-3.0}\pm 0.6)$~MeV and a width of
$(15.3^{+10.4}_{-6.1}\pm 2.5)$~MeV, evidence for another structure
with a mass of $(4274.4^{+8.4}_{-6.7}\pm1.9)$ MeV and a width of
$(32.3^{+21.9}_{-15.3}\pm 7.6)$ MeV is reported. The significance of this
structure is estimated to be approximately 3.1$\sigma$.
The $X(4140)$ was seen by CMS and D0 in both
$B^+ \to J/\psi \phi K^+$ decays with a greater than $5.0\sigma$ and $3.1\sigma$
significance, respectively~\cite{Chatrchyan:2013dma,Abazov:2013xda}.
The second peak observed in the updated analysis of $B^+ \to J/\psi \phi K^+$ by
CDF~\cite{Aaltonen:2011at} was also seen by CMS~\cite{Chatrchyan:2013dma},
but the measured mass was higher by $3.2\sigma$.

Looking for more production modes will undoubtedly help us to understand
the $X(4140)$.
The first evidence for the prompt production of $X(4140) \to \phi J/\psi$ was
presented by D0 based on 10.4~fb$^{-1}$ of $p \bar{p}$ collision data~\cite{Abazov:2015sxa}.
The measured mass and width are $(4152.5\pm1.7^{+6.2}_{-5.4})$~MeV
and $(16.3\pm5.6\pm11.4)$ MeV, respectively.
BESIII searched for the $X(4140)$ via
$e^+e^- \to \gamma \phi J/\psi$ at $\sqrt{s}=$4.23, 4.26, 4.36, and 4.60~GeV,
but no significant $X(4140)$
signal was observed in any of these data
samples~\cite{Ablikim:2014atq,Ablikim:2017cbv}. The upper limits of the
product of the cross section and branching fraction $\sigma[e^+e^-
\to \gamma X(4140)]B[X(4140)\to \phi\jpsi]$ are determined
to be 0.35, 0.28, 0.33, and 1.2~pb at $\sqrt{s} =$4.23, 4.26,
4.36, and 4.60~GeV, respectively, at the 90\% C.L.

To search for the $X(4140)$, Belle did a two-photon
analysis of $\gamma \gamma \to \phi J/\psi$~\cite{Shen:2009vs}. The
$\phi \jpsi$ mass spectrum is shown in Fig.~\ref{belle-twophoton}(b), where
the open histogram shows the experimental data,
and the shaded histogram is from normalized $\phi$ and $J/\psi$ mass sidebands.
The arrow shows the expected position of the $X(4140)$.
Instead of observation of the $X(4140)$, Belle found a
$3.2\sigma$ evidence for a
narrow $\phi J/\psi$ peak at $(4350.6\,^{+4.6}_{-5.1}\pm0.7)$ MeV with a
width of $(13^{+18}_{-9}\pm4)$ MeV~\cite{Shen:2009vs}. This structure is
called $X(4350)$. The fit results are shown in Fig.~\ref{belle-twophoton}(b)
with the solid curve for the best fit and
the dashed line for the backgrounds. It should be noted that the production
of the $X(4140)$ in two-photon fusion is forbidden by
the Landau--Young theorem if its spin is one as described below.

\begin{figure}[htbp]
\centering
\includegraphics[height=5cm]{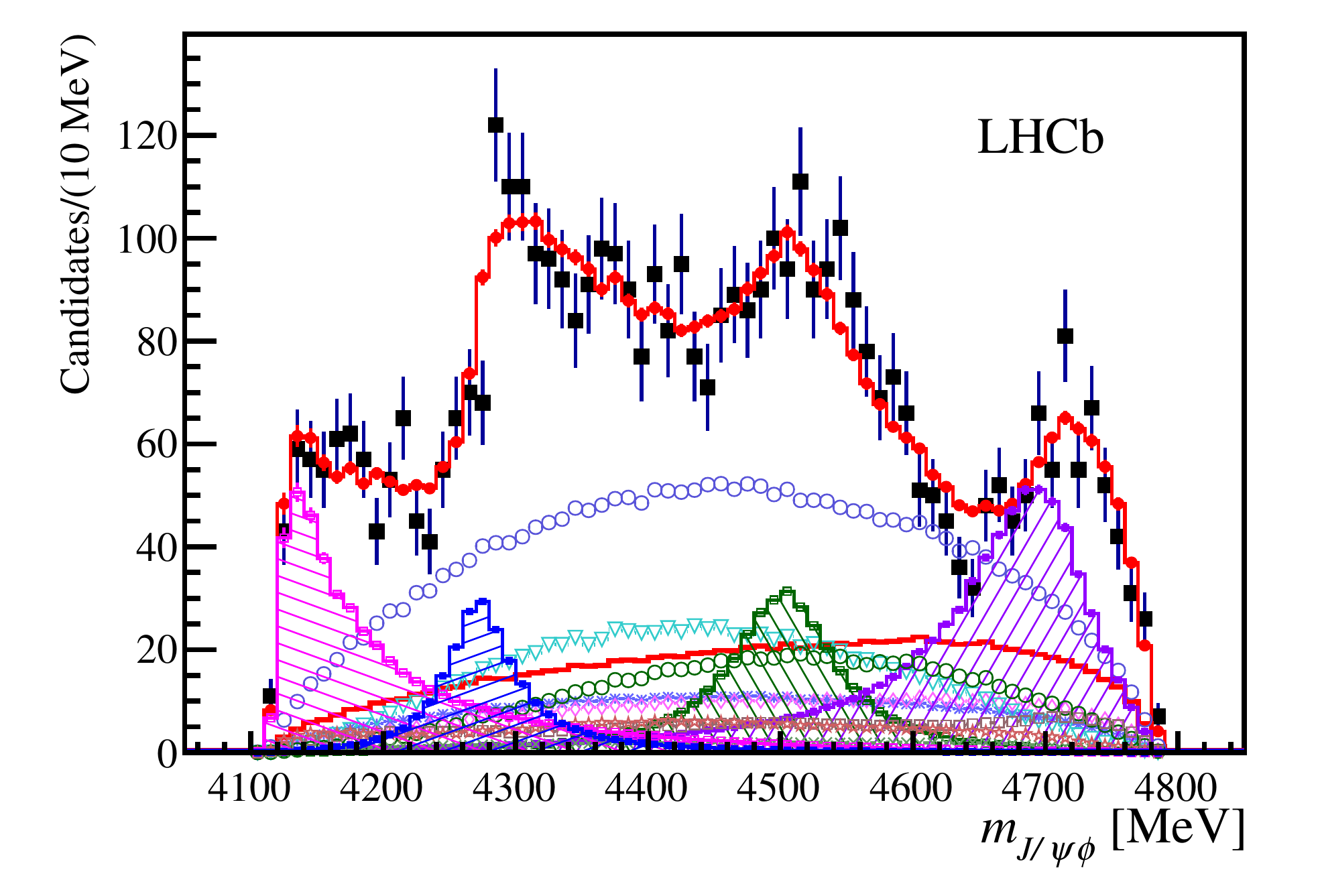}
\caption{Distributions of $\phi\jpsi$ invariant mass for the
$B^+\to J/\psi \phi K^+$ candidates (black data points) compared
with the fit containing eight $K^{*+}\to \phi K^+$ and five $X\to
J/\psi\phi$ contributions~\cite{Aaij:2016iza}. The total fit is given by the red
points with error bars. Individual fit components are also shown.
 \label{y4140-fit}}
\end{figure}

Considering the complicated structures and confusing experimental
situation concerning the $\phi J/\psi$ mass spectrum, LHCb did a full amplitude
analysis of the selected $4289\pm 151$ $B^+\to K^+ \phi J/\psi$
events using $3$~fb$^{-1}$ data collected at c.m.\ energies 7 and
8~TeV~\cite{Aaij:2016iza}.
This analysis offers the best sensitivity to study
the resonant structures in the $\phi\jpsi$ system.
The data requires not only two $J^{PC}=1^{++}$ states, the $\chi_{c1}(4140)$ aka
$X(4140)$ and $\chi_{c1}(4274)$ aka $X(4274)$ observed by CDF and CMS before, but also two new broad
$J^{PC}=0^{++}$ states, $X(4500)$ and $X(4700)$. Figure~\ref{y4140-fit} shows the
$\phi J/\psi$ invariant mass distribution for the selected signal candidates,
where four $\phi J/\psi$ structures,
$X(4140)$, $X(4274)$, $X(4500)$, and $X(4700)$,
are needed~\cite{Aaij:2016iza}. These resonance parameters including
signal significance, $J^{PC}$ values, and the measured mass and width
are listed in Table~\ref{y4140}.

\begin{figure*}[htbp]
\begin{center}
\includegraphics[height=5cm]{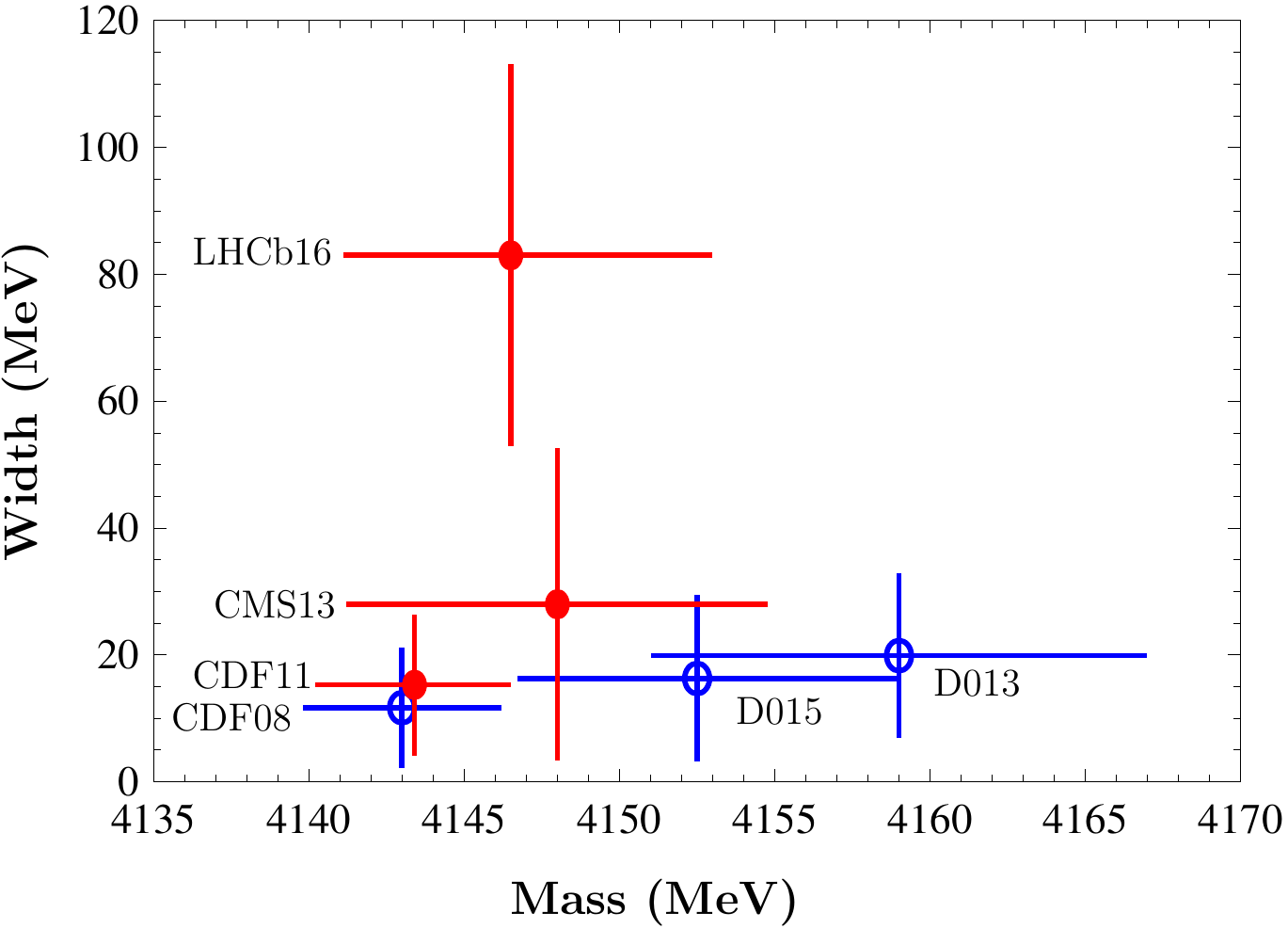}
\end{center}
\caption{A comparison of the resonant parameters of $\chi_{c1}(4140)$ aka
$X(4140)$ reported from different
measurements~\cite{Aaltonen:2009tz,Aaltonen:2011at,Chatrchyan:2013dma,Abazov:2013xda,Abazov:2015sxa,Aaij:2016iza,Aaij:2016nsc}.
The filled (open) circles indicate the significance of the $X(4140)$ signals
greater (less) than $5\sigma$.}\label{y4140-expcom}
\end{figure*}

A comparison of the resonant parameters of $X(4140)$
reported from the different
measurements~\cite{Aaltonen:2009tz,Aaltonen:2011at,Chatrchyan:2013dma,Abazov:2013xda,Abazov:2015sxa,Aaij:2016iza,Aaij:2016nsc}
is shown in Fig.~\ref{y4140-expcom}, where the filled (open)
circles indicate the significance of the $X(4140)$
signals greater (less) than $5 \sigma$.
Note that Ref.~\cite{Wang:2017mrt} finds that the LHCb data are in fact compatible with a narrow
$X(4140)$ as long as a broad $X(4160)$ state,
a candidate for a $D_s^*\bar D_s^*$ molecular state, as well as the cusp for the
$D_s^*\bar D_s^*$ threshold are included in the analysis. Confirmation,
especially of the $X(4500)$ and $X(4700)$,
from other experiments and further experimental investigation of them are needed.
In the near future, the Belle II experiment can reanalyze the
$B^+\to K^+ \phi J/\psi$ decays.

\begin{table}[htbp]
 \centering
 \begin{tabular}{ccccc}
 \hline\hline
 States & Significance & $J^{PC}$ & Mass (MeV) & Width (MeV) \\
 \hline
 $X(4140)$ & $8.4\sigma$ & $1^{++}$ & $4146.5\pm 4.5^{+4.6}_{-2.8}$
 & $83\pm 21^{+21}_{-14}$ \\
 $X(4274)$ & $6.0\sigma$ & $1^{++}$ & $4273.3\pm 8.3^{+17.2}_{-3.6}$
 & $56\pm 11^{+8}_{-11}$ \\
 $X(4500)$ & $6.1\sigma$ & $0^{++}$ & $4506\pm 11^{+12}_{-15}$
 & $92\pm 21^{+21}_{-20}$ \\
 $X(4700)$ & $5.6\sigma$ & $0^{++}$ & $4704\pm 10^{+14}_{-24}$
 & $120\pm 31^{+42}_{-33}$ \\
\hline\hline
\end{tabular}
\caption{\label{y4140} Results for significances, masses, and widths of the $\phi\jpsi$ components in $B^+ \to
J/\psi \phi K^+$ from the LHCb experiment~\cite{Aaij:2016iza}.}
\end{table}

\subsubsection{The $Y$ states}
\label{Sect:3.1.2}
Among the charmonium-like states, there are many vectors with
quantum numbers $J^{PC} = 1^{--}$ that are usually called $Y$
states, and according to the new PDG naming scheme (see Sec.~\ref{sec:namingscheme})
should be called $\psi($mass$)$, like the $\psi(4260)$ aka $Y(4260)$~\cite{Aubert:2005rm},
$\psi(4360)$ aka $Y(4360)$~\cite{Aubert:2007zz,Wang:2007ea}, and
$\psi(4660)$ aka $Y(4660)$~\cite{Wang:2007ea}.
The $Y$ states show strong coupling to hidden-charm final states in contrast to the vector charmonium
states in the same energy region, $\psi(4040)$ and $\psi(4160)$,
 which couple dominantly to the ground state open-charm meson
pairs~\cite{Tanabashi:2018oca}. These $Y$ states are good candidates for new
types of exotic particles and stimulated many theoretical
interpretations, including compact tetraquarks, molecules, hybrids, or
hadrocharmonia~(see, e.g., Ref.~\cite{Brambilla:2010cs} and references therein).

The first $Y$ state, the $Y(4260)$, was observed by BaBar
via the ISR process $\EE\to \pi^+\pi^- J/\psi$ using data samples of
211 fb$^{-1}$ at $\sqrt{s}=10.58$ GeV and 22 fb$^{-1}$ at
10.54 GeV~\cite{Aubert:2005rm}.
After that, a few $Y$ states were discovered at $B$ factories using
ISR technology. The idea of utilizing ISR from a high-mass
state to explore electron-positron processes at all energies below
that state was outlined in Ref.~\cite{Baier:1969kaa}.
With ISR events the whole hadron spectrum is visible so that the
line shape of the resonance and
fine structures can be investigated, while
the effective luminosity and detection efficiency are relatively low.
To study the $Y$ states, a method complementary to the ISR
is to take data via direct $\EE$ annihilation in the charmonium energy region.
In particular, BESIII has collected large data samples above 4~GeV in recent years~\cite{Ablikim:2009aa},
which allow accurate determinations of the cross sections for some final states.
The drawback is that since the data are taken at fixed energy points, one might miss
narrow structures.

\vspace{0.3cm}\noindent
$\bullet$ {\it The $\psi(4230) \ \mbox{aka} \ Y(4230)$ state}~\label{Sec:ppjpsi}
\vspace{0.3cm}

The process $\EE\to \pi^+\pi^- J/\psi$ via ISR at c.m.\ energies up to
5.0~GeV was first studied by BaBar, where an unexpected structure at
about 4.26 GeV was observed clearly~\cite{Aubert:2005rm}.
It is the first observed $Y$ state, which is referred to as
$Y(4260)$.
Subsequently, although the $Y(4260)$ was confirmed
by Belle in the same process,
Belle found that $Y(4260)$ alone cannot describe
the line shape satisfactorily~\cite{Yuan:2007sj}.
Besides the $Y(4260)$,
Belle also observed a broad excess near 4~GeV, called $Y(4008)$~\cite{Yuan:2007sj}.
Improved measurements with both BaBar~\cite{Lees:2012cn} and
Belle~\cite{Liu:2013dau} full data samples confirmed the existence of a
component in addition to $Y(4260)$ in $\EE\to \pi^+\pi^- J/\psi$ but the
line shape was parametrized with different models.

\begin{figure*}
\begin{center}
\includegraphics[height=5cm]{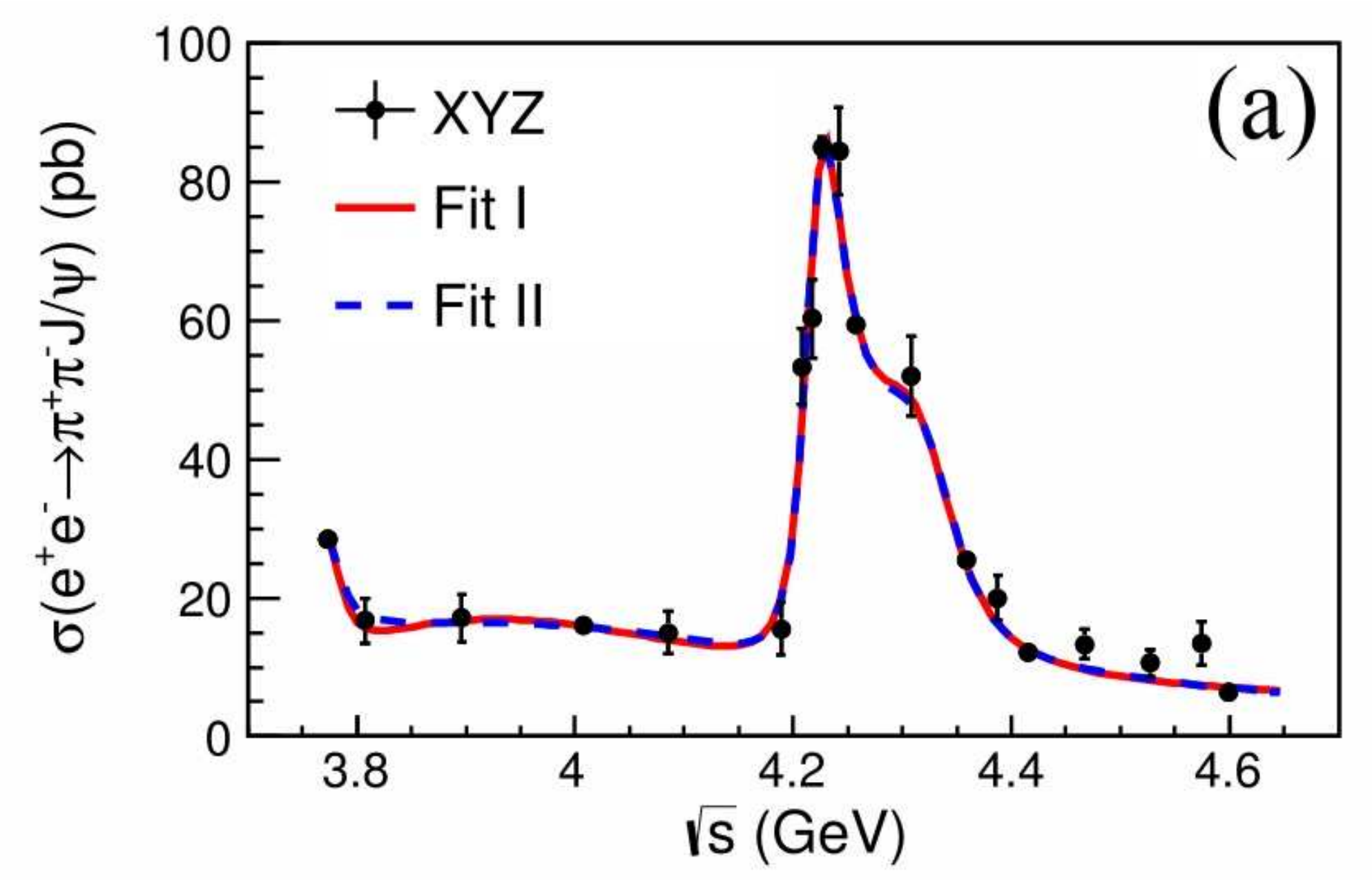}
\includegraphics[height=4.8cm]{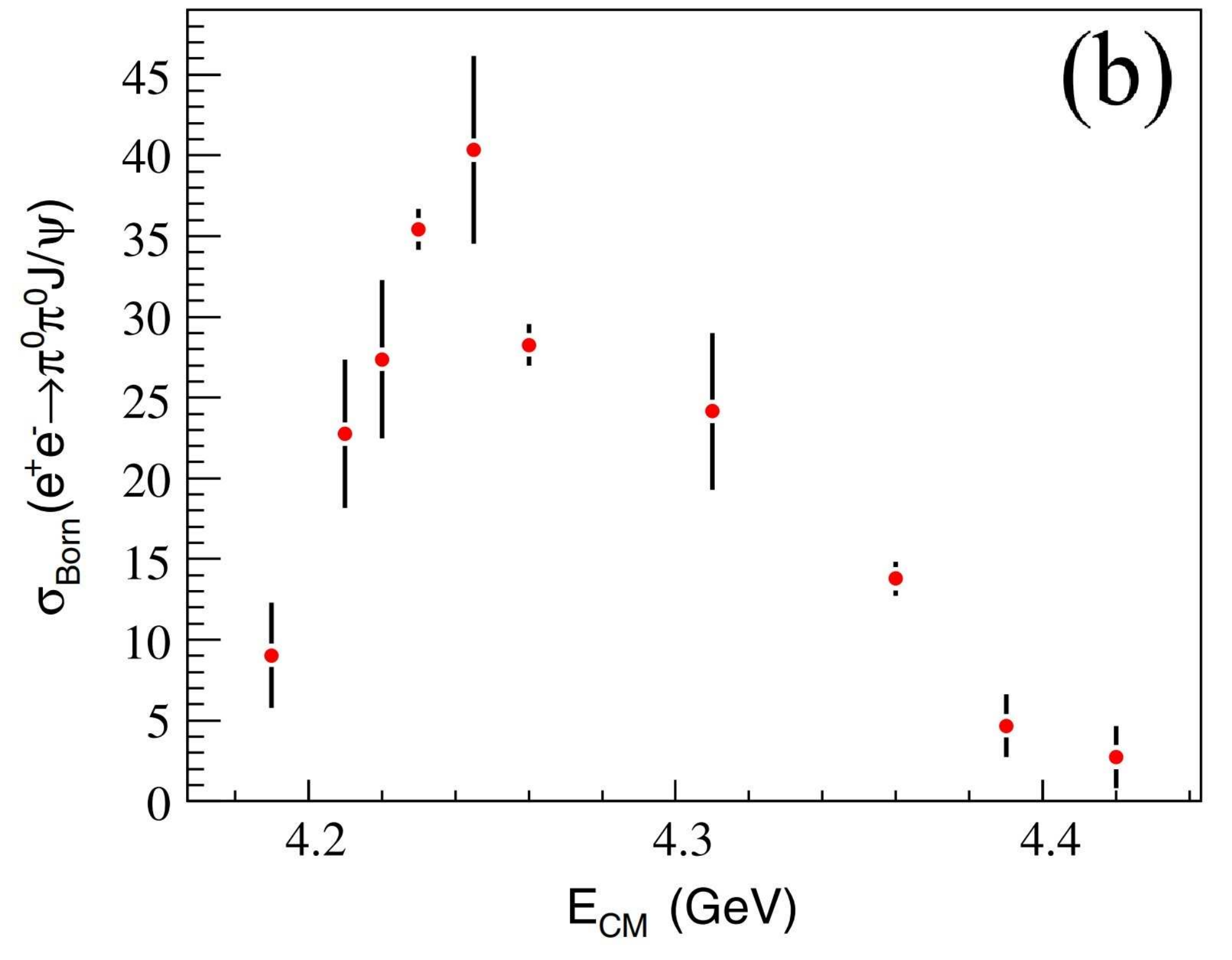}
\caption{Measured cross sections of (a) $\EE\to \ppjpsi$~\cite{Ablikim:2016qzw} and (b)
$e^+e^- \to \pi^0 \pi^0 J/\psi$~\cite{Ablikim:2014dxl}. In (a), the solid and dashed lines are
the fits with the coherent sum of three BW functions (Fit I, red
solid curve) and the coherent sum of an exponential continuum and
two BW functions (Fit II, blue dashed curve).} \label{xsec-fit}
\end{center}
\end{figure*}

BESIII reported a precise measurement of $e^+e^- \to \pi^+ \pi^- J/\psi$ cross
sections from 3.77 to 4.60~GeV using data samples with an integrated
luminosity of 9~fb$^{-1}$~\cite{Ablikim:2016qzw}. While the nature of the
events at around 4~GeV is still ambiguous, the dominant resonant
structure, the so called $Y(4260)$, was found to have a mass of
$(4222.0\pm 3.1\pm 1.4)$~MeV and a width of $(44.1\pm 4.3\pm
2.0)$~MeV. In addition, a new resonance with a
mass of around 4.32~GeV is needed to describe the high
precision data. Its mass and width are $(4320.0\pm 10.4 \pm
7.0)$~MeV and $(101.4^{+25.3}_{-19.7}\pm 10.2)$~MeV with a
statistical significance larger than $7.6\sigma$.
This resonance is called $Y(4320)$.
However, it should be mentioned that the analysis mentioned above
was performed using BW functions that generate
symmetric line shapes --- however, already the analysis of the older
data based on the molecular picture, which naturally generates asymmetric
line shapes even from a single state only,
found a mass for the $Y(4260)$ close to
4230~MeV~\cite{Cleven:2013mka}
that is also qualitatively consistent with the new data as discussed
in Sec.~\ref{Sect:4.1.5} (see Fig.~\ref{fig:Ydecays_oldandnew}).
Figure~\ref{xsec-fit}(a) shows the measured $\EE\to \pi^+\pi^- J/\psi$
cross sections, where one can see clearly the $Y(4260)$
structure observed by BaBar and Belle
experiments, but its peaking position is at around 4.22~GeV rather than
4.26~GeV from the previous fits~\cite{Lees:2012cn,Liu:2013dau}.
Since both mass and width of $Y(4320)$ are consistent with those of the
 $Y(4360)$ resonance
observed in $\EE\to \pp\psp$ by BaBar and Belle~\cite{Wang:2014hta,Lees:2012pv}, they could be the same
state, which needs to be confirmed with precise measurements of these
two resonant parameters in the future.
It is worth pointing out that the lower mass structure [called
$\psi(4230) \ \mbox{aka} \ Y(4230)$ hereafter] is the main component of the $Y(4260)$
structure with an improved measurement of the resonant parameters.
But we also note that an accurate measurement of the cross section at
4.32~GeV is very important
since its size will determine how to describe the line shape of $e^+e^- \to \pi^+ \pi^- J/\psi$ cross
sections, in particular if there is one or two resonances.

Besides the $\pi^+ \pi^- J/\psi$ final states, the neutral
process $e^+e^- \to \pi^0 \pi^0 J/\psi$ was measured by BESIII from 4.19 to
4.42~GeV
corresponding to an integrated luminosity of 2.8~fb$^{-1}$~\cite{Ablikim:2014dxl}.
The measured cross sections are shown in Fig.~\ref{xsec-fit}(b),
where the $Y(4260)$ signals are clear.
Although the covered energy range is narrower than in the charged mode,
the measured line shape is in good agreement with that in Fig.~\ref{xsec-fit}(a).
In the future, BESIII is able to improve the measurements by using more data
points.

From the BESIII data, one can see that besides the
$Y(4230)$ peak,
the $\pp\jpsi$ cross section is at the 10--15~pb level. Whether the
cross section is due to the pure continuum process or from the decays
of other charmonium or charmonium-like states is still not clear.
Once more data samples are available at BESIII or Belle II in the future, one
can try to fit the cross sections by adding coherent $\psi(4040)$ and
$\psi(4160)$
resonances to check their couplings to the $\pp\jpsi$ final state.

\begin{figure*}
\begin{center}
\includegraphics[height=5cm]{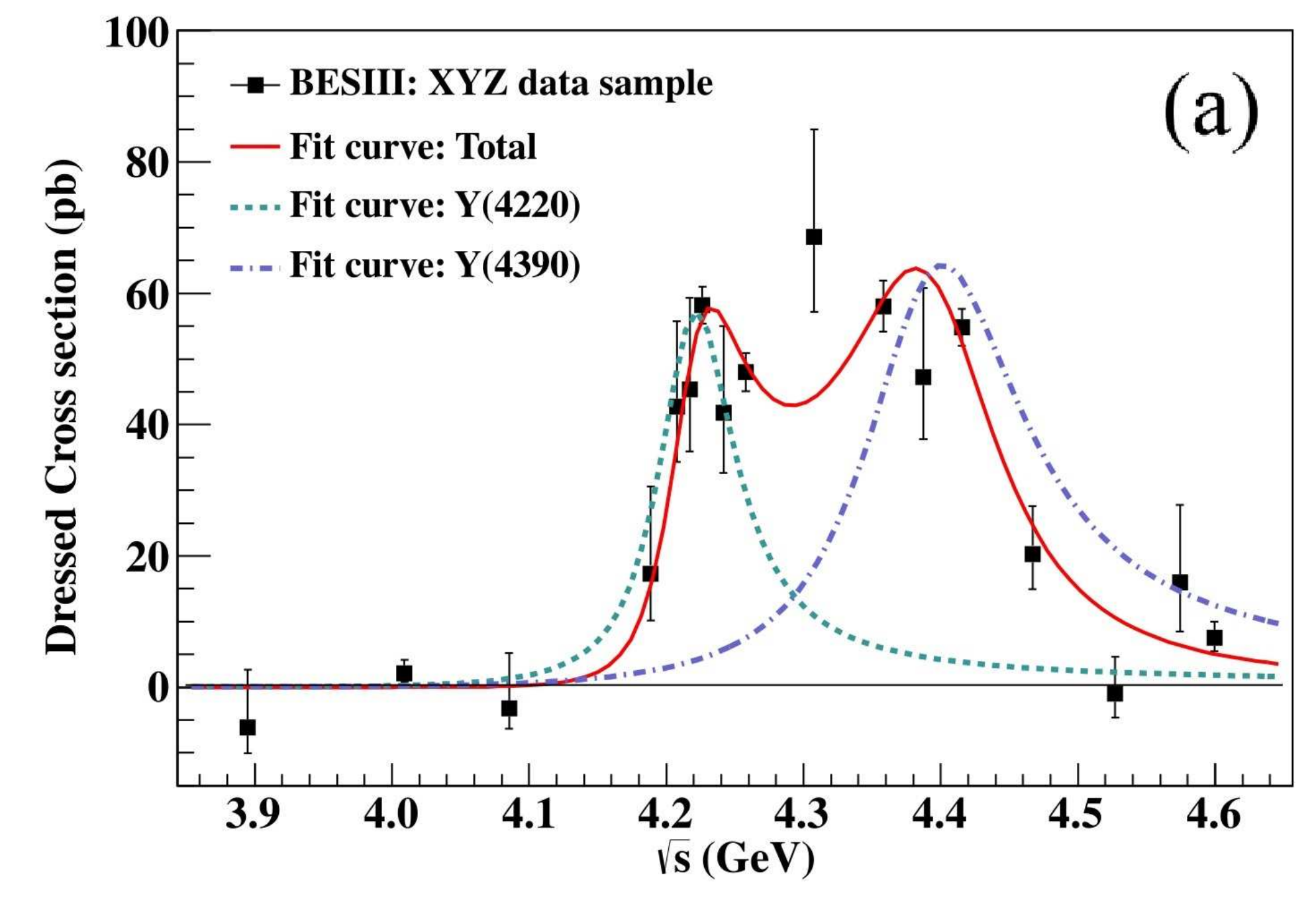}
\includegraphics[height=5.3cm]{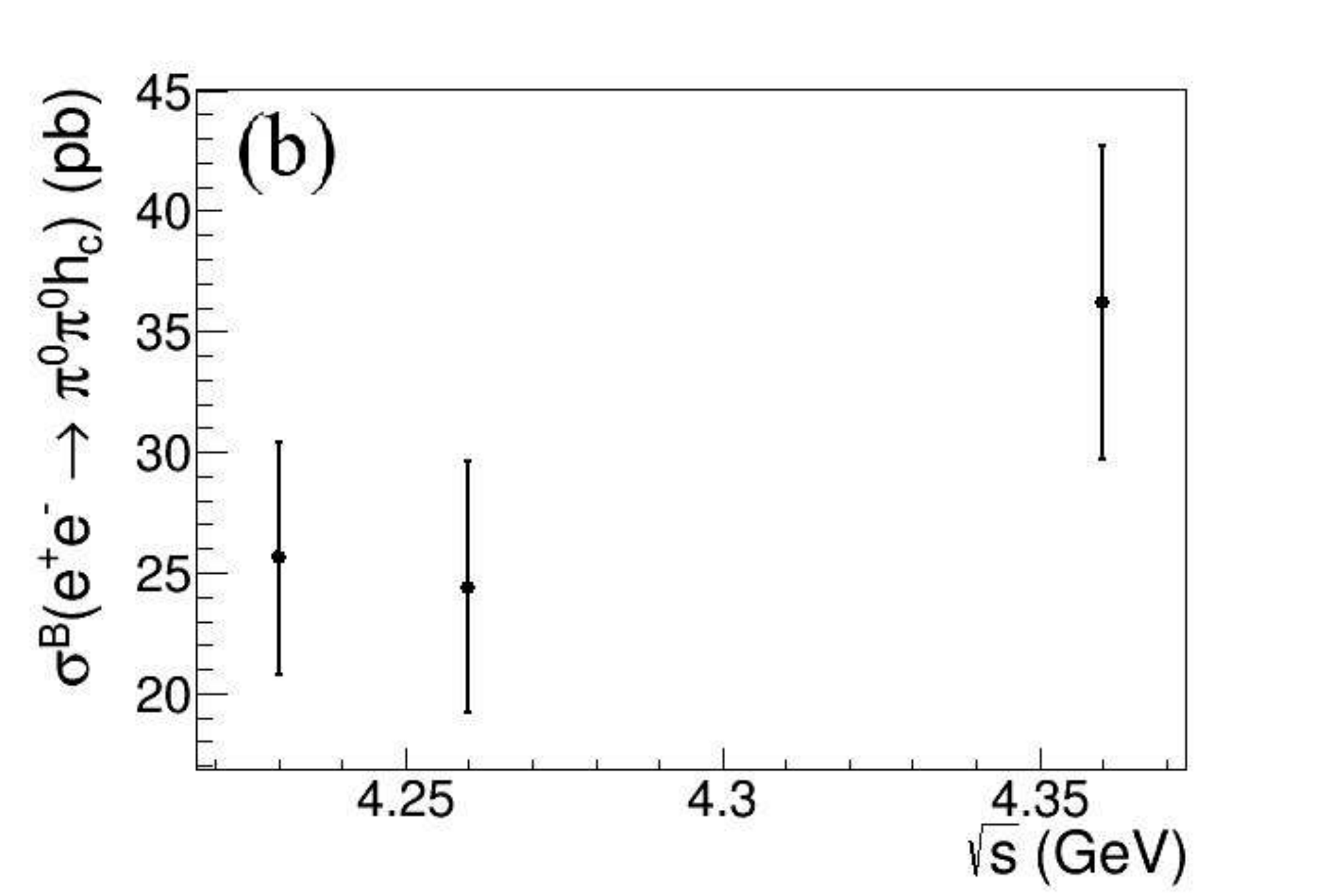}
\caption{Measured cross sections of (a) $\EE\to \pi^+ \pi^- h_c$~\cite{BESIII:2016adj} and (b)
$\EE\to \pi^0 \pi^0 h_c$~\cite{Ablikim:2014dxl}. In (a), the solid curve
is the fit with the coherent sum of two BW functions, and the dashed
and dash-dotted curves show the contributions from the two structures
$\psi(4230) \ \mbox{aka} \ Y(4230)$ and $Y(4390)$.} \label{cs}
\end{center}
\end{figure*}

Replacing the $J/\psi$ with an $h_c$, BESIII reported the cross section
measurement of $\EE\to \pphc$ at 13 c.m.\ energies from 3.9 to 4.4~GeV and
found the magnitude of the cross sections is about the same as that of
$\EE\to \ppjpsi$ but with a different line shape~\cite{Ablikim:2013wzq}.
Although no quantitative results were given in interpreting the
$\pphc$ line shape, the resonant structure at around
4.22~GeV is obvious~\cite{Ablikim:2013wzq}.
A follow-up measurement of $\EE\to \pp\hc$ cross sections at
c.m.\ energies from 3.9 to 4.6~GeV was done in 2017 with improved
precision~\cite{BESIII:2016adj}.
The cross sections of the neutral process $\EE\to \pi^0 \pi^0 h_c$
were also measured by BESIII at $\sqrt{s}=$4.23, 4.26, and
4.36 GeV~\cite{Ablikim:2014dxl}.
In all the above measurements, the $\hc$ is reconstructed via its
electric-dipole transition $\hc\to \gamma\etac$ with $\etac$ to 16 exclusive
hadronic final states. The measured cross sections are shown in Fig.~\ref{cs}.
The Born cross sections of $\EE\to \pi^0 \pi^0 h_c$ are found to be about
half of those of
$\EE\to \pi^+ \pi^- h_c$ within less than $2\sigma$.

The cross sections of $\EE\to \pp\hc$ indicate that there are possibly
two resonant structures in the studied energy range although
the error bar at around 4.3~GeV is large and it seems very important
to understand the line shape.
Assuming that there are two interfering resonances,
BESIII performed a fit to the distribution of the $\EE\to \pp\hc$ cross
sections and obtained the parameters of the two resonances:
$M=(4218.4^{+5.5}_{-4.5}\pm 0.9)$~MeV,
$\Gamma=(66.0^{+12.3}_{-8.3}\pm 0.4)$~MeV, and the product of the
electronic partial width and the decay branching fraction
$\Gamma_{\EE}\BR[Y(4230)\to \pphc] = (4.6^{+2.9}_{-1.4}\pm
0.8)$~eV for $Y(4230)$, and $M=(4391.5^{+6.3}_{-6.8}\pm
1.0)$~MeV, $\Gamma=(139.5^{+16.2}_{-20.6}\pm 0.6)$~MeV,
and $\Gamma_{\EE}\BR[Y(4390)\to \pphc]
=(11.6^{+5.0}_{-4.4}\pm1.9)$~eV for $Y(4390)$, with a relative
phase of $\phi=(3.1^{+0.7}_{-0.9}\pm 0.2)$~rad. The parameters of
the $Y(4230)$ are consistent with those of the
resonance observed in $\EE\to \ppjpsi$~\cite{Ablikim:2016qzw}
discussed above and in $\EE\to \omega\chi_{c0}$~\cite{Ablikim:2014qwy}
discussed below. The high mass structure is different from
the $Y(4360)$~\cite{Lees:2012pv,Wang:2014hta} (discussed below) and
$\psi(4415)$~\cite{Tanabashi:2018oca}.
In the future, the precision of the cross section at around 4.3~GeV needs
to be improved in order to better understand the line shape of $\EE\to \pp\hc$.

\begin{figure}[htbp]
\begin{center}
\includegraphics[height=3cm]{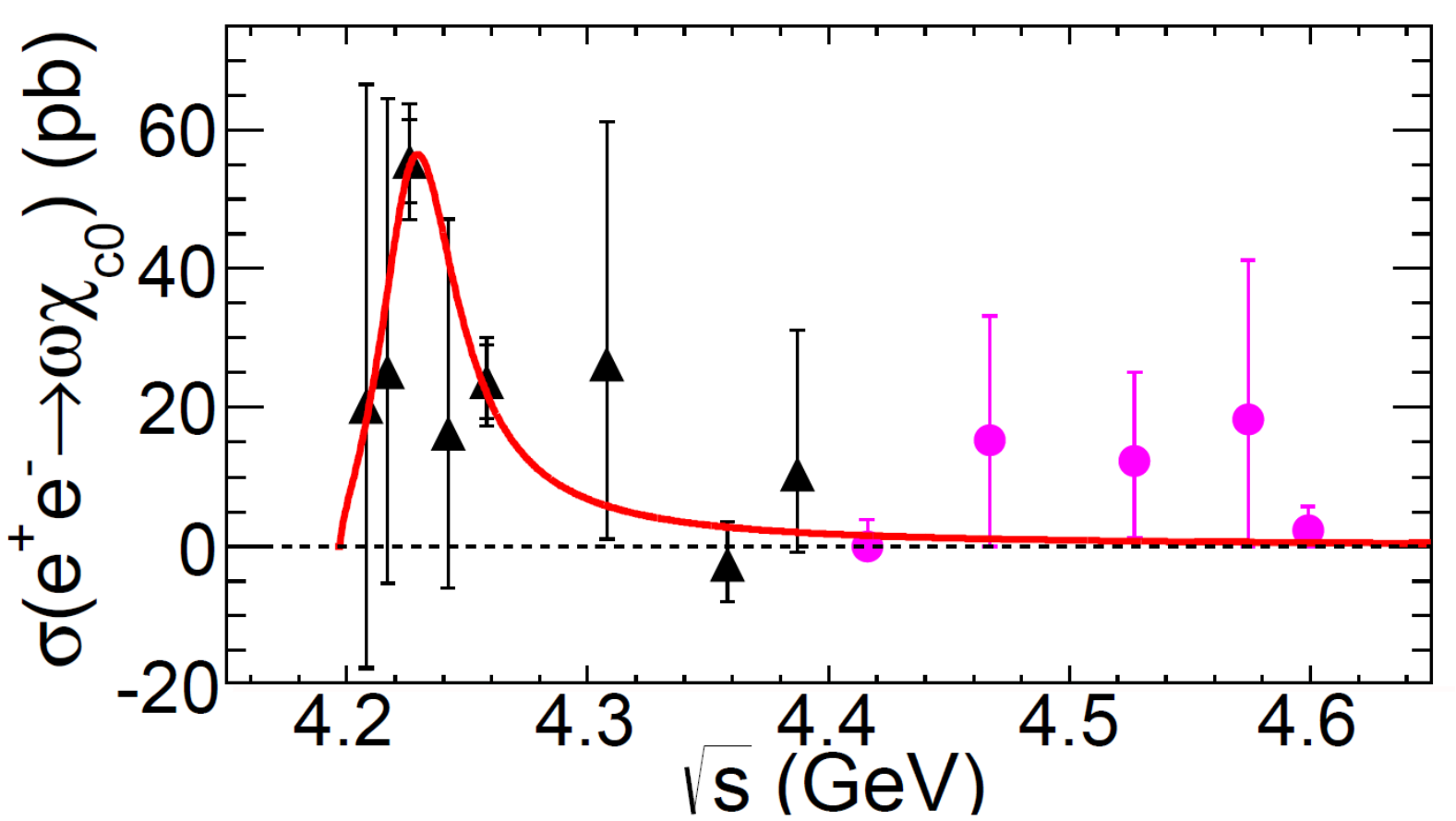}
\includegraphics[height=3cm]{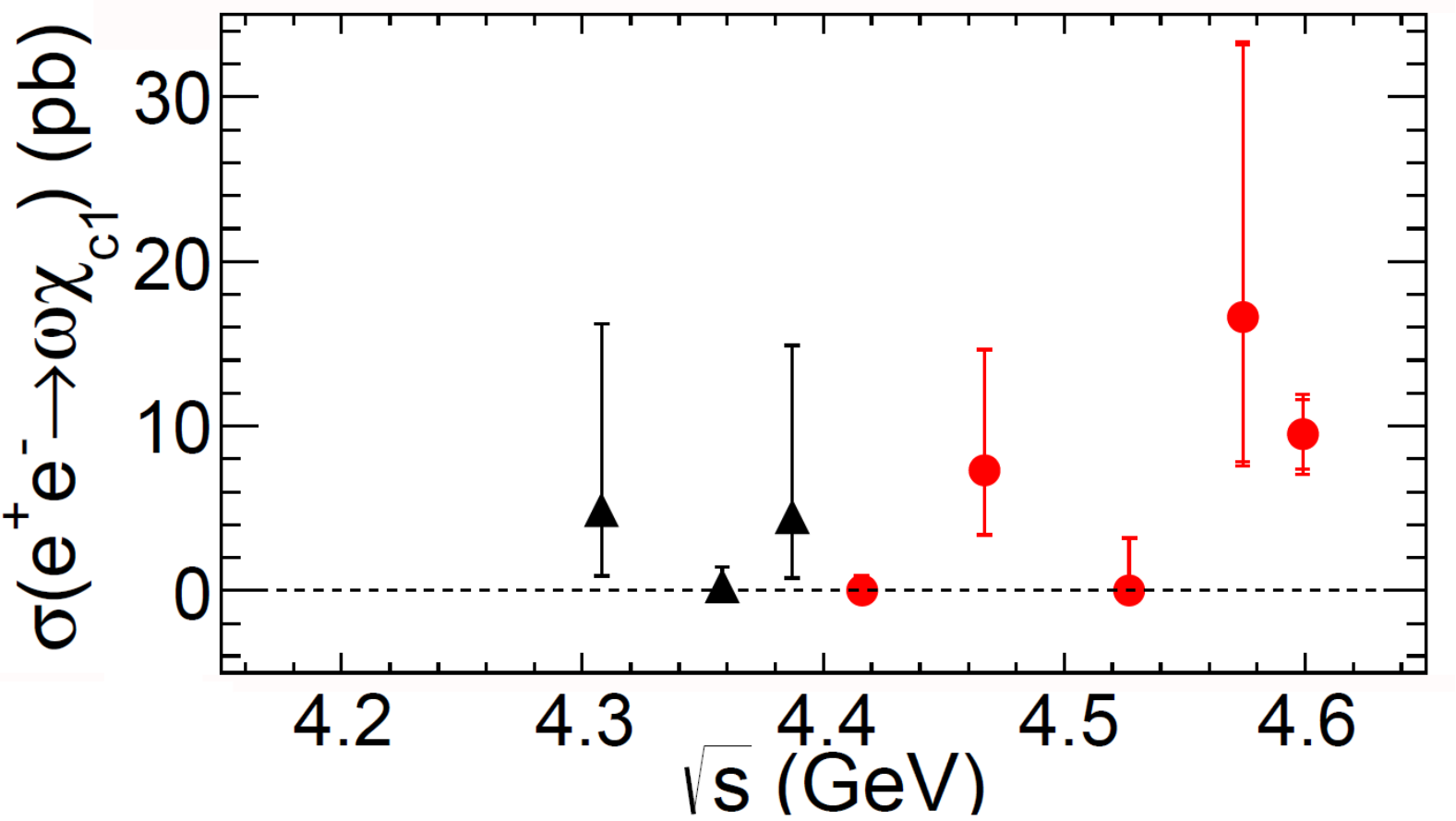}
\includegraphics[height=3cm]{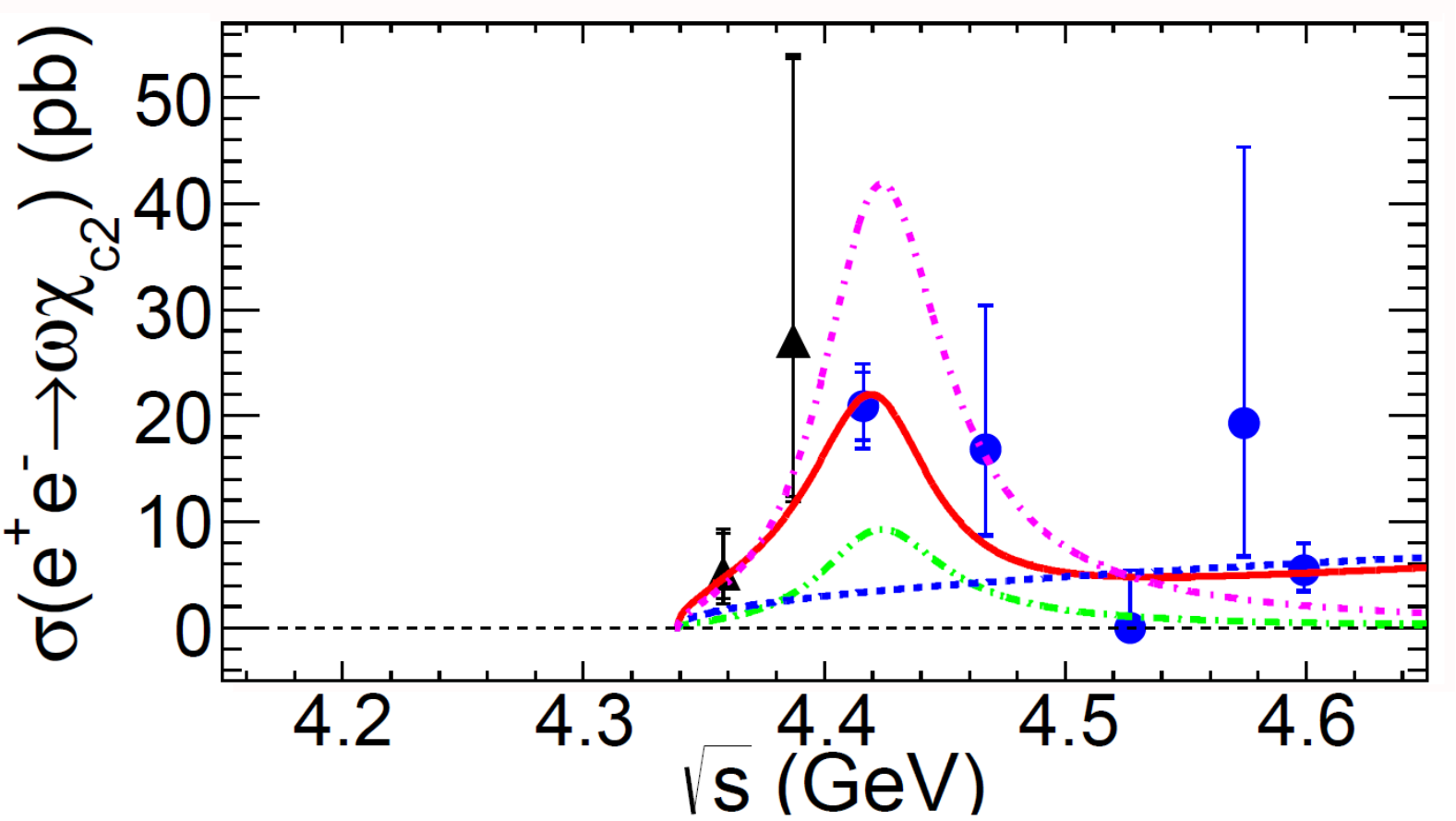}
\caption{ Measured Born cross section for $\EE \to \omega
\chi_{cJ}$ ($J=$0, 1, 2) as a function of c.m.\ energy~\cite{Ablikim:2014qwy,Ablikim:2015uix}.
The smaller error bars are
statistical and the larger error bars are the combined statistical
and systematic errors. The solid curves show the fit results.}
\label{fig:crosssectionall}
\end{center}
\end{figure}

A possible strong coupling of the $Y(4230)$ to the
$\omega\chi_{cJ}$ ($J=0,~1,~2$)
final state was proposed by a few authors~\cite{Yuan:2005dr,Dai:2012pb}.
In 2014, BESIII reported the cross section measurement of $\EE\to
\omega \chi_{c0}$ at 9 c.m.\ energies from 4.21 to 4.42~GeV, where
$\chi_{c0}$ candidates are reconstructed via a pair of
$\pp$ or $\kk$~\cite{Ablikim:2014qwy}. The Born cross sections
are measured to be $(55.4\pm 6.0\pm 5.9)$ pb and $(23.7\pm 5.3\pm
3.5)$~pb at $\sqrt{s}=4.23$ and $4.26$~GeV, respectively,
which are comparable to those of the $\ppjpsi$ process~\cite{Ablikim:2016qzw}.
Later BESIII updated the measurement of $e^+e^- \to \omega \chi_{c0}$
with higher energy data up to 4.6~GeV included~\cite{Ablikim:2015uix}.
Besides the energy points at $4.23$ and $4.26$~GeV,
no significant signals are found at other energy points, and the upper limits
on the cross sections at the 90\% C.L.\ are determined~\cite{Ablikim:2015uix}.
The cross sections are shown in Fig.~\ref{fig:crosssectionall} (left plot),
where a clear peaking structure close to the threshold is observed although
there are only two statistically significant measurements available.
Assuming the $\omega\chi_{c0}$ signals come from a single
resonance, the cross sections of $\EE \to \omega \chi_{c0}$ are
fitted with a BW function as shown in Fig.~\ref{fig:crosssectionall} (left plot).
The fitted mass and width are
$(4226\pm 8\pm 6)$~MeV and $(39\pm 12\pm 2)$~MeV, respectively,
with a statistical significance of more than $9\sigma$.

\begin{figure}[htbp]
\begin{center}
\includegraphics[height=5.5cm]{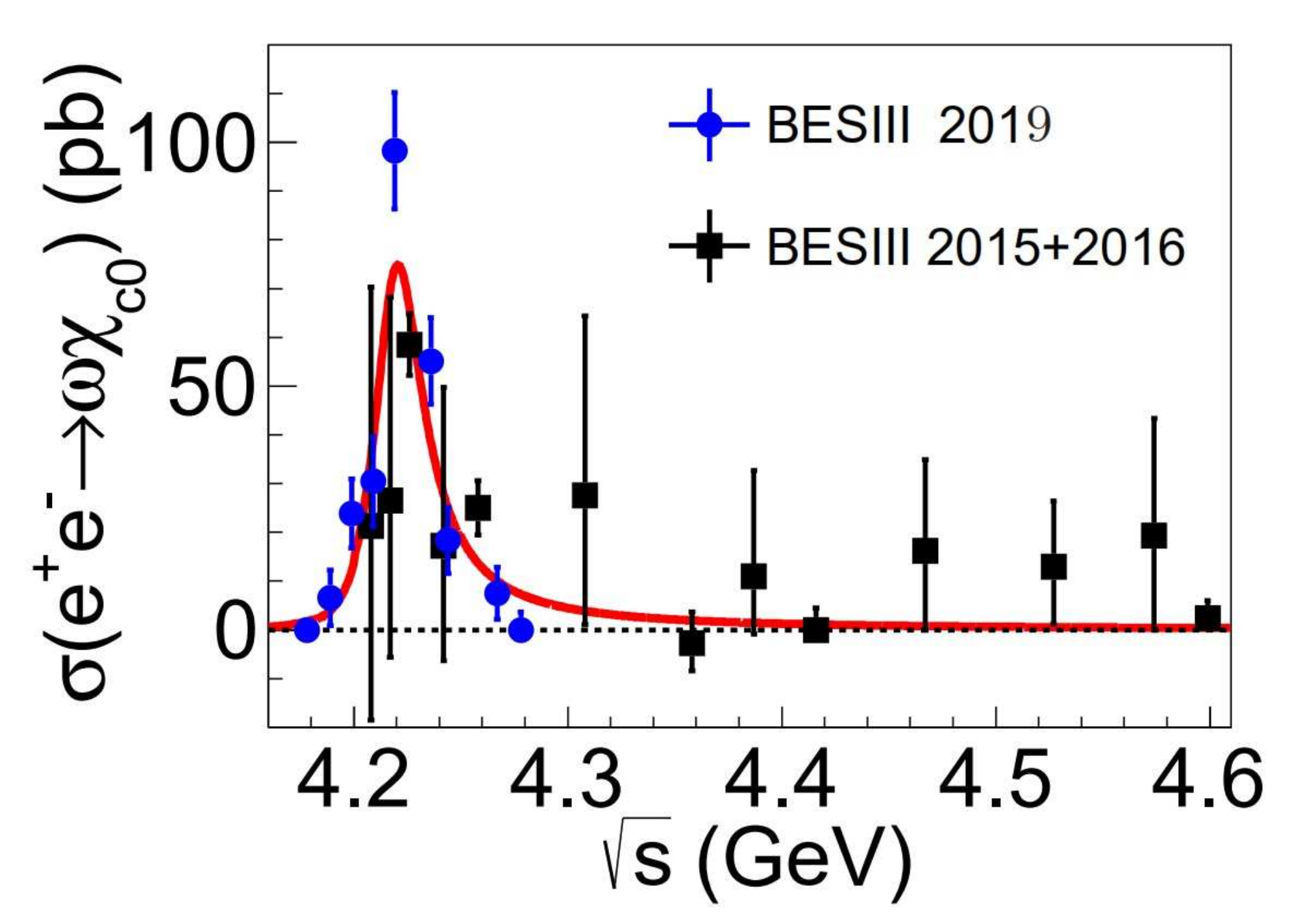}
\caption{The $e^+e^- \to \omega \chi_{c0}$ cross section as a function of the
c.m.\ energy from BESIII. The blue points are from the updated measurement~\cite{Ablikim:2019apl}
and the black square points are from previous measurements~\cite{Ablikim:2014qwy,Ablikim:2015uix}.
The red solid line is the fit result assuming all the
$\omega \chi_{c0}$ signals come from a single resonance.}
\label{fig:bes3oc2}
\end{center}
\end{figure}

Very recently the cross section of the process $e^+e^- \to \omega \chi_{c0}$
was again updated by BESIII at c.m.\ energies $\sqrt{s}$ from 4.178 to
4.278~GeV using a data sample of 7~fb$^{-1}$~\cite{Ablikim:2019apl}. The event
selection is almost the same as before~\cite{Ablikim:2014qwy,Ablikim:2015uix}.
The updated $e^+e^- \to \omega \chi_{c0}$ cross section as a function of the
c.m.\ energy is shown in
Fig.~\ref{fig:bes3oc2}, where the blue points are from the updated
measurement~\cite{Ablikim:2019apl}
and the black square points are from previous
measurements~\cite{Ablikim:2014qwy,Ablikim:2015uix}.
As expected, the $Y(4230)$ signal is observed
clearly. By assuming that all the
$\omega \chi_{c0}$ signals come from this resonance,
the fit result is shown in Fig.~\ref{fig:bes3oc2} with a solid red line and
the fitted mass and width
are $M=(4218.5\pm1.6\pm4.0)$~MeV and $\Gamma=(28.2\pm3.9\pm1.6)$~MeV. The
updated measurement
confirms and statistically improves upon the previous observation~\cite{Ablikim:2014qwy,Ablikim:2015uix}.
The parameters of this state are consistent with those of the narrow structure in the
$\EE\to \pp\hc$~\cite{Yuan:2005dr,BESIII:2016adj},
$\pi^+\pi^-J/\psi$~\cite{Ablikim:2016qzw},
$\ddpi$~\cite{Ablikim:2018vxx}, and $\pp\psp$~\cite{Ablikim:2017oaf} (discussed below)
processes.

For the $Y(4230)$, one of the big puzzles is
that there is a dip at around 4.22~GeV in the distribution of cross sections
of the charm meson pairs, instead of a peak. Therefore, whether or not this
state can couple to open charm channels is a crucial issue for
understanding its nature. Utilizing the measured cross sections of
$e^+e^-\to D^*\bar{D}^*$ and $D_s^*\bar{D}_s^*$ by
Belle~\cite{Abe:2006fj, Pakhlova:2010ek}, the authors of Ref.~\cite{Xue:2017xpu} performed
fits using $\psi(4040)$, $\psi(4160)$, and $\psi(4415)$
together with the $Y(4230)$ allowing for an interference
between them~\cite{Xue:2017xpu}. The cross sections can be well
described by these states, and the interference between the
$Y(4230)$ and the other charmonia produces a dip around 4.22~GeV in the
$e^+e^-\to D^*\bar{D}^*$ cross section line shape. However, the
errors of current Belle measurements are large especially for
$e^+e^-\to D_s^*\bar{D}_s^*$ cross sections. More precise measurements should
be done at BESIII to clarify the role played by the $Y(4230)$.

\begin{figure}[htbp]
\centering
\includegraphics[height=4.5cm]{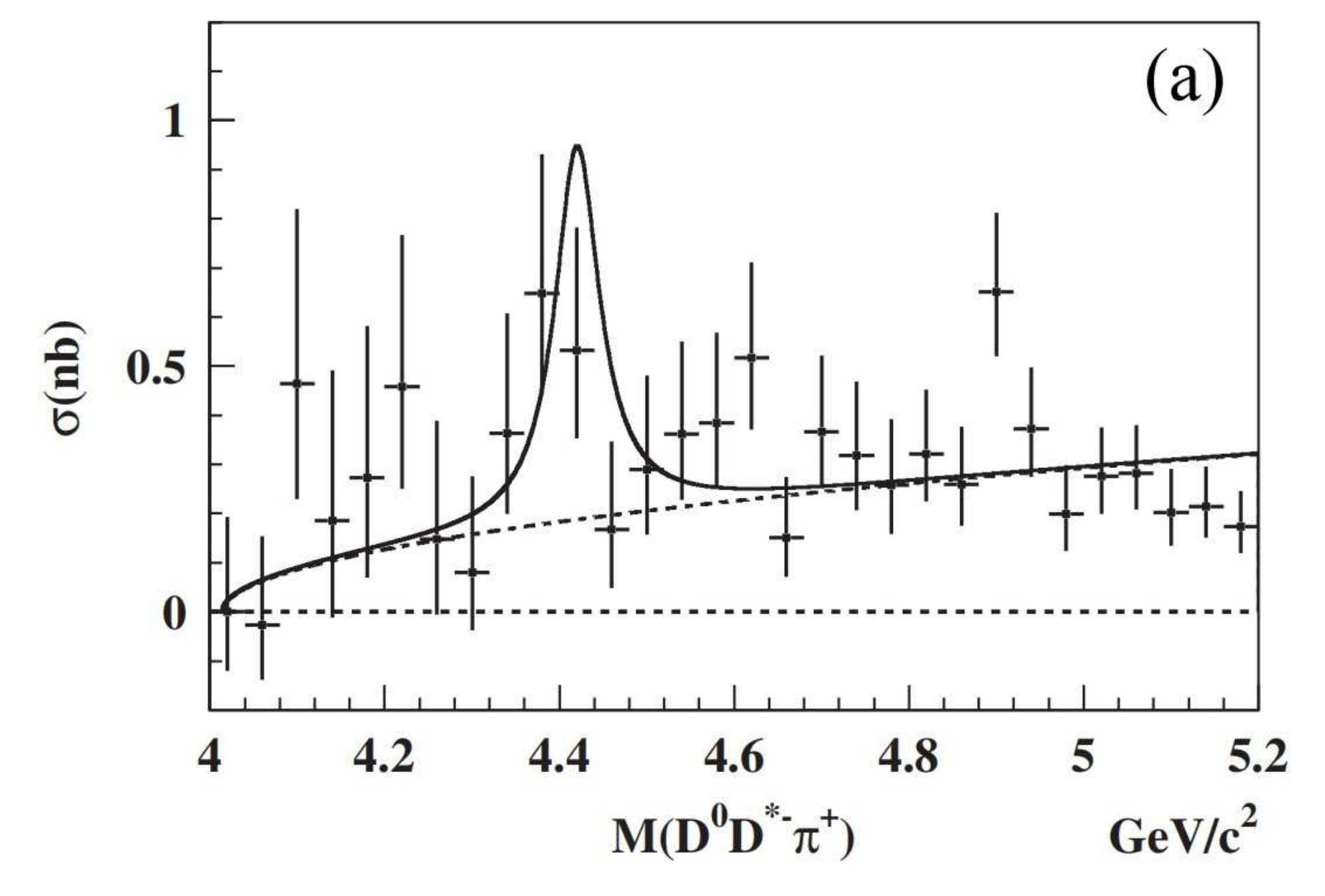}
\includegraphics[height=4.5cm]{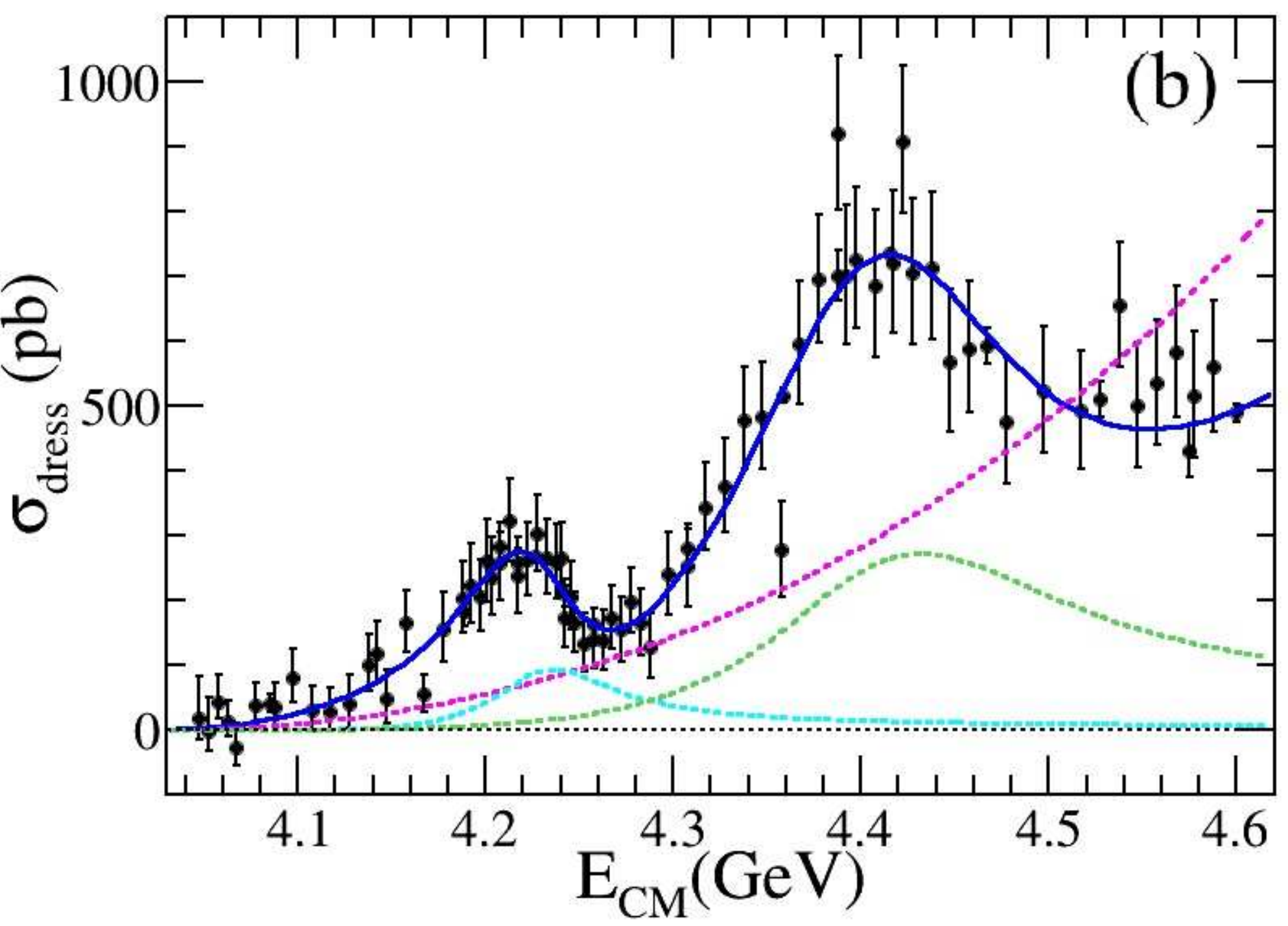}
\caption{Measured cross sections of $\EE\to \ddpi$ from (a) Belle~\cite{Pakhlova:2009jv}
and (b) BESIII~\cite{Ablikim:2018vxx}
experiments. Belle did a fit to cross sections with a $\psi(4415)$ signal
plus a threshold function, while BESIII did a fit with the coherent sum of
a direct three-body phase-space term and two BW functions for the $Y(4230)$
and $Y(4390)$. The solid lines are the total fit results. } \label{fig_cross}
\end{figure}

Motivated by the absence of open-charm decay channels for the $Y$ states,
Belle did the first measurement of the exclusive cross section for
$\EE\to \ddpi$ as a function of c.m.\ energy from the $\ddpi$ threshold to
5.2~GeV with ISR using an integrated luminosity of 695 fb$^{-1}$ data sample
at the $\Upsilon(4S)$
resonance~\cite{Pakhlova:2009jv}. The measured cross sections are shown in Fig.~\ref{fig_cross}(a).
No evidence for the $Y(4230)$, $Y(4360)$, $\psi(4415)$, $Y(4630)$, or
$Y(4660)$ was found with the limited statistics. Belle performed a likelihood fit
with a possible $\psi(4415)$ signal contribution plus a
threshold function. The fit yields $14.4\pm6.2^{+1.0}_{-9.5}$ signal events for
the $\psi(4415)$ state with a statistical significance of $3.1\sigma$.
A 90\% C.L.\ upper limit on the peak cross section for the $\EE \to \psi(4415) \to \ddpi$
process at $\psi(4415)$ nominal mass is obtained to be 0.76 nb.

Recently BESIII reported an improved measurement of the cross section of
$\EE \to \ddpi$ at c.m.\ energies from 4.05 to 4.60~GeV at 15 energy
points with integrated luminosity larger than 40~pb$^{-1}$ for each point
and 69 ``$R$-scan data" points with integrated luminosity smaller
than 20~pb$^{-1}$ for each point~\cite{Ablikim:2018vxx}, where
the $D^0$ meson is reconstructed via $D^0 \to K^-\pi^+$ and
the bachelor $\pi^+$ is also reconstructed, while the
$D^{*-}$ is inferred from energy-momentum conservation
in order to increase the statistics. The measured cross sections are shown in Fig.~\ref{fig_cross}(b).
Two resonant structures in good agreement with the $Y(4230)$ and $Y(4390)$ observed
in $\pphc$~\cite{BESIII:2016adj} are identified over a smoothly
increasing non-resonant term which can be parametrized with a
three-body phase-space amplitude. Therefore, a fit to the cross section from BESIII measurement
is performed to determine the parameters of the two resonant structures.

The total amplitude is described by the coherent sum of a direct three-body phase-space
term for $\EE\to \ddpi$ and two BW functions, representing the two
resonant structures. The fit yields a mass of $(4228.6 \pm
4.1\pm6.3)$~MeV and a width of $(77.0 \pm 6.8\pm 6.3)$~MeV for the lower
mass structure. As for the higher mass structure, there are complicated effects
which require further detailed amplitude analysis since a few excited $\psi$ states
and $Y$ charmonium-like states may overlap in this wide mass region.
Models including one additional known resonance, either
$Y(4320)$, $Y(4360)$, or $\psi(4415)$ with their masses
and widths fixed to the world average values~\cite{Tanabashi:2018oca},
can improve the fit quality. However, the statistical significance of any additional
resonance is less than $2\sigma$. Since the lower mass resonance is in
good agreement with the $Y(4230)$ observed in $\pp\jpsi$, $\pphc$,
and $\omega\chi_{c0}$ modes~\cite{Ablikim:2016qzw,BESIII:2016adj}, this
indicates the first observation of the $Y(4230)$ decays into an
open-charm final state $\ddpi$. The measured Born cross section of
$e^+e^- \to \ddpi$ at the
$Y(4230)$ peak is higher than the sum of the known
hidden-charm channels, which tells us the $\ddpi$ final state may be the dominant decay mode of the
$Y(4230)$ state.

An obvious feature in the above four channels $e^+e^-\to \omega
\chi_{c0}$, $\pphc$, $\pi^+ \pi^- J/\psi$, and $\ddpi$ from the BESIII
measurements is that there is a common structure at around
4.22~GeV, i.e., the $Y(4230)$. Considering this,
the authors of Ref.~\cite{Gao:2017sqa} did a simultaneous
fit to the cross sections of these four processes.
The measured mass and width of the $Y(4230)$ are $(4219.6 \pm 3.3 \pm
5.1)$~MeV and $(56.0 \pm 3.6 \pm 6.9)$~MeV, respectively.

The leptonic decay width for a vector state is an important
quantity for discriminating various theoretical interpretations
of its nature. As current measurements only give the product of
the leptonic decay width and the strong decay widths,
the magnitude of the leptonic decay width
determines how the strong decay widths sum up to the total width.
Smaller leptonic decay width means that the strong decay widths
will be relatively enhanced and vice versa.
The estimate of quenched\footnote{i.e., not including the effects of sea quarks and so
leading to a non-unitary theory and an unknown systematic uncertainty --- see Sec.~\ref{Sect:4.3}.}
lattice QCD for the leptonic decay width of
the $Y(4230)$ is about 40~eV~\cite{Chen:2016ejo}
assuming it to be a hybrid meson and neglecting its coupling to meson-meson
decay channels; the predicted upper limit of
the $Y(4230)$ leptonic
decay width is about 500~eV if the $Y(4230)$ is a
hadronic molecule dominated by $D\bar{D}_1(2420)$~\cite{Qin:2016spb}; the
leptonic decay width is only about 23~eV for the $\omega\chi_{c0}$ molecule
interpretation~\cite{Dai:2012pb}, where no contributions from the open
charm decay channel are included in the analysis.

By considering the isospin symmetric modes of the measured
channels, the authors of Ref.~\cite{Gao:2017sqa} estimated the
lower limit on the leptonic partial width of the
$Y(4230)$ decays
for the first time: $\Gamma_{e^+e^-}[Y(4230)]>(29.1\pm 2.5\pm 7.0)~\hbox{eV}$.
This lower limit is close to quenched lattice QCD result for a hybrid
vector charmonium state~\cite{Chen:2016ejo}. By considering other $Y(4230)$
decay modes, such as $\pp\psp$, $\eta h_c$ discussed below, the above estimation can be refined.

\begin{figure}[htbp]
\begin{center}
\includegraphics[height=5.5cm]{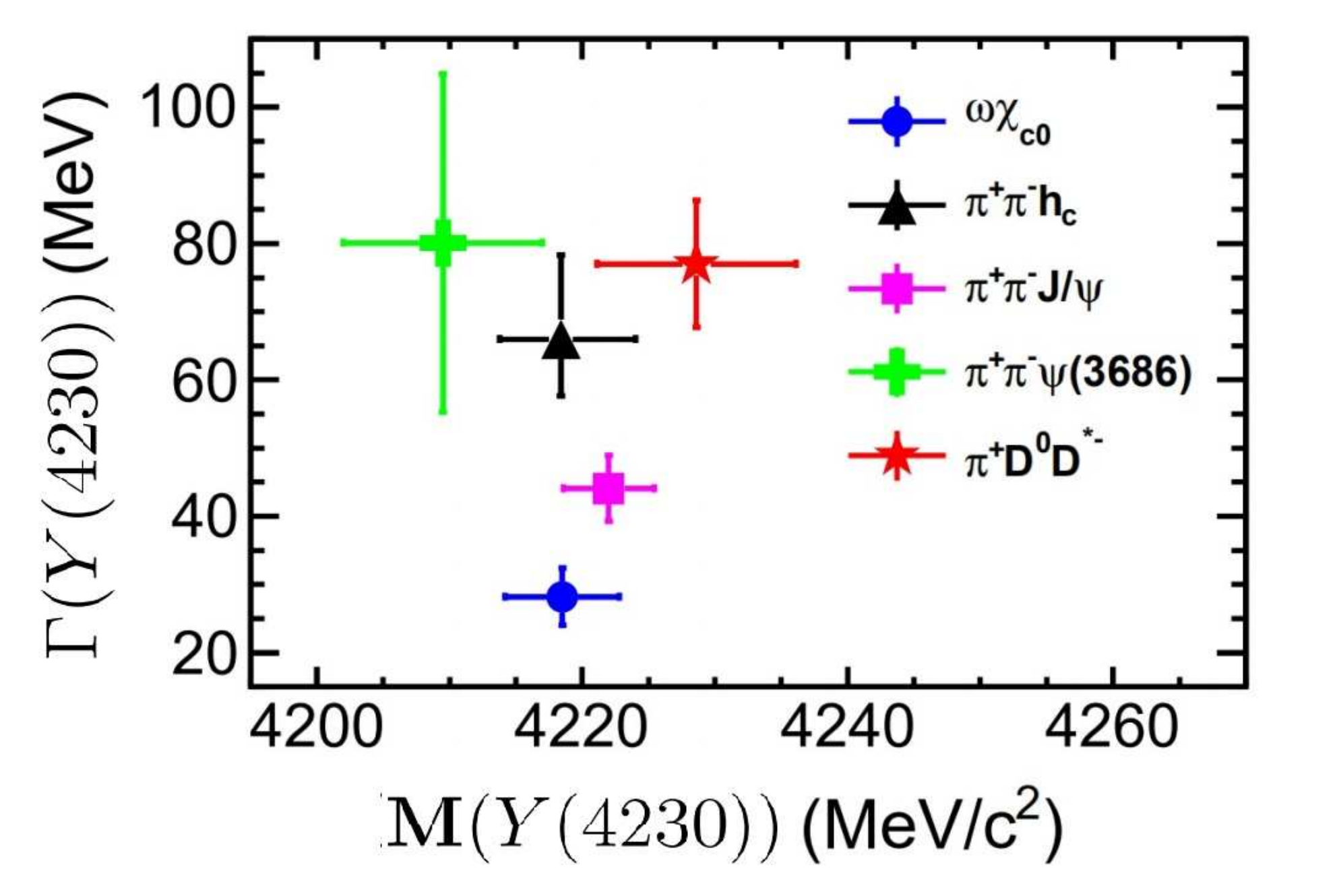}
\caption{The measured mass and width of
the $Y(4230)$ from the processes
$\pi^+\pi^-J/\psi$~\cite{Ablikim:2016qzw}, $\EE\to \pp\hc$~\cite{BESIII:2016adj},
$\ddpi$~\cite{Ablikim:2018vxx}, $\pp\psp$~\cite{Ablikim:2017oaf},
and $\omega \chi_{c0}$~\cite{Ablikim:2019apl}
from the BESIII Collaboration.}
\label{fig:ymasswidth}
\end{center}
\end{figure}

In an analysis of $\EE\to \pp\psp$, Belle found evidence for the
contribution from the $Y(4230)$~\cite{Wang:2014hta}
while high-statistics BESIII data confirmed the observation of this mode
together with a measurement of the resonant parameters~\cite{Ablikim:2017oaf}.
Figure~\ref{fig:ymasswidth} shows the measured mass and width of
the $Y(4230)$ from the processes
$\pi^+\pi^-J/\psi$~\cite{Ablikim:2016qzw}, $\EE\to \pp\hc$~\cite{BESIII:2016adj},
$\ddpi$~\cite{Ablikim:2018vxx}, $\pp\psp$~\cite{Ablikim:2017oaf},
and $\omega \chi_{c0}$~\cite{Ablikim:2019apl}
from the BESIII Collaboration. Although the measured masses are consistent
with each other, the widths from the processes
$\EE\to \pp\hc$, $\pp\psp$, and $\ddpi$
are larger than those from the processes $\EE\to \pi^+\pi^-J/\psi$ and
$\omega \chi_{c0}$. At the moment, we cannot draw a conclusion on whether
the structure observed in these processes is the same state or whether
the inconsistencies are caused by the BW parametrization. Further experimental
studies with higher statistics are needed to draw a more reliable conclusion
on the nature of this structure.

The compact tetraquark model~\cite{Faccini:2013lda} predicted a sizeable
isospin-violating process $Y(4230) \to \eta \pi^0 J/\psi$ with $Z_c^0$ decaying into $\pi^0 \jpsi$
and possibly $\eta \jpsi$. The molecular model~\cite{Wu:2013onz} predicted
a peak in the cross section of $e^+e^- \to \eta \pi^0 J/\psi$
at the $D_1 \bar{D}$ threshold and a narrow peak in the $\eta \jpsi$
mass spectrum at the $D\bar{D}^*$ threshold. Therefore, using data samples
collected at c.m.\ energies of $\sqrt{s}=$4.009, 4.226, 4.257, 4.358, 4.416,
and 4.599~GeV with a total luminosity of 4.5~fb$^{-1}$,
BESIII searched for such an isospin-violating decay
process~\cite{Ablikim:2015xfo}.
No $Y(4230)$ signal is observed, and upper limits
on the cross sections
$e^+e^- \to \eta \pi^0 J/\psi$ at the 90\% C.L.\ are determined to be 3.6, 1.7, 2.4, 1.4, 0.9, and 1.9~pb for $\sqrt{s}=$4.009, 4.226, 4.257, 4.358, 4.416,
and 4.599~GeV, respectively.

\begin{figure}[htbp]
\begin{center}
\includegraphics[height=7cm]{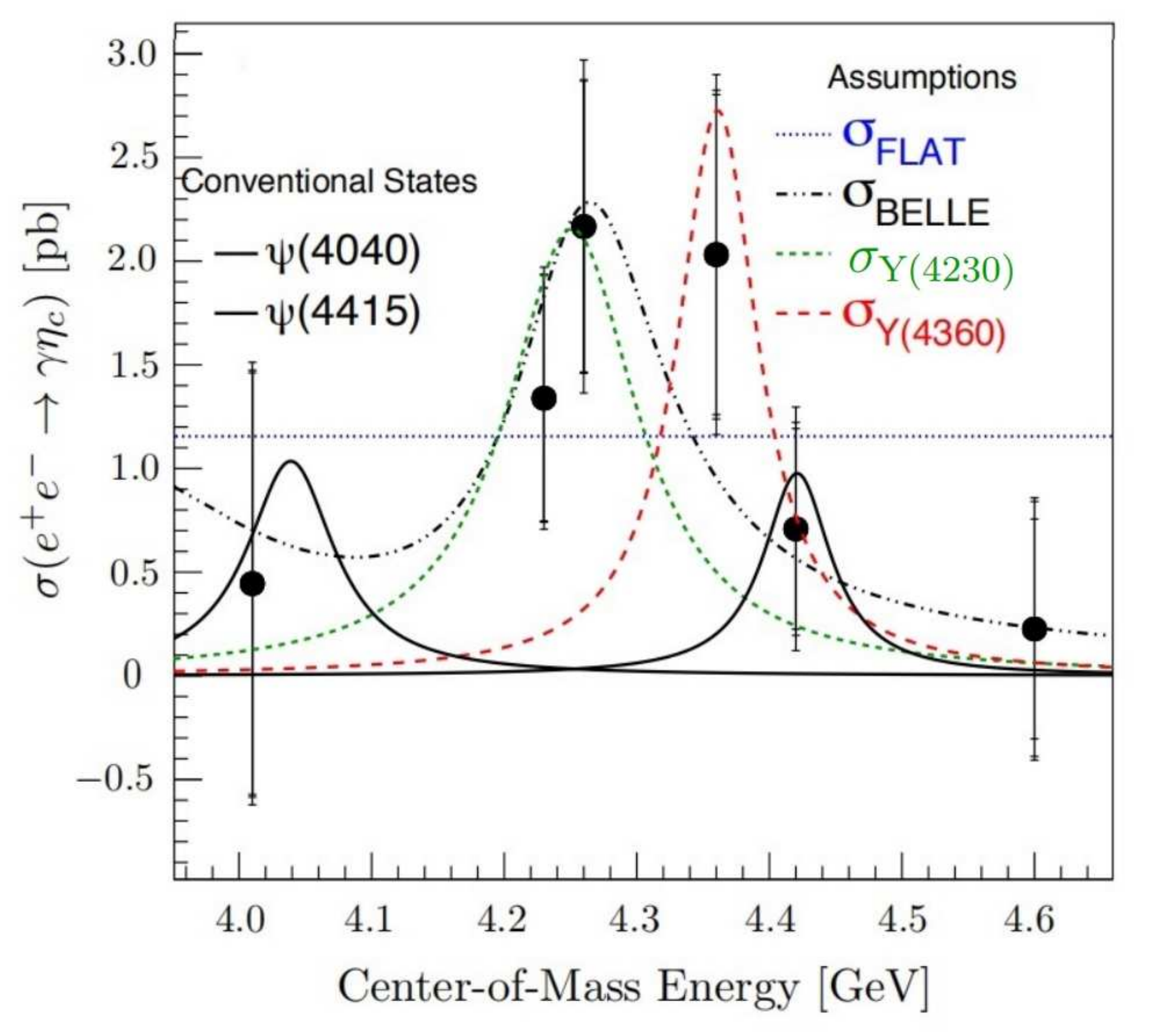}
\caption{ Measured Born cross section for $\EE \to \gamma \eta_c$
as a function of c.m.\ energy~\cite{Ablikim:2017ove} and comparison with
different assumptions about the energy-dependence of the
cross section (broken lines). The predicted cross sections for
$e^+e^- \to \psi(4040)/\psi(4415) \to \gamma \eta_c$
are shown as solid lines.}\label{fig:getac}
\end{center}
\end{figure}

Under the assumption of $Y(4230)$ being a hybrid meson,
the quenched lattice study~\cite{Dudek:2009kk} suggested that the rate of decay to $\gamma \eta_c$ may be enhanced relative to $\gamma \chi_{c0}$.
Finding evidence for $Y(4230) \to \gamma \eta_c$ could
thus give additional support to the hybrid interpretation.
BESIII searched for the process $e^+e^- \to \gamma \eta_c$
at six c.m.\ energies between 4.01 and 4.60~GeV corresponding to a total integrated
luminosity of 4.6~fb$^{-1}$~\cite{Ablikim:2017ove}.
The final distribution of $\sigma(e^+e^- \to \gamma \eta_c)$ is shown
as the points in Fig.~\ref{fig:getac}. The lines in Fig.~\ref{fig:getac} show
the different assumptions of resulting cross sections as a function of energy,
where (1) $\sigma_{\rm FLAT}$: the cross section is constant,
(2) $\sigma_{\rm BELLE}$:
the cross section follows the Belle parametrization of
$\sigma(e^+e^- \to \pi^+ \pi^- \jpsi)$~\cite{Liu:2013dau},
(3) $\sigma_{Y(4230)}$: the cross section follows a non-relativistic BW
distribution for the
$Y(4260)$ with mass and width values from the PDG~\cite{Tanabashi:2018oca}, and (4) $\sigma_{Y(4360)}$:
the cross section follows a non-relativistic BW distribution for the $Y(4360)$ with mass and
width values from the PDG~\cite{Tanabashi:2018oca}. With current statistics,
different assumptions for the energy dependence of the cross section can not
be distinguished, but the cross section is somewhat better explained by $\sigma_{Y(4230)}$.
The expected rates of $e^+e^- \to \psi(4040)/\psi(4415) \to \gamma \eta_c$ are also
shown in Fig.~\ref{fig:getac} as solid lines.
Both significances are 1.9$\sigma$.
Much larger data samples are required to confirm the existence of the $Y(4230) \to \gamma \eta_c$ process.

Some light-hadron final states are also searched for by BESIII for
the $Y(4230)$
decays including $K_{S}^{0}K^{\pm}\pi^{\mp}$~\cite{Ablikim:2018jbb},
$K_{S}^{0}K^{\pm}\pi^{\mp}\pi^0$, and $K_{S}^{0}K^{\pm}\pi^{\mp}\eta$~\cite{Ablikim:2018ddb}.
The Born cross sections of $e^+e^-$ to these final states are measured
up to 4.6~GeV. In fitting these cross sections, besides the continuum component,
a $Y(4230)$ signal is added.
No clear signal is observed, thus the 90\% C.L.\ upper limits
on the cross sections of $e^+e^- \to Y(4230)$ times the branching fractions
of the $Y(4230)$ to these final states are set.

\begin{figure}[htbp]
\begin{center}
\includegraphics[height=7cm]{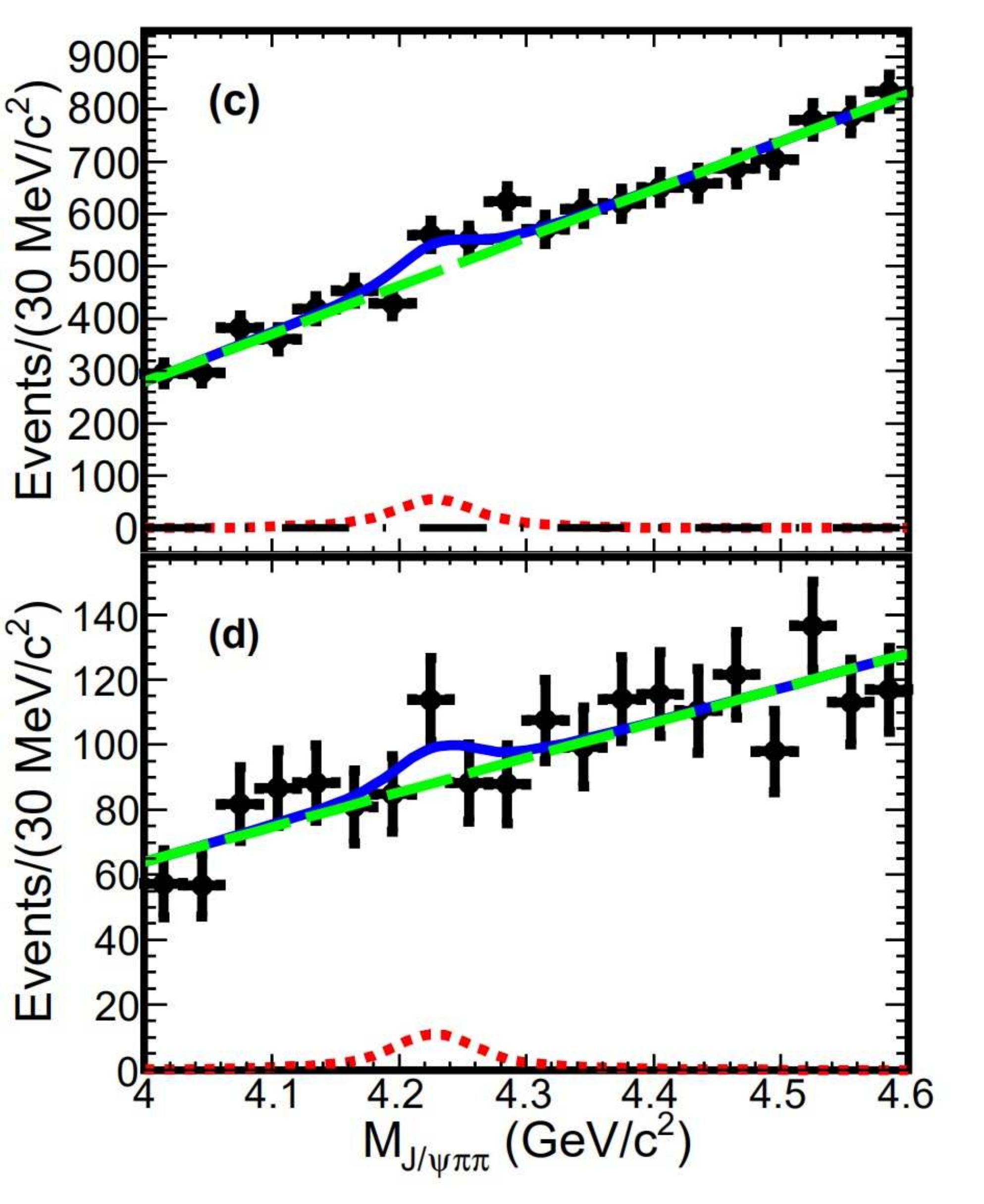}
\caption{The $\pi^+ \pi^- J/\psi$ mass spectra from charged (top) and
neutral (bottom) $B$ decays $B^+ \to K^+ \pi^+ \pi^- J/\psi$ and
$B^0 \to K^0 \pi^+ \pi^- J/\psi$~\cite{Garg:2019hor}.
The blue solid lines are the best fits,
the green dashed curves are the fitted backgrounds,
and the red dotted curves are the signals.}\label{fig:bby4260}
\end{center}
\end{figure}

The $Y(4230)$ was also searched for in the charged
$B$ decay
$B^+ \to K^+ \pi^+ \pi^- J/\psi$ by the BaBar Collaboration based on
$232\times 10^6$
$B\bar{B}$ pairs and a signal with a statistical
significance of 3.1$\sigma$ was observed~\cite{Aubert:2005zh}.
Very recently, Belle did the same search in both charged and neutral $B$ decays $B^+ \to K^+ \pi^+ \pi^- J/\psi$ and
$B^0 \to K^0 \pi^+ \pi^- J/\psi$ by fully reconstructing the final states from
$B$ decays based on $771.58$ million $B\bar{B}$ pairs~\cite{Garg:2019hor}.
The obtained $\pi^+ \pi^- J/\psi$ mass spectra after subtracting the
combinatorial background
from charged and neutral $B$ decays are shown in Fig.~\ref{fig:bby4260},
where no clear
$Y(4230)$ signal can be seen. The
signal significances of the $Y(4230)$ taking into account the
systematic uncertainties are 2.1$\sigma$ and 0.9$\sigma$
by an unbinned extended maximum-likelihood fit to these two
$\pi^+ \pi^- J/\psi$ mass spectra with a sum of two Gaussians as a $Y(4230)$ signal shape.
The blue solid lines in Fig.~\ref{fig:bby4260} are the best fits.
The upper limits on the product branching fractions
$\BR[B^+ \to K^+ Y(4230)]$$\BR[Y(4230) \to \pi^+ \pi^- J/\psi]$
and $\BR[B^0 \to K^0 Y(4230)]$$\BR[Y(4230) \to \pi^+ \pi^- J/\psi]$
at the 90\% C.L.\ are determined to be $1.4\times 10^{-5}$ and $1.7 \times 10^{-5}$, respectively.
In a search for the $\zc$ in $b$-hadron decays, the D0 experiment reported
evidence for the $Y(4230)$ production associated with its $\pi\zc$ decay
mode~\cite{Abazov:2018cyu} (discussed in Sec.~\ref{Sect:3.2}).
These studies will obviously benefit from much larger data samples.

Further studies of $Y(4230)$ decays in experiments
are needed with larger data statistics. Being well above the thresholds of
many final states with $\eta_c$, such as $\pi\rho\eta_c$, $\omega\eta_c$, and
$\phi\eta_c$, and final states like $\eta h_c$, $K\bar{K}J/\psi$,
$\eta^{(\prime)}\jpsi$, and $\eta^{(\prime)}\psp$, $Y(4230)$ may decay into such final states with substantial rates.
In addition, the decays into open charm final states other than $D \bar{D}^*\pi$ such as $D\bar{D}$,
$D\bar{D}^*+c.c.$, $D^*\bar{D}^*$, $D_s^+D_s^-$, $D_s^+D_s^{*-}+c.c.$ are also possible, even if the charmed mesons are in a relative $P$-wave.
The $Y(4230)$ is very close to the $D_s^{*+}D_s^{*-}$ threshold, a possible coupling to this mode should also be investigated.
Whether the $Y(4230)$ can decay into the light-hadron final states is also interesting.
Further information on these final states will be important for deeper understanding of the nature of the $Y(4230)$.

For the $Y(4230)$, popular theoretical
interpretations include a $D\bar{D}_1(2420)$
molecule~\cite{Cleven:2013mka,Ding:2008gr,Wang:2013cya},
a $c\bar{c}$-gluon hybrid~\cite{Zhu:2005hp,Close:2005iz,Kou:2005gt},
a compact tetraquark state~\cite{Maiani:2014aja}, a hadrocharmonium~\cite{Dubynskiy:2008mq,Li:2013ssa}.
If $Y(4230)$ is a $D\bar{D}_1(2420)$ molecule,
the binding energy is about 66~MeV which is rather large but not excluded.
If $Y(4230)$ is a $c\bar{c}$-gluon hybrid, its mass is about 180~MeV below current predictions from lattice QCD~\cite{Cheung:2016bym},
but consistent with the mass determined from combining non-relativistic EFTs and lattice QCD data
of the hybrid static energies~\cite{Berwein:2015vca,Oncala:2017hop} (see also Sec.~\ref{Sect:4.2.3bis}),
all of which neglect the resonant nature of the $Y(4230)$. 	
If it were a compact tetraquark state, it would have isospin- and SU$(3)$-multiplet
partner	states. However,  none of them has been observed in experiments so far.
If it is a hadrocharmonium, its decay rate to non-$\jpsi(h_c)$ charmonium states
should be suppressed, which is not supported by the data although a possible mixing of two
hadrocharmonia with spin 1 and spin 0 compact $\bar cc$ cores can make the scenario consistent~\cite{Li:2013ssa}.
At this point in time the nature of this lightest negative parity exotic candidate is not yet fully settled.

There have been a number of different interpretations proposed for
the $Y(4320)$ and $Y(4390)$, including a $\psi(3^3D_1)$
state~\cite{Chen:2018fsi,Bhavsar:2018umj},
a compact tetraquark state~\cite{Wang:2018ejf,Wang:2018rfw},
a hadrocharmonium~\cite{Ferretti:2018kzy},
a $D^{\ast}\bar{D}_1(2420)$ molecule~\cite{Wang:2013kra,Chen:2017abq,He:2017mbh},
a state formed from dynamical diquarks~\cite{Lebed:2017min},
a hybrid state~\cite{Oncala:2017hop}
and so on. In addition to the above explanations, a special explanation is
that the $Y(4320)$ and $Y(4390)$ may not exist at all.
The authors of Ref.~\cite{Chen:2017uof} found that the fits to the
distributions of $e^+e^- \to \pi^+ \pi^- J/\psi$ and $\pi^+ \pi^- h_c$
cross sections measured by BESIII with three
interfering resonances $Y(4230)$, $\psi(4160)$, and $\psi(4415)$
can also describe the data well with a goodness of the fit of
$\chi^2/ndf=118/153$ and $18/69$, respectively, where $ndf$ is the number
of degrees of freedom.
Based on the fit qualities, the authors argued that two newly reported
charmonium-like states, $Y(4320)$ and $Y(4390)$, are not genuine
resonances. However, it should be stressed that the analysis of
Ref.~\cite{Chen:2017uof}
needs to implement a huge amount of spin-symmetry violation as well as very
large non--resonant couplings at odds with our current understanding of the
role of the heavy quark spin symmetry in doubly-heavy systems (see also Sec.~\ref{Sect:4.2}).

\vspace{0.3cm}\noindent
$\bullet$ {\it The $\psi(4360)$ aka $Y(4360)$ and
$\psi(4660)$ aka $Y(4660)$ states}~\label{Sec:y4630}
\vspace{0.3cm}

Replacing the $J/\psi$ with a $\psp$, BaBar studied the ISR process
$e^+e^- \to \pi^+\pi^-\psp$
to search for the exotic state found previously,
namely $Y(4260)$~\cite{Aubert:2007zz}. Instead of
the $Y(4260)$, a clear structure around 4.32~GeV was
observed~\cite{Aubert:2007zz}. Subsequently, Belle checked this process and
found that there are actually
two resonant structures at 4.36 and 4.66~GeV, denoted as the
$Y(4360)$
and $Y(4660)$~\cite{Wang:2007ea}.

The BaBar experiment reported the update of the study of $\EE\to
\pp\psp$ with ISR events with the full data
sample recorded at and near the $\Upsilon(nS)$ ($n$=2,~3,~4)
resonances with an integrated luminosity of
520~fb$^{-1}$~\cite{Lees:2012pv}. The cross sections for
$\EE\to \pp\psp$ from 3.95
to 5.95~GeV have been measured. A fit to the $\pp\psp$ mass distribution
yields a mass of $(4340\pm 16\pm 9)$~MeV and a width of
$(94\pm 32\pm 13)$~MeV for the $Y(4360)$, and a mass of
$(4669\pm21\pm 3)$~MeV and a width of $(104\pm 48\pm 10)$~MeV for the
$Y(4660)$~\cite{Lees:2012pv}. The results are in good
agreement with the Belle measurement and confirm the $Y(4660)$
observed by the Belle experiment~\cite{Wang:2007ea}.
In both the Belle and BaBar measurements, the $\pi^+ \pi^-$ mass distribution
appears to differ from the phase-space expectation. For the
$Y(4660)\to \pp \psp$ decays, there is an indication of an accumulation
of events in the vicinity of the $f_0(980)$ state.

\begin{figure}[!htbp]
\centering
 \psfig{file=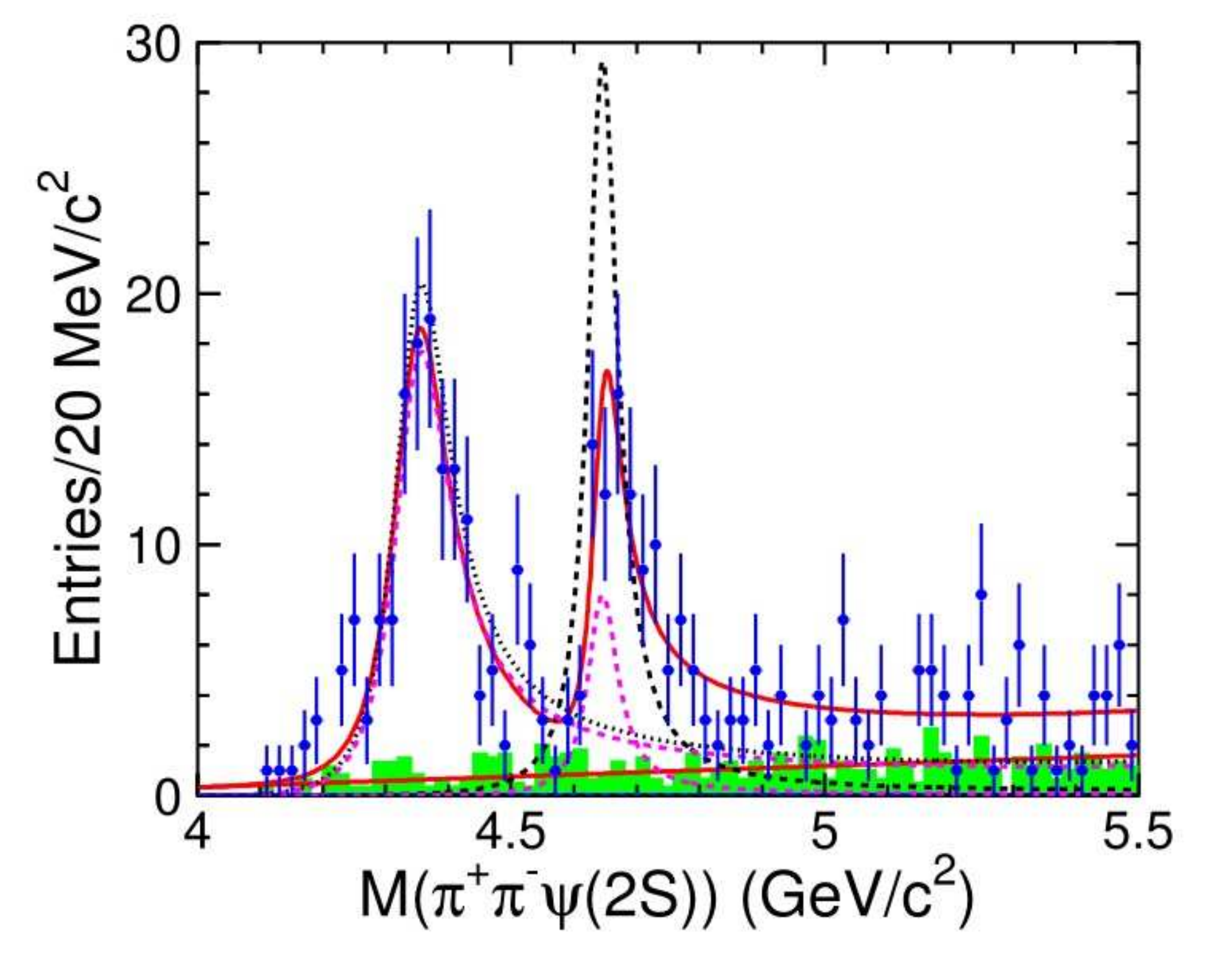, height=5.0cm}
 \psfig{file=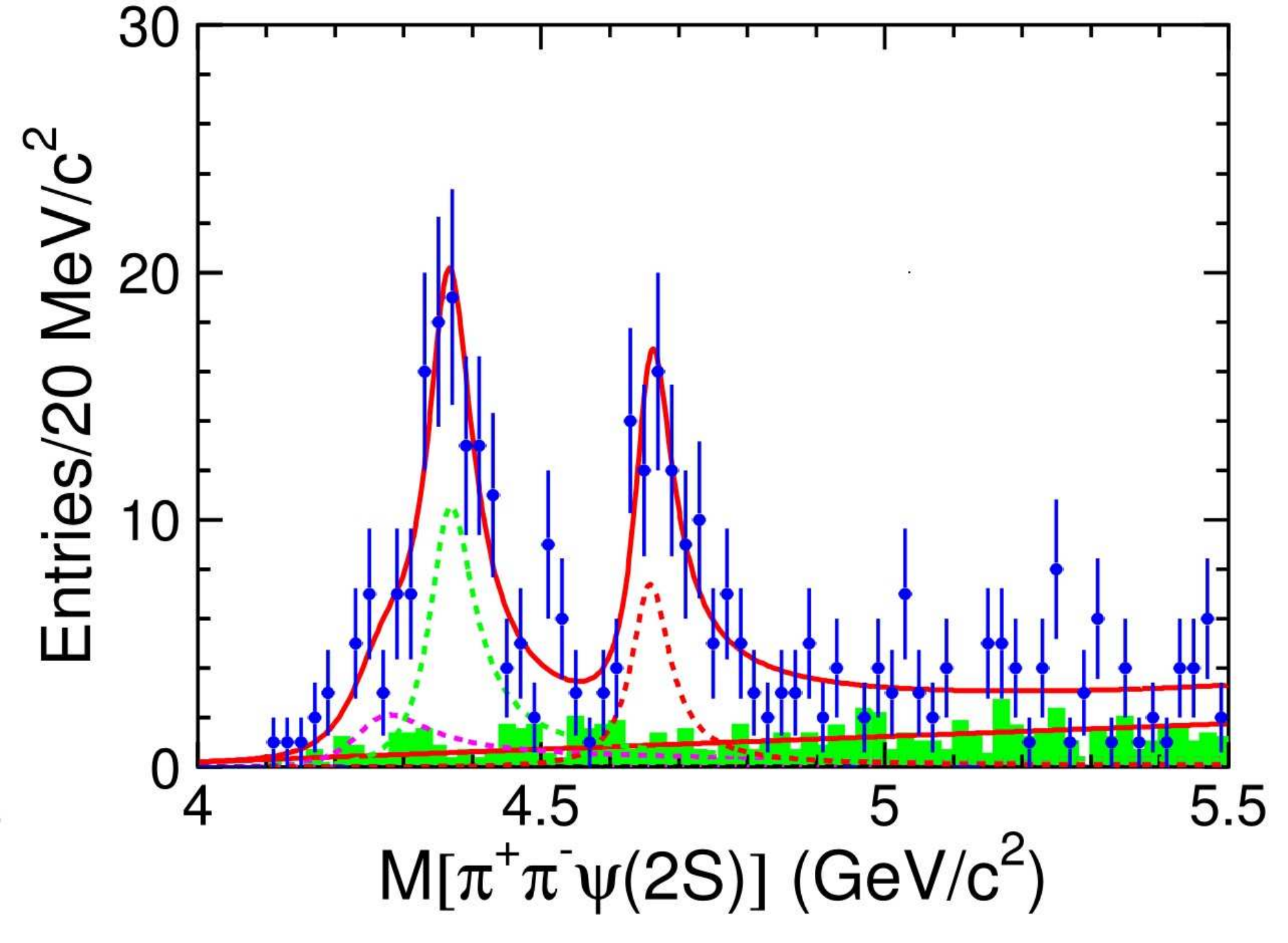, height=4.9cm}
\caption{The $\pp\psp$ invariant mass distributions from the Belle
experiment~\cite{Wang:2014hta} and the fit results with the coherent sum
of two BW functions (left panel) and with the coherent sum of three BW
functions (right panel). The points with error bars are data while
the shaded histograms are the normalized $\psp$ mass sideband backgrounds.
The curves show the best fits and the dashed curves show the
contributions from different BW components. } \label{2bwfit}
\end{figure}

The $Y(4360)$ and $Y(4660)$ parameters
were measured with improved precision with the full 980~fb$^{-1}$ data sample
of Belle~\cite{Wang:2014hta}.
Fitting the mass spectrum of $\pp\psp$ with two coherent BW functions (see
Fig.~\ref{2bwfit}~(left)), Belle obtained $M[Y(4360)] = (4347\pm
6\pm 3)$~MeV, $\Gamma[Y(4360)] = (103\pm 9\pm 5)$~MeV,
$M[Y(4660)] = (4652\pm 10\pm 8)$~MeV, and $\Gamma[Y(4660)] = (68\pm 11\pm 1)$~MeV.
Belle also noticed that there are a number of events in the vicinity of the
$Y(4230)$ mass. The fit with the $Y(4230)$
included is also performed.
In the fit, the mass and width of the $Y(4230)$ are fixed
to the latest measured values at Belle~\cite{Liu:2013dau}.
The signal significance of the $Y(4230)$ is found to be only $2.4\sigma$.
In this fit, one obtains $M[Y(4360)]=(4365\pm 7\pm
4)$~MeV, $\Gamma[Y(4360)]=(74\pm 14\pm 4)$~MeV,
$M[Y(4660)]=(4660\pm 9\pm 12)$~MeV, and
$\Gamma[Y(4660)]=(74\pm 12\pm 4)$~MeV. By comparing the fitted results in
these two fits, one can find that the resonant parameters depend strongly on whether
there is an additional $Y(4230)$.

\begin{figure}[!htbp]
\begin{center}
\includegraphics[width=0.5\textwidth]{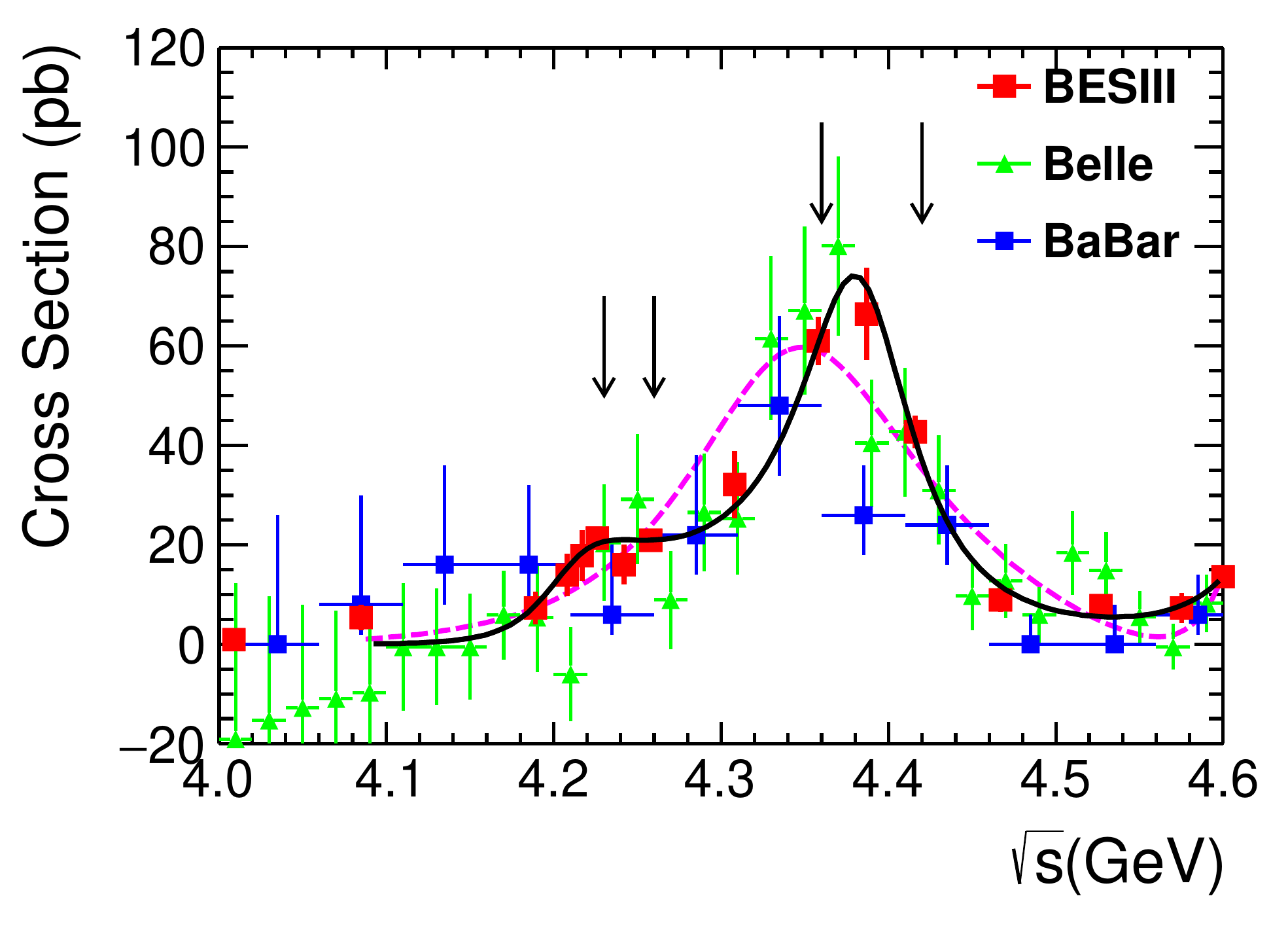}
\caption{Cross sections of $\EE\to \pp\psp$, where
the large red squares, green triangles, and small blue squares
are from BESIII~\cite{Ablikim:2017oaf},
 Belle~\cite{Wang:2014hta},
 and BaBar~\cite{Lees:2012pv} measurements, respectively. The solid (dashed)
curve is the fit to the BESIII measurement with the sum of three (two) BW
 functions. The arrows indicate the locations of four energy points
 with large integrated luminosities. } \label{comresult}
\end{center}
\end{figure}

BESIII measured $\EE\to \pp\psp$ cross sections at 16 c.m.\ energies from 4.0 to
4.6~GeV using 5.1~fb$^{-1}$ of data in total, where the $\psp$ candidates are
reconstructed via two decay modes $\psp\to \pp\jpsi$
and $\psp\to {\rm neutrals}+\jpsi$ (neutral=$\piz\piz$, $\piz$, $\eta$,
and $\gamma\gamma$)~\cite{Ablikim:2017oaf}.
After event selection criteria were applied, a prominent $\psp$
signal over a small background is observed in both $\psp$ decay modes.
The measured cross sections are shown in Fig.~\ref{comresult} together with the
comparisons from the Belle and BaBar
measurements~\cite{Lees:2012pv,Wang:2014hta}.
The results are in good consistency with former BaBar and Belle
results~\cite{Lees:2012pv,Wang:2014hta}, and have much improved precision.
A binned $\chi^2$ fit with the coherent sum of three BW amplitudes for the
$Y(4230)$, $Y(4360)$,
and $Y(4660)$ is applied to
describe the $e^+e^- \to \pp \psp$ cross section in a energy range
from 4.085 to 4.600~GeV.
As the BESIII data can only reach 4.6~GeV, the parameters of the
$Y(4660)$ are fixed to the Belle
measurement~\cite{Wang:2014hta} in the fit.
The fit results in a mass $M=(4383.8\pm 4.2\pm 0.8)$~MeV and a width
$\Gamma=(84.2\pm 12.5\pm 2.1)$~MeV for the $Y(4360)$.
By comparing to the fit with the coherent sum of two BW amplitudes,
$Y(4360)$ and $Y(4660)$, the data require
a lower-mass resonance with a mass $M=(4209.5\pm 7.4\pm 1.4)$~MeV and a
width $\Gamma=(80.1\pm 24.6\pm 2.9)$~MeV with a statistical
significance of $5.8\sigma$. This is the first observation of the
new decay mode $Y(4230)\to \pp\psp$.
The fit results with the coherent sum of three BW amplitudes and two BW
amplitudes are shown in Fig.~\ref{comresult} with solid and dashed lines.

\begin{figure}[!htbp]
\begin{center}
\includegraphics[width=3in,height=2.22in,angle=0]{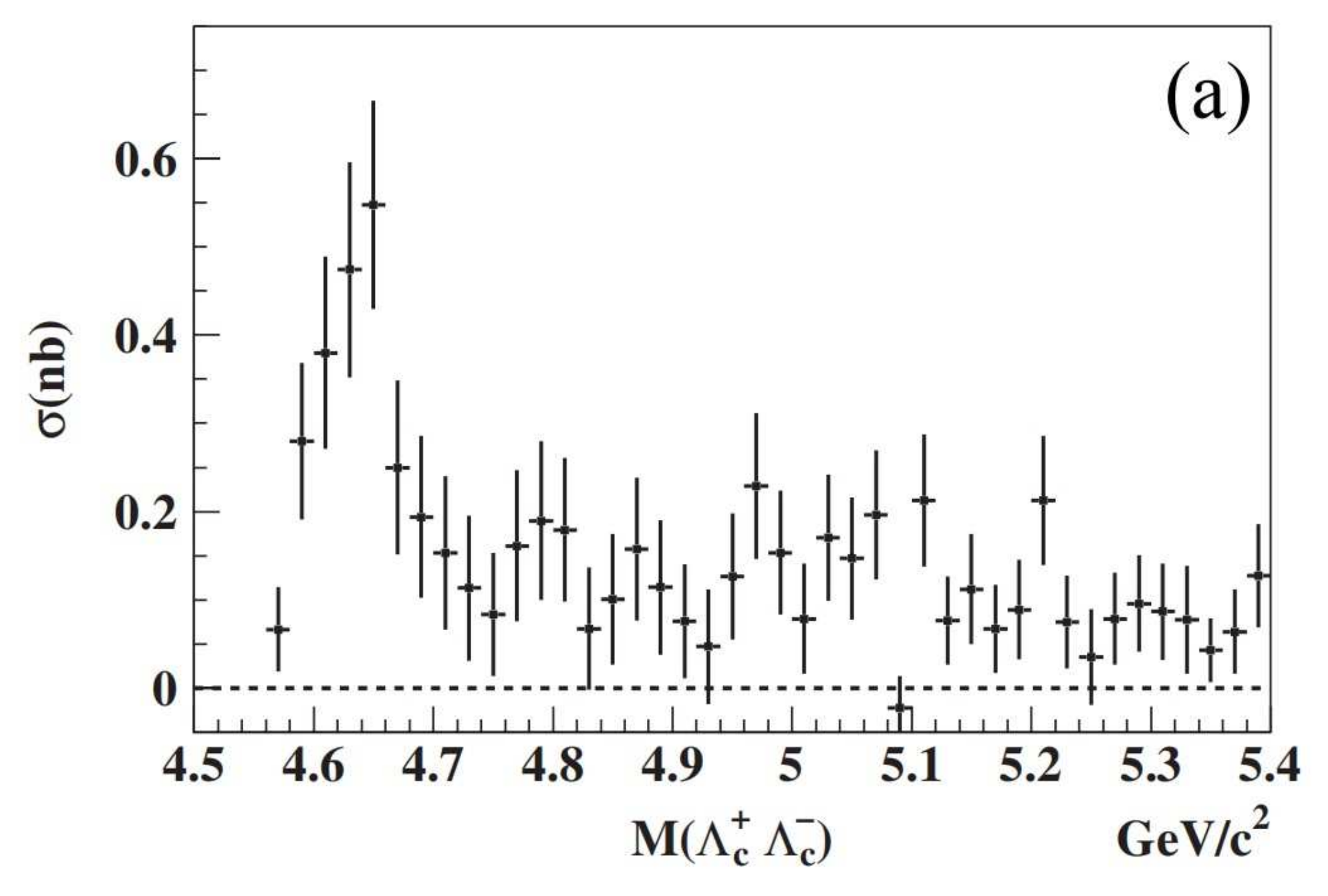}
\includegraphics[width=3in,height=2.22in,angle=0]{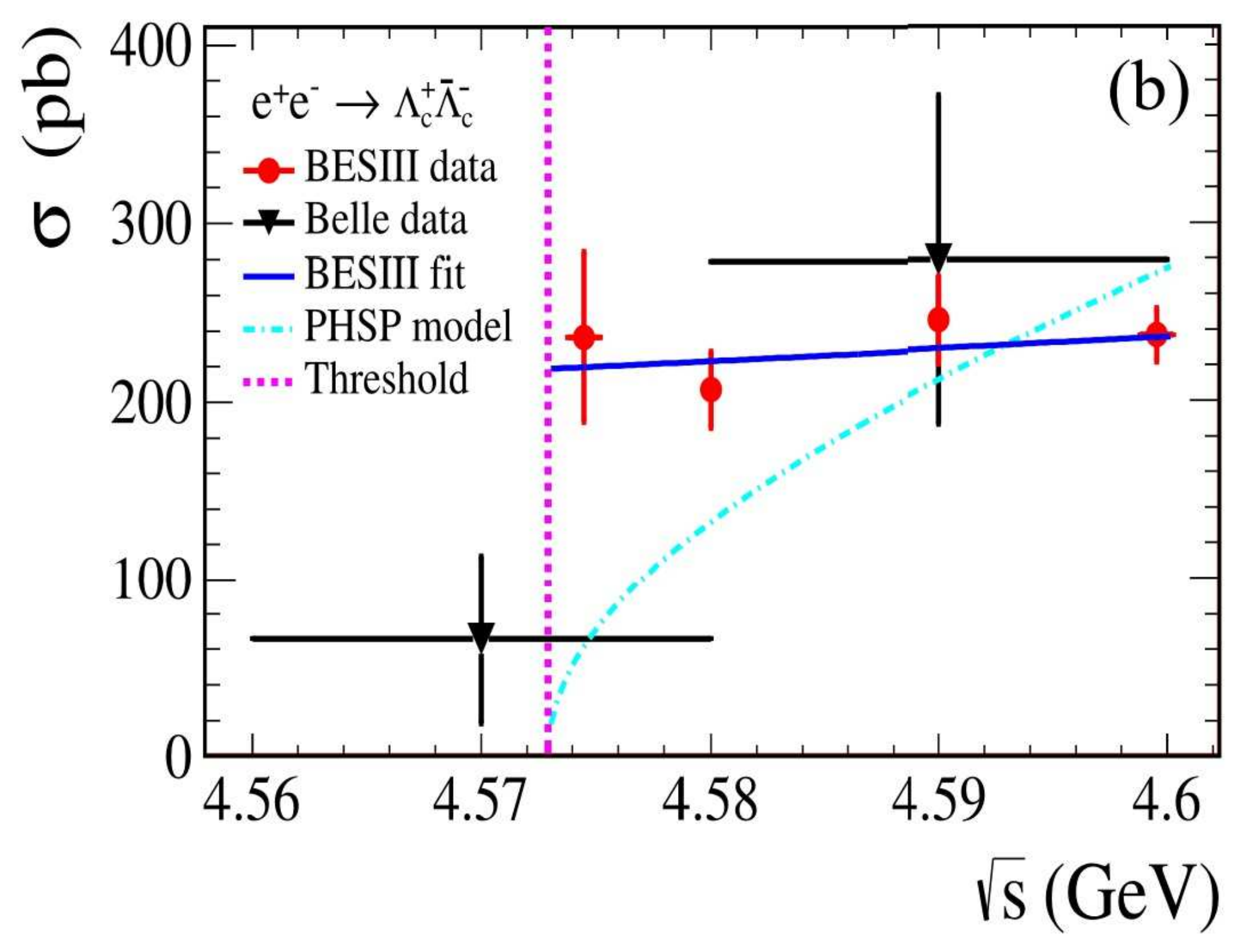}
\caption{The measured $e^+e^-\to \Lambda_c^+ \Lambda_c^-$ cross
sections from (a) Belle~\cite{Pakhlova:2008vn} and
(b) BESIII~\cite{Ablikim:2017lct}
together with those from Belle from threshold to 4.6~GeV.} \label{FitLinShp}
\end{center}
\end{figure}

The $Y(4630)$ was first observed by Belle in the $\Lambda_{c}^{+}
\bar{\Lambda}_{c}^{-}$ invariant mass distribution in the ISR process
$\EE \to \Lambda_{c}^{+} \bar{\Lambda}_{c}^{-}$~\cite{Pakhlova:2008vn}
with a measured mass of $(4634^{+8+5}_{-7-8})$ MeV and a width of
$(92^{+40+10}_{-24-21})$ MeV.
The measured cross sections are shown in Fig.~\ref{FitLinShp}(a),
where the $Y(4630)$ peak is evident near the $\Lambda_{c}^{+}
\bar{\Lambda}_{c}^{-}$ threshold with a statistical significance of $8.8\sigma$.
The measured $Y(4630)$ parameters are consistent within errors with the
mass and width of $Y(4660)$, that was found in
$\pp \psp$ decays via ISR~\cite{Wang:2007ea,Lees:2012pv,Wang:2014hta}.

Recently BESIII performed a measurement of $e^+e^-\to \Lambda_c^+ \bar{\Lambda}_c^-$ using
data samples at $\sqrt{s}=4574.5$, 4580.0, 4590.0, and 4599.5 MeV~\cite{Ablikim:2017lct}.
Figure~\ref{FitLinShp}(b) shows the measured $e^+e^-\to
\Lambda_c^+ \bar{\Lambda}_c^-$ cross sections from BESIII~\cite{Ablikim:2017lct}
together with those from Belle~\cite{Pakhlova:2008vn} near
the $\Lambda_c^+ \bar{\Lambda}_c^-$ mass threshold.
The non-zero cross section from BESIII near the
$\Lambda_c^+ \bar{\Lambda}_c^-$ production threshold is evident.
This means that when the cross sections are fitted, the effect of
the threshold should be considered, which will affect the
parameters of $Y(4630)$. Due to a large uncertainty in the Belle measurement,
such a threshold effect was not considered.
Unfortunately, the maximum energy point that BESIII can achieve is 4.6 GeV. In the future, with much
larger data samples accumulated by Belle II, the $Y(4630)$ parameters can be measured more precisely.

As the $Y(4630)$ mass is very close to that of the $Y(4660)$
observed by Belle and BaBar in the process
$e^+e^- \to \pi^+\pi^-\psp$~\cite{Wang:2007ea,Lees:2012pv,Wang:2014hta},
many theoretical explanations assume they are the same
state~\cite{Bugg:2008sk,Cotugno:2009ys,Guo:2010tk}. In
Refs.~\cite{Guo:2008zg,Guo:2009id}, where the $Y(4660)$ is
modeled as an $f_{0}(980)\psp$ bound state, the authors predicted that it
should have a spin partner---a $f_{0}(980)\eta_{c}(2S)$ bound
state denoted as the $Y_{\eta}$---with a mass and width of $(4613
\pm 4)$ MeV and around 30 MeV, respectively, and a large
partial width into $\Lambda_{c}^{+}\bar{\Lambda}_{c}^{-}$~\cite{Guo:2010tk,Guo:2009id}.

\begin{figure}[htbp]
\begin{center}
\includegraphics[width=6cm]{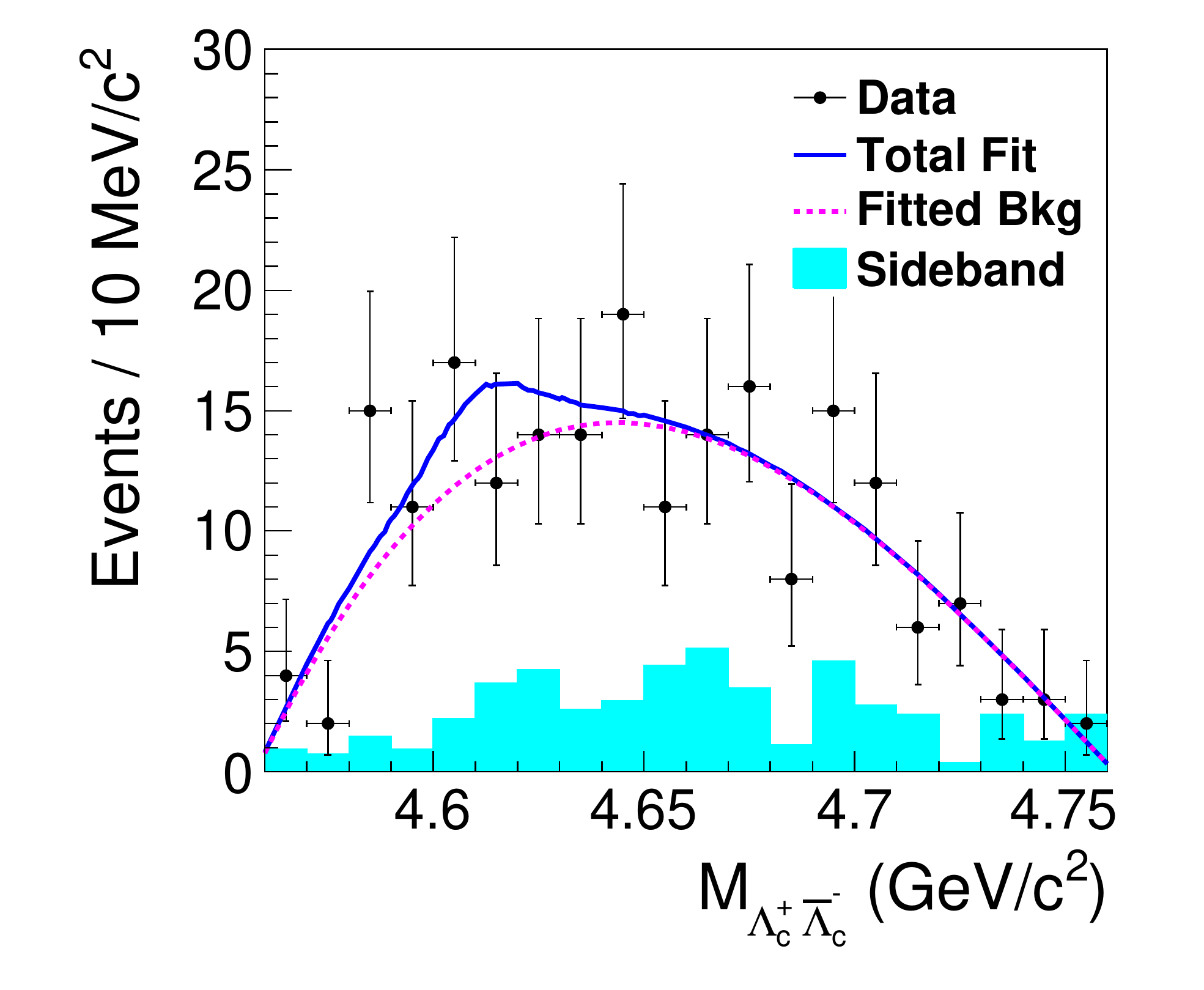}
\includegraphics[width=6cm]{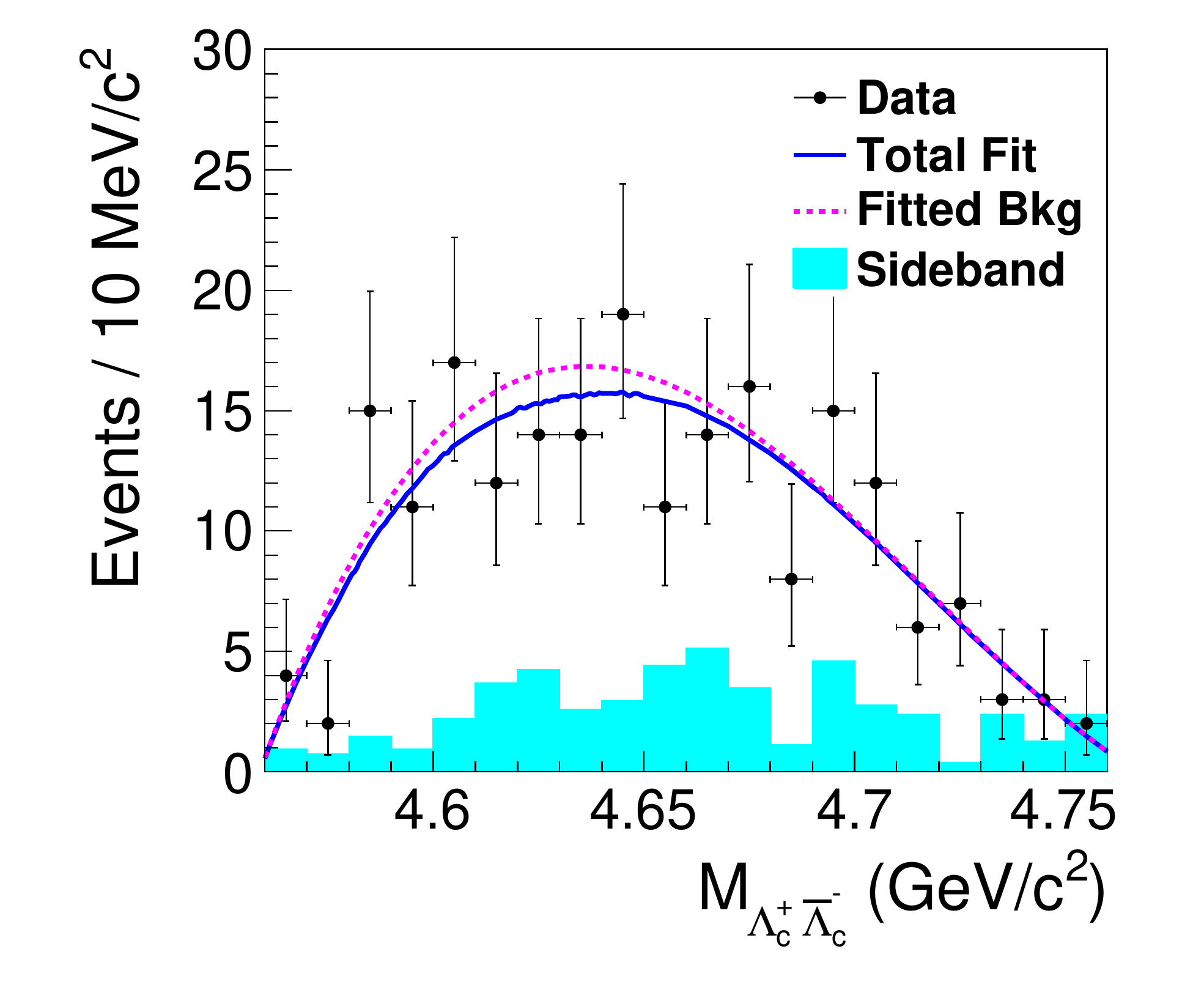}
\put(-300,120){\bf (a)}
 \put(-125,120){\bf (b)}
\caption{\label{Yplot} The $\Lambda_{c}^{+}\bar{\Lambda}_{c}^{-}$
invariant mass spectra in $B^{-}
\to K^{-} \Lambda_{c}^{+} \bar{\Lambda}_{c}^{-}$ decays from
Belle~\cite{Li:2017uvv} with (a)
$Y_{\eta}$ and (b) Y(4660) signals included in the fits. The
shaded cyan histograms are from the normalized $\Lambda_{c}^{+}$
and $\bar{\Lambda}_{c}^{-}$ mass sidebands. }
\end{center}
\end{figure}

To search for $Y_{\eta}$ in the $\Lambda_{c}^{+} \bar{\Lambda}_{c}^{-}$ system,
Belle performed an updated measurement of
$B^{-}\to K^{-} \Lambda_{c}^{+} \bar{\Lambda}_{c}^{-}$ using a sample of
$(772\pm11)\times 10^{6} B\bar{B}$ pairs~\cite{Li:2017uvv}. The
obtained $M_{\Lambda_{c}^{+}\bar{\Lambda}_{c}^{-}}$ spectrum is
shown in Fig.~\ref{Yplot}, in which no clear $Y_{\eta}$ or
$Y(4660)$ signal is seen.
The 90\% C.L.\ upper limits for the $Y(4660)$ and its theoretically predicted spin partner $Y_{\eta}$
are set to be ${\cal B}[B^- \to K^- Y(4660)]$${\cal B}[Y(4660) \to
\Lambda_{c}^{+} \bar{\Lambda}_{c}^{-}]<1.2\times 10^{-4}$ and
${\cal B}[B^- \to K^- Y_{\eta}]{\cal B}[Y_{\eta} \to \Lambda_{c}^{+} \bar{\Lambda}_{c}^{-}]<2.0 \times 10^{-4}$ respectively.
Subsequently, Belle also searched for the $Y(4660)$
and $Y_{\eta}$ in $\bar{B}^{0} \to K^{0}_{S} \Lambda_{c}^{+}
\bar{\Lambda}_{c}^{-}$ decays~\cite{Li:2018fmq}.
Similarly, no evidence of $Y_{\eta}$ or $Y(4660)$ is seen
in the $\Lambda_{c}^{+} \bar{\Lambda}_{c}^{-}$ mass spectrum.
The 90\% C.L.\ upper limits are $\BR[\bar{B}^0 \to \bar{K}^0 Y(4660)]\BR[Y(4660) \to
\Lambda_{c}^{+} \bar{\Lambda}_{c}^{-}] < 2.3 \times 10^{-4}$
and $\BR[\bar{B}^0 \to \bar{K}^0 Y_{\eta}]\BR[Y_{\eta} \to
\Lambda_{c}^{+} \bar{\Lambda}_{c}^{-}] < 2.2 \times 10^{-4}$.

\vspace{0.3cm}\noindent
$\bullet$ {\it Possible $Y$ structures in other final states}
\vspace{0.3cm}

Besides the $\EE$ annihilation final states discussed above in the study
of the $Y(4230)$, $Y(4360)$,
and $Y(4660)$
states, many other final states are studied to search for the new decay modes
of these $Y$ and the excited $\psi$ states, and new $Y$ states.

\vspace{0.3cm}\noindent
{\it 1) $e^+e^- \to K^+ K^- J/\psi$ and $K^0_S K^0_S J/\psi$}
\vspace{0.3cm}

\begin{figure}[htbp]
\begin{center}
\includegraphics[height=6cm]{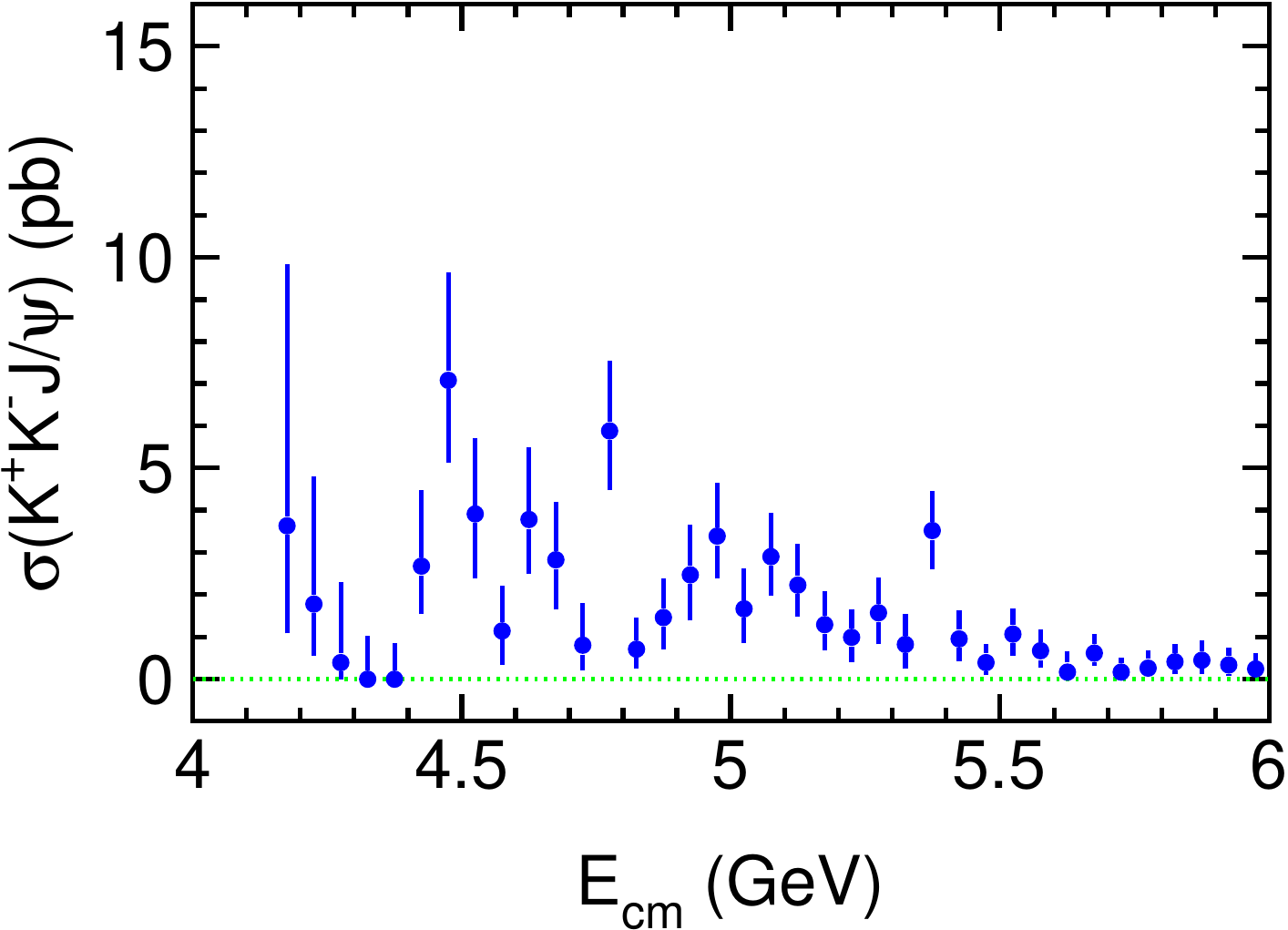}
\includegraphics[height=6cm]{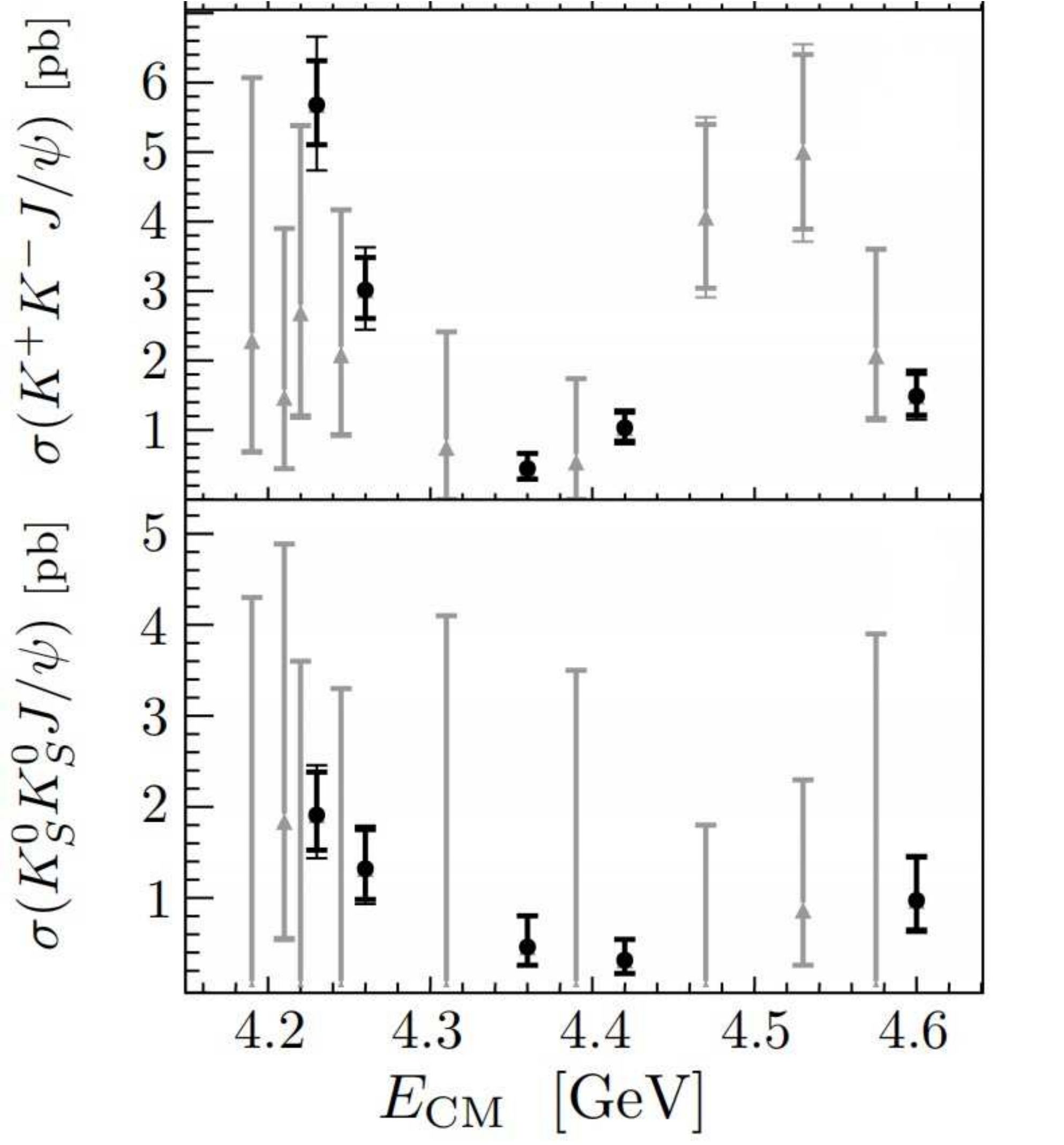}
\caption{\label{bes3-kkjpsi} The measured Born cross sections of
$e^+e^- \to K^+ K^- J/\psi$
from Belle (left panel), and those of
$e^+e^- \to K^+ K^- J/\psi$ and $K^0_S K^0_S J/\psi$ from BESIII (right panel).
For the left panel, the errors are statistical; a 7.8\% systematic error that is
common for all data points is not included.
For the right panel, the black circular points are for data sets
with high integrated luminosities; the gray triangular points
are for smaller data sets. Thicker error bars are for statistical
uncertainties only; thinner error bars are for combined statistical and
systematic uncertainties. In the bottom plot, the large error
bars with no central point are 90\% C.L.\ upper limits.}
\end{center}
\end{figure}

Evidence ($3.7\sigma$) for $\EE\to K^+ K^- J/\psi$ signal was observed by
CLEO using a 13.2~pb$^{-1}$ data sample at
$\sqrt{s}=4.26$~GeV~\cite{Coan:2006rv}. Belle measured for the first time
the cross section of $e^+e^- \to K^+ K^- J/\psi$ for c.m.\ energies between
threshold and 6.0~GeV using 673~fb$^{-1}$ of data~\cite{Yuan:2007bt}.
No significant signal for $Y(4230)\to K^+ K^- J/\psi$ is observed.
At the same time, Belle found evidence for $e^+e^- \to K^0_S K^0_S J/\psi$
in the same data sample.
A few years later, Belle updated the measurements of
the cross sections of $e^+ e^- \to K^+ K^- J/\psi$ and $K_S^0K_S^0J/\psi$ using a
data sample of 980~fb$^{-1}$~\cite{Shen:2014gdm} (Fig.~\ref{bes3-kkjpsi},
left panel).
No significant signal for $Y(4230)\to K^+ K^- J/\psi$ is observed either,
and $\BR[Y(4230)\to K^+ K^- J/\psi]\Gamma[Y(4230)\to e^+e^-]<0.85$~eV at a 90\% C.L.\ is set.
Belle tried to fit the $e^+e^- \to K^+ K^- J/\psi$ cross sections
using either a single BW function or using the $\psi(4415)$ plus a
second BW function, but found that both are inadequate to describe the data.

Very recently, BESIII measured the cross sections of the processes
$e^+e^- \to K^+ K^- J/\psi$ and $K^0_S K^0_S J/\psi$
at c.m.\ energies from 4.189 to 4.600~GeV using 4.7~fb$^{-1}$ of
data~\cite{Ablikim:2018epj}.
The measured Born cross sections of $e^+e^- \to K^+ K^- J/\psi$ and
$K^0_S K^0_S J/\psi$ are shown in the right panel of Fig.~\ref{bes3-kkjpsi}, where
the energy dependence of the cross section for
$e^+e^- \to K^+ K^- J/\psi$ obviously differs from that for
$\pi^+\pi^- J/\psi$ in the
region around the $Y(4230)$. The combined ratio of
the cross sections of $e^+e^- \to K^0_S K^0_S J/\psi$ and
$e^+e^- \to K^+ K^- J/\psi$ over all energies
is $0.370^{+0.064}_{-0.058}\pm0.018$, which is consistent with the expected value
of 0.5 according to isospin symmetry.

If we combine the measurements of $e^+e^- \to K^+ K^- J/\psi$
from Belle and BESIII experiments, it seems that there is a
contribution from the $Y(4230)$ with a peak cross section
of about 5~pb. In addition, there is evidence for a
structure at around 4.5~GeV with a width of about 100~MeV.
It is not clear whether it is a new structure, or the contribution
from the known $Y$ structures.

\vspace{0.3cm}\noindent
{\it 2) $e^+e^- \to \eta J/\psi$, $\eta^{\prime} \jpsi$, $\eta h_c$, and $\pi\rho\etac$}
\vspace{0.3cm}

\begin{figure}[htbp]
\begin{center}
\includegraphics[height=5.8cm]{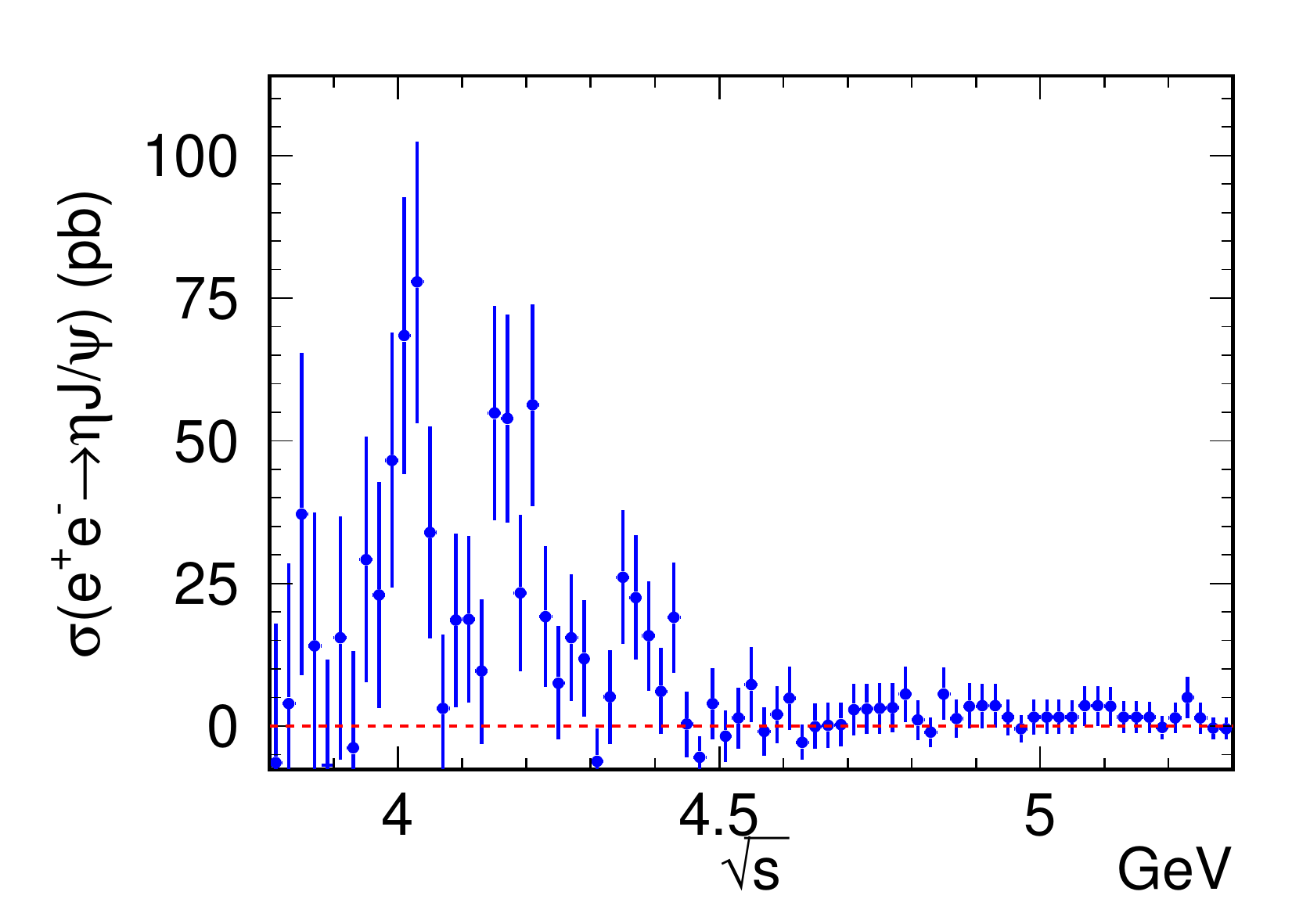}
\includegraphics[height=5.6cm]{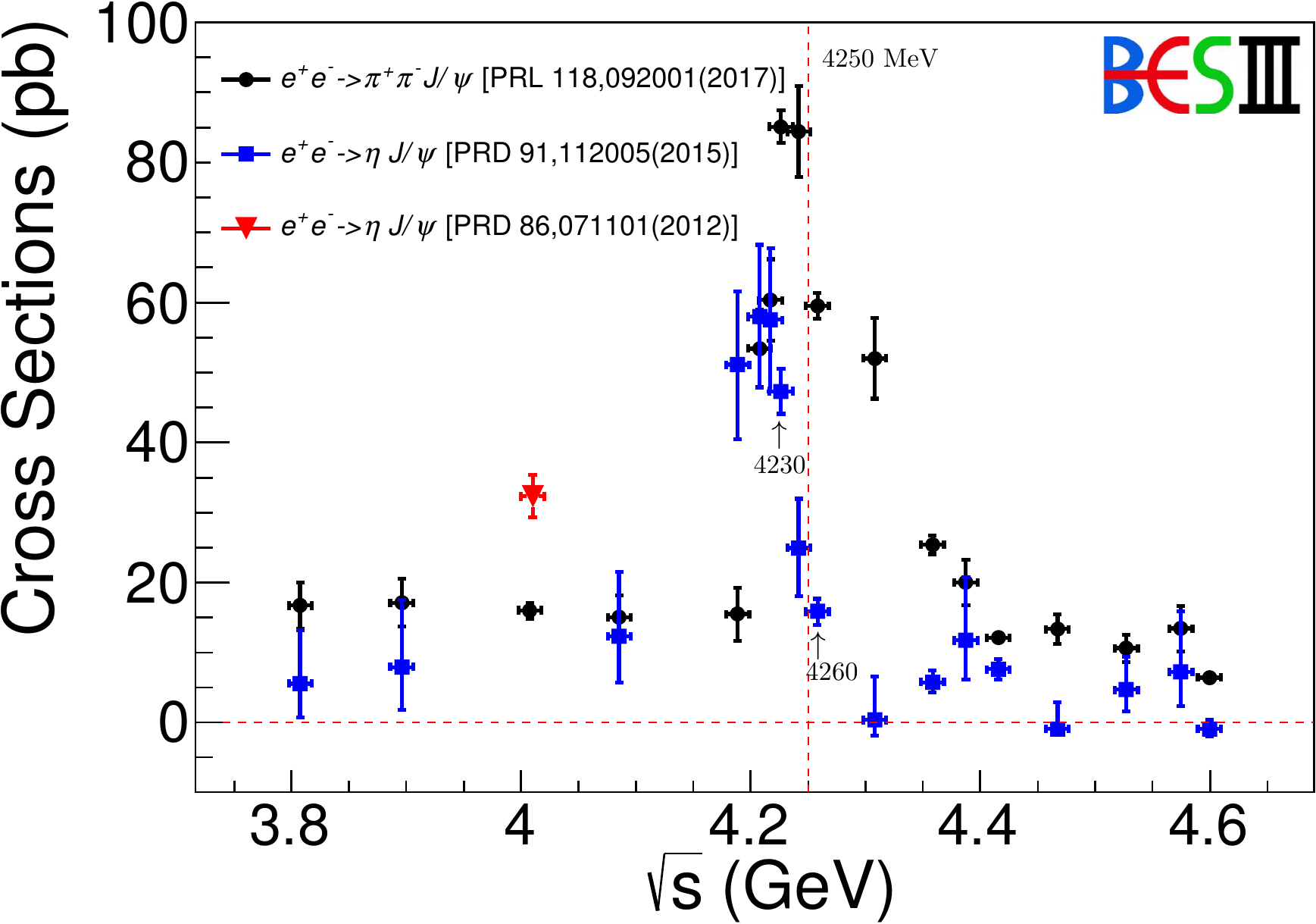}
\caption{\label{etax} The measured Born cross sections of
$e^+e^- \to \eta J/\psi$ from Belle~\cite{Wang:2012bgc} (left panel) and
BESIII~\cite{Ablikim:2012ht,Ablikim:2015xhk} (right panel) together with the cross sections
of $e^+e^- \to \pi^+\pi^- J/\psi$ from BESIII~\cite{Ablikim:2016qzw}
(right panel).
In the left panel, the errors are the
summed statistical errors of the numbers of signal and background
events. A systematic error of 8.0\% common to all the data points is not shown.
In the right panel, the errors are combined statistical and systematic errors.}
\end{center}
\end{figure}

An important way to search for $Y$ states is the investigation of
the $\eta J/\psi$ system.
Belle first reported an investigation of the $e^+e^- \to \eta J/\psi$
process using ISR events. The integrated luminosity used is
980~fb$^{-1}$~\cite{Wang:2012bgc}.
After event selection, two distinct peaks are observed in the $\eta J/\psi$
mass spectrum above 3.8~GeV, one at 4.0~GeV and the other at 4.2~GeV.
A fit to the signal events
with two coherent $P$-wave BWs for $\psi(4040)$ and $\psi(4160)$ and an
incoherent second-order polynomial
background can describe the data well. The statistical significance is
$6.5\sigma$ for $\psi(4040)$ and $7.6\sigma$ for $\psi(4160)$.
The branching fractions of $\psi(4040)\to \eta \jpsi$ and
$\psi(4160) \to \eta \jpsi$
are obtained to be $(0.56\pm0.10\pm0.17)\%$ and
$(1.30\pm0.15\pm0.24)\%$, respectively,
which correspond to about 1~MeV partial widths to $\eta \jpsi$ for these
two states.
Possible contributions from other excited charmonium-like states are examined,
including the $Y(4230)$, $Y(4360)$,
and $Y(4660)$.
None of their significance is larger than $3\sigma$, and 90\% C.L.
upper limits of their production rates in $e^+e^-$ annihilation are determined.
The measured Born cross sections are shown in the left panel of Fig.~\ref{etax}.

Using data samples at c.m.\ energies from 3.810 to 4.600~GeV, BESIII also
performed a study of $e^+e^- \to \eta J/\psi$~\cite{Ablikim:2015xhk}. The
measured cross sections are shown
in Fig.~\ref{etax} (right plot) as blue squares together with
a measurement at $\sqrt{s}=4.009$~GeV~\cite{Ablikim:2012ht}.
These measurements are compatible with the measurement by Belle, but with
a significantly improved precision at certain energies. The measured Born
cross sections are also compared to those
of $e^+e^- \to \pi^+ \pi^- \jpsi$ obtained from
BESIII~\cite{Ablikim:2016qzw} as shown in
Fig.~\ref{etax} with the dots with error bars.
Obviously the processes $e^+e^- \to \eta J/\psi$ and $\pi^+ \pi^- \jpsi$
have different line shapes including the peaking position, which indicates
the existence of a rich spectrum of $Y$ and excited $\psi$ states in this
energy region with different
coupling strengths to the various decay modes. More data samples are needed
to do an accurate measurement and describe better the shape of the cross
section.

\begin{figure}[htbp]
\begin{center}
\includegraphics[height=5cm]{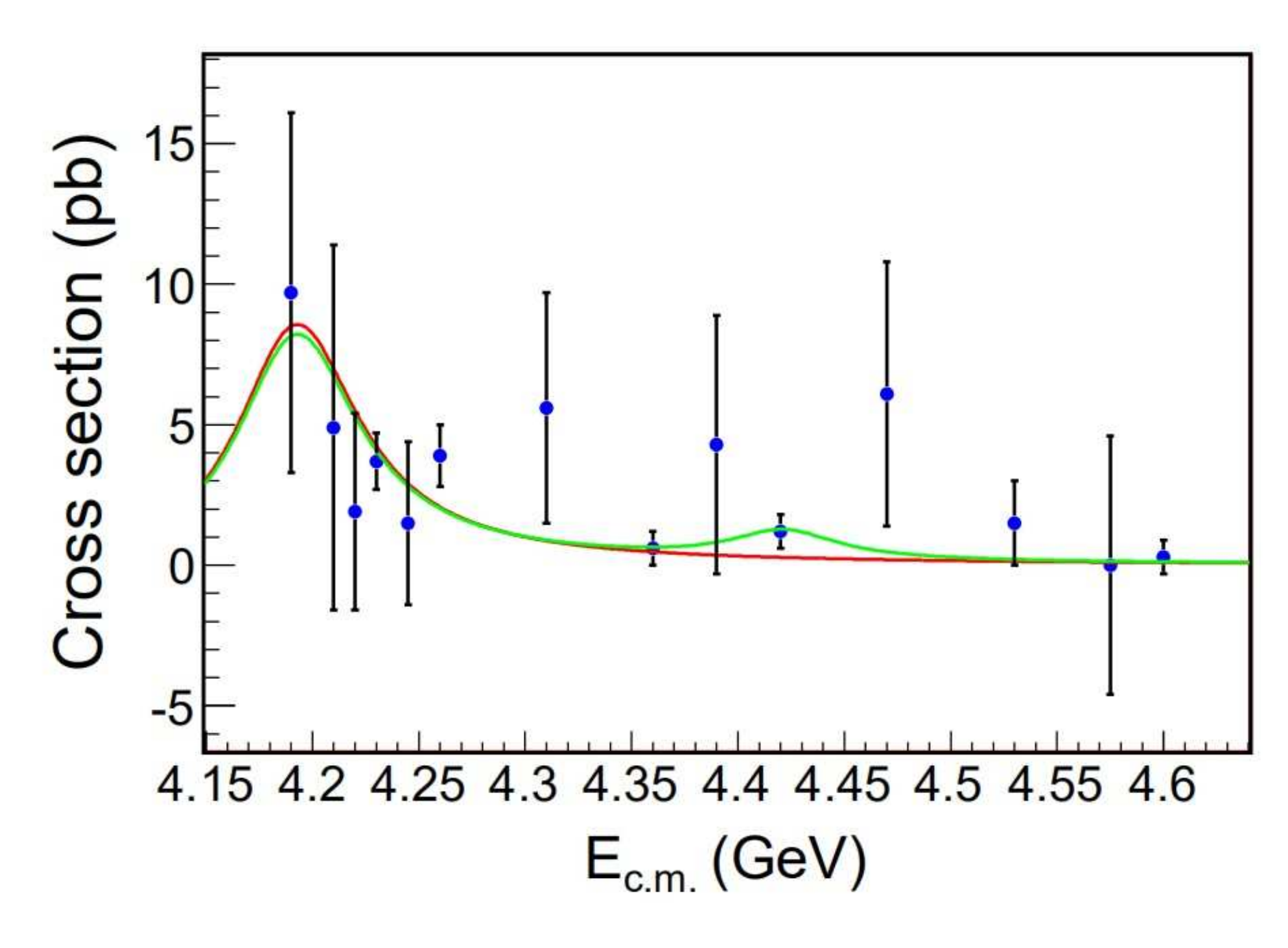}
\caption{\label{bes3-etapjpsi} The measured Born cross sections of
$e^+e^- \to \eta^{\prime} J/\psi$ by BESIII~\cite{Ablikim:2016ymr}.
The red curve is the fit with a $\psi(4160)$
resonance, and the green curve is the fit with an additional $\psi(4415)$
resonance.}
\end{center}
\end{figure}

After the observation of strong $\psi(4040)$ and $\psi(4160)$ decays into
$\eta \jpsi$, it is natural to perform the measurement of
$e^+e^- \to \eta^{\prime} \jpsi$ process.
BESIII did such a measurement to search for potential
$\eta^{\prime} \jpsi$ transitions
from charmonium and charmonium-like states using data samples of about
4.5~fb$^{-1}$
in total at c.m.\ energies from 4.189 to 4.600~GeV~\cite{Ablikim:2016ymr},
where the $\eta^{\prime}$ is reconstructed in two decay channels,
$\eta^{\prime} \to \eta \pi^+ \pi^- \to \gamma \gamma \pi^+ \pi^-$
and $\eta^{\prime} \to \gamma \pi^+ \pi^-$. Figure~\ref{bes3-etapjpsi}
shows the measured Born cross sections for $e^+e^- \to \eta^{\prime} \jpsi$.
Two alternative fits are taken to fit the cross section distribution:
one is the fit with a $\psi(4160)$ resonance with the BW parameters fixed
to PDG values~\cite{Tanabashi:2018oca},
and the other is the fit with an additional $\psi(4415)$ resonance.
The fitted results are shown in Fig.~\ref{bes3-etapjpsi}
as the red and green curves. The statistical significance of $\psi(4415)$ is
2.6$\sigma$ only.
No interesting charmonium-like states are observed with current statistics.

Some $Y$ states are observed decaying into spin-triplet charmonium states with a large rate
since the spin alignment of the $c$ and $\bar{c}$-quarks does not need to be changed between initial and final states.
However, the cross section for the spin-flip process may not be small,
for example the $Y(4230)$ and $Y(4390)$
are reported by BESIII in $e^+e^- \to \pi^+ \pi^- h_c$~\cite{BESIII:2016adj}.
This suggests a more complicated structure of the states containing both heavy-quark
spin-0 and 1 components or a milder heavy-quark spin suppression mechanism as discussed in the theory sections below.
Consequently, searching for the reaction of $e^+e^-$ to final states with an $h_c$ or $\etac$ is very important.

\begin{figure}[htbp]
\begin{center}
\includegraphics[height=4.7cm]{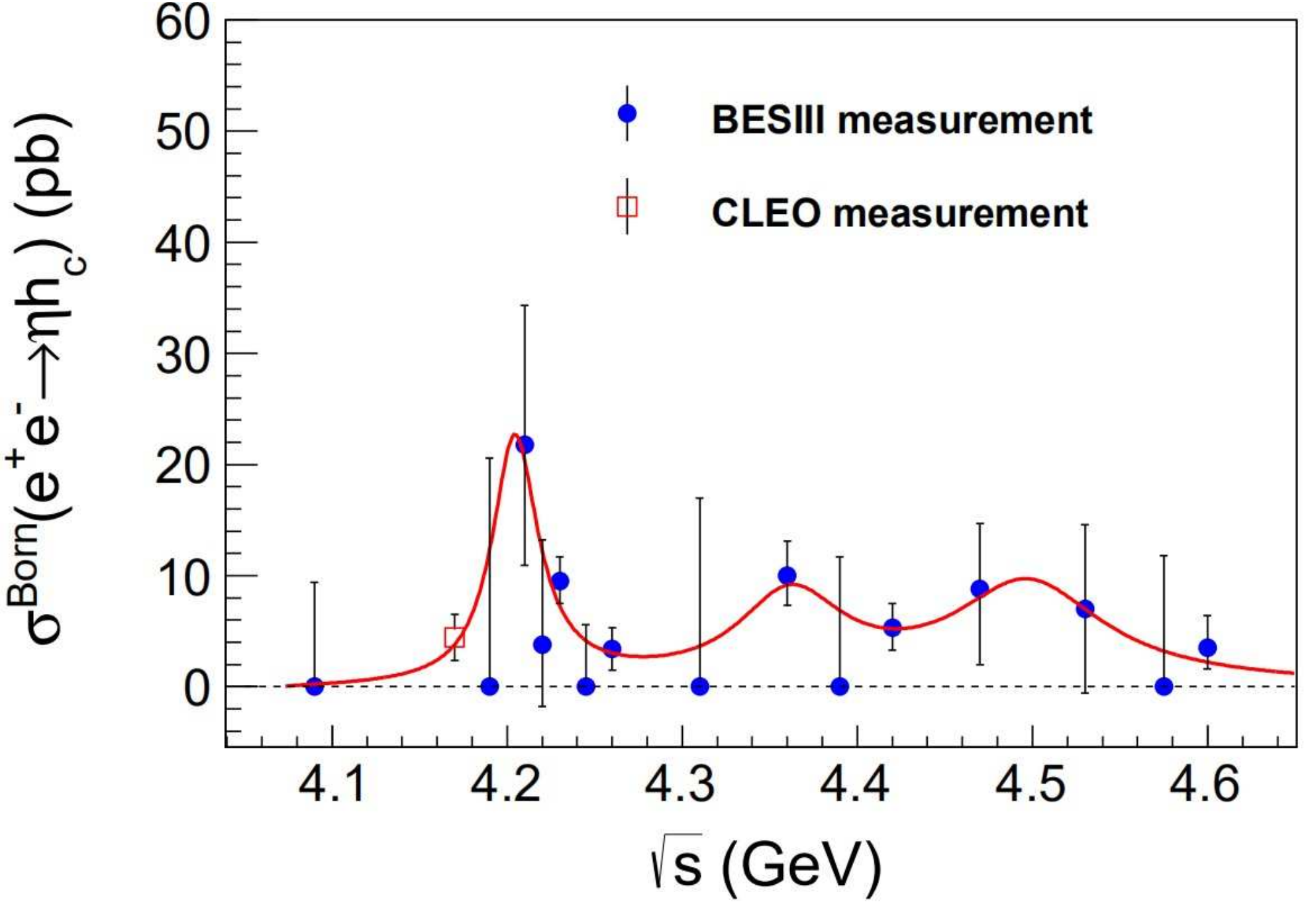}
\caption{\label{etahc} The measured Born cross sections of
$e^+e^- \to \eta h_c$ by BESIII~\cite{Ablikim:2017nkn} and
CLEO~\cite{CLEO:2011aa}.
The solid line is the fit using the coherent sum of three BW functions.}
\end{center}
\end{figure}

BESIII measured $e^+e^- \to \eta h_c \to \eta \gamma \eta_c$
with $\eta_c$ reconstructed with 16 hadronic final states using data samples
of about 4.7~fb$^{-1}$ in total at c.m.\ energies from 4.085 to 4.600~GeV~\cite{Ablikim:2017nkn}.
Figure~\ref{etahc} shows the energy-dependent $e^+e^- \to \eta h_c$ cross sections
together with the measurement at 4.17~GeV from CLEO~\cite{CLEO:2011aa}.
Clear signals and evidence for $e^+e^- \to \eta h_c$
are observed at $\sqrt{s}=4.226$ and 4.358~GeV for the first time,
and the Born cross sections are measured to be $(9.5^{+2.2}_{-2.0}\pm2.7)$
and $(10.0^{+3.1}_{-2.7}\pm2.6)$~pb, respectively.
A fit to the cross section distribution using the coherent sum of three BW
functions, the $Y(4360)$ and other two resonances with
parameters free,
is shown in Fig.~\ref{etahc} with the solid line.
The fitted parameters of the free BWs are: $M_1 =(4204\pm6)$~MeV, $\Gamma_1=(32\pm22)$~MeV,
and $M_2 = (4496\pm26)$~MeV, $\Gamma_2= (104\pm 69)$~MeV,
where the uncertainties are statistical only.
Due to the limited statistics, a further update is needed to determine the line
shape of the c.m.\ energy dependent cross section precisely.

BESIII searched for $\EE\to \pp\piz\etac$ with data at c.m.\ energies
above 4~GeV corresponding to an integrated luminosity of about 4.1~fb$^{-1}$~\cite{Ablikim:2019ipd}.
In this analysis, the $\eta_c$ is reconstructed with nine hadronic final states: $p\bar{p}$, $2(K^+K^-)$, $K^+ K^- \pi^+ \pi^-$, $K^+ K^- \pi^0$, $p \bar{p} \pi^0$,
$\ks K^\pm \pi^\mp$, $\pi^+ \pi^- \eta$, $K^+ K^- \eta$, and $\pi^+ \pi^- \pi^0 \pi^0$.
A clear signal of $\EE\to \pp\piz\etac$ is observed at $\sqrt{s}=4.23$~GeV.
From the fit to the $\eta_c$ mass spectrum, $333^{+83}_{-80}$ $\eta_c$ signal events are obtained with a statistical significance of 4.2$\sigma$.
No significant signals are observed at other c.m.\ energy points. The cross section is measured as
 \(
\sigma(\EE\to \ppp\etac)=(46^{+12}_{-11}\pm 10)~{\rm pb}
 \)
at $\sqrt{s}=4.226$~GeV. It is still not clear whether the signal is from the
$Y(4230)$ decays.

\vspace{0.3cm}\noindent
{\it 3) $\EE \to \omega \chi_{c1,2}$ }
\vspace{0.3cm}

The processes $\EE \to \omega \chi_{c1,2}$ were observed for the
first time by BESIII~\cite{Ablikim:2015uix}. Here, the
$\chi_{c1,2}$ are reconstructed via their $\gamma\jpsi$ decays.
The measured cross sections for $\EE \to \omega \chi_{c1,2}$ are shown in
Fig.~\ref{fig:crosssectionall} (middle and right plots).
A significant $\omega \chi_{c2}$ signal is found at
$\sqrt{s}=$4.42~GeV with an integrated luminosity of 1074 pb$^{-1}$,
and the cross section is measured to be $(20.9 \pm 3.2 \pm 2.5)$~pb.
With 567~pb$^{-1}$ data near $\sqrt{s}=4.6$ GeV, a clear $\omega \chi_{c1}$ signal is
seen, and the cross section is measured to be $(9.5 \pm 2.1 \pm 1.3)$~pb.
Due to low luminosity or low cross section at other energies, no significant
signals are observed. A coherent sum of the $\psi(4415)$ BW function
and a phase-space term can well describe the $\omega \chi_{c2}$ line
shape, and the branching fraction $\psi(4415)\to \omega \chi_{c2}$ is
found to be of the order of $10^{-3}$.
The solid line in Fig.~\ref{fig:crosssectionall} (right plot) shows the fit result.
Further studies based on more data samples at higher energy points
will be helpful to clarify the nature of charmonium(like) states decaying to
$\omega \chi_{cJ}$ final states.

\subsubsection{Searches for glueballs with exotic quantum numbers}
The existence of bound states of gluons (so-called ``glueballs''), with a rich spectroscopy and a complex
phenomenology, is one of the early predictions of the non-Abelian nature of QCD.
Due to the mixing between glueballs and conventional
mesons, the lack of solid information on the glueball production
mechanism, and the lack of knowledge about glueball
decay properties, none of these gluonic states have been established
unambiguously experimentally.

\begin{figure*}[htbp]
\begin{center}
\includegraphics[height=6cm,angle=-90]{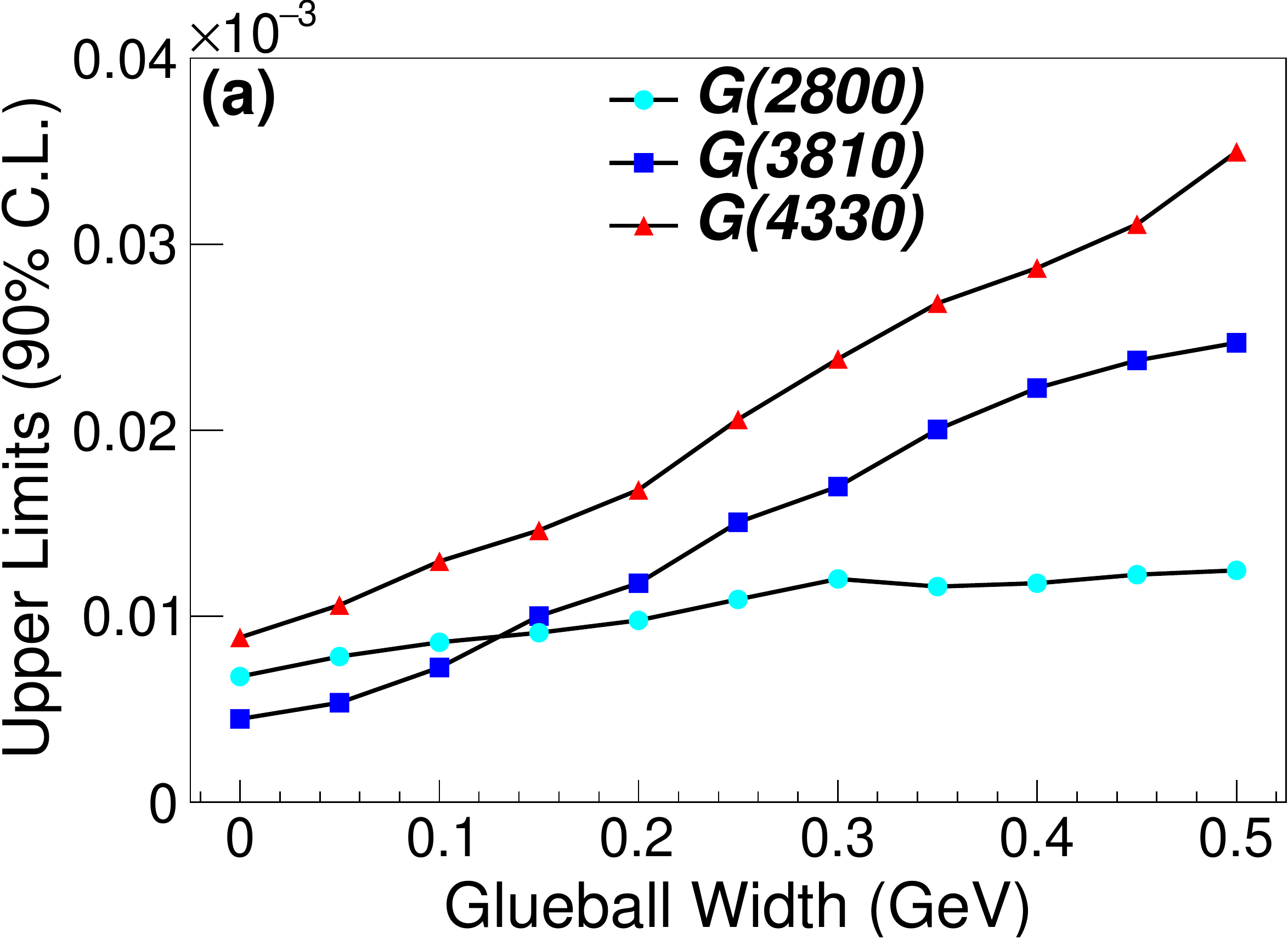}
\hspace{0.3cm}
\includegraphics[height=6cm,angle=-90]{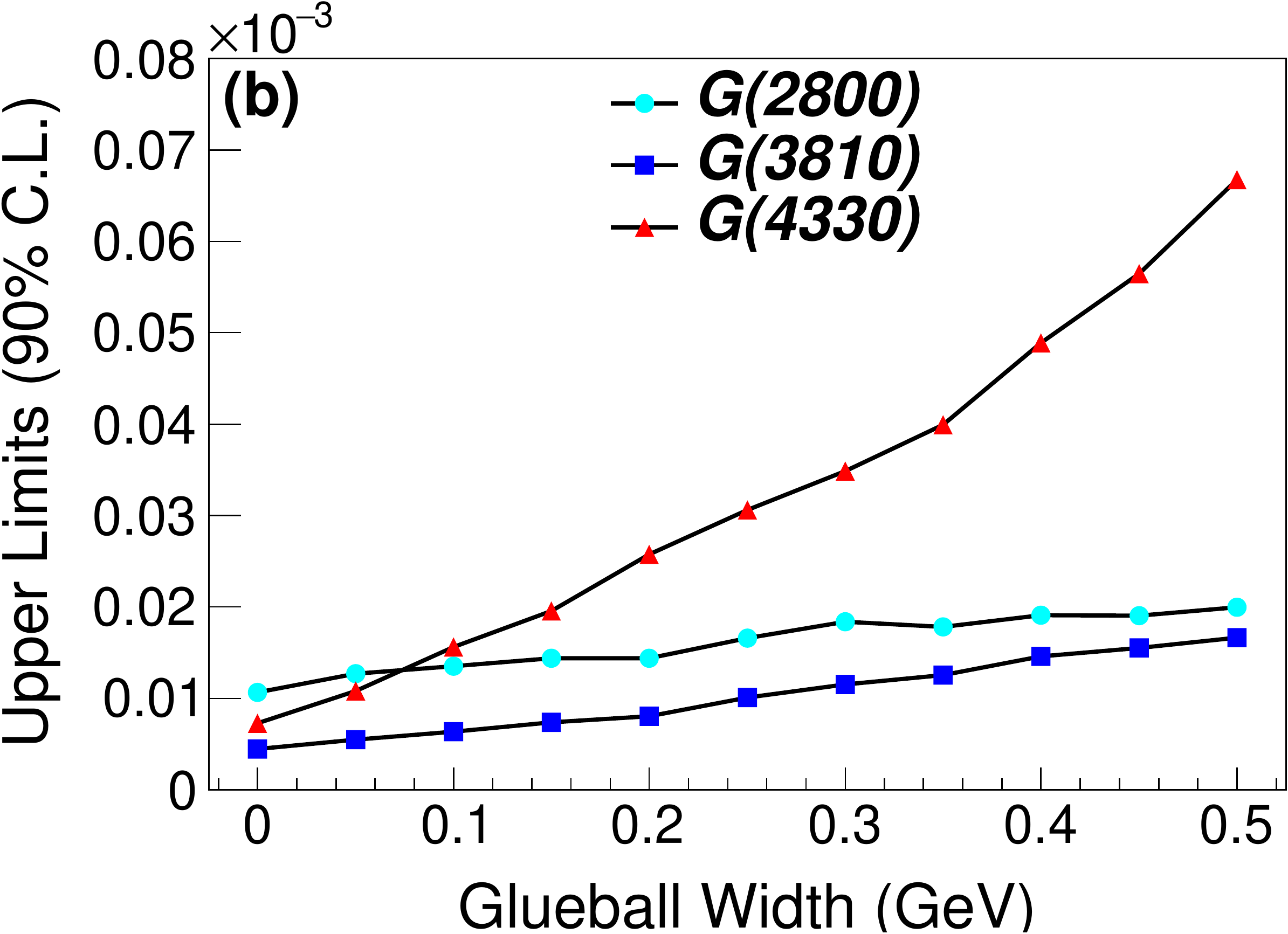}
\end{center}
\caption{The upper limits on the branching fractions for
(a) $\Upsilon(1S)\to \chi_{c1}+G_{0^{--}}$ and (b) $\Upsilon(2S)\to
\chi_{c1}+G_{0^{--}}$
as a function of the assumed $G_{0^{--}}$ width~\cite{Jia:2016cgl}.}\label{fig:final1}
\end{figure*}

Recently Belle utilized the 102M $\Upsilon(1S)$ and 158M $\Upsilon(2S)$
event data samples to search for $0^{--}$ glueballs ($G_{0^{--}}$), called
oddballs, with quantum numbers incompatible with quark-antiquark bound
states~\cite{Jia:2016cgl}. Two $0^{--}$ oddballs are
predicted using QCD sum rules~\cite{Qiao:2014vva} with masses of
$(3.81\pm0.12)$~GeV and $(4.33\pm0.13)$~GeV, while the lowest-lying state
calculated using distinct bottom-up holographic models of
QCD~\cite{Bellantuono:2015fia} has a mass of 2.80~GeV.
Belle searched for such $G_{0^{--}}$ in
$\Upsilon(1S,2S) \to \chi_{c1}+G_{0^{--}}$, $\Upsilon(1S,2S)\to
f_1(1285)+G_{0^{--}}$, $\chi_{b1} \to J/\psi+G_{0^{--}}$, and
$\chi_{b1}\to \omega+G_{0^{--}}$ processes.
No evident signal is found at the three theoretically-predicted
masses in all the studied processes, and $90\%$ C.L.\ upper limits are set on the
branching fractions for these processes.
Figure~\ref{fig:final1} shows the 90\% C.L.\ upper limits
on the branching fractions of $\Upsilon(1S)/\Upsilon(2S)\to \chi_{c1}+G_{0^{--}}$
as a function of the $0^{--}$ glueball width as an example.
Interestingly, a signal with a significance of 3.7$\sigma$ at 3.92~GeV
is observed in $\Upsilon(1S) \to f_1(1285)+G_{0^{--}}$, which will need
special attention and inspection at Belle II with much larger data samples.

\subsection{Isovector states}
\label{Sect:3.2}
While the states discussed in the previous section could, in principle, be interpreted
as conventional heavy quark-antiquark bound states distorted, e.g., by threshold or unitarization effects,
the isovector states discussed in this section clearly require going
beyond the most naive quark-antiquark picture --- at least if these states indeed exist as poles in the $S$--matrix.
Some authors claimed, however, that these states might be simply threshold cusps.
This issue is discussed at the beginning of Sec.~\ref{sect:4}.
In this section, we present the experimental evidence for structures in the
heavy quarkonium mass range with isovector quantum numbers.

\subsubsection{The $X(5568)$ state}
\label{Sect:3.2.1}
\begin{figure*}[hbt]
\begin{center}
 \includegraphics[width=0.45\linewidth]{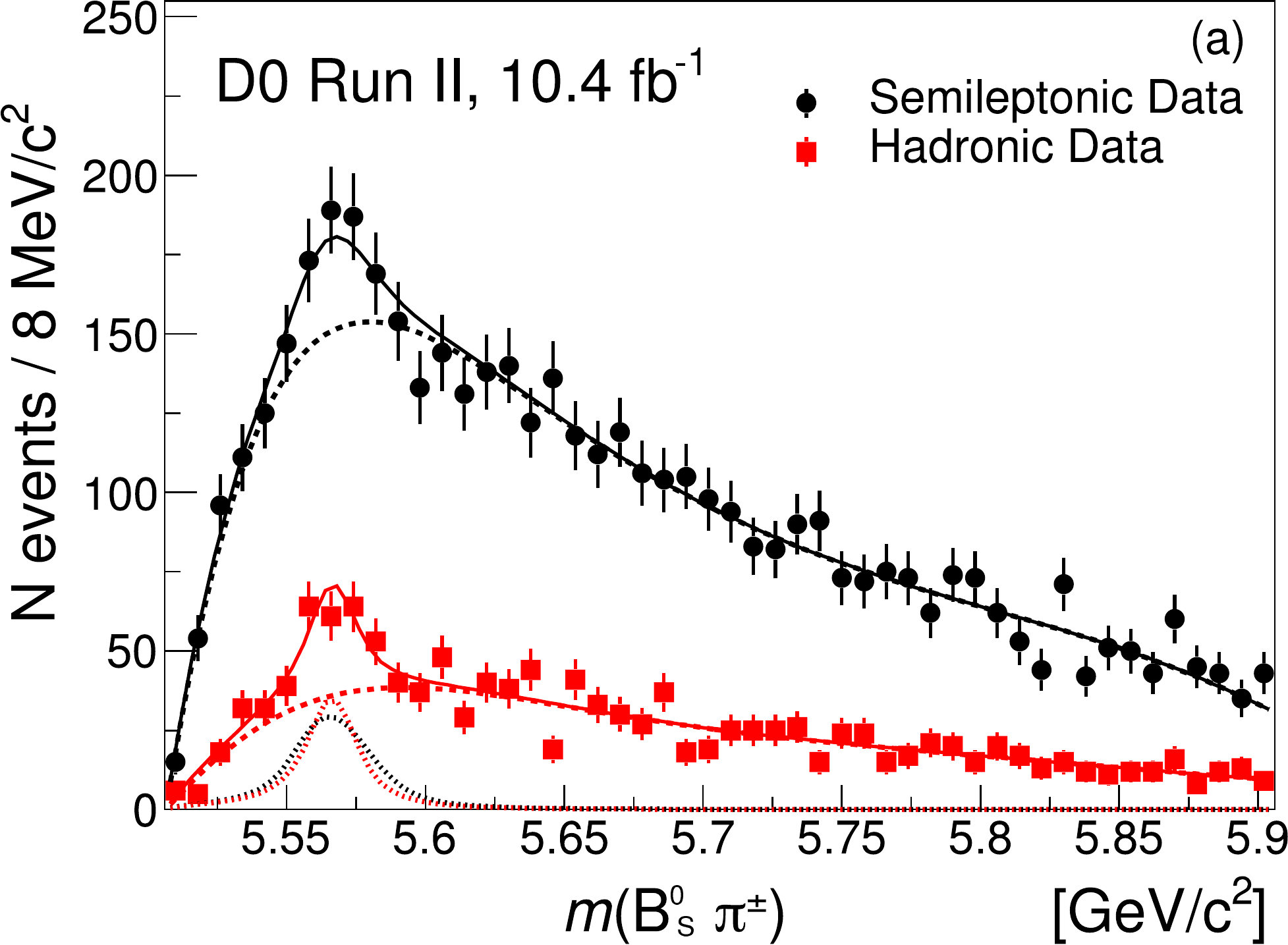}
 \includegraphics[width=0.45\linewidth]{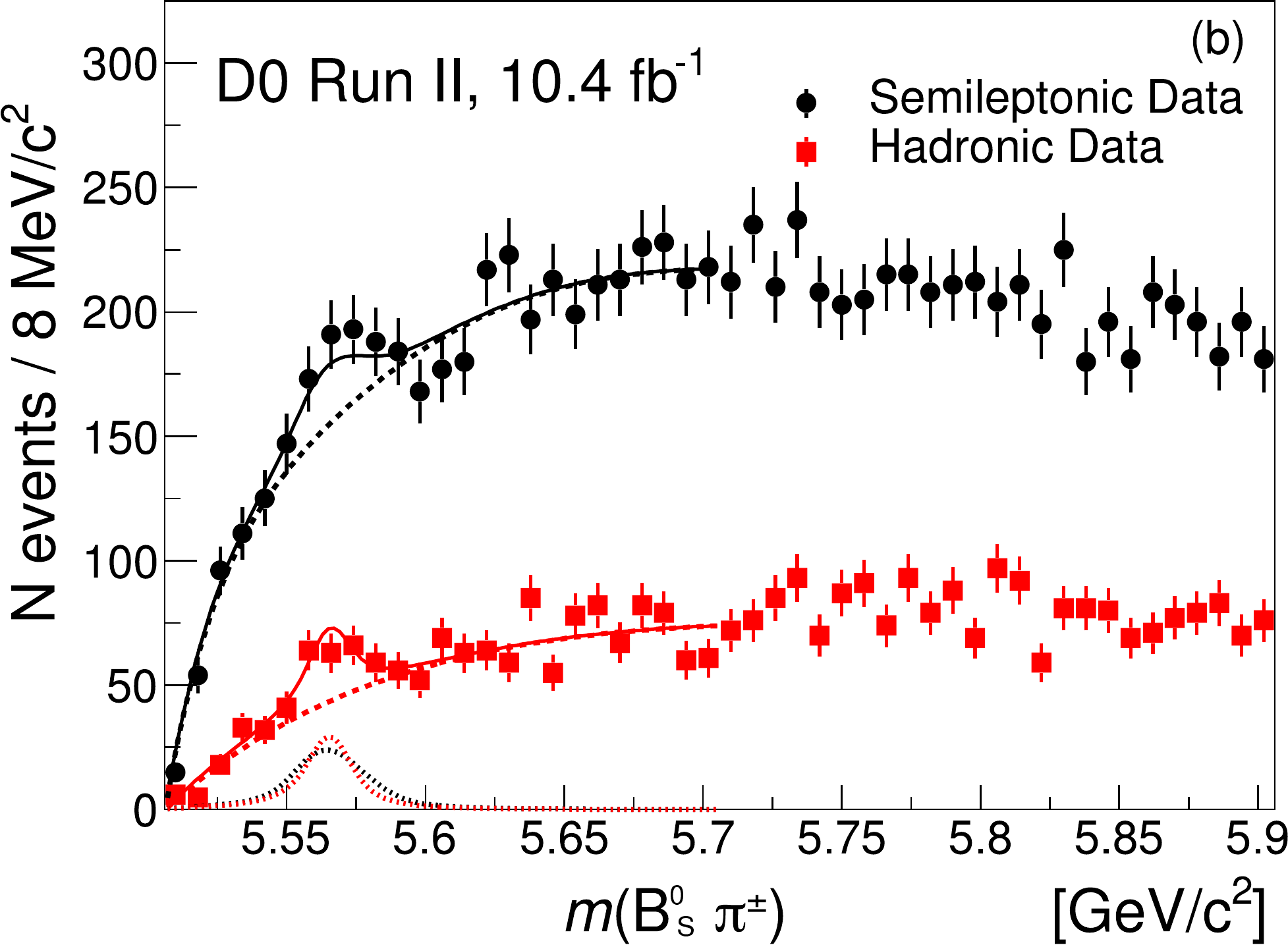}
\end{center}
\caption{\label{fig:fit_combined} The $M(B_s^0\pi^\pm)$
distributions for the hadronic (red squares) and semileptonic
(black circles) data (a) with and (b) without the cone
requirement~\cite{Abazov:2017poh}. The solid lines are the best fits. }
\end{figure*}

The D0 Collaboration reported evidence for a narrow structure,
$X(5568)$, in the hadronic decay
$X(5568) \rightarrow B_s^0 \pi^{\pm}$ with $B_s^0 \rightarrow J/\psi \phi$
with a signal significance greater than 3.9$\sigma$ in proton-antiproton
collisions at a c.m.\ energy of 1.96~TeV~\cite{D0:2016mwd}.
The measured mass and width are $M = (5567.8\pm 2.9^{+0.9}_{-1.9})$ MeV
and $\Gamma = (21.9 \pm 6.4^{+5.0}_{-2.5})$ MeV. Since the state
can be interpreted as a compact tetraquark with four different
valence quark flavors, $b,~s,~u,~d$, it has received extensive
attention from theoretical and experimental physicists.
Later the $X(5568)$ signal was confirmed by the D0 Collaboration in the
semileptonic decay of $B_s^0 \rightarrow \mu^\mp D_s^\pm \, \mathrm{X}$
with a statistical significance greater than $4.3\sigma$~\cite{Abazov:2017poh}. Figure~\ref{fig:fit_combined}
shows the $M(B_s^0\pi^\pm)$ distributions for the hadronic (red
squares) and semileptonic (black circles) data (a) with and (b)
without the cone requirement $\Delta R = \sqrt{\Delta\eta^2 +
\Delta\phi^2} <0.3$, where the cone is the angle between the
$B_s^0$ and $\pi^{\pm}$ for hadronic decay, the angle between the
$\mu^\mp D_s^\pm$ system and $\pi^{\pm}$ for semileptonic decay,
$\eta=-\ln[\tan(\theta/2)]\ $ is the pseudorapidity ($\theta$ is
the polar angle between the track momentum and the proton beam
direction), and $\phi$ is the azimuthal angle of the track. A
combined fit is performed to the selected hadronic and
semileptonic signal candidates, and the fitted results are shown
in Fig.~\ref{fig:fit_combined} as solid lines. The fitted mass and
width are $M = (5566.9 ^{+3.2+0.6}_{-3.1-1.2})$\,MeV and $\Gamma =
(18.6 ^{+7.9+3.5}_{-6.1-3.8}) $\,MeV. The signal significance with the
systematic uncertainties included is 6.7$\,\sigma$ with the cone
requirement, and 4.7$\,\sigma$ without it.

\begin{figure*}[hbt]
\begin{center}
 \includegraphics[width=0.38\linewidth]{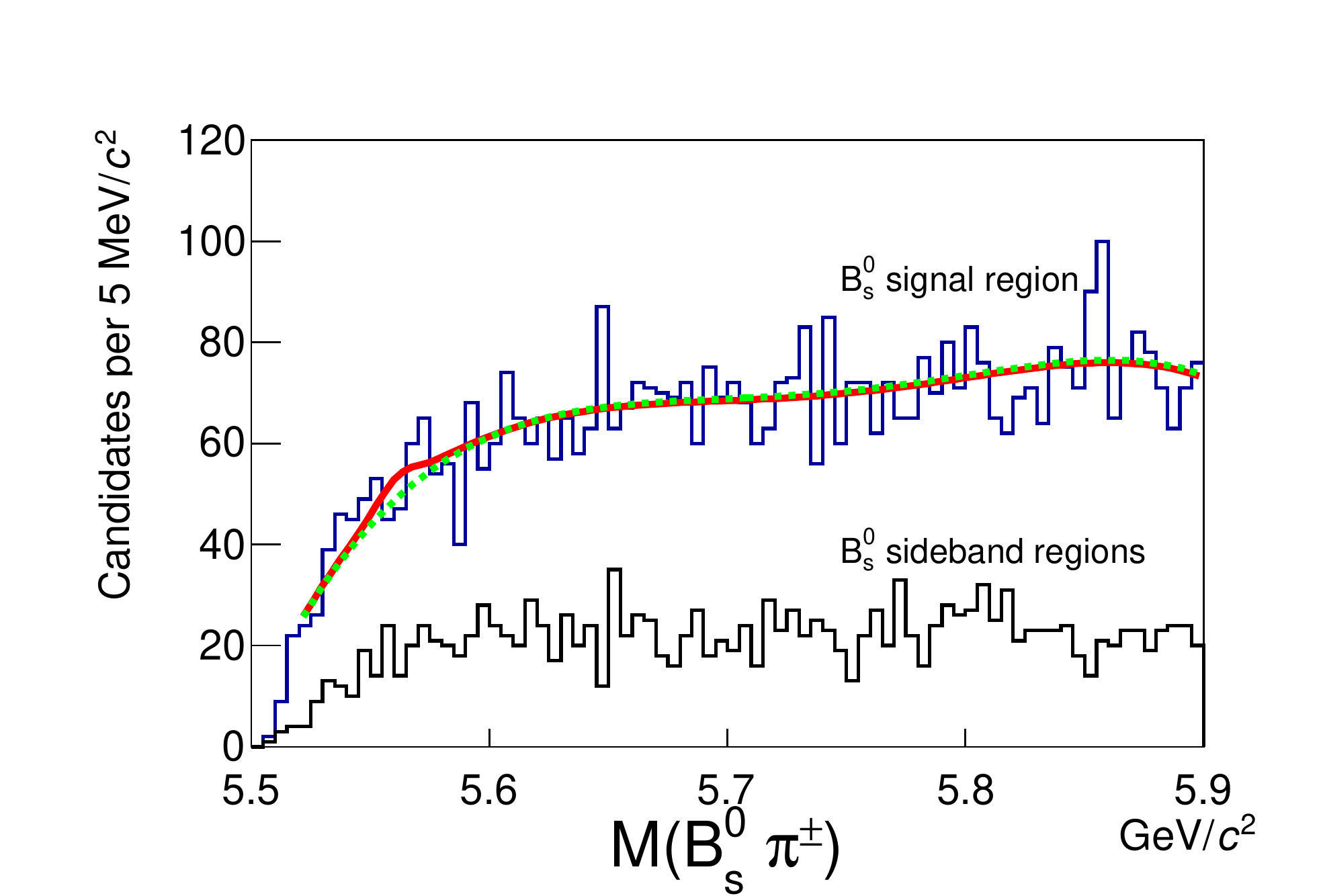}
 \includegraphics[width=0.31\linewidth]{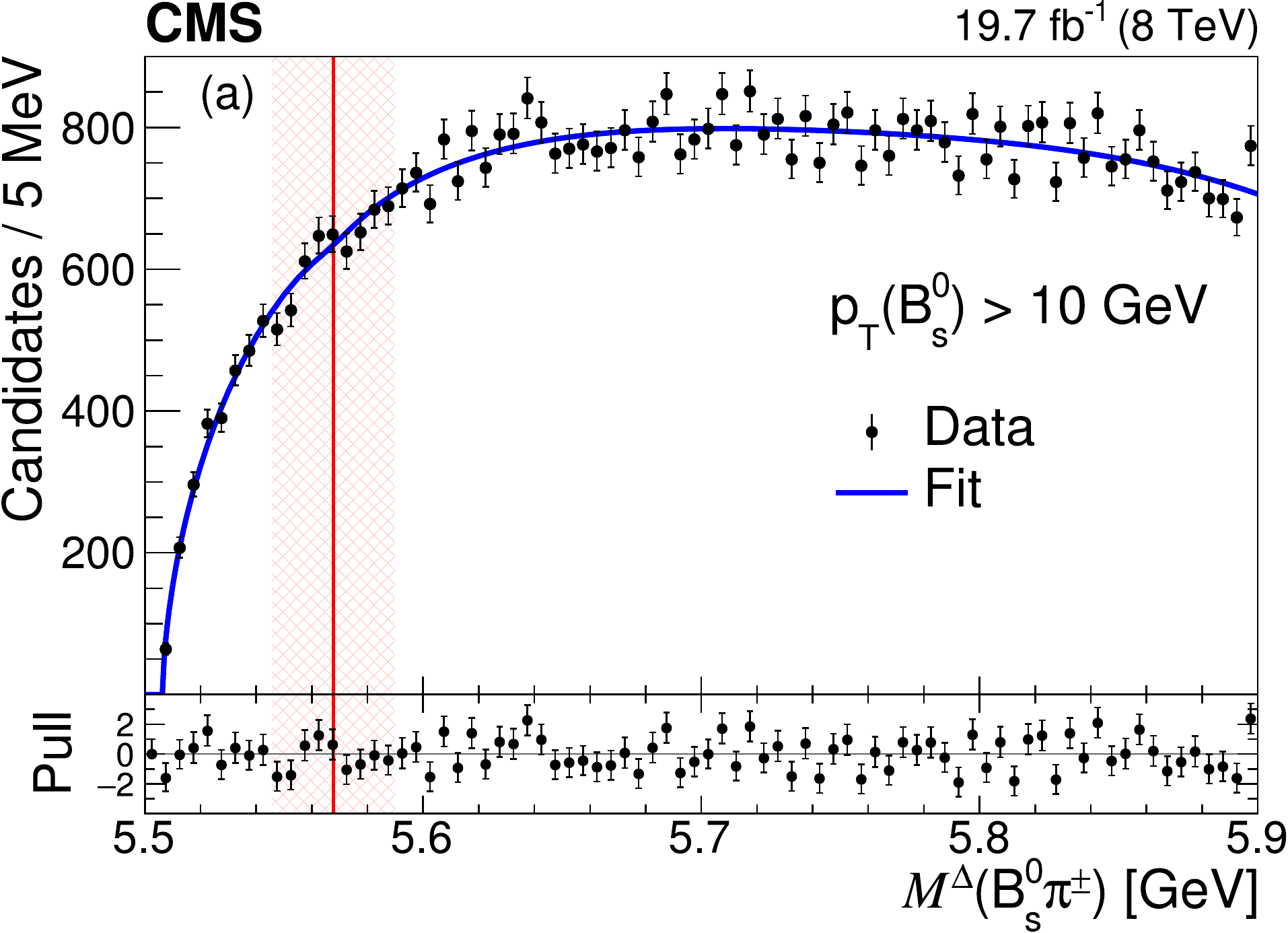}
 \includegraphics[width=0.29\linewidth]{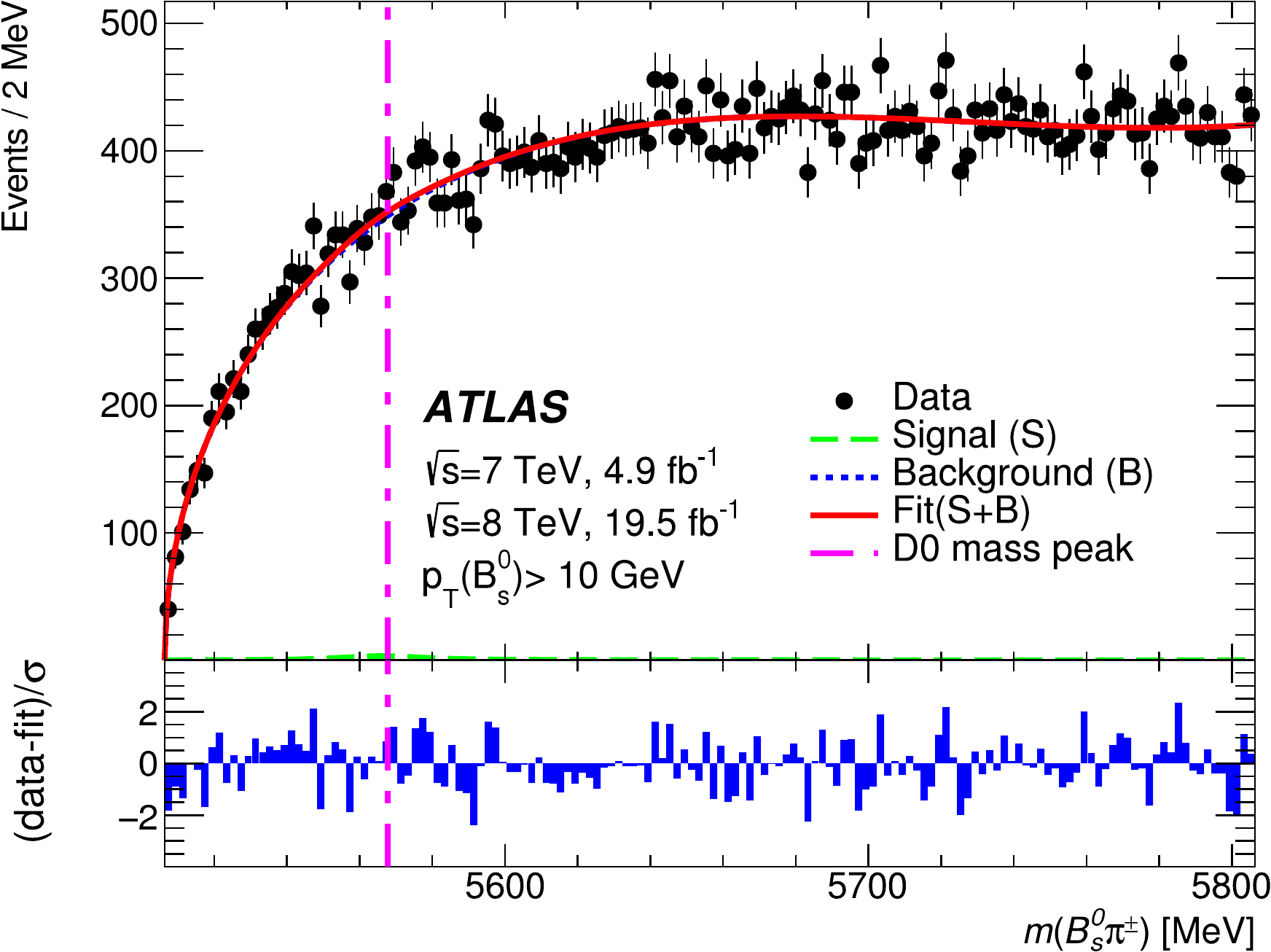}
\end{center}
\caption{\label{fig:nobspi} The $M(B_s^0\pi^\pm)$ distributions
from the CDF~\cite{Aaltonen:2017voc} (left plot),
CMS~\cite{Sirunyan:2017ofq} (middle plot), and
ATLAS~\cite{Aaboud:2018hgx} (right plot) measurements for
$B_s^0\pi^\pm \to J/\psi \phi \pi^\pm$ candidates with
$p_T(B_s^0)>10$ GeV. The solid lines are the best fits. The bottom
panels in the middle and right plots show the difference between each data
point and the fit divided by the statistical uncertainty of that
point.}
\end{figure*}

While the D0 Collaboration reported the $X(5568)$ in both
hadronic and semileptonic decays, negative results were obtained in the
LHCb~\cite{Aaij:2016iev}, CDF~\cite{Aaltonen:2017voc},
CMS~\cite{Sirunyan:2017ofq}, and ATLAS~\cite{Aaboud:2018hgx}
experiments. LHCb first provided negative results
using a 3~fb$^{-1}$ data sample of $pp$
collision data at $\sqrt{s} = 7$ and $8$~TeV~\cite{Aaij:2016iev},
where the $B_s^0$ mesons are reconstructed through the decays
$B_s^0\to D_s^{\pm}\pi^{\mp}$ and $J/\psi \phi$.
The reconstructed $B_s^0$ yield is approximately 20 times larger
than that used by D0. No significant signal is seen in the $B_s^0
\pi^{\pm}$ invariant mass distribution in the range of about 5.5
GeV to 6 GeV, and upper limits on $\sigma(pp \to X + {\rm
anything}){\cal B}(X \to B_s^0 \pi^{\pm})$ are set as a function
of the mass and width of a possible exotic meson decaying to the
$B_s^0 \pi^{\pm}$ final state. Recently, CDF, CMS, and ATLAS also
searched for the $X(5568)$ in hadronic decays using data samples of
$9.6~\textrm{fb}^{-1}$ from $p{\bar p}$ collisions at
$\sqrt{s}=1.96$ TeV, 19.7 fb$^{-1}$ from $pp$ collisions
at $\sqrt{s} = 8$ TeV, and 4.9 fb$^{-1}$ at $\sqrt{s}=7$ TeV and 19.5
fb$^{-1}$ at $\sqrt{s}=8$ TeV from $pp$ collisions,
respectively~\cite{Aaltonen:2017voc,Sirunyan:2017ofq,Aaboud:2018hgx}.
Figure~\ref{fig:nobspi} shows the reconstructed
$M(B_s^0\pi^\pm)$ distributions from the
CDF~\cite{Aaltonen:2017voc} (left plot),
CMS~\cite{Sirunyan:2017ofq} (middle plot), and
ATLAS~\cite{Aaboud:2018hgx} (right plot) measurements for
$B_s^0\pi^\pm \to J/\psi \phi \pi^\pm$ candidates with
$p_T(B_s^0)>10$ GeV, respectively. No statistically significant
peaks can be seen in either of them. Upper limits of 6.7\% from
CDF, 1.1\% from CMS, and 1.5\% from ATLAS on the fraction of
$B^0_s$ produced through the $X(5568) \rightarrow B^0_s \,
\pi^{\pm}$ process at the 95\% C.L.\ are set for
$p_T(B_s^0)>10$ GeV, respectively.

The $X(5568)$ state, if confirmed, would differ from any previously observed state,
as it must have constituent quarks with four different flavors \mbox{($b$, $s$, $u$, $d$)}.
So far, however, only the D0 experiment observed signals, while other experiments with
larger statistics reported negative results, especially the CDF
experiment from the same proton-antiproton collisions. Therefore,
the urgent task in the future is to confirm or definitely exclude the
existence of the $X(5568)$ with larger data samples and to look
for possible similar exotic states in the $B_s^{\ast} \pi$, $D_s \pi$, and $D_s^{\ast} \pi$ systems.

From the theoretical side, many explanations on the $X(5568)$ being a compact tetraquark
state~\cite{Agaev:2016mjb,Agaev:2016ijz,Agaev:2016lkl,Agaev:2016urs,Agaev:2016ifn,Wang:2016mee,Wang:2016wkj,Chen:2016mqt,Liu:2016ogz,Dias:2016dme,Stancu:2016sfd,Zhang:2017xwc}
or a meson molecule~\cite{Wang:2018jsr,Ke:2018jql} have been proposed.
At the same time, there have also been theoretical analyses that
do not support the interpretation as a compact tetraquark state~\cite{Wang:2016tsi},
as a molecular state~\cite{Albaladejo:2016eps,Kang:2016zmv,Lang:2016jpk,Chen:2016ypj,Lu:2016kxm},
or as any of the two~\cite{Chen:2016npt,Guo:2016nhb}.
In Ref.~\cite{Burns:2016gvy} all the possible interpretations, such as compact tetraquark, hadronic molecule, threshold
effect from the meson loop, and so on, have been found unable to provide a consistent explanation of the $X(5568)$.
Finally, due to the inconsistency of the interpretations in both the compact tetraquark and hadronic molecular scenarios,
the authors of Refs.~\cite{Ke:2018stp,Albuquerque:2016nlw,Esposito:2016itg} have suggested that the state might
originate from a mixing of these two scenarios.

\subsubsection{Charmonium-like charged $Z_c$ states}
\label{Sect:3.2.2}
Searching for charged charmonium-like states is one of the most
promising ways of studying exotic hadrons, since such a state
must contain at least four quarks and thus cannot be a conventional quark-antiquark meson.
Searches were performed in the combination of one charged pion/kaon/proton and a charmonium state,
like $\eta_c$, $\jpsi$, $\psp$, $\chi_c$, and $\hc$.

The first charged charmonium-like state, $Z_c(4430)^-$,
was reported in the $\pi^-\psp$ mass spectrum in $B\to K
\pi^-\psp$ decays in the Belle
experiment~\cite{Choi:2007wga,Chilikin:2013tch}, and it was confirmed
by the LHCb experiment seven years later~\cite{Aaij:2014jqa}. The
$\zc^-$ was observed in $\pi^-\jpsi$ invariant mass distribution
in the study of $\EE\to \ppjpsi$ at BESIII~\cite{Ablikim:2013mio} and
Belle~\cite{Liu:2013dau} experiments, and the $\zcp^-$ was observed
in the $\pi^-\hc$ system in $\EE\to \pphc$~\cite{Ablikim:2013wzq} only at
BESIII. There are other $Z_c$ states observed in different processes, such
as the two structures at 4050 and 4250~MeV in the
$\pi^-\chi_{c1}$ system in $B\to K \pi^-\chi_{c1}$
decays~\cite{Mizuk:2008me}; and the $Z_c(4200)^-$ in the
$\pi^-\jpsi$ invariant mass distribution in $B\to K \pi^-\jpsi$
decays~\cite{Chilikin:2014bkk} from the Belle experiment.
There is also evidence for $Z_c$ structures in the
$\pi\psp$ system at Belle~\cite{Wang:2014hta} and
BESIII~\cite{Ablikim:2017oaf} in $\EE\to \pp\psp$,
and in the $\pi \eta_c$ system at LHCb~\cite{Aaij:2018bla} in
$B^0 \to K^+ \pi^- \eta_c$.

These states indicate that a new class of hadrons has been observed.
As they are made by at least four quarks, they have been interpreted either as compact tetraquark states,
molecular states of two charmed mesons ($\bar{D}D^*$, $\bar{D}^*D^*$, $\bar{D}D_1$, $\bar{D}^*D_1$, etc.),
hadroquarkonium states, or other configurations~\cite{Chen:2016qju, Brambilla:2010cs}.

\vspace{0.3cm}\noindent
$\bullet$ {\it The $Z_c(4430)$ state}
\vspace{0.3cm}

\begin{figure*}[hbt]
\begin{center}
 \includegraphics[height=5cm]{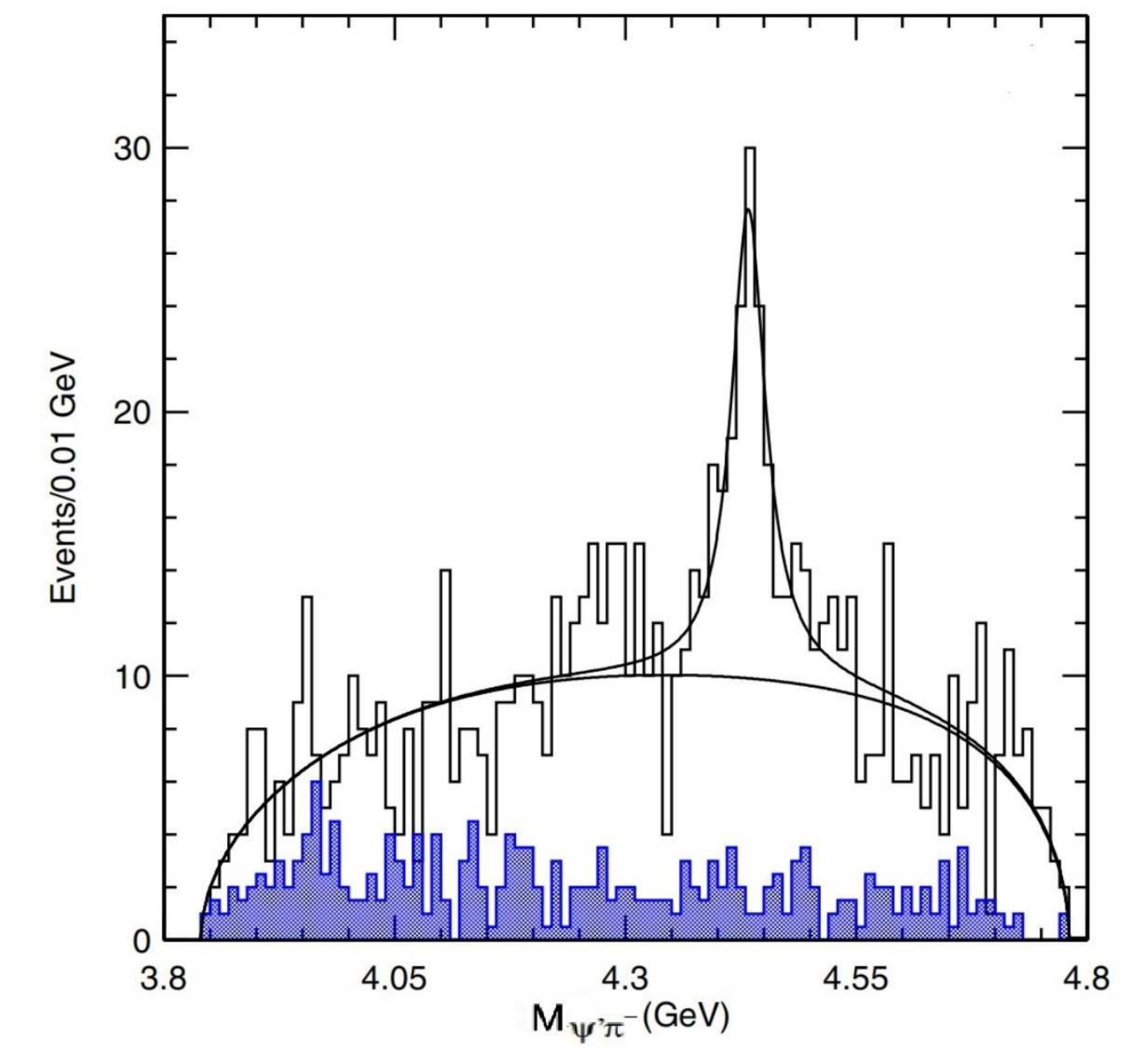}
\end{center}
\caption{\label{belle-zc4430} The $M[\pi^- \psp]$
 distribution by Belle for events in the selected $B$ signal region and with the
 $K^{\ast}(892)$ and $K_2^{\ast}(1430)$ signals veto applied~\cite{Choi:2007wga}. The
shaded histogram is from non-$B$ backgrounds.
The solid curves show the best fit results with a relativistic $S$-wave
BW function to model the peak plus a smooth phase-space-like function.}
\end{figure*}

A feature that clearly distinguishes multiquark states from hybrids or
charmonia is the possibility to
have charmo{\-}nium-like mesons with nonzero electric charge.
Thus, Belle studied exclusive $B\to K \pi^-\psp$ decays
to search for a charged charmonium-like state in the $\pi^-\psp$ system
using $657\times 10^6$ $B\bar{B}$ pairs~\cite{Choi:2007wga}.
In the Dalitz plot of $M^2(K \pi^-)$ vs. $M^2[\pi^- \psp]$,
two clear bands corresponding to $K^{\ast}(892)$ and $K_2^{\ast}(1430)$
decays to $K \pi$ final states can be seen. After vetoing these events, the
$\pi^- \psp$ mass spectrum is shown in Fig.~\ref{belle-zc4430}, where
a strong enhancement is evident near 4.43~GeV.
A fit with a relativistic $S$-wave BW function to model the peak
plus a smooth phase-space-like function yields a mass $M=(4433\pm 4\pm 2)$~MeV
and a width $\Gamma=(45^{+18+30}_{-13-13})$~MeV with a signal significance of 6.5$\sigma$.

After Belle claimed that a charged $Z_c(4430)$ particle was discovered,
BaBar analyzed the same process using an integrated luminosity of
413~fb$^{-1}$ $\Upsilon(4S)$ data~\cite{Aubert:2008aa}.
BaBar found that the $\pi \psp$ mass distribution can be well described by
the reflection of the known $K\pi$ resonances.
Although BaBar did not confirm the existence of the $Z_c(4430)$,
its results did not contradict the
Belle observation due to low statistics.
To take into account the interference effect between the
$Z_c(4430)^-$ and the $K^{\ast}$ intermediate states in $B\to K
\pi^-\psp$ decays, Belle updated their $Z_c(4430)^-$ results with a
four-dimensional (4D) amplitude
analysis~\cite{Chilikin:2013tch}. The $Z_c(4430)^-$ is observed
with a significance of $5.2\sigma$, a much larger mass of
$(4485\pm 22^{+28}_{-11})$~MeV, and a large width of
$(200^{+41+26}_{-46-35})$~MeV. The product branching fraction is
measured to be \( \BR[B^0\to Z_c(4430)^-K^+]
\BR[Z_c(4430)^-\to \pi^-\psp] = (6.0^{+1.7+2.5}_{-2.0-1.4})\times
10^{-5}\), and spin-parity $J^P=1^+$ is favored over the other
assignments by more than $3.4\sigma$.
The inconsistent results on the $Z_c(4430)^-$ between BaBar and Belle
measurements have been an open question for a very long time since there
were no new data available until recently.

\begin{figure*}[hbt]
\begin{center}
 \includegraphics[height=5cm]{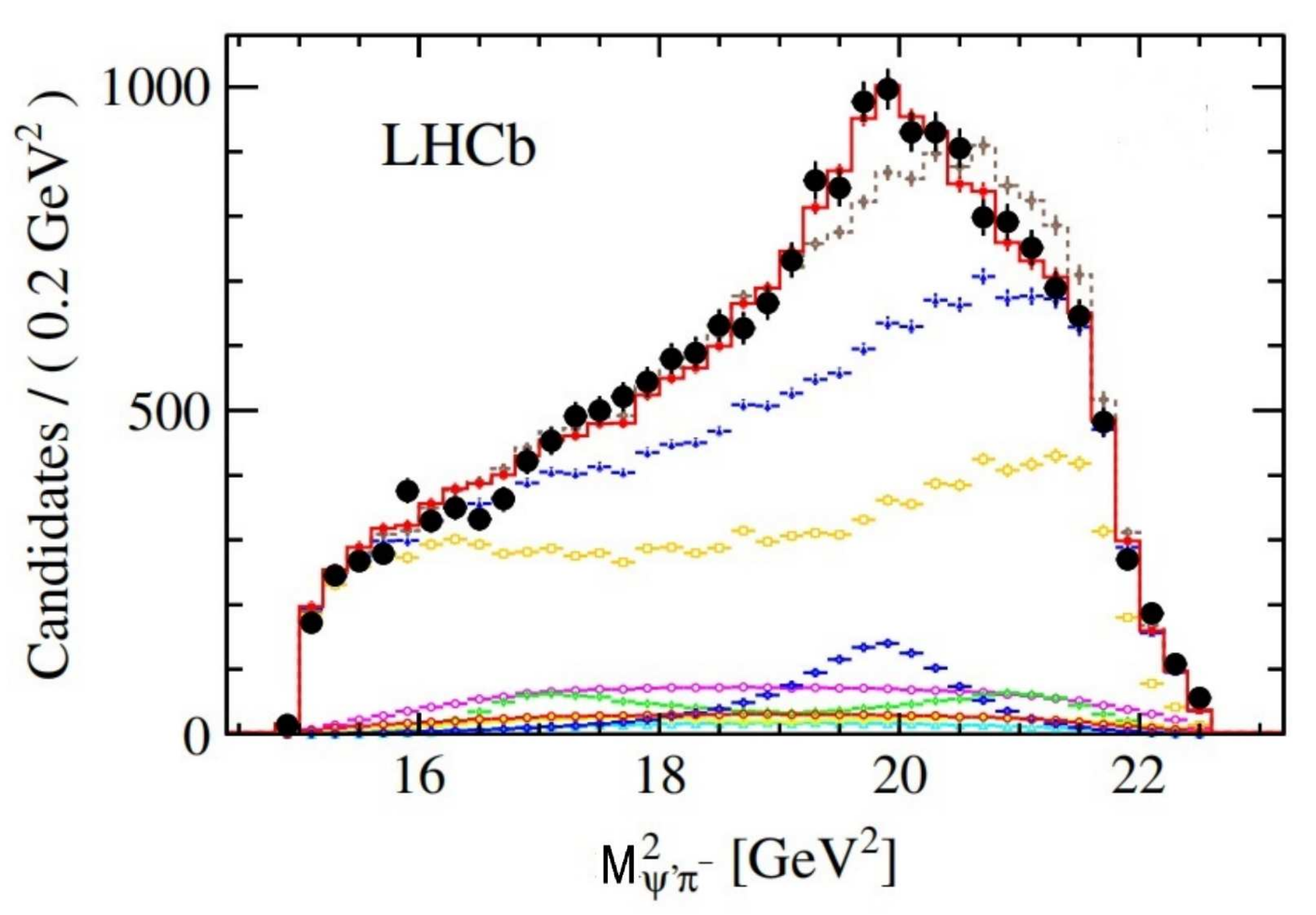}
\end{center}
\caption{\label{lhcb-zc4430} The projection of the 4D amplitude fit on the
$M_{\pi^-\psp}$~\cite{Aaij:2014jqa}.
The red solid (brown dashed) histogram represents the total amplitude
with (without) the $Z_c^-$
component, the upper (lower) blue points represent the $Z_c^-$ component
removed (taken alone),
and the orange, magenta, cyan, yellow, green, and red points represent the
$K^{\ast}(892)$, total $S$-wave, $K^{\ast}(1410)$, $K^{\ast}(1680)$, $K_2^{\ast}(1430)$,
and background terms, respectively.}
\end{figure*}

Since LHCb has large $B$-samples, the same process $B^0\to K^+\pi^-\psp$
with $\psp\to \MM$ was analyzed to search for resonant structures
using $pp$ collision data corresponding to 3 fb$^{-1}$~\cite{Aaij:2014jqa}.
After event selection, $25176\pm 174$ $B^0\to K^+\pi^-\psp \to K^+\pi^- \MM$ candidates were selected.
The order-of-magnitude increase in the signal yield over the Belle measurement~\cite{Chilikin:2013tch}
improved the sensitivity of $Z_c(4430)^-$ searches and allowed a measurement of quantum numbers and Argand plot.
Thus, a 4D model-dependent amplitude fit with
$M^2_{K^+\pi^-}$, $M^2_{\pi^-\psp}$, ${\rm cos}\theta_{\psp}$, and
$\phi$ as variables, was performed to the selected signal candidates,
where $\theta_{\psp}$ is the $\psp$ helicity
angle and $\phi$ is the angle between the $K^{\ast}$ and $\psp$ decay
planes in the $B^0$ rest frame. In the amplitude fit, all known
$K^{\ast}$ resonances, a nonresonant term, and a $Z_c^-$ amplitude
represented by a BW function are included.
The fit yields a mass of $(4475\pm
7\,{_{-25}^{+15}})$~MeV and a width of $(172\pm
13\,{_{-34}^{+37}})$~MeV, which are consistent with, but more precise
than, the Belle results~\cite{Chilikin:2013tch}.
The lowest significance for the $Z_c(4430)^-$ signal is 13.9$\sigma$.
It means the data cannot be described with $K^+ \pi^-$ resonances alone,
thus confirming the existence of the $Z_c(4430)^-$.
The projection of the 4D amplitude fit on the $M_{\pi^-\psp}$
is shown in Fig.~\ref{lhcb-zc4430}, where
the red solid (brown dashed) histogram represents the total amplitude with
(without) the $Z_c^-$
component, the upper (lower) blue points represent the $Z_c^-$ component
removed (taken alone),
and the orange, magenta, cyan, yellow, green, and red points represent the
$K^{\ast}(892)$, total $S$-wave, $K^{\ast}(1410)$, $K^{\ast}(1680)$, $K_2^{\ast}(1430)$,
and background terms, respectively.
Relative to $J^P=1^+$, the $0^-$, $1^-$, $2^+$, and $2^-$
hypotheses are ruled out by at least $9.7\sigma$, thus
the spin-parity of the $Z_c(4430)$ is established to be $1^+$ unambiguously.
In addition, LHCb measures the Argand plot of the $Z_c^-$ amplitude as
a function of $M_{\pi\psp}$, which is consistent with a rapid change of the
$Z_c(4430)^-$ phase
when its magnitude reaches the maximum, expected behavior of a
resonance. This is the first time an Argand plot is obtained for an
exotic charmonium-like state.

In the amplitude fit, LHCb also tried to add an additional resonance $Z_c^-$ and found the $p$-value of the $\chi^2$ test
improves from 12\% to 26\%, corresponding to a $6\sigma$ signal significance.
The measured mass and width of this additional $Z_c^-$ state are $(4239\pm18^{+45}_{-10})$~MeV and $(220\pm47^{+108}_{-74})$~MeV
and $0^-$ is preferred over other $J^P$ assignments by $8\sigma$.
This is the state dubbed $R_{c0}(4240)$ in the spectrum shown in Fig.~\ref{fig:barccspec}.
Although the signal significance is large for this state, Argand diagram studies are inconclusive.
Therefore, its characterization as a resonance will need further confirmation.

In a full amplitude analysis of $B^0 \to K^+ \pi^- J/\psi$ decay
based on a 711~fb$^{-1}$ $\Upsilon(4S)$ data sample at
Belle~\cite{Chilikin:2014bkk},
the $Z_c(4430)^-\to \pi^-\jpsi$ is observed as a $4.0\sigma$ signal, while
BaBar's earlier model-independent analysis of
the same mode did not find this process~\cite{Aubert:2008aa}.
Very recently, an angular analysis of $B^0 \to K^+ \pi^- J/\psi$
with $J/\psi \to \mu^+ \mu^-$ decays was performed by LHCb, although
non-$K^*$ events must exist, whether there is a contribution from the
$Z_c(4430)$ is not clear~\cite{Aaij:2019ipm}.

\vspace{0.3cm}\noindent
$\bullet$ {\it The $Z_{c1}(4050)$ and $Z_{c2}(4250)$ states}
\vspace{0.3cm}

Replacing the $\psp$, Belle subsequently checked the $\pi^- \chi_{c1}$
system in exclusive $B^0\to K^+\pi^-\chi_{c1}$ using $657\times 10^6$
$B\bar{B}$ pairs~\cite{Mizuk:2008me}.
After signal selections, the Dalitz plot distribution exhibits some distinct
features: two clear bands corresponding to $K^{\ast}(892)$ and $K^{\ast}(1430)$
decays to $K \pi$; a distinct band at $M^2(\pi^- \chi_{c1})\approx 17$ GeV$^2$
corresponding to a structure in the $\pi^- \chi_{c1}$ channel. To obtain such
structure parameters, a 6D Dalitz plot analysis was performed, where all the
known $K^+ \pi^-$ resonances below 1900~MeV
and a single exotic $\pi^- \chi_{c1}$ resonance are included.
Such a fit gives the confidence level of 0.5\%, which indicates that the shape
of the structure is not well reproduced by a single BW. Motivated by this,
an additional resonance decaying to $\pi^- \chi_{c1}$ is added to the above
fit model. The masses and widths of the two $Z_c^-$ resonances from the fit
are $M_1=(4051\pm14^{+20}_{-41})$ MeV, $\Gamma_1=(82^{+21+47}_{-17-22})$ MeV,
$M_2=(4248^{+44+180}_{-29-35})$ MeV, and $\Gamma_2=(177^{+54+316}_{-39-61})$ MeV.
They are denoted as $Z_{c1}(4050)$ and $Z_{c2}(4250)$ with signal significances
greater than $5\sigma$ for both. The confidence level for this fit is 42\%.

\begin{figure}[hbt]
\centering
 \includegraphics*[height=5cm]{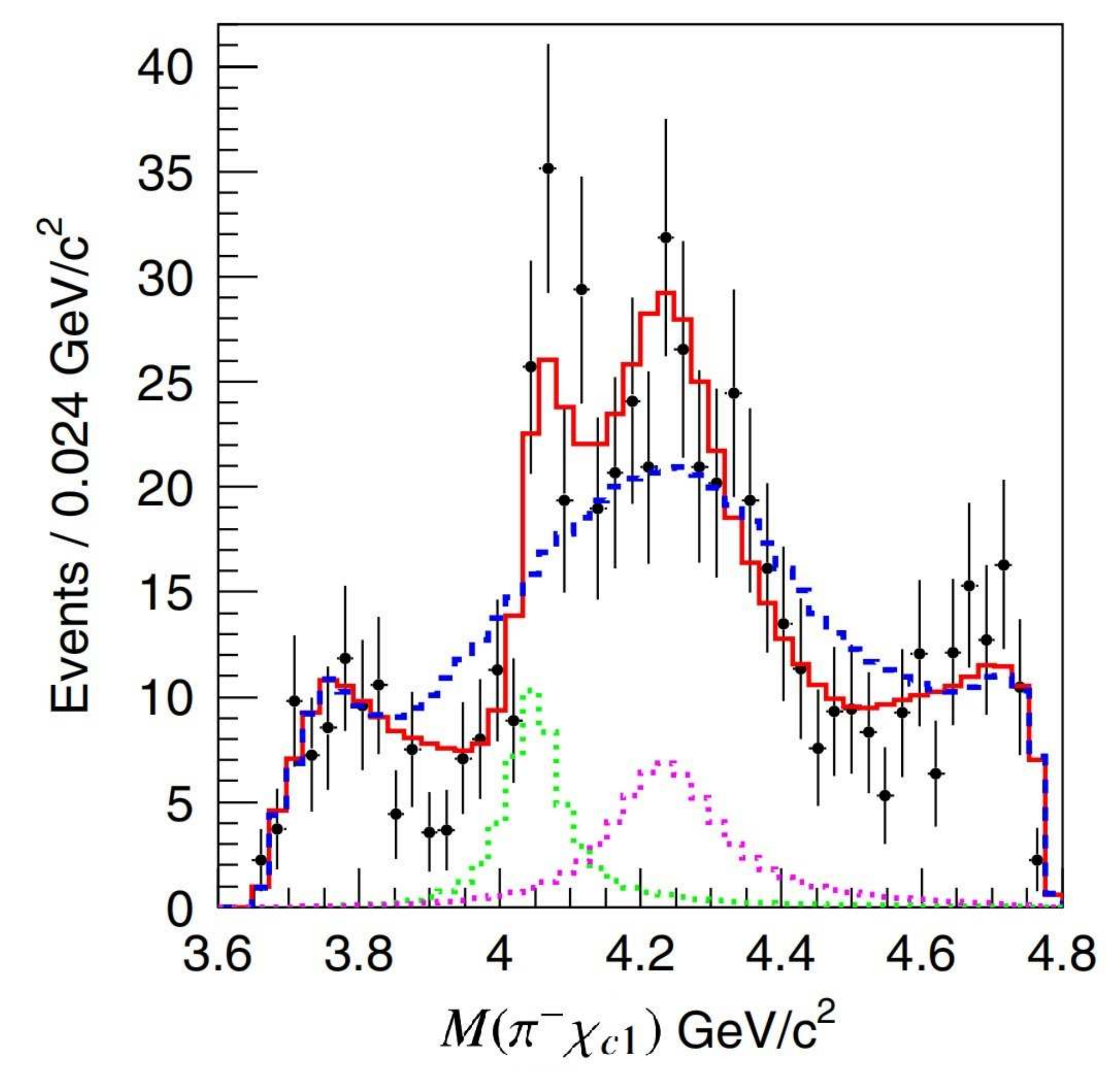}
\caption{The invariant mass distribution $M(\pi^- \chi_{c1})$ with
$1.0~{\rm GeV}^2<M^2(K^+ \pi^-)<1.75~{\rm GeV}^2$~\cite{Mizuk:2008me}.
The solid (dashed) histogram is the Dalitz plot fit result for the fit model
with all
known $K^{\ast}$ and two (without any) $\pi^- \chi_{c1}$ resonances, the dotted
histograms represent the contribution of the two $Z_{c1}(4050)$ and
$Z_{c2}(4250)$ states.\label{belle-z1z2}}
\end{figure}

The invariant mass distribution $M(\pi^- \chi_{c1})$
with $1.0~{\rm GeV}^2<M^2(K^+ \pi^-)<1.75~{\rm GeV}^2$
is shown in Fig.~\ref{belle-z1z2}, where the solid (dashed)
histogram is the Dalitz plot fit result for the fit model with all
known $K^{\ast}$ and two (without any) $\pi^- \chi_{c1}$ resonances, the dotted
histograms represent the contribution of the two $Z_{c1}(4050)$ and $Z_{c2}(4250)$ states.
These two charged resonances represent additional candidate states of
similar characteristics to the $Z_c(4430)^-$.

After Belle claimed the observation of the $Z_{c1}(4050)$ and $Z_{c2}(4250)$
states, the BaBar experiment analyzed the same process using an integrated
luminosity of 429 fb$^{-1}$~\cite{Lees:2011ik}.
In this analysis, BaBar fitted the background-subtracted and
efficiency-corrected $\pi^- \chi_{c1}$ mass distribution using the $K\pi$ mass
distribution and the corresponding normalized
$K \pi$ Legendre-polynomial moments, and found that the fit can describe
the data well without the need of any charged $Z_c$ resonance in the
$\pi^- \chi_{c1}$ system.
So their conclusion is: Neither $Z_{c1}(4050)$ nor $Z_{c2}(4250)$ is evident,
leading to 90\% C.L.\ upper limits on the branching fractions for the corresponding $B$-meson decay modes.
Given the large uncertainties, the upper limits are still compatible with the branching fraction values from the Belle measurement~\cite{Mizuk:2008me}.
Thus, further confirmation of the existence of the $Z_{c1}(4050)$ and $Z_{c2}(4250)$ states at LHCb and Belle II is needed in the future.

\vspace{0.3cm}\noindent
$\bullet$ {\it The $Z_c(4200)$ state}
\vspace{0.3cm}

\begin{figure}[hbt]
\centering
 \includegraphics*[height=5cm]{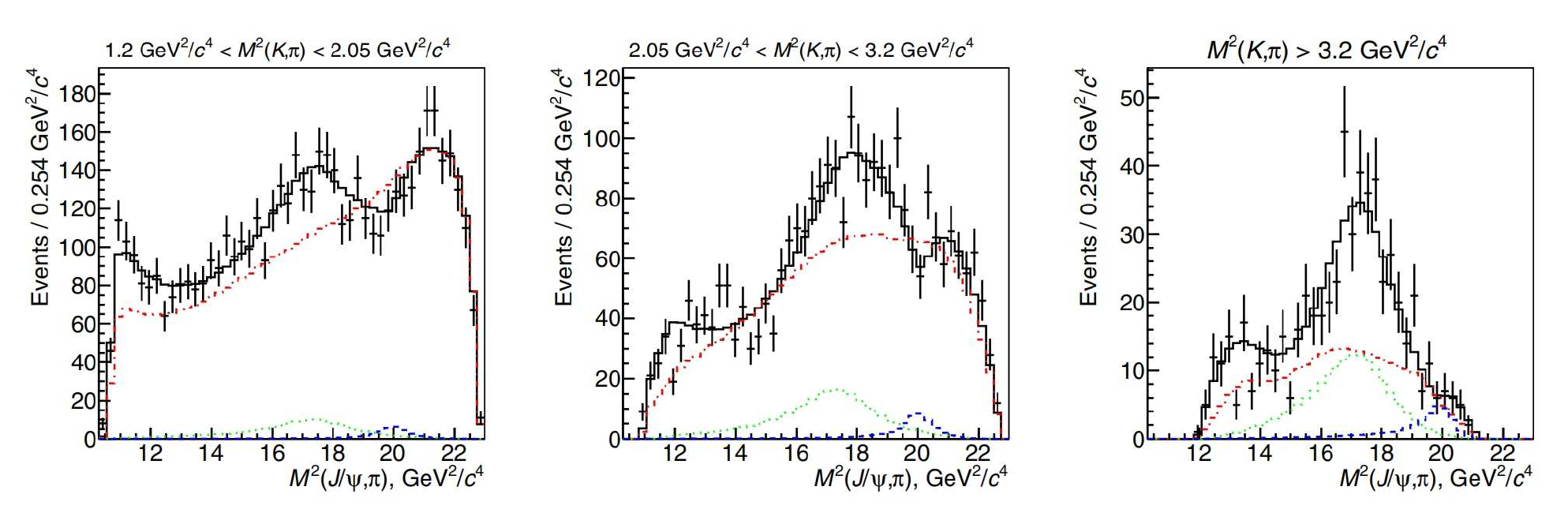}
\caption{Projections of the fit results onto the $M^2_{J/\psi \pi}$ axis
with the $Z_c(4200)$ ($J^P=1^+$) in different $K^+ \pi^-$ mass regions~\cite{Chilikin:2014bkk}.
The points with error bars are data; the solid
histograms are fit results, the dashed histograms are the $Z_c(4430)^-$
contributions, the dotted histograms are the $Z_c(4200)^-$
contributions and the dash-dotted histograms are contributions of all $K^*$
resonances.\label{belle-zc4200}}
\end{figure}

After the charged charmonium-like states of $Z_c(4430)$, $Z_{c1}(4050)$,
and $Z_{c2}(4250)$ were observed in $B$ decays by Belle, a full amplitude analysis was performed
to the ${B}^0 \to K^+ \pi^- J/\psi $ decay with $J/\psi$ decaying to a
lepton pair to search for possible charged charmonium-like
states in the $ \pi^- J/\psi$ system based on a 711~fb$^{-1}$
$\Upsilon(4S)$ data sample~\cite{Chilikin:2014bkk}.
After event selection, the total number of
signal events is $29\,990\pm190\pm50$.
For these selected signal candidates, the amplitude of the
decay ${B}^0 \to K^+ \pi^-J/\psi$ is represented as the sum of BW contributions for different
intermediate two-body states calculated using the helicity formalism in a four-dimensional parameter space,
defined as $\Phi = (M^2_{K \pi}, M^2_{J/\psi \pi}, \theta_{J/\psi}, \varphi)$, where $\theta_{J/\psi}$ is the $J/\psi$ helicity angle
[the angle between the momenta of the $(K^+ \pi^-)$ system and the $\ell^-$ in the $J/\psi$ rest frame; $\ell^-\ell^+$ is the leptonic pair from the $J/\psi$ decay]
and $\varphi$ is the angle between the planes defined by the $(\ell^+ \ell^-)$ and
$(K^+ \pi^-)$ momenta in the ${B}^0$ rest frame. The known resonances included in the
default model are $K^*_0(700)$, $K^*(892)$, $K^*(1410)$, $K^*_0(1430)$,
$K^*_2(1430)$, $K^*(1680)$, $K^*_3(1780)$, $K^*_0(1950)$, $K^*_2(1980)$,
$K^*_4(2045)$, and $Z_c(4430)^-$; a search for
additional exotic $Z_c^-$ resonances is performed.
An unbinned maximum-likelihood fit over the four-dimensional space $\Phi$ was performed.
The considered spin-parity hypotheses of a possible $Z_c^-$ resonance
are $J^P=0^-$, $1^-$, $1^+$, $2^-$ and $2^+$.
The best fit gives a global 6.2$\sigma$ significance of a $Z_c^-$
resonance with a mass of $(4196^{+31+17}_{-29-13})$ MeV
and a width of $(370^{+70+70}_{-70-132})$ MeV. Thus, a new charged $Z_c$ state, $Z_c(4200)^-$, is observed.
The preferred quantum numbers assigned to this state are $J^P=1^+$.
Projections of the fit results onto the $M^2_{J/\psi \pi}$ axis in different
$K^+ \pi^-$ mass regions for the model with the $Z_c(4200)$ ($J^P=1^+$) are shown in Fig.~\ref{belle-zc4200}.

Note that similar to the $Z_c(4430)$, BaBar's earlier model-independent
analysis of ${B}^0 \to K^+ \pi^- J/\psi $ decays did not find the $Z_c(4200)^-$
in the $\pi^- J/\psi $ system using an integrated luminosity
of 413~fb$^{-1}$ $\Upsilon(4S)$ data~\cite{Aubert:2008aa}.
BaBar made minimal assumptions about the $K_J^*$ spectrum, using
two-dimensional moments in the variables $M(K^+ \pi^-)$ and the $K^+$ helicity angle.
Within uncertainties, the $M(J/\psi \pi^-)$ spectrum in
the BaBar data was found to be adequately described using just $K_J^*$
states, without a need for exotic contribution.
The inconsistent results on the $Z_c(4200)$ between BaBar and Belle
measurements have been recently solved by LHCb as is described below.

\begin{figure}[hbt]
\centering
 \includegraphics*[height=5cm]{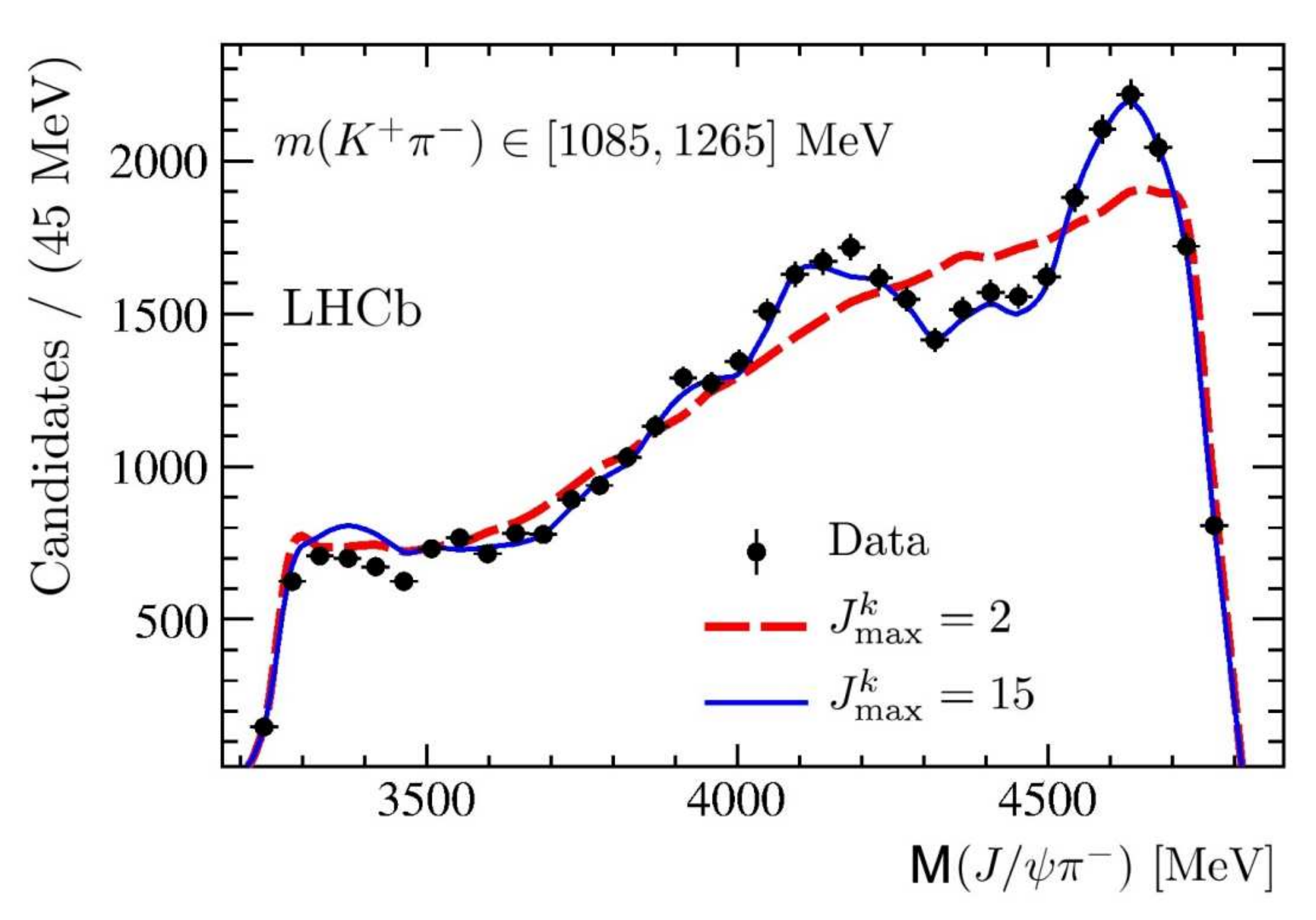}
\caption{Comparison of $M(J/\psi\pi^-)$ distributions in the
$M(K^+\pi^-)\in [1085, 1265]$~MeV
region between the background-subtracted data and simulated events
weighted by moment models
with $J^k_{\rm max}=2$ and $J^k_{\rm max}=15$~\cite{Aaij:2019ipm}.\label{lhcb-zc4200}}
\end{figure}

LHCb performed an angular analysis of $B^0 \to K^+ \pi^- J/\psi $ with
$J/\psi \to \mu^+ \mu^-$ decays using proton-proton collision data corresponding to an integrated
luminosity of 3 fb$^{-1}$~\cite{Aaij:2019ipm}.
After event selection, $554500 \pm 800$ signal candidates are
obtained in the mass region $745<M(K^+\pi^-)<1545~{\rm MeV}$.
This selected data sample is divided into 35 fine bins in $M(K^+\pi^-)$ and
a four-dimensional angular analysis is performed in each mass bin.
The four variables, $M(K^+ \pi^-)$,
$\theta_V$, $\theta_l$, and $\chi$, fully describe the decay topology,
where $\theta_V$ is the $K^+$ helicity angle
defined as the $K^+$ momentum direction in the $K^+ \pi^-$ helicity frame
with respect to the $B^0$ rest frame, $\theta_l$ is the similar lepton
helicity angle, and $\chi$ is the
the azimuthal angle between the $(\mu^+ \mu^-)$ and $(K^+ \pi^-)$ decay planes.
To maximise the sensitivity to any exotic component, an angular analysis is
performed in the $M(K^+\pi^-)\in [1085, 1445]$ MeV region, where $J^k_{\rm max}$,
the allowed spin of the highest partial wave for the $K^-\pi^+$ system
($K_J^*$), is 3.
Figure~\ref{lhcb-zc4200} shows the comparison of the $M(J/\psi \pi^-)$
distributions between the background-subtracted data and
weighted simulated events with the $J^k_{\rm max}=2$ model taken in
the $M(K^+\pi^-)\in [1085, 1265]$~MeV region.
The $J^k_{\rm max}=2$ model clearly can not describe the peaking structures
in the data around $M(\jpsi\pi^-)\approx 4200$ and 4600~MeV, which strongly indicates presence of exotic components.
The significance for exotic components is in excess of 6$\sigma$ with
systematic uncertainties included.
The structure at $M(\jpsi\pi^-)\approx 4200$~MeV is close to the exotic state
reported by Belle~\cite{Chilikin:2014bkk}.
To interpret these structures as exotic tetraquark resonances
and measure their properties will require a future model-dependent
amplitude analysis of the data.

\vspace{0.3cm}\noindent
$\bullet$ {\it The $\zc$ and $Z_c(4020)$ states}
\vspace{0.3cm}

\begin{figure}
\begin{center}
\includegraphics[height=4.3cm]{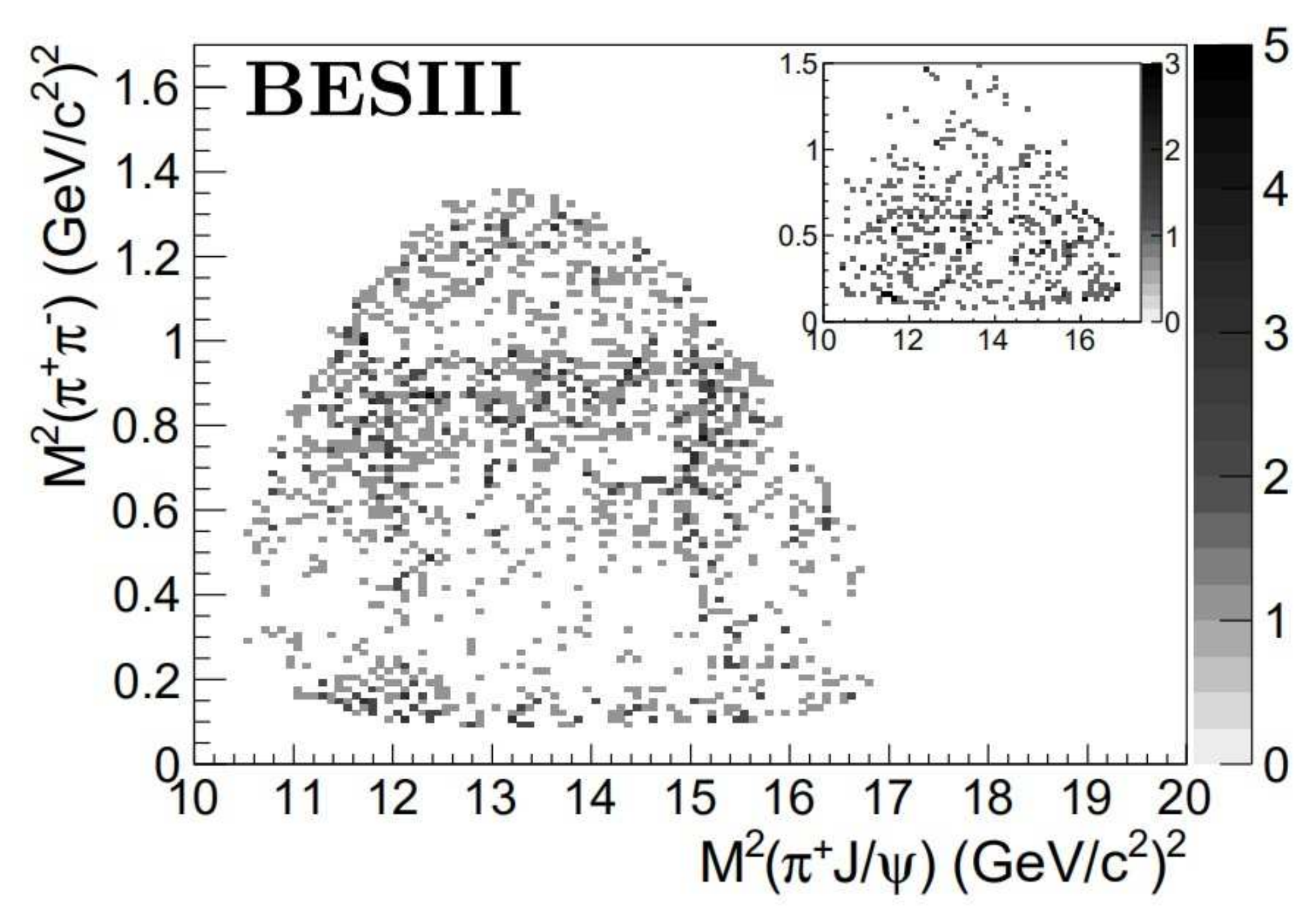}
\includegraphics[height=4.3cm]{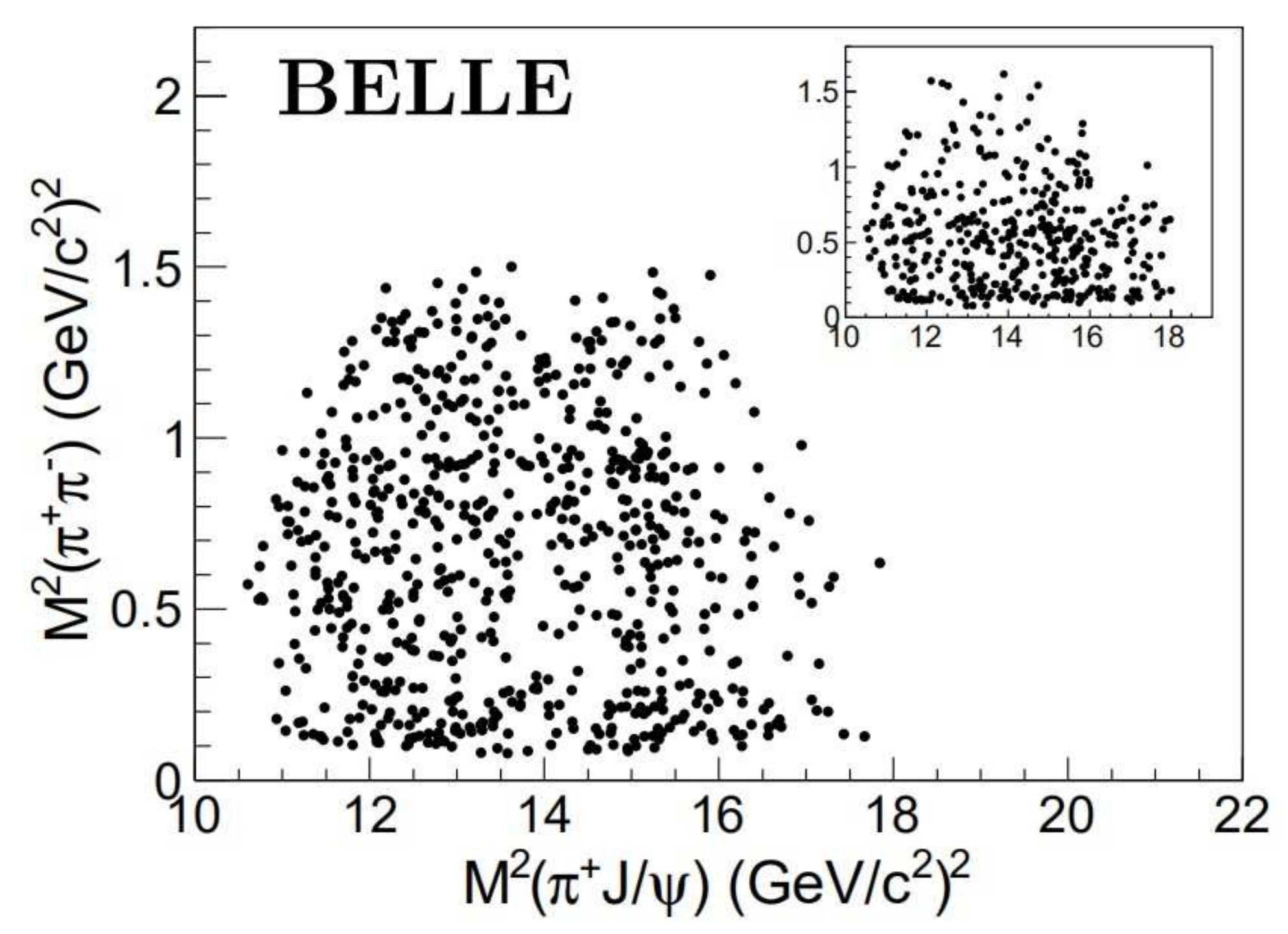}
\caption{Dalitz plots for selected $\EE\to \ppjpsi$ events in the
$\jpsi$ signal region from
BESIII~\cite{Ablikim:2013mio} and Belle~\cite{Liu:2013dau} experimental data.
The insets show background events from the $\jpsi$ mass
sidebands~(not normalized).} \label{dalitz_zc}
\end{center}
\end{figure}

To understand the intermediate states in $Y(4230)\to \pp\jpsi$,
the BESIII experiment collected a 525~pb$^{-1}$ data sample in 2013
at c.m.\ energy of 4.26~GeV, the peak of the $Y(4230)$~\cite{Ablikim:2013mio}.
The cross section of $\EE\to \ppjpsi$ at $\sqrt{s}=4.26$~GeV is measured
to be $(62.9\pm 1.9 \pm 3.7)$~pb based on the selected 1477 signal candidates, which agrees
with the previous results from the BaBar~\cite{Lees:2012cn}
and Belle~\cite{Yuan:2007sj} experiments. Using this selected signal
sample, the Dalitz plot is drawn to check for possible intermediate states,
as shown in Fig.~\ref{dalitz_zc} (left panel). At the same time, Belle updated the
measurement of the cross section of $\EE\to \ppjpsi$ from 3.8 to 5.5~GeV using the
ISR method with a 967~fb$^{-1}$ data sample~\cite{Liu:2013dau}. The Dalitz plot for events
in the $Y(4230)$ signal region
($4.15~{\rm GeV} < M(\ppjpsi) < 4.45$~GeV) is also investigated, as shown in Fig.~\ref{dalitz_zc} (right panel).

\begin{figure}
\begin{center}
\includegraphics[height=4.3cm]{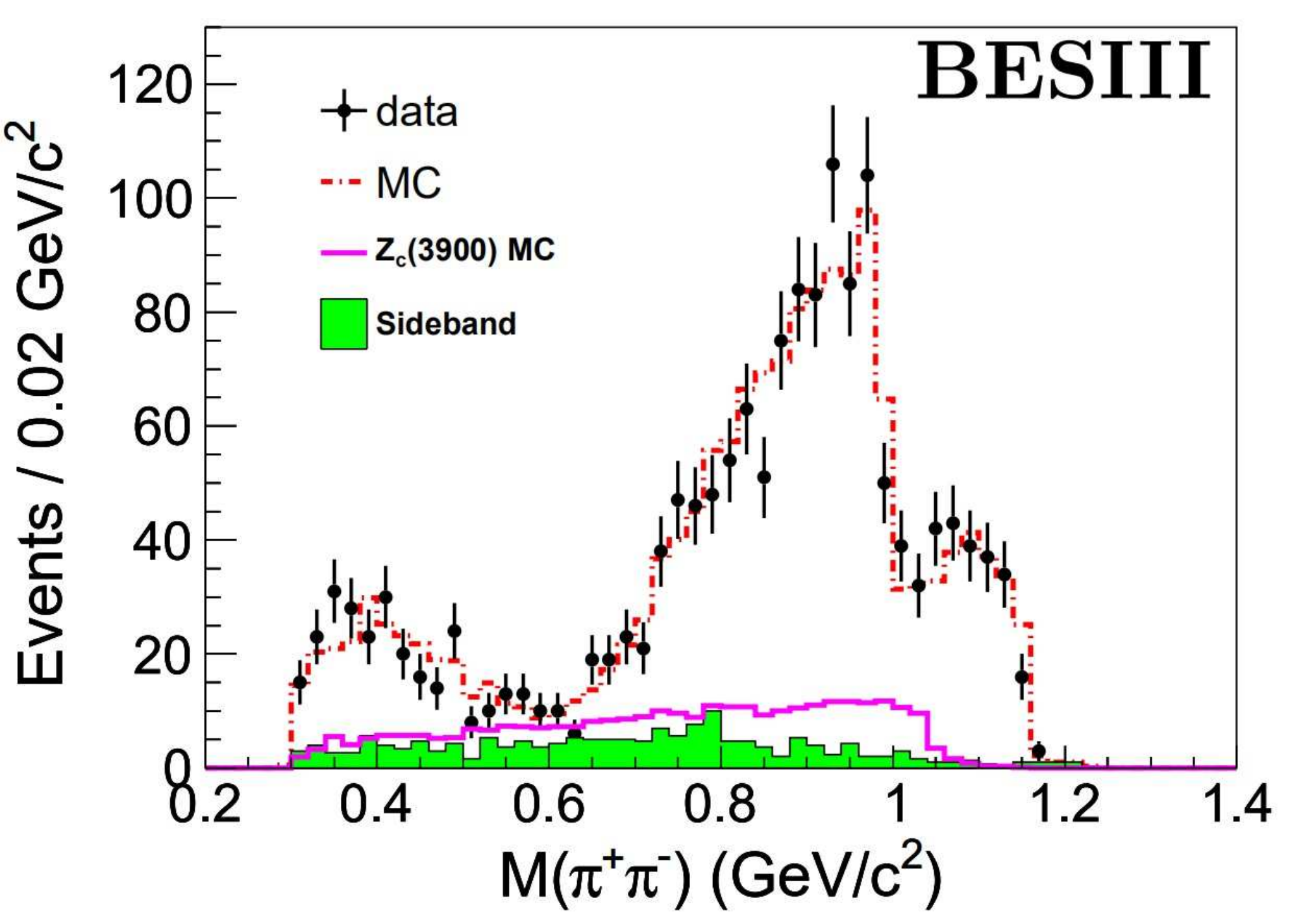}
\includegraphics[height=4.3cm]{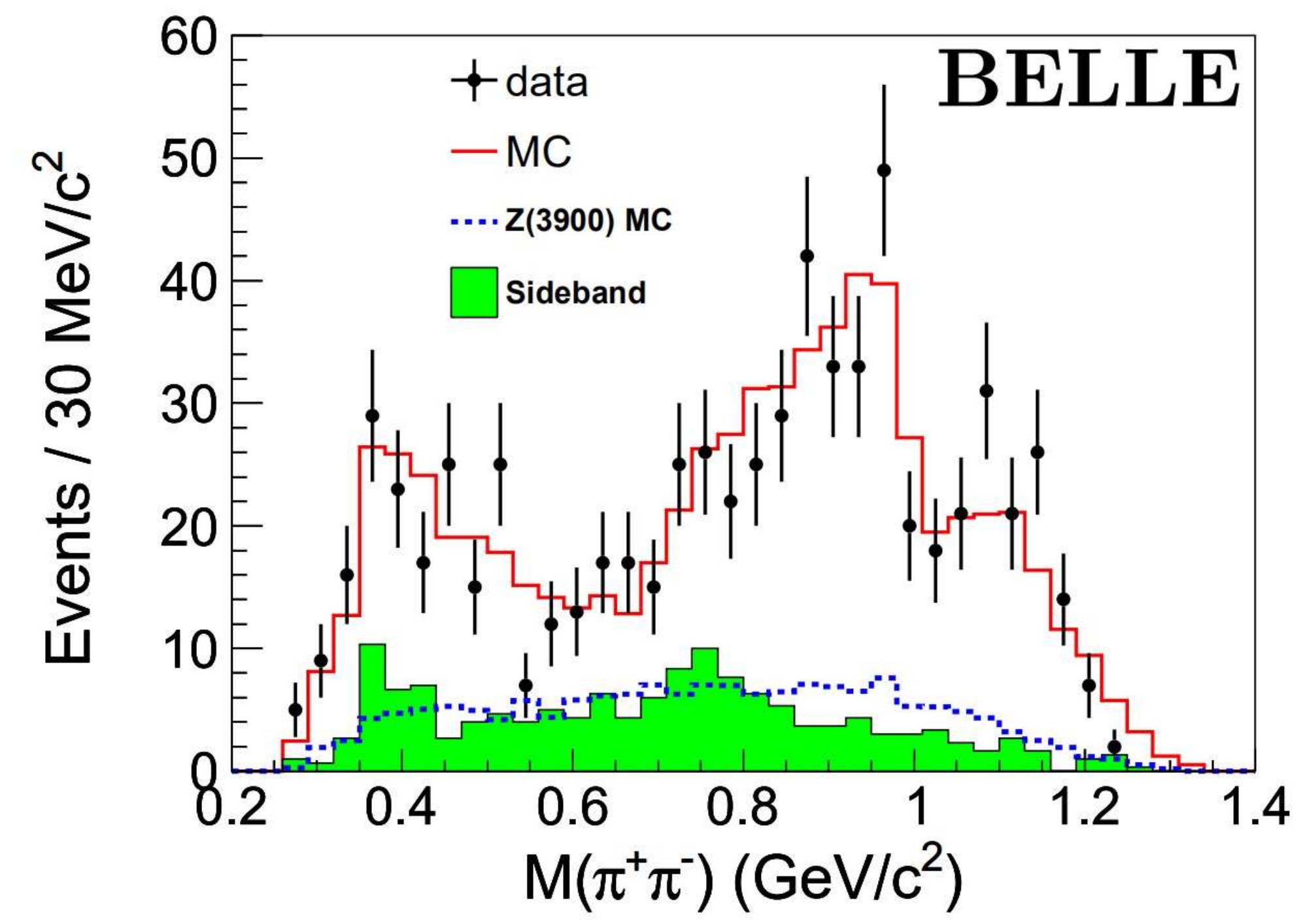}
\caption{Invariant mass distributions of $\pp$ for events in the $\jpsi$
signal region
from BESIII~\cite{Ablikim:2013mio} and Belle~\cite{Liu:2013dau} experimental
data. Points with error
bars represent data, shaded histograms are the normalized
background estimates from the $\jpsi$-mass sidebands,
red histograms represent MC simulation results from $\sigma(500)$, $f_0(980)$,
and non-resonant $\pp$ amplitudes, and lower histograms
are MC simulation results for a $\zc$ signal.} \label{zc3900-pp}
\end{center}
\end{figure}

\begin{figure}[htbp]
\centering
 \includegraphics[height=4.3cm]{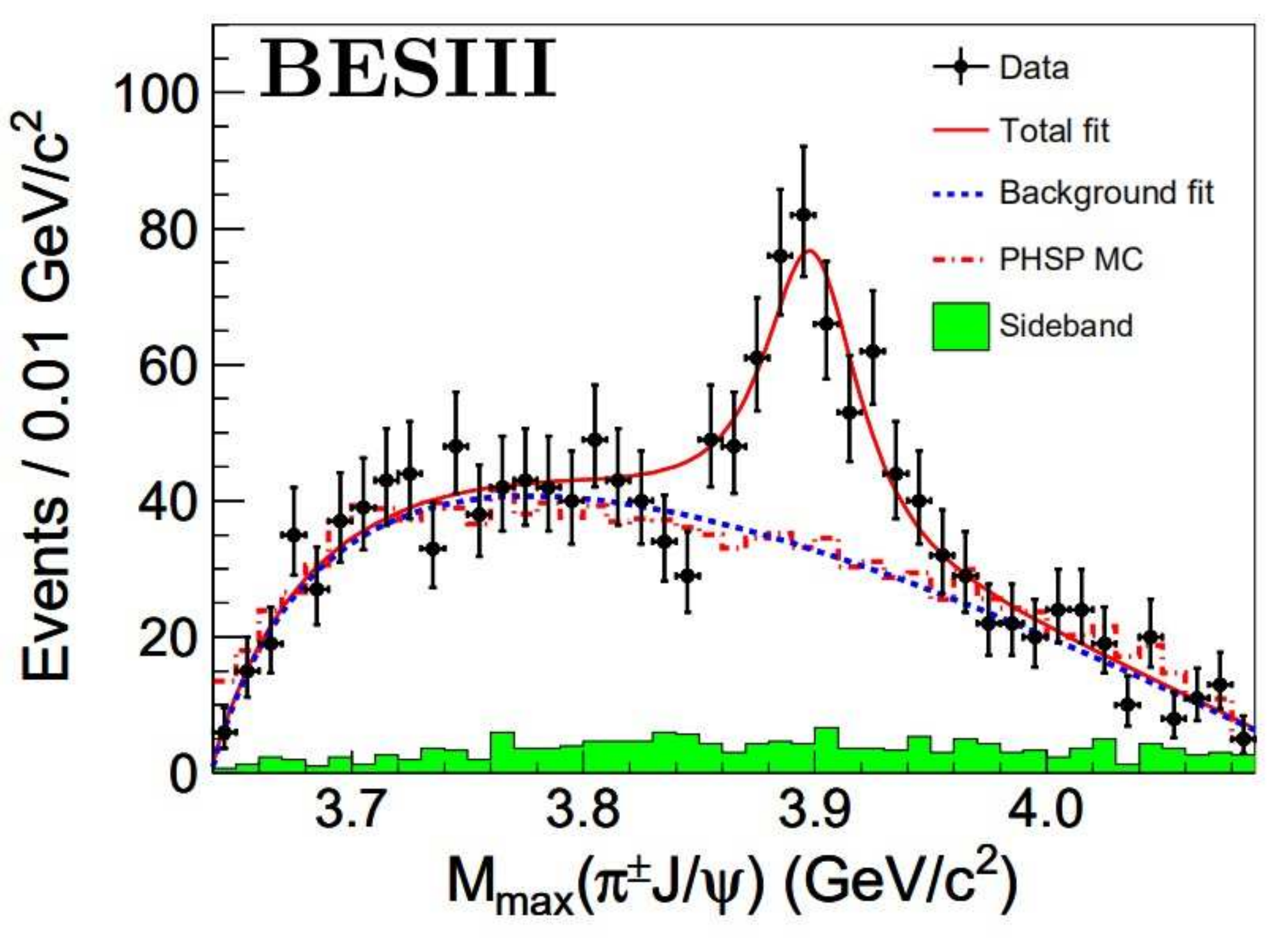}
 \includegraphics[height=4.4cm]{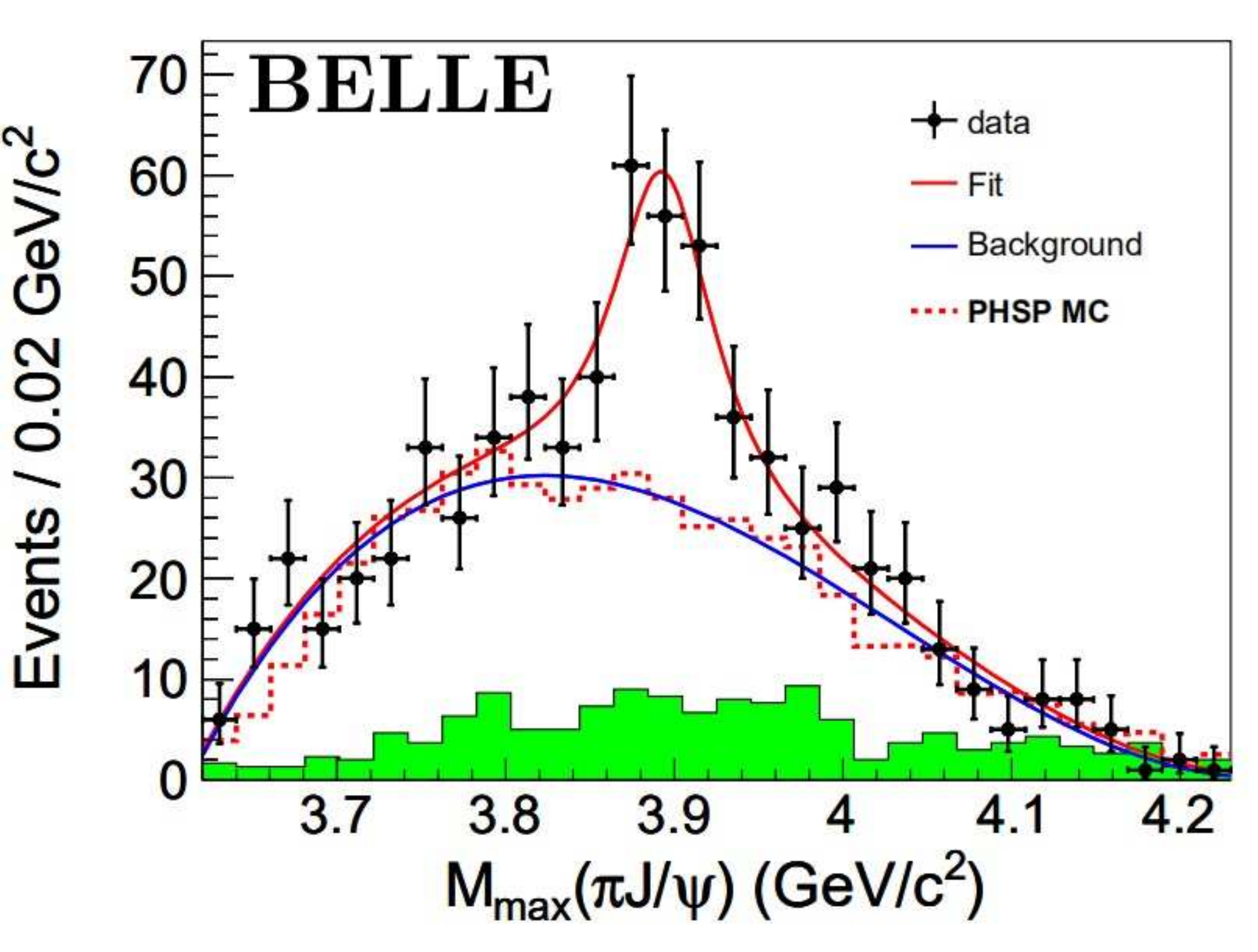}
 \caption{Unbinned maximum-likelihood fits to the
distributions of the $M_{\mathrm{max}}(\pi J/\psi)$ from
BESIII~\cite{Ablikim:2013mio} and
Belle~\cite{Liu:2013dau} experimental data. Dots with error bars are
data, the solid curves are the best fits, the dashed histograms represent
the results of the phase-space distribution, and the shaded
histograms are the normalized $\jpsi$ sideband events.}
\label{projfit}
\end{figure}

Figures~\ref{zc3900-pp} and ~\ref{projfit} show the projections of the $M(\pp)$
and $M_{\rm max}(\pi^\pm\jpsi)$ [the maximum value out of $M(\pi^+\jpsi)$ and
$M(\pi^-\jpsi)$] distributions for the signal events,
as well as the background events estimated from the normalized $\jpsi$
mass sidebands. A parametrization for the $\pp$ mass spectrum that includes $f_0(980)$, $f_0(500)$ ($\sigma$),
and a non-resonant amplitude can describe the data well, but does not
generate any peaking structure
in the $\pi^\pm\jpsi$ mass projection. The $\pp$ mass spectrum shows complicated
structures.
Unbinned maximum-likelihood fits are applied to the distributions
of $M_{\rm max}(\pi^\pm\jpsi)$ from Belle and BESIII measurements.
The signal shape is parameterized as an $S$-wave BW function
convolved with a Gaussian function with a mass resolution fixed at the MC simulated
value.
Figure~\ref{projfit} shows the fit results.
The measured masses are $(3899.0\pm 3.6\pm 4.9)~{\rm MeV}$
and $(3894.5\pm 6.6\pm 4.5)~{\rm MeV}$ and the measured widths are
$(46\pm 10\pm 20)$~MeV and $(63\pm 24\pm 26)$~MeV from the Belle
and BESIII experiments, respectively. They are consistent with each other
within the uncertainties.
The signal significance is greater than 5$\sigma$ in both measurements.
This structure is now referred to as the $\zc$.
The production ratios of $\zc$ are measured to be $$\frac{\sigma[\EE\to \pi^\pm
\zc^\mp\to \ppjpsi]} {\sigma(\EE\to \ppjpsi)}=(21.5\pm 3.3\pm
7.5)\% \ \mbox{and} \ (29.0\pm 8.9)\%$$ in the BESIII and Belle experiments,
respectively,
where the error in the Belle measurement is statistical only.
As the $\zc$ state has a strong coupling to charmonium and is charged,
it cannot be a conventional $c\bar{c}$ state. Actually, since the final
state is an isovector, the $Z_c$ state should be an isovector as well.

The $\zc$ state was confirmed shortly after with CLEO-c data at a c.m.\
energy of 4.17~GeV~\cite{Xiao:2013iha},
and the mass and width agreed very well with the BESIII and Belle
measurements. In addition, a $3.5\sigma$ evidence for $\zc^{0}$ in the CLEO-c
data was also reported in the $\EE\to \piz\piz\jpsi$
process~\cite{Xiao:2013iha}.

BESIII measured the cross sections of $e^+e^-\to \pi^0\pi^0
J/\psi$ with data in the c.m.\ energy ranges from
4.19 to 4.42~GeV~\cite{Ablikim:2014dxl}. A neutral state $\zc^0\to
\piz\jpsi$ with a significance of $10.4\sigma$ was observed, with
the mass and width measured to be $(3894.8\pm 2.3\pm
3.2)$~MeV and $(29.6\pm 8.2\pm 8.2)$~MeV, respectively,
which are close to those of $\zc^\pm$.
Thus, it is interpreted as the neutral partner of the
$\zc^\pm$. The measured production rate of $\EE\to \pi^0\zc^0$ is
about half of that for $\EE\to \pi^+\zc^-+c.c.$, which is
consistent with the expectation from isospin symmetry. This confirms that
the $\zc$ is an isovector state.

\begin{figure}[htbp]
\centering
 \includegraphics[width=6.0cm]{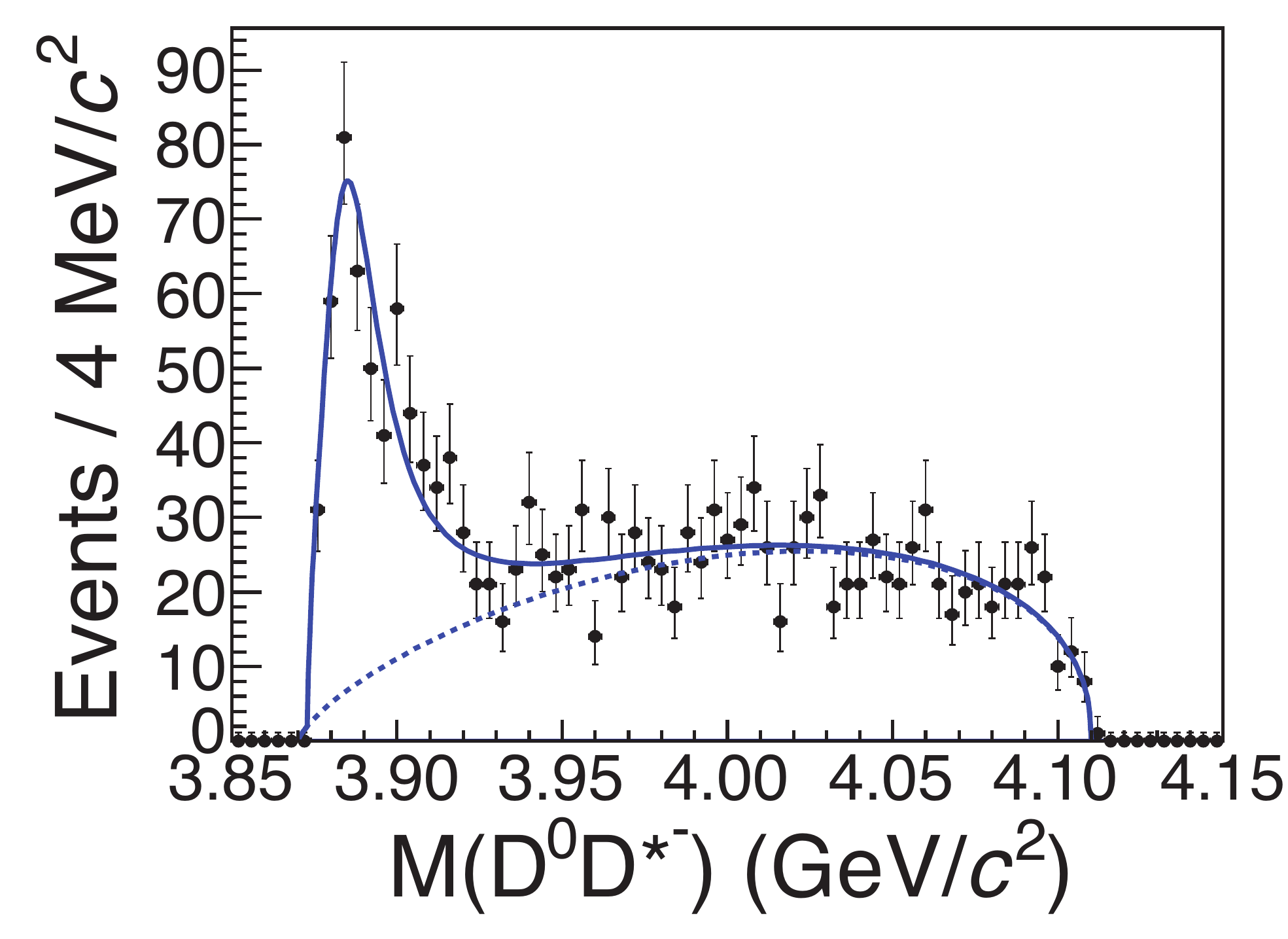}
 \includegraphics[width=6.0cm]{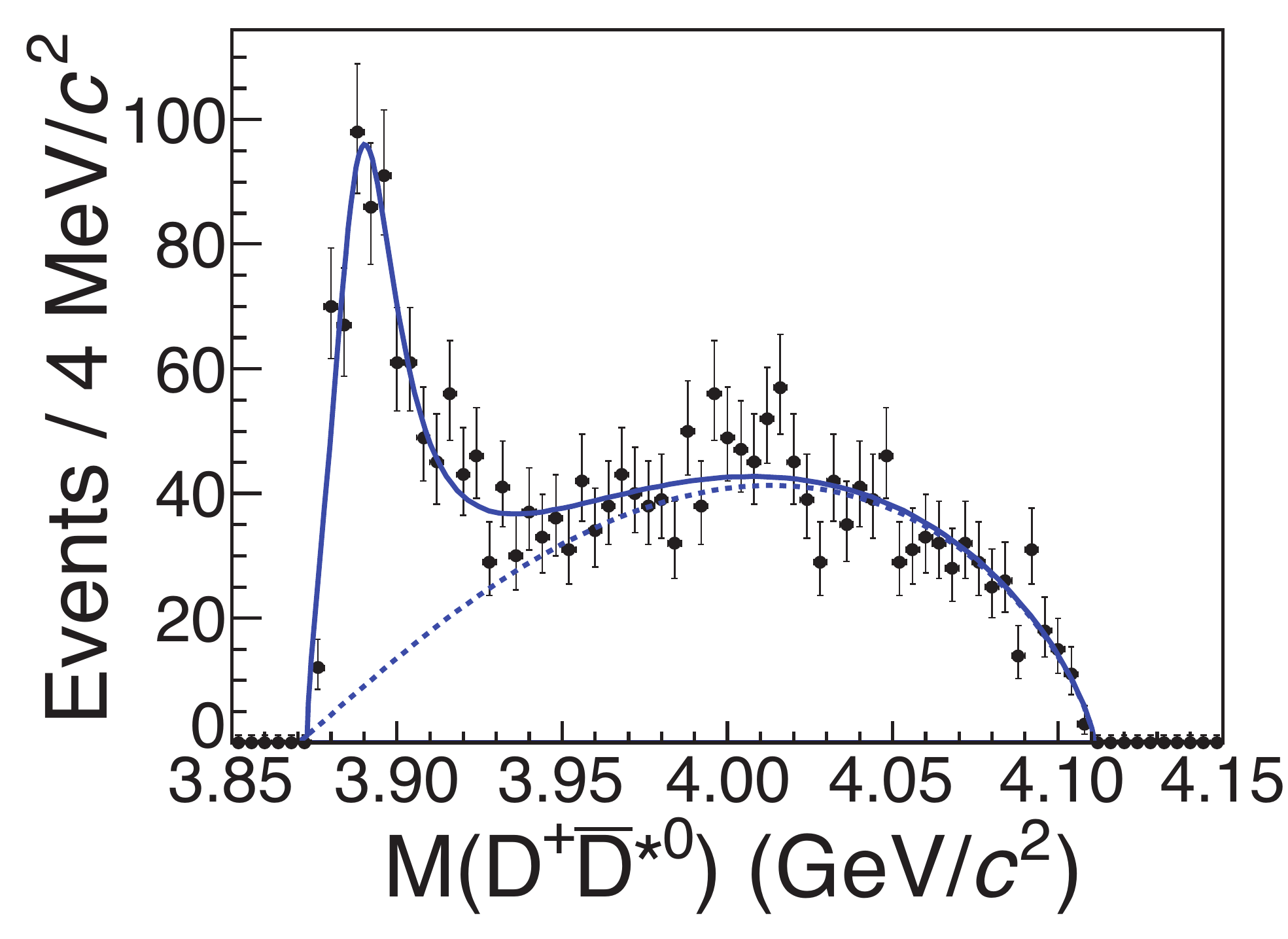}
\caption{The $M(D^0 D^{*-})$ (left) and $M(D^+\bar{D}^{*0})$
(right) distributions for selected signal candidates at $\sqrt{s}=4.26$~GeV
in the single-tag analysis~\cite{Ablikim:2013xfr}. The solid curves show
the best fits. }
\label{xuxp}
\end{figure}

As the $\zc$ mass is close to the $D\bar{D}^*$ mass threshold,
it is natural to check the $D\bar{D}^*$ mass spectrum in
$\EE\to \pi^\pm (D\bar{D}^*)^{\mp}$, which was studied by BESIII
using the data sample at $\sqrt{s}=4.26$~GeV~\cite{Ablikim:2013xfr}.
The $\EE\to \pi^{+} (D\dstrbar)^{-}+c.c.$ events are selected by a
so-called single-tag technique in which only the bachelor
$\pi^\pm$ and one final-state $D$ meson are reconstructed, and the
$\dstrbar$ is inferred from energy-momentum conservation.
In this analysis, both isospin channels $\pip D^0 D^{*-}+c.c.$ and $\pip
D^- D^{*0}+c.c.$ are studied and the $D$ mesons are reconstructed in
the $D^0\to K^-\pi^+$ and $D^+\to K^-\pip\pip$ decay channels.
As expected, a structure close to the $D\bar{D}^*$ mass threshold is
observed in the
$(D\bar{D}^*)^{\pm}$ invariant mass distributions, as shown in Fig.~\ref{xuxp}
for the selected $\EE\to \pi^{+} (D\dstrbar)^{-}+c.c.$ candidates.
Using a BW function with a mass-dependent width as a signal shape to fit the
$(D\bar{D}^*)^{\pm}$ invariant mass distributions, the pole mass and width are
determined to be $(3883.9 \pm 1.5 \pm 4.2)$~MeV and
$(24.8\pm 3.3 \pm 11.0)$~MeV, respectively.
The fit results are shown in Fig.~\ref{xuxp} with solid lines.
The production rate is measured to be $\sigma[\EE \to \pi^{\mp}
\zc^{\pm}]\BR[\zc^{\pm}\to (D\bar{D}^*)^{\pm}]=(83.5\pm
6.6 \pm 22.0)$~pb.

\begin{figure}[htbp]
\centering
 \includegraphics[width=10.0cm]{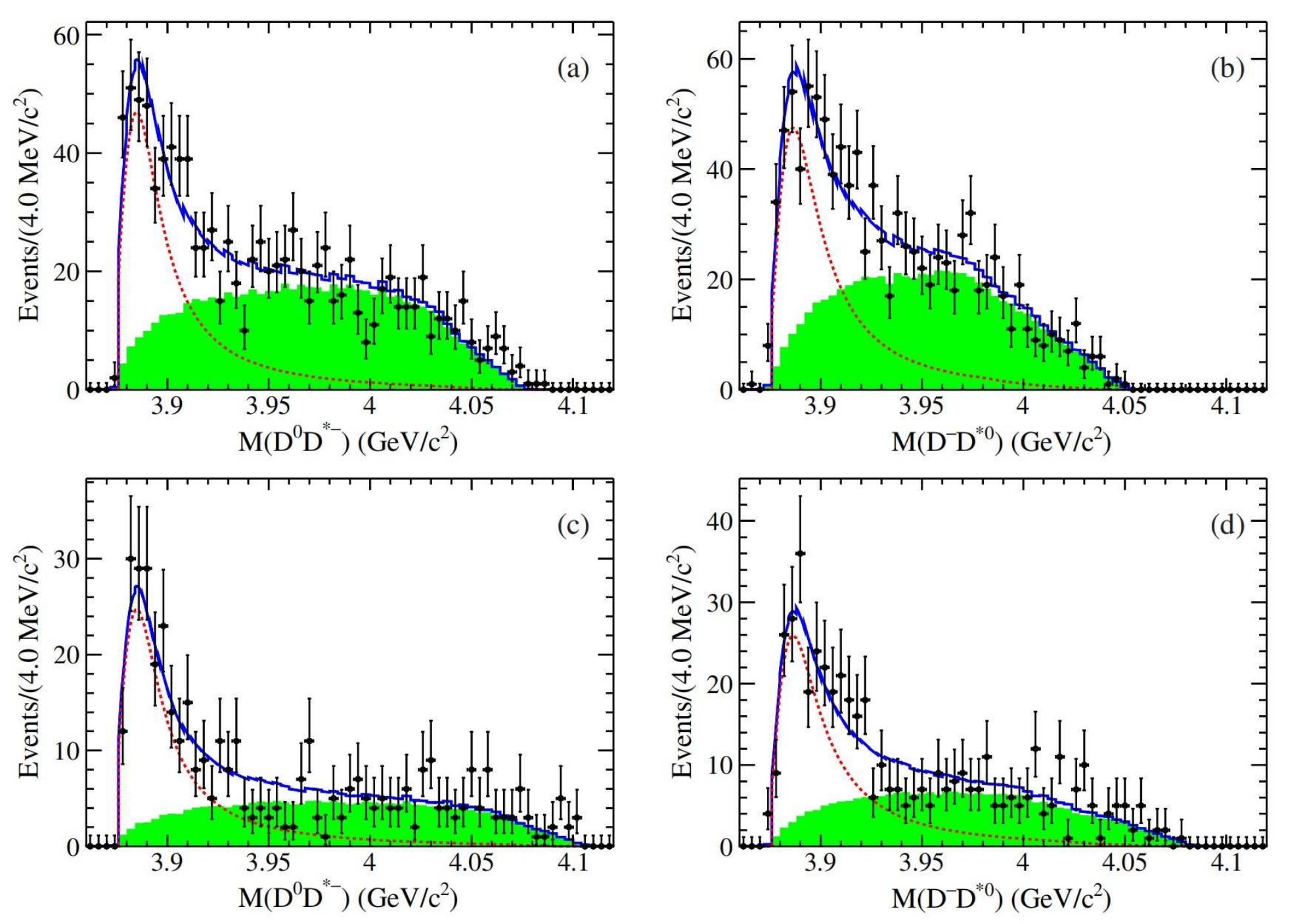}
\caption{The $M(D^0 D^{*-})$ (a,c) and $M(D^-D^{*0})$ (b,d)
distributions for selected signal candidates at $\sqrt{s}=4.23$~GeV (a,b) and
4.26 GeV (c,d) in the double-tag analysis~\cite{Ablikim:2015swa}. The solid curves show
the total fits, the dashed lines are the signal shapes, and the green shaded histograms
describe the background shapes estimated by phase space MC simulation.}
\label{bes3-ddtag}
\end{figure}

The processes $\EE \to \pip D^0 D^{*-}+c.c.$ and $\pip
D^- D^{*0}+c.c.$ are reanalyzed using a double-tag technique
with data samples at $\sqrt{s}$=4.23 and 4.26~GeV~\cite{Ablikim:2015swa}.
The bachelor $\pi^+$ and the $D$-meson pair are reconstructed, with
the $\pi$ from $D^{*-}$ and $D^{*0}$ decays inferred using
energy-momentum conservation.
The $D^0$ candidates are reconstructed in four
decay modes ($K^-\pi^+$, $K^-\pi^+\pi^0$, $K^-\pi^+\pi^+\pi^-$,
and $K^-\pi^+\pi^+\pi^-\pi^0$), and the $D^-$ in six decay modes
($K^+\pi^-\pi^-$, $K^+\pi^-\pi^-\pi^0$, $K^0_S\pi^-$,
$K^0_S\pi^-\pi^0$, $K^0_S\pi^+\pi^-\pi^-$, and $K^+K^-\pi^-$).
With both $D$ candidates reconstructed, the background level is greatly
suppressed.
Similarly, a structure close to the $D\bar{D}^*$ mass threshold is observed
in the $(D\bar{D}^*)^{\pm}$ invariant mass distributions for the two processes at
$\sqrt{s}$=4.23 and 4.26~GeV. A simultaneous fit with a BW function for the
$\zc$ signal and a phase-space distribution for the background to the
$(D\bar{D}^*)^{\pm}$ invariant mass distributions
yields a mass of $(3890.3\pm 0.8)$~MeV and a width of $(31.5\pm 3.3)$~MeV,
with a statistical significance greater than 10$\sigma$, where the errors
are statistical only. The corresponding pole mass and width are
$(3881.7\pm 1.6\pm 1.6)$~MeV and $(26.6\pm 2.0\pm 2.1)$~MeV, respectively.
The production rates are measured to be $\sigma[\EE\to
\pi^{\mp}\zc^{\pm}]\BR[\zc^{\pm}\to (DD^*)^{\pm}] =
(141.6\pm 7.9\pm 12.3)$~pb and $(108.4\pm
6.9\pm 8.8)$~pb for $\sqrt{s}$=4.23~GeV and 4.26~GeV, respectively.
The pole position of the $\zc$ and the production rate are
consistent with those from the single-tag analysis with improved precision.
Figure~\ref{bes3-ddtag} shows the $M(D^0 D^{*-})$ and $M(D^-D^{*0})$
distributions for selected signal candidates at $\sqrt{s}=4.23$~GeV and
4.26 GeV together with the projection of of the simultaneous fit.
The double-tag analysis only has $\sim$9\% events in
common with the single-tag analysis, so the two analyses are
almost statistically independent and can be combined into a
weighted average. The combined pole mass and width are $(3882.2\pm
1.1\pm 1.5)$~MeV and $(26.5\pm 1.7\pm 2.1)$~MeV,
respectively. The combined production rate $\sigma[\EE\to
\pi^{\mp}\zc^{\pm}]\BR[\zc^{\pm}\to (DD^*)^{\pm}]$ is
$(104.4\pm 4.8\pm 8.4)$~pb at $\sqrt{s}$=4.26~GeV.
In an analysis of $\EE\to \pi^0 (D\bar{D}^*)^0$, the $\zc^0\to
(D\bar{D}^*)^0$ is also observed~\cite{Ablikim:2015gda} and all the results
agree with the expectations from isospin symmetry.

Determination of the spin-parity of the $\zc$ is undoubtedly a crucial issue.
In both the single-tag and double-tag analyses of $\zc\to
D\bar{D}^*+c.c.$~\cite{Ablikim:2013xfr,Ablikim:2015swa}, by checking the angular
distribution of the $\pi$ accompanying the $\zc$, BESIII found
that the spin-parity $J^P=1^+$ of the $\zc$ is favored over
$J^P=0^-$ and $1^-$ assumptions, but $J>1$ cannot be ruled out
by simply checking one angular distribution.
The quantum numbers $J^P=0^+$ are not allowed due to spin-parity
conservation in $\zc\to \pi\jpsi$ decays.
To determine the $J^P$ values of $\zc$ precisely, BESIII performed
a PWA to the selected $\EE\to \pp\jpsi$ candidates at $\sqrt{s}=4.23$
and $4.26$~GeV~\cite{Collaboration:2017njt}.
With the same event selection as in Ref.~\cite{Ablikim:2013mio},
the numbers of selected signal events are 4154 at $\sqrt{s}=4.23$~GeV and
2447 at $\sqrt{s}=4.26$~GeV, with 365 and 272 background events,
respectively, estimated by using the normalized $\jpsi$ mass sidebands.

\begin{figure*}[!htbp]
\centering
\includegraphics[width=0.9\textwidth]{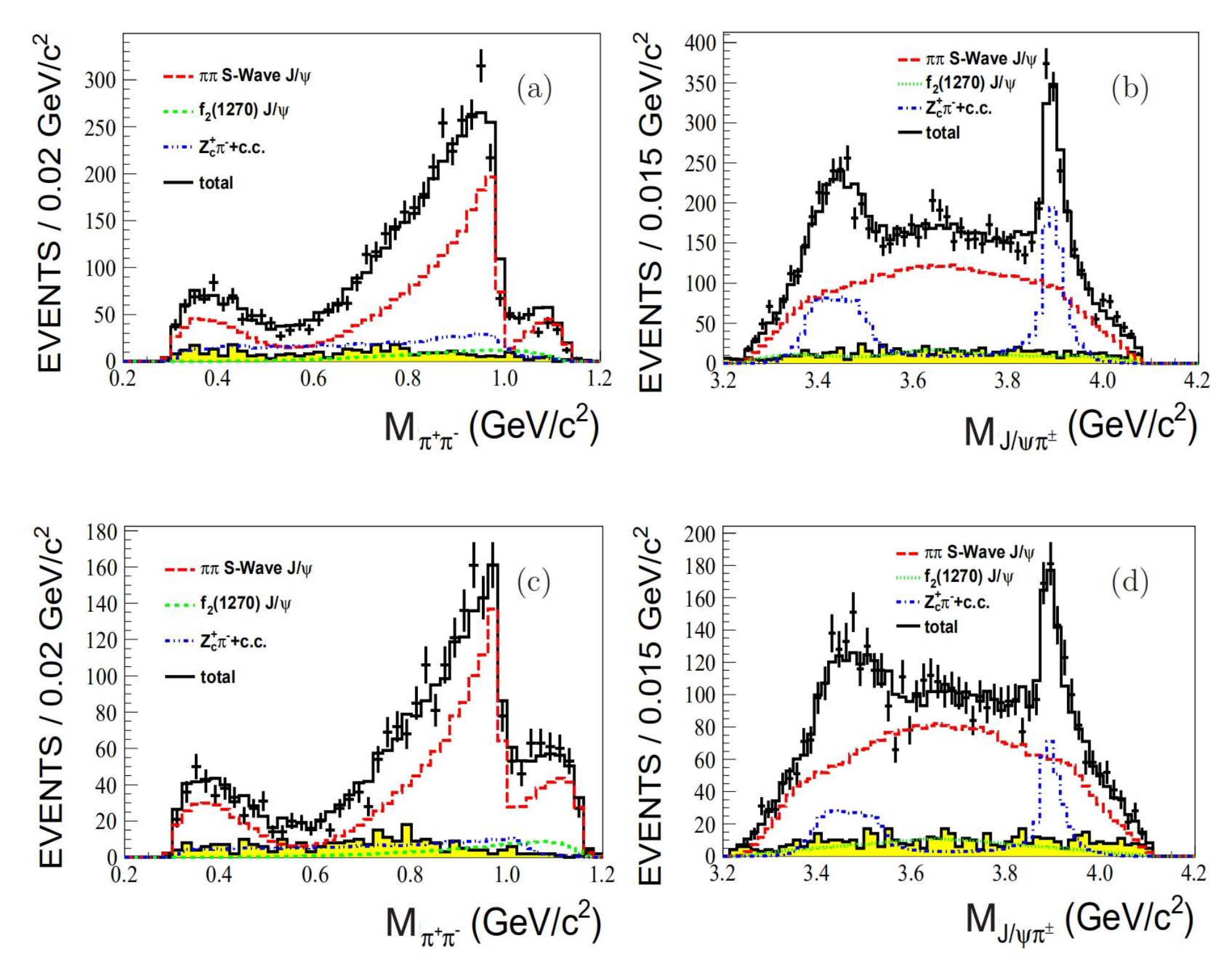}
\caption{\label{pwafitresult} Projections to $M_{\pp}$ (a,~c) and
$M_{\pi^\pm\jpsi}$ (b,~d) of the fit results with $J^P=1^+$ for
the $\zc$, at $\sqrt s=4.23$~GeV (a,~b) and $4.26$~GeV (c,~d)
from BESIII PWA~\cite{Collaboration:2017njt}. The
points with error bars are data, and the black histograms are the
total fit results. The shaded histograms denote the backgrounds. Plots
(b) and (d) are filled with two entries per event.
The contributions from fitted components are
indicated in the plots.}
\end{figure*}

Amplitudes of the PWA are constructed with the helicity-covariant
method~\cite{Chung:1997jn, Chung:1993da, Chung:2007nn}. The process
$\EE\to \pp\jpsi$ is assumed to
proceed via the $\zc$ resonance, i.e., $\EE\to \pi^\pm \zc^\mp$,
$\zc^\mp\to \pi^\mp\jpsi$, and via the non-$\zc$ decay $\EE\to
R\jpsi$, $R\to \pp$, with $R=\sigma$, $f_0(980)$, $f_2(1270)$, and
$f_0(1370)$. In the fit, the $\zc$ line shape is described with a
Flatt\'e-like formula taking into account the fact that the
$\zc^\pm$ decays are dominated by the
$(D\bar{D}^*)^\pm$~\cite{Ablikim:2013xfr,Ablikim:2015swa} and
$\pi^\pm\jpsi$~\cite{Ablikim:2013mio} final states. All processes are
added coherently to obtain the total amplitude. The fit indicates that the
spin-parity $J^P=1^+$ of the $\zc$ are
favored by more than 7$\sigma$ over other quantum numbers ($0^-$,
$1^-$, $2^-$, and $2^+$). Figure~\ref{pwafitresult} shows
projections of the simultaneous fit results with $J^P=1^+$ for the $\zc$ state
at $\sqrt{s}=4.23$ and 4.26 GeV, where contributions from fitted components
are indicated in the plots.
The pole mass and width are determined to be $(3881.2\pm 4.2\pm
52.7)$~MeV and $(51.8\pm 4.6\pm 36.0)$~MeV, respectively.
The Born cross sections for $\EE\to \pi^+\zc^-+c.c.\to \ppjpsi$ are
measured to be $(21.8\pm 1.0\pm 4.4)$~pb at $\sqrt{s}=4.23$~GeV
and $(11.0\pm 1.2\pm 5.4)$~pb at $\sqrt{s}=4.26$~GeV.
If the $\zc^\pm$ would be parametrized with a constant-width BW function,
a simultaneous fit would disfavor the BW parametrization with a significance of 6.6$\sigma$.
\vspace{0.5cm}

\begin{figure}
 \begin{center}
 \includegraphics[height=5.0cm]{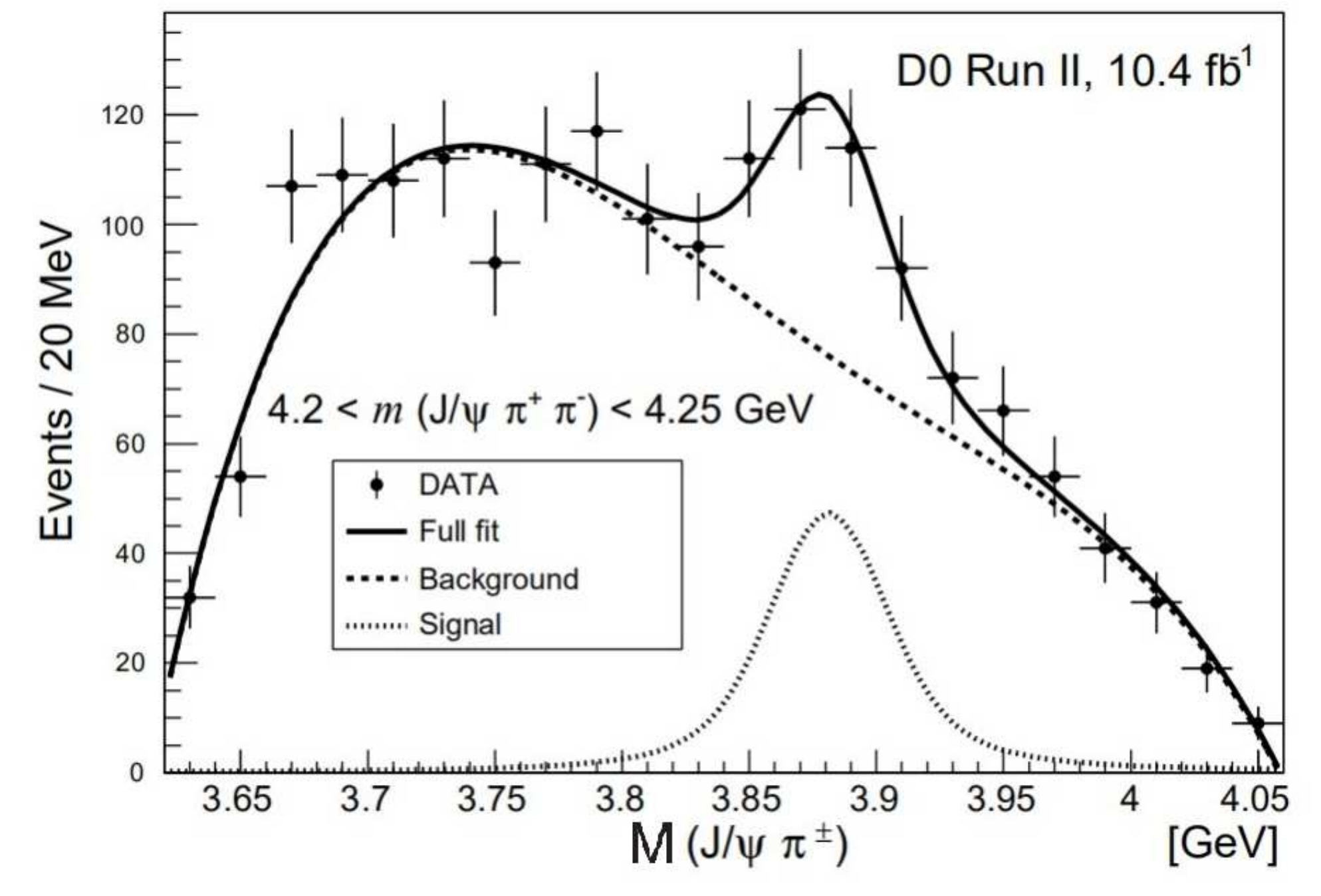}
 \includegraphics[height=5.0cm]{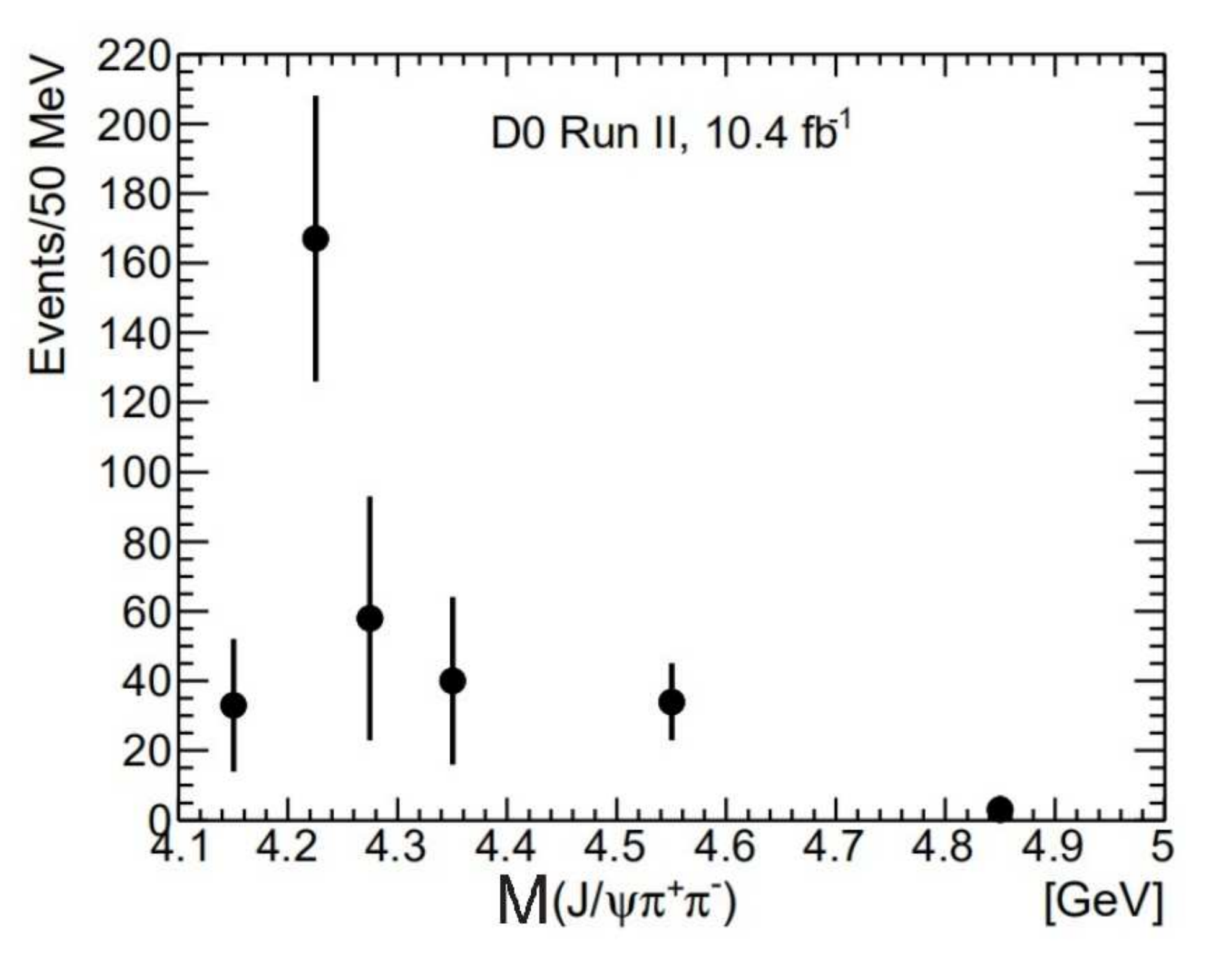}
 \end{center}
 \caption{\label{d0:zc3900}
 The fit to the invariant mass distribution of $\pi^{\pm} J/\psi$ candidates
 in $4.2<M(\pi^+ \pi^- J/\psi)<4.25$~GeV range (left plot)
 and the resulting differential distribution of the $Z_c(3900)^{\pm}$ signal yield
 versus $M(\pi^+ \pi^- J/\psi)$ (right plot)~\cite{Abazov:2018cyu}.
 }
\end{figure}

Although the $\zc$ has been clearly observed at $\EE$ colliders, it is still
necessary to check if it can be produced in other processes, such as in
$B$ decays or at hadron colliders. This will also add more experimental
information to our understanding of its inner structure. For example, the $Z_c(3900)^{\pm}$ may be produced
in $b$-hadron decays. Recently the D0 experiment looked for the $\zc$
using $10.4~\rm{fb^{-1}}$ of $p \overline p$ collision data and found evidence for the decay
$\zc^{\pm} \to \pi^\pm\jpsi$~\cite{Abazov:2018cyu}:
For the selected $\pi^+ \pi^- J/\psi$ candidates, D0 performed
binned maximum-likelihood fits to the $\pi^{\pm} J/\psi$ mass distribution
in six $\pi^+ \pi^- J/\psi $ mass intervals, (4.1$-$4.2), (4.2$-$4.25),
(4.25$-$4.3), (4.3$-$4.4), (4.4$-$4.7), and
(4.7$-$5.0)~GeV, with roughly equal numbers of signal plus background
events. Figure~\ref{d0:zc3900} (left plot) shows the fit to
the invariant mass distribution of $\pi^{\pm} J/\psi$ candidates
in the $4.2<M(\pi^+ \pi^- J/\psi)<4.25$~GeV region as an example.
A clear enhancement near the $Z_c(3900)^{\pm}$ mass can be seen, which is
consistent with the decays of
the $Y(4230)$. The significances are smaller but
nonvanishing in other $\pi^+ \pi^- J/\psi$
mass regions between 4.25 and 4.7~GeV, while no significant
signal is seen in the bins $4.1<M(\pi^+ \pi^- J/\psi)<4.2$~GeV or
$4.7<M(\pi^+ \pi^- J/\psi)<5.0$~GeV.
The measured mass is $M=(3895.0\pm5.2^{+4.0}_{-2.7})$~MeV
and the significance is 4.6$\sigma$ with systematic uncertainties included
from a fit to the data in the mass range $4.2<M(\pi^+ \pi^- J/\psi)<4.7$~GeV.
The resulting differential distribution of the $Z_c(3900)^{\pm}$ signal
yield is shown in Fig.~\ref{d0:zc3900}
(right plot). It reveals that a $Z_c(3900)^{\pm}$
signal is correlated with the $\pi^+ \pi^- J/\psi$ system
in the invariant mass range 4.2$-$4.7~GeV including
the $Y(4230)$ and $Y(4360)$ states.
There is also an indication that some $Z_c(3900)^{\pm}$ signal events
come from $b$-hadron decays to an intermediate
$J/\psi \pi^+ \pi^-$ combination with mass above that of the $Y(4360)$.
This is the first evidence for the $\zc$ production at a hadron collider.

\begin{figure}
 \begin{center}
 \includegraphics[width=220px]{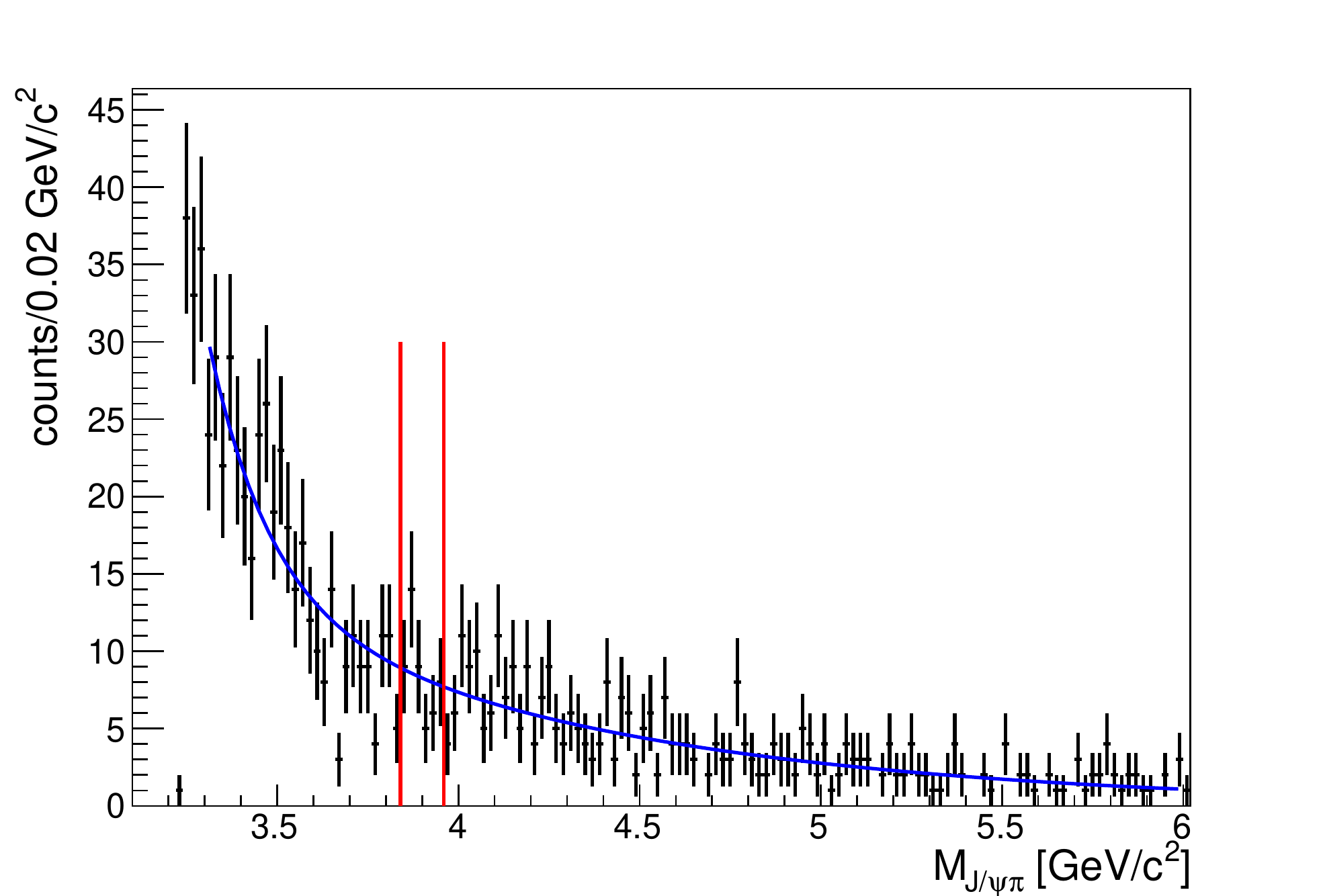}
 \end{center}
 \caption{\label{compass:zc3900} The fit to the mass spectrum of the
$\pi^{\pm} J/\psi$ candidates obtained by the COMPASS Collaboration
in the $\mu^+~N \rightarrow \mu^+ Z_c^{\pm}(3900)~N \rightarrow \mu^+ \pi^{\pm}J/\psi
N\rightarrow \mu^+\mu^+\mu^- \pi^{\pm} N$ process~\cite{Adolph:2014hba}.
The red solid lines show the expected $Z_c(3900)^{\pm}$
signal region.}
\end{figure}

Based on the vector meson dominance model, the authors of
Ref.~\cite{Lin:2013mka} predicted a sizable cross section of the reaction
$\gamma~N \rightarrow Z_c(3900)^{\pm}~N$
for $\sqrt{s_{\gamma N}}\sim 10$~GeV, where a $Z_c(3900)^{\pm}$ can be produced
by the interaction of an incoming photon with a virtual charged pion provided
by the target nucleon.
Using the data obtained by scattering positive muons of 160~GeV (2002-2010)
or 200~GeV momentum (2011) off solid $^6$LiD (2002-2004) or NH$_3$ targets
(2006-2011), COMPASS searched for the process
$\mu^+~N \rightarrow \mu^+ Z_c(3900)^{\pm}~N \rightarrow \mu^+ \pi^{\pm}J/\psi
N\rightarrow \mu^+\mu^+\mu^- \pi^{\pm} N$~\cite{Adolph:2014hba}.
The mass spectrum for the selected $\pi^{\pm} J/\psi$ candidates is shown in
Fig.~\ref{compass:zc3900}, where no statistically significant resonant
structure around 3.9~GeV can be seen. An upper limit
for the ratio $\BR[Z_c(3900)^{\pm}\rightarrow \pi^{\pm} J/\psi]\sigma[
\gamma~N \rightarrow Z_c(3900)^{\pm}~ N]/\sigma(\gamma~N \rightarrow
J/\psi~ N)$ of $3.7\times10^{-3}$ was established at the 90\% C.L.

\begin{figure}[htbp]
\begin{center}
\includegraphics[width=0.5\textwidth]{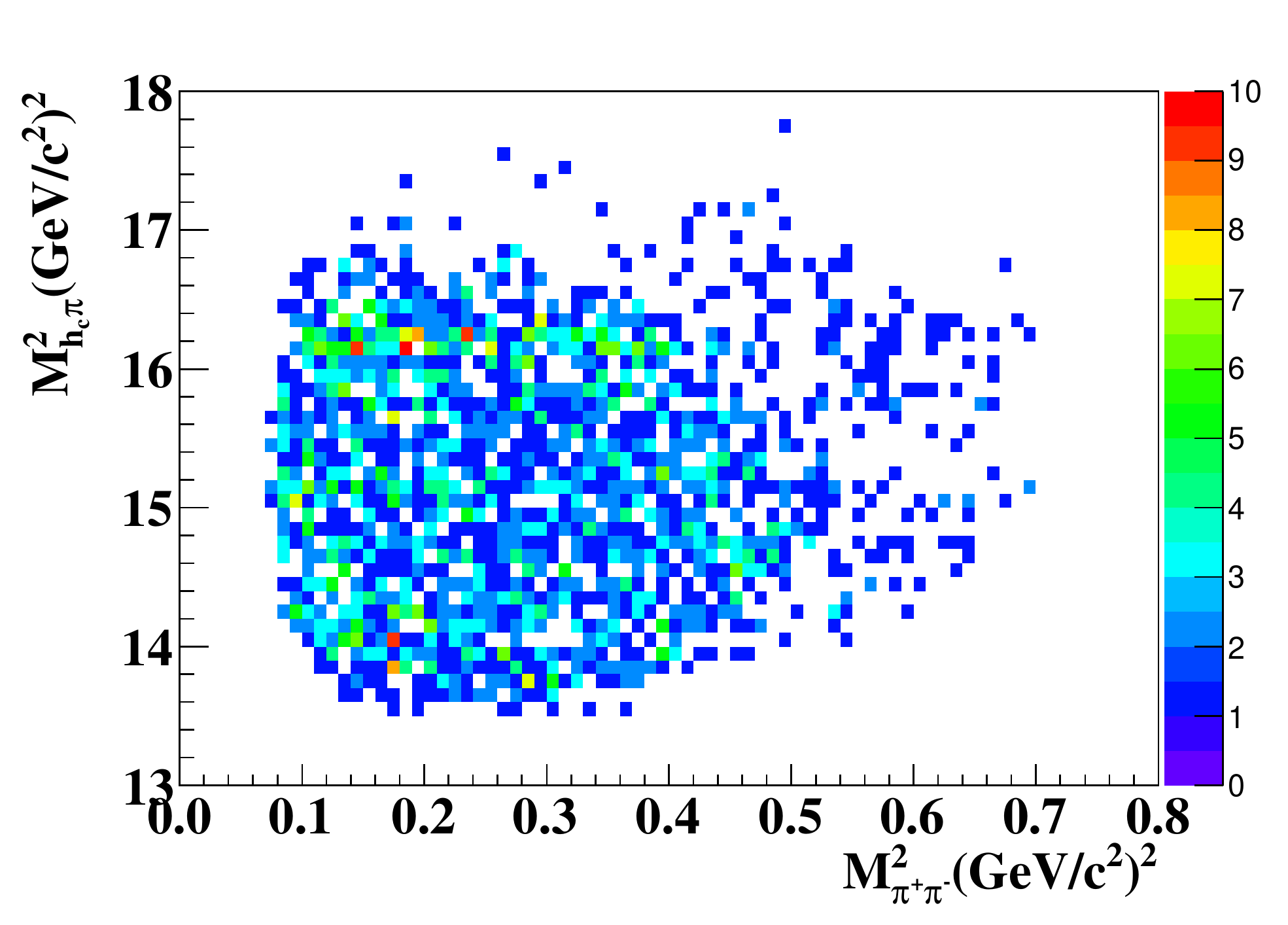}
\caption{Dalitz plot of $M^2_{\pi^+\hc}$ versus $M^2_{\pp}$ for
the selected $\EE\to \pphc$ events summed over the 4.23, 4.26, and
4.36~GeV c.m.\ energy points~\cite{Ablikim:2013wzq}.}
\label{dalitz_pphc}
\end{center}
\end{figure}

\begin{figure}[htbp]
\begin{center}
\includegraphics[width=0.45\textwidth]{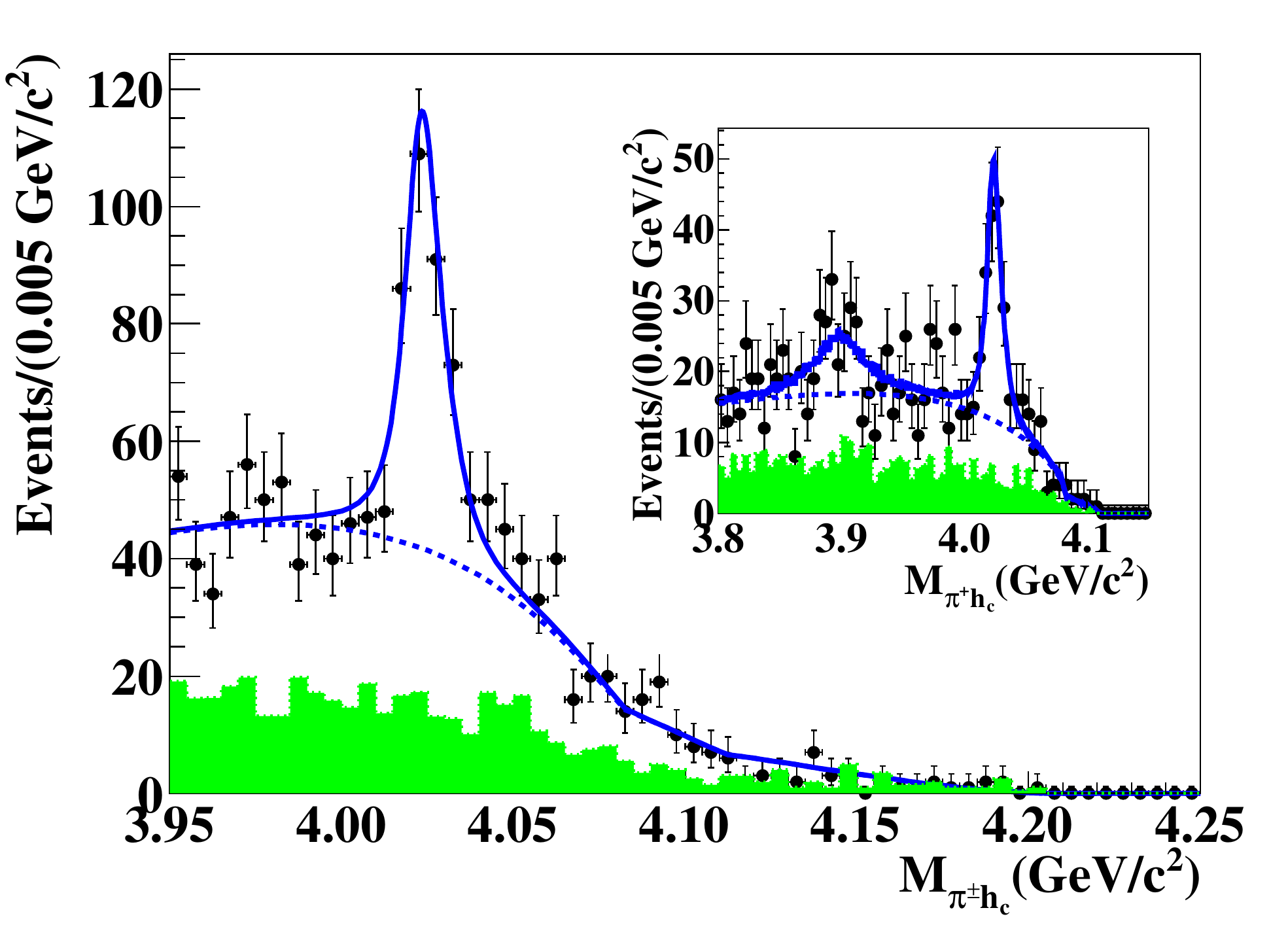}
\includegraphics[width=0.45\textwidth]{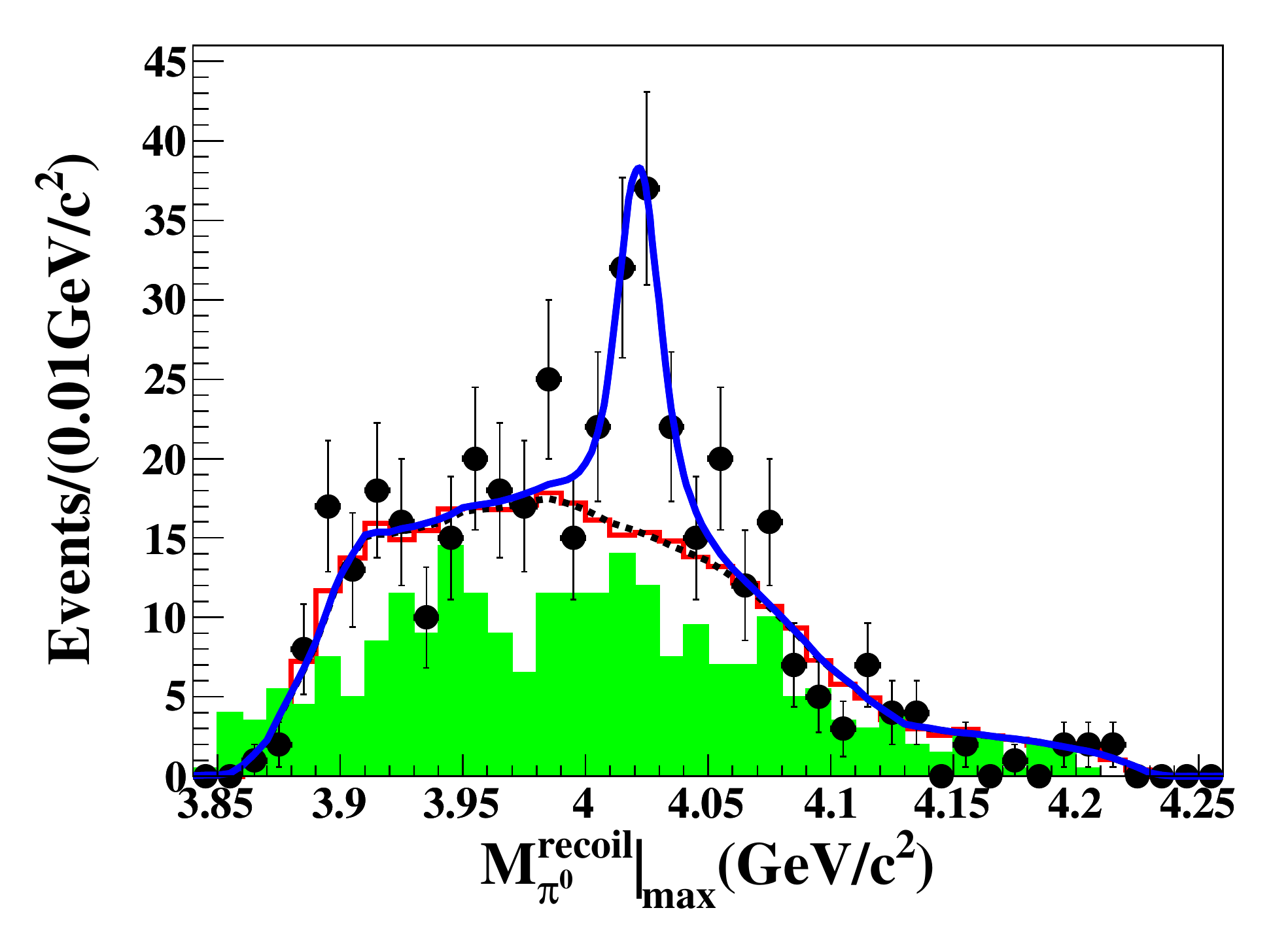}
\caption{Sum of the simultaneous fits to the $M(\pi^\pm h_c)$~\cite{Ablikim:2013wzq}
(left panel) and $M(\piz h_c)$~\cite{Ablikim:2013wzq} (right panel) distributions from
$\EE\to \pphc$ and $\piz\piz\hc$, respectively, at 4.23, 4.26, and
4.36~GeV in the BESIII data; the inset in the left panel shows the
sum of the simultaneous fit to the $M({\pi^\pm h_c})$
distributions at 4.23 and 4.26~GeV with $\zc$ and $\zcp$. Dots
with error bars are data; shaded histograms are normalized
sideband background; the solid curves show the total fits, and the
dotted curves are the backgrounds from the fits.} \label{1Dfit}
\end{center}
\end{figure}

As mentioned in Sec.~\ref{Sect:3.1.2}, BESIII measured cross sections
of $\EE\to \pphc$ at c.m.\ energies of
3.90--4.42~GeV with $\hc\to \gamma\etac$ and $\etac$ decays
into 16 hadronic final states~\cite{Ablikim:2013wzq}.
The $\hc$ signal is selected using $3.518~{\rm GeV} < M_{\gamma
\eta_c} < 3.538$~GeV, and $\pphc$ samples of 859 events at
4.23~GeV, 586 events at 4.26~GeV, and 469 events at 4.36~GeV are
obtained with purities of about $\sim$65\%.
The Dalitz plot of the selected $\pphc$ candidate
events to check for possible intermediate states is shown in Fig.~\ref{dalitz_pphc}.
There are no clear structures in the $\pp$ system, while there is
distinct evidence in the $\pi^\pm\hc$ system at about 4.02~GeV.
Figure~\ref{1Dfit}~(left) shows the projection of the $M(\pi^\pm\hc)$
(two entries per event)
distribution for the signal events summed over the three c.m.\ energy points,
as well as the background events estimated from the normalized $\hc$ mass
sidebands.
There is a significant peak at around 4.02~GeV [$\zcp$], and there are
also some events at around 3.9~GeV which could be $\zc$ as shown in
the inserted plot.

Assuming there are both $\zcp$ and $\zc$ contributions with the mass and
width of the latter fixed to the BESIII measurements~\cite{Ablikim:2013mio},
an unbinned maximum-likelihood fit is applied to the
$M(\pi^\pm\hc)$ distribution summed over the 16 $\eta_c$ decay modes.
As the $\zc$ signal overlaps with the reflection of the $\zcp$ at
$\sqrt{s}=4.36$ GeV,
the data at 4.23, 4.26, and 4.36~GeV are fitted
simultaneously to the same $\zcp$ signal function only with common mass
and width from 3.95 to 4.25~GeV,
while the data at 4.23 and 4.26~GeV are fitted
simultaneously with both of the $\zc$ and $\zcp$ signals
from 3.8 to 4.15~GeV.
The fitted results are shown in Fig.~\ref{1Dfit}~(left),
where the inset shows the sum of the simultaneous fit at 4.23 and 4.26~GeV
with $\zc$ and $\zcp$. The fit yields a mass of $(4022.9\pm 0.8\pm
2.7)~{\rm MeV}$ and a width of $(7.9\pm 2.7\pm 2.6)$~MeV,
with a statistical significance greater than $8.9\sigma$ for the $\zcp$.
The cross sections are calculated to be
$\sigma[\EE\to \pi^\pm \zcp^\mp\to \pphc] = (8.7\pm 1.9\pm 2.8\pm
1.4)$~pb at 4.23~GeV, $(7.4\pm 1.7\pm 2.1\pm 1.2)$~pb at 4.26~GeV,
and $(10.3\pm 2.3\pm 3.1\pm 1.6)$~pb at 4.36~GeV, where the first
errors are statistical, the second ones systematic, and the third
ones from the uncertainty in $\BR(\hc\to \gamma\etac)$~\cite{Tanabashi:2018oca}.
Since the statistical significance of $\zc$ is only 2.1$\sigma$,
the upper limits on the production cross sections $\sigma[\EE\to \pi^\pm \zc^\mp\to \pphc]$
are determined to be 13 pb and 11 pb at $\sqrt{s}=4.23$ and 4.26 GeV,
respectively,
at 90\% C.L., which are lower than those of $\zc\to
\pi^\pm\jpsi$~\cite{Collaboration:2017njt}.

As mentioned in Sec.~\ref{Sect:3.1.2}, BESIII measured cross sections of
$\EE\to \piz\piz\hc$ at $\sqrt{s}=$4.23,
4.26, and 4.36~GeV for the first time~\cite{Ablikim:2014dxl}.
The intermediate state of $\zcp^0$, the neutral isospin
partner of the $\zcp^\pm$, is observed in the $\piz\hc$ invariant mass
distribution, as shown in Fig.~\ref{1Dfit}~(right). This observation
indicates that there is no anomalously large isospin violation
in $\pp\hc$ and $\pi\zcp$ systems, and $\zcp$ is an isovector
state.

\begin{figure}[htbp]
\centering
\includegraphics[height=5cm]{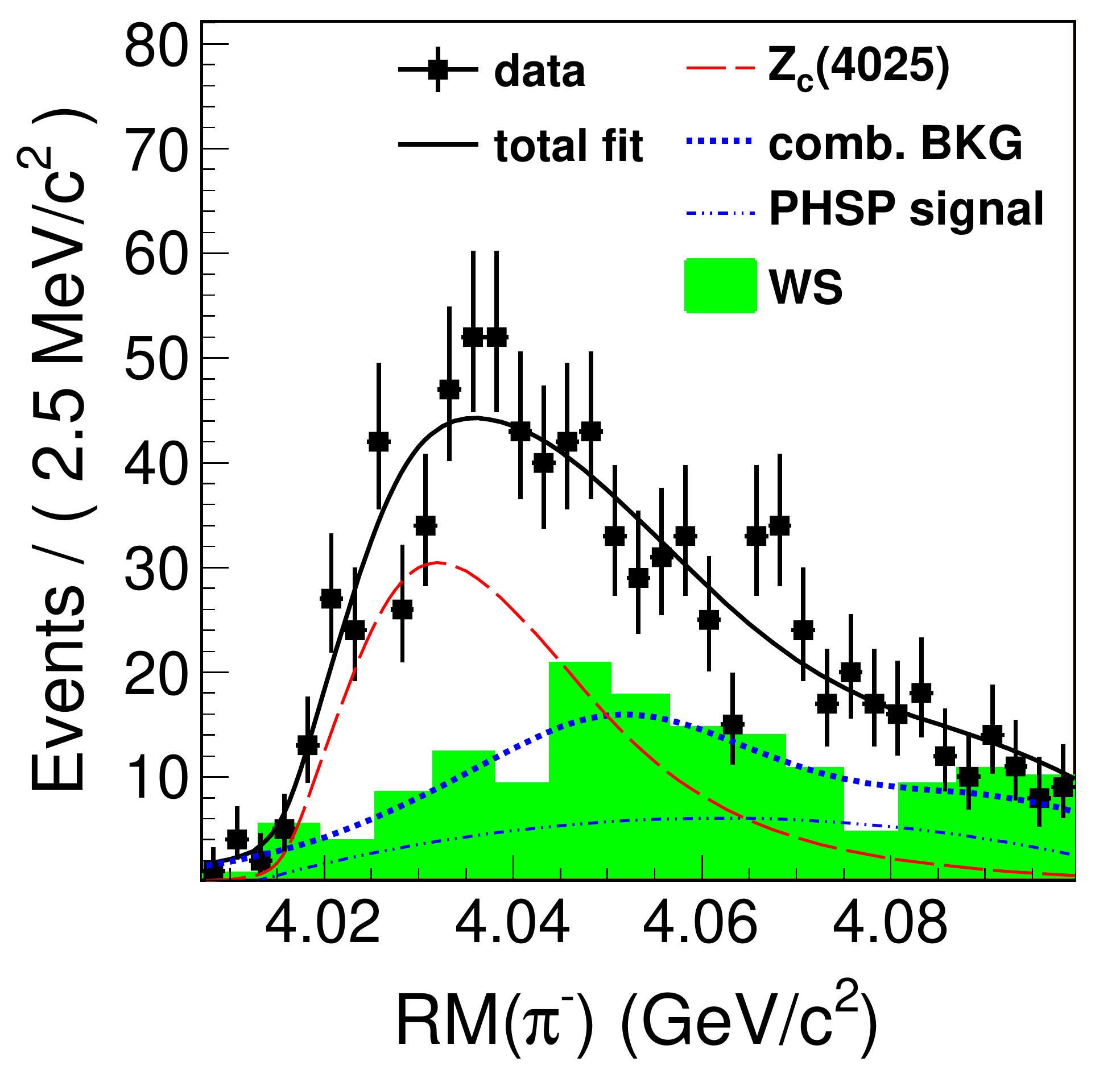}
\includegraphics[height=5cm]{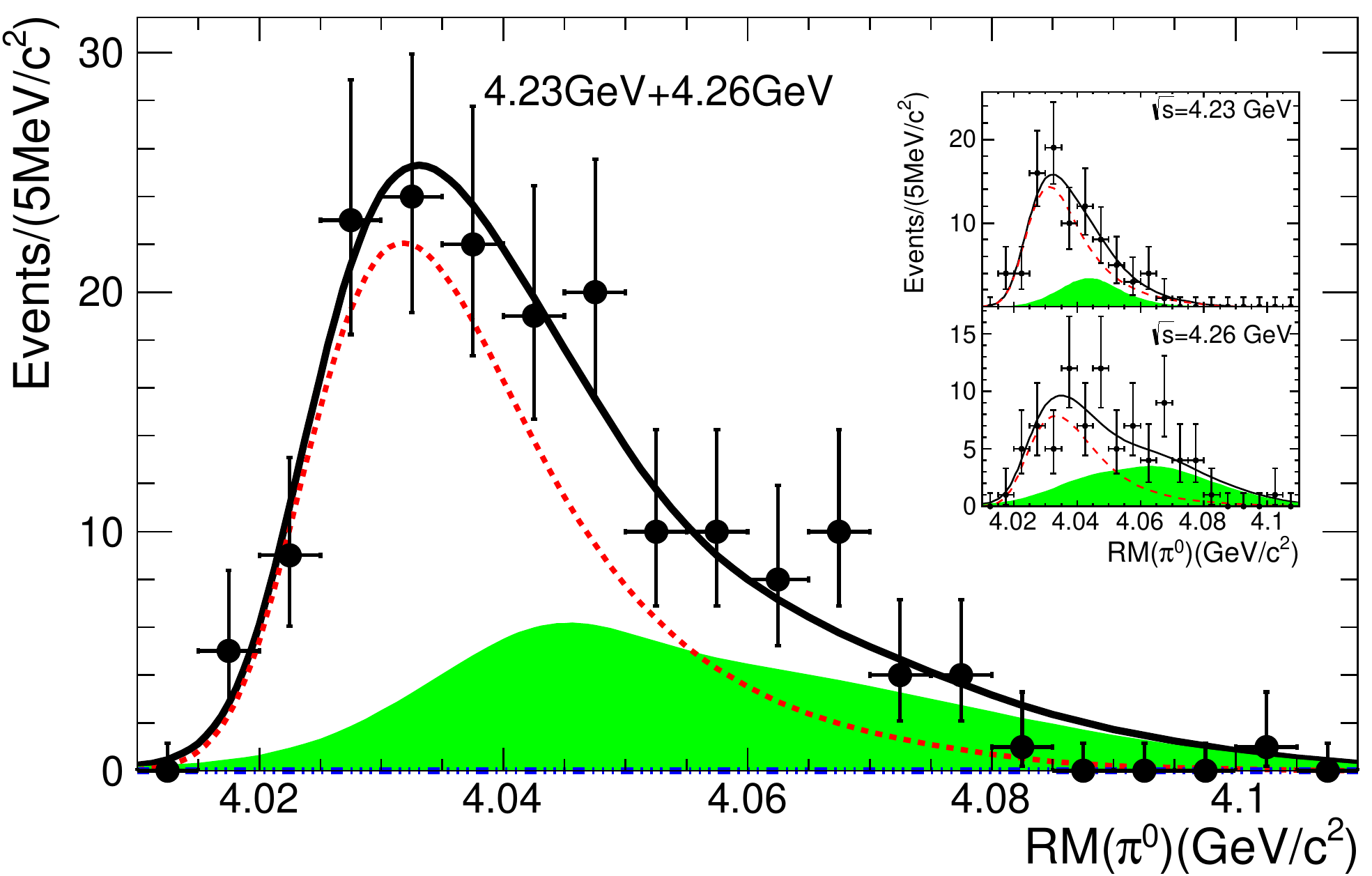}
\caption{Unbinned maximum-likelihood fits to the $\pi^\mp$ recoil
mass spectrum in $\EE\to (D^{*} \bar{D}^{*})^{\pm} \pi^\mp$
at $\sqrt{s}=4.26$~GeV~\cite{Ablikim:2013emm} (left), and to the
$\piz$ recoil mass spectrum
in $\EE\to (D^{*} \bar{D}^{*})^{0} \piz$ at
$\sqrt{s}=4.23$ and $4.26$~GeV~\cite{Ablikim:2015vvn} (right) at BESIII. }
\label{fig:fit}
\end{figure}

Since the $\zcp$ mass is close to the $D^*\bar{D}^*$ mass
threshold, it may have a strong coupling to the $D^*\bar{D}^*$
final state. The process $\EE \to (D^{*}
\bar{D}^{*})^{\pm} \pi^\mp$ [$(D^{*}
\bar{D}^{*})^{\pm}= D^{*+} \bar{D}^{*0}$ and $D^{*-} D^{*0}$] was studied
by BESIII using a 827 pb$^{-1}$ data at
$\sqrt{s}=$4.26~GeV to check the $D^*\bar{D}^*$ system~\cite{Ablikim:2013emm}.
In this analysis, a partial reconstruction technique is used, i.e.,
taking $\pi^- D^{*+} \bar{D}^{*0}$ as an example, only the bachelor $\pi^-$,
a charged $D^+$, and at least one soft $\piz$ from $D^*$ decays are
reconstructed.
By identifying the $D^+$ particle, the charges of its mother particle $D^{*+}$
and the bachelor $\pi^-$ are also unambiguously identified. The $\zcp$ state
is searched for in the bachelor $\pi^{\pm}$
recoil mass spectrum. The final $\pi^{\pm}$ recoil mass spectrum is shown in
Fig.~\ref{fig:fit}~(left),
where a structure near the $(D^* \bar{D}^*)^{\pm}$ threshold is observed
and the shaded histogram is from the combinatorial backgrounds estimated
by combining a reconstructed $D^{\pm}$ with a pion of the
wrong charge. Since the signal events from phase space (dot-dashed line) and combinatorial
backgrounds (dotted line) cannot describe the data, assuming the enhancement is due
to $\zcp$,
an unbinned maximum-likelihood fit to the $\pi^{\pm}$
recoil mass spectrum is performed with an $S$-wave BW shape to parameterize
the structure. The fit yields a mass of $(4026.3\pm 2.6\pm
3.7)$~MeV and a width of $(24.8\pm 5.6\pm 7.7)$~MeV,
with a statistical significance of 13$\sigma$. The fit results are shown in
Fig.~\ref{fig:fit}~(left). The Born
cross section of $\EE \to (D^{*}
\bar{D}^{*})^{\pm} \pi^\mp$ is measured to be $(137\pm 9\pm 15)$~pb
at $\sqrt{s}=$4.26~GeV.
From the fit results, $401\pm 47$ $\zcp$ signal events are obtained,
and the associated ratio of the production rates is determined to be
$$\frac{\sigma[\EE\to \zcp^\pm \pi^\mp \to (D^{*} \bar{D}^{*})^{\pm} \pi^\mp]}{\sigma[e^+e^-\to
(D^{*} \bar{D}^{*})^{\pm} \pi^\mp]}=0.65\pm
0.09\pm 0.06 . $$

The processes $\EE \to (D^{*0} \bar{D}^{*0})\piz$ and
$(D^{*+}{D}^{*-}) \piz$ are also studied at BESIII to search for
the neutral partner of the $\zcp$
with integrated luminosities of 1092 pb$^{-1}$ at $\sqrt{s}=$ 4.23
and 826 pb$^{-1}$ at $\sqrt{s}=$4.26~GeV~\cite{Ablikim:2015vvn}. In this
analysis, two $D$ mesons are reconstructed together with the bachelor $\piz$.
The recoil mass distribution of the bachelor $\pi^0$
is shown in Fig.~\ref{fig:fit}~(right), together with the
distributions at $\sqrt{s}=$4.23 GeV and 4.26 GeV in the
inset plots. The points with error bars are the data
and the shaded histograms represent the inclusive backgrounds.
Enhancements over the inclusive backgrounds
estimated using the inclusive MC samples for both data
samples can be seen, which cannot be explained by three-body
nonresonant processes either. A simultaneous unbinned maximum-likelihood
fit to the bachelor $\pi^0$ recoil mass spectra at $\sqrt{s}=$4.23 GeV and
4.26 GeV is performed with BW functions with a mass-dependent width as
$\zcp$ signal shapes.
The solid lines show the fit results and the dotted red lines stand
for the $Z_c(4020)^0$ signals.
The fit determines the mass and width of the $\zcp$ pole
position to be $(4025.5^{+2.0}_{-4.7}\pm 3.1)$~MeV and $(23.0\pm6.0 \pm 1.0)$~MeV, respectively.
From the simultaneous fit, $69.5\pm 9.2$ and $46.1\pm 8.5$
$\zcp^0$ signal events are obtained at 4.23 and 4.26~GeV,
respectively, with a statistical significance of $5.9\sigma$. The
Born cross section $\sigma[\EE\to \zcp^0 \piz \to(D^{*0}
\bar{D}^{*0} + D^{*+}{D}^{*-})\piz]$ is measured to be $(61.6\pm
8.2\pm 9.0)$~pb at 4.23~GeV and $(43.4\pm8.0 \pm 5.4)$~pb at 4.26~GeV.
BESIII also obtained the ratio
$$\frac{\sigma[\EE \to \zcp^0 \piz \to(D^{*}
\bar{D}^{*})^0\piz]}{\sigma[\EE\to \zcp^+ \pi^-\to(D^{*}
\bar{D}^{*})^+\pi^-]}\approx 1 \ $$
 at $\sqrt{s}=4.26$~GeV, which is expected from isospin symmetry.
This also confirms that the isospin of the $\zcp$ is one.

As mentioned before, to determine the $J^P$ values of $\zc$,
BESIII did the PWA to the selected $\EE\to \pp\jpsi$ events at
$\sqrt{s}=$4.23 and 4.26~GeV~\cite{Collaboration:2017njt}.
In the PWA, BESIII also searched for the process
$\EE\to \pi^-\zcp^+ +c.c.\to \pp\jpsi$,
with the $\zcp^\pm$ assumed to be a $1^+$ state added in the global fit.
Its mass is taken from Ref.~\cite{Ablikim:2013wzq} and width is taken
as the observed value. The fit yields a 3$\sigma$ statistical significance
for $\zcp^\pm \to \pi^\pm
\jpsi$ in the combined data. The Born cross sections are measured to be $(0.2\pm 0.1)$~pb at
$\sqrt{s}=4.23$~GeV and $(0.8\pm 0.4)$~pb at $\sqrt{s}=4.26$~GeV,
and the corresponding upper limits at the 90\% C.L.\ are estimated
to be $0.9$ and $1.4$~pb, respectively.

Although the observations of the $\zc$ and $Z_c(4020)$ indicate that they
are not conventional mesons consisting of a quark-antiquark pair, their
exact quark configuration is still unknown. There are many models
developed to interpret their inner structure, including loosely bound
hadronic molecules of two charmed mesons, compact
tetraquarks, hadroquarkonium, and so on. Therefore, how to discriminate
between the molecule and compact tetraquark scenarios experimentally is an important research
topic that will be discussed in some depth in various subsections of Sec.~\ref{sect:4}.
It has recently been suggested that the relative decay rate of
$\zc\to \rho \eta_c$ to $\pi \jpsi$ [or $\zcp \to \rho \eta_c$ to $\pi h_c$]
can be used to discriminate them~\cite{Esposito:2014hsa}.
In the compact tetraquark scenario, the predicted ratio of
$\BR[\zc\to \rho\etac]/\BR[\zc\to \pi\jpsi]$
is 230 or 0.27, depending on whether or not the spin-spin interaction
outside the diquarks is kept~\cite{Esposito:2014hsa,Li:2014pfa,Ke:2013gia,Ma:2014zva,Ma:2015nmy,Voloshin:2018pqn}.
In the molecular framework, on the other hand, this ratio is only 0.046.
Similarly, the predicted ratio of
$\BR[\zcp\to \rho\etac]/\BR[\zcp\to \pi\jpsi]$ is 6.6 in the compact tetraquark model,
but only 0.01 in the meson molecule model~\cite{Esposito:2014hsa}.
Therefore, a search for the decays
$\zc/\zcp \to \rho \eta_c$ offers an important opportunity to understand the
internal structure of the $\zc$ and $\zcp$.

\begin{figure*}[htbp]
\centering
\includegraphics[height=5cm]{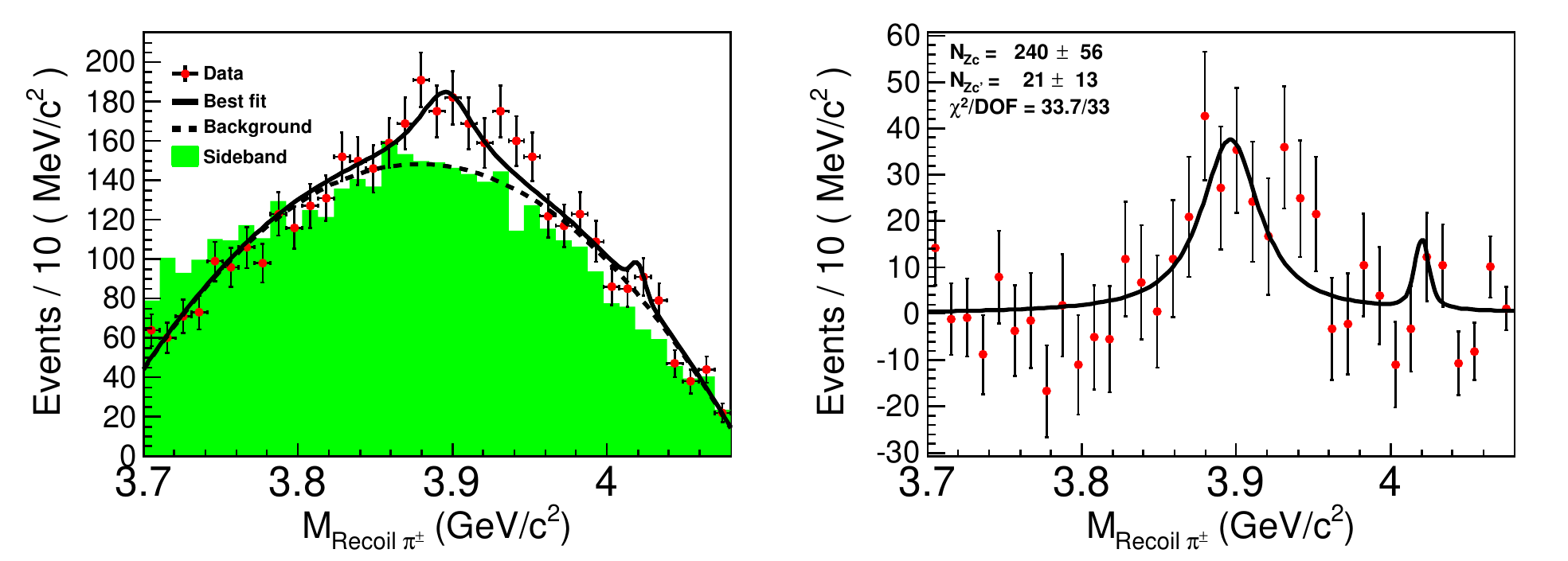}
\caption{The $\pi^\pm$ recoil mass distribution
summed over 9 $\eta_c$ decay channels in $\EE\to
\pi^{\pm}\rho^{\mp} \etac$ at $\sqrt{s}=4.23$~GeV and fit with
$Z_c(3900/4020)^{\pm}$ signals (left panel), and the same plot
with background subtracted (right panel)~\cite{Ablikim:2019ipd}.
Dots with error bars are
data, the shaded histogram is from $\etac$ mass sidebands, normalized to
the number of backgrounds from the fit, the solid lines are the total
fits, and the dotted line is the background. \label{fit_Zc_data_4230}}
\end{figure*}

BESIII searched for $\EE\to \pp\piz\etac$ and intermediate states
decaying into $\rho\etac$ with data at c.m.\ energies above 4~GeV
corresponding to an integrated luminosity of about 4.1~fb$^{-1}$~\cite{Ablikim:2019ipd}.
In this analysis, $\eta_c$ is
reconstructed with 9 hadronic final states: $p\bar{p}$, $2(K^+
K^-)$, $K^+ K^- \pi^+ \pi^-$, $K^+ K^- \pi^0$, $p \bar{p} \pi^0$,
$\ks K^\pm \pi^\mp$, $\pi^+ \pi^- \eta$, $K^+ K^- \eta$, and
$\pi^+ \pi^- \pi^0 \pi^0$.
A clear signal of $\EE\to \pp\piz\etac$ is observed at $\sqrt{s}=4.23$~GeV.
From the fit to the $\eta_c$ mass spectrum, $333^{+83}_{-80}$ $\eta_c$ signal events are obtained with a statistical
significance of 4.2$\sigma$. No significant signals are observed at other
c.m.\ energy points. The $Z_c(3900/4020)^{\pm}\to \rho^\pm \etac$
signals are examined after requiring that the invariant mass of
an $\etac$ candidate is within the $\etac$ signal region
[2.95, 3.02] GeV and the invariant mass of $\pi^{\pm}\pi^{0}$ is within the $\rho^\pm$
signal region [0.675, 0.875] GeV. The recoil mass of the remaining $\pi^\mp$
(equivalent to the invariant mass of $\rho^\pm\etac$) is shown in
Fig.~\ref{fit_Zc_data_4230} for the data at $\sqrt{s}=4.23$~GeV,
together with the contribution from the $\eta_c$ mass sideband events (the shaded histogram).
In Fig.~\ref{fit_Zc_data_4230}, the $\zc^\pm$ signal is apparent,
but there is no statistically significant $\zcp^{\pm}$ signal.
The $\rho^\pm\etac$ invariant mass distribution
is fitted with the contributions from $\zc$ and $\zcp$ together
with a smooth background. In the fit, a possible interference between
the signal and the background is neglected.
The solid line in the left plot of Fig.~\ref{fit_Zc_data_4230} shows the best fit,
while the right plot is the background-subtracted distribution.
The total $\zc^{\pm}$ signal yield is $240^{+56}_{-54}$ events with a statistical significance of 4.3$\sigma$,
and that of the $\zcp^{\pm}$ is $21^{+15}_{-11}$ events with a statistical significance of 1.0$\sigma$.
The $\zc$ signals at other c.m.\ energies and the $\zcp$ signals at all the c.m.\
energies are not statistically significant.
The cross section is measured as
 \(
\sigma[\EE\to \pi^\mp \zc^\pm\to \pi^\mp\rho^\pm\etac]=(48 \pm 11
\pm 11)~{\rm pb}
 \)
at $\sqrt{s}=4.23$~GeV. This result is equal within errors to the
cross section of $\EE\to \ppp\etac$, which is $(46^{+12}_{-11}\pm
10)\rm \,pb$. This indicates that the $\EE\to \ppp\etac$ process
is saturated by the process $\EE\to \pi^\mp \zc^\pm\to
\pi^\mp\rho^\pm\etac$. No signal is observed at $\sqrt{s}=$4.26,
4.36, 4.42, and 4.6~GeV, and the upper limits of the production cross sections
at the 90\% C.L.\ are determined to be 62 pb, 36 pb, 44 pb, and 14 pb,
respectively.
The upper limits of the production cross section
$\sigma[\EE\to \pi^\mp \zcp^\pm\to \pi^\mp\rho^\pm\etac]$
at the 90\% C.L.\ are determined to be 14 pb, 6 pb, 14 pb, 11 pb, and 21 pb,
respectively, at $\sqrt{s}=4.23$, 4.26, 4.36, 4.42, and 4.6~GeV.

Using the results from Refs.~\cite{Ablikim:2013wzq,Collaboration:2017njt}, the ratio
$R_{\zc}=\BR[\zc\to \rho\etac]/\BR[\zc\to \pi\jpsi]$ is $2.2\pm0.9$ at
$\sqrt{s}=4.23$~GeV and less than 5.6 at $\sqrt{s}=4.26$~GeV at the 90\% C.L., while the ratio
$R_{\zcp}=\BR[\zcp^\pm\to \rho^{\pm} \etac]/\BR[\zcp^\pm\to \pi^{\pm} \hc]$ is less than 1.6, 0.9, and 1.4 at c.m.\
energies of 4.23, 4.26, and 4.36~GeV, respectively, at the 90\% C.L.\ For
the $R_{\zc}$, the current result seems to favor the compact tetraquark interpretation,
while $R_{\zcp}$ is somewhat more consistent with the molecular one.
A clear discrimination is not yet possible because of both the experimental and theoretical uncertainties.

From the above studies, we conclude that the $\zc$ has been observed at
both $\EE$ and hadron colliders. It is an isovector state with positive
$G$-parity and spin-parity $J^P=1^+$. It
decays into $\pi\jpsi$ and $D\dstrbar$, and it may also decay into
$\rho\etac$ and $\pi\hc$ final states. The neutral $\zc$ has a
negative $C$-parity. Although the reported masses and widths are consistent
with each other in many measurements, in
most of the cases the possible interference between the $\zc$ and other
amplitudes is neglected.
A PWA to the selected $\EE\to \ppjpsi$ events has been done by BESIII~\cite{Collaboration:2017njt},
giving a pole mass of $(3881.2\pm 4.2\pm
52.7)$~MeV and a width of $(51.8\pm 4.6\pm 36.0)$~MeV,
where the large errors are dominated by the uncertainties in the
parametrization of the $\pp$ $S$-wave amplitudes.
In this PWA, a drawback is the assumption of $\zc$ decaying dominantly
into $\pi\jpsi$ and $D\dstrbar$
modes, which may introduce bias, as the $\rho\etac$ mode has been
observed with a larger decay rate than $\pi\jpsi$. There could
be other decay modes such as $\pi\hc$, $\pi\psp$ and so on.
This needs to be checked with larger data samples in the future.
The reported production cross section for $\EE\to \pi^+\zc^-+c.c.$
suffers from the same problems mentioned above.
The only reliable measurement is the product
cross section of $\EE\to \pi^+\zc^-+c.c.\to \ppjpsi$ determined
from PWA which is $(21.8\pm 1.0\pm 4.4)$~pb at $\sqrt{s}=4.23$~GeV
and $(11.0\pm 1.2\pm 5.4)$~pb at
$\sqrt{s}=4.26$~GeV~\cite{Collaboration:2017njt}.
The former at 4.23~GeV is twice as big as the latter at 4.26~GeV.
It would be very important to measure the $\zc$ production cross sections
at other c.m.\ energies, to check if the line shape of $\EE\to
\pi\zc$ is the same as that of $\EE\to \ppjpsi$. This will be an
important piece of information in understanding the nature of the
$\zc$ and the production mechanism.

As for the $\zcp$, we conclude that it is an
isovector state with positive $G$-parity, similar to the $\zc$
state. The spin-parity quantum numbers of the $\zcp$ are not
measured yet, but $J^P=1^+$ are assumed in all the analyses.
The currently observed decay modes are $\pi\hc$ and $D^*\dstrbar$.
With current statistics at BESIII, the possible decay modes of
$\pi\jpsi$ and $\rho\etac$ have not been observed yet.
The neutral $\zcp$ has a negative $C$-parity.
The reported masses and widths in both analyses of
$\pphc$ and $D^*\dstrbar$ final states may have bias,
since the reported values depend on different
assumptions on the signal shape and the possible
interference between the $\zcp$ and other amplitudes is neglected.
In the analyses of $\EE \to \pi D^*\dstrbar$~\cite{Ablikim:2013emm,Ablikim:2015vvn}, the
$\zcp^\pm$ and $\zcp^0$ are parameterized with different line
shapes and the pole mass and width are reported in the latter
case. In addition, the fractions of non-$\zcp$ events in $\EE\to
\pi D^*\dstrbar$ are quite different in charged and neutral modes.
All these suggest that improved measurements of $\EE\to \pi
D^*\dstrbar$, both charged and neutral modes, using more $D$-tag
modes and data at other c.m.\ energies are necessary.
The most reliable measurement is probably from $\EE\to \pphc$
mode~\cite{Ablikim:2013wzq} due to the very narrow width.
The mass and width are measured to be
$(4022.9\pm 0.8\pm 2.7)~{\rm MeV}$ and $(7.9\pm 2.7\pm 2.6)$~MeV,
respectively, in this mode. Of course, a combined
analysis of $\pphc$ and $\pi D^*\dstrbar$ modes will give
more reliable measurements of the resonant parameters.
The production cross section for $\EE\to \pi\zcp$ suffers from the
same problems mentioned in the $\zcp$ mass and width
determination. It would be very important to measure the $\zcp$
production cross sections as a function of the c.m.\ energy with
PWA, to check if the $\EE\to \pi\zcp$ process is from continuum
production or from decays of some resonant structures, such as the
$Y(4230)$ and
$Y(4390)$ observed in $\EE\to \pphc$~\cite{BESIII:2016adj}.
This will be an important piece of information in understanding
the nature of the $\zcp$ and the production mechanism.

\vspace{0.3cm}\noindent
$\bullet$ {\it The $Z_c(4050)$ state}
\vspace{0.3cm}

\begin{figure}[htbp]
\centering
\includegraphics[height=5cm]{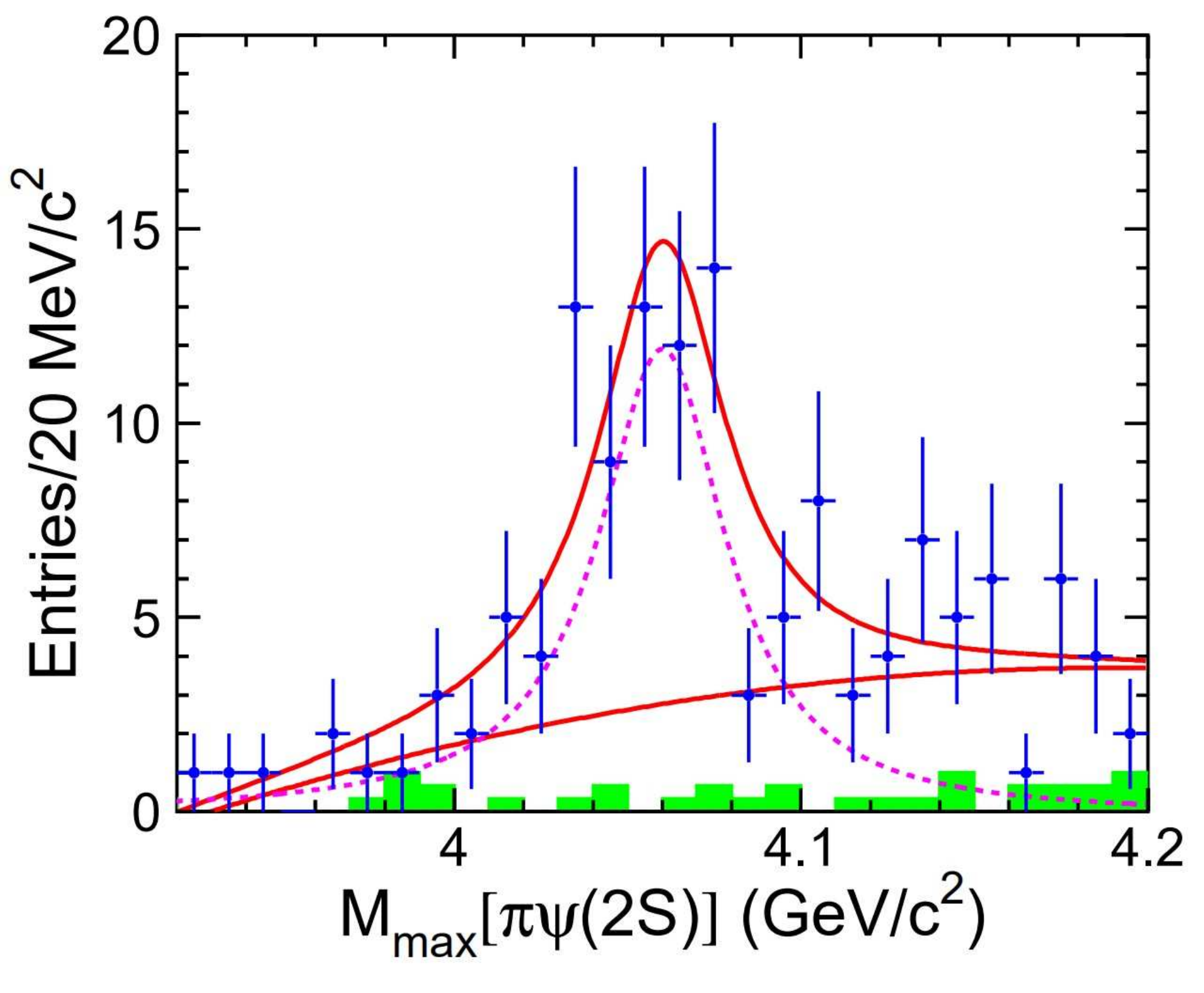}
\includegraphics[height=5cm]{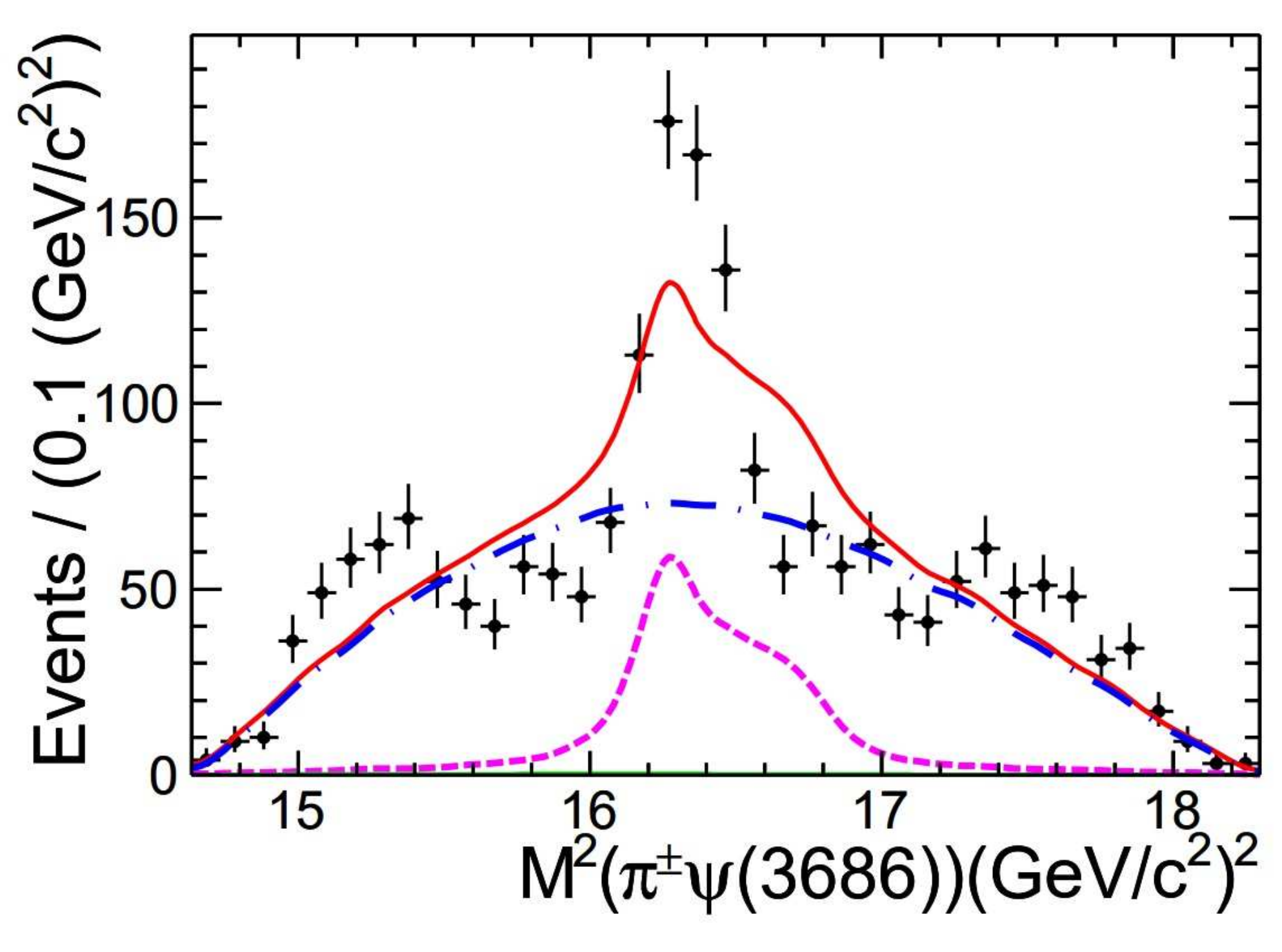}
\caption{The distributions of (a) $M_{\rm max}[\pi^{\pm}\psp]$ from
$\psi(4360)$ aka $Y(4360)$ decays from Belle~\cite{Wang:2014hta} and (b) $M^2[\pi^\pm \psp]$
at $\sqrt{s}$=4.416 GeV (two entries per event) from BESIII~\cite{Ablikim:2017oaf}.
The points with error bars represent the data; the shaded histogram is from
the $\psp$ mass sidebands, the solid curves are the best fits,
and the dashed curves show the shapes of the intermediate states. } \label{mppsp-fit}
\end{figure}

Belle updated the measurement of $\EE\to \pp\psp$ via ISR using the
980~fb$^{-1}$ full data sample~\cite{Wang:2014hta}.
Two distinct resonances, $Y(4360)$ and $Y(4660)$, are observed.
Possible charged charmonium-like structures in
$\pi^{\pm}\psp$ final states from the $Y(4360)$ or
$Y(4660)$
decays are searched for by checking the Dalitz plot of the selected
candidate events.
Figure~\ref{mppsp-fit} (left) shows the $M_{\rm max}[\pi^{\pm}\psp]$ distribution,
the maximum of $M[\pim\psip]$ and $M[\pip\psip]$, in $Y(4360)$
decays ($4.0<M[\pp\psp]<4.5$~GeV), where an excess evidence at around 4.05~GeV
can be seen. An unbinned maximum-likelihood fit is performed on
the distribution of $M_{\rm max}[\pi^{\pm}\psp]$.
The excess is parameterized with a BW function and the
non-resonant non-interfering background with a second-order
polynomial function. The fit yields a mass of $(4054\pm 3\pm 1)$~MeV
and a width of $(45\pm 11\pm 6)$~MeV, as shown in Fig.~\ref{mppsp-fit} (left).
The statistical significance of the signal is $3.5\sigma$ with systematic
uncertainties included.
Due to limited statistics in $Y(4660)$ decays
($4.5< M_{\pp\psp} < 4.9$~GeV), no significant structure in the
$\pi^{\pm} \psp$ system is observed.

As mentioned before, BESIII studied the process $\EE\to \pp\psip$ using 5.1~fb$^{-1}$
of data at c.m.\ energies from 4.0 to 4.6~GeV~\cite{Ablikim:2017oaf}.
Intermediate states are investigated in the data samples that have
large integrated luminosity. For data at $\sqrt{s}=$4.416 GeV, a prominent
narrow structure is observed around 4030 MeV in the $M[\pi^{\pm} \psp]$ spectrum,
as shown in Fig.~\ref{mppsp-fit} (right). For data at $\sqrt{s}=$4.358 GeV,
there is no obvious structure observed in the $M[\pi^{\pm} \psp]$ spectrum.
For data at $\sqrt{s}=$4.258 GeV, the possible structures with
masses of 3900 and 4030 MeV in the $M[\pi^{\pm} \psp]$ spectrum
have kinematic reflections at each other mass positions. For data
at $\sqrt{s}=$4.226 GeV,
no structure is clearly seen, which is very different from the behavior
at the nearby energy point of 4.258 GeV. From the above, we see the
structures in the $M[\pi^{\pm} \psp]$ spectrum
are correlated with $M(\pi^+ \pi^-)$ and $M[\pp \psp]$. A dedicated PWA
is needed to extract reliable resonance parameters.

To characterize the structure observed in the
$M[\pi^{\pm} \psp]$ spectrum for data at $\sqrt{s}=4.416$ GeV,
an unbinned maximum-likelihood fit to the Dalitz plot is applied.
Assuming an intermediate $Z_c$ state in the $\pi\psp$
system with spin-parity $J^P=1^{+}$, the Dalitz plot is
parameterized by the coherent sum of the process $\EE \to \pi Z_c\to
\pp\psp$ and the direct process $\EE\to \pp\psp$. The fit yields a
mass of $M = (4032.1\pm 2.4)$~MeV and a width of $\Gamma$ =
$(26.1\pm 5.3)$~MeV for the intermediate state with a significance
of $9.2\sigma$. The fit projection on $M^{2}[\pi^{\pm}\psip]$ is shown in
Fig.~\ref{mppsp-fit} (right). Since the overall fit curve does
not match the peaking structure and the corresponding C.L.\ of the fit is
8\% only, the reported errors are statistical only for the parameters of
the $Z_c$ structure.

The authors of Ref.~\cite{bondar_charm2018} reported their preliminary
PWA results on $\EE\to \pp\psp$ at the Charm 2018 meeting by using
BESIII published results. A similar PWA method used in the analysis of
$\Upsilon(5S)\to \pi Z_b \to \pi^+ \pi^- \Upsilon(nS)$ ($n=1,~2,~3$)~\cite{Belle:2011aa}
is taken to refit $\EE\to \pp\psp$ events from BESIII at $\sqrt{s}=4.416$~GeV~\cite{Ablikim:2017oaf}.
In the fit, the
interference effect between the $\pp$ amplitude and the $Z_c$
amplitude is taken into account properly. The fit quality is much improved.
It is found that the structure can be
described well with a charged state with a mass of $(4019.0\pm
1.9)$~MeV and a width of $(29\pm 4)$~MeV, or the $\zcp$ state
observed in $\pphc$ final state~\cite{Ablikim:2013wzq}.
The ratio $\sigma[\EE\to \pi^+\zcp^-+c.c.\to \pp\psp]/\sigma[\EE\to\pp\psp]$
from the fit is $(12.0\pm 3.7)\%$, corresponding to
$\sigma[\EE\to \pi^+\zcp^-+c.c.\to \pp\psp]=(5.1\pm
1.6)$~pb. If such PWA results are confirmed in the future,
it means a new decay mode of $\zcp \to \pi \psp$ is found.
Thus, both $\pp\psp$ and $\pphc$ final states need to be further investigated
to understand the intermediate structures.

\vspace{0.3cm}\noindent
$\bullet$ {\it The $Z_c(4100)$ state}
\vspace{0.3cm}

Motivated by a series of discovered $Z_c$ states, and a prediction
from the diquark model
~\cite{Maiani:2004vq} of a possible exotic state decaying to the $\pi\eta_c$
system, the LHCb Collaboration recently performed a Dalitz plot analysis of
$B^0 \to K^+ \pi^-\eta_c $ using data samples of
4.7 fb$^{-1}$ $pp$ collisions at c.m.\ energies of $\sqrt{s}=7,~8,$
and $13$~TeV~\cite{Aaij:2018bla}.

\begin{figure}[htbp]
\centering
\includegraphics[height=5cm]{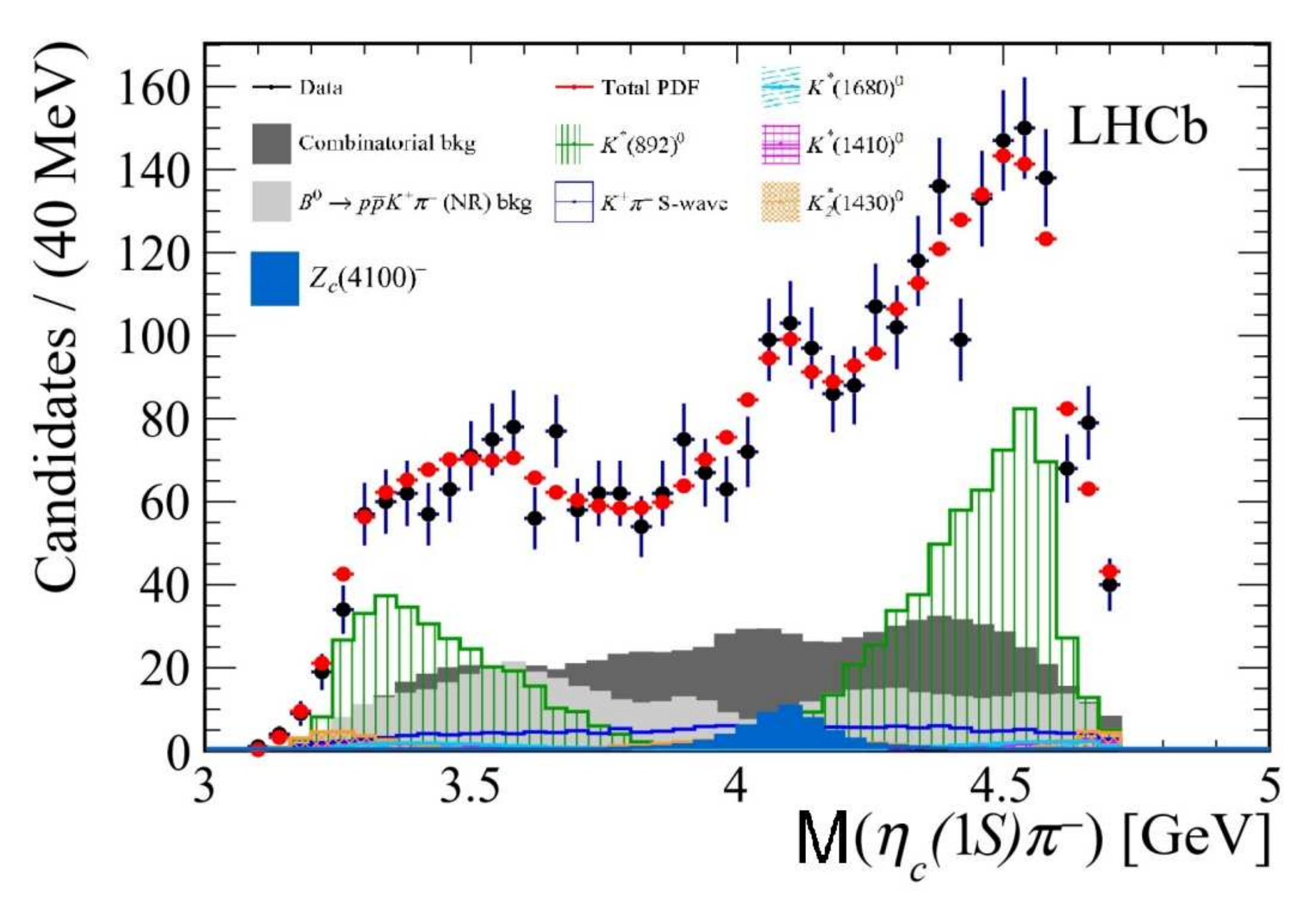}
\caption{The projections of the data and amplitude fit onto the $\pi^-\eta_c$
system from a Dalitz plot analysis of $B^0 \to K^+ \pi^-\eta_c$ performed by the LHCb Collaboration~\cite{Aaij:2018bla},
where the contributions from each included component are shown and
a charged charmonium-like resonance is needed for a satisfactory description
of the data
with a significance of $3.2\sigma$ with systematic errors considered.} \label{fig:lhcbzc4100}
\end{figure}

After event selection, there are $2105\pm75$ $B^0 \to K^+ \pi^-\eta_c $
signal candidates, and
the measured branching fraction $\BR(B^0 \to \eta_c K^+ \pi^-)$
is $(5.73 \pm 0.24 \pm 0.13 \pm 0.66) \times 10^{-4}$,
where the first uncertainty is statistical, the second systematic, and the
third is due to the limited knowledge of the external branching fractions.
For these selected signal candidates, an isobar model is used to perform the
Dalitz plot analysis, where seven components [$K^{*}(892)^0$, $K^{*}(1410)^0$,
$K^{*}_0(1430)^0$, $K^{*}_2(1430)^0$, $K^{*}(1680)^0$, $K^{*}_0(1950)^0$, and non-resonant] states
are taken to describe the $K^+ \pi^-$ system and an
additional exotic $Z_c^- \to \eta_c \pi^-$ component is assumed.
Satisfactory description of the data is achieved when including a contribution
representing an exotic $\pi^-\eta_c$ resonant state with
$\chi^2/{ndf}=1.3$.
The parameters of such a charged charmonium-like resonance
are $M_{Z_c^-} = (4096 \pm 20~ ^{+18}_{-22})$ MeV and
$\Gamma_{Z_c^-}= (152 \pm 58~^{+60}_{-35})$ MeV
with a significance of $3.2\sigma$ with systematic errors considered.
The projections of the data and amplitude fit onto the $\pi^-\eta_c $
system are shown in Fig.~\ref{fig:lhcbzc4100}, where the contributions from
each included component are shown. The spin-parity
assignments $J^P=0^+$ and $J^{P}=1^-$ are both consistent with the
data.

Some phenomenological models have been developed to explain the nature
of the charged $Z_c(4100)$, including an effect
arising from an $S$-wave $D^*\bar{D}^*$
rescattering with $I^GJ^{PC}=1^-0^{++}$, a resonance produced by the
$P$-wave $D^*\bar{D}^*$
interaction of which the neutral partner has exotic quantum numbers
$I^GJ^{PC}=1^-1^{-+}$~\cite{Zhao:2018xrd},
a four-quark state of the hadrocharmonium type~\cite{Voloshin:2018vym},
and a compact tetraquark state~\cite{Wu:2018xdi}.
However, by constructing a diquark-antidiquark current in QCD sum rules,
the predicted mass
in Ref.~\cite{Wang:2018ntv}
is $(4.24\pm 0.1)$ GeV, which agrees with the $Y(4230)$
mass and disfavors
assigning the $Z_c(4100)$ as a compact tetraquark state.
Due to low signal significance of $Z_c(4100)$ in LHCb data, more data will be
required to confirm this state. The $Z_c(4100)^-$ and its neutral partner can be
searched for at Belle and Belle II in $B \to K \pi\eta_c $ to conclusively
determine the nature of the $Z_c(4100)$ resonance.

\vspace{0.3cm}\noindent
$\bullet$ {\it Search for $Z_c$ pair production}
\vspace{0.3cm}

Considerable efforts in theory have been devoted to interpret the charged charmonium-like states as compact tetraquarks,
molecules, or hadrocharmonia~\cite{Brambilla:2010cs,Brambilla:2014jmp,Yuan:2014rta}.
To distinguish among these explanations, more experimental or theoretical input is needed. A
new idea in this respect is presented in Refs.~\cite{Brodsky:2016uln,Brodsky:2015wza},
where for double $Z^{\pm}_{c}$ production in
$e^{+}e^{-} \to Z^{+}_{c}Z^{-}_{c}$ the dependence on $s$
(the $e^+ e^-$ c.m.\ energy squared) of the electromagnetic
form factor, $F_{Z_{c}^{+}Z_{c}^{-}}$ is claimed to be related to the number of active constituents
in the states. However, it remains unclear from which values of $s$ onwards this scaling is applicable.

Belle searched for doubly charged charmonium-like state production with
$102$ million $\Upsilon(1S)$ events,
$158$ million $\Upsilon(2S)$ events, $89.5$~fb$^{-1}$ data collected at
$\sqrt{s} = 10.52$~GeV, $711.0$~fb$^{-1}$ data collected at $\sqrt{s}
=10.58$~GeV, and $121.4$~fb$^{-1}$ data collected at $\sqrt{s} =10.867$~GeV.
No significant signals are observed in any of the studied modes, and the 90\% C.L.\ upper limits
on ${\cal B}[\Upsilon(1S,2S)\to Z^{+}_{c}Z^{(\prime) -}_{c}]{\cal B}(Z^{+}_{c}\to\pi^{+}+c\bar c)$ [$c\bar c=J/\psi$, $\chi_{c1}(1P)$, $\psi(2S)$] and
$\sigma[e^+e^- \to Z^{+}_{c}Z^{(\prime) -}_{c}]{\cal B}(Z^{+}_{c} \to \pi^+ +c\bar c)$ at $\sqrt{s}$ = 10.52, 10.58, and 10.867 GeV are
in the range of $(1.8-45.5)\times 10^{-6}$ and $(1.3-143.9)$ fb, respectively.
Here, $Z_{c}$ refers to the
$Z_{c}(3900)$ and $Z_{c}(4200)$ observed in the $\pi J/\psi$ final
state, the $Z_{c1}(4050)$ and $Z_{c2}(4250)$ in the
$\pi\chi_{c1}(1P)$ final state, and the $Z_{c}(4050)$ and
$Z_{c}(4430)$ in the $\pi\psi(2S)$ final state.

\subsubsection{Bottomonium-like charged $Z_b$ states}
\label{Sect:3.2.3}
\begin{figure*}[htbp]
\begin{center}
\includegraphics[height=4.5cm]{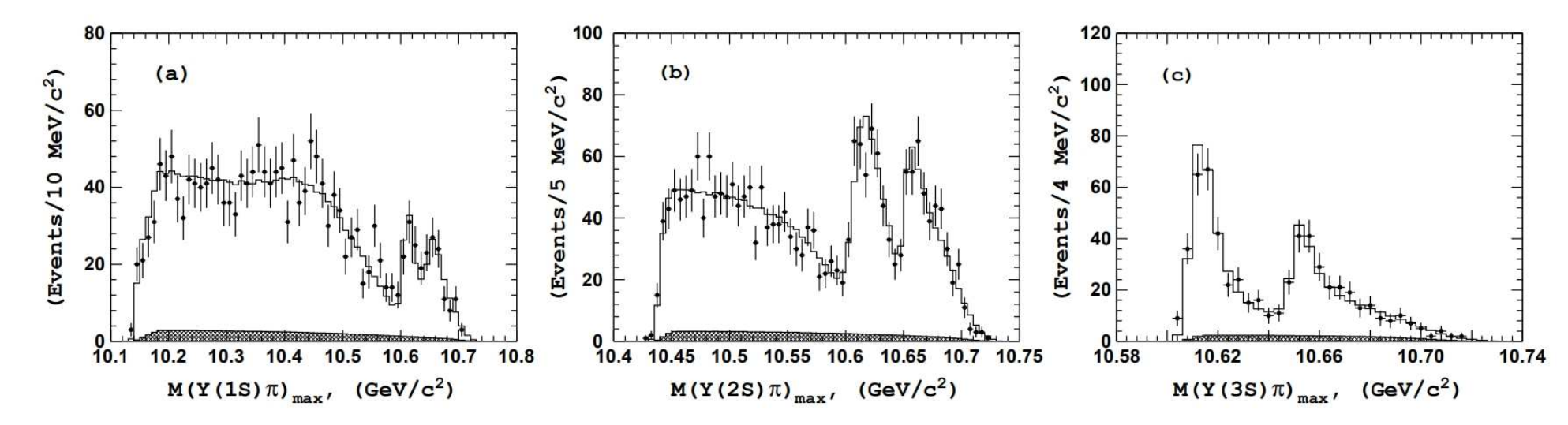}
\end{center}
\caption{Comparison of fit results (open histogram) with
 experimental data (points with error bars) for events in the
 (a) $\Upsilon(1S)$, (b) $\Upsilon(2S)$, and (c) $\Upsilon(3S)$
 signal regions~\cite{Belle:2011aa}. The hatched histogram shows the background component.}
\label{belle-y5s-zb}
\end{figure*}

\begin{figure}[htbp]
\begin{center}
\includegraphics[height=4.5cm]{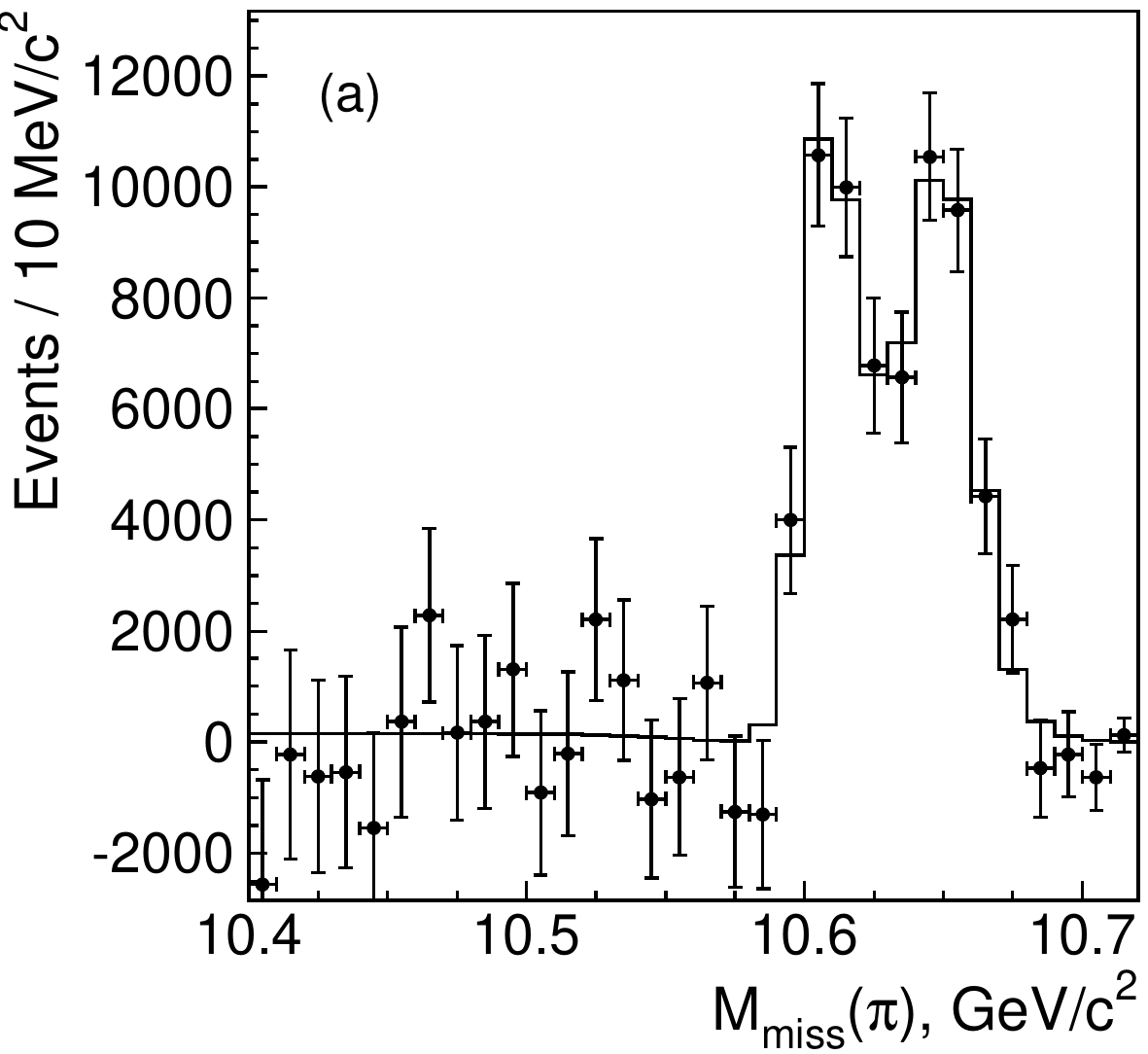}
\includegraphics[height=4.5cm]{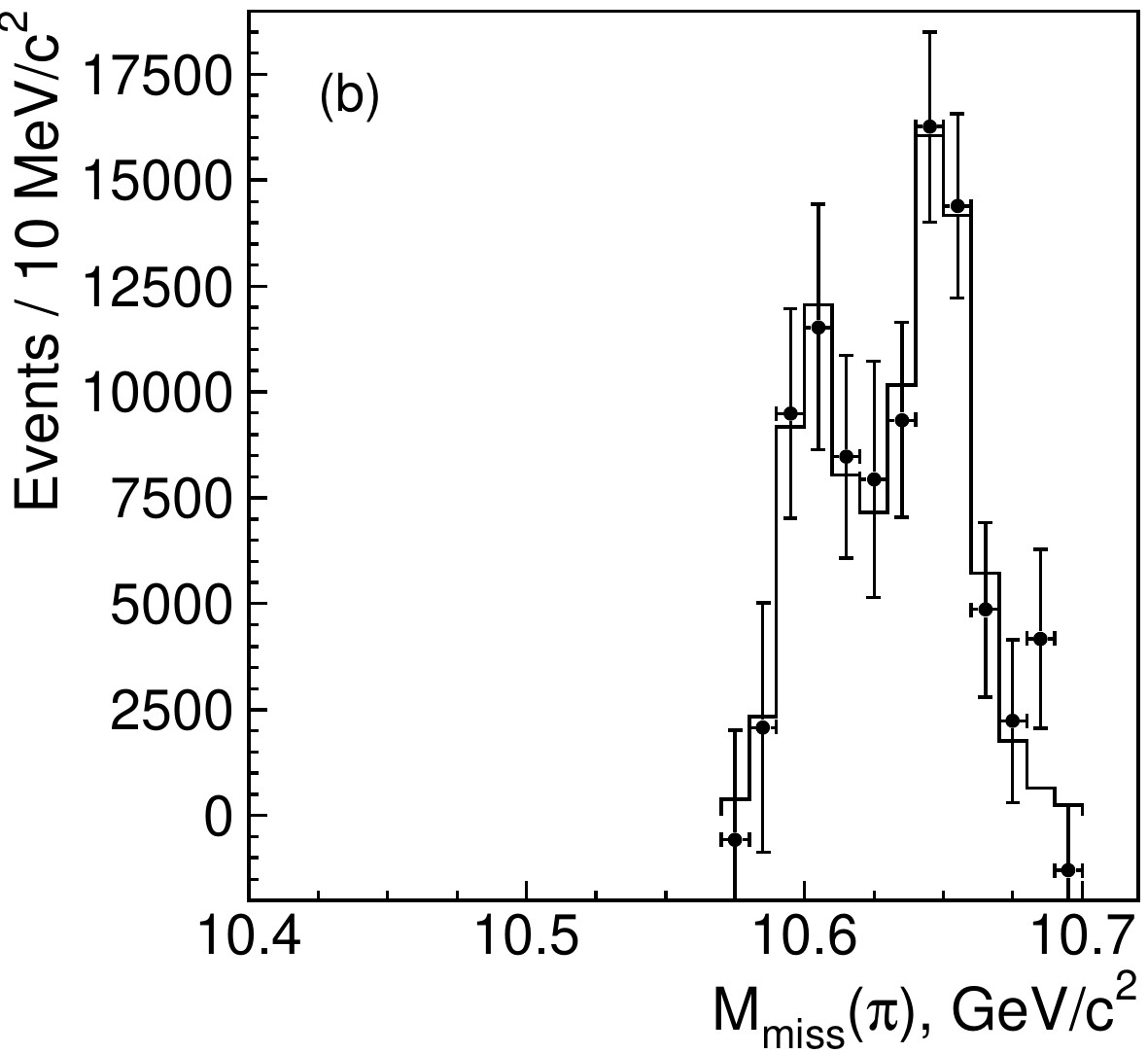}
\end{center}
\caption{The yields of (a) $\pi^+\pi^-h_b(1P)$ and (b) $\pi^+\pi^-h_b(2P)$ as a function of $\pi$ missing mass
(points with error bars) and results of the fit (histogram).}
\label{fig:mhbpi}
\end{figure}

Although the processes $\Upsilon(5S)\to \pp\Upsilon(nS)$ ($n=1$,
$2$, $3$) were observed~\cite{Abe:2007tk} at Belle in 2008 with
21.7~fb$^{-1}$ data collected near the $\Upsilon(5S)$ peak, the
Dalitz plots of the processes were only investigated in 2012
together with the processes $\Upsilon(5S)\to \pp h_b(mP)$ ($m=1$,
$2$) with 121.4~fb$^{-1}$ data collected in the vicinity of the
$\Upsilon(5S)$ resonance~\cite{Belle:2011aa}.
Amplitude analyses of the three-body $\Upsilon(5S)\to \pp\Upsilon(nS)$ decays with
$\Upsilon(nS) \to \mu^+ \mu^-$ are
performed by means of unbinned maximum likelihood fits to two-dimensional
$M^2[\pi^+\Upsilon(nS)]$ versus $M^2[\pi^-\Upsilon(nS)]$ Dalitz distributions~\cite{Belle:2011aa}.
One-dimensional invariant mass projections for
events in the $\Upsilon(nS)$ signal regions are shown in Fig.~\ref{belle-y5s-zb},
where two peaks are observed in the $\pi\Upsilon(nS)$ system near $10.61$ GeV
and $10.65$ GeV [named as $Z_b(10610)$ and $Z_b(10650)$].
The combined statistical significance of the two peaks exceeds $10\,\sigma$
for all $\pi^+\pi^-\Upsilon(nS)$ channels.
The yields of $\Upsilon(5S)\to \pi^+\pi^-h_b(1P)$ and $\pi^+\pi^-h_b(2P)$ as a function of $\pi$ missing mass
are shown in Fig.~\ref{fig:mhbpi}, where the $Z_b(10610)$ and $Z_b(10650)$ signals are clear
although the available phase-space is smaller for $\pi h_b(2P)$.
The histograms in Fig.~\ref{fig:mhbpi} are the fit results with interfering
$Z_b(10610)$ and $Z_b(10650)$ signals.
Analyses of charged pion angular distributions favor the $J^P=1^+$ spin-parity
assignment for both the $Z_b(10610)$ and $Z_b(10650)$.
Their $\piz$ transition modes were
measured to confirm their isospin~\cite{Krokovny:2013mgx}, and the
open bottom decay modes were studied to understand the couplings to
various final states~\cite{Garmash:2015rfd}.
Note that their decay modes fix the isospin of the $Z_b$ states to one.

\begin{figure}[htbp]
\includegraphics[height=3.0cm]{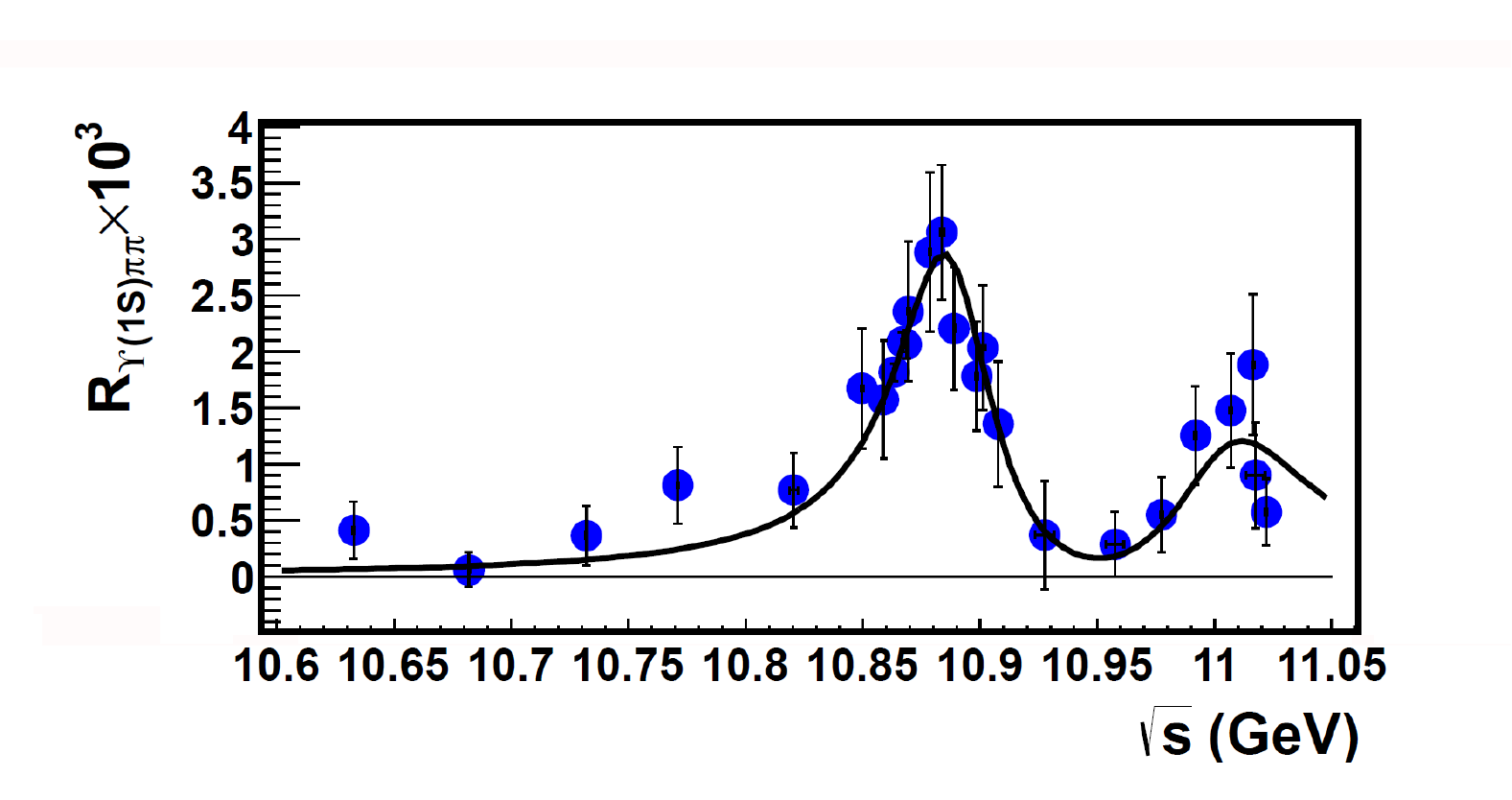}
\includegraphics[height=2.6cm]{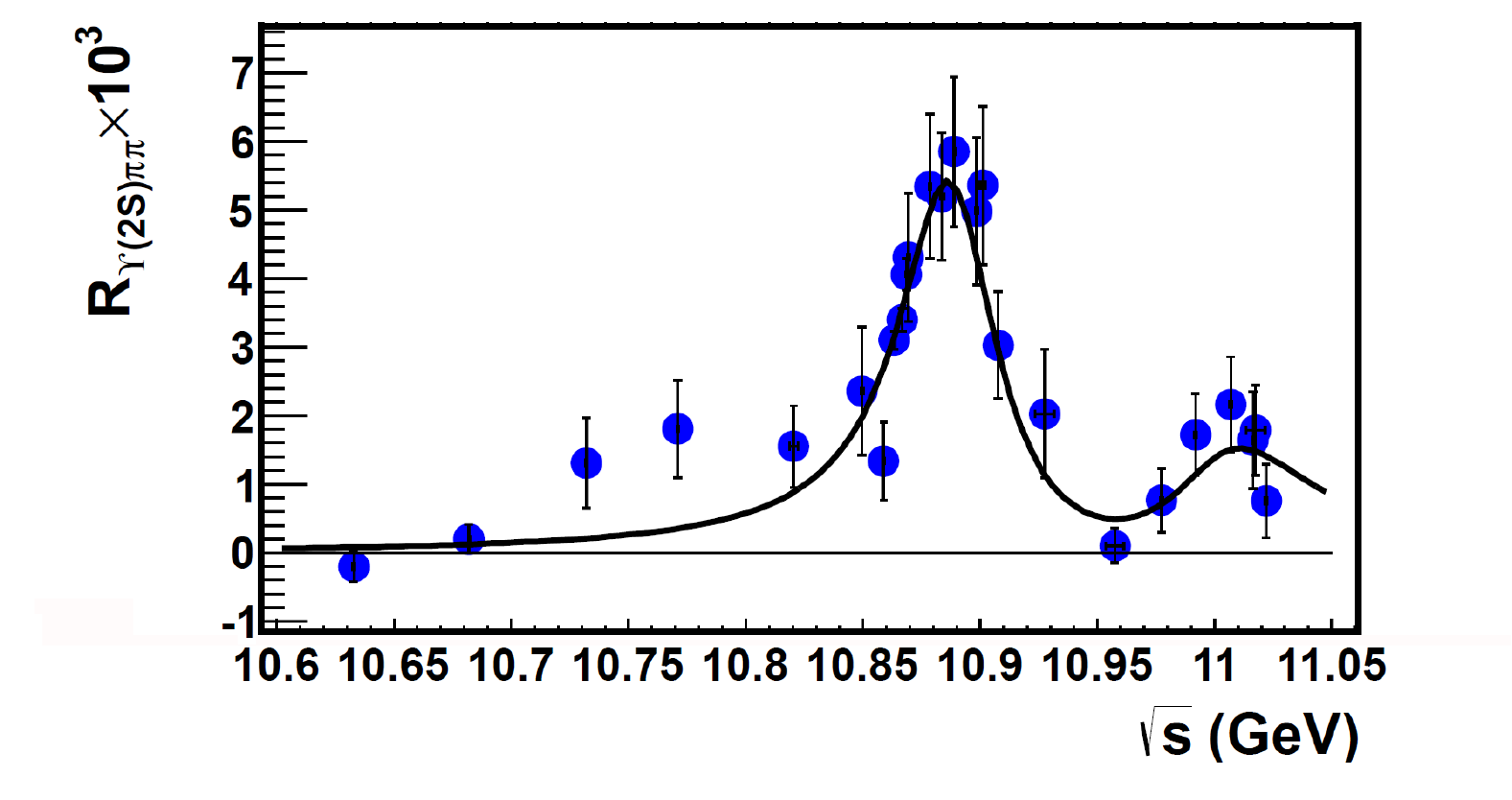}
\includegraphics[height=3.1cm]{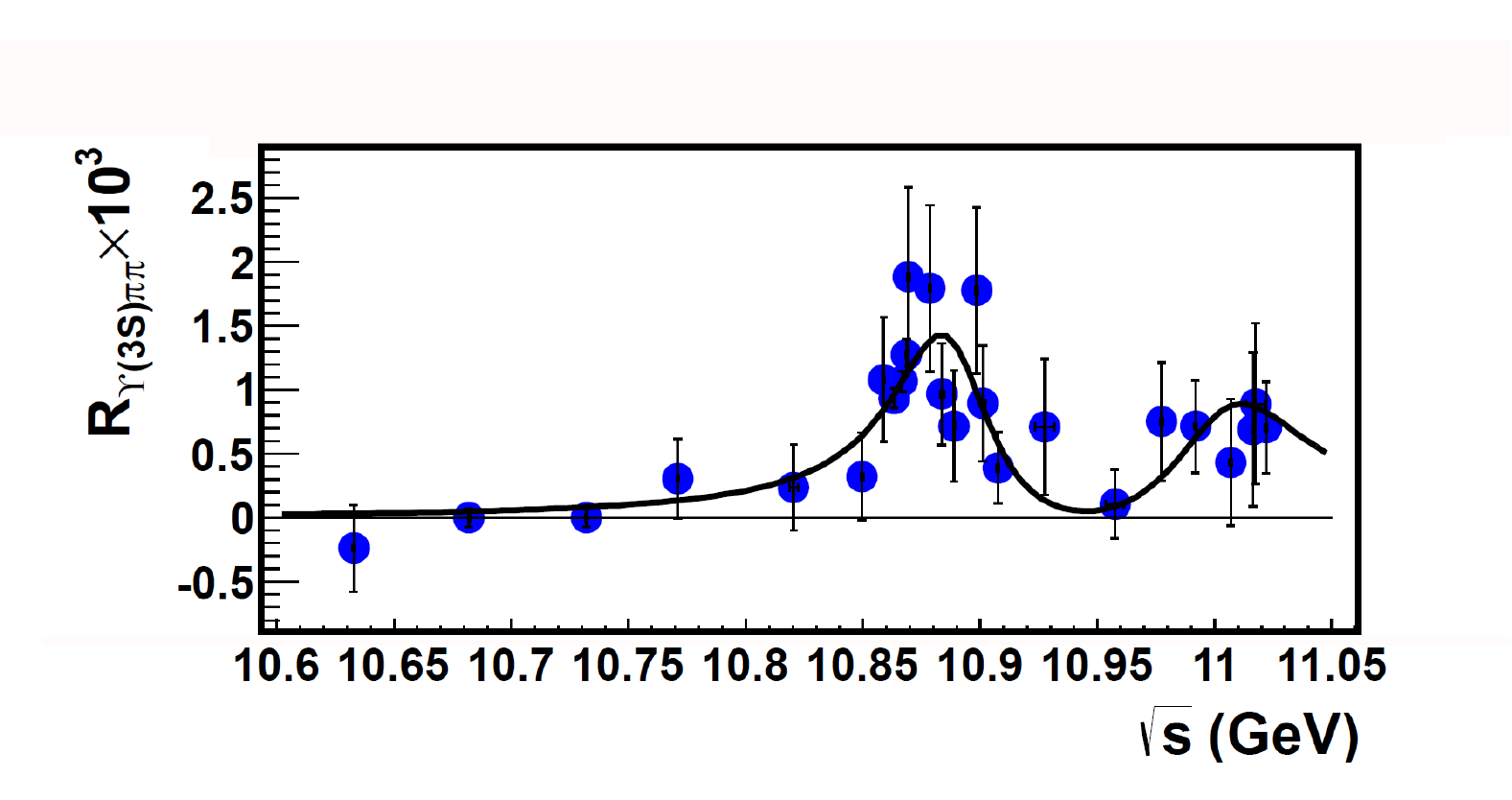}
\caption{Cross sections of $\EE\to \pp\Upsilon(1S)$ (left),
$\pp\Upsilon(2S)$ (middle), and $\pp\Upsilon(3S)$ (right)~\cite{Santel:2015qga}, and
the fits with a coherent sum of two BW functions. Error
bars are statistical only.} \label{pipins}
\end{figure}

The cross sections of $\EE\to \pp\Upsilon(nS)$ ($n$ = 1, 2, 3) at the
$\Upsilon(5S)$ peak and 22 energy points between 10.63 and
11.02~GeV with approximately 1~fb$^{-1}$ of the collected luminosity each are
measured by Belle~\cite{Santel:2015qga}.
Candidate $\pi^+\pi^-\Upsilon(nS)(\to \mu^+\mu^-)$ events are
selected for the measurement. Figure~\ref{pipins} shows $R_{\pp\Upsilon(nS)}\equiv \sigma[e^+e^-\to
\pp\Upsilon(nS)]/\sigma^0(\EE\to \MM)$. The cross sections are fit
for the masses and widths of the $\Upsilon(5S)$ and $\Upsilon(6S)$
resonances. It is found that $\EE\to \pp\Upsilon(nS)$ is dominated
by the two resonances. With $\EE\to \pp\Upsilon(nS)$, Belle
measured $M_{\Upsilon(10860)}=(10891.1\pm 3.2^{+0.6}_{-1.7})$~MeV and
$\Gamma_{\Upsilon(10860)} = (53.7^{+7.1}_{-5.6}\,^{+0.9}_{-5.4})$~MeV, and
reported the first measurements
$M_{\Upsilon(11020)}=(10987.5^{+6.4}_{-2.5}\,^{+9.0}_{-2.1})$~MeV,
$\Gamma_{\Upsilon(11020)} = (61^{+9}_{-19}\,^{+2}_{-20})$~MeV, and the
relative phase $\phi_{\Upsilon(11020)}-\phi_{\Upsilon(10860)} = (-1.0\pm
0.4\,^{+1.0}_{-0.1})$ rad.

The large statistics at the $\Upsilon(5S)$ peak make a study of
the intermediate states of $\EE\to \pp\Upsilon(nS)$
possible~\cite{Belle:2011aa}. After event selection, 1905, 2312,
and 635 candidate events are left for the $\pp\Upsilon(1S)$,
$\pp\Upsilon(2S)$, and $\pp\Upsilon(3S)$ final states,
respectively. Belle performed a full amplitude analysis of
three-body $\EE\to \pp\Upsilon(nS)$ transitions and determined the
relative fractions of various quasi-two-body components of the
three-body amplitudes as well as the spin and parity of the two
observed $Z_b$ states. The favored quantum numbers are $J^P=1^+$
for both $Z_b(10610)$ and $Z_b(10650)$ states, and the
alternative $J^P=1^-$ and $J^P=2^\pm$ combinations are rejected at
confidence levels exceeding six standard deviations.
Results of the amplitude analysis are summarized in
Table~\ref{tab:results}, where fractions of individual
quasi-two-body modes, masses and widths of the two $Z_b$ states,
the relative phase ($\phi_Z$) between the two $Z_b$ amplitudes, and
fraction $c_{Z_b(10610)}/c_{Z_b(10650)}$ of their amplitudes are
given.

\begin{table*}[htbp]
\centering
 \begin{tabular}{lccc} \hline
Parameter & ~~~~$\pp\Upsilon(1S)$~~~~ &
 ~~~~$\pp\Upsilon(2S)$~~~~ &
 ~~~~$\pp\Upsilon(3S)$~~~~
\\ \hline
 $f_{Z^\mp_b(10610)\pi^\pm}$, \% &
 $4.8\pm 1.2^{+1.5}_{-0.3}$ &
 $18.1\pm 3.1^{+4.2}_{-0.3}$ &
 $30.0\pm 6.3^{+5.4}_{-7.1}$
\\
 $Z_b(10610)$ mass, MeV &
 ~~$10608.5\pm 3.4^{+3.7}_{-1.4}$~~ &
 ~~$10608.1\pm 1.2^{+1.5}_{-0.2}$~~ &
 ~~$10607.4\pm 1.5^{+0.8}_{-0.2}$~~
\\
 $Z_b(10610)$ width, MeV &
 ~~$18.5\pm 5.3^{+6.1}_{-2.3}$~~ &
 ~~$20.8\pm 2.5^{+0.3}_{-2.1}$~~ &
 ~~$18.7\pm 3.4^{+2.5}_{-1.3}$~~
\\
 $f_{Z^\mp_b(10650)\pi^\pm}$, \% &
 $0.87\pm 0.32^{+0.16}_{-0.12}$ &
 $4.05\pm 1.2^{+0.95}_{-0.15}$ &
 $13.3\pm 3.6^{+2.6}_{-1.4}$
\\
 $Z_b(10650)$ mass, MeV &
 ~~$10656.7\pm 5.0^{+1.1}_{-3.1}$~~ &
 ~~$10650.7\pm 1.5^{+0.5}_{-0.2}$~~ &
 ~~$10651.2\pm 1.0^{+0.4}_{-0.3}$~~
\\
 $Z_b(10650)$ width, MeV &
 ~~$12.1^{+11.3+2.7}_{-4.8-0.6}$~~ &
 ~~$14.2\pm 3.7^{+0.9}_{-0.4}$~~ &
 ~~$ 9.3\pm 2.2^{+0.3}_{-0.5}$~~
\\
 $\phi_{Z}$, degrees &
 $67\pm 36^{+24}_{-52}$ &
 $-10\pm 13^{+34}_{-12}$ &
 $-5\pm 22^{+15}_{-33}$
\\
 $c_{Z_b(10650)}/c_{Z_b(10610)}$ &
 $0.40\pm 0.12^{+0.05}_{-0.11}$ &
 $0.53\pm 0.07^{+0.32}_{-0.11}$ &
 $0.69\pm 0.09^{+0.18}_{-0.07}$
\\
 $f_{\Upsilon(nS) f_2(1270)}$, \% &
 $14.6\pm 1.5^{+6.3}_{-0.7}$ &
 $4.09\pm 1.0^{+0.33}_{-1.0}$ &
 $-$
\\
 $f_{\Upsilon(nS)(\pp)_S}$, \% &
 $86.5\pm 3.2^{+3.3}_{-4.9}$ &
 $101.0\pm 4.2^{+6.5}_{-3.5}$ &
 $44.0\pm 6.2^{+1.8}_{-4.3}$
\\
 ~~~$f_{\Upsilon(nS) f_0(980)}$, \% &
 $6.9\pm 1.6^{+0.8}_{-2.8}$ &
 $-$ &
 $-$
\\
\hline
\end{tabular}
\caption{Summary of results of fits to $\Upsilon(5S)\to\pp\Upsilon(nS)$
events in the signal regions.} \label{tab:results}
\end{table*}

The processes $\Upsilon(5S)\to \piz\piz \Upsilon(nS)$ are observed
at Belle with the $121.4\,{\rm fb}^{-1}$ data sample
too~\cite{Krokovny:2013mgx}, and the measured cross sections,
 $\sigma[\EE\to \piz\piz \Upsilon(1S)]=(1.16\pm 0.06\pm 0.10)\, {\rm pb}$,
 $\sigma[\EE\to \piz\piz \Upsilon(2S)]=(1.87\pm 0.11\pm 0.23)\, {\rm pb}$, and
 $\sigma[\EE\to \piz\piz \Upsilon(3S)]=(0.98\pm 0.24\pm 0.19)\, {\rm pb}$,
are consistent with the expectations from isospin conservation
based on $\sigma[\EE\to \pp \Upsilon(nS)]$ measured at the
same c.m.\ energy.
The Dalitz analyses of the processes $\EE\to \piz\piz
\Upsilon(2S)$ and $\piz\piz \Upsilon(3S)$ indicate that the
neutral partner of the $Z_b(10610)$ is observed in its
$\piz\Upsilon(2S)$ and $\piz\Upsilon(3S)$ decay modes with a
significance of $6.5\sigma$ including systematic
uncertainties~\cite{Krokovny:2013mgx}. Its
measured mass, $(10609\pm 4\pm 4)$~MeV, is consistent with
the mass of the corresponding charged state, the $Z_b(10610)$. The
$Z_b(10650)$ signal is not significant in any of the $\piz\piz
\Upsilon(nS)$ channels with the limited statistics.

\begin{figure}[htbp]
\centering
\includegraphics[height=7cm]{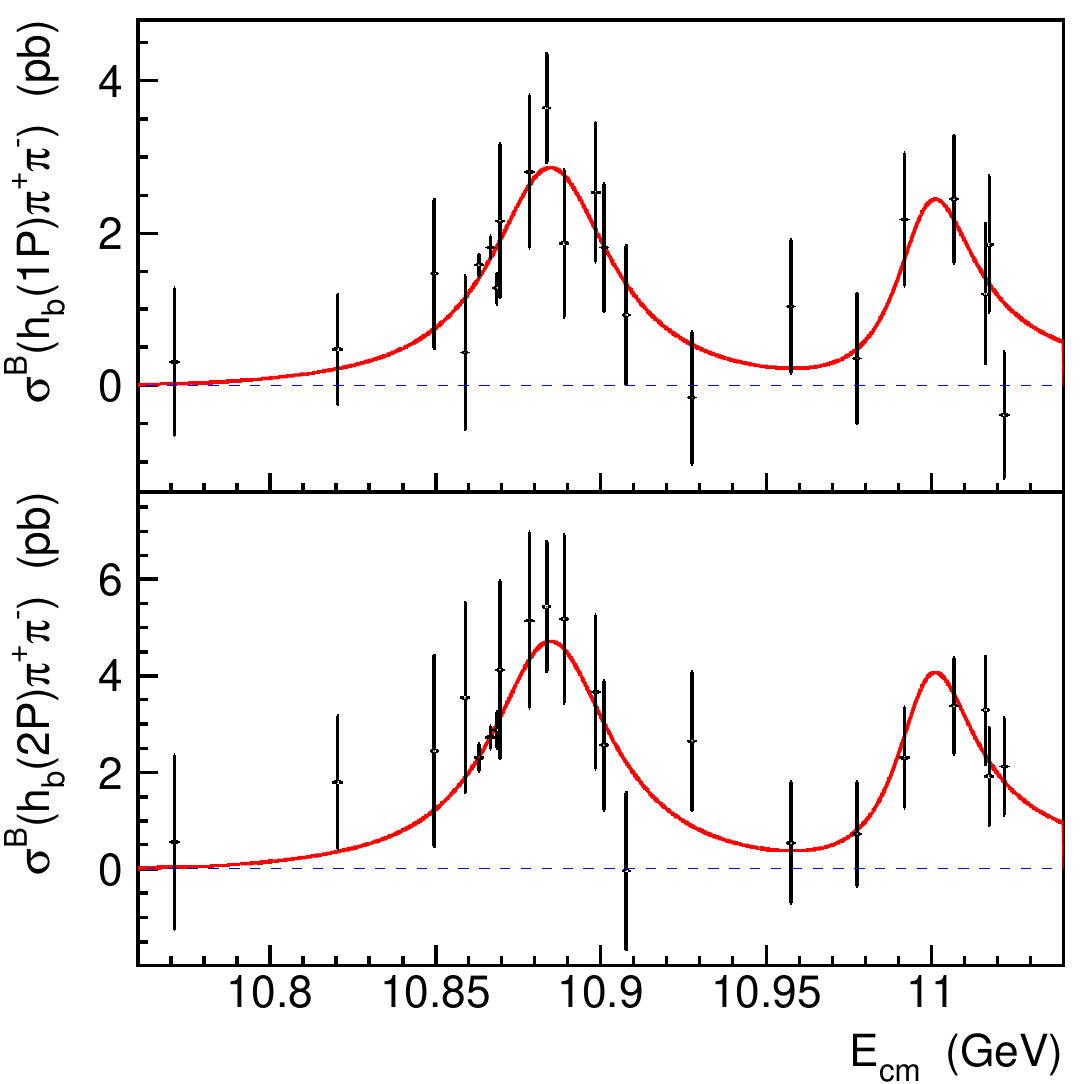}
\caption{Cross sections for the $\EE\to \pp h_b(1P)$ (top) and
$\EE\to \pp h_b(2P)$ (bottom) processes as a function of c.m.\
energy measured by Belle~\cite{Abdesselam:2015zza}. Points with error bars
are the data and red solid curves are
the fit results. } \label{hbpp_born_simfit}
\end{figure}

Belle measured $\EE\to \pp h_b(nP)$ ($n=$1, 2) with on-resonance
$\Upsilon(5S)$ data of 121.4~fb$^{-1}$, and energy scan data in the
range from about 10.77 to 11.02~GeV taken at 19 points of about
1~fb$^{-1}$ each~\cite{Abdesselam:2015zza}.
The processes $\EE\to \pp h_b(np)$ are reconstructed inclusively
using the missing mass of $\pp$ pairs, $M^{\rm
miss}_{\pp}=\sqrt{(\sqrt{s}-E_{\pp}^*)^2-p_{\pp}^{*2}}$, where
$E^*_{\pp}$ and $p^*_{\pp}$ are the energy and momentum of the
$\pp$ pair measured in the c.m.\ frame. The resulting cross sections are shown in
Fig.~\ref{hbpp_born_simfit}. Belle performed a simultaneous fit to
the energy dependence of the $\EE\to \pp h_b(nP)$ $(n=1,2)$ cross
sections. The fit function is a coherent sum of two BW amplitudes
and (optionally) a constant with an energy
continuum contribution:
\begin{equation*}
A_n\;f(s)\; \Bigl| BW(s,M_5,\Gamma_5) +a\,e^{i\,\phi}BW(s,M_6,\Gamma_6)
+b\,e^{i\,\delta} \Bigr|^2, \label{eq:fit_fun}
\end{equation*}
where $f(s)$ is the phase space function, which is calculated
numerically taking into account the measured $Z_b$ line shape,
$BW(s,M,\Gamma)$ is a BW amplitude
$BW(s,M,\Gamma)=M\Gamma/(s-M^2+iM\Gamma)$. The parameters $A_1$,
$A_2$, $M_5$, $\Gamma_5$, $M_6$, $\Gamma_6$, $a$, $\phi$ and
(optionally) $b$, $\delta$ are floated in the fit.
The significance of the non-resonant continuum
contribution from the fit is found to be $1.5\sigma$ only.
Thus the default fit function
does not include the continuum contribution. The fit results for
the default model are given in Table~\ref{fitpipihnp}.

\begin{table}[htbp]
\centering
\begin{tabular}{lc}
\hline
 Parameter & Results \\ \hline
$M_5$~(MeV) & $10884.7^{+3.2}_{-2.9}{^{+8.6}_{-0.6}}$ \\
$\Gamma_5$~(MeV) & $44.2^{+11.9}_{-7.8}{^{+2.2}_{-15.8}}$ \\
$M_6$~(MeV) & $10998.6\pm6.1{^{+16.1}_{-1.1}}$ \\
$\Gamma_6$~(MeV) & $29^{+20}_{-12}{^{+2}_{-7}}$ \\
$A_1 / 10^3$ & $4.8^{+2.7}_{-0.8}$ \\
$A_2 / 10^3$ & $8.0^{+4.6}_{-1.3}$ \\
$a$ & $0.64^{+0.37}_{-0.11}{^{+0.13}_{-0}}$ \\
$(\phi/\pi)$ & $0.1^{+0.3}_{-0.5}$ \\ \hline
\end{tabular}
\caption{Fit results to the measured $\EE\to \pp h_b(nP)$ $(n=1,2)$ cross
sections~\cite{Abdesselam:2015zza}.} \label{fitpipihnp}
\end{table}

\begin{figure*}[htb]
\centering
\includegraphics[height=5cm]{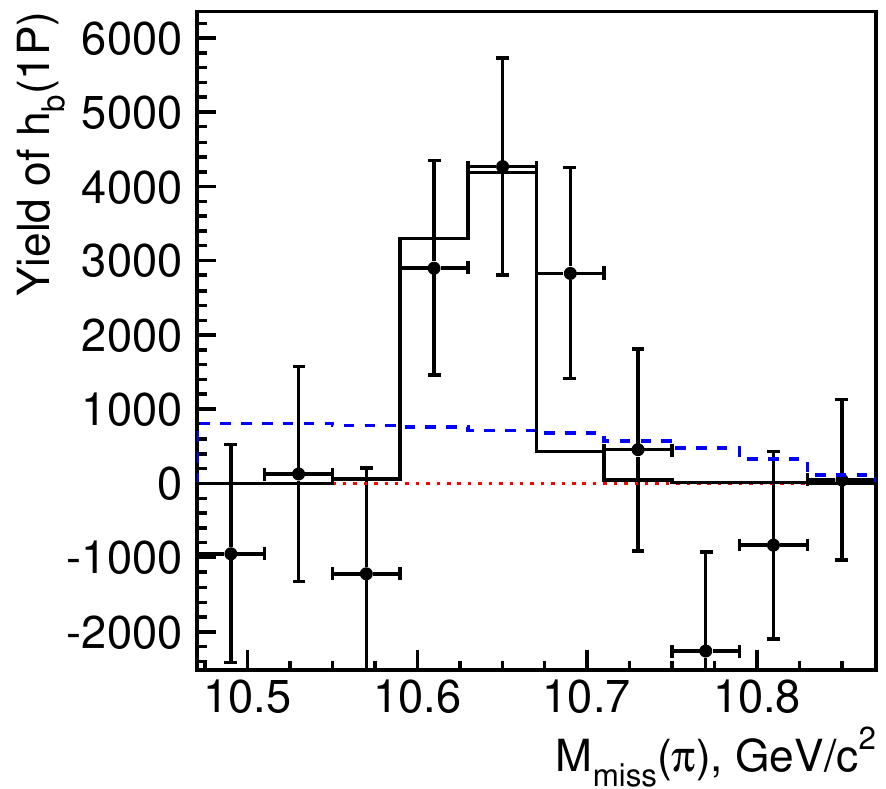}
\includegraphics[height=5cm]{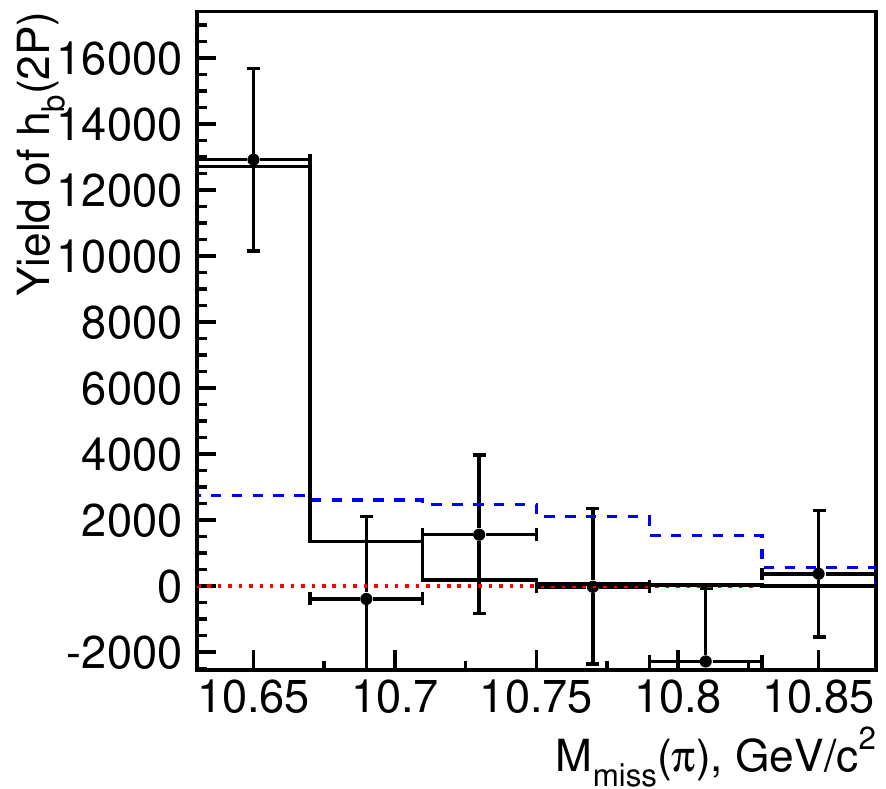}
\caption{The efficiency-corrected yields of $\pp h_b(1P)$ (left)
and $\pp h_b(2P)$ (right) as a function of $\pi$ missing mass~\cite{Abdesselam:2015zza}.
Points represent data, the black solid histogram represents the
fit result with the shape fixed from the $\Upsilon(5S)$ analysis,
the blue dashed histogram is the result of the fit to the
phase-space distribution. } \label{hb_vs_mmp}
\end{figure*}

The efficiency-corrected $\pi h_b(1P)$ and $\pi h_b(2P)$ invariant
mass distributions
are shown in Fig.~\ref{hb_vs_mmp}~\cite{Abdesselam:2015zza}.
The data do not follow a
phase-space distribution, but populate the mass region of the
$Z_b(10610)$ and $Z_b(10650)$ states. Belle found that the
transitions are dominated by the intermediate $Z_b(10610)$ and
$Z_b(10650)$ states, but the limited statistics do not allow a
measurement of the contribution from each mode.

\begin{figure}[htbp]
\centering
 \includegraphics[width=0.45\textwidth]{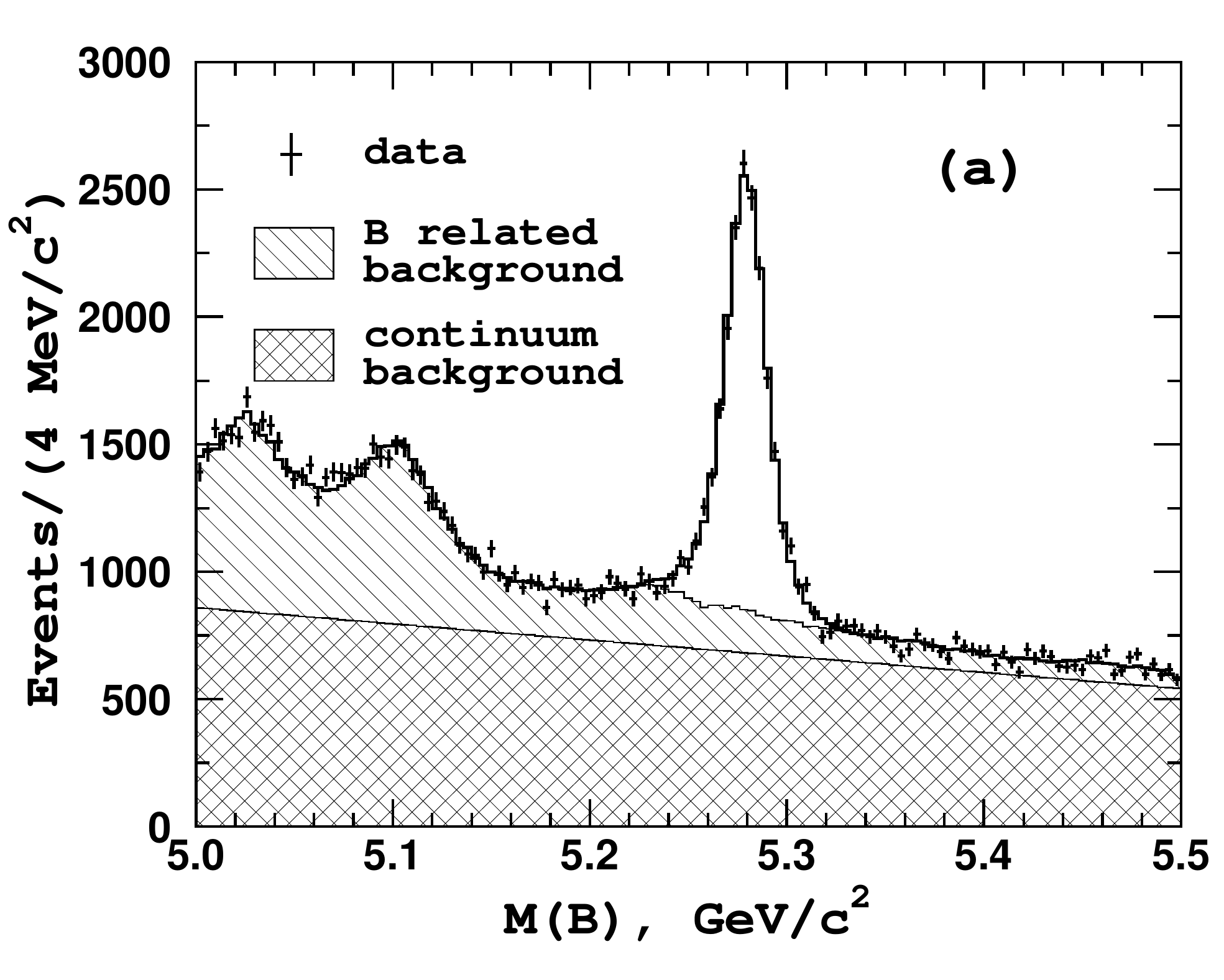}
 \includegraphics[width=0.45\textwidth]{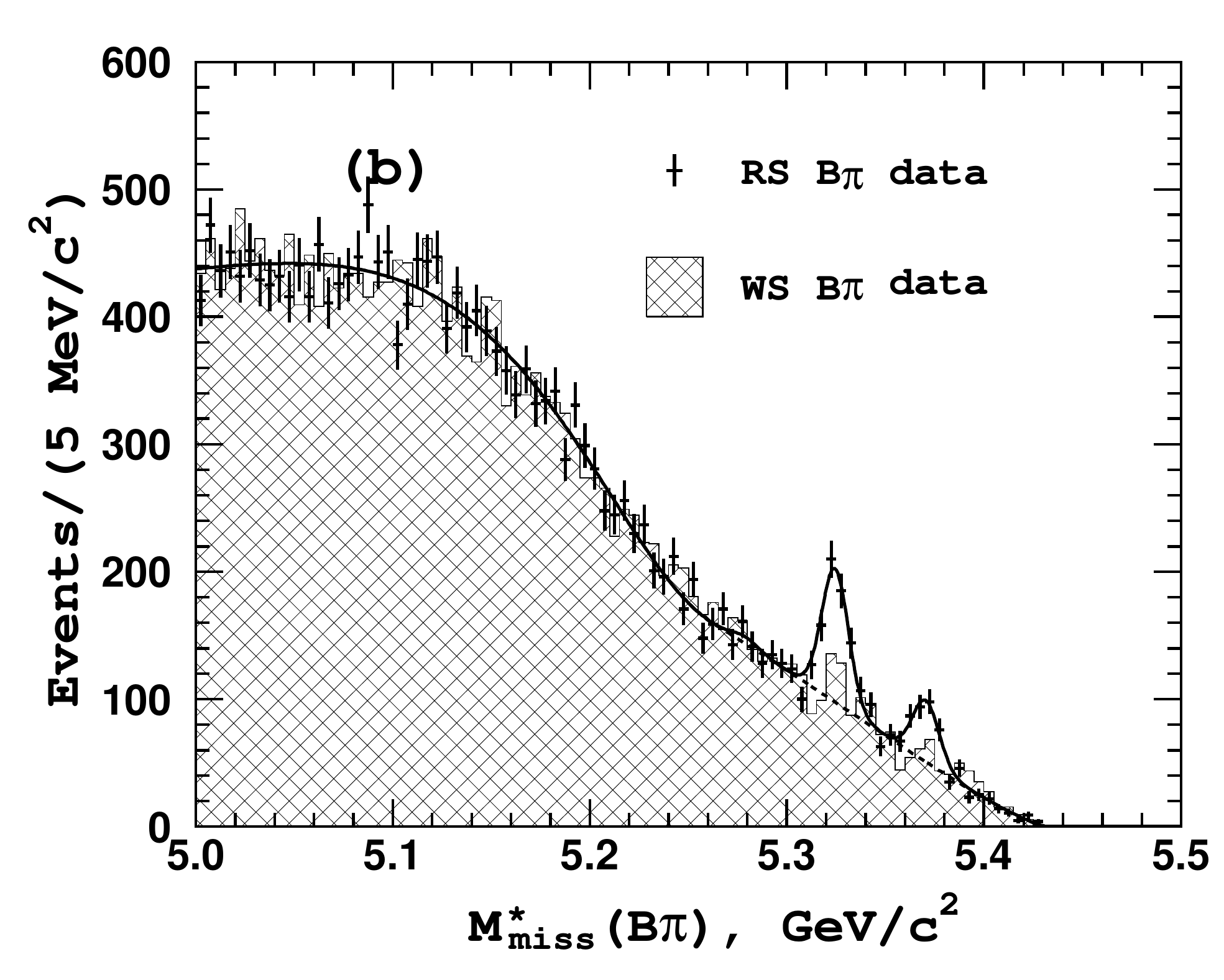}
\caption{The invariant mass (a) and $M^*_{\rm miss}(B\pi)$ (b)
distributions for $B$ candidates in the $B$ signal region
from the analysis of $\Upsilon(5S)\to B^{(*)}\bar{B}^{(*)}\pi$ decay
by Belle~\cite{Garmash:2015rfd}. Points
with error bars are data. The open histogram in (a) is the fit to
data. The solid line in (b) is the fit to the $B\pi$ missing mass;
the dashed line is the background level.}
 \label{bbpi}
\end{figure}

Evidence for the three-body $\Upsilon(5S)\to B\bar{B}^*\pi$ decay
has been reported previously by Belle, based on a data sample of
$23.6$~fb$^{-1}$~\cite{Drutskoy:2010an}. The analysis is updated with
a data sample of $121.4$~fb$^{-1}$ on the
$\Upsilon(5S)$ resonance~\cite{Garmash:2015rfd}. For brevity,
$B^+\bar{B}^0\pi^-$ and $B^-B^0\pi^+$ final states are referred to
as $B\bar{B}\pi$; $B^+\bar{B}^{*0}\pi^-$, $B^-B^{*0}\pi^+$,
$B^0B^{*-}\pi^+$, and $\bar{B}^0B^{*+}\pi^-$ final states are
referred to as $B\bar{B}^*\pi$; and $B^{*+}\bar{B}^{*0}\pi^-$ and
$B^{*-}B^{*0}\pi^+$ final states are denoted as $B^*\bar{B}^*\pi$. The
inclusion of the charge-conjugate mode is implied here.
The $B$ mesons are reconstructed in 18 decay channels: $B^+\to J/\psi
K^{(*)+}$, $B^+\to \bar{D}^{(*)0}\pi^+$, $B^0\to J/\psi K^{(*)0}$, and
$B^0\to D^{(*)-}\pi^+$. The $B$ candidates are identified by their
reconstructed invariant mass $M(B)$ as shown in Fig.~\ref{bbpi}(a).
The requirement $P(B)<1.35$~GeV, where $P(B)$ is the momentum of $B$ candidates in the c.m.\ frame,
is applied to retain $B$ mesons produced in both two-body and multibody
processes.

Reconstructed $B^+$ or $\bar{B}^0$ candidates are combined with
$\pi^-$'s and the missing mass, $M_{\rm miss}(B\pi)$, is
calculated as $M_{\rm miss}(B\pi) =
\sqrt{(\sqrt{s}-E_{B\pi})^2-P^2_{B\pi}}$, where $E_{B\pi}$ and
$P_{B\pi}$ are the measured energy and momentum of the
reconstructed $B\pi$ combination. Signal $\EE\to B\bar{B}^*\pi$ events
produce a narrow peak in the $M_{\rm miss}(B\pi)$ spectrum around
the nominal $\bar{B}^*$ mass, while $\EE\to B^*\bar{B}^*\pi$ events produce a
peak at $m_{\bar{B}^*}+\Delta m_{B^*}$, where $\Delta
m_{B^*}=m_{B^*}-m_{B}$, due to the missed photon from the $B^*\to
\gamma B$ decay. To remove the correlation between $M_{\rm
miss}(B\pi)$ and $M(B)$ and to improve the resolution, $M_{\rm
miss}^*=M_{\rm miss}(B\pi)+M(B)-m_B$ instead of $M_{\rm
miss}(B\pi)$ is used. The $M_{\rm miss}^*$ distribution is shown
in Fig.~\ref{bbpi}(b), where peaks corresponding to the $B\bar{B}^*\pi$
and $B^*\bar{B}^*\pi$ signals are evident. The fit to the $M_{\rm
miss}^*$ distribution yields $N_{B\bar{B}\pi}=13\pm 25$,
$N_{B\bar{B}^*\pi}=357\pm 30$, and $N_{B^*\bar{B}^*\pi}=161\pm 21$ signal
events. The statistical significances of the observed $B\bar{B}^*\pi$
and $B^*\bar{B}^*\pi$ signals are $9.3\sigma$ and $8.1\sigma$,
respectively.

\begin{figure}[htbp]
\centering
 \includegraphics[height=7cm]{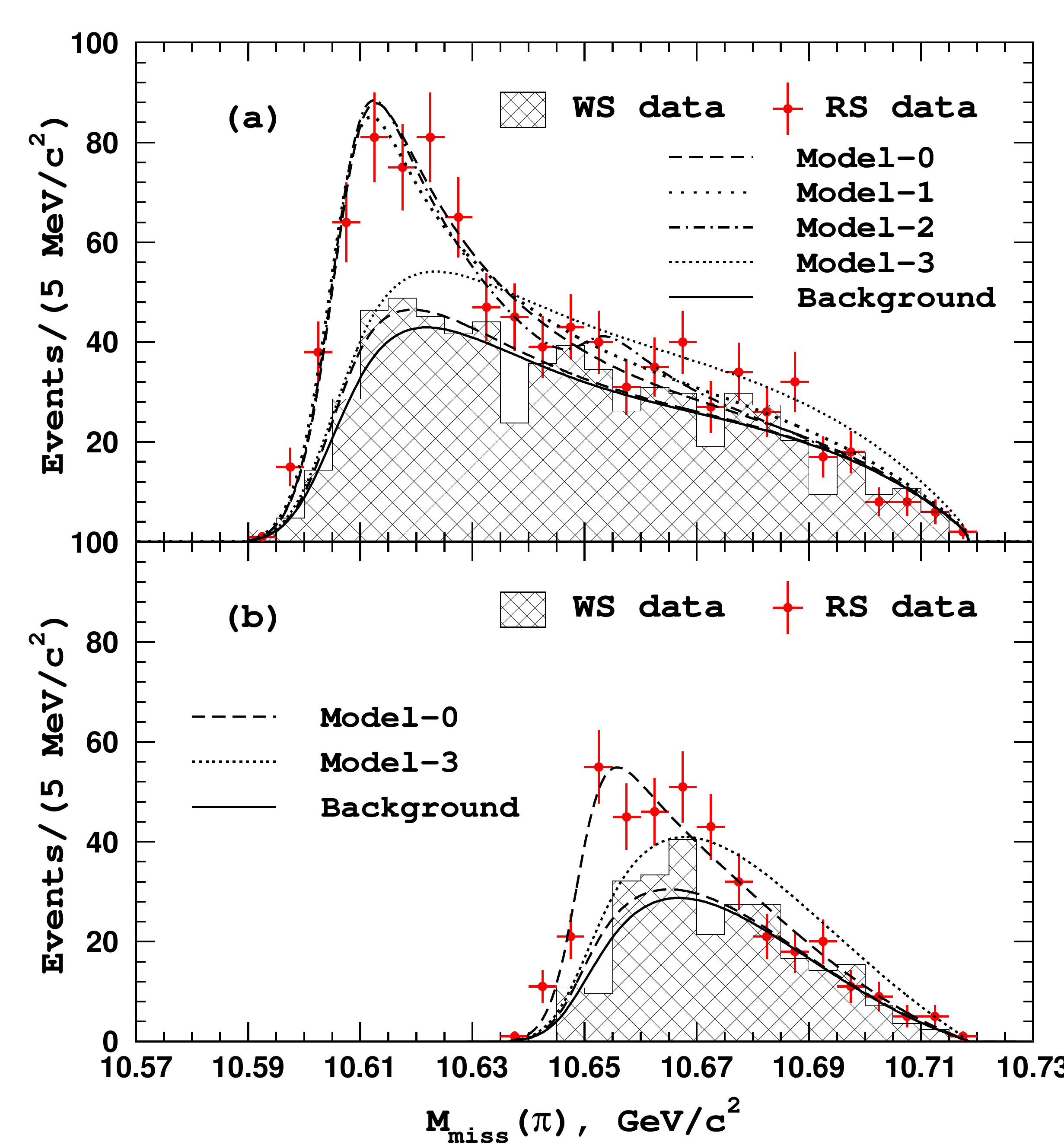}
\caption{The $M_{\rm miss}(\pi)$ distributions for the (a) $B\bar{B}^*\pi$
and (b) $B^*\bar{B}^*\pi$ candidate events from the analysis of
$\Upsilon(5S)\to B^{(*)}\bar{B}^{(*)}\pi$ decay
by Belle~\cite{Garmash:2015rfd}.}
 \label{bbpi_zb}
\end{figure}

The $\pi$ missing mass $M_{\rm miss}(\pi) =
\sqrt{(\sqrt{s}-E_\pi)^2-P^2_{\pi}}$ distributions are shown in
Fig.~\ref{bbpi_zb} for selected $B\bar{B}^*\pi$ ($|M_{\rm
miss}^*-m_{\bar{B}^*}|<15$~MeV/$c^2$) and $B^*\bar{B}^*\pi$ candidate events
[$|M_{\rm miss}^*-(m_{\bar{B}^*}+\Delta m_B)|<12$~MeV/$c^2$ with $\Delta
m_B=m_{B^*}-m_B$]. Here $E_{\pi}$ and $P_{\pi}$ are the
reconstructed energy and momentum, respectively, of the charged
pion in the c.m.\ frame. It is found that $\EE\to B\bar{B}^*\pi$ is
dominated by $Z_b(10610)$ and $\EE\to B^*\bar{B}^*\pi$ is dominated by
$Z_b(10650)$. The fits to the $\pi$ missing mass distributions with
different models parameterizing the line shapes are also shown in
Fig.~\ref{bbpi_zb}, where Model-0 is the fit with only
$Z_b(10610)$ for $B\bar{B}^*\pi$ events and $Z_b(10650)$ for $B^*\bar{B}^*\pi$ events,
Model-1 is the fit with an additional non-resonant component for
$B\bar{B}^*\pi$ events,
Model-2 is the fit with a combination of two $Z_b$ amplitudes for $B\bar{B}^*\pi$ events,
and Model-3 is the fit with only a pure non-resonant amplitude for both of
$B\bar{B}^*\pi$ and $B^*\bar{B}^*\pi$ events.

The relative fractions for $Z_b$ decays, assuming that they are
saturated by the already observed $\pi\Upsilon(nS)$, $\pi h_b(mP)$,
and $B^{(*)}\bar{B}^*$ channels, are listed in Table~\ref{br_zb}.
However, this assumption needs to be further checked.
For example, the decay mode of $Z_c(3900)\to \rho\eta_c$ is observed with a similar rate
as $Z_c(3900)\to \pi J/\psi$ in the BESIII experiment;
hence $Z_b\to \rho\eta_b$, as its counterpart in the bottom sector,
is expected to happen with a possible rate of a few percent.

\begin{table}[htbp]
 \centering
 \begin{tabular}{lcc} \hline \hline
 ~Channel~ & \multicolumn{2}{c}{Fraction, \%} \\
 & ~~~~~~~~$Z_b(10610)$~~~~~~~~ & ~~$Z_b(10650)$~~ \\
\hline
 $\pi^+\Upsilon(1{\rm S})$ & $0.54^{+0.16+0.11}_{-0.13-0.08}$ & $0.17^{+0.07+0.03}_{-0.06-0.02}$ \\
 $\pi^+\Upsilon(2{\rm S})$ & $3.62^{+0.76+0.79}_{-0.59-0.53}$ & $1.39^{+0.48+0.34}_{-0.38-0.23}$ \\
 $\pi^+\Upsilon(3{\rm S})$ & $2.15^{+0.55+0.60}_{-0.42-0.43}$ & $1.63^{+0.53+0.39}_{-0.42-0.28}$ \\
 $\pi^+ h_b(1{\rm P})$ & $3.45^{+0.87+0.86}_{-0.71-0.63}$ & $8.41^{+2.43+1.49}_{-2.12-1.06}$ \\
 $\pi^+ h_b(2{\rm P})$ & $4.67^{+1.24+1.18}_{-1.00-0.89}$ & $14.7^{+3.2+2.8}_{-2.8-2.3}$ \\
 $B^+\bar{B}^{*0}+\bar{B}^0B^{*+}$
 & $85.6^{+1.5+1.5}_{-2.0-2.1}$ & $-$ \\
 $B^{*+}\bar{B}^{*0}$ & $-$ & $73.7^{+3.4+2.7}_{-4.4-3.5}$ \\
\hline \hline
 \end{tabular}
\caption{Branching fractions for the $Z_b(10610)$ and $Z_b(10650)$
decays~\cite{Garmash:2015rfd}. The first uncertainties are statistical and
the second systematic.} \label{br_zb}
\end{table}

While the mass determinations reported above for both the $Z_b(10610)$ and $Z_b(10650)$
are above the nearby open-flavor thresholds,
more refined analyses seem to prefer lower masses --- typically below the threshold~\cite{Guo:2016bjq}.
When, however, the pion dynamics is included fully dynamically like in Ref.~\cite{Wang:2018jlv},
this leads to an additional energy dependence that pushes the $Z_b(10650)$ pole a little above the $B^*\bar B^*$ threshold.
These remarks illustrate that amplitudes consistent with unitarity and analyticity should be used to
extract parameters of very near-threshold states reliably (see discussion in Sec.~\ref{Sect:4.1.1}).

\subsection{Isospin-half states}
\label{sect:3.3}

\subsubsection{Searches for multi-quark  states with strangeness}
\label{sect:3.3.1}
After the discovery of the $Z_c$ states, it is natural to search also for their possible strange partners, $Z_{cs}$,
by looking at $K^{\pm}$ associated with a charmonium state.
The lowest $J^P=1^+$ $Z_{cs}$ state has been predicted to have a mass of $(3.97 \pm 0.08)$~GeV~\cite{Lee:2008uy}
and its decay widths to $K^+ J/\psi$, $K^{\ast+}\eta_c$, $D_s^+ \bar{D}^{\ast0}$, and $\bar{D}^{0} D_s^{\ast+}$
have been calculated using QCD sum rules~\cite{Dias:2013qga}.
The $Z_{cs}$ has also been predicted in the single-kaon emission model~\cite{Chen:2013wca,Chen:2011cj}.

Similar to the process $e^+e^- \to Y(4230)\to \pi^+\pi^-J/\psi$ in which the $Z_c(3900)$ is observed~\cite{Ablikim:2013mio, Liu:2013dau},
the $e^+e^- \to K^+ K^- J/\psi $ process is a suitable process to search for the $Z_{cs}$.
Belle has updated the cross sections of $e^+ e^- \to K^+ K^-J/\psi $ via ISR at c.m.\ energies between the threshold and 6.0~GeV
using a data sample of 980~fb$^{-1}$~\cite{Shen:2014gdm}.
Possible intermediate states for the selected $e^+e^- \to K^+ K^-J/\psi$ events have been investigated by examining the Dalitz plot,
but no clear structure has been observed in the $K^{\pm} \jpsi$ system.
A larger data sample seems therefore necessary to obtain more information about possible structures in the $K^+ K^- J/\psi $ and $K^{\pm} J/\psi$ systems.

\subsubsection{Pentaquark states}
\label{sect:3.3.2}
\begin{figure*}
\begin{center}
\includegraphics[height=6cm]{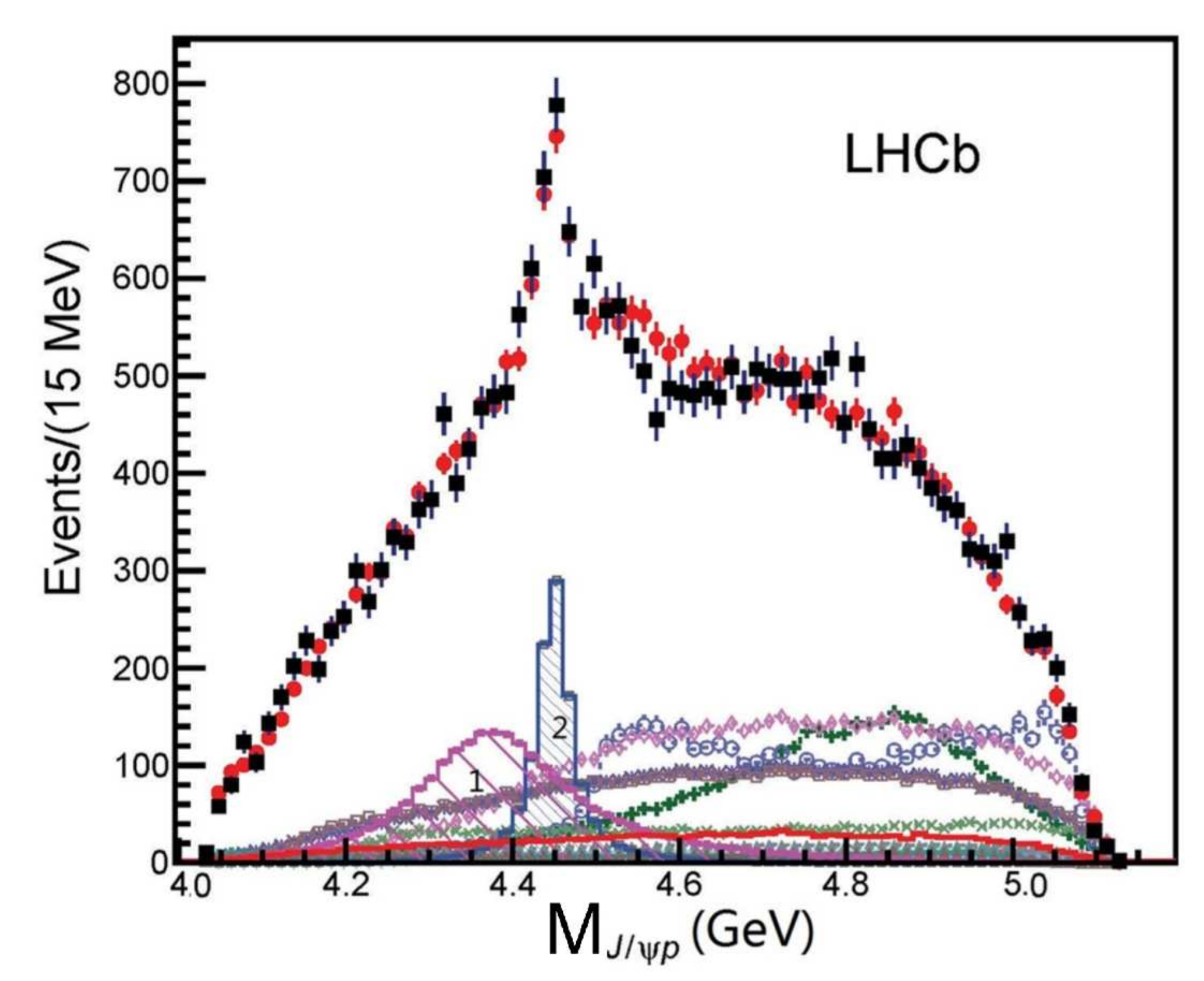}
\end{center}
\caption{Distribution of the $J/\psi p$ invariant mass and the fit
with two $P_c^+$ states~\cite{Aaij:2015tga}. The purple
hatched histogram is the $P_c(4380)^+$ and the blue hatched histogram is the $P_c(4450)^+$ signal.}
\label{lhcb-pc}
\end{figure*}

In 2015, LHCb reported the observation of two exotic structures, named as $P_c(4380)^+$ (9$\sigma$ significance)
and $P_c(4450)^+$ (12$\sigma$ significance), in the $J/\psi p$ system in $\Lambda_b^0 \to J/\psi K^-p$~\cite{Aaij:2015tga}.
The $P_c(4380)^+$ has a mass of $(4380\pm 8\pm 29)$~MeV and a width of $(205\pm 18\pm 86)$~MeV, while the
$P_c(4450)^+$ is much narrower with a mass of $(4449.8\pm 1.7\pm 2.5)$~MeV and a width of $(39\pm 5\pm 19)$~MeV.
The results of the fit to the $J/\psi p$ invariant mass distribution are shown in Fig.~\ref{lhcb-pc},
where the purple hatched and blue hatched histograms, labeled as ``1'' and ``2'',
are from the $P_c(4380)^+$ and $P_c(4450)^+$ contributions, respectively.
Since the valence structure of $J/\psi p$ is $c\bar{c}uud$, the newly discovered
particles consist of at least five quarks.

Considering the complicated $\Lambda^{*}$ spectroscopy in the amplitude
analysis of $\Lambda_b^0 \to J/\psi K^-p$, LHCb checked the
level of consistency of the data with minimal assumptions about
the spin and lineshape of possible $\Lambda^{*}$ contributions.
It is demonstrated that at more than 9 standard deviations
$\Lambda_b^0 \to J/\psi K^-p$ decays cannot be described with $K^- p$
contributions alone, and that $J/\psi p$ contributions play a dominant role.
These model-independent results support the existence of the
$P_c(4380)^+$ and $P_c(4450)^+$~\cite{Aaij:2016phn}.
Based on the measured fractions of the $P_c(4380)^+$ and $P_c(4450)^+$~\cite{Aaij:2015tga}
and the measured $\BR(\Lambda_b^0 \to J/\psi K^-p)$~\cite{Aaij:2015fea},
the branching fractions $\BR(\Lambda_b^0\to P_c^+ K^-)\BR(P_c^+\to J/\psi p)$ are
determined to be $(2.66\pm0.22\pm1.33^{+0.48}_{-0.38})\times10^{-5}$
for $P_c(4380)^+$ and $(1.30\pm0.16\pm0.35^{+0.23}_{-0.18})\times10^{-5}$
for $P_c(4450)^+$~\cite{Aaij:2015fea}, respectively, where the first uncertainty is statistical, the second is systematic,
and the third is due to the systematic uncertainty on $\BR(\Lambda_b^0 \to J/\psi K^-p)$.

Many theoretical interpretations of these two $P_c$ states have been
developed, including a pentaquark doublet, hadronic molecules
composed of an anticharm meson and a charm baryon, a $\chi_{c1} p$ resonance and so on.
The structure of these two states is still an open question, and none of the
explanations in the literature have been accepted
unanimously~\cite{Chen:2016qju}.

Very recently, the LHCb Collaboration updated the same analysis with
more data~\cite{Aaij:2019vzc} and discovered a new state
$P_c(4312)^+$ in the $J/\psi p$ invariant mass spectrum. At
the same time they found that the $P_c(4450)^+$ is composed of two narrow
overlapping structures, $P_c(4440)^+$ and $P_c(4457)^+$, with the following resonant parameters:
\begin{eqnarray}
\nonumber P_c(4312)^+:&& M = (4311.9 \pm 0.7 ^{+6.8}_{-0.6}) \mbox{ MeV} \, , \,\,
 \Gamma = (9.8 \pm 2.7 ^{+3.7}_{-4.5}) \mbox{ MeV} \, ,
\\ \nonumber P_c(4440)^+:&& M = (4440.3 \pm 1.3 ^{+4.1}_{-4.7}) \mbox{ MeV} \, , \,\,
 \Gamma = (20.6 \pm 4.9 ^{+8.7}_{-10.1}) \mbox{ MeV} \, ,
\\ \nonumber P_c(4457)^+:&& M = (4457.3 \pm 0.6 ^{+4.1}_{-1.7}) \mbox{ MeV} \, , \,\,
 \Gamma = (6.4 \pm 2.0 ^{+5.7}_{-1.9}) \mbox{ MeV} \, .
\end{eqnarray}
All these values are obtained by fitting the $J/\psi p$ invariant mass spectrum
directly without doing PWA, so the numbers are subject to change in a more
sophisticated analysis and the spin-parities are not determined.
It should be noticed that the $P_c(4312)$, $P_c(4440)$, and $P_c(4457)$
are just below the $\Sigma_c^+ \bar D^0$ and $\Sigma_c^+ \bar D^{*0}$ thresholds of 4318 and 4460~MeV, respectively,
so that it is natural to interpret them as bound
states composed of charmed baryons and anticharmed mesons.
On the other hand, the two-peak structure of the
$P_c(4450)$ was predicted in Ref.~\cite{Eides:2015dtr} after LHCb reported the discovery of the two first pentaquarks.
In the interpretation of Ref.~\cite{Eides:2015dtr}, however, the $P_c(4450)$ is a bound state of a charmonium $\psi(2S)$
and a nucleon consisting of two almost degenerate pentaquark states with $J^P=(1/2)^-$ and $J^P=(3/2)^-$.

There should exist other processes leading to the production of pentaquark
states that can be accessed by experiments.
Moreover, if pentaquark states exist, we should be able to observe more than just a few of them.
One would expect more states with either different spin-parity, or different flavor content.
To confirm the existence of the found pentaquark states and search for new ones, LHCb did the following analyses:
(1) A full amplitude analysis of $\Lambda_b^0 \to J/\psi p \pi^-$ was performed
using an integrated luminosity of $3$~fb$^{-1}$~\cite{Aaij:2016ymb}.
Besides the nucleon excitations, the $P_c(4380)^+\to J/\psi p$,
$P_c(4450)^+\to J/\psi p$, and $Z_c(4200)^-\to \jpsi\pi^-$ states are
also needed for a better description of the data with a significance of
more than 3$\sigma$.
(2) The pentaquark states containing a single $b$ (anti)quark decaying weakly
into four specific final states $J/\psi K^+ \pi^- p$, $J/\psi K^- \pi^- p$,
$J/\psi K^- \pi^+ p$, and $J/\psi \phi p$
were searched for, using an integrated luminosity of 3 fb$^{-1}$~\cite{Aaij:2017jgf}.
No evidence for these decays was found. Upper limits at 90\% C.L.\ on the
ratios of the production cross sections of these states times the branching
fractions into the searched modes, with respect to the production and
decay of the $\Lambda_b^0$ baryon in the mode $J/\psi K^- p$, were set at the $10^{-3}$ level.
(3) The Cabibbo-suppressed decay $\Lambda^0_b\rightarrow\psi(2S)p\pi^-$ was observed for the first time
and the $\psi(2S) p$ and $\psi(2S) \pi^-$ mass spectra were investigated,
but no evidence for contributions from exotic states was found~\cite{Aaij:2018jlf}.
With a larger data sample, a detailed amplitude analysis of this decay could be performed
to search for possible exotic states.
(4) Clear signals of $\Lambda_b^0 \to \psi(2S) p K^-$, $J/\psi \pi^+ \pi^- p K^-$~\cite{Aaij:2016wxd},
$\chi_{c1} p K^-$, and $\chi_{c2} p K^-$~\cite{Aaij:2017awb} were observed,
but no intermediate states were investigated. For these processes, LHCb may be
conducting further analyses of intermediate states.

One of the most promising ways to independently confirm the pentaquarks
is the $J/\psi$ photoproduction off a nucleon using the
$\gamma p \to J/\psi p$ reaction, as suggested in Ref.~\cite{Wang:2015jsa}.
Similar ideas were also presented in
Refs.~\cite{Kubarovsky:2016whd, Karliner:2015voa, Paryev:2018fyv, Meziani:2016lhg, Blin:2016dlf, Kubarovsky:2015aaa, Fernandez-Ramirez:2017gzc}.
$J/\psi$ photoproduction can benefit from the experimental fact that the pentaquark states are coupled to $J/\psi p$.
Note that $S$-channel pentaquark productions via high energy photon-nucleon scattering can be understood in the framework of the vector
meson dominance model for photon-$J/\psi$ coupling. The strong production and decay of the pentaquarks can be connected to their couplings to $J/\psi p$.

\begin{figure*}[htbp]
\begin{center}
\includegraphics[height=6cm]{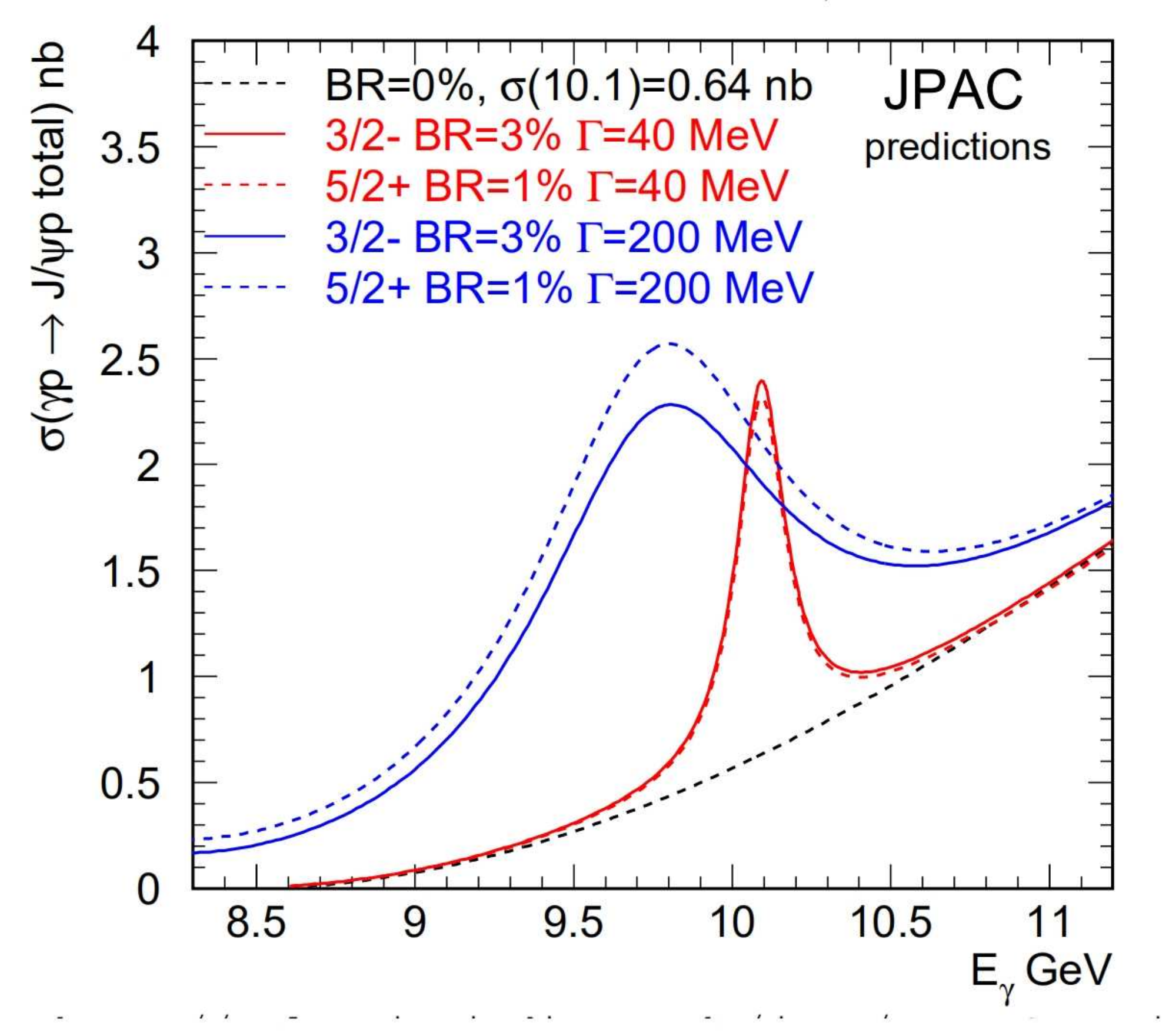}
\includegraphics[height=6.4cm]{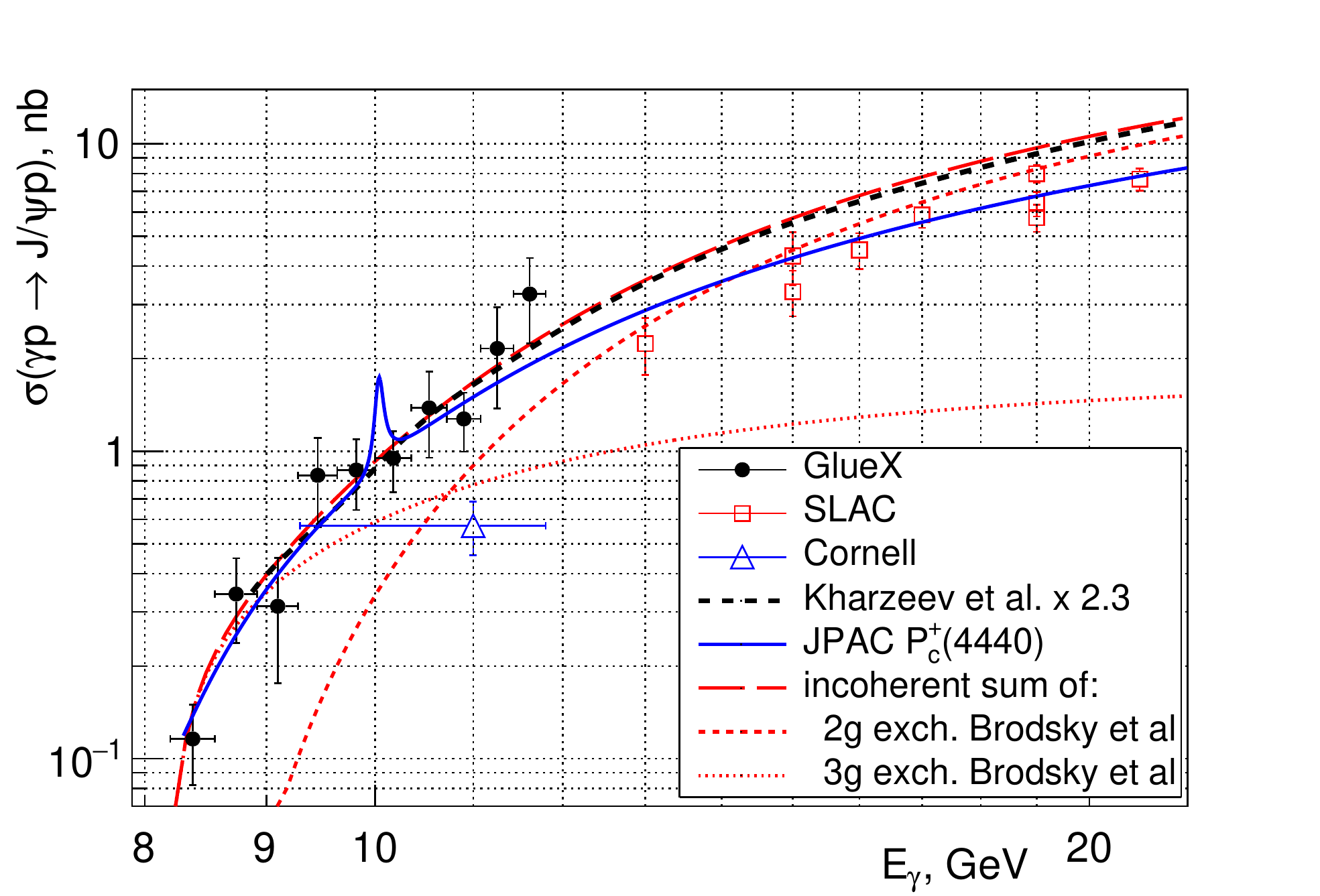}
\end{center}
\caption{The theoretical predictions on the cross sections of
$\gamma p \to P_c \to J/\psi p$ with different assumptions of $J^P$ values,
$\BR(P_c \to J/\psi p)$,
and width for $P_c$ states~\cite{Blin:2016dlf} (left plot), and the measured
cross sections of $\gamma p \to J/\psi p$ by GlueX~\cite{Ali:2019lzf} together with the
comparison to the Cornell \cite{Gittelman:1975ix} and SLAC~\cite{Camerini:1975cy} data,
the theoretical predictions~\cite{Brodsky:2000zc,Kharzeev:1999jt}, and the JPAC model~\cite{Blin:2016dlf}
corresponding to $\mathcal{B}(P_c(4440)^+ \to J/\psi p)=1.6\%$ for the $J^P=3/2^-$ case
as discussed in the text.}\label{gluex-pc}
\end{figure*}

If a pentaquark state $P_c$ does exist, the $P_c$ production would manifest itself as a peak in the cross sections of
$\gamma p \to J/\psi p$ [$\sigma(E_\gamma)$].
Figure~\ref{gluex-pc} (left) shows the expected line shapes and cross sections
of pentaquark states with different assumptions for the $J^P$ values,
$\BR(P_c \to J/\psi p)$,
and width from the Joint Physics Analysis Center (JPAC) predictions~\cite{Blin:2016dlf}.
Recently the cross sections of $\gamma p \to J/\psi p$
were measured by GlueX at $8.2<E_\gamma<11.8$ GeV using
about $25\%$ of the total data accumulated by the GlueX experiment
from 2016 to date, where clear $J/\psi$ signals are observed~\cite{Ali:2019lzf}.
The measured total cross section in bins of beam energy is shown in Fig.~\ref{gluex-pc} (right),
and compared to the earlier measurements at Cornell~\cite{Gittelman:1975ix} and SLAC~\cite{Camerini:1975cy}.
The measured cross sections do not favor either pure two- or three-hard-gluon exchange
predicted by the Brodsky {\it et al.} model~\cite{Brodsky:2000zc}, and a combination of the two processes is
required to fit the data adequately as shown in Fig.~\ref{gluex-pc} (right) assuming no interference between the two contributions.
The total cross section calculations of Kharzeev {\it et al.}~\cite{Kharzeev:1999jt}
multiplied by a factor 2.3 agree well with the measurements implying
large gluonic contribution to the nuclear mass and are shown in Fig.~\ref{gluex-pc} (right).
The $P_c(4312)^+$, $P_c(4440)^+$, and $P_c(4457)^+$ states
are expected to appear as structures at
$E_\gamma = 9.44$, $10.04$, and $10.12$~GeV in the cross-section results
shown in Fig.~\ref{gluex-pc} (right). Thus, no evidence for the $P_c$ state can be seen.
The expected $P_c(4440)^+$ contribution from the JPAC model-predicted yield
with assumption of $\BR(P_c(4440)^+ \to J/\psi p)=1.6\%$ is also shown in
Fig.~\ref{gluex-pc} (right).
GlueX is planning to analyze the full data sample including data taken in
2018, therefore a higher sensitivity is expected in the near future.

Many theoretical models precisely predicted the masses and decay
widths of the multiquark states, but the internal structure of these states
is still uncertain.
The magnetic moments of a particle are fundamental parameters, which are
directly related to the charge and current distributions in the hadrons
and these parameters are directly connected to the spatial
distributions of quarks and gluons inside the hadrons.
Therefore, the magnetic moments provide copious information
on the underlying quark structure and can be used to distinguish the
preferred configuration from various theoretical models and deepen our
understanding of the underlying dynamics. However,
works on the magnetic moments of the exotic particles $XYZ$ are rare. Recently,
under the assumption of $Z_c(3900)$ being a compact tetraquark state, the magnetic moment of the
$Z_c(3900)$ ($\mu_Z$) was calculated to be $|\mu_Z|=(0.35^{+0.24}_{-0.19})~\mu_N$
or $(0.47^{+0.27}_{-0.22})~\mu_N$
based on different sets of input parameters with the QCD sum rules~\cite{Wang:2017dce},
and $|\mu_Z|=(0.67\pm 0.32)~\mu_N$ with the light-cone QCD sum rules~\cite{Ozdem:2017jqh},
where $\mu_N$ is the nucleon magneton.
Also in Refs.~\cite{Wang:2016dzu,Ozdem:2018qeh} the electromagnetic multipole moments
of pentaquarks were calculated exploiting the flavor structure or with the light-cone QCD sum
rule mentioned above, respectively, by assuming, e.g., a diquark-diquark-antiquark and $\bar D^*\Sigma_c$
molecular structure with different quantum numbers $J^P = \frac{3}{2}^-$.
The obtained values of the electromagnetic multipole moments
under the two different assumptions show large differences from each other.
The magnetic moments of the $Z_c(3900)$ and $P_c(4380)^+$ can be measured in future experiments,
for example the electromagnetic multipole moments of $P_c(4380)^+$ can be
extracted through the process $\gamma^{(*)}p\rightarrow P_c(4380)^+ \rightarrow
P_c(4380)^+\, \gamma \rightarrow J/\psi\, p\, \gamma $ in the GlueX experiment.
Measurements and comparisons with predictions may provide
valuable knowledge on the internal structure of these multiquark states.

To understand the pentaquark states better, further experimental research
should be pursued with the currently available and the forthcoming
experimental data. There have been many suggestions on the
discovery channels for these and other exotic pentaquarks,
such as (1) in $B$ decays: $B^0 \to p \bar{p} K^0$,
$\bar{B}^0 \to D^0 p \bar{p}$; (2) in baryon decays:
$\Lambda_c^+ \to p K^0 \bar{K}^0$, $\Lambda_b^0 \to K^- \chi_{c1} p$;
(3) in quarkonium decays: $\Upsilon(nS)\to J/\psi p+X$, $\chi_{cJ} p+X$,
and $D^{(*)-} p + X$ ($n=1$, 2, 3); (4) in $e^+e^-$ continuum
processes: $e^+e^- \to J/\psi p + X$, $\chi_{cJ} p+X$,
$D^{(*)-} p + X$, and $D^{(*)} \Lambda + X$. It is
clear that a systematic search for baryon-meson resonances should
be pursued in various processes, where the baryon could be $p$,
$\Lambda$, $\Sigma$, $\Xi$, $\Omega$, $\Sigma_c$, ..., and the
meson $\pi$, $\eta$, $\omega$, $\phi$, $K$, $D$, $J/\psi$, $\chi_{cJ}$, ...~.

\section{Theoretical foundations}
\label{sect:4}
Quantum Chromodynamics is the sector of the Standard Model that describes the strong interaction.
It depends on only one coupling, $\alpha_s$.
Once renormalized, the coupling becomes small at high energies --- a phenomenon known as asymptotic freedom~\cite{Gross:1973id,Politzer:1973fx}.
Asymptotic freedom allows weak-coupling perturbative calculations of high-energy processes in terms of quarks and gluons,
the fundamental degrees of freedom of QCD.
Equivalently, instead of $\alpha_s$, one may use the intrinsic energy scale, $\Lambda_{\text{QCD}}$.
The precise value of $\Lambda_{\text{QCD}}$ depends on the adopted renormalization scheme:
in the $\overline{\text{MS}}$ scheme with three flavors, $\Lambda_{\overline{\rm MS}} = (332 \pm 17)$~MeV~\cite{Tanabashi:2018oca}.

At low energies, i.e., at energies of the order of $\Lambda_{\text{QCD}}$ or
smaller, the coupling becomes large and weak-coupling perturbative calculations
are no longer an available tool for computing observables.
Also, the right low-energy degrees of freedom of the strong interaction
are hadrons rather than quarks and gluons, as the former are detected in the experiments and not the latter.
This phenomenon goes under the name of color confinement, because hadrons are
color singlets, while quarks and gluons transform as triplet and octet
representations, respectively, under the color SU(3) group underlying QCD.
QCD is the only sector of the Standard Model that exhibits a strongly-coupled low-energy regime.

Since non relativistic bound states are characterized by a momentum transfer and a binding energy that is much
smaller than the mass, even the study of hadrons made of heavy quarks requires at some point the use of methods for QCD at low energies.
Because low-energy QCD cannot be addressed with weak-coupling perturbation theory, alternative approaches have been developed.
Most of the approaches can be grouped in the following categories:
phenomenological approaches, which will be discussed in Sec.~\ref{Sect:4.1},
effective field theories, which will be presented in Sec.~\ref{Sect:4.2},
and lattice QCD numerical computations, whose main results will be listed in Sec.~\ref{Sect:4.3}.
A full description of each of these approaches would deserve a review by
itself and would go far beyond the scope of the present one.
Hence we will present them with a focus mostly on their applications to the $XYZ$ states.

The connection of phenomenological approaches to QCD is at most plausible,
but they provide a framework for the classification of the states,
offer some valuable physics insights for observables where more rigorous alternatives are not available
and have the potential to pave the way towards more systematic approaches.
Sum rules supply rigorous results, but are limited in their applications,
accuracy and also in the insights they offer into the physical problem.
A detailed discussion of such phenomenological approaches lies beyond the scope of the present paper, especially given that there exist recent reviews dedicated to all of them --- in each subsection devoted to a particular approach we provide the corresponding reference together with a brief introduction to the subject.
On the contrary, in this review we put most emphasis on the discussion of effective field theories and lattice QCD,
which are rigorous and systematic approaches.
Their common feature is that the calculations have a sound connection to QCD
and can be improved systematically.
Of the highest relevance for the subject of this report are heavy quark
effective theory --- see Sec.~\ref{Sect:4.2.1} --- and potential non-relativistic QCD
--- see Secs.~\ref{Sect:4.2.3} and~\ref{Sect:4.2.3bis} --- both operating
fundamentally at the quark-gluon level, and chiral perturbation theory and its non-perturbative version
--- see Sec.~\ref{Sect:4.2.4} --- operating at the hadron level.

\subsection{Threshold effects without new states}
\label{sec:cusps}
Before examining the different approaches in detail, it is necessary,
however, to comment on alternative explanations for the structures found
in the various data sets, namely those that try to describe them without
nearby poles and accordingly without any states.
In these explanations the structures are usually generated by the
branch points that
emerge whenever the energy crosses a threshold related to a new channel. As
will be shown below those are of particular importance for thresholds of two
(near-)stable particles in a relative $S$-wave, for then they can induce pronounced
non-analyticities of the production amplitudes.
Clearly this kind of explanation is quite appealing, since not only does it
provide a reason why many of the additional
structures are located right at the thresholds, but also, if confirmed,
would remove many states from the particle listings.
There are also attempts to explain some structures as interference
phenomena. Both concepts will now be examined critically.

\begin{figure}[t!]
\begin{center}
 \includegraphics[width=0.5\linewidth]{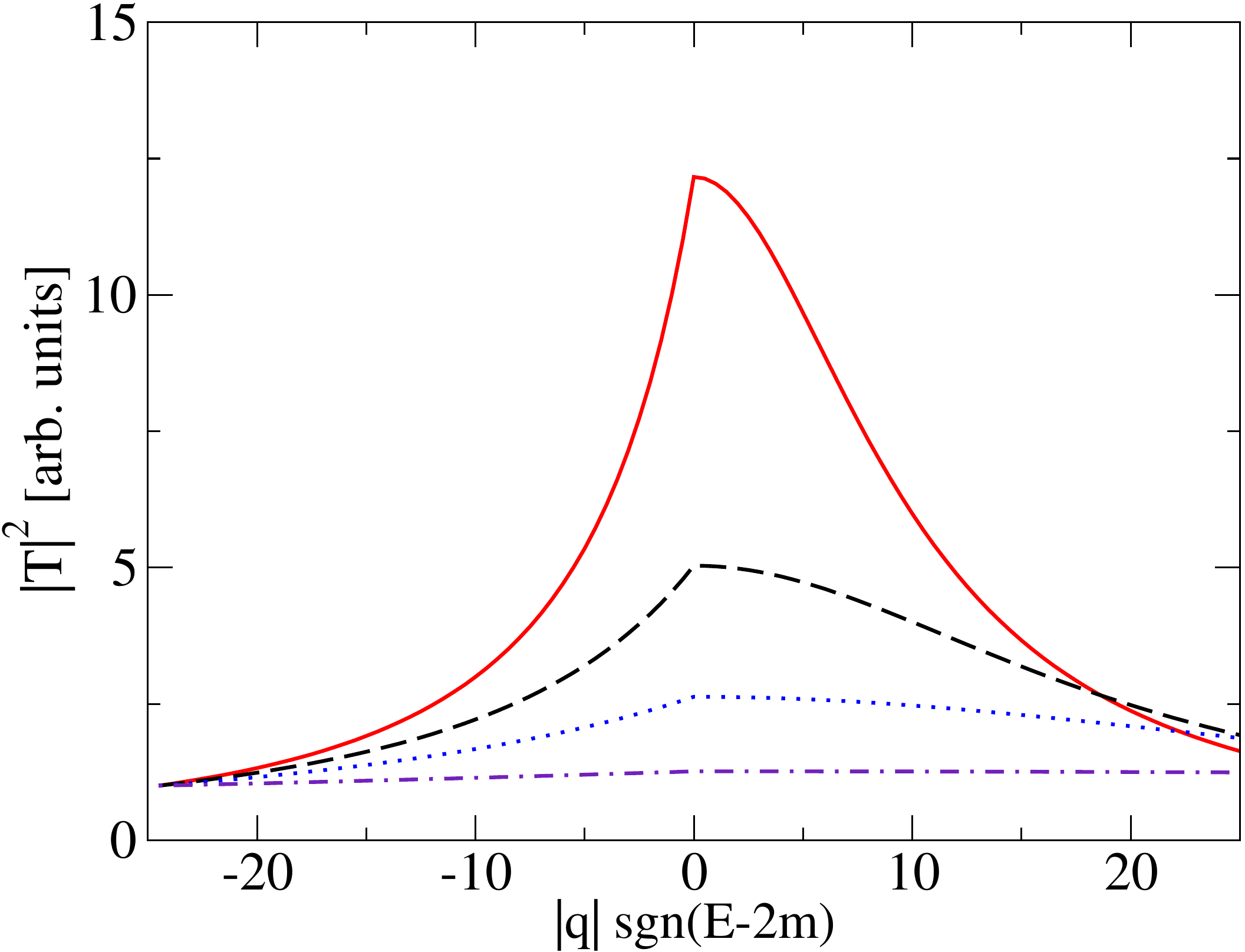}
 \caption{Threshold structure of the $T$-matrix of Eq.~\eqref{eq:Tthr} as a
function of the scattering length $a$.
 Shown is the absolute square of the amplitude for $a=20$ fm (red solid),
10 fm (black dashed), 5 fm (blue dotted),
 and 1 fm (magenta dot-dashed). Note that in the sign convention employed here
 positive scattering lengths refer to virtual states.}
\label{fig:cuspofa}
\end{center}
\end{figure}

We begin with the former. It should be stressed that whenever the total energy
(or a subsystem energy) or, more appropriately for relativistic systems, the
Mandelstam variable $s$ crosses an $N$-particle threshold, the
corresponding scattering amplitude develops a branch point in the energy plane.
The corresponding phase space scales as $(s-M_{\rm thr.}^2)^{(3N-5)/2}$, where the
threshold mass is denoted by $M_{\rm thr.}=\sum_{i=1}^N m_i$ with $m_i$ for the
mass of particle $i$. The ``-5'' term comes from the four-dimensional delta
function present in the expression for the
phase space acknowledging that energy integration counts as two powers of the
momentum. As a consequence of unitarity
the phase space not only enters the expressions for cross sections but also
scales the imaginary parts
of the corresponding loop diagrams and in this way enters the amplitudes.
This allows for a pronounced non-analyticity at threshold, driven by a
diverging, discontinuous derivative,
of the amplitude only for $N=2$. Accordingly we will now focus on two-body
reactions.
We may write the partial-wave-projected elastic $2\to 2$ $T$-matrix
generically for small momenta as
\begin{equation}
T(s)\propto \frac{q^{2L}F^2(s)}{1/a - i q^{(2L+1)}F^2(s)+{\cal O}\left(q^{4}\right)}\,,
\label{eq:Tthr}
\end{equation}
where $q$ is the particle momentum in the rest frame of the system --- e.g.,
for $m_1=m_2=m$ we may write $q=\sqrt{s/4-m^2+i\epsilon}$.
Here the term $+i\epsilon$ ensures that the square root is chosen positive
for real valued $s>4m^2$, which refers to the physical sheet. Accordingly
one may move to the unphysical sheet by choosing the $-i\epsilon$-prescription,
or, equivalently, the negative square root.
In Eq.~(\ref{eq:Tthr}) $L$ denotes the angular
momentum of the particle pair and $a$ is the scattering length.
With the sign convention employed in Eq.~(\ref{eq:Tthr}), positive and
negative values of $a$ lead to poles on the unphysical and the physical sheet,
respectively. The former are called virtual states, the latter bound states.
To allow for resonances one would need to add an effective range term,
$+1/2\, rq^2$, to the denominator, however, this is not necessary for
the kind of reasoning we are after here~\footnote{For a detailed discussion of the
connection of effective range parameters and pole locations we refer to Ref.~\cite{Hanhart:2014ssa}.}.
As done in many analyses, in addition, a form factor, denoted here as $F(s)$
with $F(0)=1$, is introduced
for convenience. It mimics the internal structure of interacting particles and
becomes operative as soon as the momentum transferred is large enough
($q\gtrsim\Lambda$ with a suitable $\Lambda$ depending on the particular system at hand) to resolve the latter.
From Eq.~(\ref{eq:Tthr}) one reads off immediately that the mentioned
pronounced structure emerging from a discontinuity in the first derivative of the scattering amplitude ---
called a cusp --- appears only for $L=0$.
A cusp is always present when a threshold related to a pair of stable particles
in a relative $S$-wave opens, however, one also sees that its significance
depends very sensitively on the value of the scattering length $a$. In this
discussion let us focus on positive values of $a$ since bound states
leave a clear imprint in observables and are connected to physical states
beyond any doubt.
The dependence of the cusp on the value of $a$ is shown in
Fig.~\ref{fig:cuspofa} where, for the illustration purpose, we employ
\begin{equation}
F(s)= \exp(-q^4/\Lambda^4)\,,
\end{equation}
with $\Lambda=0.33$~GeV. The result is given for $a=20$, $10$, $5$, and
$1$~fm, respectively. One can clearly see that a pronounced
structure appears only for the two larger values while, for example, for $a=1$
fm the amplitude shows only a very small kink
that cannot generate any structure in data. Neglecting the effect of $F(s)$,
the value of $a$ can be translated directly into the location of the corresponding pole below
the threshold on the second sheet: One finds $E=-E_v=-1/(2\mu a^2)$.
Using $2\mu=M_D=1865$~MeV for illustration, one finds $E_v\approx 0.05$, $0.2$,
$1$, and $20$ MeV for $a=20$, $10$, $5$, and $1$~fm, respectively.
Thus, the presence of a pole on the unphysical sheet very close to the
threshold is necessary to
have a pronounced cusp effect. Since a pole is to be identified with a state,
a pronounced cusp in the data is also a signature of a near-threshold state.

In view of the reasoning of the previous paragraph the claim of Refs.~\cite{Bugg:2004rk,Bugg:2011jr,Chen:2011pv,Chen:2011xk,Chen:2011pu,Chen:2013coa,Chen:2013wca,Swanson:2014tra} that
various near-threshold states are simply cusp effects without any nearby pole
does not look justified.
However, in all these works production reactions were studied with final
states different from those which generate the cusps [for example, the
$Z_c(3900)$ is claimed to be a $D\bar D^*$ cusp, but studied only
in the $\pi J/\psi$ final state in Ref.~\cite{Swanson:2014tra}]
--- thus in this case the transition to the final state is independent of
the interaction in the nearby channel. It was first pointed out in Ref.~\cite{Guo:2014iya}
that the observable most sensitive to the pole structure of the $S$-matrix is the final state that matches the nearby threshold.
This transition (up to possible contributions from left-hand cuts) is proportional to the
$T$-matrix in that channel. As argued above, however,
a pronounced cusp in the $T$-matrix occurs only if there is a nearby pole.
Based on this, the authors of Ref.~\cite{Guo:2014iya}
argue that at least $Z_c(3900)$, $Z_c(4020)$, $Z_b(10610)$, $Z_b(10650)$,
and $\chi_{c1}(3872)$ aka $X(3872)$ should exist as states
since they are all connected to signals in the nearby channels.\footnote{
The logic of the presentation of Ref.~\cite{Guo:2014iya} is somewhat different:
It is shown there that, as soon as one fits the data in the nearby channel with a one-loop diagram,
the size of the interaction in that channel calls for a resummation, which
then necessarily generates a pole.}
The statements of Ref.~\cite{Guo:2014iya} were criticised in
Ref.~\cite{Swanson:2015bsa}.
However, in the latter work, the signal in the nearby channel is not
generated by the cusp itself, but by the vertex form factor,
while the structures in the other channels are still assumed to be driven by
the cusp.
Moreover, if this explanation were correct it would imply that there should be
an enhancement at every threshold, since the vertex factors should be universal.

In some cases there should be a triangle mechanism at work that can in
principle mimic a resonance structure (for a recent discussion
of the issue see, for example, Refs.~\cite{Wang:2013hga,Szczepaniak:2015eza}).
However, it is shown in Refs.~\cite{Wang:2013hga,Gong:2016jzb}
for the case of the $Z_c(3900)$ that the triangle singularity alone cannot accommodate the data available at various energies.

Another proposal for the origin of the $XYZ$ structures without nearby poles
is that they emerge as interference effects
of various nearby quarkonia. For example, in Ref.~\cite{Chen:2017uof} it is
argued that an interference of the quark states $\psi(4160)$ and $\psi(4415)$ together with a non-resonant
background and only one additional resonance located at 4220~MeV can
explain all available data for $e^+e^-\to \pi^+\pi^-J/\psi$,
$\pi^+\pi^-h_c$, and $\pi^+D^0D^{*-}$. It is even shown
that already without this additional resonance a reasonable
fit to the data can be achieved. However, this description
comes at the price that the coupling of the photon to the $\pi\pi J/\psi$
channel is smaller than that to the $\pi\pi h_c$ channel
although the latter is, contrary to the former, forbidden by spin symmetry.
Moreover, in all channels studied a fine tuning of the non--resonant terms vs.
the resonant contributions is necessary
in order to diminish the appearance of structures near the resonance
positions. In this sense the results of
Ref.~\cite{Chen:2017uof} appear not to be a very natural explanation.
In Ref.~\cite{vanBeveren:2010mg} the Belle and BaBar data for
$e^+e^-\to \pi^+\pi^- J/\psi$ in the mass range
from 3.8 to 4.7~GeV are also explained as an interference of charmonium
states with some very broad structure.
However, the resulting line shape appears to be inconsistent with the new
high-accuracy data by BESIII (see Fig.~\ref{xsec-fit}).
Moreover, it is unclear how the $\pi\pi h_c$ final state can be understood within this picture.

From here on we will assume that at least most of the structures established in the last
fifteen years correspond, indeed, to states.
We will, therefore, discuss the theoretical implications of their existence in some depth.

\subsection{The quark model}
\label{Sect:4.1.1}
The quark model lies at the basis of the foundation of QCD.
It allows one to identify hadrons as compound objects consisting of quarks and antiquarks and provides a framework to describe their dynamics
inside hadrons employing the rich and well-developed apparatus of quantum mechanics. In particular, this framework is given by potential models that capture different aspects of this dynamics.
Nowadays, the quark model has been replaced in many aspects by more rigorous
treatments following directly from QCD.
These are effective field theories of QCD, which we will discuss
in Sec.~\ref{Sect:4.2}, and lattice QCD, which we will present in Sec.~\ref{Sect:4.3}.
Nevertheless, the quark model is still essential for
{\em (i)} classifying hadrons and
{\em (ii)} describing some features of near- and above-open-flavor-threshold states, where more rigorous alternatives are missing.
An important disclaimer is in order here. The presentation of the material in this subsection pretends neither to be chronological nor comprehensive.
An interested reader is advised to check dedicated review papers on the quark model which are supposed to meet these criteria, for example the review
\cite{Richard:2012xw}. On the contrary, as it has just been stated, in this review, the quark model is only used for classification purposes as well as a benchmark for the properties of hadrons containing heavy quarks, especially those regarded as candidates to exotic states. For this reason, the presentation of the basic features of the quark model contained in this subsection is rather schematic. When its predictions are exemplified, versions of the model which appear to be most recent, rather well-developed, and provide a unified description for both charmonium and bottomonium spectrum are preferred among many others. As such representative examples we choose (i) a relativistic potential model augmented with a heavy-light meson pair creation mechanism \cite{Ferretti:2013faa,Ferretti:2013vua} and (ii) a relativistic quasipotential model of Refs.~\cite{Galkin:1992ry,Ebert:1999xv,Ebert:2002pp,Ebert:2013iya}. Obviously, for charmonium and especially for bottomonium systems, relativistic effects are supposed to provide only little corrections, so using the relativistic form of the dynamics may be justified by a closer connection with the light-quark sector where the same models may provide a reasonable description of the properties of hadrons.

\vspace{0.3cm}\noindent
$\bullet$ {\it Potential models and classification}
\vspace{0.3cm}

The classification of mesons in terms of bound states of a constituent (or valence) quark and antiquark, follows the same patterns as for atoms.
Baryons are understood as bound states of three quarks. In this review, following the commonly accepted wisdom, we identify as exotic those hadrons which do not fit into the quark model classification of mesons and baryons. Such
exotic hadrons either involve valence gluons (these are glueballs, which
are bound states made of gluons only, or hybrids, which are bound states of quarks and gluons) or are made from more
than three valence quarks (these are multiquark states).
Nevertheless, exotic hadrons can also have the quantum numbers of ordinary hadrons and can, therefore, mix with the latter.

The quark model describes the interaction
of the constituent quarks in terms of potentials which are usually phenomenologically motivated, but may also be
inferred from lattice QCD computations.
This description works fairly well even for light quarks (with the exception of
the would-be Goldstone bosons of the chiral symmetry breaking, {\em in primis}
the pions), but it is only for heavy quarks that potential models can be
justified from first principles. We will show this rigorously in Sec.~\ref{Sect:4.2.3}.
The physical reason is however clear: the interaction between constituents
may be described by a potential, i.e., an instantaneous interaction,
only if the typical time scale is much larger than the average distance between the constituents.
This happens if the typical energy of the constituents is much smaller
than their relative momentum, i.e., if the quarks move non relativistically.
For the latter to be true their masses should be much larger than any
other scale in the system, in particular,
$\Lambda_{\text{QCD}}$, that qualifies the quarks as heavy.
The heavy quark masses, $m_h$, being much larger than any other scale
justifies organizing the potential as an expansion in $1/m_h$.

In this section, we will restrict our discussion of the quark model only to
quark--antiquark mesons (or quarkonia). Then a typical potential-model Hamiltonian for a quark--antiquark system has the form
\begin{equation}
H_0 = T + V,
\label{H0}
\end{equation}
where $T$ is the kinetic energy of the constituents, and $V$ is the potential.
For heavy quarks the kinetic energy can be non relativistically expanded but it is sometimes treated exactly, see, for instance,
Ref.~\cite{Godfrey:1985xj}.
The potential is usually also expanded in powers of $1/m_h$. If relativistic corrections are also treated exactly in $V$
one speaks of a relativistic quark model, examples of which will be discussed later in this section.
In general, potential quark models reduce the study of quark bound states to a
quantum mechanical problem with the entire QCD dynamics encoded in the potential.

\begin{figure}[ht]
\begin{center}
\includegraphics[width=0.6\linewidth]{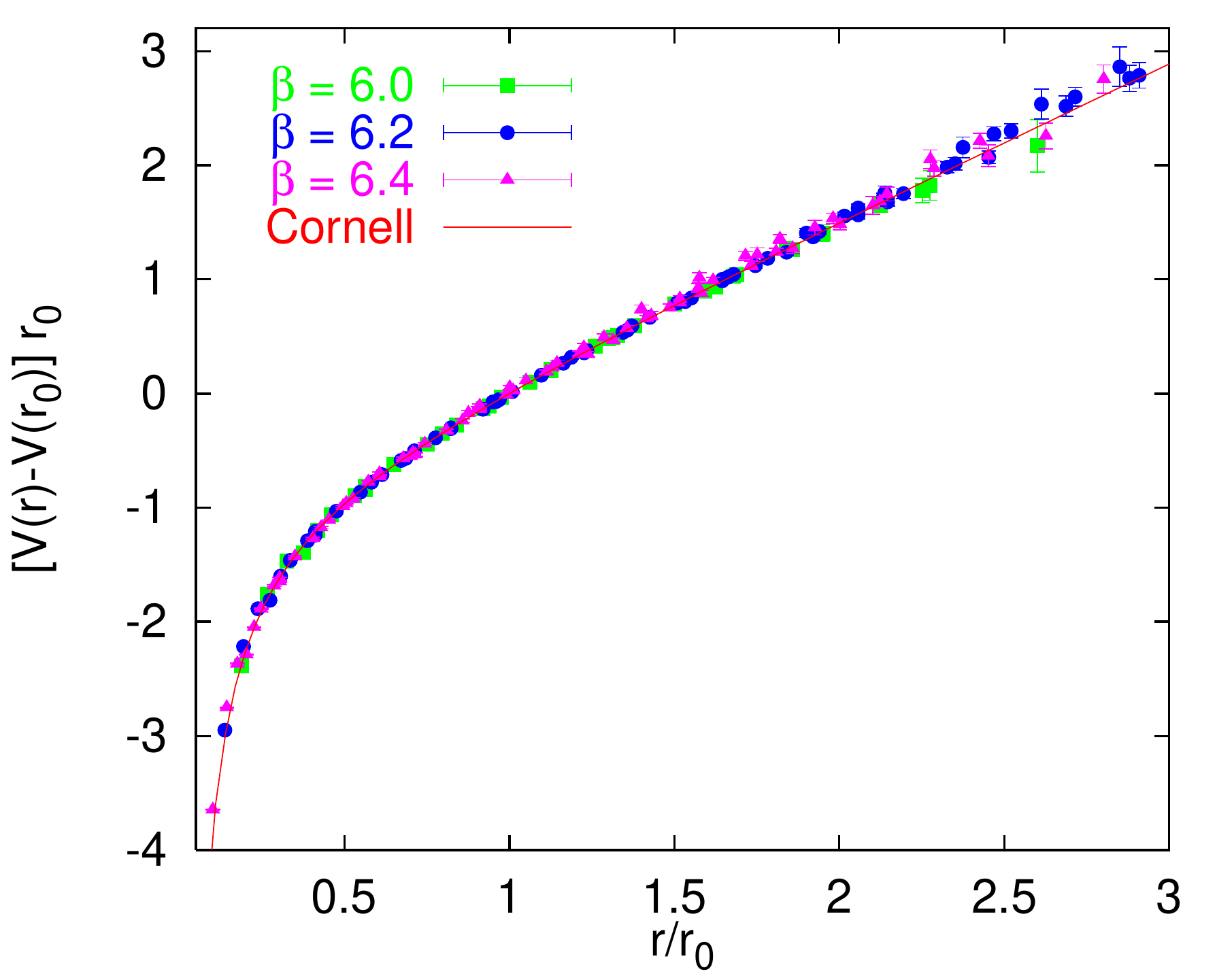}
\caption{The static quark-antiquark energy from quenched lattice QCD for
different couplings $\beta=3/(2\pi\alpha_s)$ as a function of the
quark-antiquark distance $r$.
The lattice scale is $r_0\approx 0.5$~fm. The red solid line corresponds to
the Cornell potential of Eq.~\eqref{Cornell}. Adapted from~\cite{Bali:2000gf}.}
\label{fig:V0}
\end{center}
\end{figure}

In the meson case, the potential $V$ is a function of the distance $r$ between the quark and the antiquark, their masses, spin, momenta
and angular momenta.
If it is organized as an expansion in the inverse of the masses, $m_h$, the
leading-order term is the static potential, $V^{(0)}$, which is only a function of $r$.
At short distances, $r \ll 1/\Lambda_{\text{QCD}}$, asymptotic freedom
constrains the potential to be Coulombic. At large distances,
$r \gg 1/\Lambda_{\text{QCD}}$, the potential is confining.
This is confirmed by lattice simulations of the static quark-antiquark energy,
see Fig.~\ref{fig:V0}; for a more recent determination in 2+1 flavor lattice
QCD see, for instance, Ref.~\cite{Bazavov:2009bb}.
Lattice simulations also indicate that the confining part of the static
potential grows linearly with the distance $r$.
A simple parameterization of the static potential consists, then, in the sum
of a Coulombic and a linear potential,
\begin{equation}
V^{(0)}(r) = -\frac{\kappa}{r} + \sigma r + C.
\label{Cornell}
\end{equation}
This parameterization goes under the name of Cornell potential~\cite{Eichten:1978tg}.
The parameter $\kappa$ would be $4 \alpha_s/3$ plus higher-order corrections
at very short distances, if the linear term was subtracted from the lattice
data; the coefficient $4/3$ is the Casimir of the fundamental SU(3)
representation and the coupling $\alpha_s$ is the strong coupling of QCD.
The parameter $\sigma$, whose typical value varies in the range
0.16-0.2~GeV$^2$, is known as the string tension.
It may be understood as a constant force that prevents color charges from separating.
The emergence of a string tension at long distances is suggestive of a
description of the long-range interaction
between heavy quark-antiquark pairs in terms of the effective string theory~\cite{Nambu:1978bd}.
The constant $C$ is a normalization constant of the dimension of an energy;
as we will discuss later in the section, it may also serve to mimic the effect of hadronic loops.
These parameters can be determined from fits to lattice QCD data, see e.g.~Fig.~\ref{fig:V0}, as well as from phenomenological approaches.

At higher orders in the $1/m_h$ expansion, the potential $V$ exhibits also
spin- and momentum-dependent terms.
Like in Eq.~\eqref{Cornell} for the static potential,
also higher-order potentials in the $1/m_h$ expansion may be parameterized
by summing a short distance part with a long distance confining one.
The short-distance part may be computed in perturbation theory --- see Eqs.~\eqref{V1pNRQCD}-\eqref{V2S12pNRQCD} below for the
leading-order expressions.
The long-distance part may be derived from the effective string theory,
see, for instance, Ref.~\cite{Brambilla:2014eaa}.
Spin-dependent terms show up at order $1/m_h^2$.
In the quark model this just reflects the smallness of the fine and hyperfine splittings.
We will see in Sec.~\ref{Sect:4.2} that, in the framework of non relativistic effective field theories,
the suppression in $1/m_h$ of the spin-dependent potentials is a consequence
of the heavy quark spin symmetry.
The first derivation of the spin-dependent potentials goes back to
Refs.~\cite{Eichten:1980mw,Gromes:1984ma,Barchielli:1988zp} and played a
crucial role in establishing the nature of the confining interaction in QCD.
Finally, we remark that kinetic energy and potential in the
Hamiltonian \eqref{H0} depend on the heavy quark masses.
In the quark model, masses are just free parameters like $\sigma$ and $\kappa$.
They are sometimes called constituent masses. Their values are related to the
constant $C$ and fixed through fits to the data (typically, to the mass spectrum).
If the Hamiltonian is derived and defined in the context of the effective
field theory of Sec.~\ref{Sect:4.2.3}, then the mass acquires a precise
meaning. If no constant is added/subtracted from the static potential, it is
the pole mass, whereas different constants define masses in different
subtraction schemes, see Sec.~\ref{Sect:4.2.1}.

A review of the determination of the potential in lattice QCD from the year
2000 is given in Ref.~\cite{Bali:2000gf}.
More recent determinations can be found, for example,
in Refs.~\cite{Koma:2006si,Koma:2006fw,Koma:2007jq,Bazavov:2009bb}.
We discuss these lattice results, along with analogous results for the quark-quark potential and the quark-antiquark potential in the presence of a gluonic excitation, later in this review --- see Secs.~\ref{Sect:4.2.3}, \ref{Sect:4.2.3bis} and \ref{Sect:4.3}.
The potential may also be computed through models of the QCD low-energy dynamics and
the agreement/disagreement of the model predictions with data or lattice
determinations will then provide useful insight in the QCD low-energy dynamics.
Most of the models share the basic features of the Cornell model discussed above.
In the short range, they agree with perturbative QCD, as a result of
asymptotic freedom, and, in the long range, they agree with the expectations
of the effective string theory.
In the intermediate range they show different behaviours.
Examples of model determinations of the quarkonium potential can be found in
Refs.~\cite{Simonov:1988mj,Brambilla:1996aq,Baker:1996mu,Brambilla:1999ja,DiGiacomo:2000irz,Badalian:2008sv}.

Finally, we add a few comments on the quantum numbers of quark-antiquark
states. In practical applications, the Hamiltonian of the quark model is
diagonalized in the $J^{PC}$ basis, where $J$ is the total angular momentum,
$\veJ=\veL+\veS$, $\veL$ the total orbital angular momentum, $\veS$ the total
spin, while $P=(-1)^{L+1}$ and $C=(-1)^{L+S}$ are the spatial and charge parity,
respectively, of the corresponding quark-antiquark state.
While the Hamiltonian commutes with $\veJ$, it does not commute neither with
$\veS$ nor with $\veL$ separately,
that is, generally speaking, neither $\veS$ nor $\veL$ are good quantum
numbers for describing a quark-antiquark system.
The same is true also for atomic systems.

The case $J=L$ is trivial, as the angular wave function is simply given by
the spherical harmonic $Y_{JM}(\ven)$ or by the spherical vector
${\bm Y}_{JLM}(\ven)$ for a spin-singlet [$P=(-1)^{J+1}$, $C=(-1)^{J}$] or
a spin-triplet [$P=(-1)^{J+1}$, $C=(-1)^{J+1}$], respectively.
Here $M$ is the magnetic quantum number and $\ven$ is the unit vector in the
direction of the radius $\ver$ or momentum $\vep$,
depending on whether the eigenstate problem is formulated in coordinate or momentum space.
Details of the formalism of the spherical harmonics and vectors can be found,
for instance, in the book~\cite{Varshalovich:1988ye}.

The case $L=J\pm 1$ [$P=(-1)^J$, $C=(-1)^J$] is more involved (with the
obvious exception of the $0^{++}$ scalar meson case, where $J=0$ and $L=1$).
The reason is that, because of the spin-dependent potential, the wave function
mixes terms proportional to ${\bm Y}_{J(J-1)M}(\ven)$ and ${\bm Y}_{J(J+1)M}(\ven)$.
For example, a vector state $1^{--}$ is a mixture of $^3S_1$ and $^3D_1$ wave
functions, with the coefficients which depend on the dynamics.
It is commonly accepted, however, that some quarkonium vector states are
predominantly $^3S_1$ states [for example, $\psi(3686)=\psi(2S)$],
while others are predominantly $^3D_1$ states [for example, $\psi(3770)$].
Sometimes the relative coefficient between the two wave functions can be
fixed from additional symmetry-based arguments.
For example, if chiral symmetry is effectively restored (i.e., becomes
manifest) in the spectrum of highly excited quarkonia \cite{Glozman:2003bt,Glozman:2007ek},
then the latter fills approximate multiplets of the
$\mbox{SU(2)}_L\times \mbox{SU(2)}_R$ chiral group.
As a consequence, there must exist two types of vectors,
\begin{eqnarray}
 \displaystyle |\bar{q}\gamma^i q \rangle&=&\sqrt{\frac23}|{}^3S_1\rangle+\sqrt{\frac13}|{}^3D_1\rangle\,,
 \label{veccur}\\
 \displaystyle |\bar{q}\sigma^{0i} q \rangle&=&\sqrt{\frac13}|{}^3S_1\rangle-\sqrt{\frac23}|{}^3D_1\rangle\,,
\end{eqnarray}
where the interpolating currents are written in the kets on the left-hand
side, and the mixing coefficients between the states on the right-hand side
are completely fixed by chiral symmetry.\footnote{
 Notice that, since the photon mediates electromagnetic interactions
described by a vector current,
 the $S$-$D$ wave function decomposition of a photoproduced
fermion-antifermion pair should be given by Eq.~\eqref{veccur}.
 Indeed, as one can readily verify, in the ultrarelativistic limit,
the differential cross section of $e^+e^-$ annihilation into a pair of
fermions shows an angular dependence consistent with the
decomposition \eqref{veccur}~\cite{Glozman:2009bt}:
 \be
 \frac{d\sigma}{d\Omega}\propto 1+\cos^2\theta =\frac{16\pi}{3}\left|\sqrt{\frac23} {\bm Y}_{101}(\ven)+\sqrt{\frac13}{\bm Y}_{121}(\ven)\right|^2.
 \ee
}
Here $q$ is a generic quark field, $\sigma^{\mu\nu}=(i/4)[\gamma^\mu\gamma^\nu]$
and the $\gamma$'s are the Dirac matrices.
The corresponding chiral basis for all quantum numbers is built in
Ref.~\cite{Glozman:2007at}.

$S$-$D$ mixing takes place not only between quark-antiquark states with
the same radial quantum numbers, but also between states with different
ones~\cite{Heikkila:1983wd}.
Moreover, since states with similar masses mix more strongly it may happen
that higher radially excited $S$-wave states play an important role in the
mixing with $D$-states that are heavy due to the large orbital angular momentum.
There exist claims in the literature that such a mixing may allow, for
instance, to get more realistic values of the leptonic widths of vector
charmonium and bottomonium states
and even to treat some $Y$ states [in particular, $Y(4230)$ and $Y(4360)$]
as generic $\bar{c}c$ quarkonia --- see
Refs.~\cite{LlanesEstrada:2005hz,Shah:2012js,Badalian:2017nyv,Fu:2018yxq}
and references therein.
It is important to notice, however, that the underlying mechanism for this
mixing involves creation of heavy-light meson pairs, and, as such, lies
beyond the scope of the potential models discussed so far.
The effect of hadronic loops and above-threshold states on the quarkonium
spectrum will be discussed below. Besides that, although the suggested mixing mechanism is interesting by itself, its reliable implementation requires, {\em inter alia}, a better understanding of the mixing patterns between states with different angular momenta within the quark model scheme.

\vspace{0.3cm}\noindent
$\bullet$ {\it Coupling to an open-flavor threshold}
\vspace{0.3cm}

The quark model discussed above relies upon the assumption that only the
heavy constituent quark and antiquark of the quarkonium are relevant
degrees of freedom at scales comparable with the typical momentum transfer
and binding energy inside the bound state.
This is not the case anymore, however, when the quark-antiquark distance
becomes so large
that the energy stored in the string formed between the heavy quark-antiquark
pair exceeds twice the mass of the lightest heavy-light, $(\bar{Q}q)$,
meson ($\sigma r\gtrsim 2M_{\bar{Q}q}$, where $Q$ and $q$ denote the heavy and
the light quarks, respectively).
At this distance it is energetically favorable for the string to break, which
means that the heavy quark-antiquark pair decays strongly as
$(\bar{Q}Q)\to({\bar Q}q)+({\bar q}Q)$.
The open-flavor mesons created in this way are clearly degrees of freedom
not accounted for by the naive potential models discussed so far.

The phenomenon of the string breaking is sometimes incorporated effectively
into potential models through the flattening of the static quark-antiquark
potential at relatively large quark-antiquark
distances~\cite{Bernard:2002sb,Duncan:2000kr,Bali:2005fu,Badalian:2002xy}.
Moreover, the constant parameter $C$ in Eq.~\eqref{Cornell} can be related
to the real part of the hadronic loops.
If in this adaptation of the Cornell potential the constant $C$ is taken as a universal parameter of the model, the accuracy of predictions appears to be rather low.
Some attempts to relate the constant $C$ with the light-quark content of
the meson were made in Ref.~\cite{Kalashnikova:2001ig},
while in Ref.~\cite{Simonov:2001iv} this parameter was directly related
to the quark selfenergies and evaluated using the field correlator method~\cite{DiGiacomo:2000irz}.

The effects of hadronic loops were studied in many theoretical works --- see, for example,
Refs.~\cite{Eichten:1975ag,Eichten:1975bk,Eichten:1978tg,Eichten:1979ms,vanBeveren:1979bd,Tornqvist:1979hx,vanBeveren:1982qb,Ono:1983rd,Heikkila:1983wd,
Tornqvist:1984fy,Zenczykowski:1985uh,Geiger:1989yc,Geiger:1991qe,Geiger:1991ab,Geiger:1992va,Morel:2002vk,vanBeveren:2003af,vanBeveren:2003jv,
vanBeveren:2003kd,vanBeveren:2004ve,vanBeveren:2004bz,Hwang:2004cd,Eichten:2004uh,Amsler:2004ps,Kalashnikova:2005ui,Swanson:2005rc,vanBeveren:2005ha,Eichten:2005ga,
Rupp:2006sb,vanBeveren:2006st,vanBeveren:2006ih,Hanhart:2007yq,Kalashnikova:2007qz,Pennington:2007xr,Lu:2017hma,Lu:2016mbb} to mention some.
These studies were triggered by the discovery of the narrow charm-strange
mesons $D_{s0}^*(2317)$~\cite{Aubert:2003fg} and $D_{s1}(2460)$ \cite{Besson:2003cp},
which appear to have much lower ($\sim100$ MeV) masses than predicted
by quark models. Such a discrepancy was explained by a strong coupling of
these states to the $DK$~\cite{Godfrey:1986wj,vanBeveren:2003kd} and
$D^*K$ channel \cite{vanBeveren:2003kd,Close:2005se}, respectively.
Thus, one has to proceed beyond the scope of the simple constituent quark,
potential model and to treat open-flavor decays properly.
Sometimes, this is described as ``unquenching'' the quark model.

A systematic approach to hadronic-loop effects on the spectrum of
quark-antiquark states is developed in Ref.~\cite{Barnes:2007xu} and a number
of loop theorems are formulated and proved there.
In particular, to incorporate hadronic-loop effects in the quark model a
physical hadron is modeled as a bare valence state $|{\rm A}\rangle$
augmented by two-hadron continuum components, $|{\rm BC}\rangle$,
\be
|\Psi\rangle = |{\rm A}\rangle + \sum_{\rm \{BC\}} \psi_{\rm BC} |{\rm BC}\rangle\,,
\label{ABCloop}
\ee
and it is assumed that the Hamiltonian consists of a valence Hamiltonian $H_0$ [the quark model Hamiltonian \eqref{H0}]
and an interaction term $H_I$ that couples the valence and continuum sectors,
\be
H=H_0+H_I\,.
\ee
The shift in the hadron A mass due to its coupling to the BC continuum
channels can be evaluated using the second-order quantum-mechanical
perturbation theory formula written in the rest frame of the decaying particle A,
\be
\Delta M_{\rm A}^{({\rm BC})}=\int d^3p \;\frac{|\langle {\rm BC}|H_I|{\rm A}\rangle|^2}{M_{\rm A}-E_{\rm BC}+i0}\,,\qquad E_{\rm BC}=\sqrt{\vep^2+M_{\rm B}^2}+\sqrt{\vep^2+M_{\rm C}^2}\,,
\label{massshift}
\ee
where $\vep=\vep_{\rm B}=-\vep_{\rm C}$ and a particular microscopic model
needs to be invoked to provide a form factor which regulates the integral.

If $M_{\rm A}>M_{\rm B}+M_{\rm C}$, then $\Delta M_{\rm A}^{({\rm BC})}$ acquires an
imaginary part which is trivially related to the partial decay width,
\be
\Gamma({\rm A}\to {\rm BC})=-2{\rm Im}(\Delta M_{\rm A}^{({\rm BC})}).
\label{ABCwidth}
\ee
The total hadronic shift and the total decay width for a given meson A
are given by the sum over all open-flavor channels $\{{\rm BC}\}$,
\be
\Delta M_{\rm A}=\sum_{\{{\rm BC}\}}\Delta M_{\rm A}^{({\rm BC})}\,, \qquad \Gamma_{\rm A}=\sum_{\{{\rm BC}\}}\Gamma({\rm A}\to {\rm BC})\,.
\label{DG}
\ee

\begin{figure}[ht]
\begin{center}
\includegraphics[width=0.3\linewidth]{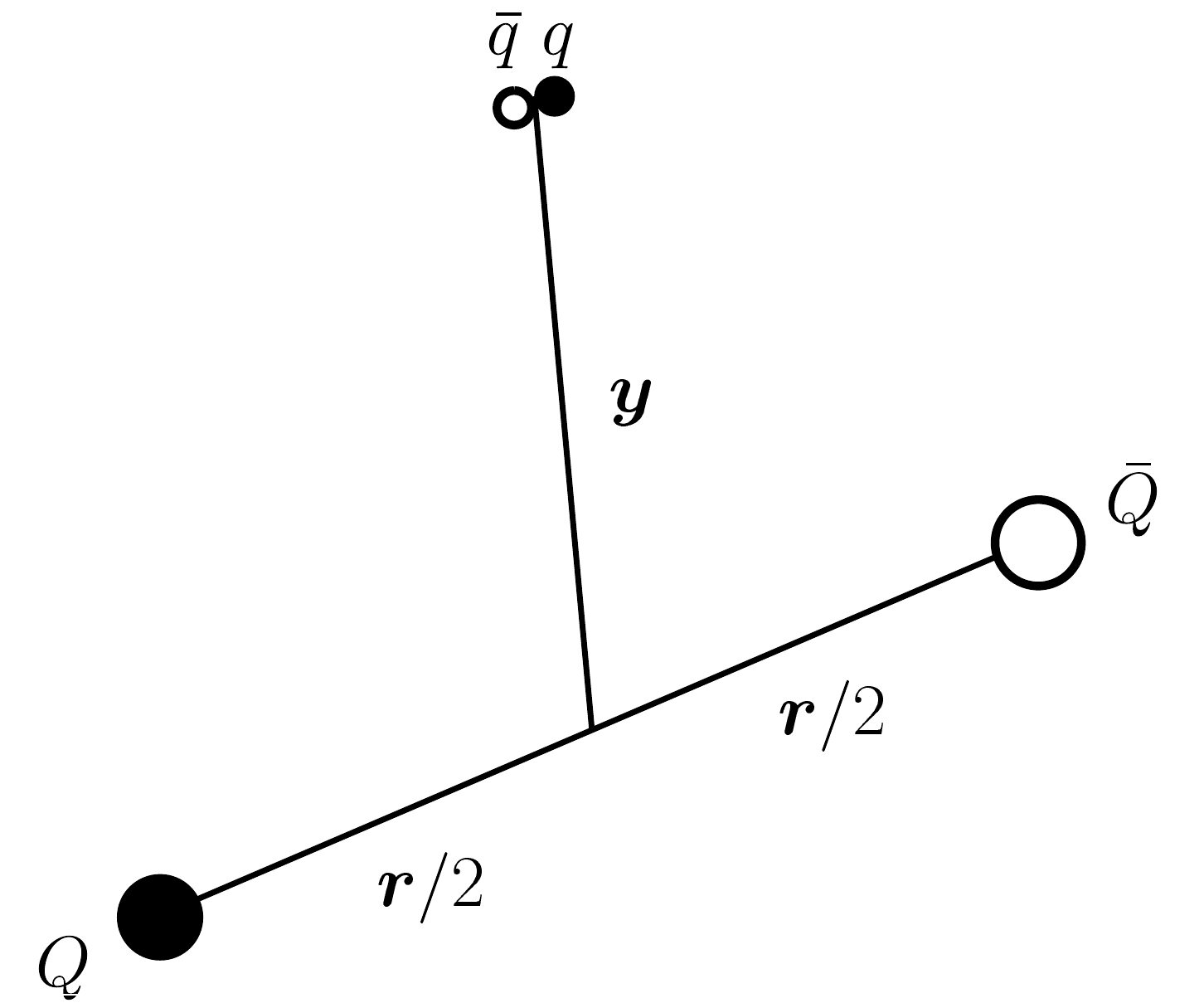}
\caption{Graphical representation of the $(q\bar{q})$ pair creation
mechanism. Adapted from Ref.~\cite{Hammer:2016prh}.}\label{fig:3P0}
\end{center}
\end{figure}

In order to proceed, one has to solve the eigenstate problem for the
quark model Hamiltonian $H_0$ and to devise a suitable mechanism for the strong decays described by the term $H_I$.
It was suggested long ago~\cite{Micu:1968mk,LeYaouanc:1972vsx,LeYaouanc:1973ldf,LeYaouanc:1974cvx}
that such a mechanism could be provided by the creation
from the vacuum of light-quark pairs with vacuum quantum numbers $J^{PC}=0^{++}$, that is, $^3P_0$ pairs. For this
reason, this model is known nowadays as the $^3P_0$ model.\footnote{
 An alternative approach, based on the $^1S_0$ pair creation, is also known
in the literature --- see, for example, Refs.~\cite{Alcock:1983gb,Kumano:1988ga}.
 However, this model is not supported by experimental data~\cite{Geiger:1994kr}.}
See also Refs.~\cite{Blundell:1995ev,LeYaouanc:1977fsz,LeYaouanc:1977gm,Page:1995rh,Godfrey:1985xj,Kokoski:1985is,Ackleh:1996yt} for further details.
Since a quark-antiquark pair is created by the operator $b^\dagger d^\dagger$
(here $b$ and $d$ are the quark and antiquark annihilation operators),
 the $^3P_0$ model can be formulated with the help of the interaction
Hamiltonian (see, for example, Refs.~\cite{Barnes:1996ff,Ackleh:1996yt})
\be
H_I=g\int d^3 x \;\bar{\psi}\psi,
\label{H3P0}
\ee
where $g$, or, equivalently, the dimensionless quantity $\gamma_q=g/(2m_q)$,
is a free parameter of the model.
A graphical representation of the quark-pair creation is in Fig.~\ref{fig:3P0}.
The values of $\gamma_q$ typically lie in the range $0.3$-$0.5$ depending on
the light-quark flavor and on the values taken by other parameters of the
model --- see, for example, Refs.~\cite{Kalashnikova:2005ui,Ferretti:2012zz}.

For states $\{{\rm A}\}$ belonging to a given $\{n,L\}$ multiplet, the
following three theorems about hadronic loops hold~\cite{Barnes:2007xu}:
\begin{enumerate}
\item The total mass shifts [see Eq.~(\ref{DG})] are equal.
\item The total decay widths [see Eq.~(\ref{DG})] are equal.
\item The mixing amplitude between any two valence basis states $i$ and $f$ vanishes if $L_i\neq L_f$ or $S_i \neq S_f$.
\end{enumerate}
The above conclusions hold if all members of the $(Q\bar{Q})$ multiplet
have equal masses and the heavy-light mesons in the loops also have
identical masses.
In addition, the final-state interaction in the continuum channels is neglected.
Although in reality these conditions cannot be met exactly, the results of
numerical calculations performed in Ref.~\cite{Barnes:2007xu}
show that the claims of the theorems are fulfilled with a quite high accuracy
in the spectrum of charmonium.
Moreover, those numerical calculations also show that the total hadronic
shifts for low-lying $\bar{c}c$ states are quite large, of the order of
400-500~MeV.
Somewhat lower shifts, of the order of 200~MeV, have been obtained for the
same charmonia in a similar approach in Ref.~\cite{Kalashnikova:2005ui}.
The conclusion drawn by the authors of Ref.~\cite{Barnes:2007xu} is that,
although hadronic-loop effects may be largely renormalized away,
their accurate description calls for further developments in the
understanding of the valence-continuum coupling.

As a representative example of a typical quarkonium spectrum calculation in
the framework of a quark model coupled to the open-flavor threshold,
we present the results of Refs.~\cite{Ferretti:2013faa,Ferretti:2013vua}.
In those references, the authors use a Hamiltonian $H_0$ with relativistic
kinetic energies and a Cornell-type potential that includes a static and a
spin-dependent part (no spin-independent part, however, beyond the static potential).
Moreover, they include quark-antiquark pair creation through the $^3P_0$ model
described above --- see Eq.~\eqref{H3P0}.
In addition, the creation point of the light-quark pair is smeared with a
form factor of the size $r_q$, which takes values in the range
0.25-0.35~fm~\cite{SilvestreBrac:1991pw,Geiger:1996re,Geiger:1991qe,Geiger:1991ab,Geiger:1992va,Ferretti:2013faa}.
Further details of the model, as well as the values of the parameters used
in the actual calculations,
can be found in Refs.~\cite{Ferretti:2013faa,Ferretti:2013vua}, while here
we only quote the results obtained for the charmonium and bottomonium mass
spectra --- see Table~\ref{tab:uqm}.

\begin{table*}
\begin{center}
{\small
\begin{tabular}{cccccccccccccccc}
\hline
\hline \\
State & $J^{PC}$ & $D\bar D$ & $\bar DD^*$ & $\bar D^*D^*$ & $D_s \bar D_s$ & $D_s \bar D_s^*$ & $D_s^* \bar D_s^*$ & $\eta_c\eta_c$ & $\eta_cJ/\psi$ & $J/\psi J/\psi$
& $\Sigma(M_0^2)$ & $M_0$ & $M$ & $M_{\text{exp.}}$ \\
 & & & $D \bar D^*$ & & & $\bar D_s D_s^*$ & & & & & &
 & & & \\ \\
\hline \\
$\eta_c(1^1S_0)$ & $0^{-+}$ & -- & -34 & -31 & -- & -8 & -8 & -- & -- & -2 & -83 & 3062 & 2979 & 2984 \\
$J/\psi(1^3S_1)$ & $1^{--}$ & -8 & -27 & -41 & -2 & -6 & -10 & -- & -2 & -- & -96 & 3233 & 3137 & 3097 \\
$\eta_c(2^1S_0)$ & $0^{-+}$ & -- & -52 & -41 & -- & -9 & -8 & -- & -- & -1 & -111 & 3699 & 3588 & 3638 \\
$\psi(2^3S_1)$ & $1^{--}$ & -18 & -42 & -54 & -2 & -7 & -10 & -- & -1 & -- & -134 & 3774 & 3640 & 3686 \\
$h_c(1^1P_1)$ & $1^{+-}$ & -- & -59 & -48 & -- & -11 & -10 & -- & -2 & -- & -130 & 3631 & 3501 & 3525 \\
$\chi_{c0}(1^3P_0)$ & $0^{++}$ & -31 & -- & -72 & -4 & -- & -15 & 0 & -- & -3 & -125 & 3555 & 3430 & 3415 \\
$\chi_{c1}(1^3P_1)$ & $1^{++}$ & -- & -54 & -53 & -- & -9 & -11 & -- & -- & -2 & -129 & 3623 & 3494 & 3511 \\
$\chi_{c2}(1^3P_2)$ & $2^{++}$ & -17 & -40 & -57 & -3 & -8 & -10 & 0 & -- & -2 & -137 & 3664 & 3527 & 3556 \\
$h_c(2^1P_1)$ & $1^{+-}$ & -- & -55 & -76 & -- & -12 & -8 & -- & -1 & -- & -152 & 4029 & 3877 & -- \\
$\chi_{c0}(2^3P_0)$ & $0^{++}$ & -23 & -- & -86 & -1 & -- & -13 & 0 & -- & -1 & -124 & 3987 & 3863 & 3860 \\
$\chi_{c1}(2^3P_1)$ & $1^{++}$ & -- & -30 & -66 & -- & -11 & -9 & -- & -- & -1 & -117 & 4025 & 3908 & 3872 \\
$\chi_{c2}(2^3P_2)$ & $2^{++}$ & -2 & -42 & -54 & -4 & -8 & -10 & 0 & -- & -1 & -121 & 4053 & 3932 & 3927 \\
$\eta_{c2}(1^1D_2)$ & $2^{-+}$ & -- & -99 & -62 & -- & -12 & -10 & -- & -- & -1 & -184 & 3925 & 3741 & -- \\
$\psi(3770)(1^3D_1)$ & $1^{--}$ & -11 & -40 & -84 & -4 & -2 & -16 & -- & 0 & -- & -157 & 3907 & 3750 & 3773 \\
$\psi_2(3823)(1^3D_2)$ & $2^{--}$ & -- & -106 & -61 & -- & -11 & -11 & -- & -1 & -- & -190 & 3926 & 3736 & 3823 \\
$\psi_3(3842)(1^3D_3)$ & $3^{--}$ & -25 & -49 & -88 & -4 & -8 & -10 & -- & -1 & -- & -185 & 3936 & 3751 & 3842 \\ \\
\end{tabular}
\begin{tabular}{cccccccccccccccccc}
\hline
\hline \\
State & $J^{PC}$ & $B \bar B$ & $B\bar B^*$ & $B^*\bar B^*$ & $B_s \bar B_s$ & $B_s \bar B_s^*$ & $B_s^* \bar B_s^*$ & $B_c \bar B_c$ & $B_c \bar B_c^*$ & $B_c^* \bar B_c^*$ & $\eta_b
\eta_b$ & $\eta_b \Upsilon$ & $\Upsilon \Upsilon$ & $\Sigma(M_0^2)$ & $M_0$ & $M$ & $M_{\text{exp.}}$ \\
 & & & $\bar BB^*$ & & & $\bar B_s B_s^*$ & & & $\bar B_c B_c^*$ & &
 & & & & & & \\ \\
\hline \\
$\eta_b(1^1S_0)$ & $0^{-+}$ & -- & -26 & -26 & -- & -5 & -5 & -- & -1 & -1 & -- & -- & 0 & -64 & 9455 & 9391 & 9399 \\
$\Upsilon(1^3S_1)$ & $1^{--}$ & -5 & -19 & -32 & -1 & -4 & -7 & 0 & 0 & -1 & -- & 0 & -- & -69 & 9558 & 9489 & 9460 \\
$\eta_b(2^1S_0)$ & $0^{-+}$ & -- & -43 & -41 & -- & -8 & -7 & -- & -1 & -1 & -- & -- & 0 & -101 & 10081 & 9980 & 9999 \\
$\Upsilon(2^3S_1)$ & $1^{--}$ & -8 & -31 & -51 & -2 & -6 & -9 & 0 & 0 & -1 & -- & 0 & -- & -108 & 10130 & 10022 & 10023 \\
$\eta_b(3^1S_0)$ & $0^{-+}$ & -- & -59 & -52 & -- & -8 & -8 & -- & -1 & -1 & -- & -- & 0 & -129 & 10467 & 10338 & -- \\
$\Upsilon(3^3S_1)$ & $1^{--}$ & -14 & -45 & -68 & -2 & -6 & -10 & 0 & 0 & -1 & -- & 0 & -- & -146 & 10504 & 10358 & 10355 \\
$h_b(1^1P_1)$ & $1^{+-}$ & -- & -49 & -47 & -- & -9 & -8 & -- & -1 & -1 & -- & 0 & -- & -115 & 10000 & 9885 & 9899 \\
$\chi_{b0}(1^3P_0)$ & $0^{++}$ & -22 & -- & -69 & -3 & -- & -13 & 0 & -- & -1 & 0 & -- & 0 & -108 & 9957 & 9849 & 9859 \\
$\chi_{b1}(1^3P_1)$ & $1^{++}$ & -- & -46 & -49 & -- & -8 & -9 & -- & -1 & -1 & -- & -- & 0 & -114 & 9993 & 9879 & 9893 \\
$\chi_{b2}(1^3P_2)$ & $2^{++}$ & -11 & -32 & -55 & -2 & -6 & -9 & 0 & -1 & -1 & 0 & -- & 0 & -117 & 10017 & 9900 & 9912 \\
$h_b(2^1P_1)$ & $1^{+-}$ & -- & -66 & -59 & -- & -10 & -9 & -- & -1 & -1 & -- & 0 & -- & -146 & 10393 & 10247 & 10269 \\
$\chi_{b0}(2^3P_0)$ & $0^{++}$ & -33 & -- & -85 & -4 & -- & -14 & 0 & -- & -1 & 0 & -- & 0 & -137 & 10363 & 10226 & 10233 \\
$\chi_{b1}(2^3P_1)$ & $1^{++}$ & -- & -63 & -60 & -- & -9 & -10 & -- & -1 & -1 & -- & -- & 0 & -144 & 10388 & 10244 & 10255 \\
$\chi_{b2}(2^3P_2)$ & $2^{++}$ & -16 & -42 & -72 & -2 & -6 & -10 & 0 & 0 & -1 & 0 & -- & 0 & -149 & 10406 & 10257 & 10269 \\
$h_b(3^1P_1)$ & $1^{+-}$ & -- & -18 & -73 & -- & -11 & -10 & -- & -1 & -1 & -- & 0 & -- & -114 & 10705 & 10591 & -- \\
$\chi_{b0}(3^3P_0)$ & $0^{++}$ & -4 & -- & -160 & -6 & -- & -15 & 0 & -- & -1 & 0 & -- & 0 & -186 & 10681 & 10495 & -- \\
$\chi_{b1}(3^3P_1)$ & $1^{++}$ & -- & -25 & -74 & -- & -11 & -10 & -- & 0 & -1 & -- & -- & 0 & -121 & 10701 & 10580 & 10512 \\
$\chi_{b2}(3^3P_2)$ & $2^{++}$ & -19 & -16 & -79 & -3 & -8 & -12 & 0 & 0 & -1 & 0 & -- & 0 & -138 & 10716 & 10578 & -- \\
$\Upsilon_2(1^1D_2)$ & $2^{-+}$ & -- & -72 & -66 & -- & -11 & -10 & -- & -1 & -1 & -- & -- & 0 & -161 & 10283 & 10122 & -- \\
$\Upsilon(1^3D_1)$ & $1^{--}$ & -24 & -22 & -90 & -3 & -3 & -16 & 0 & 0 & -1 & -- & 0 & -- & -159 & 10271 & 10112 & -- \\
$\Upsilon_2(1^3D_2)$ & $2^{--}$ & -- & -70 & -68 & -- & -10 & -11 & -- & -1 & -1 & -- & 0 & -- & -161 & 10282 & 10121 & 10164 \\
$\Upsilon_3(1^3D_3)$ & $3^{--}$ & -18 & -43 & -78 & -3 & -8 & -11 & 0 & -1 & -1 & -- & 0 & -- & -163 & 10290 & 10127 & -- \\ \\
\hline
\hline
\end{tabular}}
\end{center}
\caption{Selfenergies due to coupling to the meson-meson continuum (meson
loops) (in MeV) for various charmonium and bottomonium states as evaluated
in Refs.~\cite{Ferretti:2013faa,Ferretti:2013vua}.
 The masses $M_0$ are the quarkonium masses without hadronic-loop effects,
i.e., the quarkonium masses computed from the valence Hamiltonian $H_0$.
 The masses $M = M_0 + \Sigma(M_0^2)$ are the final masses including
hadronic-loop effects.
 The masses $M_{\text{exp.}}$ are the experimental masses from Ref.~\cite{Tanabashi:2018oca}.
 }
\label{tab:uqm}
\end{table*}

We have just seen that unquenching the quark model reveals new effects related to
hadronic loops.
In particular, if a quarkonium state lies above an open-flavor threshold,
not only its mass is shifted, as a result of its coupling to the open-flavor
channel, but, in addition, the bound-state pole moves away from the real axis
into the complex plane and the mass acquires an imaginary part, conventionally identified with the width of the state. The mass (energy) distribution that describes such a state is the famous BW distribution. To arrive at it let us parametrize the
$S$-matrix element for the elastic scattering process $(\bar{Q}q)+(\bar{q}Q)\to (\bar{Q}q)+(\bar{q}Q)$ as
\be
S=1+2iA.
\ee
Then the unitary condition for the amplitude $A$,
\be
AA^{\dagger}=\frac{1}{2i}(A-A^{\dagger})\,,
\label{unitarity}
\ee
is fulfilled automatically for
\be
A=K(1-iK)^{-1}\,,
\label{Aampl}
\ee
if the quantity $K$ is real (in case of multichannel dynamics the
corresponding $K$-matrix must be Hermitian, $K^{\dagger}=K$).
If the scattering $(\bar{Q}q)+(\bar{q}Q) \to (\bar{Q}q)+(\bar{q}Q)$ proceeds
through the formation of the state $(\bar{Q}Q)$ of mass $M$ then
\be
K=G(s)\frac{1}{M^2-s}G(s)=\frac{\Gamma(s)\sqrt{s}}{M^2-s},\quad \Gamma(s)=\frac{G^2(s)}{\sqrt{s}}\,,
\label{K1}
\ee
which gives the amplitude
\be
A = \frac{\Gamma(s)\sqrt{s}}{M^2-s-i\Gamma(s)\sqrt{s}} = G(s)\frac{1}{M^2-s-\Sigma(s)}G(s)\,.
\label{BW1}
\ee
The amplitude \eqref{BW1} describes all possible rescatterings
$(\bar{Q}q)(\bar{q}Q) \to (\bar{Q}Q) \to (\bar{Q}q)(\bar{q}Q) \to(\bar{Q}Q) \dots$ in the intermediate state
that result in the appearance of the selfenergy (hadronic loop) operator
$\Sigma(s)$ in the denominator.
The selfenergy modifies the propagator of the bare $(\bar{Q}Q)$ state.
Since the amplitude is an analytic function of $s$, the operator $\Sigma(s)$
is defined both above and below the corresponding open-flavor threshold
$(\bar{Q}q)(\bar{q}Q)$, and, therefore, it contributes in general to both the real and imaginary parts of the denominator.
Therefore, indeed, the bare pole of the amplitude, $s_{\rm bare}=M^2$, moves from its
original location both along the real axis (mass shift) and the imaginary axis,
that is, the pole position acquires also an imaginary part --- the width.

In Eq.~\eqref{BW1} we recognise the relativistic form of the BW
distribution for an isolated resonance.
If the form factor $G(s)$ in Eq.~\eqref{K1} weakly depends on $s$ for
$s\approx M^2$ ($G(s)\approx G(M^2)\equiv g$)
and $\Gamma\equiv\Gamma(M^2)=g^2/\sqrt{M}\ll M$, then the amplitude (\ref{BW1}) has a pole at
\be
s_{\rm pole}= \left(M-\frac{i}{2}\Gamma\right)^2 \approx M^2-iM\Gamma\ .
\label{s0}
\ee
Therefore, the larger the coupling $g$ of the quarkonium state to the
continuum channel, the larger is its width and, consequently, the more the
corresponding pole in the amplitude moves away from the real axis.
Nevertheless, the single-resonance amplitude \eqref{BW1} is unitary by
construction.
Multiresonance systems, instead, have to be treated with care, since
unitarisation effects may play an important and sometimes quite unexpected
role.
In particular, it has been argued for a long time that, when building quark
models with coupled channels, unitarisation cannot be neglected,
and may be even responsible for the appearance of extra, dynamically
generated resonances --- see, for example,
Refs.~\cite{vanBeveren:1979bd,Tornqvist:1995ay,Boglione:1997aw,vanBeveren:2003vs,vanBeveren:2006ua,vanBeveren:2008rs,Rupp:2015taa,Rupp:2012py,Wolkanowski:2015lsa,Hammer:2016prh} and references therein.

In Ref.~\cite{Hammer:2016prh}, the effects of unitarisation have
been studied in a simple toy model in which a scalar field $\varphi$
[viewed as a $(q\bar{Q})$ meson] interacts with a set of scalar
quark-antiquark fields $R_n$ ($n \in \lbrace 1, \ldots, N \rbrace$)
associated with the $n$-th radial excitations of a $(Q\bar{Q})$ system.
The number of excitations has been assumed to be large, $N\gg 1$.
The fields $\varphi$ and $\bar{\varphi}$ can be produced from $R_n$
through some strong open-flavor decay mechanism controlled by a coupling~$g$.
The behaviour of the poles as $g$ is varied in a wide range can be summarized as follows:
\begin{itemize}
\item For small values of the coupling $g$, unitarisation effects are
negligibly small, so that all poles are independent of each other,
and they deviate only slightly from their respective positions,
$s_n^{(0)}=M_n^2$, found for $g=0$.
Their behaviour complies, therefore, with the one described by Eq.~\eqref{s0}.
\item For growing values of the coupling $g$, as soon as the selfenergy
due to the loop operator becomes as large as the level spacing
[$|\Sigma_n|\simeq M^2(R_{n+1})-M^2(R_n)$],
unitarisation effects become important and cannot be treated perturbatively
anymore. In particular, for most states the pole trajectories bend and their
widths decrease. A similar observation that quark states may become narrower
as the coupling to the continuum grows was made in
Ref.~\cite{vanBeveren:2006ua}.
Such a nontrivial behaviour disproves an old prejudice that, because of the large phase space available,
the states high up in the spectrum acquire large widths and, therefore, cannot be observed.
However, a large width inevitably implies strong unitarisation effects,
which, as a back reaction, tame it.
\item At least one state possesses a pole trajectory that spans a wide (compared to the level spacing) range
and acquires contributions from multiple bare poles.
This behaviour has been interpreted in Ref.~\cite{Hammer:2016prh} as a sort of collective phenomenon.
\end{itemize}

The conclusion is that unitarisation effects may substantially affect the
properties of hadronic states,
such as the real and imaginary parts of the corresponding poles in the
complex plane (conventionally interpreted as the mass and the width of the state),
and, for strongly-coupled systems, result in some quite unexpected behaviour of
the system, such as a decrease of the width with the increase of the
coupling and a sort of collective phenomenon involving multiple poles.
Such effects may manifest themselves in the line shapes of the hadronic
resonances, which can depart from the simple symmetric BW form.

\vspace{0.3cm}\noindent
$\bullet$ {\it Quasipotential model}
\vspace{0.3cm}

In relativistic quark models, classes of relativistic corrections are included in the potential.
The inclusion is not systematic and may miss some equally important
corrections, nevertheless, the aim of these models is to capture some
features of the relativistic dynamics.
Hence, one possible use of relativistic quark models is to provide a checking
ground for the convergence of the non relativistic expansion.
This is particularly relevant for not so heavy quarks, like the charm quark.

Among the relativistic quark models, in the following we will consider
the quasipotential approach of Refs.~\cite{Galkin:1992ry,Ebert:1999xv}.
In this approach a meson is described as a bound state of a quark of mass
$m_1$ and an antiquark of mass $m_2$ with the wave function satisfying
a quasipotential equation~\cite{Logunov:1963yc} of the Schr\"odinger type~\cite{Martynenko:1986ug},
\begin{equation}
\label{quas}
{\left(\frac{b^2(M)}{2\mu_{R}}-\frac{\vep^2}{2\mu_{R}}\right)\Psi_{M}(\vep)} =\int\frac{d^3 q}{(2\pi)^3} V(\vep,\veq;M)\,\Psi_{M}(\veq)\,,
\end{equation}
where
\begin{equation}
\mu_{R}=\frac{E_1E_2}{E_1+E_2}=\frac{M^4-(m^2_1-m^2_2)^2}{4M^3},\quad E_1=\frac{M^2-m_2^2+m_1^2}{2M}, \quad E_2=\frac{M^2-m_1^2+m_2^2}{2M},\quad M=E_1+E_2\,,
\end{equation}
and the c.m. relative momentum squared on mass shell reads
\begin{equation}
{b^2(M) } = \frac{[M^2-(m_1+m_2)^2][M^2-(m_1-m_2)^2]}{4M^2}.
\end{equation}
The kernel $V(\vep,\veq;M)$ in Eq.~(\ref{quas}) is a quasipotential operator
that describes the interaction between the quark and the antiquark,
\begin{equation}
\label{qpot}
V(\vep,\veq;M)=\bar{u}_1(p)\bar{u}_2(-p){\mathcal V}(\vep, {\bm{q}};M)u_1(q)u_2(-q)\,,
\end{equation}
where the $u$'s are Dirac spinors, and the interaction,
\be
{\mathcal V}(\vep,\veq;M) = -\frac{16\pi}{3}\alpha_sD_{ \mu\nu}({\bm{k}})\gamma_1^{\mu}\gamma_2^{\nu} +V^V_{\rm conf}(\vek)\Gamma_1^{\mu} \Gamma_{2;\mu}+V^S_{\rm conf}(\vek)\,,
\qquad \vek=\vep-\veq\,,
\ee
consists of a gluon exchange and a confining term [somehow providing a
Lorentz-covariant version of the Cornell potential \eqref{Cornell}].
The gluon propagator in Coulomb gauge reads
\begin{equation}
D^{00}(\vek) = \frac{1}{\vek^2}\,, \qquad D^{ij}(\vek) = \frac{1}{k^2}\left(\delta^{ij}-\frac{k^ik^j}{\vek^2}\right)\,, \qquad D^{0i} = D^{i0} = 0\,,
\end{equation}
whereas the confining term consists of a scalar and a vector part:
\be
\label{vlin}
V_{\rm conf}^V(r)=(1-\varepsilon)Ar+B\,, \qquad V_{\rm conf}^S(r)=\varepsilon Ar\,, \qquad V_{\rm conf}(r)=V_{\rm conf}^S(r)+V_{\rm conf}^V(r)=Ar+B\,.
\ee
Finally, the effective long-range vector vertex is given by
\begin{equation}
\label{kappa}
\Gamma_{\mu}(\vek)=\gamma_{\mu} + \frac{i\kappa}{2m}\sigma_{\mu\nu}k^{\nu}\,,
\end{equation}
where $\kappa$ is taken as a constant and is a parameter of the model.

\begin{table}[ht]
\begin{center}
\begin{tabular}{|c|c||c|c|c||c|c|c|}
\hline
$n\,{}^{2S+1}L_J$ \vphantom{$\int_0^1$}&$J^{PC}$&$\bar{c}c$ state& Theory & Exp & $\bar{b}b$ state & Theory & Exp\\
\hline
$1\,{}^1S_0$& $0^{-+}$ & $\eta_c(1S)$ & 2981 & 2983.9$\pm$0.5 &$\eta_b(1S)$ & 9398 & 9399.0$\pm$2.3\\
$2\,{}^1S_0$& $0^{-+}$ & $\eta_c(2S)$ & 3635 & 3637.6$\pm$1.2 &$\eta_b(2S)$ & 9990 & 9999$\pm$4\\
$1\,{}^3S_1$& $1^{--}$ & $J/\psi(1S)$ & 3096 & 3096.900$\pm$0.006 &$\Upsilon(1S)$ & 9460 & 9460.30$\pm$0.26\\
$2\,{}^3S_1$& $1^{--}$ & $\psi(2S)$ & 3685 & 3687.097$\pm$0.025 &$\Upsilon(2S)$ & 10023& 10023.26$\pm$0.31\\
$3\,{}^3S_1$& $1^{--}$ & $\psi(4040)$ & 4039 & 4039$\pm$1 &$\Upsilon(3S)$ & 10355& 10355.2$\pm$0.5\\
$4\,{}^3S_1$& $1^{--}$ & $\psi(4415)$ & 4427 & 4421$\pm$4 &$\Upsilon(4S)$ & 10586& 10579.4$\pm$1.2\\
$5\,{}^3S_1$& $1^{--}$ & & 4837 & &$\Upsilon(10860)$ & 10869& $10889.9_{-2.6}^{+3.2}$\\
$6\,{}^3S_1$& $1^{--}$ & & 5167 & &$\Upsilon(11020)$ & 11088& $10992.9_{-3.1}^{+10.0}$\\
$1\,{}^3P_0$& $0^{++}$ & $\chi_{c0}(1P)$ & 3413 & 3414.71$\pm$0.30 &$\chi_{b0}(1P)$ & 9859 & 9859.44$\pm$0.42$\pm$0.31\\
$2\,{}^3P_0$& $0^{++}$ & $\chi_{c0}(2P)$ & 3870 & $3862_{-35}^{+50}$($*$) &$\chi_{b0}(2P)$ & 10233& 10232.5$\pm$0.4$\pm$0.5\\
$1\,{}^3P_1$& $1^{++}$ & $\chi_{c1}(1P)$ & 3511 & 3510.67$\pm$0.05 &$\chi_{b1}(1P)$ & 9892 & 9892.78$\pm$0.26$\pm$0.31\\
$2\,{}^3P_1$& $1^{++}$ & $\chi_{c1}(2P)$ & 3906 & 3871.69$\pm$0.17($**$) &$\chi_{b1}(2P)$ & 10255& 10255.46$\pm$0.22$\pm$0.50\\
$3\,{}^3P_1$& $1^{++}$ & $\chi_{c1}(3P)$ & 4319 & &$\chi_{b1}(3P)$ & 10541& 10512.1$\pm$2.3\\
$1\,{}^3P_2$& $2^{++}$ & $\chi_{c2}(1P)$ & 3555 & 3556.17$\pm$0.07 &$\chi_{b2}(1P)$ & 9912 & 9912.21$\pm$0.26$\pm$0.31\\
$2\,{}^3P_2$& $2^{++}$ & $\chi_{c2}(2P)$ & 3949 & 3927.2$\pm$2.6 &$\chi_{b2}(2P)$ & 10268& 10268.65$\pm$0.22$\pm$0.50\\
$1\,{}^1P_1$& $1^{+-}$ & $h_c(1P)$ & 3525 & 3525.38$\pm$0.11 &$h_b(1P)$ & 9900& 9899.3$\pm$0.8\\
$2\,{}^1P_1$& $1^{+-}$ & $h_c(2P)$ & 3926 & &$h_b(2P)$ & 10260& 10259.8$\pm$1.2\\
$1\,{}^3D_1$& $1^{--}$ & $\psi(3770)$ & 3783 & 3773.13$\pm$0.35 & & 10154&\\
$2\,{}^3D_1$& $1^{--}$ & $\psi(4160)$ & 4150 & 4191$\pm$5 & & 10435&\\
$1\,{}^3D_2$& $2^{--}$ & $\psi_2(3823)$ & 3795 & 3822.2$\pm$1.2 &$\Upsilon_2(1D)$ & 10161& 10163.7$\pm$1.4\\
\hline
\end{tabular}
\end{center}
\caption{Charmonium and bottomonium mass spectra (in MeV) obtained in
Ref.~\cite{Ebert:2002pp} and updated in Ref.~\cite{Ebert:2013iya} in the
framework of the relativistic quark model~\cite{Galkin:1992ry,Ebert:1999xv}.
 Predictions of the model are compared with the experimental data from
the PDG~\cite{Tanabashi:2018oca}.
 Only states with well established quantum numbers are considered,
with the only exception made for the charmonium $\chi_{c0}(2P)$ state
(marked with an asterisk)
 reported recently by the Belle Collaboration~\cite{Chilikin:2017evr},
which fits well into the quark model scheme.
 The charmonium $\chi_{c1}(2P)$ state (marked with a double asterisk)
is identified with the $X(3872)$.
 However, for this state the effect of the coupling to the $D\bar{D}^*$
threshold would be very large and result in a shift downwards of the mass
of the state, see discussion in the following paragraphs.
 The effects of the coupling to open-flavor thresholds have not been
included in the results of the table.
 Further predictions for other $c\bar{c}$ and $b\bar{b}$ quarkonia can
be found in Refs.~\cite{Ebert:2002pp,Ebert:2013iya}.}
\label{candb}
\end{table}

Expressions of the quasipotential for heavy quarkonia, expanded
non relativistically (i.e., in the inverse of the heavy quark masses and for small heavy quark momenta)
without and with retardation corrections to the confining potential, can be
found in Refs.~\cite{Galkin:1992ry} and~\cite{Ebert:1999xv}, respectively.
The spin-dependent interaction reduces in this way to the spin-dependent
potential of the non relativistic quark model.
In particular, it agrees with the Cornell-type parameterization of Ref.~\cite{Eichten:1980mw}.

The model is completely fixed by setting
\be
A=0.18~\mbox{GeV}^2,\quad B=-0.16~\mbox{GeV},\quad \varepsilon=-1,\quad \kappa=-1,\quad m_b=4.88~\mbox{GeV},\quad m_c=1.55~\mbox{GeV}\,,
\ee
where $A$ can be identified with the string tension in the Cornell potential
and $m_b$ and $m_c$ are the bottom and charm mass parameters, respectively.
Predictions of the model for the mass spectrum of $\bar{c}c$ and $\bar{b}b$
quarkonia are collected in Table~\ref{candb}.

\vspace{0.3cm}\noindent
$\bullet$ {\it Beyond the quark model: exotic states and the $\chi_{c1}(3872)$ aka $X(3872)$}
\vspace{0.3cm}

\begin{figure}[ht]
\begin{center}
\includegraphics[width=0.8\linewidth]{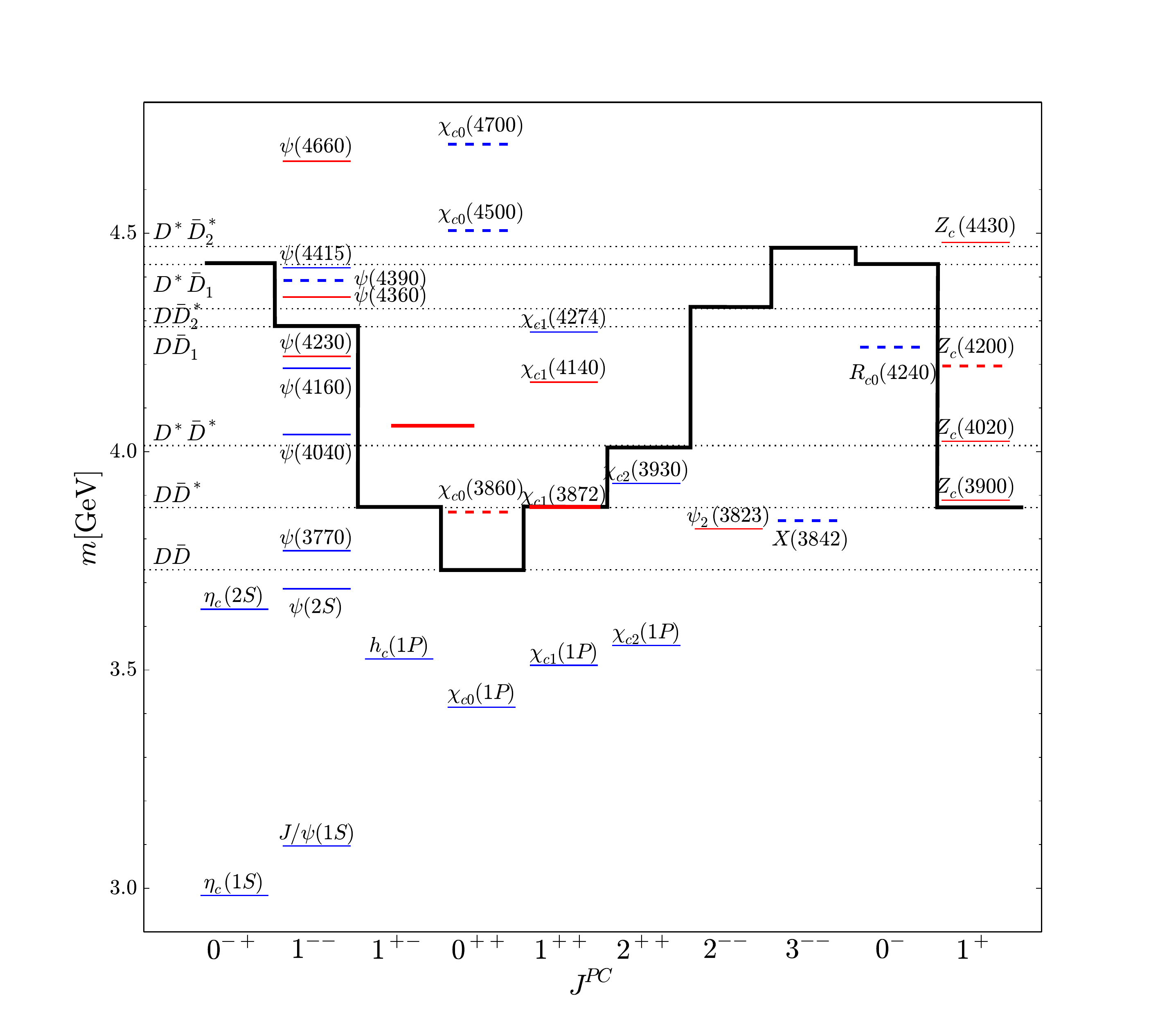}
\caption{The spectrum of states in the $c\bar{c}$ sector as of July 2019 (see the caption of Fig.1 for the notations and further details). The red lines show exotic candidates with established quantum numbers and discussed in this review.
The blue lines correspond to the experimental
charmonium spectrum listed in Table~\ref{candb}.
In each $J^{PC}$ channel the thick black line indicates the lowest threshold energy where a pair of narrow resonances can couple to the given quantum number in an S wave. Note that the parity of the channels that contain the ground-state $D$ and $D^*$ in S wave can only be positive, while by replacing one of the mesons by either a $D_1$ or a $D_2$ gives a negative parity. Higher partial waves change the parity by the additional factor $(-1)^L$, with $L$ being the angular momentum in the pair.}
\label{fig:cbarc2}
\end{center}
\end{figure}

It is instructive to put the charmonium states predicted by the quark model
in a single plot (for definiteness we use the masses listed in
Table~\ref{candb};
using the values from Table~\ref{tab:uqm} would result in a quite similar plot)
--- see Fig.~\ref{fig:cbarc2}.
In addition, a few observed vector states not present in the table are shown in red.
It is clearly seen from Table~\ref{candb}
that, while most of the charmonium states are well described by the quark model, still there are some
(shown in red and named $X$'s and $Y$'s) that do not fit into the quark model scheme.
Furthermore, the charged charmonium and bottomonium states observed experimentally cannot fit into the $Q\bar{Q}$ scheme as a matter of principle,
since their minimal quark content is four quarks.
As detailed in Sec.~\ref{Sect:3.2}, charged states are the $Z_c(3900)^{\pm}$, $Z_c(4020)^{\pm}$, and $Z_c(4430)^{\pm}$
in the charmonium spectrum, and the $Z_b(10610)^{\pm}$ and $Z_b(10650)^{\pm}$ in the bottomonium spectrum.
Other additional states claimed in the literature, but not yet confirmed, have been discussed in Sec.~\ref{Sect:3}.
All these states are exotic states with respect to the quark model.

We conclude by adding a few comments on the spectral identification of
the state $X(3872)$, which is the first and best known
of the possible charmonium-like exotica.
Since the properties of this state were discussed in detail in
Sec.~\ref{Sect:3.1.1}, we only mention here that it was observed by the
Belle Collaboration in 2003~\cite{Choi:2003ue}
and that its quantum numbers, $J^{PC}=1^{++}$, were unambiguously determined
by the LHCb Collaboration only 10 years later, in
2013~\cite{Aaij:2013zoa,Aaij:2015eva}.
The most up-to-date value of its mass given by the
PDG~\cite{Tanabashi:2018oca} is quoted in the last column of
Table~\ref{tab:X3872}.

\begin{table*}[ht]
\begin{center}
\begin{tabular}{|c|c|c|c|c|c|c|}
\hline
Reference&\cite{Godfrey:1985xj}&\cite{Ferretti:2013faa}&\cite{Kalashnikova:2005ui}&\cite{Ebert:2013iya}&\cite{Danilkin:2010cc}&\cite{Tanabashi:2018oca}\\
\hline
Mass, MeV&3960&3908&3990&3906&3954&$3871.69\pm 0.17$\\
\hline
\end{tabular}
\end{center}
\caption{Predictions of some quark models for the mass of the genuine
$\chi_{c1}(2P)$ (or $\chi_{c1}'$) charmonium.
The last column contains the experimental mass of the exotic charmonium-like state $\chi_{c1}(3872)$ aka $X(3872)$.}\label{tab:X3872}
\end{table*}

From the point of view of the quark model, the best $c\bar{c}$ candidate for
the $X(3872)$ is the radially excited $\chi_{c1}(2P)$
charmonium, since it possesses the right quantum numbers and lies nearby.
However, such an identification meets certain problems.
In particular, the quark model is unable to explain the low mass of the $X$.
Indeed, predictions of quark models for this state --- see, for example,
those collected in Table~\ref{tab:X3872} --- are in the range 3900-4000~MeV,
that is, from a few dozen to more than a hundred MeV higher than needed.
Furthermore, it is clearly seen from Fig.~\ref{fig:cbarc2} that the mass of
the $X(3872)$ is extremely close (within just 1 MeV!) to the neutral $D\bar{D}^*$ threshold
--- a fact that cannot be explained by the quark model other than by an accident.
Such a proximity to an open-flavor threshold implies that this threshold strongly impacts the properties of the state,
and that coupled channel effects need to be taken into account --- see, for example, Refs.~\cite{Kalashnikova:2005ui,Danilkin:2010cc}.
In particular, it is shown in Ref.~\cite{Kalashnikova:2005ui} that coupling of the bare $2^3P_1$
charmonium state to the $D\bar{D}^*$ channel generates,
together with a quarkonium of mass 3990 MeV (see Table~\ref{tab:X3872}),
a very shallow virtual state at the $D\bar{D}^*$ threshold.
This virtual state has a very small admixture of $c\bar{c}$ charmonium in the wave function
and is suggested to be identified with the $X(3872)$.
In Ref.~\cite{Danilkin:2010cc} the peak at the neutral $D\bar{D}^*$ threshold is argued to be due to a BW resonance shifted by
channel coupling from its original position of 3954~MeV (see Table~\ref{tab:X3872}).

In order for the $X(3872)$ to be unambiguously identified with or distinguished from the $\chi_{c1}(2P)$
on spectroscopic grounds, an interesting observation was made in Ref.~\cite{Lebed:2017yme}.
The spin-dependent potential induces a splitting within spin-triplet states
of the same radial excitation $n$ and angular momentum $L$.
The center-of-gravity mass of these states is defined as
\be
M^{\text{c.o.g.}}_{n{^3}L_J} =\frac{1}{3(2L+1)} \left( (2L-1) M(n{}^3 L_{J = L-1})+(2L+1) M(n{}^3 L_{J = L})+(2L+3) M(n{}^3 L_{J = L+1})\right)\,.
\label{Deltanl}
\ee
It is well known~\cite{Vairo:2009tn} that the hyperfine splitting between
the mass of the spin-singlet heavy quarkonium state $h$ ($J^{PC}=1^{+-}$)
and the centre-of-gravity mass of the corresponding spin triplet
$\chi$ ($J^{PC}=J^{++}$, $J=0,1,2$),
\be
\Delta_{n,1}=M_{n{^1}P_1}- {M}^{\text{c.o.g.}}_{n{^3}P_J}\,,
\label{hfscog}
\ee
is very small both in the charm and bottom sector. Indeed, the experimental
values read~\cite{Tanabashi:2018oca}
\be
\Delta_{1,1}(c\bar{c})=0.08(13)~\text{MeV}\,,\qquad \Delta_{1,1}(b\bar{b})=-0.57(88)~\text{MeV}\,, \qquad \Delta_{2,1}(b\bar{b})=-0.44(1.31)~\text{MeV}\,.
\ee
Thus, it was suggested in Ref.~\cite{Lebed:2017yme} that the splitting
$\Delta_{n,L}$ could be used as a benchmark for the quark content of hadronic states.
A large $\Delta_{n,L}$ would signal that at least one state in the quartet
contains a significant non-$Q\bar{Q}$ component and, therefore, should be
regarded as an exotic state.
In particular, the experimental measure of the mass of the $h_c(2P)$
would allow to test the entire charmonium $2P$ quartet against the
identification of the $X(3872)$ with the $\chi_{c1}(2P)$,
under the assumption that $\chi_{c0}(2P)$ and $\chi_{c2}(2P)$ have been identified correctly.

\subsection{Phenomenological approaches to exotic hadrons}
\label{Sect:4.1}
In this section various phenomenological approaches are discussed, which
are often used in the literature to classify mesons, including candidates for exotic states.
As mentioned above, we define the latter to be states which demonstrate properties at odds with the
predictions of the simplest quark model describing mesons as bound states of a quark and an antiquark.
Amongst the exotic states there are some that carry quantum numbers that are
incompatible with the simple quark model --- for example,
the isovector quarkonium-like states introduced in Secs.~\ref{Sect:3.2.1}, \ref{Sect:3.2.2} and~\ref{Sect:3.2.3}
--- and others that have quantum numbers allowed for a quark-antiquark state,
but show properties at odds with the quark model predictions.
Here the most famous example is the $\chi_{c1}(3872)$ aka $X(3872)$ introduced in Sec.~\ref{Sect:3.1.1}.

\begin{figure}[t!]
\begin{center}
 \includegraphics[width=0.7\linewidth]{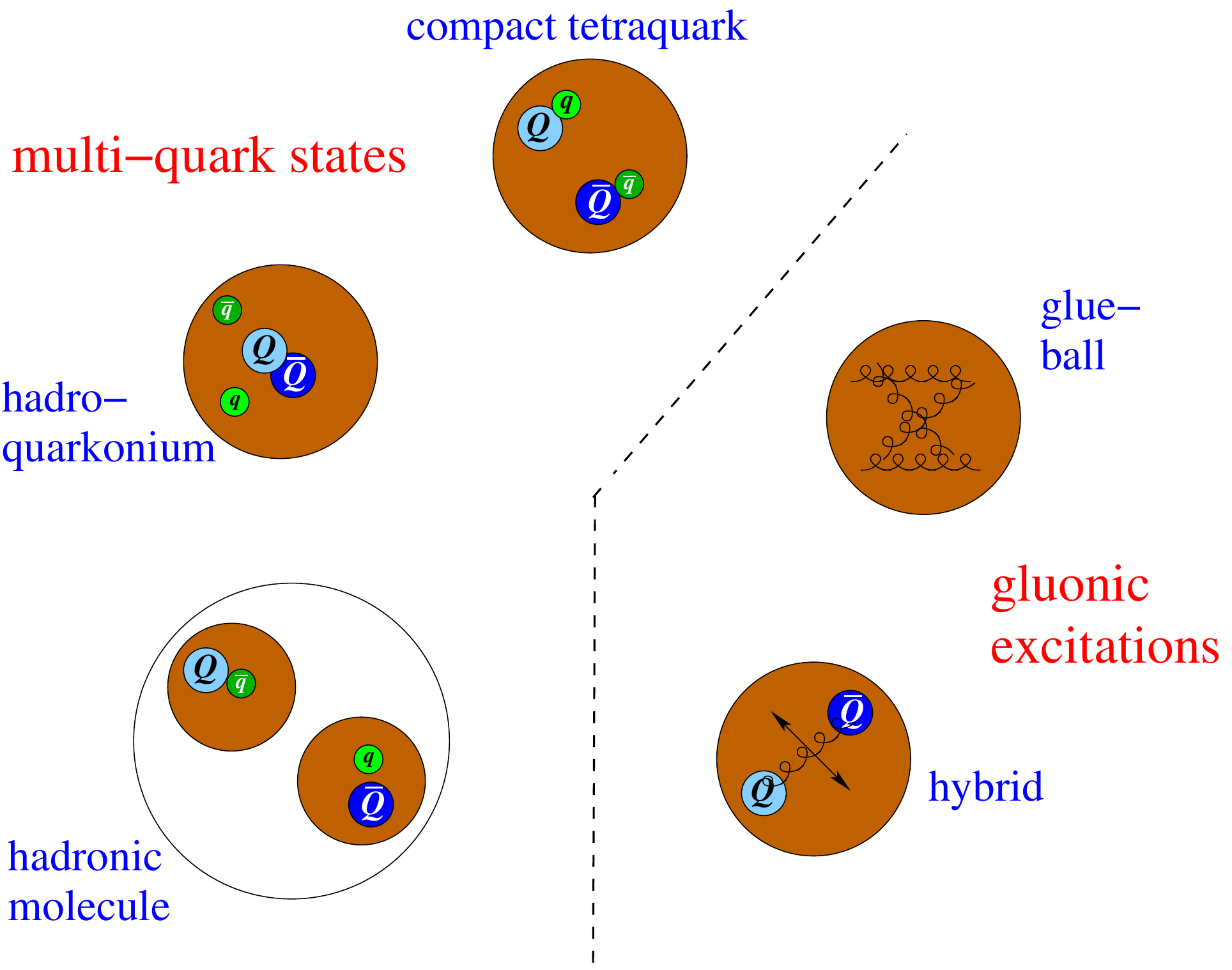}
\caption{Graphic illustration of the most common proposals for the structure
of exotic states. Details are given in the text.}
\label{fig:exotics}
\end{center}
\end{figure}

Amongst the scenarios relevant for exotic states in the spectrum of heavy
quarks, that is those containing a heavy $Q\bar{Q}$ pair,
one may distinguish states with an active gluonic degree of freedom (hybrid mesons) and multiquark configurations that are
distinguished on the basis of their internal quark clustering --- a graphical
representation of the most common scenarios for exotics is shown in Fig.~\ref{fig:exotics}.
It is fair to note here that Gell-Mann proposed the existence of multiquark
states already in 1964, when introducing his famous classification model
for mesons and baryons \cite{GellMann:1964nj}.

\subsubsection{Hybrids}
\label{Sect:4.1.2}
Due to its non-Abelian nature, QCD allows for more colorless hadrons
than just quark-antiquark mesons and three-quark baryons.
Of relevance for the present review are quarkonium hybrids or simply hybrids,
which constitute the subject of the present section, and multi-quarks, which will be discussed in the next sections.
With the word ``hybrid'' we identify conventionally quark-antiquark mesons
with excited gluonic degrees of freedom.
Gluons bring a new type of excitation to the system, in addition to the
rotational and radial motion of the quarks in ordinary $(Q\bar{Q})$ systems
discussed in the previous section.

Predictions for various properties of hybrids with heavy quarks found in the vast literature
on the subject rely on different approaches.
In particular, the interested reader can find predictions from the flux-tube model in
Ref.~\cite{Close:1994hc,Barnes:1995hc},
from the Coulomb-gauge QCD approach in
Refs.~\cite{Cotanch:2001mc,General:2006ed,LlanesEstrada:2000hj},
from a constituent gluon model for gluelumps (short-distance hybrids) in
Ref.~\cite{Abreu:2005uw},
from the constituent gluon model in
Refs.~\cite{Horn:1977rq,Iddir:1998yc} ---
see also the discussion below --- and
from the QCD string model in
Refs.~\cite{Simonov:1999qj,Simonov:2001rn,Kalashnikova:2002tg,Buisseret:2007ed,Kalashnikova:2008qr,Kalashnikova:2016bta}.
An effective field theory description based on non-relativistic effective
field theories has also been developed.
This is the Born--Oppenheimer effective field theory
of Refs.~\cite{Braaten:2013boa,Braaten:2014ita,Braaten:2014qka,Berwein:2015vca,Oncala:2017hop,Brambilla:2017uyf,Soto:2017one,Brambilla:2018pyn}
discussed in more detail in Sec.~\ref{Sect:4.2.3bis},
An independent source of information about the masses and splittings of hybrids comes from lattice QCD
calculations~\cite{Michael:2003xg,Liu:2005rc,Luo:2005zg,Liu:2012ze,Cheung:2016bym,Juge:1999ie,Liao:2001yh}.

Models differ substantially in the way they describe hybrids.
For example, in the flux tube model~\cite{Isgur:1984bm}, hybrids are
described as phonon-type excitations, while in the constituent gluon
model~\cite{Horn:1977rq,Iddir:1998yc} they carry both color and spin.
Nevertheless, there is some consensus about few basic facts.


In addition, the excitation of the gluonic degrees of freedom contributes at least with
approximately 1~GeV to the mass of the system, so that the mass of the ground-state hybrid can be roughly
estimated to be $2m_h+1~\mbox{GeV}$, where $m_h$ is the mass of the heavy quark.
This gives about 4~GeV for the mass of the lowest charmonium hybrid and 11~GeV for the one of the lowest bottomonium hybrid.
Results obtained by different theoretical approaches, as well as
determinations from lattice QCD (mostly in the quenched approximation, but
Ref.~\cite{Cheung:2016bym} is a 2+1 flavor dynamical lattice simulation,
with a pion mass of about 240~MeV, discussed in Sec.~\ref{Sect:4.3:charmonia}) for the lowest charmonium hybrid are
collected in Table~\ref{tab:cchybryd}.
We see that, indeed, all predictions are generally consistent with each other
and with the simple estimate made above for what concerns the value of the
mass of the lowest-lying hybrid.
The situation for bottomonium hybrids looks similar.
Indeed, the most recent calculations place the mass of the lowest
bottomonium hybrid around 11~GeV, also in agreement
with the simple estimate made above --- see
Refs.~\cite{Braaten:2014ita,Berwein:2015vca,Kalashnikova:2016bta} for more details.

Clearly, as the table shows, there are also considerable differences between the different approaches.
First, not all of them identify the same hybrid state as the lowest-lying one.
Moreover, also the (approximate) degeneracy pattern of the different hybrid multiplets differ.
These disagreements may serve to select among different approaches.

\begin{table}[ht]
\begin{center}
\begin{tabular}{|c|c|c|c|}
\hline
Model& Reference& Quantum numbers & Mass, GeV\\
\hline
Flux-tube&\cite{Barnes:1995hc}& --- & $4.1-4.2$\\
\hline
Gluelump& \cite{Abreu:2005uw} & $2^{++}$& 4.12\\
\hline
QCD string& \cite{Buisseret:2006sz} & $1^{-+}$ & $4.2 \pm 0.2$\\
 &\cite{Kalashnikova:2008qr,Kalashnikova:2016bta}& $1^{--}$, $(0,1,2)^{-+}$ & $4.3-4.5$\\
\hline
BOEFT&\cite{Berwein:2015vca}& $1^{--}$, $(0,1,2)^{-+}$ & $4.15 \pm 0.15$\\
\hline
Lattice& \cite{Cheung:2016bym,Michael:2003xg} & $1^{-+}$ & $4.3-4.4$\\
&\cite{Cheung:2016bym,Liu:2005rc,Luo:2005zg} & $0^{-+}$ & $4.3-4.4$\\
\hline
\end{tabular}
\end{center}
\caption{Predictions of various models, effective field theory and lattice
QCD for the mass of the lowest charmonium hybrid.
 BOEFT stands for the Born--Oppenheimer effective field theory discussed in Sec.~\ref{Sect:4.2.3bis}.}\label{tab:cchybryd}
\end{table}

\vspace{0.3cm}\noindent
$\bullet$ {\it Constituent gluons and potential models}
\vspace{0.3cm}

The quark model may be extended to describe hybrids by allowing, besides
constituent quarks and antiquarks, also constituent gluons.
In the simplest realisation, the hybrid is a quark-antiquark pair augmented by a single gluonic excitation.
The constituent gluon plays the role of an additional gluonic degree of
freedom that binds with the heavy quark-antiquark pair to form the hybrid.
It can be given an effective mass of about 1~GeV. As was stated above,
this is the typical energy gap between the heavy quark-antiquark levels and
their first gluonic excitations.
Because in this picture the gluon, like the other constituents, is a
massive mode, it is possible to describe the interaction
among the constituents in terms of a potential and expand it in the inverse of the masses.
The valence gluon, located at $\ver_g$, is connected to the quark, located
at $\ver_Q$, and the antiquark, located at $\ver_{\bar{Q}}$,
by two strings with the same string tension $\sigma$ as in the quark-antiquark potential of Eq.~\eqref{Cornell}.
Then, for a hybrid in the constituent gluon picture, the long-range, confining, static potential is
\be
V_{\rm conf}[Q\bar{Q}g]=\sigma |\ver_Q-\ver_g|+\sigma |\ver_{\bar Q}-\ver_g|\,,
\label{h0}
\ee
and the leading-order short-range static potential reads
\be
V_{\rm Coul}[Q\bar{Q}g]=-\frac{3\alpha_s}{2|\ver_Q-\ver_g|}-\frac{3\alpha_s}{2|\ver_{\bar Q}-\ver_g|}+\frac{\alpha_s}{6|\ver_Q-\ver_{\bar Q}|}\,.
\label{coulomb}
\ee
Note that at leading order the short-range potential is just the sum of
three two-body Coulomb potentials in the suitable SU(3) color representation of the interacting pair:
the gluon-heavy-quark pair in the fundamental representation, the gluon-heavy-antiquark pair
in the complex conjugate of the fundamental representation, and the heavy quark-antiquark pair in the octet representation.
The different factors in front of the Coulomb potentials are due to the different representations;
the octet potential is repulsive, hence at short distances the hybrid is
kept together by the bindings of the quark and antiquark with the constituent gluon alone.
At higher orders in the inverse of the masses, one may establish also Cornell-like potentials similar to the static potential.
In particular, one gets a spin-dependent potential of the same form as the
spin-dependent potential for ordinary quarkonia --- see, for example, Ref.~\cite{Kalashnikova:2008qr}.
As we will see in Sec.~\ref{Sect:4.2.3bis} --- and, in particular, with
Eq.~\eqref{sdm2} --- the naive constituent gluon potential model misses,
however, even the leading, only $1/m_h$ suppressed, spin-dependent potential in the hybrid system.
This happens because this potential has no analog in ordinary quarkonium.
As a consequence, the spin symmetry is broken more strongly in hybrids than in quarkonia.

The spatial and charge-conjugation parity of a hybrid read
\be
P=(-1)^{L + K}, \qquad C=(-1)^{L+S+1}\,,
\label{JPCmag}
\ee
for a chromomagnetic gluon ($L_g = K$), and
\be
P=(-1)^{L+K+1},\qquad C=(-1)^{L+S+1}\,,
\label{JPCel}
\ee
for a chromoelectric gluon ($L_g = K \pm 1$), where $L_g$ is the relative
angular momentum between the heavy quark-antiquark pair and the gluon,
$K$ is the total angular momentum of the gluon, $L$ is the orbital angular momentum
of the quark-antiquark pair, and $S$ is the spin of the quark-antiquark pair.
Hybrids with a chromoelectric gluon couple strongly to pairs of $S$-wave
heavy-light mesons and, as a result, they become very broad and cannot be
observed as resonances~\cite{Iddir:1998yc}.
In the following, we will stick, therefore, to hybrids with a chromomagnetic gluon.

For hybrids with a chromomagnetic gluon (that is, a $1^{+-}$ gluon),
if the heavy quark-antiquark pair is in an $S$-wave spin-singlet state ($S = L = 0$)
and $L_g = K = 1$, then, according to Eq.~\eqref{JPCmag}, the quantum numbers
are $1^{--}$, that is, the hybrid is a vector state.
If, on the contrary, the $Q\bar{Q}$ pair is in an $S$-wave spin-triplet state,
then the hybrid may have the quantum numbers $J^{PC}=J^{-+}$ with $J=0,1,2$;
hence the quantum numbers include the exotic $1^{-+}$.
These states constitute the so-called $H_1$ multiplet and are (approximately)
degenerate --- see Table~\ref{tab:spin_multiplet}.
This pattern of (approximate) degeneracies has been confirmed by model
calculations, effective field theory analyses, and lattice simulations.
Similarly, the binding of a chromomagnetic gluon with a $P$-wave heavy
quark-antiquark pair ($L=1$) with the quantum numbers
$\{1^{+-}, (0, 1, 2)^{++} \}$ (corresponding to different spin states)
gives hybrids with the quantum numbers of the $H_2 \cup H_3 \cup H_4$
multiplets --- see Table~\ref{tab:spin_multiplet}.
This (approximate) degeneracy pattern is specific of the constituent gluon
picture~\cite{Braaten:2014qka,Berwein:2015vca}.
Different (approximate) degeneracy patterns show up in different approaches.
We will discuss the degeneracy pattern emerging in the Born--Oppenheimer
effective field theory approach in Sec.~\ref{Sect:4.2.3bis}.

Due to the symmetry of the wave function, it was found that a selection rule exists that
suppresses the decay of the vector hybrid with a chromomagnetic gluon into a pair of $S$-wave heavy-light mesons in the final
state~\cite{Iddir:1998yc,Kalashnikova:1993xb,Kou:2005gt,Isgur:1985vy}.
Similarly to the case of ordinary quarkonia, but in the constituent gluon model,
the relative strength of the strong decays of hybrids into one $S$-wave and
one $P$-wave heavy-light meson has been estimated in the heavy-quark limit
and the corresponding spin-recoupling coefficients have been computed~\cite{Kalashnikova:2008qr}.
Although the heavy-quark symmetry is not exact for the $c$ and $b$ quarks and,
therefore, its predictions cannot be treated as exact results,
nevertheless these recoupling coefficients set a particular pattern for the open-flavor decays of the hybrids.
Transitions from hybrids to quarkonia have been considered in non-relativistic
effective field theories that by construction implement the heavy quark
symmetry~\cite{Oncala:2017hop}; these transitions will be reviewed in Sec.~\ref{Sect:4.2.3bis}.

A comment on the $P$-wave positive-parity $D_J$ ($B_J$) mesons is in order here.
In the strict heavy-quark limit, the quadruplet of such $P$-wave heavy-light
mesons turns to a pair of doubly degenerate states, $P_j$, with $j=1/2$ or
$j=3/2$ being the light quark total angular momentum.
Since the $P_{1/2}$ and $P_{3/2}$ mesons decay via pion emission to lower-lying
heavy-light mesons in the $S$-wave and in the $D$-wave, respectively,
the $P_{1/2}$ mesons appear to be much broader than the $P_{3/2}$ ones.
This makes it hardly feasible to identify experimentally either of the two
$P_{1/2}$ quadruplet members ($D_0$ and $D_1$ in the $c$-sector and $B_0$ and $B_1$ in the $b$-sector)
in the final state, while this is possible for the other two members of
the $P$-wave quadruplet ($D_1$ and $D_2$ in the $c$-sector and $B_1$ and $B_2$ in the $b$-sector).
Then, since the most prominent decay channels of the $D_1(2420)$ and
$D_2(2460)$ are the $D^{(*)}\pi$ modes,
the open-flavor decays of the vector charmonium hybrid should be saturated by $D^*\bar{D}^{(*)}\pi$ final states.
The situation with bottomonium hybrids is similar.
At this point, it is instructive to compare the production of a
heavy-light meson from the $P_{3/2}$ doublet accompanied by an $S$-wave meson
in the decays of ordinary quarkonia and hybrids.
Such a production is only possible if the produced light-quark pair has total angular momentum equal to 1.
This condition cannot be fulfilled in the vector quarkonium decay, as in this case $j_{q\bar{q}}=0$.
Therefore, the amplitude for the quarkonium decay into such a final state
vanishes in the strict heavy-quark limit~\cite{Eichten:1978tg,Li:2013yka}.
Instead, open-flavor decays of $(Q\bar{Q}g)$ hybrids proceed through the
gluon conversion into a light quark-antiquark pair, which, therefore,
carries the angular momentum of a vector, $j_{q\bar{q}}=1$.
This implies that there is no suppression of the amplitude for the vector
hybrid decay into one $S$-wave and one $P_{3/2}$-wave open-flavor meson.

The heavy-quark limit is not exact in QCD.
Beyond this limit, one needs to take into account corrections controlled by
the ratio $\Lambda_{\rm QCD}/m_h$, which may be sizeable in the charm sector.
In particular, the physical $P$-wave heavy-light mesons come not as pure $P_{1/2}$ and $P_{3/2}$ states,
but as mixtures governed by a mixing angle $\theta$.
The probability of the decay of an ordinary quarkonium into a final state
containing one $S$-wave and one narrow $P$-wave open-flavor meson
is proportional to $\sin^2\theta$, while the same probability for the hybrid is proportional to $\cos^2\theta$.
The heavy-quark selection rule manifests itself in the data if
$\theta \ll 1$. Recent estimates of the mixing angle $\theta$ in the charm
and bottom systems can be found, for instance, in
Ref.~\cite{Kalashnikova:2016bta}.
They show that, at least in the bottomonium sector, the heavy-quark selection
rule may indeed allow one to distinguish hybrids from conventional quarkonia.

The heavy quark-antiquark pair ($Q\bar{Q}$) in a hybrid is dominantly in a color-octet state,
while in an ordinary quarkonium it is dominantly in a color-singlet state.
This observation leads to one more prediction for the decays of hybrids:
since the color-octet quark-antiquark pair cannot annihilate into a photon,
which would eventually convert into a lepton pair,
the leptonic width of hybrids has to be small and, in particular, hybrids
should not be seen as an enhancement in the $R$-ratio scan.

For what concerns hidden-flavor decays of hybrids, they can happen according to the following scheme:
\be
(Q\bar{Q})_8g
\genfrac{}{}{0pt}{}{\nearrow}{\searrow}\raisebox{0mm}{$
\begin{array}{l}
(Q\bar{Q})_1(gg)\to (Q\bar{Q})+\mbox{light hadrons}\\[3mm]
(ng)\to\mbox{light hadrons},\quad n\geqslant 2
\end{array}
$},
\label{2ways}
\ee
that is, the quark-antiquark pair inside the hybrid, either converts from the color-octet state into a color-singlet one by emitting a gluon, $g$,
or it annihilates completely into gluons.
Then the gluons convert into light hadrons.
In the former case, the hybrids hidden-flavor decays populate final states
with hidden flavor, which may provide a clear experimental signal~\cite{Close:1994zq}.
In the latter case, final states not containing heavy quarks are enhanced~\cite{Dunietz:1996zn}.
See also Ref.~\cite{Close:1997wp} for further details on the experimental signatures and search strategies for charmonium hybrids in $B$-meson decays.
Various estimates of the hybrid production probability in such decays can be found in Refs.~\cite{Close:1997wp,Chiladze:1998ti}.
In particular, it is argued that the hybrid production probability is
$\approx 10^{-3}$-$10^{-2}$, which lies in the same ballpark as the production probability of conventional quarkonia.

\vspace{0.3cm}\noindent
$\bullet$ {\it The $\psi(4230)$ aka $Y(4230)$}
\vspace{0.3cm}

\begin{figure*}[htbp]
\centering
\includegraphics[width=0.48\textwidth]{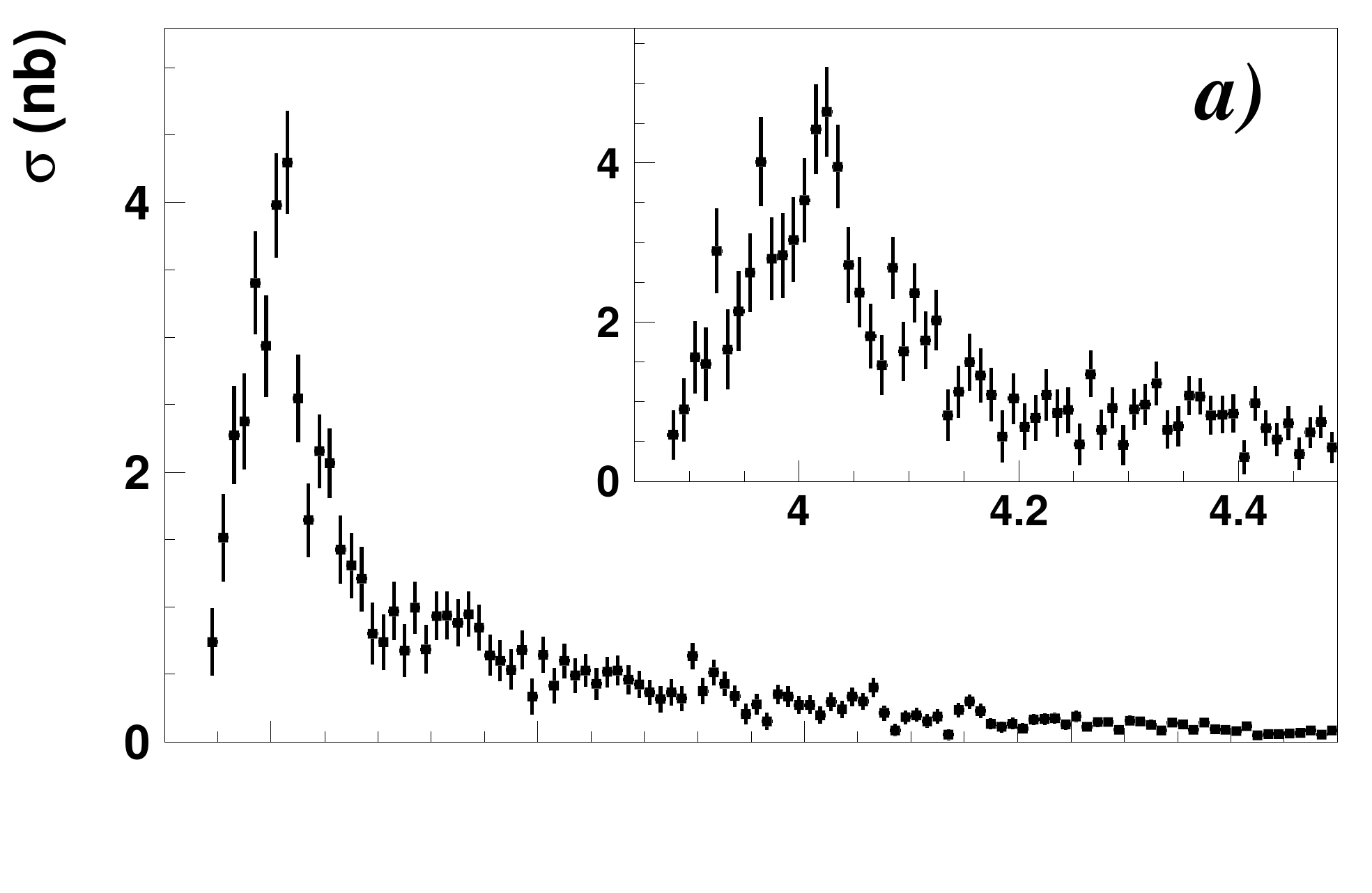} \\[-1.1cm]
\includegraphics[width=0.48\textwidth]{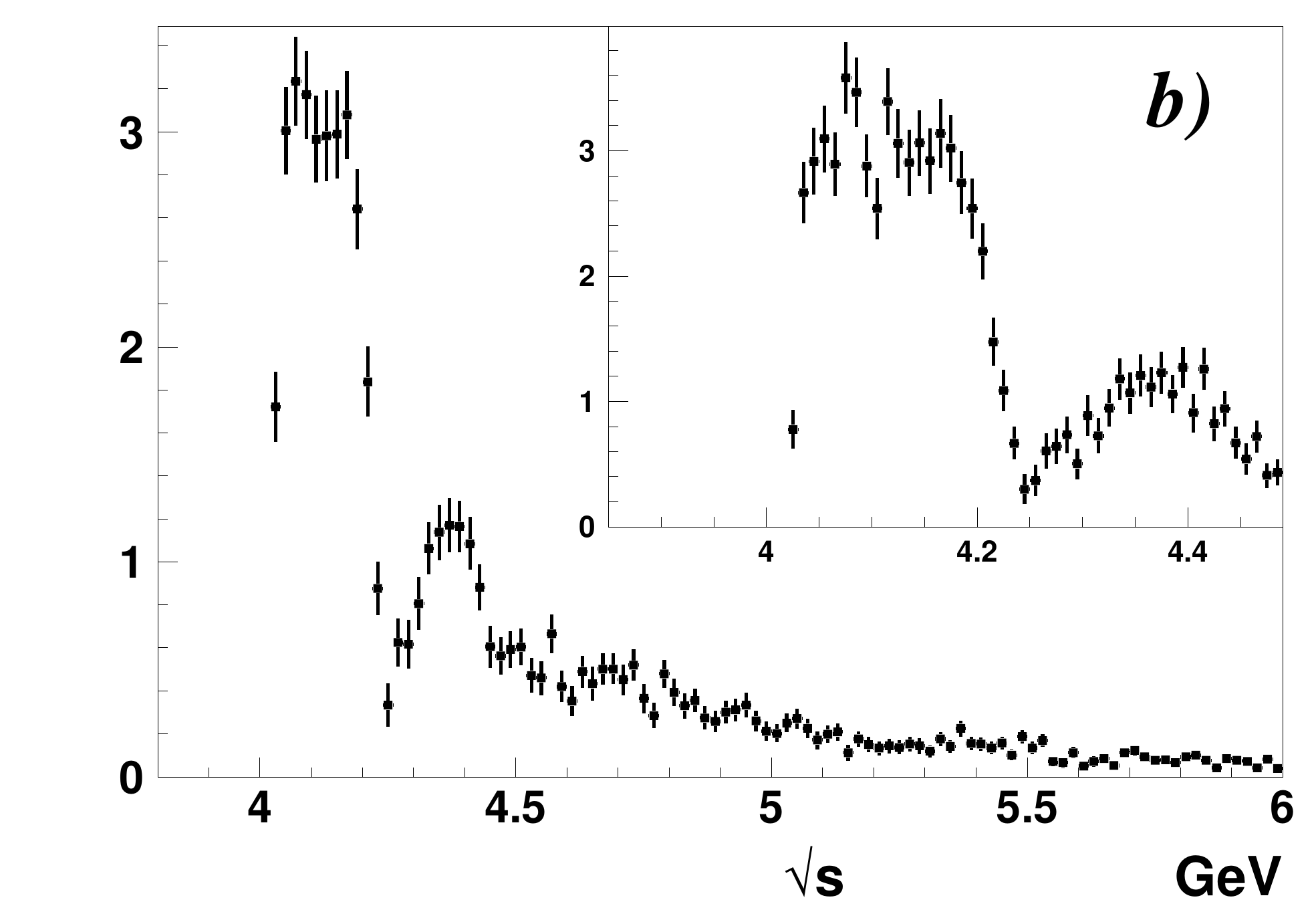}
\caption{Recent results from the Belle Collaboration~\cite{Zhukova:2017pen}
on the $e^+e^-$ annihilation into the $D\bar{D}^*$~(a) and
$D^*\bar{D}^*$~(b) final states
via ISR using a 951~fb$^{-1}$ data sample.}
\label{fig:DDstDstDst}
\end{figure*}

Among the exotic $XYZ$ states the most prominent candidate for a hybrid is the $Y(4230)$
[the literature often refers to $\psi(4260)$ aka $Y(4260)$, however, the most recent BESIII data shows that
the mass of the state needs to be lower --- compare the discussion
in Sec.~\ref{Sect:3.1.2}], as it shows some features expected for a charmonium hybrid.
To begin with, it has a mass~\cite{Tanabashi:2018oca},
\be
M=(4230\pm 8)~\mbox{MeV}\,,
\label{M4260}
\ee
close to the theoretical predictions discussed above --- see Table~\ref{tab:cchybryd}
and the discussion in Sec.~\ref{Sect:4.2.3bis}, in particular Fig.~\ref{fig:exp101016} ---
and, what is more important, there are indications that it has a decay pattern
that is not typical of conventional quarkonia but specific of hybrids.
Indeed, the ratio $R=\sigma(e^+e^-\to\mbox{hadrons})/\sigma(e^+e^-\to\mu^+\mu^-)$ has a dip in the energy region around the mass 4230 MeV.
Besides, measurements of the electronic width of the $Y(4230)$ give small values~\cite{Tanabashi:2018oca} in
agreement with the expectations for hybrids.
Furthermore, one can see a pronounced dip in the cross sections of the
$e^+e^-$ annihilation into open-charm final states --- see,
for example, Fig.~\ref{fig:DDstDstDst} for the most recent data from the
Belle Collaboration~\cite{Zhukova:2017pen}.
Also, the expectation, discussed above, that the open-flavor decays of the
charmonium hybrid should populate the three-body $D^*\bar{D}^{(*)}\pi$
final states is in line with the recent BESIII data --- see Fig.~\ref{fig_cross}.
Thus, the $Y(4230)$ could be a hybrid charmonium with a
spin 1~\cite{Close:2005iz,Berwein:2015vca} or spin 0~\cite{Kou:2005gt,Kalashnikova:2008qr} $c\bar{c}$ core, or a mixture of both.

It should be mentioned, however, that there exist alternative
explanations for the established properties of the $Y(4230)$.
In particular, nontrivial structures in the shape of the inclusive and
exclusive cross section can be explained by unitarisation effects --- see, for example, Ref.~\cite{Uglov:2016orr}.
Also, the proximity of the $Y(4230)$ mass to the $D\bar{D}_1$
threshold hints to a large admixture of the molecular component in the wave function.
Then, the experimentally observed enhancement in the three-body final state
$D\bar{D}^*\pi$ just above the c.m. energy of $4.2$~GeV --- see Fig.~\ref{fig_cross} ---
is naturally explained by the dominating decay chain for the molecule, $Y(4230)\to D\bar{D}_1\to D\bar{D}^*\pi$.
A detailed discussion of the experimental situation is given in Sec.~\ref{Sect:3.1.2}, and further details of
the molecular model for the $Y(4230)$ are collected in Sec.~\ref{Sect:4.1.5}.
Finally, it is possible that a realistic picture of the $Y(4230)$ might require that both hybrid and quarkonium (short-range),
and molecular (long-range) components coexist in the wave function.

\subsubsection{Hadroquarkonium}
\label{Sect:4.1.3}
Triggered by the experimental fact that all the candidates for heavy exotic
states were discovered in final states that contain a quarkonium and
light quarks, in Ref.~\cite{Voloshin:2007dx} it was proposed that those states
could be viewed as a compact quarkonium core surrounded by an excited light-quark cloud.
Later the idea was worked out in depth in Ref.~\cite{Dubynskiy:2008mq}.
This structure naturally explains that
in the decay the compact core simply stays intact while the light-quark cloud gets
radiated off in terms of multipion states.
The core and the cloud are held together by the QCD analogue of the
Van der Waals force known from molecular physics (for recent studies, see~\cite{Brambilla:2017ffe,TarrusCastella:2018php}).
In Ref.~\cite{Dubynskiy:2008mq} this picture is applied to the
$\psi(4260)$ aka $Y(4260)$~\footnote{Although there is now striking evidence that
the $Y(4260)$ is to be identified with the $Y(4230)$, in this paragraph we stick to the old mass value
to keep closer contact to the connected publications.},
$\psi(4360)$ aka $Y(4360)$, $\psi(4660)$ aka $Y(4660)$ and $Z_c(4430)$.
In Ref.~\cite{Voloshin:2013dpa} the hadrocharmonium picture is contrasted with the
molecular picture for the $Z_c(3900)$. This mechanism provides a
natural explanation why states like $Y(4260)$ do not decay
to $D^{(*)}\bar D^{(*)}$ as the lower-lying
vector states. Note that only the latter decay pattern is in line with the
predictions from the quark model. The hadroquarkonium picture makes distinct
predictions for both decay patterns and spin-partner states.
Because of the heavy-quark symmetry, states with the same light-quark cloud, but the heavy-quark spin coupled differently,
should be degenerate, up to terms suppressed by powers of $\Lambda_{\rm QCD}/m_h \ll 1$, where $m_h$ denotes either the charm or the bottom mass.
This also implies that the spin of the heavy quarks is conserved in decays.
For a more detailed discussion of the heavy-quark symmetry as well as the symmetry breaking terms we refer to Sec.~\ref{Sect:4.2.1}.

The effective Hamiltonian underlying the concept of hadroquarkonium is given by
the leading term of the QCD multipole expansion:
\begin{equation}
H_{\rm eff}= -\frac12 \alpha^{(\bar QQ)} E_i^a E_i^a \ ,
\end{equation}
where $\vec E^a$ is the chromoelectric field with color index $a$ and $\alpha^{(\bar QQ)}$
is the chromo-electric polarizability of the $\bar QQ$ state. One may write
\begin{equation}
\alpha^{(\bar QQ)}=\frac1{16}\langle (\bar QQ)|\xi^a r_i {\cal G} r_i \xi^a| (\bar QQ)\rangle \ ,
\end{equation}
where $\xi^a$ denotes the difference of the color generators acting on the quark and the antiquark, and $r_i$ denotes
their relative position. The Green's function $\cal G$ describes the propagation of the heavy quark
pair in the color octet.
As pointed out in Ref.~\cite{Dubynskiy:2008mq} this structure suggests that
the most likely hadroquarkonia should emerge for higher charmonium resonances such as $\psi(2S)$ and $\chi_{cJ}$
in conjunction with some excited hight quark cloud.

To investigate the phenomenological implications of the Hamiltonian quoted above,
we start with discussing the properties of $Z_c(3900)$ that would emerge if
this state were a hadrocharmonium, in particular a $J/\psi$ core surrounded by a light-quark cloud.
Here we largely follow Ref.~\cite{Voloshin:2013dpa}.
In this picture the $Z_c(4430)$ seen in the $\psi(2S)\pi$ final state can be
viewed as a radial excitation of the $Z_c(3900)$ in the same way as the $\psi(2S)$ is the first excited state of the $J/\psi$.
Spin interactions scale with the inverse of the heavy-quark mass such that the spin of heavy quarks is
conserved at least in the limit of infinitely heavy quarks.
For the system at hand this implies that the $Z_c(3900)$ should be found
only in final states with heavy-quark spin equal to 1.
This is in conflict with the presence of a $Z_c(3900)$ signal in the
$\pi h_c$ final state, see inset in the left panel of Fig.~\ref{1Dfit}.
Moreover, in this picture one should expect the decay of the $Z_c(3900)$ into
open-flavor two-meson states to be suppressed
compared to the decay into $\pi J/\psi$. However, this is also not the case.
Thus, one needs to conclude that the $Z_c(3900)$ does not qualify as a state
with a dominant $\pi J/\psi$ component.

\begin{figure}[t!]
\begin{center}
 \includegraphics[width=0.7\linewidth]{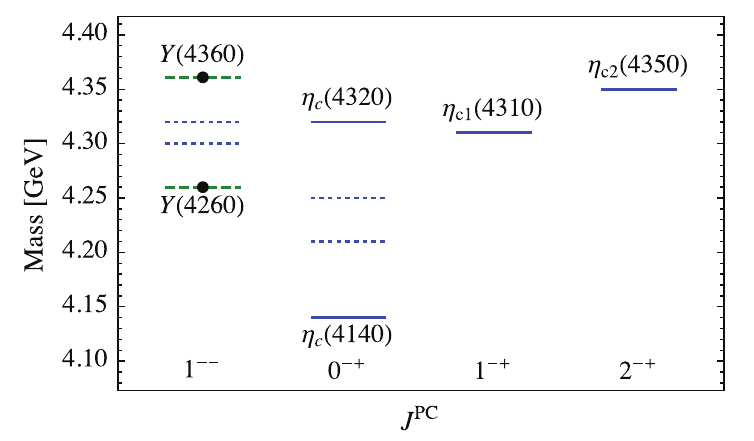}
\caption{Family of spin partners of the $\psi(4260)$ aka $Y(4260)$ and
$\psi(4360)$ aka $Y(4360)$ that should
exist~\cite{Cleven:2015era}, if both states qualify
as mixed hadrocharmonia as proposed in Ref.~\cite{Li:2013ssa}. The short
dashed lines show the seed states and the (green) long dashed lines show the
masses used as input. The (blue) solid lines then
show the predictions that emerge from imposing spin symmetry.}
\label{fig:hadrocharm_spinpartners}
\end{center}
\end{figure}

The second example we want to discuss is the $Y(4260)$
and $Y(4360)$.
Originally they were proposed to be
hadrocharmonium states with a scalar-isoscalar cloud and a $J/\psi$ and
$\psi(2S)$ core, respectively~\cite{Dubynskiy:2008mq}, since they were
observed in the corresponding $c {\bar c} 2\pi$ final states.
A model calculation performed in Ref.~\cite{MartinezTorres:2009xb},
where a state near the mass of the $Y(4260)$ emerges from
non-perturbative $f_0(980) J/\psi$ interactions\footnote{In this work, the
system is treated as a three-body system with the $\pi\pi-K\bar K$ system interacting with the $J/\psi$.}
supports this picture.
However, recently the two states were also seen in the $\pi\pi h_c$ final
state with a comparable rate~\cite{BESIII:2016adj} (c.f. Fig.~\ref{cs}),
which would imply a significant amount of spin-symmetry violation if the
picture mentioned above were right. To explain the simultaneous observation
of $Y(4260)$ and $Y(4360)$ in final states
with heavy-quark spin equal to 1
and 0, in Ref.~\cite{Li:2013ssa} it was proposed that both states are mixed
hadrochamonia from two different seeds: One that contains a $J^{PC}=1^{--}$
charmonium core [to be identified with $J/\psi$ or $\psi(2S)$ or a
mixture thereof with the latter being prominent] surrounded by a $0^{++}$
light-quark cloud and one that contains a $1^{+-}$ charmonium core [to be
identified with $h_c(1P)$] surrounded by a $0^{-+}$ light-quark cloud.
In Ref.~\cite{Li:2013ssa} it is shown that such an assignment allows one to at
least qualitatively describe the data
in the $\pi\pi J/\psi$, the $\pi\pi \psi(2S)$ as well as the $\pi\pi h_c$ final
states, if the seed states are located
at 4.30 and 4.32 GeV, respectively. While this mass difference is quite
small, it is stressed in Ref.~\cite{Li:2013ssa} that
there is nothing that prohibits such a near degeneracy.
In Ref.~\cite{Cleven:2015era} it was pointed out that the mixing scheme
just described allows one to predict four spin-partner states of
$Y(4260)$ and $Y(4360)$: One simply has
to replace the cores of the seed states by their respective spin-partner states.
Accordingly, $\psi(2S)$ needs to be replaced by $\eta_c(2S)$ giving rise
to a state with $\eta_c$ quantum numbers when being paired with the $0^{++}$
cloud and $h_c$ needs to be replaced by the triplet of
$\{\chi_{c0},\chi_{c1}, \chi_{c2}\}$ giving rise to states with quantum numbers
of $\eta_c$, $\eta_{c1}$ and $\eta_{c2}$. Naturally, the two $\eta_{c}$ states
will mix in the same way as their vector partners. The pattern that emerges
is shown in Fig.~\ref{fig:hadrocharm_spinpartners}.
As a special feature, the scheme predicts with $\eta_c(4140)$ a relatively light $\eta_c$ state
that should not decay to $D^*\bar D$ but predominantly $\pi\pi \eta_c$ and $\pi\pi\chi_{c0}$.
This highly nontrivial prediction awaits its experimental confirmation or refutation.
Note that in the molecular as well as in the compact tetraquark picture the lightest
$0^{-+}$ state is located in mass above the $Y(4260)$ as will be explained in detail in
Secs.~\ref{Sect:4.1.4} and ~\ref{Sect:4.1.5}.

The
$Y(4660)$ with quantum numbers $1^{--}$ is seen
in the $\pi\pi\psi(2S)$ final state and is therefore a candidate for a
hadrocharmonium with a $0^{++}$ light-quark cloud surrounding a
$\psi(2S)$ core~\cite{Dubynskiy:2008mq}. In addition, the mass of the
$Y(4660)$ is very close to the threshold for the production
of $f_0(980)\psi(2S)$ which led the authors of
Ref.~\cite{Guo:2008zg} to propose that the state is a $\psi(2S)-f_0(980)$
molecular state --- note that in all previous
cases the quarkonium--light-quark resonance threshold was far below
the mass of the state such that in those cases a molecular assignment cannot
be applied. The implications of spin symmetry are the same
as before: If the $Y(4660)$ has a prominent
$f_0(980)\psi(2S)$ substructure, there should exist a spin
partner, $Y_\eta$, with similar properties as the $Y(4660)$
and $f_0(980)\eta_c(2S)$ as a prominent component~\cite{Guo:2009id} ---
this should be true in both pictures if we call the state a molecule or a
hadrocharmonium, however, so far experimental searches for this state were
not successful --- see the discussion in Sec.~\ref{Sect:3.1}.
Note that the enhancement seen in $\Lambda_c\bar \Lambda_c$ at 4630 MeV
is consistent with originating from the
$Y(4660)$~\cite{Cotugno:2009ys,Guo:2010tk,Dai:2017fwx}.

Very recently, Ref.~\cite{Voloshin:2018vym} proposed that the two charged
states $Z_c(4100)$ and $Z_c(4200)$ claimed in Refs.~\cite{Aaij:2018bla} and
\cite{Chilikin:2014bkk}, respectively, are hadrocharmonia (the experimental
evidence is shown and discussed in Sec.~\ref{Sect:3.2.2}). The analysis
presented in this work is very similar to the one discussed in the previous
paragraph, only that here at least evidence for the pair of spin-partner
states is already found: The $Z_c(4100)$ is proposed to be a hadrocharmonium
with an $\eta_c$ core and the $Z_c(4200)$
to have a $J/\psi$ core --- both with a light-quark cloud with pion quantum numbers.
A typical prediction that emerges from this assignment is for the productions (around the same invariant mass as the resonances)
\begin{equation}
\frac{\Gamma [B^0\to Z_c(4100)^- K^+]}{\Gamma [B^0\to Z_c(4200)^- K^+]}\approx
\left.\frac{\Gamma [B^0\to \eta_c\pi^- K^+]}{\Gamma [B^0\to J/\psi \pi^- K^+]}\right|_{M(c\bar c\pi)\approx M(Z_c)}
\end{equation}
and for the decays
\begin{equation}
\Gamma [Z_c(4100)\to \eta_c(2S)\pi]\simeq \Gamma [Z_c(4200)\to \psi(2S)\pi] \ ,
\end{equation}
up to spin-symmetry-violating corrections. If this prediction were confirmed
experimentally, it would provide strong evidence for the mentioned
hadrocharmonium assignment of both $Z_c(4100)$ and $Z_c(4200)$.

\subsubsection{Compact tetraquarks}
\label{Sect:4.1.4}
Sometimes the term tetraquark is used generically for all states containing four constituent quarks irrespective of their clustering.
{In the older literature multi-quark states containing diquarks (a quark--quark pair) and anti-diquarks  (an antiquark--antiquark pair),
 as subclusters were sometimes called
``diquonium''~\cite{Chan:1978nk,Gavela:1978hq}.
However,
in line with the recent literature on the XYZ states, in this review, we use the term ``compact tetraquark'' to refer to
multiquark states that show this kind of clustering.} It should be stressed that
 so far no mechanism is identified in QCD that would lead to such clusters --- this issue is discussed a little
more in the following paragraph; for a more general discussion of four-quark states we refer to Ref.~\cite{Richard:2018yrm} and references
therein.
The first microscopic model working with diquark degrees of freedom was
presented by Jaffe in 1976~\cite{Jaffe:1976ig} at that time focussing on light quarks.
For a pedagogical introduction to the subject of diquark clustering we refer to Ref.~\cite{Selem:2006nd} and for a recent book see Ref.~\cite{Ali:2019roi}.
Heavy-light diquarks were first discussed in Ref.~\cite{Maiani:2004vq}. A comprehensive discussion of further developments of these ideas and their application to the spectrum of the $XYZ$ states can be found in a recent review \cite{Esposito:2016noz}.
In this review we will focus on the results that were later developed on the basis of this work.
Results for compact tetraquarks obtained in the quasipotential quark model described
in Sec.~\ref{Sect:4.1.1} can be found in Refs.~\cite{Ebert:2005nc,Ebert:2008se,Ebert:2007rn,Ebert:2008kb,Ebert:2010af}.

A pair of quarks can be either in the $\mathbf{[\bar 3]}$ or in the
$\mathbf{[6]}$ representation of the color group.
There is phenomenological evidence that the former configuration is more
tightly bound --- for example, the one-gluon exchange is attractive in the $\mathbf{[\bar 3]}$ but repulsive
in the $\mathbf{[6]}$. Accordingly, the latter is neglected in many works,
although the $\mathbf{[6]}$ may play an important role considering the findings of, e.g., Refs.~\cite{Ader:1981db,Czarnecki:2017vco}
for multiheavy systems, which put into question the emergence of
compact diquarks as building blocks of compact hadrons.

\begin{figure}[t!]
\begin{center}
 \includegraphics[width=0.4\linewidth]{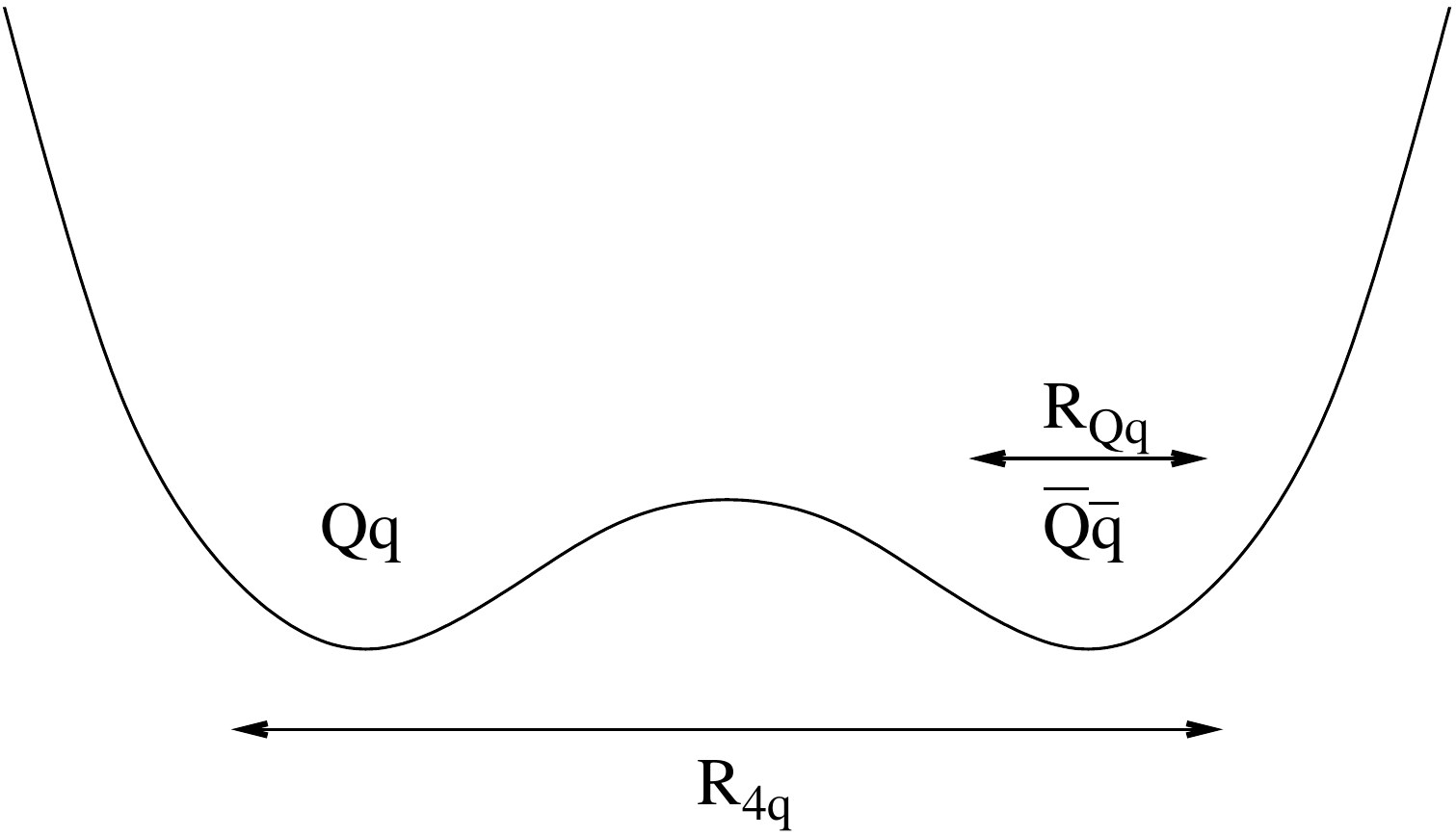}
\caption{Proposed potential between a diquark and an antidiquark forming a
compact tetraquark~\cite{Maiani:2017kyi}. Here $R_{Qq}$
denotes the size of the diquark while $R_{4q}$ stands for the size of the
whole tetraquark.}
\label{fig:tetraquarkpot}
\end{center}
\end{figure}

Usually it is found that the spin-zero diquarks are more
tightly bound in both the light-quark
sector~\cite{Jaffe:1976ig}\footnote{Nevertheless also the spin-1 diquarks are
found to contribute considerably, e.g., to the binding of light
baryons~\cite{Ishii:1993rt,Hanhart:1995tc}.}
as well as in the heavy-quark sector~\cite{Ali:2017wsf}.
In the latter case the two spin configurations
form a spin doublet in analogy to, e.g., $D$ and $D^*$ or $B$ and $B^*$. It
is therefore reasonable that the mass difference between spin-1 and spin-0
$cq$-diquarks, where $c$ ($q$) denotes the
charm (a light) quark, is within 120 MeV~\cite{Maiani:2014aja}, fixed to the
mass difference of $Z_c(3900)$ and $Z_c(4020)$, found to be of the order of the $D$-$D^*$ mass difference.
In Ref.~\cite{Maiani:2014aja} it is argued that this phenomenology follows from the condition that the spin-spin interaction
is operative predominantly inside the diquarks (driving the mentioned mass difference) and not between the light quarks as assumed originally~\cite{Maiani:2004vq}.
For this to happen it appears necessary that the diquark and the antidiquark do not come too close.
Therefore in Ref.~\cite{Maiani:2017kyi} an idea put forward in
Ref.~\cite{Selem:2006nd} was adapted to quarkonium-like states, namely that
they are sitting in distinct wells of a double-well potential
as shown in Fig.~\ref{fig:tetraquarkpot}. The emerging structure is
characterized by two length scales, the diquark radius, $R_{Qq}$, as well as
the tetraquark radius, $R_{4q}$. In Ref.~\cite{Maiani:2017kyi}
it is claimed that a sensible phenomenology emerges when $R_{Qq}\ll R_{4q}$ --- as possible values $R_{Qq}\sim 0.7$ fm and $R_{4q}\sim 2$ fm are advocated.
Besides providing a reason for the hierarchy of the spin interactions, the potential of Fig.~\ref{fig:tetraquarkpot} can also provide a reason
why exotics typically prefer to decay into pairs of open-flavor mesons instead of a quarkonium
and a light meson as will be discussed in a little more detail below. So far, however, no mechanism was proposed
that could explain the emergence of a potential of the type of Fig.~\ref{fig:tetraquarkpot}. Its existence
emerges as a phenomenological necessity to justify the diquark picture. An alternative
approach trying to justify the emergence of compact diquark building blocks
in compact tetraquarks was proposed in Ref.~\cite{Brodsky:2014xia} and applied in Refs.~\cite{Lebed:2017min}.
However, since the emerging phenomenology of the two approaches is similar, we are reviewing here the research focussed around the work \cite{Maiani:2017kyi} by Maiani et al.

The assumption that the spin-spin interactions are active only within the diquarks and not between the light quarks influences significantly the decays of compact tetraquarks.
In particular, since the heavy-quark spins within the compact tetraquarks turn out not to be correlated with each other,
decays into both spin-1 and spin-0 quarkonia are allowed~\cite{Maiani:2014aja} --- in line with observations.
The same mechanism is at work in the molecular model described in the next section.

Furthermore it turns out that a spin-tensor force needs to be included into the Hamiltonian to avoid the appearance of the light high-spin states found in Ref.~\cite{Cleven:2015era}.
The Hamiltonian then reads~\cite{Ali:2017wsf}:
\begin{equation}
 2m_{\cal Q}+\frac{B_{\cal Q}}{2}\bm{L}^2+2a_Y\bm{L}\cdot \bm{S}+\frac{b_Y}{4}S_{12}+2\kappa_{Qq}\left(\bm{S_q}\cdot \bm{S_Q} + \bm{S_{\bar q}}\cdot \bm{S_{\bar Q}}\right) \ ,
\label{eq:tetraquarkM}
\end{equation}
where $S_{12}$ is the tensor operator $S_{12}/4 = 3({\bm{S}_1}\cdot \hat{{\bm{r}}})({{\bm{S}}}_2\cdot \hat{{\bm{r}}})-({{\bm{S}}}_1\cdot {{\bm{S}}}_2)$,
${{\bm{S}}}={{\bm{S}}}_1+{{\bm{S}}}_2$, with ${{\bm{S}}}_1$ and ${{\bm{S}}}_2$ the spins of the diquark and the antidiquark, respectively,
$\hat{\bm{r}}={\bm{r}}/r$ is a unit vector pointing along the radius vector, $Q$ is the heavy quark flavor and $q$ the light quark one.
The tensor operator gives a nonvanishing contribution only for states with $S_2=S_1=1$.
Moreover, the terms that contain the angular momentum operator contribute only for $L\neq 0$.

The parameters in Eq.~(\ref{eq:tetraquarkM}) are fitted to the data.
To do so, one has to establish which observed exotic candidate (see Sec.~\ref{Sect:3}) should be identified with which compact tetraquark.
The positive-parity compact tetraquarks that emerge from the picture sketched above are discussed in some depth in Ref.~\cite{Maiani:2014aja}.
In particular, using the mass difference of $Z_c(3900)$ and $Z_c(4020)$ as input allows the authors to fix $\kappa_{Qq}=67$ MeV.
In this paper, also states with $0^{++}$ and $2^{++}$ are predicted, however, those are almost 100 MeV away from the states observed with these quantum numbers.

\begin{figure}[t!]
\begin{center}
 \includegraphics[width=0.7\linewidth]{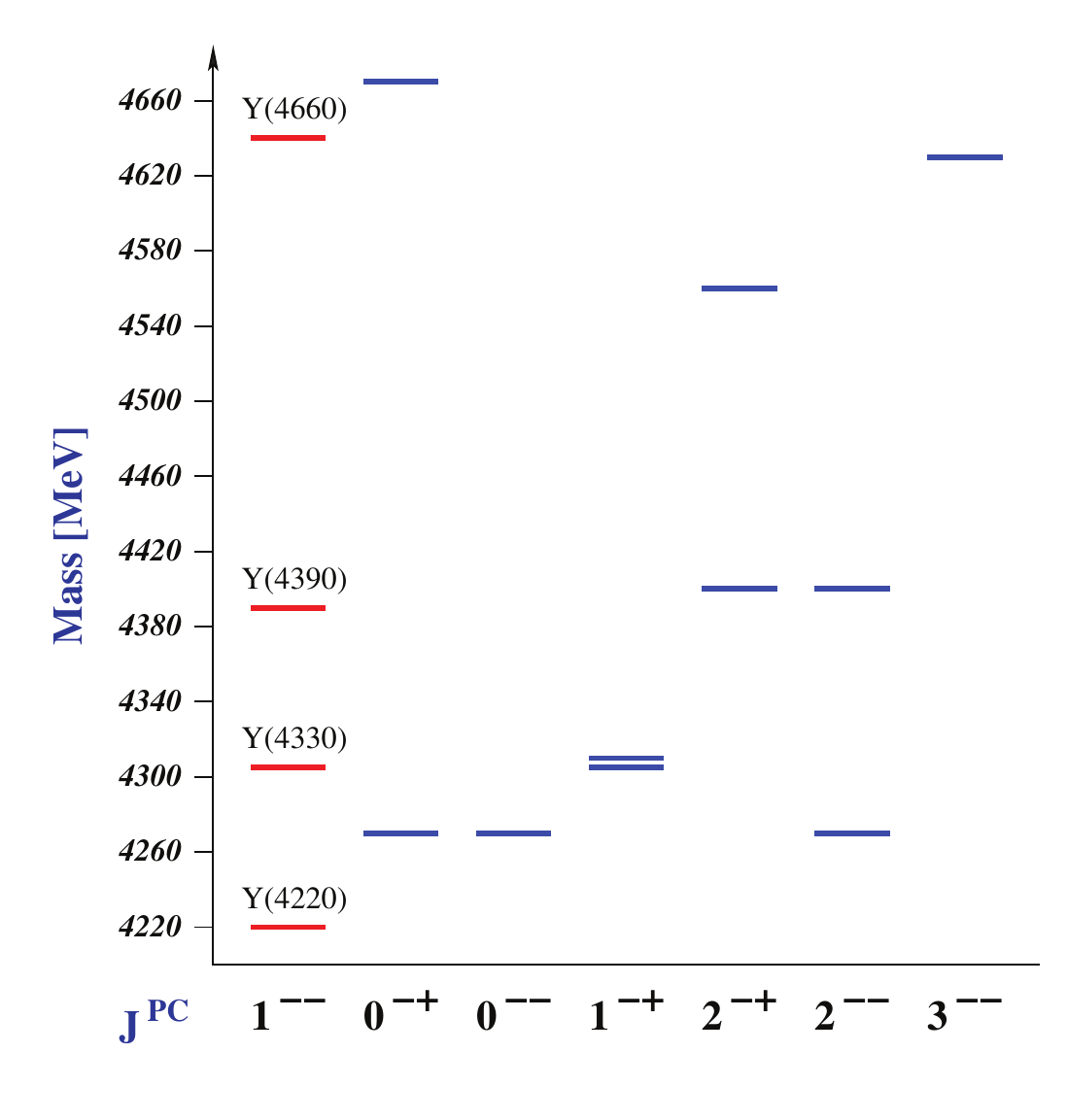}
\caption{Spectrum of negative-parity compact tetraquarks according to the preferred
solution of Ref.~\cite{Ali:2017wsf}. Note that $0^{--}$ and $1^{-+}$ are exotic quantum numbers
that cannot be generated for $Q\bar Q$ states.}
\label{fig:tetraquarkspec}
\end{center}
\end{figure}

Very interesting are the negative-parity states.
In the compact tetraquark model those arise from two diquarks in a $P$-wave that have their constituents in a relative $S$-wave.
Since the diquarks have either spin 1 or spin 0, in total four ground states are possible with $J^{PC}=1^{--}$,
namely $[0,0]_0$, $[1,0]_1$, $[1,1]_0$, and $[1,1]_2$, where the diquark spins are given in the bracket and the
subscript denotes the total spin that resulted from their coupling.
Note that $[1,1]_1$ would produce the wrong $C$-parity.
The preferred assignments of Ref.~\cite{Ali:2017wsf} for those four states
are\footnote{We use here the names currently adopted by
the Particle Data Group instead of those quoted in the original
experimental works and in Ref.~\cite{Ali:2017wsf}.}
$Y(4220)$, $Y(4330)$,
 $Y(4390)$, and $Y(4660)$~\footnote{As in the previous
section also here we stick to the mass values used in the original publication although there
is evidence that $Y(4330)$ and $Y(4390)$ could be realisations of $Y(4360)$.},
since this pattern is consistent with the spin-spin interaction determined in
the even-parity sector as well as an analysis of $\Omega_c$ states~\cite{Ali:2017wsf}.
This is in contrast with the analysis of Ref.~\cite{Maiani:2014aja}, where the $Y(4008)$
is taken as a genuine structure, which is not confirmed by the recent BESIII data, although
the existence of a very broad state below 4.2 GeV can not be excluded (see Sec.~\ref{Sect:3.1.2}).
As will be explained below, the hadronic-molecule picture for the
negative-parity states allows for only
three ground states --- thus a proper mapping of the singularity structure
of the $S$--matrix should allow us to eventually decide how QCD structures
its non-$Q\bar Q$ heavy mesons. Moreover, since now all
parameters of Eq.~(\ref{eq:tetraquarkM}) are fixed, it is straightforward
to make predictions for negative-parity states with $J\neq 1$. The resulting
spectrum is shown in Fig.~\ref{fig:tetraquarkspec}.
Clearly, Eq.~(\ref{eq:tetraquarkM}) leads to a very rich and specific
spectrum that contains both quantum numbers
that are allowed in the quark model and those that are not, like
$0^{--}$ or $1^{-+}$. This spectrum will be compared
to what is expected for molecular states in the next section.

The compact tetraquark model is also capable of explaining some unexpected decay
chains. For example, in the coupling scheme outlined above both the
$Y(4360)$ as well as $X(3872)$ have
the same heavy-quark spin structure
$\sim (\bar s=0, s=1) + (\bar s=1,s=0)$, where $\bar s$ and $s$ denote the
spin of the antidiquark and the diquark, respectively,
and thus a transition $Y(4360)\to \gamma X(3872)$ appears naturally as electric
dipole transition~\cite{Ali:2017wsf}. On the other hand, this structure implies
that the $Y(4360)$ has purely heavy-quark spin 1 and as such
should not appear in the $\pi\pi h_c$ final state. Present data does not yet
allow one to decide how many states there are in the mass region $4200-4400$~MeV with quantum numbers
$1^{--}$ (see Sec.~\ref{Sect:3.1.2}), however, upcoming data with improved
statistics should eventually allow one to conclude
whether the pattern presented is realised in nature.

Before closing this section, a few comments should be made about the decays of the compact tetraquarks.
As outlined above, the candidates for exotic states typically decay to both
open-flavor channels as well as hidden-flavor
channels --- e.g. $X(3872)$ is seen in $D^0\bar D^{*0}$ and amongst others in $\pi\pi J/\psi$, $3\pi J/\psi$,
$Z_c(3900)$ is seen in $D\bar D^*$ as well as in $\pi J/\psi$,
$Y(4230)$ is seen in $D\bar D^*\pi$ as well as in $\pi\pi J/\psi$,
$Z_b(10610)$ is seen in $B\bar B^*$ as well as in $\pi\Upsilon(nS)$,
and $\pi h_b(mP)$ and $Z_b(10650)$ is seen in $B^*\bar B^*$ as well as in $\pi\Upsilon(nS)$ and $\pi h_b(mP)$.
Common to all these examples is that, although the open-flavor channels have a lot less
phase space available, still the states decay predominantly into those. This is commonly taken as a
strong indication for a molecular nature of these states, however, it may also be understood from the
proposed diquark--antidiquark potential shown in Fig.~\ref{fig:tetraquarkpot},
in which case a constituent of a diquark must tunnel through the barrier to enable the decay.
Since the tunnelling for heavy quarks is exponentially suppressed compared to that of the light quarks,
the pattern described appears naturally~\cite{Maiani:2017kyi}.

In Ref.~\cite{Chen:2019mgp} it was found from an analysis of the $\pi\pi$
spectra from $e^+e^-\to \pi^+\pi^- J/\psi$ in the 4260~MeV mass
range, that the source of the pion pair must contain a sizeable flavor-octet component.
This observation is in line with a prominent molecular or compact tetraquark interpretation of the vector states in this mass
range but poses a problem for their conventional $c\bar c$ or hybrid interpretation.
The results of Ref.~\cite{Chen:2019mgp} could be confirmed if data became available also for $e^+e^-\to K {\bar K} J/\psi $.

An interesting ansatz to study compact tetraquarks is via the large $N_c$ expansion first introduced by 't Hooft~\cite{tHooft:1973alw}.
Under the assumptions that confinement survives the large $N_c$ limit and that
the strong coupling constant scales
as $1/\sqrt{N_c}$ which provides a smooth limit for $N_c\to\infty$,
rigorous conclusions about features of the QCD spectrum become accessible ---
e.g. in this limit there is an infinite tower of stable quark-antiquark pairs generated from an infinite sum of planar QCD diagrams.
Until a few years ago there was the belief that all configurations that contain four or more quarks and survive the large $N_c$ limit
are just given by noninteracting multimeson states~\cite{Coleman:1985rnk}.
However, in 2013 this was questioned in Ref.~\cite{Weinberg:2013cfa}: There it was argued that $if$ there are compact tetraquark
poles present in the connected piece of the amplitude, they will also survive the large $N_c$ limit, in particular
with a width that decreases as $N_c$ increases. This work initiated a series of publications --- see, e.g., Refs.~\cite{Knecht:2013yqa,Lebed:2013aka}.
In all those works the focus is still on planar diagrams.
In contrast to this, Refs.~\cite{Maiani:2016hxw,Maiani:2018pef} argue
that non--planar diagrams should be favored for providing the origin of very narrow compact tetraquarks.
A recent critical reflection on the subject is in Refs.~\cite{Lucha:2017gqq,Lucha:2018dzq,Sazdjian:2018rpt}.

Recently the interest into $QQ\bar q\bar q$ tetraquarks, first studied in Ref.~\cite{Ader:1981db}, has revived.
There exist now studies from QCD sum rules~\cite{Du:2012wp} (see also Sec.~\ref{Sect:4.sumrules}), lattice QCD~\cite{Bicudo:2015kna,Francis:2016hui,Bicudo:2016ooe} (see also Sec.~\ref{Sect:4.3})
as well as phenomenology~\cite{Cui:2006mp,Luo:2017eub,Karliner:2017qjm,Eichten:2017ffp}.
Especially the last works employ the observation of doubly-heavy baryons to make predictions for doubly-heavy tetraquarks.
The connection between these systems might be most compactly illustrated by the mass formula~\cite{Eichten:2017ffp}
\begin{equation}
{m(QQ\bar q\bar q) - m(QQ q) \approx m(\bar Q\bar q\bar q)-m(\bar Q q)} \ ,
\label{doublyheavy}
\end{equation}
which is based on the quark-diquark symmetry, see also Sec.~\ref{Sect:4.2.1}.
This symmetry is realized in nature, if heavy diquarks form compact substructures in hadrons, for this would allow
one to perform a systematic expansion in $r_{QQ}/r_{qq}$, where $r_{qq}$ denotes the size of the light-quark cloud that
may be estimated as $1/\Lambda_{\rm QCD}$.
If $r_{QQ}/r_{qq}$ is a small parameter, one may safely assume the $QQ$
diquark to be in a color--antitriplet configuration, since for heavy quarks the $QQ$ interaction should be largely
governed by the one-gluon exchange, which is attractive only in this channel.
Then Eq.~(\ref{doublyheavy}) follows naturally.
After the recent observation of the first doubly-heavy baryon~\cite{Aaij:2017ueg},
Eq.~(\ref{doublyheavy}) might be also used the other way around
to argue that if this (approximate) equality is not realized in nature,
then QCD does not favor doubly-heavy compact diquarks.
Therefore the experimental search for tetraquark structures should be performed with high priority at, e.g. Belle II and LHCb.
In the bottomonium sector, the studies mentioned above find typically a deeply bound $bb\bar u\bar d$
system with $J^P=1^+$ 100-200 MeV below the $B^-B^{* 0}$ threshold.
The issue of multiquarks with two heavy quarks (in contrast to a heavy quark
and its antiquark) is picked up again in Sec.~\ref{Sect:4.2.1}
from the effective field theory perspective.

\subsubsection{Hadronic molecules}
\label{Sect:4.1.5}

Hadronic molecules are compound states of two hadrons. For a recent review
on the subject we refer to Ref.~\cite{Guo:2017jvc}.
The concept is nothing else but a generalisation of nuclei
to systems of mesons. Accordingly one may
derive some properties to be expected for hadronic molecules from those of
light nuclei --- because of this we start this section with a short review of few-nucleon systems.

It is widely accepted that nuclei are bound states of protons and neutrons.
The lightest nucleus is the deuteron with a binding energy of $E_b({\rm deuteron})=2.22$ MeV,
where $E_b(X)=m_1+m_2-M_X$, with $m_i$ and $M_X$ denoting the masses of the constituents and the mass of the state $X$, respectively.
An important quantity for hadronic molecules is the so-called binding momentum,
\begin{equation}
\gamma = \sqrt{2\mu E_b},\quad \mu=m_1m_2/(m_1+m_2),
\end{equation}
which defines the size of the molecule since the large distance behaviour of a molecular state scales as $\exp(-\gamma r)$.
At the same time $\gamma$ is a measure of the typical momentum within the bound state.
In the case of the deuteron one finds $\gamma=45.7$ MeV.
In other nuclear systems the corresponding number can be as small as 13 MeV --- the $\Lambda$
separation energy of hypertriton is only $E_b=130$~keV --- and as large as
the almost 200 MeV that is necessary to kick a single nucleon our of $^4$He.
Accordingly we should expect a similar range of binding momenta also for bound states of two mesons.

It should be noted, however, that while the analogy between nuclei and
mesonic molecules is apparent within, e.g., potential models, it does not
survive the large $N_c$ limit: While nuclei should continue to exist in the large $N_c$ limit~\cite{Coleman:1985rnk},
since a baryon contains $N_c$ quarks and this adds to the multiplicity of the possible diagrams,
mesonic molecules will most probably not~\cite{Pelaez:2015qba}.

In general, the internucleon interaction is the strongest in the $S$-waves,
since $S$-waves do not have a centrifugal barrier (see, for example, the discussion in the beginning of section~\ref{sect:4}).
Indeed, the deepest nuclear bound states that appear (if allowed by the Pauli principle) are $S$-wave ones. Therefore,
in this section we will mainly focus on mesonic systems in the $S$-wave. However,
also $P$-wave bound states are possible. For example\footnote{C.H. is grateful to Andreas
Nogga for providing him with valuable information on nuclear level schemes.}: $^6$Li is located
1.47~MeV below the deuteron-$^4$He threshold. It may thus be viewed as a
deuteron-$^4$He bound state. The first negative-parity excitation
of $^6$Li is located 18~MeV above the ground state~\cite{Tilley:2002vg} ---
a number of the order of the energy difference between different nuclear shells. Having this said,
it is certainly not straightforward to learn something about possible $P$-wave states in mesonic systems from their appearance in nuclei.
We come back to the proposed $P$-wave hadronic molecules in the $XYZ$-family near the end of this section.

That heavy mesons should form bound states was proposed in Ref.~\cite{Voloshin:1976ap} already in 1976.
The assumed binding mechanism was the exchange of vector mesons. Later,
in Ref.~\cite{Tornqvist:1993ng} similar conclusions were drawn from calculations using one-pion exchange for the potential.
In particular, both calculations predicted that, for example, the
$D\bar D^*$ system, where the $X(3872)$ was found in 2003, should bind.

The role of one-pion exchange to molecular binding is widely discussed in the literature.
It was argued in Ref.~\cite{Suzuki:2005ha} that the pion cut, if kinematically accessible, should be kept in the calculations.
This issue is further elaborated in Ref.~\cite{Baru:2011rs}.
In Ref.~\cite{Baru:2015nea}, it is stressed that the one-pion exchange is well
defined only when being accompanied with a local counterterm.
Thus it is fair to state that at present
little is known about the binding potential of hadronic molecules,
if they exist\footnote{Note that there are indications
that one-pion exchange leaves some imprint in the line shapes of the
$Z_b$ states~\cite{Wang:2018jlv} as is detailed below.}.
However, as it will become clear below, many statements can already be made without detailed knowledge of the potential.

There is an intense discussion in the literature whether a bound state
as shallow as the $X(3872)$ could be produced
copiously in reactions with a large momentum transfer. One can find works
that conclude from the observed
rates that the $X(3872)$ cannot be a shallow bound
state~\cite{Suzuki:2005ha,Bignamini:2009sk,Esposito:2015fsa,Esposito:2017qef},
while others argue that as soon as final-state interactions are taken
into account and proper momentum ranges are considered, there is no contradiction between a
molecular nature of the $X(3872)$ and its production
rates~\cite{Artoisenet:2009wk,Artoisenet:2010uu,Guo:2013ufa,Albaladejo:2017blx,Wang:2017gay,Braaten:2018eov}.
In this review we only state that at present there is no full understanding
of the production rates of shallow molecular states.

Implications of the molecular scenario are contrasted with different quark model approaches in Ref.~\cite{Hanhart:2019isz}.

\vspace{0.3cm}\noindent
$\bullet$ {\it The Weinberg criterion and its implications}
\vspace{0.3cm}

In order to discuss the implications of the molecular nature for a given state, we first need to define the notion of a molecule.
The most popular definition goes back to Weinberg who set up a scheme that allowed him to quantify the molecular component of the deuteron~\cite{Weinberg:1965zz}.
In particular, in his work Weinberg showed that the residue of a bound-state pole located close to a threshold
could be written as\footnote{A simple straightforward derivation of this relation in a somewhat
less general form was obtained eralier by L. Landau \cite{Landaumolecule}.}, using the notation of Ref.~\cite{Guo:2017jvc},
\begin{equation}
\frac{g_{\rm eff}^2}{4\pi}= 4M^2\left(\frac{\gamma}{\mu}\right)(1-\lambda^2),
\label{eq:geffdef}
\end{equation}
where $\lambda^2$ denotes the probability to find a non-molecular component in the bound-state
wave function, $\mu$ is the reduced mass of the nearby two-hadron channel, and $M$ is the mass
of the bound state. Weinberg also showed that $\lambda^2$ could be identified
with the wave function renormalisation constant.
In particular, the effective coupling of the bound state to the continuum gets
maximal when the state is a pure molecule ($\lambda^2=0$).
Since, for a shallow bound state, the low-energy scattering amplitude
parameterised, for example,
in terms of the scattering length and the effective range can be expressed
in terms of $g_{\rm eff}$, this coupling can be regarded an observable.

It is important to keep in mind, when it comes to identifying hadronic molecules, that only those observables
are useful that are sensitive to the effective coupling defined in Eq.~(\ref{eq:geffdef}).
If, for example, a reaction is sensitive to the short-ranged part of the wave function, no
statement about the molecular admixture of the given state is possible.
This observation, relevant for both the production of $\xx$ in large momentum transfer
reactions as well as its radiative decays, will be discussed in more detail in Sec.~\ref{Sect:4.2.4}.

Equation (\ref{eq:geffdef}) acquires corrections of the order of
$(\gamma/\beta)$, where $\beta$ is the mass scale
of either the next higher channel or the inverse of the range of forces.
Accordingly, the value of $g_{\rm eff}$ can only be used for shallow bound states.
However, starting from Eq.~(\ref{eq:geffdef}) it appears justified to assume that a state with $\lambda^2\to 0$ is characterised
by a large effective coupling to the two-hadron channel that forms the molecule.
As will be illustrated below, the prominence of these two-hadron channels
has important phenomenological implications, even in cases where the
quantitative connection of Eq.~(\ref{eq:geffdef}) is lost because of a too large binding momentum $\gamma$.
Weinberg's argument was generalised to resonances in
Refs.~\cite{Baru:2003qq,Sekihara:2014kya}.
Coupled channels are discussed in Ref.~\cite{Gamermann:2009uq}.
Some related ideas for inferring that a resonance in coupled-channel scattering is dominated by
molecular configurations, based on the presence and position of poles on
unphysical sheets, are discussed in Refs.~\cite{Morgan:1992ge,Morgan:1993td,Baru:2004xg}.

\begin{figure}[t!]
\begin{center}
\includegraphics[width=0.4\linewidth]{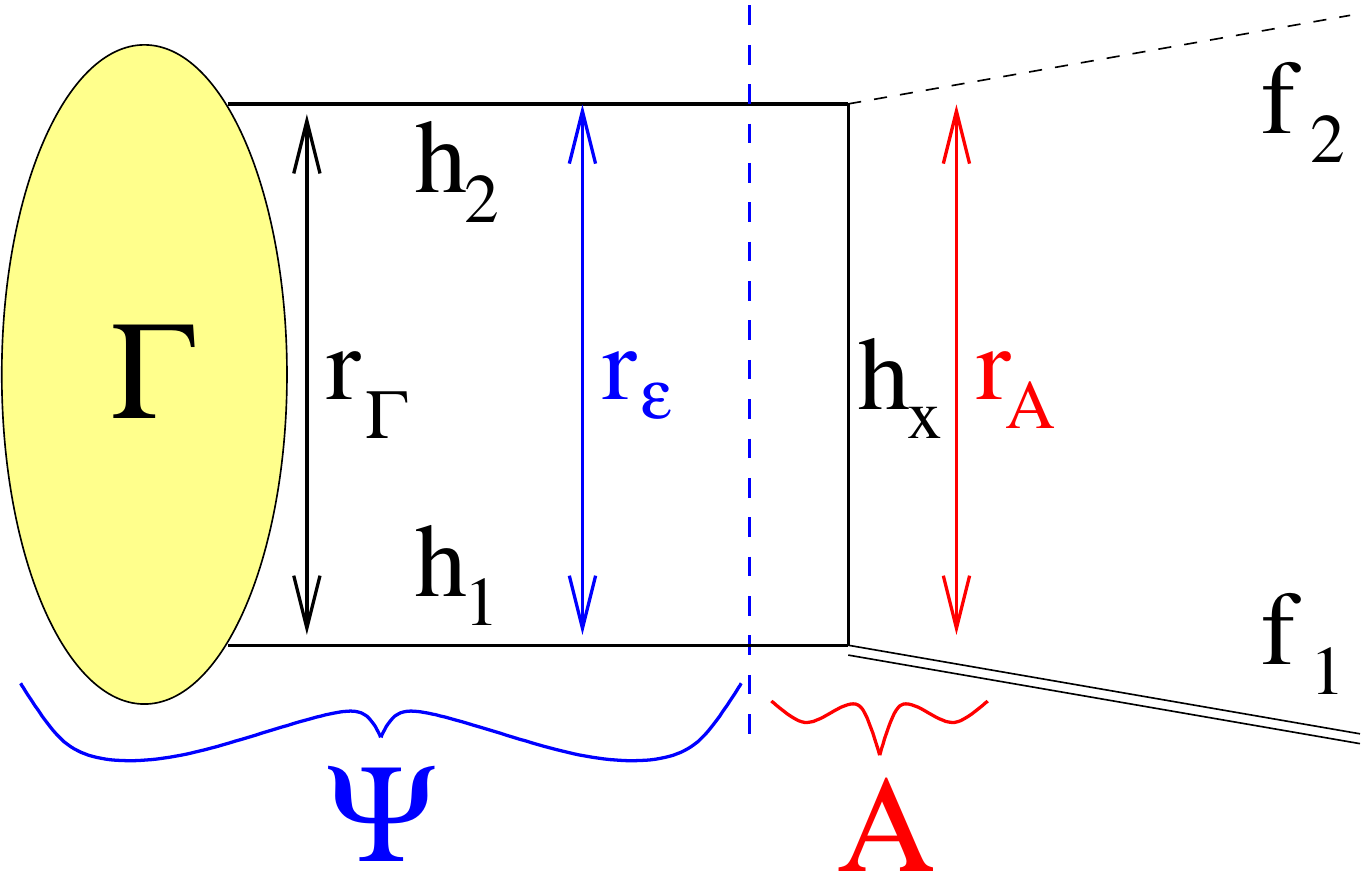}
\caption{Illustration of the strong decay of a hadronic molecule,
composed of two hadrons $h_1$ and $h_2$, into $f_1$ and $f_2$.
The vertex for the transition of the molecule into its constituents is
denoted by $\varGamma$, the corresponding
wave function by $\Psi$, and the annihilation potential by $A$. Solid lines
denote the propagation of the $h_1$, $h_2$ and the
intermediate hadron $h_x$, whereas the double line and the dashed line
denote the decay products $f_1$ and $f_2$.}
\label{fig:scales_decay}
\end{center}
\end{figure}

The observation that in a molecular state the coupling to its constituents
gets large has direct implications on how decays of molecules are calculated:
The most natural decay chain runs via its constituents. Accordingly, it appears natural to
calculate these decays in analogy to those of the positronium via a
factorisation into the wave function at the origin times an on-shell
annihilation amplitude. However, this ansatz is in general
not justified~\cite{Hanhart:2007wa} and a careful study of the scales involved appears to be necessary.
Those are illustrated in Fig.~\ref{fig:scales_decay}: The length scale of the
decay vertex is typically controlled by the mass of the lightest exchange
meson that contributes to the molecular binding --- thus we have
$r_\Gamma\sim 1/\beta$ --- and the size of the wave function is controlled
by the binding momentum, as was explained above, $r_\epsilon\sim 1/\gamma$.
The length scale $r_A$ is controlled by either the cut $f_2h_xh_1$ or the cut $f_1h_xh_2$.
For example, the former may be written as
$$
r_A \sim \frac{1}{M-E_{f_2}-E_{h_x}-E_{h_1}} = \frac{1}{E_{f_1}-E_{h_x}-E_{h_1}},
$$
where $M$ denotes the mass of the molecular state and $E_n$ denotes the energy of the state $n$.
The aforementioned factorisation of the decay amplitude is justified only if $r_A\ll r_\epsilon$ and $r_A\ll r_\Gamma$.
Such a hierarchy of scales holds, for example, for the positronium, where
$r_\Gamma\sim r_\epsilon \sim 1/(\alpha_{\mbox{\tiny QED}} m_e)$ and $E_{f_1}\simeq E_{h_x}\simeq E_{h_1}\simeq m_e$,
such that $r_A/r_\epsilon\simeq \alpha_{\mbox{\tiny QED}}$, where
$\alpha_{\mbox{\tiny QED}}=1/137$ is the electromagnetic fine structure constant.
However, if the decay $f_1\to h_xh_1$ is (nearly) allowed kinematically,
$r_A$ can become very large. In fact, by assumption we have a large
$r_\epsilon$ since we study shallow bound states, and therefore
$r_A\simeq r_\epsilon$ can even imply the proximity of a so-called
triangle or Landau singularity~\cite{Landau:1959fi} which can lead to a
significant enhancement of the transition rates. We come back to this discussion in section~\ref{Sect:4.2.3}.
In a possible decay $X(3872)\to \pi^0\chi_{cJ}$ which is of direct
relevance for this review, the factorisation method is contrasted with the full evaluation of the hadronic loop in Ref.~\cite{Mehen:2015efa}.
The role of the triangle singularities in the decays of heavy
quarkonium-like states is discussed in Ref.~\cite{Wang:2013hga}.

\vspace{0.3cm}\noindent
$\bullet$ {\it General considerations}
\vspace{0.3cm}

Before discussing examples of molecular candidates and some model descriptions
for the $XYZ$-states, some general remarks are in order.
First of all, it should be stressed that relatively narrow, shallow
molecular states can only be formed from similarly narrow constituents,
since the width of the molecular state is strongly correlated with the width
of the constituents~\cite{Filin:2010se}. Stated differently, the constituents
must be sufficiently long-living to allow the molecular
state to form~\cite{Guo:2011dd}. Focusing on states
containing a $c\bar c$ pair and no strange quarks, this implies that only
the ground-state spin doublet $\{D,D^*\}$, with the
quantum numbers $0^-$ and $1^-$, respectively, and the excited
spin doublet $\{D_1,D_2\}$, with the quantum numbers $1^+$ and $2^+$, respectively, are of interest.
The latter pair has a light quark with $j_l^P=3/2^-$ that allows for a decay
into $D^{(*)}\pi$ in the $D$ wave only, which is the reason for their narrow
widths of the order of 30~MeV. The other low-lying pair with positive parity,
$\{D_0,D_1\}$ with quantum numbers $0^+$ and $1^+$, can decay to $D\pi$ in the $S$-wave, which explains their
large widths of the order of 300~MeV. This width is too broad to allow $\{D_0,D_1\}$ to form molecular states.
Accordingly we expect that molecular states with positive parity are significantly lighter than those with negative parity.
The lightest negative-parity molecule can only be a $D_1\bar D$ bound state with quantum numbers $J^P=1^-$.
The nominal $D_1\bar D$ threshold is at $2420+1866=4286$ MeV and indeed, the lowest in mass candidate for an exotic vector state is
$Y(4230)$, located only 50 MeV below this threshold --- this connection was exploited for the first time in Ref.~\cite{Wang:2013cya}.
An additional consequence of assuming a molecular structure for the
negative-parity exotic candidates is that the lightest exotic pseudoscalar
needs to be heavier than the lightest vector by about the mass difference $M_{D^*}-M_D=140$ MeV,
since $J=0$ can only be formed from $D_1\bar D^*$ and not from $D_1\bar D$. The resulting pattern of relevant thresholds is illustrated in Fig.~\ref{fig:cbarc2}.
This figure clearly shows that, while all states below the lowest $S$-wave threshold for a the given quantum number
show properties in line with the quark model, most of the exotic candidates occur close to or above the threshold.

\begin{figure}[t!]
\begin{center}
 \includegraphics[width=0.8\linewidth]{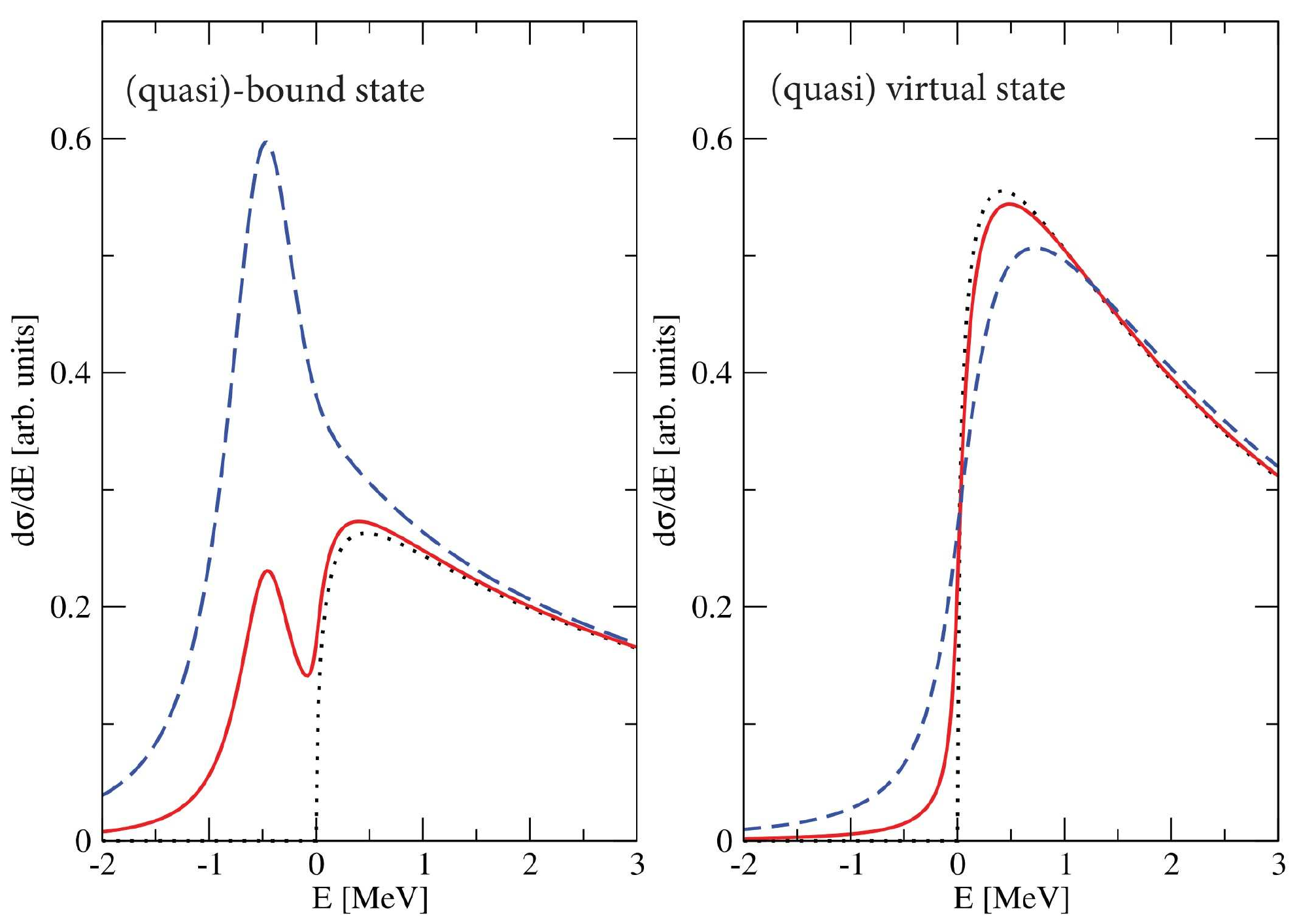}
\caption{Line shapes for a quasibound (left panel) and virtual state
(right panel) for different widths of one of the constituents: 0, 0.1, and
1 MeV as the dotted, solid, and dashed line in order. The mass of the state
is 0.5 MeV below the threshold and an inelastic width of 1.5 MeV is added
into the resonance propagator.}
\label{fig:lineshapes}
\end{center}
\end{figure}

The width of a constituent not only provides a natural scale for the width of the molecular state but it also has
an impact on the line shapes of the molecule. Consider a decay chain of
the resonance $R$ containing an unstable constituent $A$,
$R\to AB\to [cd]B$, and study the line shape in the channel $cdB$.
The effect of the finite width of a molecular constituent on the resulting
line shapes is exemplified in Fig.~\ref{fig:lineshapes},
where different line shapes are plotted for a pole located on the real axis
0.5 MeV below the $AB$ threshold on the first (left panel) or second (right panel) sheet.
The plots are generated for an additional width of 1.5~MeV, due to other decay channels of the resonance,
employing the expressions of Ref.~\cite{Braaten:2007dw}, which are applicable to narrow
states (the formalism was generalised to allow for larger widths in Ref.~\cite{Hanhart:2010wh}).
The width assumed for the state $A$ is 0, 0.1, and 1 MeV for the solid, dotted and dashed line, in order.
A pole on the unphysical sheet can in general only show up as a structure at the nominal $AB$ threshold
or above. However, a pole on the physical sheet generates in addition
to the $AB$ continuum a structure in the data near the pole location,
if sufficient strength from the finite width of the resonance $A$ leaks there.
In the inelastic channels, a pole on the physical $AB$ sheet
leads to a structure near the pole location, while a pole on the unphysical
sheet gives also in this case a signal only at the nominal $AB$ threshold.

The mass parameters of all $Z$ states extracted in the experimental papers
are above the corresponding continuum thresholds. However,
this may in part be a consequence of parameter extractions using symmetric and energy-independent BW functions.
For example, in Ref.~\cite{Cleven:2011gp} it was shown that the data existing at the time for the $Z_b$ states
were consistent with them being bound states. The more refined study of Ref.~\cite{Guo:2016bjq}
identifies both states as virtual states. The inclusion of pion exchange non-perturbatively in the
analysis leaves the pole of the $Z_b(10610)$ untouched within errors,
while the pole related to the $Z_b(10650)$ might even be located slightly above the threshold~\cite{Wang:2018jlv}.
Recent theoretical studies for the pole location of the $Z_c(3900)$ find
that current data are also consistent with an above-threshold resonance~\cite{Albaladejo:2015lob}
or a virtual state~\cite{Pilloni:2016obd}, however, in both cases the pole
location is not well constrained.

\vspace{0.3cm}\noindent
$\bullet$ {\it Remarks on the role of the heavy-quark spin}
\vspace{0.3cm}

Four of the established charged states in the charmonium sector are prime
candidates for
hadronic molecules as was quickly recognised by many
authors~\cite{Bondar:2011ev,Mehen:2011yh,Wang:2013cya,Guo:2013sya,Dias:2014pva,Aceti:2014uea}.
Indeed, they are located very close to $S$-wave thresholds of narrow
resonances: The $Z_c(3900)$ lies at the $D^*\bar D$ threshold, the
$Z_c(4020)$ is close to the $D^*\bar D^*$ one,
the $Z_b(10610)$ and $Z_b(10650)$ reside near the $B^*\bar{B}$ and
$B^*\bar{B}^*$ thresholds, respectively.
For the $Z_b$ states it is intriguing that they both are seen in
three $\pi\Upsilon$ and two $\pi h_b$ channels with
a very similar strength in the two-pion transitions from the $\Upsilon(10860)$
bottomonium, although the initial state
as well as the $\Upsilon$'s in the final states have a heavy-quark spin of 1
while the $h_b$'s has a heavy-quark spin 0. This fact appears to be very
unexpected since in heavy-quark systems the spin of the heavy quarks should be conserved given that spin-dependent
interactions are suppressed as $(\Lambda_{\rm QCD}/m_h)$, where $m_h$ denotes the heavy-quark mass (see Sec.~\ref{Sect:4.2.1}).
In addition, the interference patterns between the $Z_b(10610)$ and $Z_b(10650)$
appear to be different in the two types of channels: While in the former
there appears to be a destructive interference between the two structures,
it is constructive in the latter.
Both features find a natural explanation if $Z_b(10610)$ and $Z_b(10650)$
are assumed to be $B^*\bar B$ and $B^*\bar B^*$ molecular
states~\cite{Bondar:2011ev}, for then one finds
\begin{eqnarray} \nonumber
Z_b(10610)\sim B^*\bar B-B\bar B^* =\frac{1}{\sqrt{2}}\left(0^-_{\bar bb}\otimes 1^-_{\bar qq} - 1^-_{\bar bb}\otimes 0^-_{\bar qq}\right), &&\\
Z_b(10650)\sim \quad~B^*\bar B^*\quad~=\frac{1}{\sqrt{2}}\left(0^-_{\bar bb}\otimes 1^-_{\bar qq} + 1^-_{\bar bb}\otimes 0^-_{\bar qq}\right).&&
\end{eqnarray}
For both states the first and second term in parentheses provide the coupling
to the final state that contains $h_b$ and $\Upsilon$, respectively.
An additional support for the molecular interpretation of all four $Z$'s comes
from the fact that they all decay, with a branching fraction larger than 80\%,
into the nearby open-flavor channel despite a very small phase space
available in these decays while the phase space is wide open for their
hidden-flavor channels.
For a comparison of a molecular \emph{versus} hadrocharmonium
interpretation of the $Z_c(3900)$ we refer to Ref.~\cite{Voloshin:2013dpa}.

If one admits that the charged states are good candidates for isovector
hadronic molecules, it appears natural that there are also isoscalar hadronic molecules.
Indeed, let us assume that the binding potential is provided by meson exchanges
and that amongst those the most prominent one is an isovector meson exchange
(this is an established picture for two-nucleon systems with the exchanged particle being the pion).
For the exchange of such an isovector particle between two isospin doublets
the ratio of isospin factors of the isoscalar to the isovector channel
is -3. An additional minus sign appears when switching the
$C$-parity (clearly, for isovector states the $C$-parity is defined only for
the neutral component). In particular, the picture just drawn would be
consistent with a prominent isovector exchange potential
if molecular states existed simultaneously in the $D^*\bar D$ system with $I=0$ and $J^{PC}=1^{++}$, and with $I=1$ and $J^{PC}=1^{+-}$.
Furthermore, in this scenario there should not be any isovector (isoscalar) $1^{++}$ ($1^{+-}$) states.
This naive pattern is in line with experimental observations.
This is clearly different from the compact tetraquark picture,
where each isoscalar state should be accompanied by its nearly degenerate isovector partner.
However, this reasoning does not include that, e.g., pion exchange naturally provides large $S$-wave to $D$-wave transitions via its tensor force ---
as soon as those are included one-pion exchange can provide binding also in, e.g., isovector $1^{++}$ channels
(see Refs.~\cite{Baru:2017gwo,Baru:2019xnh} and the discussion in Sec.~\ref{Sect:4.2.4}).
In addition, the channel coupling driven by the tensor force of the one pion exchange also generates
a transition of the $Z_b(10650)$ to the $B\bar B^*$ channel, which, however, is not seen in the data.
It is shown in Ref.~\cite{Wang:2018jlv} that this transition is largely absorbed into a formally subleading counter
term that needs to be promoted to leading order to render the calculation cut--off independent --- this is discussed
in some detail in Sec.~\ref{Sect:4.2.4}.

The feature described above for the $Z_b$ states, namely that molecules
in the intermediate states evade the rule that the heavy-quark spin
remains unchanged in a hadronic transition, is quite general. For example,
in Ref.~\cite{Li:2013yka} the following decompositions
of the above mentioned two-hadron states with the quantum numbers $1^{--}$
are given as
\begin{eqnarray}
(D_1\bar D - \bar D_1 D)&:& \frac{1}{2\sqrt{2}} \psi_{11}+\frac{\sqrt{5}}{2\sqrt{2}} \psi_{12}+\frac{1}{2} \psi_{01} , \\
(D_1\bar D^* - \bar D_1 D^*)&:& \frac{3}{4} \psi_{11}-\frac{\sqrt{5}}{4} \psi_{12}+\frac{1}{2\sqrt{2}} \psi_{01} , \\
(D_1\bar D - \bar D_1 D)&:& \frac{\sqrt{5}}{4} \psi_{11}+\frac{1}{4} \psi_{12}-\frac{\sqrt{5}}{2\sqrt{2}} \psi_{01} ,
\end{eqnarray}
where $\psi_{1J}=1_{Q\bar Q}^{--}\otimes J_{q\bar q}^{++}$ and $\psi_{01}=0_{Q\bar Q}^{-+}\otimes 1_{q\bar q}^{+-}$,
where as before $Q$ ($q$) denotes a heavy (light) quark.
Accordingly, if there exist resonances generated as two-hadron states in
these channels, one should expect a similar population for the final states
with the total heavy-quark spin equal to 0
or 1. And indeed, in the mass range covered by the three corresponding
thresholds, where at least three vector states\footnote{Some analyses
claim that even more are needed --- see chapter~\ref{Sect:3.1}.},
$Y(4230)$, $Y(4360)$ and $\psi(4415)$,
coexist,
one finds the $e^+e^-\to \pi\pi h_c$ cross section of the order of 60 pb and
the $e^+e^-\to \pi\pi J/\psi$ cross section of the order of 80 pb ---
cf. Figs.~\ref{xsec-fit} and \ref{cs}.
It should be stressed that those total cross sections have only a very small
contribution from the $Z_c$ states.
Stated differently, the invariant mass spectrum that shows the $Z_c$ states
in the $\pi h_c$ final state
(cf. Fig.~\ref{1Dfit}) shows also a strong non-resonant contribution.
This is in contrast to the data on the $Z_b$'s
in the $\pi h_b$ channels where the signal is resonant only
which in the molecular picture is understood as a consequence of the
large mass gap between the $\Upsilon(10860)$ and the first relevant two heavy-hadron S-wave
threshold, $B_1\bar B$ at 11004 MeV.

\begin{figure}[t!]
\begin{center}
 \includegraphics[width=0.7\linewidth]{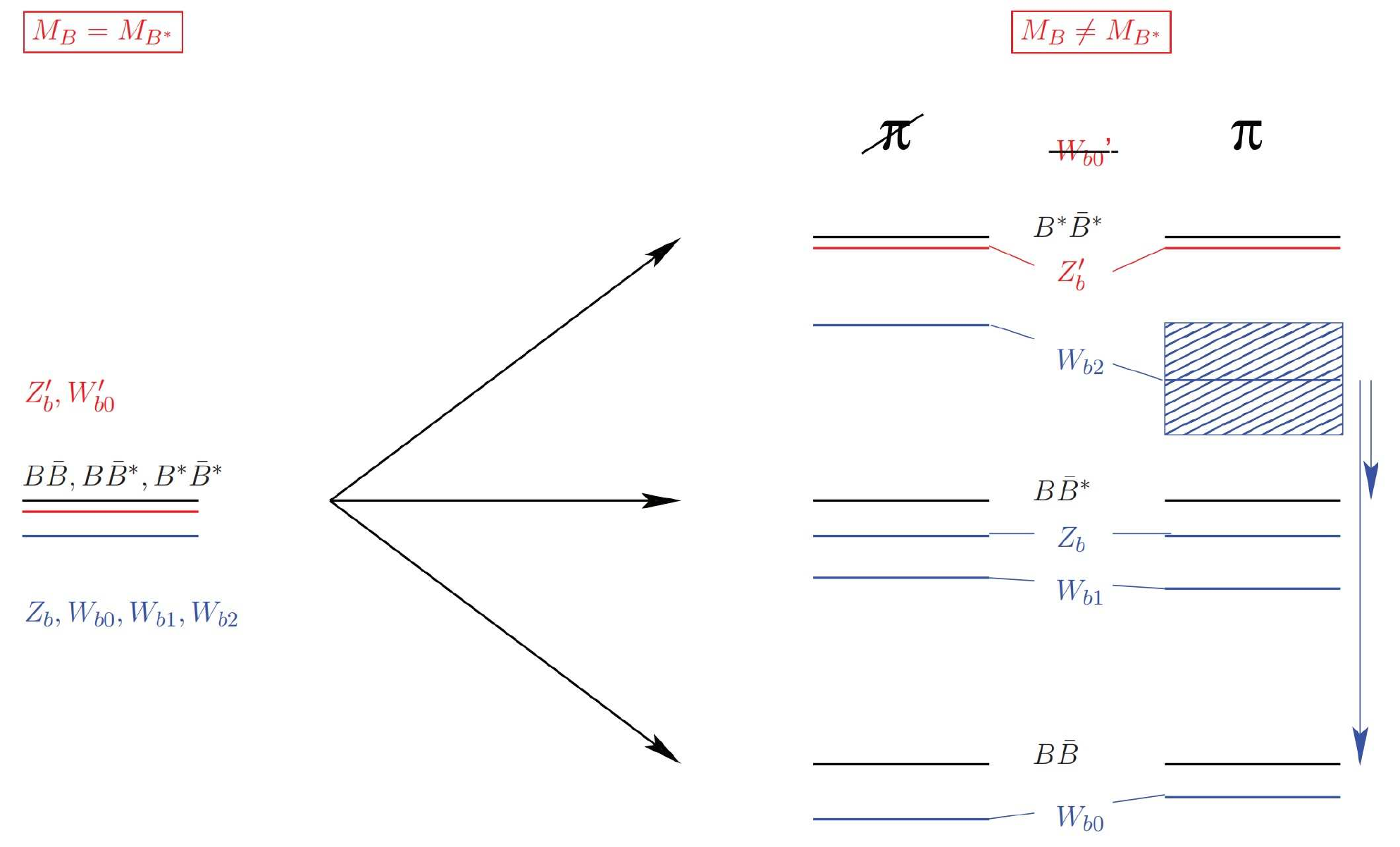}
\caption{Sketch of the mass spectrum of $Z_b(10610)$ and $Z_b(10650)$ and their spin
partner states in the spin-symmetric limit (left) and when the $B^*$-$B$ mass
splitting is included (right). In the latter case calculations
with (marked by ${\mathbf{ \pi}}$) and without one-pion exchange (marked by ${\mathbf{ \pi \!\!\!/}}$)
 in the potential are compared as detailed
in the text. The results underlying the figure are from
Ref.~\cite{Baru:2017gwo} based on the input masses for the $Z_b$ states
from Ref.~\cite{Cleven:2011gp}.}
\label{fig:Zbspinpartners}
\end{center}
\end{figure}

Since in QCD the spin of the heavy quarks decouples from the interaction up to corrections of the order of $\Lambda_{\rm QCD}/m_h$,
one can also predict spin partner states for hadronic molecules.
It was stressed in Ref.~\cite{Cleven:2015era} that the pattern of spin-symmetry
breaking is very different between different models for the $XYZ$ states.
For example, to leading order in the $1/m_b$ expansion the $Z_b$-states
and their spin partners form two multiplets: One doublet which contains
the $J^{PC}=1^{+-}$ state $Z_b(10650)$ and a $0^{++}$, called $W_{b0}'$, and
one quartet, which contains the $J^{PC}=1^{+-}$ state $Z_b(10610)$ as well as the
$0^{++}$ , $1^{++}$, and $2^{++}$ states called $W_{b0}$, $W_{b1}$, and $W_{b2}$,
respectively~\cite{Voloshin:2011qa,Mehen:2011yh,Baru:2017gwo}. If one allows
for spin-symmetry violation, whose leading effect for hadronic molecules is
in this case driven by the $B^*$-$B$ mass difference, the members of the
multiplets follow the splitting of the thresholds --- see Fig.~\ref{fig:Zbspinpartners}.
As soon as one--pion exchange is included, many additional
partial waves need to be included in the calculation~\cite{Baru:2016iwj} since the tensor force provides strong $S$-$D$ transitions.
As a result of this, the spin-2 state acquires a width from the now allowed transition
from $B^*\bar B^*(\mbox{$S$-wave})\to B\bar B^{(*)}(\mbox{$D$-wave})$.
Because of this additional transition there also appears a sizeable mass shift once pions are included.
It was shown in Ref.~\cite{Baru:2017gwo} that the spectrum of spin partner
states is very sensitive to the input masses used for the $Z_b$ states.
A more refined study using as input the masses determined directly from a fit to the $Z_b$ line
shapes is presented in Ref.~\cite{Wang:2018jlv} and will be discussed in more detail in Sec.~\ref{Sect:4.2.4}.
Moreover, the impact of the pions on the pole positions and the line shapes is important also for the spin partner states --- in particular, while both $Z_b$'s and their siblings $W_{bJ}$ appear to be virtual states in the pionless theory, in the full EFT, including pions, most of these states turn to above-threshold resonances --- see Ref.~\cite{Baru:2019xnh} for details.

A similar pattern of spin partner states is expected to also emerge in
case of the isoscalar states in both the charm and the bottom sector.
Thus, if the molecular picture is correct, then, in the strict
spin-symmetry limit, the $1^{++}$ state $X(3872)$
should be a member of a multiplet together with one $1^{-+}$, one $0^{++}$,
and one $2^{++}$ state. In addition, in this limit there could be a second multiplet with a $1^{-+}$ state and another $0^{++}$ state~\cite{Nieves:2012tt}.
As mentioned above, the leading spin-symmetry-violating effect comes from the mass difference of the ground-state
pseudoscalar and vector states, such that the interactions may be assumed spin symmetric.
Still, to constrain all masses of the members of those multiplets, at least two states are needed as input.
In Ref.~\cite{Nieves:2012tt} the masses of the $X(3872)$ and of the $X(3915)$, assumed to be a scalar $D^*\bar D^*$ molecular state,
were employed, and, based on this, using a model with contact interactions only, in total four additional states were predicted.
Note that an uncertainty estimate of these predictions is in general very difficult,
since the $X(3915)$ is located 100~MeV below the relevant threshold and corrections to the pure contact interactions
should be expected. However, it is interesting to observe that the spin partner of the $X(3872)$
with $J^{PC}=2^{++}$ is generated from an interaction identical to
that of the $X(3872)$ --- thus, the small binding energy
of this state with respect to the neutral $D\bar D^*$ threshold
should directly translate to a small binding energy of the spin-2 state with respect to the $D^*\bar D^*$ threshold.
This observation was confirmed by the explicit numerical
calculations of Ref.~\cite{Nieves:2012tt}.
Once one-pion exchange is included, transitions from the $D^*\bar D^*$
$S$-wave with spin 2 to the $D\bar D$ $D$-wave become possible which can
lead to widths of a few MeV~\cite{Albaladejo:2015dsa} to tens of MeV~\cite{Baru:2016iwj}.

It is interesting to note that Refs.~\cite{Gamermann:2006nm,Nieves:2012tt,HidalgoDuque:2012pq} all find a
resonance with a mass about 3700 MeV that couples to $D\bar D$ strongly
and is interpreted as a $D\bar D$ molecular state. Various reactions are proposed to look for this state,
e.g. $e^+e^-\to J/\psi D\bar D$~\cite{Gamermann:2007mu} and radiative decays of heavier vector states~\cite{Xiao:2012iq}.
While there are indications that this state might have left traces in the data, so far
the analyses are not fully conclusive.

An SU(3)-flavor generalisation of the research of Ref.~\cite{Nieves:2012tt} is provided in Ref.~\cite{HidalgoDuque:2012pq}.
Here the authors constrain the parameters by also investigating the decays
of $X(3872)$ into $\rho J/\psi$ and $\omega J/\psi$,
which constrains the isovector interaction, and propose that $X(4140)$ is a
$0^{++}$ state (at the time of this work the $Z_c$ states were not yet found).
From this input they find that there should neither be a $1^{++}$ nor a $2^{++}$ isoscalar state with hidden strangeness.
However, since now the $X(4140)$ is established to be a $1^{++}$ state, this analysis should be redone. In contrast to Ref.~\cite{HidalgoDuque:2012pq},
Ref.~\cite{Molina:2009ct} finds a $2^{++}$ $D_s^*\bar D_s^*$ bound state at 4160~MeV, based on a similar construction but
using vector-meson exchanges for the potential. Here more data, especially
for states with different quantum numbers are needed before a clear theoretical understanding can emerge.

\vspace{0.3cm}\noindent
$\bullet$ {\it $\chi_{c1}(3872)$ aka $X(3872)$ from a molecular perspective}
\vspace{0.3cm}

The probably most prominent example of a molecular state is the
$X(3872)$, not only because its existence was predicted already in Refs.~\cite{Voloshin:1976ap,Tornqvist:1993ng}
but also because its mass is incredibly close to the $D^0\bar D^{0\, *}$ threshold. For a review with focus on
the $X(3872)$ as a molecular state see Ref.~\cite{Kalashnikova:2018vkv}.
Originally, the $X(3872)$ was claimed to be (predominantly)
a bound state of $D^0\bar D^{0*}$~\cite{Close:2003sg,Braaten:2004fk}, which translates to an equal
admixture of an isospin-1 and an isospin-0 component.
However, this is at odds with experiments.
From the first glance one might think that this is in line with the
experimental observation that it decays at near equal rates into the
$\rho^0 J/\psi $ and $\omega J/\psi$ final states (see
section~\ref{Sect:3.1.1}), the former having isospin 1, the latter isospin 0.
However, this can only be understood, if the $X(3872)$ is predominantly an isoscalar state,
since the formal $\rho^0 J/\psi $ threshold is located below the
mass of the $X$ while that of the $\omega J/\psi$ channel is 7 MeV above,
which is almost as large as the width of the $\omega$.
Thus, the decay into the latter channel is heavily suppressed kinematically compared to that into
the former, especially in light of the large width of the $\rho^0$ of the order of 150 MeV. Thus,
had the $X(3872)$ been to 50\% an isovector state it would have almost exclusively decayed into $\rho^0 J/\psi $.
Moreover, it was shown in Ref.~\cite{Gamermann:2009fv}, and more recently in Ref.~\cite{Zhou:2017txt},
that one can understand the apparent isospin violation in the $X(3872)$ decay quantitatively,
if one considers in the decay the large isospin violation provided by the meson loops
induced by a large coupling of the molecular state to the two--meson continuum according to Eq.~(\ref{eq:geffdef}):
Meson loops in a relative $S$-wave show a strong mass dependence in form of
a pronounced cusp structure exactly at the two--meson threshold,
if there is a pole nearby (see Sec.~\ref{sec:cusps}).
In isospin-violating transitions the loops in the neutral and the charged channel enter with a relative minus
sign and equal coupling strength. Thus, if the masses of the meson pair in the two channels are
equal, they cancel exactly as required by isospin symmetry. However, in
reality the $D^\pm D^{*\mp}$ threshold is located 8~MeV above the
$D^0\bar D^{*0}$ threshold, which results in an enhanced isospin violation.
A second argument in favor of a predominant isoscalar nature of the
$X(3872)$ is given in Ref.~\cite{Mehen:2015efa},
where it is shown that if there had been a sizeable $I=1$ component in the
wave function, then the $X(3872)$ should have copiously
decayed into the $\pi^0\chi_{cJ}$ channels --- with the partial widths
larger than the current upper limit of the $X(3872)$ total width.
Finally, in Ref.~\cite{Aceti:2012cb} it is shown that a charged component
in the $X(3872)$ is necessary to get a quantitative
understanding of its radiative decays.

At present it is not known on which sheet the pole of the $X(3872)$ is located:
It might be a shallow bound state (similar to the deuteron and hypertriton mentioned at the beginning of this chapter),
which would imply a pole on the physical sheet, or a virtual state like the two-neutron state (or isovector nucleon-nucleon state),
which would imply a pole on the unphysical sheet.
It is generally believed that the latter scenario unambiguously points at a
two-meson nature of the corresponding state.
Unfortunately, an analysis of the currently existing data showed consistency with both
scenarios~\cite{Hanhart:2007yq,Braaten:2007dw}. The latter reference
stressed that in case of a bound state it is necessary to consider the width
of the $D^*$ meson and also provided analytic expressions for the resulting line shapes that can be used for narrow constituents.
Since the width estimated for the neutral $D^*$ is of the order
of 0.1 MeV and the $X(3872)$ binding energy is 0.2 MeV
or lower, one expects a structure below the $D^0\bar D^{*0}$ threshold in the $D^0\bar D^-\pi^+$ channel, if the
$X(3872)$ is a bound state. This
structure should be absent if the $X(3872)$ were a
virtual state (cf. Fig.~\ref{fig:lineshapes} and the corresponding discussion).
If the transition rate $p\bar p\to X(3872)$ is large enough, there is hope
that the PANDA experiment planned at the FAIR facility (see Sec.~\ref{sect:6.1}) could provide a direct measurement of this line shape.
An alternative method to get high accuracy information of the pole location of $X(3872)$ has been provided recently based
on the interplay of a triangle singularity and the corresponding pole~\cite{Guo:2019qcn} --- this idea is discussed in more detail in Sec.~\ref{Sect:4.2.4}.

\vspace{0.3cm}\noindent
$\bullet$ {\it $\psi(4230)$ aka $Y(4230)$ as hadronic molecule }
\vspace{0.3cm}

\begin{figure}[t!]
\begin{center}
 \includegraphics[width=0.8\linewidth]{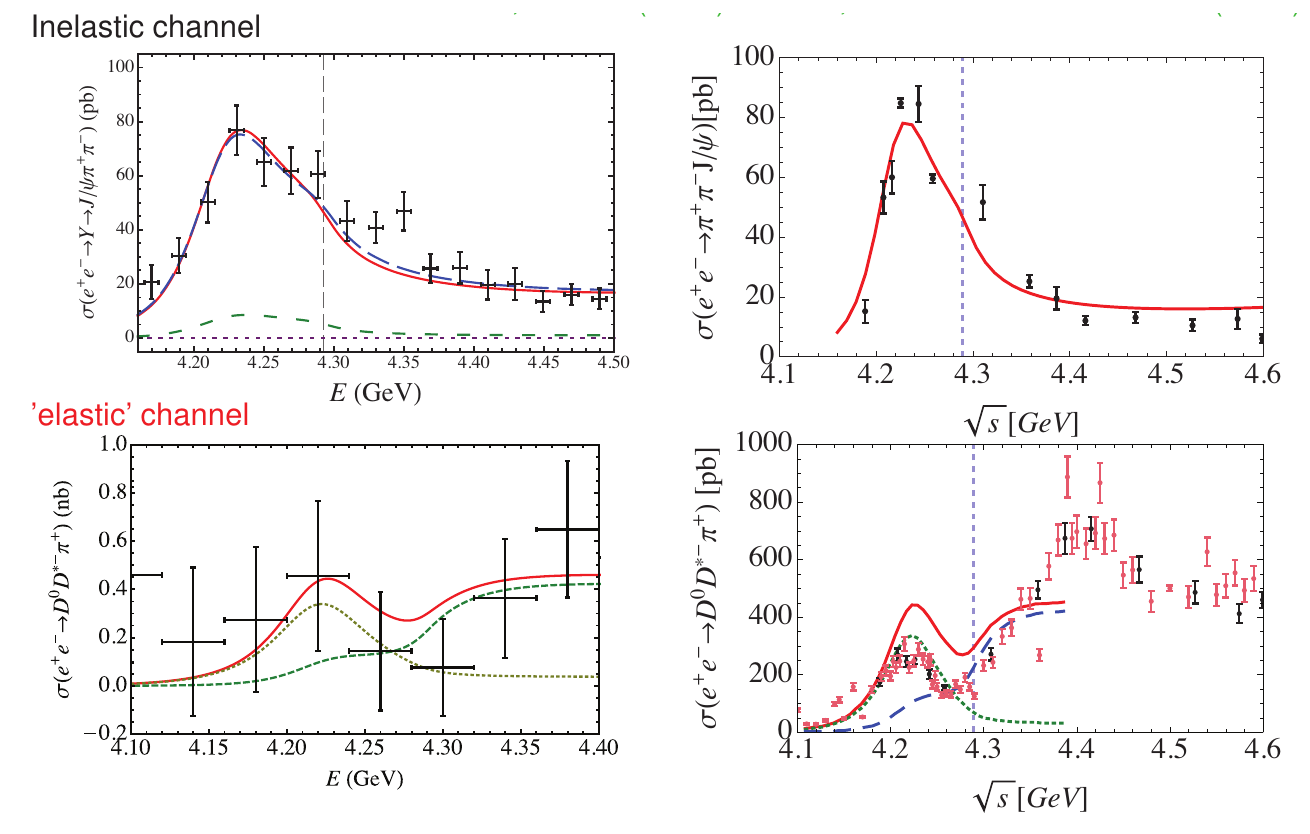}
\caption{Lineshapes of the $\psi(4230)$ aka $Y(4230)$ in the $\pi\pi J/\psi $ channel (upper panel) and the $D\bar D^*\pi$ channel (lower panel).
 The left column shows the data available before 2014 from Belle~\cite{Pakhlova:2009jv} (lower panel) and~\cite{Yuan:2007sj} (upper panel)
 and the right column shows the very recent data from BESIII~\cite{Ablikim:2016qzw,Ablikim:2018vxx}.
 In all panels the red lines show the results of the full model of Ref.~\cite{Cleven:2013mka} with its parameters fixed in 2013.}
\label{fig:Ydecays_oldandnew}
\end{center}
\end{figure}

Next we come back to the $Y(4260)$ and list some more
properties that point at a $D_1\bar D$ molecular nature of this state.
The reasoning is based on the fact that a state that has a nearby two-hadron channel as its prominent molecular component will couple to it strongly.
Accordingly, it appears natural in the molecular scenario for the $Y(4260)$ that the nominal $D_1 \bar D$
threshold leaves an imprint, for example, on its line shapes.
This is shown in Fig.~\ref{fig:Ydecays_oldandnew}.
In all panels the red solid line denotes the results of the full model of Ref.~\cite{Cleven:2013mka}.
All model parameters were fixed in 2013 to the spectra and angular distributions existing at that time.
The results for the $\pi\pi J/\psi$ channel clearly show that the molecular model naturally generates asymmetric line shapes from just a single state.
In contrast to this, the experimental analysis of the new data (upper right panel) generate the asymmetry
from two BW resonances --- see the discussion in Sec.~\ref{Sect:3.1} above.
One conclusion from the molecular picture is that the structure known as $Y(4260)$ should be identified with the $Y(4230)$.
We will use this name in the rest of this section.

Since the natural decay of a molecular state is into its constituents, and the strongest decay channel
of the $D_1(2420)$ is $D^*\pi$, a $D_1\bar D$ molecular state must leave a strong imprint in the $\bar{D}D^*\pi$ final state.
At the time when the model parameters of Ref.~\cite{Cleven:2013mka} were fixed, no high quality data in
this channel were available and the results shown in
the lower right panel of Fig.~\ref{fig:Ydecays_oldandnew} came as a prediction.
Also this calculation contains only one state --- still the data and the
calculation contain in addition to the peak of the $Y(4230)$
a second structure above the nominal $D_1\bar D$ threshold located at
4280~MeV --- here a similarity of the red curve and the typical line shapes for molecules with unstable constituents shown in
Fig.~\ref{fig:lineshapes} should be emphasized.
The missing strength to account for the data above 4350 MeV might come from
the $\psi(4415)$ or $D_2\bar D^*$ channel missing in the model.

\begin{figure}[t!]
\begin{center}
 \includegraphics[width=0.8\linewidth]{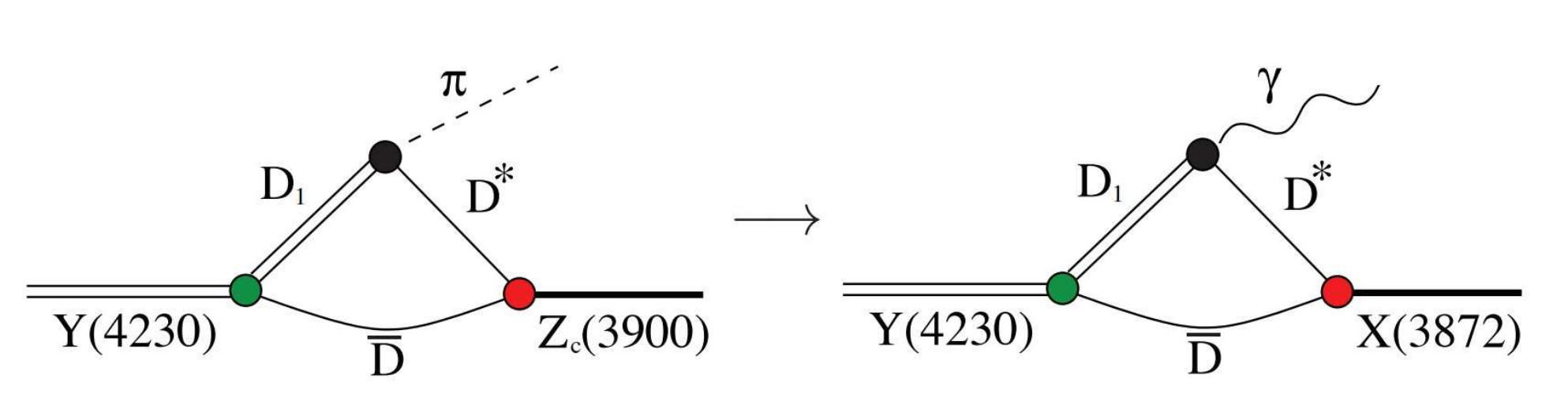}
\caption{Triangle mechanisms driving both $Y(4230)\to\pi Z_c(3900)$ and $Y(4230)\to \gamma X(3872)$ decay in the molecular picture.}
\label{fig:Y2XandY2Z}
\end{center}
\end{figure}

A $D_1\bar D$ molecular nature for the $Y(4230)$ also predicts peculiar decay patterns of this state.
For example, it was stressed in Ref.~\cite{Wang:2013cya} that within this scenario it appears
natural that the $Z_c(3900)$ envisioned as
$D\bar D^*$ bound state gets excited in the $Y(4230)$ decay,
since the prime decay channel of the $D_1(2420)$ is $\pi D^*$
and, accordingly, the triangle mechanism shown in the left half of
Fig.~\ref{fig:Y2XandY2Z} provides an efficient source of low-energy $D^*\bar D$ pairs leading to a copious
production of the related isovector molecular state.
Moreover, the quark model predicts that the $D_1(2420)$ also decays into $D^*\gamma$
--- thus, the very same triangle mechanism, shown in the right panel of Fig.~\ref{fig:Y2XandY2Z}, should again
provide $D^*\bar D$ pairs, however, now with the opposite $C$-parity since $\pi^0$ and $\gamma$ have $C=+$ and $C=-$, respectively.
Accordingly, if the above explanation for the observation of the $Z_c(3900)$ in the decay
$Y(4230)\to\pi Z_c(3900)$ is correct, necessarily $X(3872)$ must also be produced
in the $Y(4230)\to \gamma X(3872)$ reaction \cite{Guo:2013nza}. This prediction
was confirmed experimentally shortly after its publication --- see Fig.~\ref{fit-mx}.

As already mentioned above, the findings of Ref.~\cite{Chen:2019mgp}, where the
$\pi\pi$ spectra from $e^+e^-\to \pi^+\pi^- J/\psi$ in the 4260~MeV mass range were studied,
that the source contains a sizeable flavor-octet component, are fully in line with
a molecular as well as compact tetraquark interpretation of those states.

\vspace{0.3cm}\noindent
$\bullet$ {\it Closing remarks on hadronic molecules}
\vspace{0.3cm}

So far in this section we have focused on general properties of molecular
states. To go beyond this and predict, for example, partner
states that should exist given the symmetries of QCD, dynamical equations
need to be solved. Those need as input
phenomenological models like meson exchange models --- see, for example,
Refs.~\cite{Zhang:2006ix,Ding:2009vj,Li:2012ss,Zhao:2014gqa,He:2017mbh,Liu:2019stu}; many
additional references are also provided in the review
articles~\cite{Chen:2016qju,Liu:2019zoy}. Alternatively, one can construct the scattering
potentials in an effective field theory framework. The latter method will be
detailed in Sec.~\ref{Sect:4.2.4} below.
Sometimes flavor SU(4) is used to constrain scattering potentials --- see,
for example, Refs.~\cite{Molina:2009ct,Dias:2014pva,Aceti:2014uea} ---
however, it should be stressed that such studies
necessarily come with uncontrolled systematic uncertainties, since
flavor SU(4) is not a symmetry of QCD (cf. chapter 16.4. of Ref.~\cite{Georgi:1999wka}).
Note that some studies constructed on the basis of flavor SU(4) are in fact
only sensitive to the flavor SU(3) subgroup and as such
show a connection to QCD symmetries --- see, e.g., the discussion in Sec. IIA of Ref.~\cite{Sakai:2017avl}, where it is argued that
in practice what is often employed is only the well established SU(2) symmetry.

From the molecular perspective, the $Z_c(4430)$ is the most problematic, since
it is not close to any $S$-wave threshold\footnote{This state is located basically at the $D_2\bar D^*$ threshold,
but here an $S$-wave calls for a negative parity at odds with the quantum
numbers of the $Z_c(4430)$ determined by LHCb~\cite{Aaij:2014jqa}.} --- see
Fig.~\ref{fig:cbarc2}. The $Z_c(4430)$ could be a signature of
the triangle singularity as originally proposed in Ref.~\cite{Uglov:2016nql}.
If this were a proper explanation of these resonances, there should be no signal of the $Z_c(4430)$ in the $D^*\bar D$ channel,
according to Schmid's theorem (see Sec.~\ref{Sect:4.2.4} for a more detailed discussion), which could be tested experimentally.
A different set of triangle diagrams to explain both the $Z_c(4200)$ and the $Z_c(4430)$ was proposed in
 Ref.~\cite{Nakamura:2019btl}. Here the latter (former) state emerges from the triangle
 $\bar B^0\to \bar K^*(892)Y(4260)\to [K^-\pi^+]Y(4260)\to \pi^+\psi$ ($\bar B^0\to \bar K^*_2(1430)\psi(3770)\to [K^-\pi^+]\psi(3770)\to \pi^+\psi$),
 where the $\psi$ in the final state could be either $J/\psi$ or $\psi(2S)$. For these transitions to be prominent
 there should be significant rates observable experimentally for $\bar B^0\to \bar K^*(892)Y(4260)$ and
 $\bar B^0\to \bar K^*_2(1430)\psi(3770)$, respectively, making also this prediction testable experimentally.

Since a state like $Z_c(4430)$ cannot be an $S$-wave bound state of a pair of ground-state mesons, Ref.~\cite{Barnes:2014csa}
employs a one-pion exchange model to test
the hypothesis whether this state could be a molecule of the $D^*\bar D(2S)-D^*(2S)\bar D$ system
with the relevant thresholds assumed at 4509 MeV and 4590 MeV, respectively.
The authors conclude that with their standard parameters they cannot get an
isovector bound state in this system, however, they find significantly more deeply bound isoscalar states. If the pion
coupling in the model is increased to a strength sufficient to get the $Z_c(4430)$ as a bound state,
the isoscalar partner states appear to be at least 100 MeV more bound.
Those states have not been seen experimentally.

Alternatively, also employing a one-pion exchange model, in Ref.~\cite{He:2017mbh} the $Z_c(4430)$ is proposed to be a $P$-wave
$\bar D_1(2420)D^*$ resonance. In the same work $Y(4390)$ is explained as its isoscalar $S$-wave partner.
In this calculation the excitation from the $S$- to the $P$-wave costs about
40 MeV. If this picture is correct, there should be a signature of the
$Z_c(4430)$ in the $D^*\bar D\pi$ channel --- a prediction that can be tested experimentally.
Another candidate for a $P$-wave bound state, this time in the $D^*\bar D^*$ channel, could be the $Z_c(4100)$
if it has negative parity \cite{Zhao:2018xrd}.

To summarise, the molecular model provides natural explanations for various observables, such as the masses, line shapes, and decay patterns, of various candidate exotic states.
Based on the arguments presented above, for example, $X(3872)$, $Z_c(3900)$, $Z_c(4020)$, $Y(4230)$, and $Y(4360)$
in the charmonium sector, and $Z_b(10610)$ and $Z_b(10650)$ in the bottomonium sector are good candidates for states having a prominent molecular component.
However, it should not remain unmentioned that also the molecular picture suffers from a few unresolved issues:
\begin{itemize}
\item At present it does not explain the relative strengths of the $Z_c(3900)$ and $Z_c(4020)$ at the various beam energies in the $\pi h_c$ and $\pi J/\psi$ final states.
 While the anticipated production mechanism via a triangle might play an important role here as proposed in Refs.~\cite{Wang:2013cya,Wang:2013hga},
 full understanding of all available data is still lacking.
\item The state $Z_c(4430)$ is not located close to any $S$-wave threshold and thus calls for an alternative explanation.
 It is our understanding that so far no convincing explanation for this state in a two-hadron scenario has been provided.
\item In addition, the interplay of the molecular states with compact quark states ($Q \bar{Q}$ quarkonia, $Q\bar{Q}g$ hybrids and so on) is not understood yet,
 although some efforts have already been taken in this direction \cite{Cincioglu:2016fkm}.
\end{itemize}

\subsubsection{Sum rules}
\label{Sect:4.sumrules}
Another non-perturbative approach often applied to both generic and exotic
hadrons is the QCD sum rules method~\cite{Shifman:1978bx,Shifman:1978by}.
A thorough description of the method and a comprehensive list of references
on the subject can be found in the recent review~\cite{Albuquerque:2018jkn},
which is specifically devoted to the sum rules approach to exotic XYZ hadrons.

The cornerstone of the method is the quark--hadron duality conjecture,
which allows one to relate observables for hadronic states with the properties
of the non-perturbative QCD vacuum encoded in various local condensates.
To this end one considers a correlation function of currents with the
quantum numbers of the studied resonance
and evaluates it in terms of local condensates of different dimensions using
the operator product expansion.
For example, for the two-point correlation function one has
\be
\Pi(Q^2) = i\int d^4 x \; e^{iqx} \; \langle 0|T[j(x)j^\dagger(0)]|0\rangle = \Pi^{\rm pert}(Q^2) + C_3(Q^2)\langle\bar{q}q\rangle + C_4(Q^2)\langle g^2\mbox{Tr}(F_{\mu\nu}F^{\mu\nu})\rangle + \dots\,,
\label{Piexp}
\ee
where $j(x)$ is an interpolating current, $Q^2=-q^2$, $\Pi^{\rm pert}(Q^2)$ is the perturbative contribution to the correlation function,
and $\langle\bar{q}q\rangle$, $\langle g^2\mbox{Tr}(F_{\mu\nu}F^{\mu\nu}))\rangle$ and so on are the quark, gluon and so on local condensates characterising the QCD vacuum.
Where to truncate the series in the right-hand side of Eq.~\eqref{Piexp} depends also on the resonance under study --- typically the quark and the gluon condensates are sufficient to describe generic quarkonia,
while one has to include higher-order condensates to provide a good
convergence of the series for exotic states.
The values of the condensates are assumed to be known and taken as input parameters.
For example, the chiral condensate $\langle\bar{q}q\rangle$ is related with the light quark masses and the pion decay constant $f_\pi$ through the Gell-Mann--Oakes--Renner relation~\cite{GellMann:1968rz}.
One can also extract them from lattice QCD calculations of the same correlation functions in Euclidean spacetime --- for recent determinations of the gluon condensate see Refs.~\cite{Bali:2014sja,DelDebbio:2018vhr}.
Finally, the numerical values of the condensates can be extracted from the sum rules applied to hadronic states with well established properties, such as the masses and the decay widths.
It has to be remarked, however, that only a few low-dimension condensates
are known and that additional assumptions need to be invoked to treat
higher-dimension condensates.
In particular, a factorisation hypothesis is usually employed, which implies
that higher-dimension condensates are reduced to suitable powers of the
lower-dimension ones.
This results in a further error, difficult to assess precisely, that contributes to the overall uncertainty of the method.
The Wilson coefficients $C_d(Q^2)$ of the expansion \eqref{Piexp} can be expressed in terms of particular sets of Feynman diagrams and, as such, can be computed straightforwardly in perturbative QCD.
Finally, the correlator $\Pi(Q^2)$ can be represented as a dispersive integral,
\be
\Pi(Q^2)=\int_{s_{\rm min}}^\infty ds\; \frac{\rho(s)}{s-Q^2}\,, \qquad \rho(s)=\frac{1}{\pi}\mbox{Im}[\Pi(s)]\,.
\ee
This completes the first stage of the calculation based directly on QCD,
and which involves only its fundamental degrees of freedom, i.e., quarks
and gluons.

Alternatively, according to the quark--hadron duality conjecture, the
correlation function \eqref{Piexp} can be evaluated in terms of low-energy
effective degrees of freedom, i.e., hadrons.
In particular, this amounts to constructing the spectral density $\rho(s)$
as a suitable function that depends on the masses, coupling constants, and
other parameters of the physical hadronic states.
Generically, all hadronic states $h$, for which the matrix element does not
vanish, contribute to $\langle h|j(0)|0\rangle$ and, therefore, to the spectral density $\rho$.
However, the contribution of high-lying broad hadronic states is usually taken into account effectively through a smooth continuum function added to the
sharp contributions from the low-lying narrow states.
This smooth function is expected to vanish below a certain threshold $s_0$.

To improve the convergence of the operator product expansion series by
suppressing higher-order contributions,
it is common to follow the original works~\cite{Shifman:1978bx,Shifman:1978by} and to employ a Borel transformation depending on a parameter $M$ called the
Borel mass.
The typical value of this parameter is set by the mass scale of the hadronic
system under study.
The results can be regarded as reliable only if they are robust with respect
to a variation of the parameter $M$ in a sufficiently wide range --- called
the Borel window.
Choosing a Borel parameter that is too small amounts to truncating the
operator product expansion series before accounting for all contributions
necessary to describe the hadronic system with enough precision.
On the other hand, too large values of $M$ would allow the poorly controlled
continuum to contribute.
Thus, the appropriate choice of the Borel parameter is a crucial point in
the approach.
Furthermore, consistency requires the lower bound of the continuum
contribution $s_0$ to depend on the value of the Borel parameter
$M$~\cite{Lucha:2007pz}.
Indeed, it was noticed a long time ago that
choosing a constant threshold
$s_0$ results in an insufficient accuracy of the
calculation~\cite{Pascual:1984zb}.

Studying exotic hadronic states in the sum rules approach is more difficult
than considering just generic quarkonia.
On the one hand, as was mentioned above, the operator product expansion
series has to be truncated at higher orders to provide a sufficient level
of convergence.
This implies that higher-order condensates need to be retained in the series.
The evaluation of such higher-order condensates requires additional
approximations, such as the factorisation assumption.
For example, the quartic quark condensate is expressed as $\langle\bar{q}q\bar{q}q\rangle=\rho\langle\bar{q}q\rangle^2$, where the coefficient $\rho$ is
expected to be of order one.
However, choosing the best value of $\rho$ and estimating the corresponding
uncertainty requires further theoretical assumptions.
On the other hand, the interpolating current $j(x)$ entering the correlation
function needs to be constructed in such a way that it overlaps with all
relevant, ordinary and exotic, components of the hadronic state,
hence, not only quark-antiquark components, but also hybrid, molecular,
and compact tetraquark ones.
The mixing angles between the different components in the interpolating current are additional parameters.

\begin{table}[ht]
\begin{center}
\begin{tabular}{|c|c|c|} \hline
 state & structure & $J^{PC}$ \\ \hline
 $\chi_{c1}(3872)$ aka $X(3872)$ & mixed $\chi_{c1}-D\bar{D}^*$ & $1^{++}$ \\
 $Z_c(3900)^+$ & $D\bar{D}^*$ & $1^{+-}$ \\
 $X(3940)$ & mixed $\chi_{c0}-D^*\bar{D}^*$ & $0^{++}$ \\
 $Z_c(4020)^+$ & $D^*\bar{D}^*$ & $1^{+-}$ or $2^{++}$ \\
 $Z_c(4100)^+$ & $D_0^*\bar{D}_0^*$ & $0^{++}$ \\
 $X(4160)$ & $D_s^*\bar{D}_s^*$ & $2^{++}$ \\
 $Z_c(4200)^+$ & $[cs][\bar{c}\bar{s}]$ & $1^{+}$ \\
 $X(4250)^+$ & $D\bar{D}_1$ & $1^-$ \\
 $\psi(4260)$ aka $Y(4260)$ & mixed $J/\psi-[cq][\bar{c}\bar{q}]$ & $1^{--}$ \\
 $\psi(4360)$ aka $Y(4360)$ & $[cq][\bar{c}\bar{q}]$ & $1^{--}$ \\
 $\psi(4660)$ aka $Y(4660)$ & $[cs][\bar{c}\bar{s}]$ & $1^{--}$ \\
\hline
 \end{tabular}
 \caption{Structures and quantum numbers of several charmonium-like states obtained within the sum rules approach (adapted from Ref.~\cite{Albuquerque:2018jkn}).
 Only those states that appear in the listings of the PDG with the
assumed or unknown quantum numbers are kept and listed using their
official names.
 \label{tabfinal}}
 \end{center}
 \end{table}

The sum rule approach can be used not only to calculate masses but also to study three--point functions like for magnetic moments.
For the evaluation of magnetic moments of the $Z(3900)$ state see, e.g., Refs.~\cite{Wang:2017dce,Ozdem:2017jqh}.

To summarize, the sum rules approach to exotic hadrons is a strong
analytical tool aimed at relating the properties of the hadronic states
under study directly with intrinsic features of the QCD vacuum encoded in various local condensates.
Shortcomings of this approach are the necessity of additional assumptions
and approximations that can make systematic improvements and reliable
uncertainty estimates difficult, especially for exotic near-threshold states.
Predictions of the sum rules for the structures and quantum numbers of $XYZ$ states in the charmonium spectrum are collected in Table~\ref{tabfinal}, taken from the recent review~\cite{Albuquerque:2018jkn}.
There the interested reader can find further details of the method and
its application to exotic hadrons.
Results from sum rules for charmonium and bottomonium hybrids have been
briefly summarized also in Sec.~\ref{Sect:4.2.3bis}.
For recent developments on the foundation of sum rules for compact tetraquarks and
their specific differences with respect to sum rules for ordinary quarkonia we refer to Ref.~\cite{Lucha:2019pmp}.

\subsection{Effective field theories for QCD}
\label{Sect:4.2}
In a physical system, the long-distance (or low-energy) dynamics usually
does not depend on the details of the physics at short distances (or
high energies). For example, the details of atomic physics are irrelevant
to describe planetary motion; the only feature of the nucleus relevant to
chemistry is its charge. Effective field theories (EFTs) are a theoretical
tool that allows to put this generic observation into a rigorous framework.
In particular, under some conditions, EFTs are able to describe systems
characterized by several energy scales just in terms of a few relevant degrees of freedom.
This description is improvable in a systematic way.
It has, however, a limited, but well defined, range of applicability.

A necessary condition for the construction of an effective field theory
is the existence of a hierarchy amongst the energy scales of the system.
For instance, in the case of a system with two scales, $\Lambda$ and $Q$, we can construct an EFT if $\Lambda \gg Q$.
The effective field theory Lagrangian is then organized as an expansion in $Q/\Lambda$.
Each term in the expansion is made of the fields describing the system at the low-energy scale $Q$, also called effective degrees of freedom.
It can be any term as long as it is consistent with the symmetry principles.
In turn, the resulting scattering matrix will be the most general one consistent with analyticity,
perturbative unitarity, cluster decomposition and the symmetry principles~\cite{Weinberg:1978kz}.

Analytic terms in the expansion parameter $Q/\Lambda$ are accounted for by the operators of the EFT.
Non-analytic terms, carrying the contributions of the high-energy modes in the original, fundamental theory, which are no longer dynamical in the EFT,
are encoded in the parameters multiplying the EFT operators.
These parameters are the Wilson coefficients of the EFT, also called low-energy constants in hadronic EFTs like Chiral Perturbation Theory.
Hence, EFTs automatically factorize, for any observable, high-energy from low-energy contributions.
The Wilson coefficients of the EFT Lagrangian are fixed by matching to the fundamental theory,
i.e. by requiring the EFT and the fundamental theory to describe the same physics
(observables, Green functions, scattering matrices, ...) at any given order of the expansion parameter $Q/\Lambda$.
Alternatively they can also be fixed from data. In case of hadronic EFTs they are fixed either from experimental
data or from matching to results from lattice QCD.

To allow for controlled calculations based on the effective Lagrangian,
operators, as well as the quantum corrections, are organized according to their expected importance.
Operators in the Lagrangian are counted in powers of the small expansion
parameter $Q/\Lambda$, whereas quantum corrections are either computed exactly or counted in powers of the coupling constant.
EFTs are, in general, non-renormalizable, however, they are at each order in the expansion parameter.
This leads eventually to finite results, whose different terms scale according to the power counting.
The power counting may or may not be obvious.
Nevertheless, once a power counting has been established, effective field theories prove to be very useful, sometimes the only tool,
to compute in a systematic and rigorous fashion observables in multiscale systems.

One of the strengths of EFTs is that they enjoy at leading order a larger symmetry than the fundamental theory,
as a consequence of the fact that some of the original degrees of freedom have been integrated out. For example, the heavy-quark symmetry of the
Heavy Quark Effective Theory (see Sec.~\ref{Sect:4.2.1}) is a hidden, approximate symmetry of the physical system
that is made manifest by the EFT at leading order.
Although higher-order terms in the EFT restore the symmetry of the full theory, having a more symmetric zeroth-order solution
contributes to the predictive power of the EFT by reducing the number of its low-energy parameters.

The fundamental theory for hadronic physics is QCD. Its Lagrangian density reads (for simplicity gauge fixing,
ghost fields and a possible $\theta$-term are omitted)~\cite{Fritzsch:1973pi}
\begin{equation}
{\cal L}_{\rm QCD}=\sum_{i=1}^{n_f} \bar q_i\left(i\gamma_\mu D^\mu - m_i\right)q_i -\frac{1}{4}F_{\mu\nu}^aF^{\mu\nu\;a}\ ,
\end{equation}
where the quark fields are denoted by $q_i$, the sum runs over the $n_f$
quark flavors, $m_i$ are the quark masses,
the covariant derivative is $D^\mu = \partial^\mu - i g A^\mu$,
with $A^\mu = A^{\mu\,a}T^a$ the gluon fields and $T^a$ the SU(3) generators,
$F_{\mu\nu}^a$ is the field strength tensor, and $g$ is the gauge coupling,
related to the strong coupling via $\alpha_s = g^2/(4\pi)$.
Of particular relevance in hadronic physics are the EFTs that can be constructed from QCD in the limit of small and large quark masses.

For light quarks we can expand the QCD Lagrangian for small masses and write at leading order
\begin{equation}
{\cal L}_{\rm QCD} = \sum_{i=1}^{n_\ell} \bar q_{i\,L}\left(i\slashed{\partial} + g \slashed{A}^a T^a\right) q_{i\,L}
+ \bar q_{i\,R}\left(i\slashed{\partial} + g \slashed{A}^a T^a\right) q_{i\,R} - \frac{1}{4}F_{\mu\nu}^aF^{\mu\nu\;a},
\end{equation}
where $q_{i\,R}$ ($q_{i\,L}$) are the right- (left-) handed components of the $n_\ell$ light-quark fields.
In this limit, the Lagrangian is U(1)$_V$ $\times$SU$(n_\ell)_V$ $\times$U$(1)_A$ $\times$SU($n_\ell)_A$ invariant ($V$ stands for vector, $A$ for axial vector).
However, the SU($n_\ell)_V \times$SU($n_\ell)_A$ symmetry is spontaneously broken to SU($n_\ell)_V$, which has
as a consequence that, at low energies, the effective degrees of freedom
of QCD are the Goldstone bosons that emerge from the symmetry breaking rather than quarks and gluons.
The effective field theory that describes QCD at low energies is called
Chiral Perturbation Theory (ChPT)~\cite{Gasser:1983yg,Leutwyler:1993iq}.
Its effective degrees of freedom are the Goldstone bosons of the
spontaneously broken SU($n_\ell)_A$ symmetry. For $n_\ell=2$ these are pions.
The exploited hierarchy of scales is $\Lambda_\chi \gg M_\pi$, where $M_\pi$ is the pion mass and $\Lambda_\chi\sim 4\pi f_\pi\sim 1$~GeV is the hadronic scale associated
with the spontaneous breaking of the chiral symmetry; $f_\pi$ is the pion decay constant.
Hence ChPT is constructed as a perturbative expansion in powers of $M_\pi/\Lambda_\chi$.
Because ChPT is an EFT for degrees of freedom that live below the hadronic
scale, its matching to QCD is non-perturbative.
Chiral Perturbation Theory plays a crucial role in describing nuclei and hadronic molecules, see Refs.~\cite{Epelbaum:2008ga} and~\cite{Guo:2017jvc} and references therein.
We will discuss it in this context in Sec.~\ref{Sect:4.2.4}.

For hadrons made of one heavy quark, like heavy-light mesons and baryons,
the proper EFT is called Heavy Quark Effective Theory (HQET).
It exploits the hierarchy of scales $m_h \gg \Lambda_{\rm QCD}$, where $m_h$ is
the heavy quark mass and $\Lambda_{\rm QCD}$ is the typical, hadronic scale, of the same order as
$\Lambda_\chi$, relevant for these systems.
The condition $m_h \gg \Lambda_{\rm QCD}$ is fulfilled by the charm, bottom,
and top quarks.\footnote{
Top quarks will not play any role in the following, as this review focuses
on hadrons and the top quark has no time
to form any before decaying weakly into a $b$ quark.
}
In a sense, the HQET describes QCD in the opposite limit of ChPT, however,
the HQET is not the large mass limit of QCD.
We will discuss the HQET in the following Sec.~\ref{Sect:4.2.1}.
If we consider systems made of more than one heavy quark, like quarkonia or
quarkonium-like states or doubly-heavy baryons, then more scales become relevant.
These are the scales of the heavy-quark-heavy-(anti)quark interaction, i.e., the
typical momentum transfer, $m_h v$, and the typical binding energy $m_h v^2$,
where $v$ is the relative velocity of the heavy quarks.
At each of these scales one can construct an EFT, specifically, NRQCD at
the scale $m_h v$, which will be discussed in Sec.~\ref{Sect:4.2.2}, and potential NRQCD (pNRQCD)
at the scale $m_h v^2$, which will be discussed in Sec.~\ref{Sect:4.2.3}.
The version of pNRQCD suited to deal with systems made of two or more heavy
quarks bound with some light degrees of freedom, like light quarks or gluons,
is called Born--Oppenheimer EFT and will be presented in Sec.~\ref{Sect:4.2.3bis}.
Finally, effective field theories at the hadron level are described in Sec.~\ref{Sect:4.2.4}.

\subsubsection{Heavy Quark Effective Field Theory}
\label{Sect:4.2.1}
The HQET is the EFT suited to describe hadrons made of one heavy particle and light degrees of freedom,
also called heavy-light hadrons~\cite{Isgur:1989vq,Isgur:1989ed,Isgur:1991wq,Eichten:1989zv}
(for an early review see, for instance, Ref.~\cite{Neubert:1993mb}, for a textbook see Ref.~\cite{Manohar:2000dt}).
The heavy particle was originally designated to be a heavy quark.
However, under some circumstances, it can be also a composite particle made by more than one heavy quark;
this is the case when the internal modes of the composite heavy particle may be ignored.
The light degrees of freedom are made by quarks and gluons. Among the light
quarks we may distinguish between valence quarks and
sea quarks, where the first ones are those that establish, together with
the heavy degrees of freedom, the quantum numbers of the heavy-light hadron.

The HQET exploits the hierarchy $m_h \gg \Lambda_{\rm QCD}$ that characterizes heavy-light hadrons made by a heavy quark of mass $m_h$.
High-energy degrees of freedom that live at the energy scale $m_h$ are integrated out from QCD.
The resulting EFT is made of low-energy degrees of freedom living at the scale $\Lambda_{\rm QCD}$.
These are the low-energy modes of the heavy quark (antiquark), described by a Pauli spinor $\psi$ ($\chi$) that
annihilates (creates) the heavy quark (antiquark), and low-energy gluons and light quarks.
The HQET is constructed as an expansion in $1/m_h$: the heavy quark expansion.
Matrix elements of operators of dimension $d$ are of order $\Lambda_{\rm QCD}^d$, hence the higher the dimension of the operator the higher the suppression in $\Lambda_{\rm QCD}/m_h$.
In the rest frame of the heavy-light hadron, the HQET Lagrangian density for a heavy quark reads up to order $1/m_h^2$
(the HQET Lagrangian including $1/m_h^4$ terms has been derived in Refs.~\cite{Gunawardana:2017zix,Kobach:2017xkw})
\begin{align}
{\cal L}_{\rm HQET} = &\, \psi^\dagger \Biggl\{ i D_0 + \, \frac{{\bm{D}}^2}{2 m_h} - c_F\, \frac{\bm{\sigma} \cdot g{\bm{B}}}{2 m_h}
- c_D \, \frac{ \left[{\bm{D}} \cdot, g{\bm{E}} \right]}{8 m_h^2}
- i c_S \, \frac{\bm{\sigma} \cdot \left[{\bm{D}} \times, g{\bm{E}} \right] }{8 m_h^2} \Biggr\} \psi
\nonumber\\
&- \frac{1}{4} F^a_{\mu \nu} F^{a\,\mu \nu} + \frac{d_2}{m_h^2} F^a_{\mu \nu} D^2 F^{a\,\mu \nu} - \frac{d_3}{m_h^2} g f_{abc}F^a_{\mu\nu} F^b_{\mu\alpha} F^c_{\nu\alpha}
+ \sum_{\ell=1}^{n_\ell} \bar q_\ell\left(i\gamma_\mu D^\mu - m_\ell\right)q_\ell ,
\label{eq:sect4:HQET}
\end{align}
where $[{\bm{D}} \cdot, g{\bm{E}}] = {\bm{D}} \cdot g{\bm{E}} - g {\bm{E}} \cdot {\bm{D}}$ and $[{\bm{D}} \times, g{\bm{E}}]={\bm{D}} \times g{\bm{E}} - g{\bm{E}} \times {\bm{D}}$,
${\bm{E}}^i = F^{i0}$ is the chromoelectric field, ${\bm{B}}^i = -\epsilon_{ijk}F^{jk}/2$ the chromomagnetic one with the totally antisymmetric tensor
$\epsilon_{ijk}$ ($\epsilon_{123}=1$), and $\bm{\sigma}$ are the Pauli matrices.
The fields $q_\ell$ stand for $n_\ell$ light-quark fields, i.e., the mass $m_\ell$ is much smaller than $\Lambda_{\rm QCD}$ and may be set to zero in many applications.
On the other hand, $m_h$ has to be understood as the heavy quark pole mass, hence not the mass in the QCD Lagrangian.
The coefficients $c_F$, $c_D$, $c_S$, $d_2$, and $d_3$ are Wilson coefficients of the EFT.
They encode the contributions of the high-energy modes that have been integrated out from QCD. Since the high-energy scale, $m_h$, is larger than
$\Lambda_{\rm QCD}$, the Wilson coefficients may be computed in perturbation theory and organized as an expansion in $\alpha_s$ (at a typical scale of order $m_h$).
The coefficients $c_F$, $c_D$, and $c_S$ are 1 at leading order, while the perturbative series of the coefficients $d_2$ and $d_3$ starts at order $\alpha_s$.
The one-loop expression of the coefficients may be found in Ref.~\cite{Manohar:1997qy}.
Some of the coefficients are known far beyond one loop. For instance, the Fermi coefficient $c_F$, which plays a crucial role in the spin splittings,
is known up to three loops~\cite{Grozin:2007fh}. Not all the coefficients are independent.
For instance, Poincar\'e invariance of QCD relates $c_F$ and the spin-orbit coefficient $c_S$: $c_S = 2 c_F-1$~\cite{Luke:1992cs,Manohar:1997qy,Brambilla:2003nt}. This relation is exact.
The HQET Lagrangian for a heavy antiquark may be obtained from Eq.~\eqref{eq:sect4:HQET} by charge conjugation.
In Eq.~\eqref{eq:sect4:HQET} we have not considered $1/m_h^2$ suppressed operators involving light quarks, since their impact is negligible in most hadronic observables.
They have been considered first in Ref.~\cite{Balzereit:1998am}; for a recent calculation see Ref.~\cite{Moreno:2017sgd}.

The impact of the HQET on the physics involving heavy-light hadrons and, in particular, their weak decays has been enormous.
The reason is that the leading-order HQET Lagrangian (${\cal L}_{\rm HQET} = \psi^\dagger i D_0 \psi - F^a_{\mu \nu} F^{a\,\mu \nu}/4$) makes manifest a hidden symmetry of heavy-light hadrons.
This symmetry is the heavy-quark symmetry and stands for invariance with respect to the heavy-quark flavor and spin.
Moreover, the leading-order HQET Lagrangian is exactly renormalizable.
Higher-order operators in Eq.~\eqref{eq:sect4:HQET} break this symmetry (and exact renormalizability), however, they do it in a controlled, perturbative way.
Hence, observables computed up to some order in the HQET expansion depend on fewer and more
universal non-perturbative matrix elements than they would in a full QCD calculation.
This makes the heavy quark expansion more predictive than a full QCD calculation.
We will mention in the following a few implications of the heavy quark
expansion for the spectrum of heavy-light hadrons.

Heavy-light meson masses, expressed in the HQET as an expansion up to order $1/m_h$ in the inverse of the heavy quark mass~\cite{Falk:1992wt}, read
\begin{equation}
M_{H^{(*)}} = m_h + \bar{\Lambda} + \frac{\mu_\pi^2}{2m_h} - d_{H^{(*)}} \frac{\mu_G^2(m_h)}{2m_h} + \mathcal{O}(1/m_h^2),
\label{eq:HQETmesonmass}
\end{equation}
where $M_{H^{(*)}}$ is the spin singlet (triplet) meson mass, $m_h$ the heavy
quark pole mass, $\bar{\Lambda}$ the binding energy in the static limit,
of order $\Lambda_{\rm QCD}$, ${\mu_\pi^2}/{2m_h}$ the kinetic energy of the heavy quark ($\mu_\pi^2$ is the matrix element of
$\psi^\dagger{\bm{D}}^2\psi$), of order $\Lambda_{\rm QCD}^2/m_h$, $d_{H^{(*)}}$ is 1 for $H$ and $-1/3$ for $H^*$,
and $d_{H^{(*)}} {\mu_G^2(m_h)}/{2m_h}$ is the matrix element of $c_F \,\psi^\dagger\bm{\sigma}\cdot g{\bm{B}}/(2m_h)\psi$, of order $\Lambda_{\rm QCD}^2/m_h$.
The heavy quark symmetry manifests itself through the universality of the leading term $M_{H^{(*)}} - m_h \approx \bar{\Lambda}$,
as well as of the matrix elements $\mu_\pi^2$ and $\mu_G^2(m_h)/c_F(m_h)$, which depend neither on the heavy quark flavor nor on the heavy quark spin.
The flavor dependence of $\mu_G^2(m_h)$ comes entirely from the Wilson coefficient $c_F$, which depends on $m_h$ through the running of the strong coupling.

A primary use of Eq.~\eqref{eq:HQETmesonmass} is to determine the heavy quark
masses from the (measured) meson masses.
Since the relation between the $\overline{\text{MS}}$ mass (or any ultraviolet
mass) and the pole mass is given by a poorly convergent perturbative series
(at present, this relation is known up to four loops~\cite{Marquard:2015qpa,Marquard:2016dcn})
\footnote{For model-dependent estimates of the five- and six- loop contributions
see, for instance, Ref.~\cite{Kataev:2018mob}.}, an intermediate step is necessary.
This consists in rewriting Eq.~\eqref{eq:HQETmesonmass} in a scheme that preserves the power counting of the EFT,
which excludes a direct use of the $\overline{\text{MS}}$ scheme, and defines a mass that is related to the $\overline{\text{MS}}$ mass through a convergent perturbative series.
There are many possibilities for such a scheme~\cite{Uraltsev:1996rd,Beneke:1998rk,Hoang:1999ye,Pineda:2001zq,Hoang:2008yj,Brambilla:2017hcq}.
The approach favored by the recent lattice determination of Ref.~\cite{Bazavov:2018omf} is the minimal renormalon subtraction (MRS) scheme~\cite{Brambilla:2017hcq}.
The obtained heavy quark masses from the $D_s$ and $B_s$ masses read
\begin{align}
\overline{m}_c &= 1273 (4)_\text{stat} (1)_\text{syst} (10)_{\alpha_s} (0)_{f_{\pi,\text{PDG}}}~\text{MeV} = 1273(10)~\text{MeV},\\
\overline{m}_b &= 4201 (12)_\text{stat} (1)_\text{syst} (8)_{\alpha_s} (1)_{f_{\pi,\text{PDG}}}~\text{MeV} = 4201(14)~\text{MeV} ,
\end{align}
where $\overline{m}_h$ is the $\overline{\text{MS}}$ mass of the quark $h$
at the scale of its $\overline{\text{MS}}$ mass,
and the uncertainties are statistical, systematic, due to uncertainties in $\alpha_s$ and in the pion decay constant $f_\pi$ from the PDG
used to set the lattice scale.

The quantities $\bar{\Lambda}$, ${\mu_\pi^2}$ and ${\mu_G^2(m_h)}$ are non-perturbative. The lattice determination of Ref.~\cite{Bazavov:2018omf} finds
\begin{align}
 \bar{\Lambda}_{\text{MRS}} &= 555 (25)_\text{stat} (8)_\text{syst} (16)_{\alpha_s} (1)_{f_{\pi,\text{PDG}}}~\text{MeV},\\
 \mu_\pi^2 	 &= 0.05 (16)_\text{stat} (13)_\text{syst} (06)_{\alpha_s} (00)_{f_{\pi,\text{PDG}}}~\text{GeV}^2, \\
 \mu_G^2(m_b) &= 0.38 (01)_\text{stat} (01)_\text{syst} (00)_{\alpha_s} (00)_{f_{\pi,\text{PDG}}}~\text{GeV}^2,
\end{align}
where $\bar{\Lambda}$ is in the MRS scheme and the quantity $\mu_G^2$ has been evaluated for the $b$ quark.
At higher orders in the $1/m_h$ expansion, more matrix elements have to be determined, a fact that increasingly limits the predictive power of the HQET.

Clearly, Eq.~\eqref{eq:HQETmesonmass} can be immediately extended to heavy-light baryons.
What will change is the explicit value of the non-perturbative matrix elements, as the light degrees of freedom are different from the mesonic case.
More interesting, from a conceptual point of view, is the case of doubly-heavy baryons.
These systems are characterized by more energy scales than heavy-light hadrons.
One of these scales is the inverse of the typical distance between the two heavy quarks.
If it is much larger than $\Lambda_{\rm QCD}$, a case more likely to be realized by doubly bottomed baryons than by doubly charmed baryons, then one can integrate out this scale.
At the low-energy scale of the resulting EFT, one does not resolve the two heavy quarks anymore and the resulting EFT is just the HQET (for the heavy antiquark)
with the role of the heavy antiquark given to an effective heavy diquark field
with the same quantum numbers~\cite{Savage:1990di,Bardeen:2003kt,Brambilla:2005yk,Fleming:2005pd,Mehen:2006vv,Ma:2015cfa,Mehen:2017nrh,Ma:2017nik,Cheng:2018mwu,Mehen:2019cxn}.
This is an appealing picture with predictive power, as it links the doubly-heavy baryon observables to the much better known heavy-light meson ones.
Nevertheless, it has been also challenged by lattice data that, at least in the doubly charmed baryon spectrum, do not
seem to fully support the diquark picture~\cite{Padmanath:2015jea}.
It is anyway clear that for any doubly-heavy baryon, and even more so for doubly charmed baryons, the heavy quark-quark dynamics
provides a contribution that cannot be entirely neglected and that a diquark picture is theoretically justified
only if the typical heavy quark distance is much smaller than $1/\Lambda_{\rm QCD}$.
Doubly charmed baryons have attracted more attention since the LHCb discovery of a resonance in
the $\Lambda_c^+K^-\pi^+\pi^-$ mass spectrum at a mass of $(3621.40\pm0.78)$~MeV that is consistent with a $\Xi^{++}_{cc}$ baryon~\cite{Aaij:2017ueg,Aaij:2018wzf}.
Earlier observations by the SELEX experiment remained unconfirmed by other experiments~\cite{Mattson:2002vu,Ocherashvili:2004hi} and show
a huge isospin violation difficult to understand from an EFT point of view~\cite{Brodsky:2011zs}.

The heavy quark symmetry may be also applied to link doubly-heavy tetraquarks (tetraquarks made of two heavy quarks
and two light anti-quarks)
with heavy-light baryons sharing the same light-quark content~\cite{Eichten:2017ffp,Mehen:2017nrh}.
Even more relevant now that we have an experimental determination of a doubly-heavy baryon.
In Ref.~\cite{Eichten:2017ffp}, using the mass formula in Eq.~\eqref{eq:HQETmesonmass} and experimental input for heavy-light baryon and meson masses,
the authors show evidence, at least from the HQET, that there are doubly bottomed tetraquark states $bb\bar{u}\bar{d}$, $bb\bar{u}\bar{s}$, and $bb\bar{d}\bar{s}$
that are stable against strong decays
(these are states for which the decays into two heavy-light meson pairs, and a doubly-heavy baryon and a light antibaryon turn out to be kinematically forbidden),
while the doubly charmed tetraquark $cc\bar{q}_\ell\bar{q}_{\ell'}$, mixed tetraquark $bc\bar{q}_\ell\bar{q}_{\ell'}$,
and heavier doubly bottomed tetraquark states dissociate into pairs of heavy-light mesons.
The existence of a stable doubly bottomed $I = 0$ tetraquark has been predicted by the quark model of Ref.~\cite{Karliner:2017qjm},
the lattice QCD calculations in Refs.~\cite{Francis:2016hui,Bicudo:2016ooe},
and it is supported by the arguments of Ref.~\cite{Bicudo:2016ooe}.
In Ref.~\cite{Mehen:2017nrh} it has been pointed out that if the mass of this stable tetraquark is smaller than $10405$~MeV
(this is the case for the prediction in Ref.~\cite{Karliner:2017qjm}),
then the lowest-lying doubly bottomed tetraquark with quantum numbers $J^P = 1^+$ and $I = 1$ would also likely be stable against strong decays.

\begin{figure}[h]
\begin{center}
\includegraphics[width=0.55\linewidth]{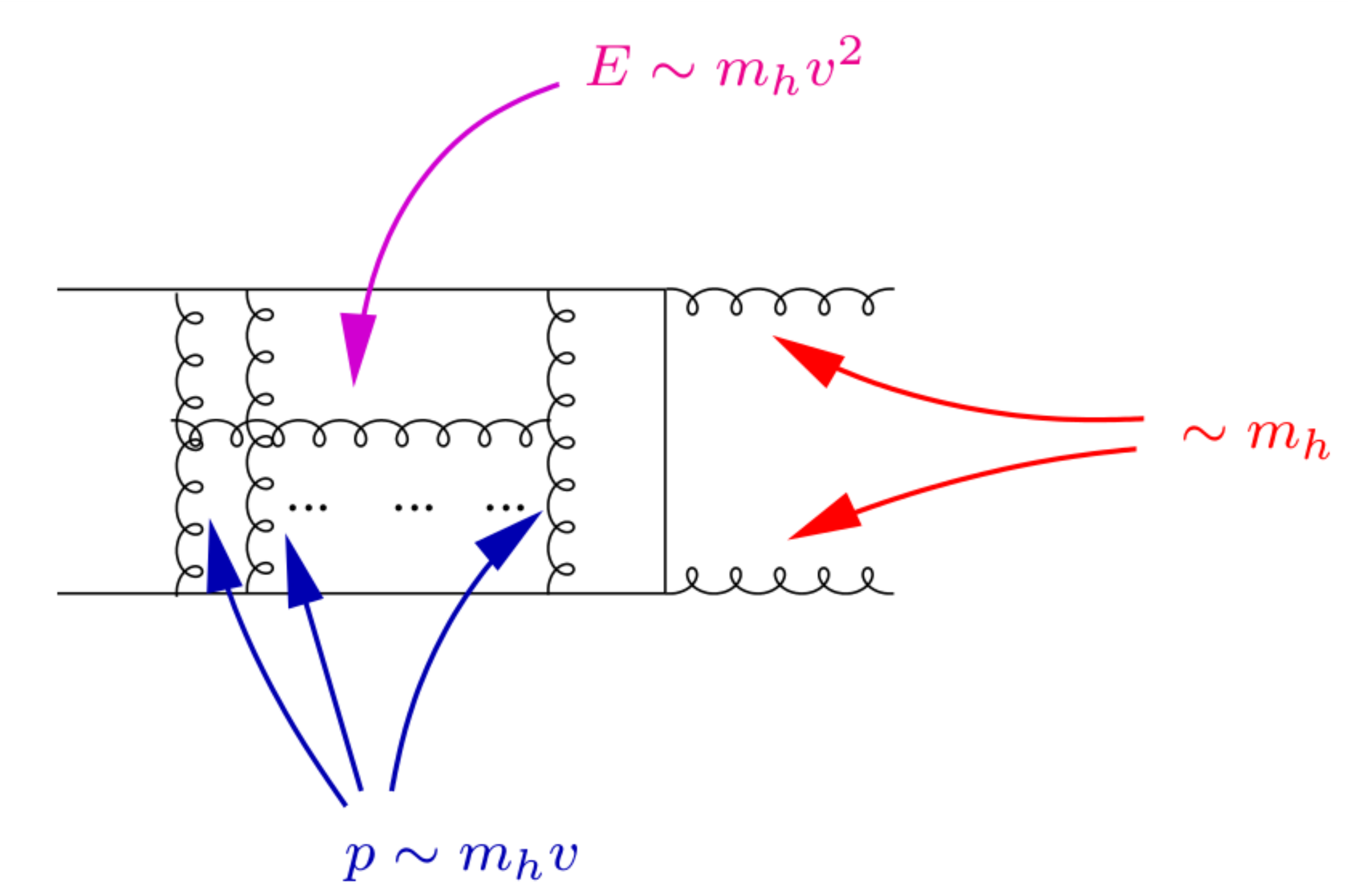}\includegraphics[width=0.45\linewidth]{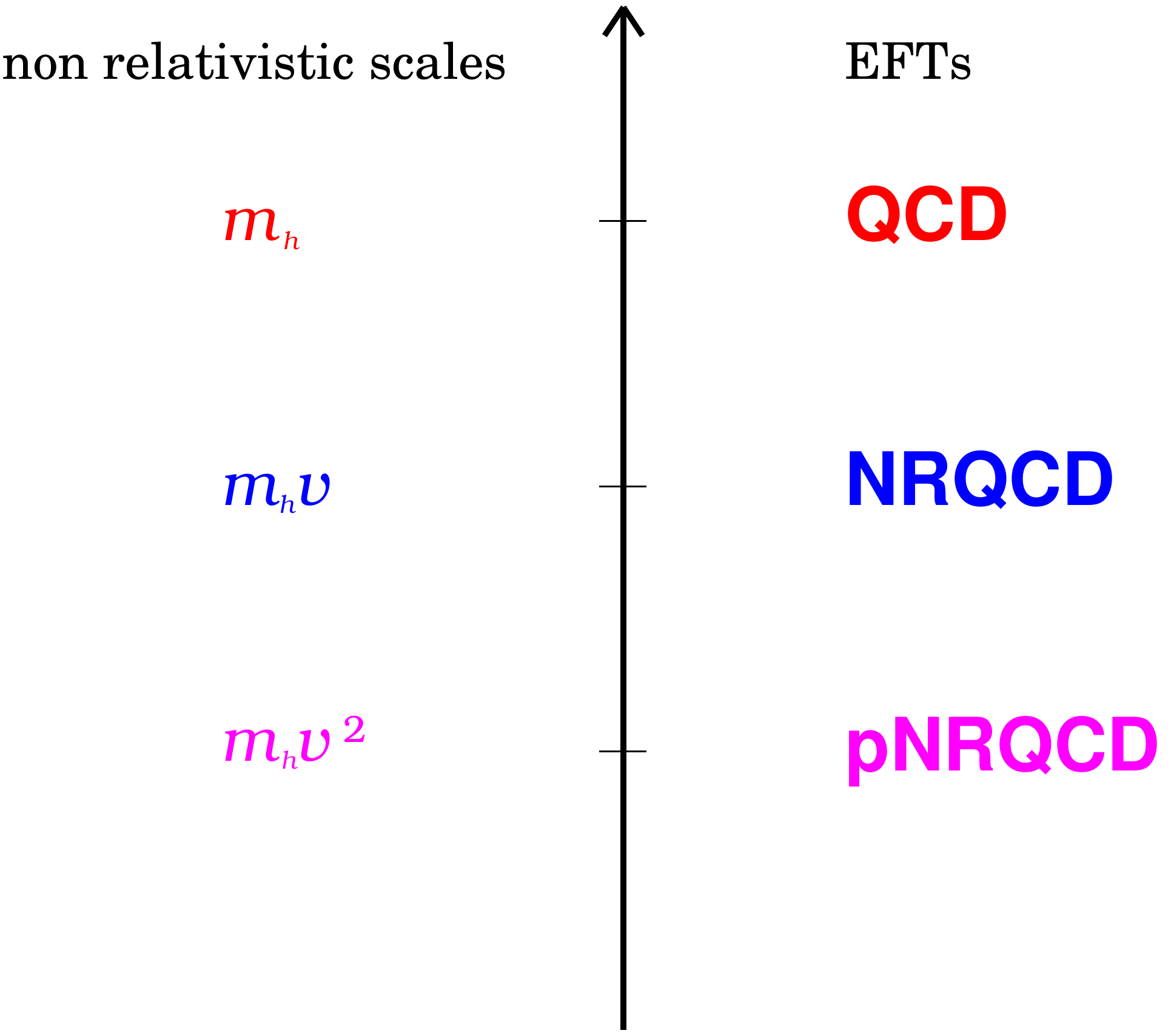}
\caption{Left: Typical energy scales entangled in a quarkonium annihilation. Right: Hierarchy of energy scales and corresponding EFTs.}
\label{fig:scalesNR}
\end{center}
\end{figure}

\subsubsection{Non Relativistic QCD}
\label{Sect:4.2.2}
Systems made of two or more non-relativistic particles, like two heavy quarks or a heavy quark and a heavy antiquark,
are characterized by more energy scales than systems where just one particle is non-relativistic.
These energy scales are the inverse radius (or radii) defining the size of the heavy particle system and
the energy of the excitations of the heavy particle system. In the case of heavy quarks of mass $m_h$, the inverse
size of the heavy quark system is of order $m_hv$ and the energy of its excitations is of order $m_hv^2$,
where $v\ll 1$ is the velocity of the heavy quark relative to the c.m. frame (for simplicity, we do not distinguish here between heavy quarks of different flavor).
This is analogous to what happens in the hydrogen atom, whose inverse Bohr radius is $m_e \alpha$ and whose
energy levels are of order $m_e \alpha^2$, $m_e$ being the mass of the electron and $\alpha$ the fine structure constant.
The fine structure constant also provides the relative velocity of the electron in the atom.
In QCD, the velocity $v$ of the heavy quark may (for weakly-coupled bound states) or may not (for strongly-coupled bound states)
be proportional to the strong coupling~$\alpha_s$. Regardless of this, the relative velocity of the heavy quark is much smaller than the velocity of light,
and may serve to hierarchically order the non-relativistic energy scales:
\begin{equation}
 m_h \gg m_h v \gg m_h v^2.
\label{sect4:scales}
\end{equation}
The non-relativistic energy scales are also correlated.
The tower of hierarchically ordered energy scales in Eq.~\eqref{sect4:scales} calls for the construction of a tower of non-relativistic EFTs~\cite{Brambilla:2004jw},
see Fig. \ref{fig:scalesNR}.
In the last twenty years, the development of non-relativistic EFTs of QCD has been the major theoretical breakthrough
in the description of quarkonium and quarkonium-like systems~\cite{Brambilla:2004wf,Brambilla:2010cs,Brambilla:2014jmp}.
For a more historical perspective, see Ref.~\cite{Vairo:2009rs}.

\begin{figure}[h]
\begin{center}
\includegraphics[width=1\linewidth]{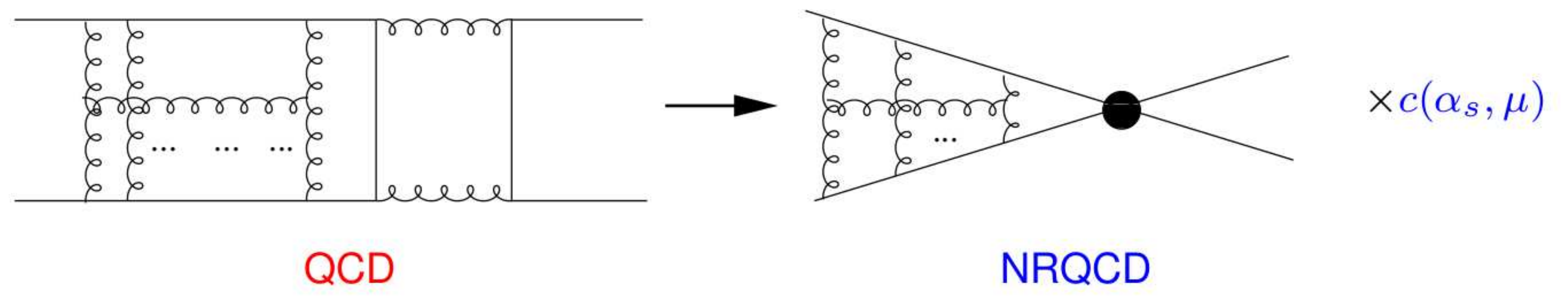}
\caption{Schematic matching of NRQCD; $c$ is a Wilson coefficient of NRQCD, $\mu$ is the infrared renormalization scale of NRQCD.}
\label{fig:matchingNRQCD}
\end{center}
\end{figure}

The EFT that follows from QCD by integrating out the energy scale of the heavy quark mass, $m_h$, which is the largest scale in Eq.~\eqref{sect4:scales},
and that is suited to describe systems made of heavy quarks and heavy antiquarks, is NRQCD~\cite{Caswell:1985ui}, see Fig. \ref{fig:matchingNRQCD}.
Although the Lagrangian of the theory is identical to the one of the HQET in the two-fermion and gauge sectors, nevertheless the power counting is different.
One consequence of this is that the leading-order NRQCD Lagrangian includes the kinetic energy operators,
$\psi^\dagger \bm{\nabla}^2/(2 m_h)\psi - \chi^\dagger \bm{\nabla}^2/(2 m_h)\chi$.
Therefore, differently from the HQET leading-order Lagrangian, it is not renormalizable.

Because of this seeming difficulty, NRQCD has been used first for lattice QCD calculations involving heavy quarks~\cite{Thacker:1990bm,Lepage:1992tx}.
The advantage there is that, once the heavy quark mass has been integrated out, the lattice spacing, $a$, is no more constrained to be smaller than $1/m_h$,
which would amount to requiring a very fine lattice if the quark is very heavy. In lattice NRQCD the constraint is relaxed to $a < 1/(m_h v)$.
Since at the same time the lattice size has to be large enough to include distances of the order of $1/\Lambda_{\text{QCD}}$ for quenched calculations and
$1/M_\pi$ for full calculations, simulations with very heavy quarks in full QCD are so far beyond reach for the required computational effort.
Indeed, still nowadays, lattice NRQCD is the only way to compute non-perturbatively observables involving bottom quarks in full QCD
(see, for instance, Refs.~\cite{Colquhoun:2014ica,Hughes:2015dba,Colquhoun:2015fuw,Hughes:2017spc,Lytle:2018ugs}).

Only after the development of the HQET, NRQCD has been systematically used for analytical calculations of quarkonium observables.
In particular, NRQCD is well suited to describe heavy quark-antiquark annihilation processes.
These happen at the energy scale $m_h$, which is exactly the energy scale that has been integrated out from QCD.
At the low-energy resolution of NRQCD no annihilation is possible and the number of heavy degrees of freedom is conserved.
All the information about the annihilation goes instead into the (imaginary part) of the four-fermion Wilson coefficients of NRQCD, which may be computed in perturbative QCD,
while the low-energy dynamics of the heavy quark-antiquark bound state is factorized into the matrix elements of the NRQCD operators.
Processes involving heavy quark-antiquark annihilations are quarkonium decays~\cite{Bodwin:1992ye,Bodwin:1994jh} and productions~\cite{Bodwin:1994jh}.
The large amount of data about quarkonium production in hadron and lepton colliders, together with the predictive power of NRQCD
and its success in most of the predictions, has established NRQCD as a standard tool for studying quarkonium annihilation processes~\cite{Brambilla:2004wf,Brambilla:2010cs,Brambilla:2014jmp}.

One novel feature of NRQCD with respect to the HQET is the presence of four-fermion operators in the Lagrangian.
These are essential to describe annihilation processes, but also, more generally, to describe correctly the short-distance interaction between the heavy particles.
Because also four-fermion operators projecting on color octet quark-antiquark states are possible,
NRQCD naturally allows for production and decay of quark-antiquark states in a color octet configuration.
Quark-antiquark states in a color octet configuration constitute a suppressed, in $v$, component of the Fock state describing a physical quarkonium.
Four-fermion color octet matrix elements are necessary in the quarkonium phenomenology~\cite{Brambilla:2004wf,Brambilla:2010cs,Brambilla:2014jmp}.
They are also necessary to cancel infrared divergences in quarkonium decay and production observables and provide
eventually finite, physical results~\cite{Bodwin:1992ye,Bodwin:1994jh}.
It should be noted, however, that the NRQCD factorization has been rigorously proved only for quarkonium decay but not yet for quarkonium production
(for progress in this direction see Refs.~\cite{Nayak:2005rw,Nayak:2005rt,Kang:2014tta,Ma:2014svb,Kang:2014pya}).
This unsolved issue may be related with the persisting difficulty in reconciling quarkonium polarization data with naive NRQCD expectations.

A last breakthrough in establishing NRQCD as a valuable tool for analytical
calculations came when it was shown that the computation of the Wilson coefficients of NRQCD
in dimensional regularization requires expanding in the heavy quark mass to avoid integrating over the high momentum region.
The result is that, even if the power countings of NRQCD and the HQET are different, the matching to QCD proceeds in the same way,
leading to the same Lagrangian in the two-fermion and gauge sectors~\cite{Manohar:1997qy}.

The NRQCD Lagrangian density for systems made of a heavy quark and a heavy antiquark of equal masses $m_h$ up to order $1/m_h^2$,
and including the $1/m_h^3$ kinetic operator, is given by
\begin{align}
{\cal L}_{\rm NRQCD} = &\, \psi^\dagger \Biggl\{ i D_0 + \frac{{\bm{D}}^2}{2 m_h} + \frac{{\bm{D}}^4}{8m_h^3}
- c_F\, \frac{\bm{\sigma} \cdot g{\bm{B}}}{2 m_h}
- c_D \, \frac{ \left[{\bm{D}} \cdot, g{\bm{E}} \right]}{8 m_h^2}
- i c_S \, \frac{\bm{\sigma} \cdot \left[{\bm{D}} \times, g{\bm{E}} \right] }{8 m_h^2} \Biggr\} \psi
\nonumber\\
& + \chi^\dagger \Biggl\{ i D_0 - \frac{{\bm{D}}^2}{2 m_h} - \frac{{\bm{D}}^4}{8m_h^3}
+ c_F\, \frac{\bm{\sigma} \cdot g{\bm{B}}}{2 m_h}
- c_D \, \frac{ \left[{\bm{D}} \cdot, g{\bm{E}} \right]}{8 m_h^2}
- i c_S \, \frac{\bm{\sigma} \cdot \left[{\bm{D}} \times, g{\bm{E}} \right] }{8 m_h^2} \Biggr\} \chi
\nonumber\\
& + \frac{f_1(^1S_0)}{m_h^2} \psi^{\dagger} \chi \chi^{\dagger} \psi
+ \frac{f_1(^3S_1)}{m_h^2} \psi^{\dagger} \bm{\sigma} \chi \cdot \chi^{\dagger} \bm{\sigma} \psi
+ \frac{f_8(^1S_0)}{m_h^2} \psi^{\dagger} {\rm T}^a \chi \chi^{\dagger} {\rm T}^a \psi
+ \frac{f_8(^3S_1)}{m_h^2} \psi^{\dagger} {\rm T}^a \bm{\sigma} \chi \cdot \chi^{\dagger} {\rm T}^a \bm{\sigma} \psi
\nonumber\\
& - \frac{1}{4} F^a_{\mu \nu} F^{a\,\mu \nu} + \frac{d_2}{m_h^2} F^a_{\mu \nu} D^2 F^{a\,\mu \nu} - \frac{d_3}{m_h^2} g f_{abc}F^a_{\mu\nu} F^b_{\mu\alpha} F^c_{\nu\alpha}
+ \sum_{\ell=1}^{n_\ell} \bar q_\ell\left(i\gamma_\mu D^\mu - m_\ell\right)q_\ell ,
\label{eq:sect4:NRQCD}
\end{align}
where, as in the HQET case, $m_h$ has to be understood as the pole mass.
The first line of Eq.~\eqref{eq:sect4:NRQCD} is the two-fermion part of the NRQCD Lagrangian.
As discussed above, it coincides, both in the operators and in the Wilson coefficients, with the two-fermion part of the HQET Lagrangian shown in Eq.~\eqref{eq:sect4:HQET}.
The second line of Eq.~\eqref{eq:sect4:NRQCD} is the two-antifermion part of the NRQCD Lagrangian and it is the charge conjugated of the two-fermion part.
The two-(anti)fermion sector of the HQET/NRQCD Lagrangian is known up to
terms of order $1/m_h^4$~\cite{Gunawardana:2017zix,Kobach:2017xkw}.
The third line of Eq.~\eqref{eq:sect4:NRQCD} is made of all possible four-fermion operators of dimension 6. The corresponding Wilson coefficients
are $f_1(^1S_0)$, $f_1(^3S_1)$, $f_8(^1S_0)$, and $f_8(^3S_1)$. The operator associated to the first (second) Wilson coefficient projects
on a heavy quark-antiquark pair in a color singlet configuration with quantum numbers $^1S_0$ ($^3S_1$), whereas
the operator associated to the third (fourth) Wilson coefficient projects
on a heavy quark-antiquark pair in a color octet configuration with quantum numbers $^1S_0$ ($^3S_1$).
The four-fermion Wilson coefficients have been computed in Refs.~\cite{Bodwin:1994jh,Pineda:1998kj}.
They have a real part that starts at order $\alpha_s$ for $f_8(^3S_1)$ and at order $\alpha_s^2$ for the other coefficients,
and they have also an imaginary part, coming from one loop or higher annihilation diagrams, which is of order $\alpha_s^2$ for
$\text{Im}\,f_1(^1S_0)$, $\text{Im}\,f_8(^1S_0)$, and $\text{Im}\,f_8(^3S_1)$, and of order $\alpha_s^3$ for $\text{Im}\,f_1(^3S_1)$.
A list of imaginary parts of four-fermion Wilson coefficients in NRQCD and related bibliography can be found in Ref.~\cite{Vairo:2003gh}.
The four-fermion sector of the NRQCD Lagrangian has been derived up to order $1/m_h^4$ (complete) and orders $1/m_h^5$ and $1/m_h^6$ (partial)
in Refs.~\cite{Brambilla:2006ph,Brambilla:2008zg,Berwein:2018fos}.
Like for the Wilson coefficients in the two-fermion sector, also the Wilson coefficients in the four-fermion sector
are not all independent, but some of them are related by Poincar\'e invariance~\cite{Brambilla:2008zg,Berwein:2018fos}.
Sometimes it may be useful to isolate the electromagnetic component of the four-fermion operator and of its corresponding Wilson coefficient.
This is the case when computing electromagnetic decay widths and photoproduction cross sections in NRQCD.
The electromagnetic operators are obtained by projecting on an intermediate QCD vacuum state, $|0\rangle$, e.g.,
$\psi^{\dagger} \chi \chi^{\dagger} \psi \to \psi^{\dagger} \chi \, |0\rangle \langle 0 |\, \chi^{\dagger} \psi$.
Finally, the fourth line of Eq.~\eqref{eq:sect4:NRQCD} contains the gauge sector of the EFT and the light quark Lagrangian.
It coincides with the equivalent second line of the HQET Lagrangian in Eq.~\eqref{eq:sect4:HQET}.
As in the HQET case, in Eq. \eqref{eq:sect4:NRQCD} we have not considered $1/m_h^2$ suppressed operators involving light quarks
(either with two light quark and two heavy quark fields or with four light quark fields, see Ref.~\cite{Brambilla:2004jw}).

Differently from the HQET, the power counting of NRQCD is not unique. The reason is that, while the HQET is a one-scale theory,
its only dynamical scale being $\Lambda_{\text{QCD}}$, NRQCD is a multiscale theory. The dynamical scales of NRQCD are, at least,
$m_hv$, $m_hv^2$, and $\Lambda_{\text{QCD}}$. In more complicated settings, even more scales can be relevant.
Hence, one can imagine different power countings, some more conservative, like assuming that the matrix elements scale according to the largest
dynamical scale, i.e., $m_hv$, see, e.g., Ref.~\cite{Brambilla:2002nu}, some less conservative or closer to a perturbative counting, see, e.g.,
Ref.~\cite{Bodwin:1994jh}.
What all the power countings have in common is that the kinetic energy scales like the binding energy and that therefore
$\psi^\dagger i\partial_0 \psi$ is of the same order as $\psi^\dagger \bm{\nabla}^2/(2 m_h)\psi$, and analogously for the antiquark.
This follows from the virial theorem and is an unavoidable consequence of the dynamics of a non-relativistic bound state.
Therefore, the leading-order NRQCD Lagrangian reads $ {\cal L}_{\rm NRQCD} = \psi^\dagger \{ i D_0 + \bm{\nabla}^2/(2 m_h) \} \psi
+ \chi^\dagger \{ i D_0 - \bm{\nabla}^2/(2 m_h) \} \chi - F^a_{\mu \nu} F^{a\,\mu \nu}/4$, which is different
from the leading-order HQET Lagrangian for a heavy quark and a heavy antiquark.
In particular, the leading-order NRQCD Lagrangian
violates the heavy-quark flavor symmetry, which means, for instance, that the bottomonium binding energy is different, even at leading order, from the charmonium one.
In the power counting of Ref.~\cite{Bodwin:1994jh} one further assumes: $D_0 \sim m_hv^2$ (when acting on $\psi$ or $\chi$),
${\bm{D}} \sim m_hv$ (when acting on $\psi$ or $\chi$), $g{\bm{E}} \sim m_h^2v^3$, and $g{\bm{B}} \sim m_h^2v^4$.
A consequence is that the heavy-quark spin symmetry is a symmetry of the leading-order NRQCD Lagrangian.
Because of the mentioned counting, we have added the operator $\psi^\dagger {\bm{D}}^4/(8m_h^3) \psi$ and its charge conjugated to the Lagrangian in Eq.~\eqref{eq:sect4:NRQCD}.
Indeed, this operator is of the same order as the operators on its right.
Matrix elements of octet operators on quarkonium states are further suppressed by the fact that they project on a subleading component of the quarkonium state,
the one made of a heavy quark-antiquark pair in a color octet configuration and gluons.
The amount of suppression depends again on the adopted power counting.

\subsubsection{Potential Non-Relativistic QCD}
\label{Sect:4.2.3}
NRQCD is well suited to describe quarkonium annihilation.
However, it is still a complicated theory to describe the quarkonium spectrum (masses, transitions, widths, ...).
In this context it is only used in lattice calculations.
The reason is that the non-relativistic bound state involves scales, $m_h v$, $m_h v^2$, and $\Lambda_{\text{QCD}}$, that
are still dynamical and entangled in NRQCD (as an illustration, see the non-annihilation part of the diagram on the left of Fig.~\ref{fig:scalesNR}).
A consequence of this is that, although the equations of motion that follow from the NRQCD Lagrangian as shown in Eq.~\eqref{eq:sect4:NRQCD} resemble
a Schr\"odinger equation for non-relativistic bound states, they are not quite that.
They involve gauge fields and do not supply a field theoretical definition and derivation of the potential that would appear in a Schr\"odinger equation.
Nevertheless, we expect that, in some non-relativistic limit, a Schr\"odinger equation describing the quantum
mechanics of the non-relativistic bound state should emerge from field theory,
since field theory may be understood as an extension of quantum mechanics that includes relativistic and radiative corrections.
In particular, in the case of quarkonium, the Schr\"odinger equation describing the bound state should be
the non-Abelian equivalent of the one describing positronium.
Another consequence is that, as already remarked in the previous section,
the power counting of NRQCD is not unique.

Since the scales $m_h v$ and $m_h v^2$ are hierarchically ordered, they may be disentangled by systematically integrating out modes
associated with scales larger than the smallest scale, $m_h v^2$, and matching to a lower energy EFT, where only degrees of freedom resolved at distances
of order $1/(m_h v^2)$ are left dynamical~\cite{Brambilla:2004jw}. This EFT is pNRQCD~\cite{Pineda:1997bj,Brambilla:1999xf}.
Because the scale $m_h v$ has been integrated out, the power counting of pNRQCD is less ambiguous than the one of NRQCD.
In situations where we can neglect the hadronic scale $\Lambda_{\text{QCD}}$,
the power counting of pNRQCD is indeed unique, as its only dynamical scale is $m_h v^2$.
Having integrated out the scale $m_h v$, which is the scale of the inverse of the distance $r$
between the heavy quark and antiquark in any matrix element of quarkonium wave functions, implies that pNRQCD is constructed as an expansion in $r$.
This is analogous to how the HQET and NRQCD are constructed.
There, having integrated out the heavy quark mass, $m_h$, the EFTs are organized as expansions in $1/m_h$,
with the Wilson coefficients encoding the non-analytic contributions, typically under the form of logarithms of $m_h$.
Here, having integrated out the dynamical scale $m_h v$, pNRQCD is organized as an expansion in $r$,
with the Wilson coefficients encoding non-analytic contributions in $r$.
Some of the Wilson coefficients of pNRQCD may be identified with the potentials in the Schr\"odinger equation of quarkonium.

The specific form of pNRQCD depends on the scale $\Lambda_{\text{QCD}}$.
If $\Lambda_{\text{QCD}}\lesssim m_h v^2$, then one deals with weakly-coupled bound states and the EFT is called weakly-coupled pNRQCD.
At distances of the order of or smaller than $1/(m_h v^2)$,
one may still resolve colored degrees of freedom (gluons, quarks, and antiquarks),
as color confinement has not yet set in. Hence gluons, quarks, and antiquarks are the degrees of freedom of weakly-coupled pNRQCD.
Weakly-coupled pNRQCD is well suited to describe tightly bound quarkonia, like the bottomonium and (to a less extent) charmonium
ground states, the $B_c$ ground state, and threshold effects in $t\bar{t}$ production.
If $\Lambda_{\text{QCD}}\gtrsim m_h v^2$, then one deals with strongly-coupled bound states and the EFT is called strongly-coupled pNRQCD.
At distances of the order of $1/(m_h v^2)$, confinement has set in and the only available degrees of freedom are color singlets.
These are, in principle, all, ordinary and exotic, heavy and light, hadrons that we might have in the spectrum.
Strongly-coupled pNRQCD is suited to describe higher states in the bottomonium and charmonium spectra, as well as quarkonium exotica.
If $m_h v \gg \Lambda_{\text{QCD}} \gg m_h v^2$, the matching to pNRQCD may be done in two steps, first integrating out (perturbatively) $m_h v$ then (non-perturbatively) $\Lambda_{\text{QCD}}$.
The advantage is that contributions coming from these two scales will be automatically factorized in the pNRQCD observables.

\begin{figure}[ht]
\begin{center}
\includegraphics[width=1\linewidth]{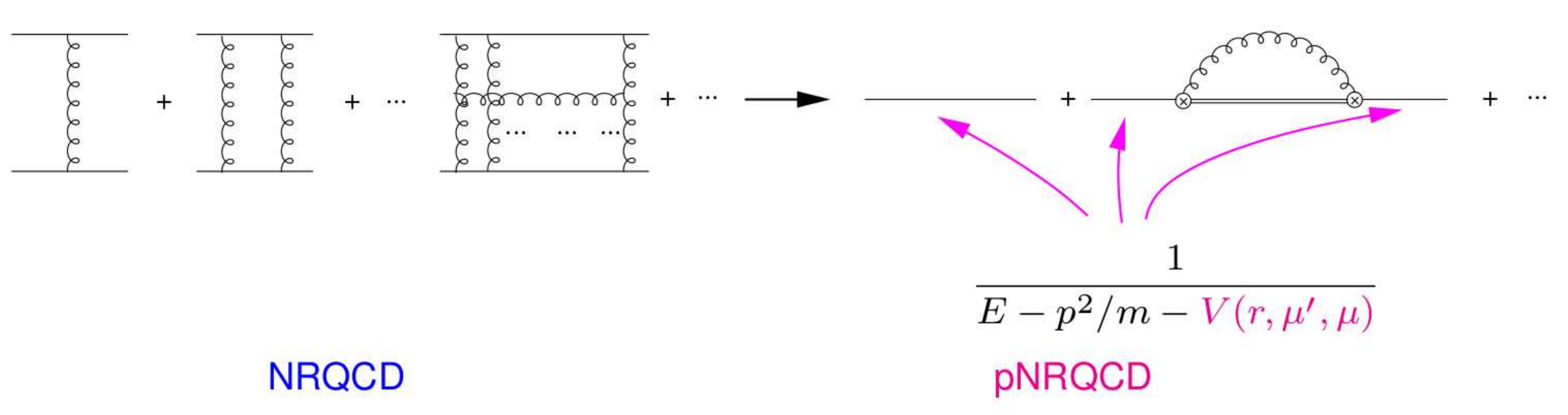}
\caption{Schematic matching of pNRQCD; $V$ is a Wilson coefficient of pNRQCD,
$\mu$ and $\mu'$ are the ultraviolet and infrared renormalization scales of pNRQCD, respectively.
In the pNRQCD Feynman diagrams a single line stands for a singlet propagator, a double line for an octet propagator,
curly lines for ultrasoft gluons, and a circle with a cross for a chromoelectric dipole interaction.}
\label{fig:matchingpNRQCD}
\end{center}
\end{figure}

\vspace{0.3cm}\noindent
$\bullet$ {\it Weakly-coupled pNRQCD}
\vspace{0.3cm}

The degrees of freedom of weakly-coupled pNRQCD are heavy quarks and antiquarks of momentum $m_h v$ and energy $m_h v^2$,
gluons of momentum and energy $m_h v^2$ (sometimes called ultrasoft gluons), and light quarks of momentum and energy $m_h v^2$.
Weakly-coupled pNRQCD follows from integrating out gluons of energy or momentum $m_h v$ (sometimes called soft gluons) from NRQCD.
Because at the scale $m_h v^2$ we cannot resolve the single heavy quark and antiquark, it may be useful to cast heavy quark and antiquark
fields into bilocal fields that depend on time, $t$, the c.m. coordinate ${\bm{R}}$ and the relative coordinate ${\bm{r}}$.
We call singlet, S, the color singlet component of the quark and antiquark field and octet, O, its color octet component, normalized as
$\mathrm{S}= \mathbbm{1}_{3\times 3} S/\sqrt{3}$ and $\mathrm{O}=\sqrt{2}O^aT^a$.
The distance $r$ typically scales like $1/(m_h v)$, while the c.m. coordinate, $R$, and the time, $t$,
typically scale like $1/(m_h v^2)$, as the quark-antiquark pair may only recoil against ultrasoft gluons.
To ensure that gluons are ultrasoft in the pNRQCD Lagrangian, gauge fields are multipole expanded in ${\bm{r}}$.
Hence gauge fields in the pNRQCD Lagrangian only depend on time and the c.m. coordinate.
The matching of pNRQCD to NRQCD is schematically illustrated in Fig.~\ref{fig:matchingpNRQCD}.
The pNRQCD Lagrangian is organized as a double expansion in $1/m_h$ and $r$.
At order $r$ in the multipole expansion and at leading order in $1/m_h$,
the weakly-coupled pNRQCD Lagrangian density has the form~\cite{Pineda:1997bj,Brambilla:1999xf}
\begin{align}
 \mathcal{L}^{\text{weak}}_{\text{pNRQCD}} =& \int d^3r\,\mathrm{Tr}\left\{ \mathrm{S}^\dagger(i\partial_0 - h_s)\mathrm{S}
+ \mathrm{O}^\dagger( iD_0{O} - h_o)\mathrm{O}\right\}
\nonumber\\
& - V_A {\rm Tr} \left\{ {\rm O}^\dagger {\bm{r}} \cdot g{\bm{E}} \,{\rm S} + {\rm S}^\dagger {\bm{r}} \cdot g{\bm{E}} \,{\rm O} \right\}
- \frac{V_B}{2} {\rm Tr} \left\{ {\rm O}^\dagger {\bm{r}} \cdot g{\bm{E}} \, {\rm O} + {\rm O}^\dagger {\rm O} {\bm{r}} \cdot g{\bm{E}} \right\}
\nonumber\\
& - \frac{1}{4} F_{\mu \nu}^{a} F^{\mu \nu \, a} + \sum_{\ell=1}^{n_\ell} \bar q_\ell\left(i\gamma_\mu D^\mu - m_\ell\right)q_\ell,
\label{eq:sect4:pNRQCDweak}
\end{align}
where (now writing also $1/m_h$, $1/m_h^2$, and $1/m_h^3$ terms in the kinetic energy and potentials)
\begin{equation}
h_s = \frac{{\bm{p}}^2}{m_h} + \frac{{\bm{P}}^2}{4 m_h} - \frac{{\bm{p}}^4}{4m_h^3} + \dots + V_s, \qquad
h_o = \frac{{\bm{p}}^2}{m_h} + \frac{{\bm{P}}^2}{4 m_h} - \frac{{\bm{p}}^4}{4m_h^3} + \dots + V_o, \qquad
\label{sect4:hsho}
\end{equation}
$iD_0 {\rm O} = i \partial_0 {\rm O} + g [A_0({\bm{R}},t),{\rm O}]$,
${\bm{P}} = -i{\bm{D}}_{\bm{R}}$ is the c.m. momentum, ${\bm{p}} = -i\bm{\nabla}_{\bm{r}}$ is the relative momentum,
and $h_s$ and $h_o$ may be interpreted as the Hamiltonian for the color singlet and color octet heavy quark-antiquark fields.
The dots in Eq.~\eqref{sect4:hsho} stand for higher-order kinetic energy terms.
The trace in Eq.~\eqref{eq:sect4:pNRQCDweak} is understood both in spin and in color.

The quantities $V_s$, $V_o$, $V_A$, and $V_B$ are Wilson coefficients of pNRQCD.
They encode contributions from the soft gluons that have been integrated out from NRQCD.
Because under the hierarchy of weakly-coupled pNRQCD the soft scale, $m_h v$, is larger than $\Lambda_\text{{QCD}}$,
the Wilson coefficients may be computed in perturbation theory, order by order in $\alpha_s$.
The Wilson coefficients are, in general, functions of ${\bm{r}}$, as well as of the spin and momentum.
At leading order, $V_A$ and $V_B$ are 1; they get possible first corrections at order $\alpha_s^2$~\cite{Brambilla:2009bi}.
The Wilson coefficients $V_s$ and $V_o$ may be identified with the color singlet and octet potentials, respectively.
Indeed, at leading order $V_s^{(0)}=-4\alpha_s/(3r)$ and $V_o^{(0)}= \alpha_s/(6r)$,
which are the Coulomb potentials in the color SU(3) fundamental and adjoint representation, respectively.
The potentials $V_s$ and $V_o$ contain, however, also momentum- and spin-dependent corrections.
To isolate the different corrections, it is useful to expand the potentials in powers of $1/m_h$.
For the singlet case (the octet case is analogous) we can write up to order $1/m_h^2$:
\begin{equation}
V_s = V_s^{(0)}(r) + \frac{V_s^{(1)}(r)}{m_h} + \frac{V_\text{SI}^{(2)}}{m_h^2} + \frac{V_\text{SD}^{(2)}}{m_h^2} \,,
\label{sec4.2.3.Vexpansion}
\end{equation}
where, at order $1/m_h^{2}$ we have distinguished between spin-independent (SI) and spin-dependent (SD) terms.
In turn, they can be organized as
\begin{align}
V_\text{SI}^{(2)} &= V_r^{(2)}(r) + \frac{1}{4} V_{p^2,\text{CM}}^{(2)}(r) {\bm{P}}^2 + V_{L^2,\text{CM}}^{(2)}(r)\, ({\bm{r}}\times {\bm{P}})^2
 + \frac{1}{2} \left\{V_{p^2}^{(2)}(r),{\bm{p}}^2 \right\} + V_{L^2}^{(2)}(r)\, {\bm{L}}^2 \,,
 \label{potentialSI2}\\
V_\text{SD}^{(2)} &= \frac{1}{2} V_{LS,\text{CM}}^{(2)}(r)\,({\bm{r}}\times {\bm{P}}) \cdot ({\bm{S}}_1 - {\bm{S}}_2)
+ V_{LS}^{(2)}(r)\, {\bm{L}} \cdot {\bm{S}}
 + V_{S^2}^{(2)}(r)\, {\bm{S}}^2 + V_{S_{12}}^{(2)}(r)\, S_{12} \,,
 \label{potentialSD2}
\end{align}
where ${\bm{S}}={\bm{S}}_{1}+{\bm{S}}_{2}=(\bm{\sigma}_{1}+\bm{\sigma}_{2})/2$ is the total spin
(${\bm{S}}_i = \bm{\sigma}_{i}/2$ is the spin of the particle $i$),
${\bm{L}}={\bm{r}}\times {\bm{p}}$ is the relative orbital angular momentum, and
$S_{12}=3(\hat{\bm{r}}\cdot\bm{\sigma}_{1})(\hat{\bm{r}}\cdot\bm{\sigma}_{2})-\bm{\sigma}_{1}\cdot\bm{\sigma}_{2}$;
$\{\;,\;\}$ stands for the anticommutator.
The potential $V_s^{(0)}$ is the static potential, the potential proportional to $V_{LS}^{(2)}$ may be identified with the spin-orbit
potential, the potential proportional to $V_{S^2}^{(2)}$ with the spin-spin potential and the potential proportional to $V_{S_{12}}^{(2)}$
with the spin tensor potential.
The above potentials read at leading (non-vanishing) order in perturbation theory (see, e.g., Ref.~\cite{Brambilla:2004jw}):
\begin{align}
&
V^{(1)}(r) = -\frac{2 \alpha_s^2}{r^2} \,,
\label{V1pNRQCD} \\
&
V_r^{(2)}(r) = \frac{4\pi}{3} \alpha_s \delta^{(3)}({\bm{r}}\,) \,,
\hspace*{0.90cm}
V_{p^2}^{(2)}(r) = -\frac{4\alpha_s}{3r} \,,
\hspace*{0.90cm}
V_{L^2}^{(2)}(r) = \frac{2\alpha_s}{3 r^3} \,,
\label{V2p2pNRQCD} \\
&
V_{LS}^{(2)}(r) = \frac{2\alpha_s}{r^3} \,,
\hspace*{1.00cm}
V_{S^2}^{(2)}(r) = \frac{16\pi \alpha_s}{9} \delta^{(3)}({\bm{r}}\,) \,,
\hspace*{1.00cm}
V_{S_{12}}^{(2)}(r) = \frac{\alpha_s}{3 r^3} \,.
\label{V2S12pNRQCD}
\end{align}
Beyond leading order, the static potential is known up to three-loop accuracy~\cite{Brambilla:1999qa,Anzai:2009tm,Smirnov:2009fh},
and also subleading logarithms, showing up at four loops, have been computed~\cite{Brambilla:2006wp};
the $1/m_h$ potential is known up to order $\alpha_s^3$~\cite{Kniehl:2001ju} and $1/m_h^2$ potentials up to order $\alpha_s^2$
(these potentials have a long history, see Ref.~\cite{Kniehl:2002br} and references therein).
We have assumed that the heavy quark and antiquark have equal masses;
for the case of a quark and an antiquark of different masses, we refer, for instance,
to Refs.~\cite{Brambilla:2001xk,Brambilla:2003nt,Brambilla:2004jw,Peset:2015vvi}.
The Wilson coefficients of pNRQCD inherit the Wilson coefficients of NRQCD.
Hence, some of the couplings appearing in the expansion of the Wilson coefficients are naturally computed at the
scale of NRQCD, $m_h$, while others, encoding the soft gluons, are naturally computed at the soft scale, $m_h v$.
In weakly-coupled pNRQCD, because the leading potential is the Coulomb potential, the Bohr radius
is proportional to $1/(m_h \alpha_s)$ and $v \sim \alpha_s$.
Finally, like in any non relativistic EFT, also the Wilson coefficients of pNRQCD are related through constraints
imposed by the relativistic invariance of the underlaying fundamental theory, QCD.
These constraints fix, for instance, the coefficients of the kinetic terms appearing in Eq.~\eqref{sect4:hsho}
to be the ones coming from expanding the relativistic kinetic energies of a free quark and antiquark.
Relations have been also found relating the potentials, e.g.,
${V_{LS,{\rm CM}}} = -{V_s^{(0)\prime}}/(2r)$ (where $V^{\prime}=dV/dr$),
$V_{{\bm{L}}^2,{\rm CM}} + {r \, V_s^{(0)\prime}}/2 = 0$, $V_{p^2,{\rm CM}} + V_{L^2,{\rm CM}} + V_s^{(0)}/2 = 0$
and many others~\cite{Gromes:1984ma,Barchielli:1988zp,Brambilla:2001xk,Brambilla:2003nt,Berwein:2018fos}.
These relations are exact, i.e., valid at any order in perturbation theory and, when applicable, also non-perturbatively.

From Eq.~\eqref{eq:sect4:pNRQCDweak} we see that the relative coordinate ${\bm{r}}$ plays the role of a continuous parameter labeling different fields.
The dynamical coordinates of the Lagrangian density are the time $t$ and the coordinate ${\bm{R}}$, which, in the case of the
fields S and O, is the c.m. coordinate.
Having written the Lagrangian in terms of singlet and octet fields has made each term in Eq.~\eqref{eq:sect4:pNRQCDweak} explicitly gauge invariant.

The power counting of pNRQCD is straightforward. We have already mentioned that $r\sim 1/(m_h v)$ and $t$, $R \sim 1/(m_h v^2)$.
Momenta scale like $p \sim m_h v$ and $P \sim m_h v^2$.
Gluon fields and light quark fields are ultrasoft and scale like $m_h v^2$ or $\Lambda_{\text{QCD}}$ to their dimension.
The leading-order Hamiltonian, ${\bm{p}}^2/m_h + V_s^{(0)}$, scales like $m_h v^2$ (and analogously in the octet case), which is the
order of the Bohr levels. The potentials listed in Eqs.~\eqref{V1pNRQCD}-\eqref{V2S12pNRQCD} scale like $m_hv^4$.
The first correction to a pure potential picture of the quarkonium interaction comes from the chromoelectric dipole
interaction terms in the second line of Eq.~\eqref{eq:sect4:pNRQCDweak}.
These operators are of order $g (m_h v^2)^2/(m_h v) \sim g m_h v^3$.
However, in order to project on singlet states the chromoelectric dipole interaction may enter only in loops
(see the diagram in the right-hand side of Fig.~\ref{fig:matchingpNRQCD}). Such a loop diagram is of order $g^2(m_hv^2)^3/(m_hv)^2 \sim g^2 m_h v^4$.
The coupling $g^2$ is computed at the ultrasoft scale. Hence, if $\Lambda_{\text{QCD}} \ll m_h v^2$, the coupling is perturbative and the
chromoelectric dipole loop diagram is suppressed with respect to the contributions coming from the potentials in Eqs.~\eqref{V1pNRQCD}-\eqref{V2S12pNRQCD}.
Elsewhere, if $\Lambda_{\text{QCD}} \sim m_h v^2$, it is of the same order.

At leading order in the multipole expansion, the equation of motion for the singlet field reads
\begin{equation}
i\partial_0\mathrm{S} = h_s \mathrm{S},
\label{pNRQCDSchr-equation}
\end{equation}
which is the Schr\"odinger equation that in quantum mechanics describes the evolution of a non-relativistic bound state.
Potential NRQCD provides therefore a QCD foundation of the potential picture underlaying many phenomenological quark models,
a rigorous QCD definition and derivation of the potentials and a range of validity for the Schr\"odinger equation itself
(for the somehow paradoxical history that has led from the Lorentz covariant formulation of the bound state problem
to the field theoretical foundation of the Schr\"odinger equation, see Ref.~\cite{Vairo:2009rs}).
As we have seen, ultrasoft gluons start contributing, and therefore correcting the potential picture,
for the spectrum, at order $m_h v^4$ or $m_h v^5$, in dependence of $\Lambda_{\text{QCD}}$.
The potentials are Wilson coefficients of an EFT, they are regularized, undergo renormalization and satisfy renormalization group equations that allow to resum potentially large logarithms
in their expressions~\cite{Brambilla:1999qa,Kniehl:1999ud,Brambilla:1999xj,Pineda:2000gza,Pineda:2001ra,Brambilla:2009bi,Pineda:2011aw,Peset:2018jkf,Anzai:2018eua}.
The proper renormalization of the potentials, highly non-trivial,
as it has to account for correlated renormalization scales originating in NRQCD and pNRQCD, guarantees, however,
that the final physical results are finite and scheme independent at any order in the expansion parameters of the EFT.
This is not the case for phenomenological models.

Weakly-coupled pNRQCD requires the condition $\Lambda_{\text{QCD}}\lesssim m_h v^2$ to be fulfilled.
This condition is realized if the size of the quarkonium is smaller and the revolution time not larger than the typical size of a heavy-light hadron.
Examples are the bottomonium ground state, the ground state of the $B_c$ system, and, to a somewhat lesser extent,
the charmonium ground state, and the first bottomonium excited states.
We recall that in a Coulombic system the size is proportional to the inverse of the mass and to the principal quantum number.
The precise dependence on the latter follows from the precise definition
of the size.
A review on applications of weakly-coupled pNRQCD to several tightly bound quarkonia can be found in Ref.~\cite{Pineda:2011dg}.
Weak-coupling determinations of the bottomonium ground state masses are typically used to extract the charm and
bottom masses~\cite{Brambilla:2001fw,Brambilla:2001qk,Recksiegel:2002za,Pineda:2001zq,Ayala:2014yxa,Beneke:2014pta,Kiyo:2015ufa,Ayala:2016sdn,Mateu:2017hlz,Peset:2018ria}.
Hence, they provide alternative observables for the extraction of the heavy quark masses to the heavy-light meson masses discussed in Sec.~\ref{Sect:4.2.1}.
The results are consistent with the ones presented in Sec.~\ref{Sect:4.2.1}, but with some different systematics,
although they face some similar issues, like the use of a proper subtraction
scheme for the mass.
Nowadays the precision is N$^3$LO, the determination of the bottom mass includes the effects due to a finite charm mass,
and the observables used are not only the masses of the vector states, $\Upsilon(1S)$ and $J/\psi$, but also the masses of the
pseudoscalar states, $\eta_b$ following the BaBar discovery~\cite{Aubert:2008ba} and $\eta_c$, and the $B_c$ ground state following the CDF discovery~\cite{Abe:1998wi}
and the latest LHCb precise measurements~\cite{Aaij:2012dd,Aaij:2013gia,Aaij:2014asa,Aaij:2016qlz}.
In Ref.~\cite{Mateu:2017hlz} the bottom and charm masses have been extracted from a global fit up to $n=3$ bottomonium states.
On the other hand, once the heavy quark masses have been established on one set of spectroscopy observables, they can be used for others
like the $B_c$ mass or the $B_c$ spectrum (see Ref.~\cite{Brambilla:2000db} for an early reference and Ref.~\cite{Peset:2015vvi}
for a status of the art calculation at N$^3$LO).
Fine and hyperfine splittings of charmonium and bottomonium have been computed perturbatively in Refs.~\cite{Recksiegel:2003fm,Brambilla:2004wu}
and at to NLL accuracy in Ref.~\cite{Kniehl:2003ap}, similarly for the $B_c^*$-$B_c$ hyperfine splitting in Ref.~\cite{Penin:2004xi}.
After an effort that lasted more than one decade the whole perturbative heavy quarkonium spectrum
has been computed at N$^3$LO~\cite{Brambilla:1999xj,Penin:2002zv,Penin:2005eu,Beneke:2005hg,Beneke:2007gj,Kiyo:2013aea,Kiyo:2014uca}.
Recently, this result has been further improved reaching N$^3$LL accuracy up to a missing contribution of the two-loop soft running~\cite{Peset:2018jkf,Anzai:2018eua}.
The N$^3$LL order represents the presently achievable precision of these calculations. Going beyond this precision will require a major computational effort, like
the four-loop determination of the static potential, that appears beyond near reach.
Electromagnetic decays of the bottomonium lowest levels have been computed including N$^2$LL corrections in Refs.~\cite{Penin:2004ay,Pineda:2006ri}.
A different power counting that includes at leading order the exact static potential has been used for these quantities in Ref.~\cite{Kiyo:2010jm}.
Corrections to the wave function and leptonic decay width of the $\Upsilon(1S)$ at N$^3$LO have been computed in Refs.~\cite{Beneke:2007pj,Beneke:2014qea}.
Non-perturbative corrections in the form of condensates have been included in Refs.~\cite{Pineda:1996uk,Rauh:2018vsv}.
Radiative quarkonium decays have been analyzed
in Refs.~\cite{Bauer:2001rh,Fleming:2002rv,Fleming:2002sr,GarciaiTormo:2004jw,GarciaiTormo:2005ch,GarciaiTormo:2005bs}.
Radiative and inclusive decays of the $\Upsilon(1S)$ may also serve as a determination of $\alpha_s$ at the bottom mass scale~\cite{Brambilla:2007cz}.
Radiative transitions, M1 and E1, at relative order $v^2$ in the velocity expansion have been computed in various power countings
in Refs.~\cite{Brambilla:2005zw,Brambilla:2012be,Pineda:2013lta,Segovia:2018qzb}. Noteworthy, pNRQCD may explain the tiny $\Upsilon(2S)\to \gamma\,\eta_b(1S)$
branching fraction measured by BaBar~\cite{Aubert:2009as}. Finally, the photon line shape in the radiative transition $J/\psi \to \gamma\,\eta_c(1S)$ has been
studied in Ref.~\cite{Brambilla:2010ey}.

\vspace{0.3cm}\noindent
$\bullet$ {\it Strongly-coupled pNRQCD}
\vspace{0.3cm}

When the hierarchy of scales is $\Lambda_{\text{QCD}}\gg m_hv^2$, then the theory enters the strong-coupling regime.
Such a regime may be appropriate to describe higher quarkonium states, and quarkonium exotica.
Strongly-coupled pNRQCD is obtained by integrating out the hadronic scale $\Lambda_{\text{QCD}}$, which means that all colored degrees of freedom are
absent~\cite{Brambilla:2000gk,Pineda:2000sz,Brambilla:2001xy,Brambilla:2002nu,Brambilla:2003mu,Brambilla:2004jw}.
Such an EFT may be constructed, in principle, for any hadron made of a heavy quark-antiquark pair,
hence for both ordinary quarkonia and exotic states where the heavy quark-antiquark pair binds with valence light quarks or gluons.

\begin{figure}[ht]
\begin{center}
 \includegraphics[width=0.5\linewidth]{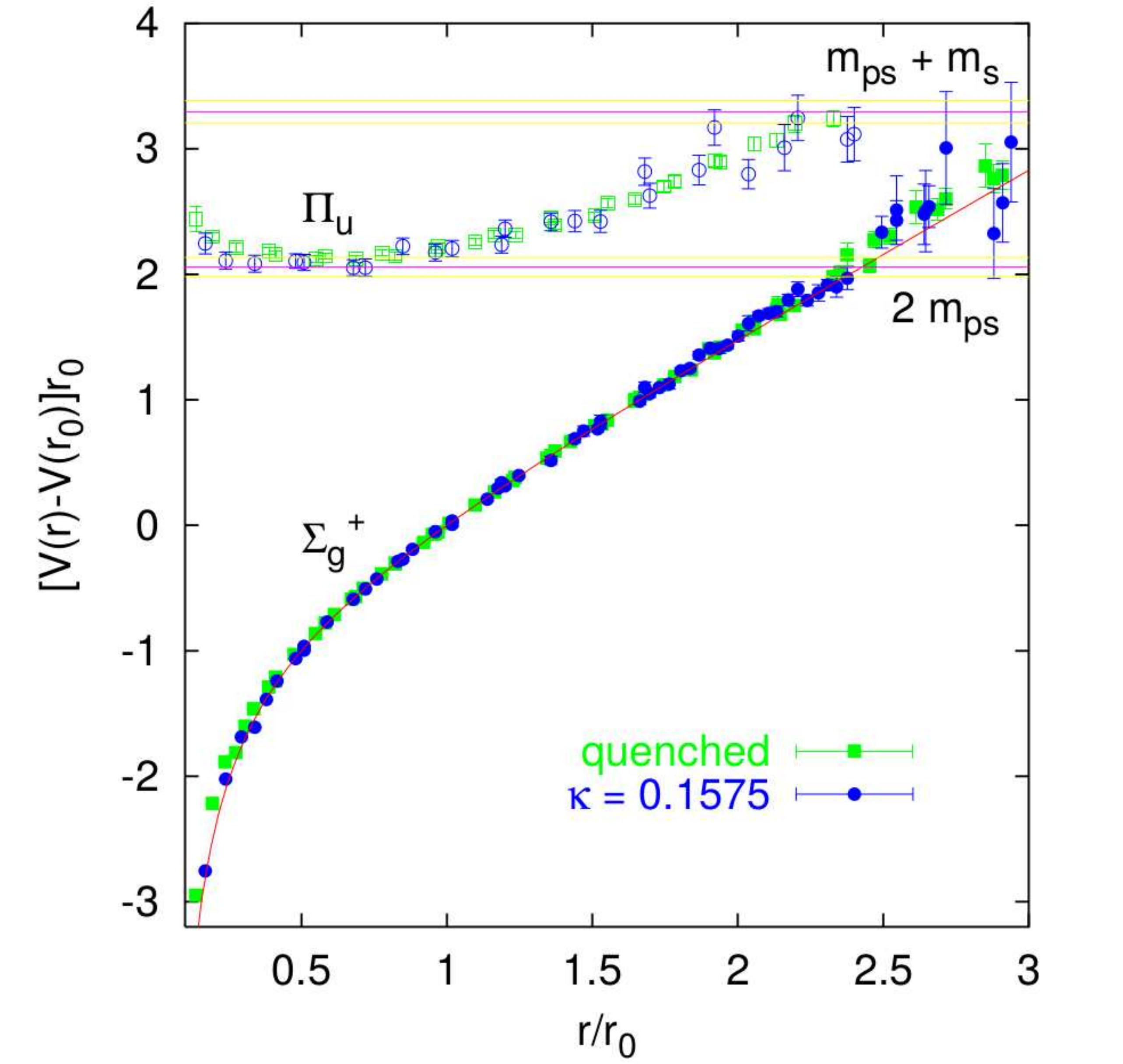}
\caption{The quarkonium static energy (labeled $\Sigma_g^+$) vs the energy of the first gluonic excitation between a static quark-antiquark pair
 (labeled $\Pi_u$) as a function of the distance between the quark and the antiquark.
 The quenched (green) and two-flavor (blue) lattice data are from~\cite{Bali:2000vr}. The unit $r_0$ is about $0.5$~fm.}
\label{fig:SESAM00}
\end{center}
\end{figure}

Let us consider in this section the simplest case of strongly-coupled pNRQCD for ordinary quarkonia.
Lattice QCD shows evidence that the quarkonium static energy is separated by a gap of order $\Lambda_{\text{QCD}}$ from the energies of the gluonic excitations
between the static quark-antiquark pair, see Fig.~\ref{fig:SESAM00}.
If this condition is also fulfilled by the binding energies, i.e.,
if the binding energies of the states that can be constructed out of the quarkonium static energy are separated by a gap of order $\Lambda_{\text{QCD}}$
from the binding energies of the states that can be constructed out of the static energies of the gluonic excitations, and from open-flavor states,
then one can integrate out all these latter higher energy states.
The resulting EFT is just made of a quark-antiquark color singlet field, whose modes are the quarkonium states, and light hadrons.
The coupling of quarkonia with light hadrons has been considered in the framework of pNRQCD in Ref.~\cite{Brambilla:2015rqa}.
It impacts very mildly spectral properties (masses, widths) of quarkonia that lie well below the open-flavor threshold.
If such quarkonia are our main concern, then we may neglect their coupling with light hadrons;
under these circumstances the pNRQCD Lagrangian density assumes the particularly simple form:
\begin{align}
\mathcal{L}_{\text{pNRQCD}}^{\text{strong}}=\int d^3r\,\text{Tr}\left\{\mathrm{S}^{\dagger}\left(i\partial_0-h_{S}\right)\mathrm{S}\right\}\,.
\label{eq:sect4:pNRQCDstrong}
\end{align}
The Hamiltonian, $h_{S}$, has the same form, given by Eqs.~\eqref{sect4:hsho}-\eqref{potentialSD2}, as in weakly-coupled pNRQCD.
The equation of motion is the Schr\"odinger equation~\eqref{pNRQCDSchr-equation},
which provides a field theoretical foundation of potential models also in the strong-coupling regime.
It also allows us to go systematically beyond the potential picture by including couplings with light degrees of freedom.

All the complications of the non-perturbative dynamics are encoded in the potentials, which at order $1/m_h$ is $V_s^{(1)}$
and at order $1/m_h^2$ are the spin-independent and spin-dependent terms identified in Eqs.~\eqref{potentialSI2} and~\eqref{potentialSD2}.
The crucial points about the EFT treatment that distinguish it from phenomenological potential models can be summarized as follows.
{\em (i)}~The potentials can be written as the products of Wilson coefficients, factorizing contributions from the high-energy scale, $m_h$,
and low-energy matrix elements, encoding contributions coming from the scales $m_h v$ and $\Lambda_{\text{QCD}}$.
The exact expressions follow from matching pNRQCD with its high-energy completion, which is NRQCD.
{\em (ii)}~The high-energy Wilson coefficients of pNRQCD are inherited from NRQCD.
These are the Wilson coefficients listed up to order $1/m_h^2$ in the NRQCD Lagrangian~\eqref{eq:sect4:NRQCD}.
As we discussed there, they are known expansions in the strong coupling (up to the computed orders).
Because the NRQCD Wilson coefficients have a real part and an imaginary part,
also the pNRQCD potentials develop a real part, responsible for the quarkonium binding, and an imaginary part, responsible for the quarkonium annihilation.
At higher orders, also contributions coming from the scale $\sqrt{m_h \Lambda_{\text{QCD}}}$ may become relevant;
these contributions can be computed in perturbation theory as $\sqrt{m_h \Lambda_{\text{QCD}}} \gg \Lambda_{\text{QCD}}$~\cite{Brambilla:2003mu}.
{\em (iii)}~The low-energy matrix elements are non-perturbative. Their field-theoretical expressions, relevant for potentials up to order $1/m_h^2$, are known.
The static potential is equal to $\displaystyle \lim_{T\to\infty} i\, \ln W/T$, where $W$ is the expectation value of a rectangular Wilson loop
of spatial extension $r$ and temporal extension $T$~\cite{Wilson:1974sk,Susskind:1976pi,Fischler:1977yf,Brown:1979ya}.
Similarly, the low-energy real parts of the other potentials can be expressed in terms of Wilson loops and field insertions on them~\cite{Eichten:1980mw,Brambilla:2000gk,Pineda:2000sz}.
These Wilson loops may be computed in weak-coupling QCD giving back the weak-coupling potentials listed in Eqs.~\eqref{V1pNRQCD}-\eqref{V2S12pNRQCD}~\cite{Peset:2017wef}.
More relevant from the phenomenological point of view is, however, their numerical non-perturbative determination via lattice QCD.
Indeed, the computation of these potentials has a long history that begins with the inception of lattice QCD.
Their most recent determinations can be found in Refs.~\cite{Koma:2006si,Koma:2006fw,Koma:2007jq}.
One should remark that lattice calculations of the quarkonium potentials have not included so far $1/m_h^2$ momentum- and spin-independent potentials, and have been limited to the pure SU(3) gauge theory.
Potentials that have not been computed on the lattice may be estimated in the long distance using the effective string theory,
which has also proved to be very successful in interpreting the existing long-distance data~\cite{Kogut:1981gm,PerezNadal:2008vm,Brambilla:2014eaa,Peset:2017wef}.
The low-energy contributions to the imaginary parts of the potential are matrix elements of the NRQCD four-fermion operators.
Hence they are local terms proportional to $\delta^3({\bm{r}})$ or derivatives of it.
Non-perturbative contributions are encoded into constants that may be expressed in terms of momenta of correlators of chromoelectric and/or chromomagnetic fields~\cite{Brambilla:2001xy,Brambilla:2002nu}.
They may be fixed on data or computed with lattice QCD, but lattice calculations have not been intensively pursued, so far, for these quantities.
{\em (iv)}~Finally, because pNRQCD retains the correct ultraviolet behaviour of the low-energy EFT, it is renormalizable order by order in the expansion parameters in both its weak-coupling and strong-coupling versions.
In particular, quantum-mechanical perturbation theory can be implemented at any order without incurring into uncanceled divergences.

Applications of strongly-coupled pNRQCD have been limited so far to the computation of quarkonium decay widths,
in particular for charmonium and bottomonium $P$-waves~\cite{Brambilla:2001xy,Brambilla:2002nu,Brambilla:2003mu,Brambilla:2004wf}.
The advantage with respect to the NRQCD approach is that, while the NRQCD four-fermion matrix elements depend on the quarkonium state,
their pNRQCD expression factorizes all the quarkonium dependence into the wave function at the origin (or its derivatives) squared.
The wave function at the origin squared gets multiplied by momenta of correlators of field-strength tensors, $F$, that are universal, quarkonium independent, constants.
Schematically, one obtains for the expression of a generic NRQCD four-fermion matrix element in pNRQCD that
$\displaystyle \langle \text{4-fermion}\rangle \sim |\text{wave-function}(0)|^2 \times \int dt \, \langle F(t) \cdots F(0) \rangle$.
This leads to a significant reduction in the number of non-perturbative parameters and allows to use information gained in the charmonium sector to make predictions in the bottomonium one.
Electroproduction of quarkonium may be treated in pNRQCD in the same way.
Hadroproduction, instead, has not been formulated yet in the pNRQCD language,
owing in part to the difficulty to state and prove rigorously the NRQCD factorization theorem in that context, see discussion and references in the previous section. Finally, very recently pNRQCD conbined with the multipole expansion has been used to compute quarkonium hadronic transitions~\cite{Pineda:2019mhw}.

\vspace{0.3cm}\noindent
$\bullet$ {\it pNRQCD for systems other than quarkonia}
\vspace{0.3cm}

Weakly-coupled pNRQCD, strongly-coupled pNRQCD or a mixture of the two can be used to describe systems with three valence quarks,
two of them heavy~\cite{Brambilla:2005yk,Fleming:2005pd,Brambilla:2009cd,Brambilla:2013vx,Mehen:2019cxn}.
The key observation is that the non-relativistic hierarchy of scales given in Eq.~\eqref{sect4:scales}, where $v$ is the relative heavy-quark velocity, is preserved.
This allows one to systematically integrate out these scales to describe eventually the baryon with a suitable EFT.
If the heavy quark-quark distance is of the order of $1/\Lambda_{\text{QCD}}$, then the valence light-quark affects the quark-quark potential.
Elsewhere, if the heavy quark-quark distance is smaller than $1/\Lambda_{\text{QCD}}$, then we may disentangle the quark-quark dynamics, described by a perturbative quark-quark potential,
from the coupling of the heavy-quark pair with the light quark. Since in this last case, the light quark sees the heavy-quark pair as a pointlike particle,
its interaction with the heavy-quark pair is described by the HQET.
Processes involving light hadrons in final or intermediate states may be described in the framework of the heavy-baryon chiral effective theory~\cite{Jenkins:1990jv}.
Similarly, one can devise EFTs for describing low-energy modes of baryons made of three heavy quarks. These states have not been discovered yet in experiments,
but they offer a unique tool to study confinement and the transition region from the Coulomb regime to the confined one in a non-trivial geometrical setting~\cite{Brambilla:2009cd}.
These issues are already addressed by lattice computations of the three-quark static potential~\cite{Takahashi:2002it,Takahashi:2004rw,Koma:2017hcm}.
Higher-order potentials in the $1/m_h$ expansion have been defined in terms of Wilson loops and field insertions on them, as in the quarkonium case, and may be eventually computed
on the lattice providing, for instance, first principle determinations of the heavy baryon fine and hyperfine splittings~\cite{Brambilla:2009cd}.

Possible bound states made of two quarkonia or of a quarkonium and a nucleon (hadroquarkonium, see Sec.~\ref{Sect:4.1.3}) may be characterized by even lower energy scales
than those characterizing the binding in quarkonia or baryons made of at least two heavy quarks.
These lower energy scales are those associated with the pion exchanges responsible for the long-range interaction.
One can treat these systems in an EFT framework by starting from the pNRQCD description of the quarkonium and the heavy-baryon chiral effective theory description of the nucleon.
The long-range pion exchange interaction sets the scale of the typical size of the system to be of the order of $1/M_\pi$, i.e., much larger than
the size of the quarkonium and even larger than its typical time scale, which is of the order of the inverse of the binding energy.
Once modes associated with the quarkonium binding energy and $M_\pi$ have been integrated out,
the quarkonium-quarkonium or the quarkonium-nucleon interaction is described by a potential that, in this way, has been systematically computed from QCD.
The coupling of quarkonium with the pions is encoded in a Wilson coefficient that may be identified with the quarkonium chromoelectric polarizability.
In the quarkonium-quarkonium system, the lowest energy EFT describing modes of energy and momentum of order $m_\pi^2/(2 m_h)$
is called van der Waals EFT (WEFT)~\cite{Brambilla:2015rqa,Brambilla:2017ffe}. The resulting potential is, in fact, the van der Waals potential.
In the quarkonium-nucleon system, the lowest energy EFT describing modes of energy and momentum of order $M_\pi^2/(2 \Lambda_\chi)$
has been dubbed potential quarkonium-nucleon EFT (pQNEFT)~\cite{TarrusCastella:2018php}.
Interest in these systems has been renewed recently after the discovery by the LHCb Collaboration
of the pentaquark states $P_c(4380)^+$ and $P_c(4450)^+$, with valence quark content $P_c^+ = \bar{c}cuud$,
as intermediate resonant states in the weak decay process $\Lambda^0_b \to J/\psi K^- p$~\cite{Aaij:2015tga,Aaij:2016phn}, see Sec.~\ref{sect:3.3.2}.

In the next section, we will deal with higher excitations of a heavy quark-antiquark pair due to gluons (hybrids) and,
on a more qualitative level, due to light quark pairs (generic tetraquarks). The framework will be that one of strongly-coupled pNRQCD.
The specific application of strongly-coupled pNRQCD to these systems will require some extra assumptions that allow for further expansions in the EFT and, therefore, in the observables.
In the case of molecular physics (i.e., QED), the leading-order term in the expansion corresponds to the Born--Oppenheimer approximation.
Although this is no more the case in QCD, we will nevertheless call the resulting EFT, the Born--Oppenheimer effective field theory (BOEFT).

\subsubsection{Born--Oppenheimer EFT}
\label{Sect:4.2.3bis}
\begin{figure}[ht]
\begin{center}
\includegraphics[width=0.4\linewidth]{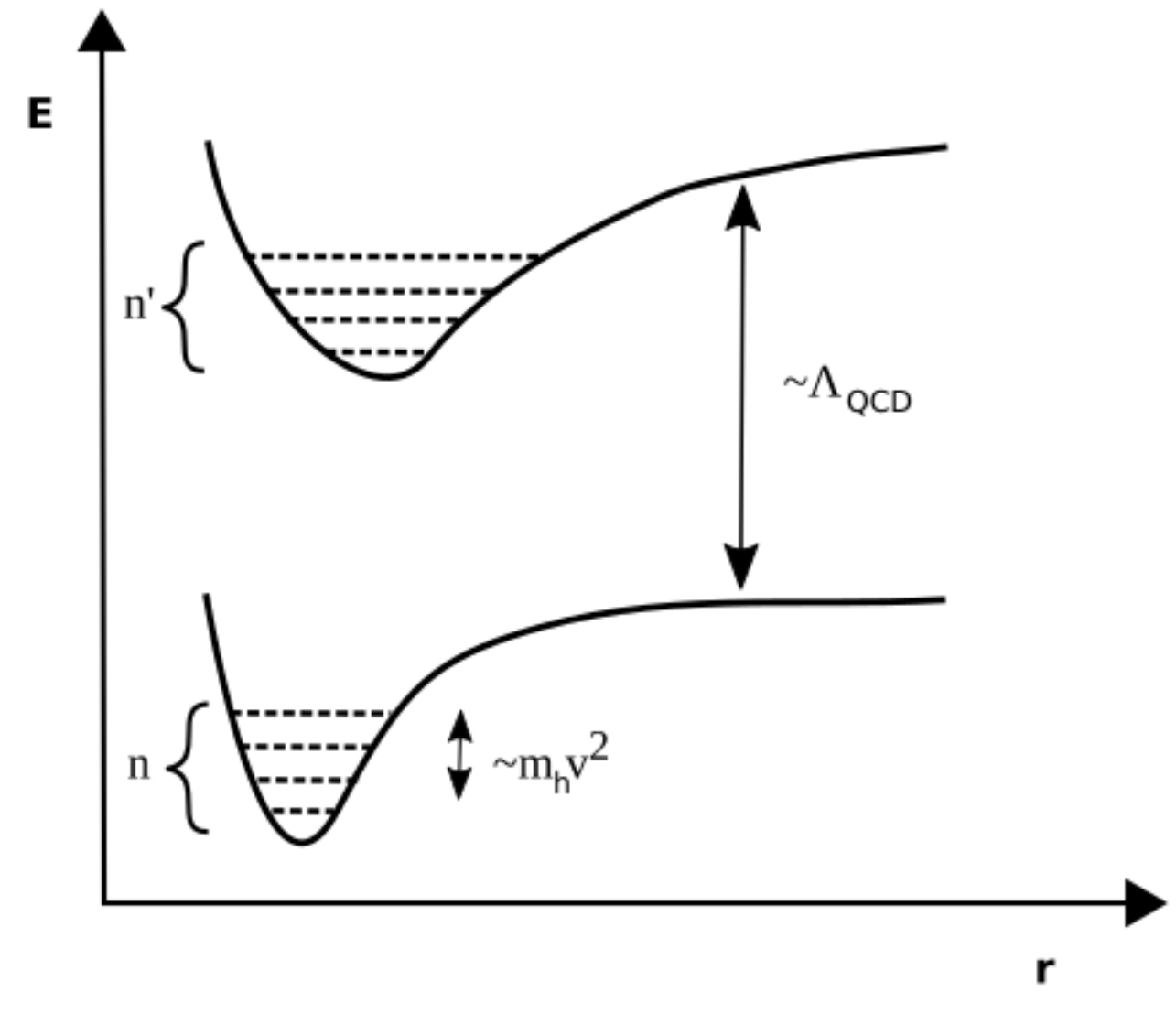}
\caption{Schematic distribution of the hybrid or tetraquark energy levels in the Born--Oppenheimer picture.}
\label{fig:hybridsBO}
\end{center}
\end{figure}

In the previous section, we have studied strongly-coupled quarkonia under the assumption that the quarkonium energy levels develop a mass gap of
order $\Lambda_{\text{QCD}}$ with respect to the spectrum of quarkonium hybrids, quarkonium plus glueballs and other exotic states.
This assumption allows one to integrate out all degrees of freedom with the exception of the heavy quark-antiquark pair and cast their contribution into a potential.
We may call this the Born--Oppenheimer assumption.
If we extend it to hybrid (tetraquark) heavy quarkonium states,
we may picture each hybrid (tetraquark) state as being a vibrational mode of a heavy quark-antiquark pair inside a potential.
The hybrid (tetraquark) potential may be identified with the energy of a gluonic (light-quark pair) excitation between the heavy quark-antiquark pair with given quantum numbers.
Ideally, gluonic (light-quark pair) excitations are separated from each other by a gap of order $\Lambda_{\text{QCD}}$,
while vibrational modes of a heavy quark-antiquark pair inside a given potential have a typical energy of order $m_h v^2$,
which is assumed to be much smaller than $\Lambda_{\text{QCD}}$.
An illustration of the distribution of vibrational modes for the different quarkonium and hybrid (or tetraquark) potentials, under the above assumptions, is shown in Fig.~\ref{fig:hybridsBO}.
A Born--Oppenheimer picture to describe quarkonium hybrids has been suggested in Refs.~\cite{Juge:1999ie,Braaten:2014qka,Berwein:2015vca}.
An EFT for quarkonium hybrids in the framework of strongly-coupled pNRQCD under the Born--Oppenheimer assumption has been developed in Ref.~\cite{Brambilla:2017uyf} under the name of Born--Oppenheimer EFT (BOEFT).
A similar approach is in Refs.~\cite{Oncala:2017hop,Soto:2017one}.
An extension to include excitations due to light quarks, i.e., tetraquarks, has been suggested in Refs.~\cite{Brambilla:2008zz,Braaten:2013boa},
and more explicitly worked out in Refs.~\cite{Braaten:2013boa,Braaten:2014ita,Braaten:2014qka,TarrusCastella:2019rit}. For recent related work see Ref.~\cite{Cai:2019orb}.

\vspace{0.3cm}\noindent
$\bullet$ {\it Hybrids}
\vspace{0.3cm}

\begin{figure}[ht]
\begin{center}
\includegraphics[width=0.525\linewidth]{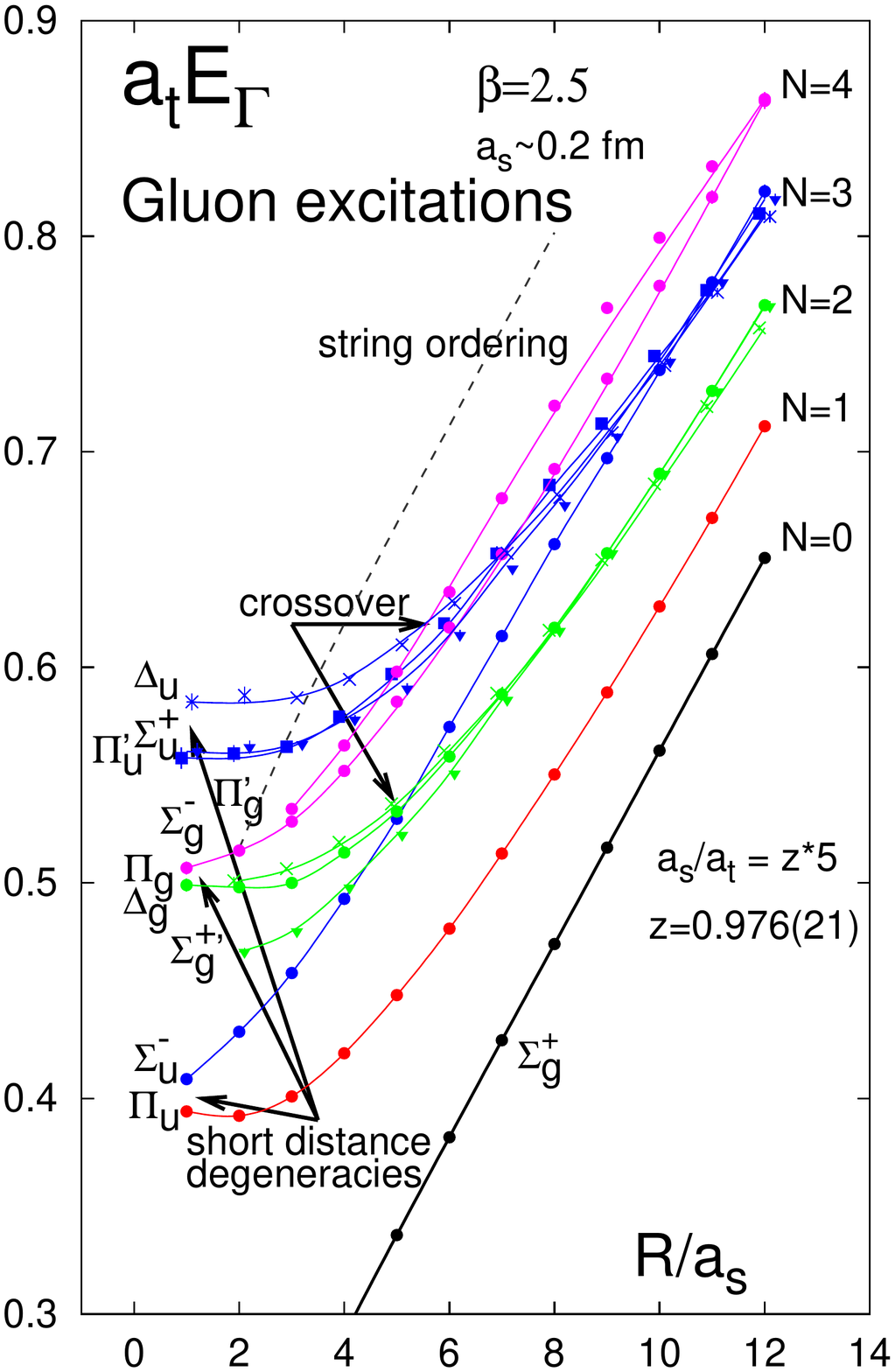}
\caption{Hybrid static energies in the lattice units $a_s$ and $a_t$ from the quenched lattice simulations of Ref.~\cite{Juge:2002br} at the lattice coupling $\beta = 2.5$.}
\label{fig:hybridsJKM}
\end{center}
\end{figure}

Hybrid potentials have been computed in the pure SU(3)-color gauge theory on the lattice for almost three decades~\cite{Perantonis:1990dy,Juge:1997nc,Foster:1998wu,Juge:2002br,Bali:2003jq,Capitani:2018rox}.
The lowest static hybrid energy has been shown in Fig.~\ref{fig:SESAM00} together with the heavy quark-antiquark static energy.
Higher static hybrid energies are shown in Fig.~\ref{fig:hybridsJKM}.
The gluonic static energies, $E_\Gamma$, are classified according to representations of the symmetry group $D_{\infty\,h}$, which is the symmetry group of diatomic molecules.
They are labeled $\Lambda_\eta^\sigma$. The quantum number $\Lambda$ is equal to $|\lambda|$, where $\lambda$ are the eigenvalues of $\hat{\bm{r}}\cdot{\bm{K}}$,
i.e., the angular momentum of the gluons, ${\bm{K}}$, projected on the unit vector, $\hat{\bm{r}}$, along the heavy-quark-antiquark axis.
$\Lambda$ may assume the values $0,1,2,\dots$; these numbers are usually indicated with capital greek letters: $\Lambda = \Sigma, \Pi, \Delta, \dots$.
The quantum number $\eta$ is the CP eigenvalue ($+1\equiv g$ and $-1 \equiv u$),
and $\sigma$ is the eigenvalue of reflection with respect to a plane passing through the heavy quark-antiquark axis.
The quantum number $\sigma$ is relevant only for $\Sigma$ states.
The lowest state, with quantum numbers $\Sigma_g^+$, describes a static quark-antiquark pair in the color singlet representation.
In general, there is more than one state for each irreducible representation of $D_{\infty\,h}$: higher states are denoted by primes, e.g., $\Pi_u$, $\Pi_u'$, $\Pi_u'', \dots$.
In the long range, the ordering of the hybrid static energies may be understood in terms of the effective string theory, see discussion in Ref.~\cite{Juge:2002br}.
In the limit $r\to 0$, the group $D_{\infty\,h}$ becomes the more symmetric group O(3)$\times$C.
This means that several different $\Lambda_\eta^\sigma$ representations reduce to the same $J^{PC}$ representation in that limit
and the corresponding static energies become degenerate~\cite{Foster:1998wu}. Hybrid states in the $r\to 0$ limit are often called gluelumps.
In particular, the gluelump multiplets ($\Sigma_u^-$, $\Pi_u$), ($\Sigma_g^{+\prime}$, $\Pi_g$), ($\Sigma_g^-$, $\Pi_g^\prime$, $\Delta_g$), ($\Sigma_u^+$, $\Pi_u^\prime$, $\Delta_u$) are degenerate.
Finally, in the very short range, $r \Lambda_{\text{QCD}} \ll 1$, all hybrid potentials behave like the Coulomb color-octet potential,
which is the short-distance component of the potential between a static quark-antiquark pair in the adjoint representation.
The symmetry O(3)$\times$C is made manifest in pNRQCD. Hence pNRQCD is the suitable EFT framework for studying hybrids in the short range~\cite{Brambilla:1999xf}.

Figure~\ref{fig:hybridsJKM} shows that close to their minimum the hybrid potentials tend, indeed, to cluster into the gluelump multiplets, i.e., according to the short-distance symmetry group O(3)$\times$C.
Hence, assuming that the physical hybrid states live close to their potential minimum, the simplified Born--Oppenheimer picture of Fig.~\ref{fig:hybridsBO} has to be modified
by replacing single potentials with multiplets of nearly degenerate potentials.
Because of the near degeneracy, vibrational modes associated to these potentials may have a significant mixing.
Different multiplets are separated by a mass gap of order $\Lambda_{\text{QCD}}$.

Let us consider, in the following, hybrids that are vibrational modes of the lowest-lying static energies.
According to Fig.~\ref{fig:hybridsJKM}, these are the static energies with quantum numbers $\Pi_u$ and $\Sigma_u^-$.
In the $r \to 0$ limit $\Pi_u$ and $\Sigma_u^-$ are degenerate and correspond to a gluonic operator with quantum numbers~$1^{+-}$.
The Born--Oppenheimer EFT Lagrangian density describing hybrids that are vibrational modes of the static potentials $\Pi_u$ and $\Sigma_u^-$
is a generalization of the Lagrangian for strongly-coupled pNRQCD, Eq.~\eqref{eq:sect4:pNRQCDstrong}, for nearly degenerate states.
It reads~\cite{Berwein:2015vca,Brambilla:2017uyf}:
\begin{equation}
 {\mathcal L}_{\text{BOEFT for }1^{+-}} =
 \int d^3r \, \sum_{\lambda\lambda'} \mathrm{Tr}\left\{ \Psi^{\dagger}_{1^{+-}\lambda}
 \left(i\partial_0 - V_{1^{+-}\lambda\lambda'}(r) + \hat{r}^{i\dagger}_{\lambda}\frac{\bm{\nabla}^2_r}{m_h}\hat{r}^i_{\lambda'}\right)\Psi_{1^{+-}\lambda'} \right\} \,.
\label{lag:hyb}
\end{equation}
An equivalent Lagrangian is in Ref.~\cite{Oncala:2017hop}.
The quantum number $\lambda$ (and $\lambda'$) may assume the values $\pm1$ and $0$.
The projectors to the eigenstates of the angular momentum of the gluons along the heavy-quark-antiquark axis are
$\hat{r}^i_0=\hat{r}^i$ and $\hat{r}^i_{\pm1}=\mp\left(\hat{\theta}^i\pm i\hat{\phi}^i\right)/\sqrt{2}$,
where $\hat{\bm r} = (\sin\theta\cos\phi,\,\sin\theta\sin\phi\,,\cos\theta)$, $\hat{\bm \theta} =$ $(\cos\theta\cos\phi,$ $\,\cos\theta\sin\phi\,,-\sin\theta)$
and $\hat{\bm \phi} = (-\sin\phi,\,\cos\phi\,,0)$.
The fields $\Psi_{1^{+-}\lambda}$ depend on time, the c.m. coordinate $\bm{R}$, and the relative coordinate $\bm{r}$.
The modes of the fields $\Psi_{1^{+-}\lambda}$ are the hybrid eigenstates that may be built out of the hybrid potentials $V_{1^{+-}\lambda\lambda'}$.
The hybrid potentials may be organized, like the quarkonium potentials, as an expansion in the inverse of the heavy quark mass, $1/m_h$.
The static potential, $V_{1^{+-}\lambda\lambda'}^{(0)} = \delta_{\lambda\lambda'} V_{1^{+-}\lambda}^{(0)}$, coincides with the hybrid static energy: $V_{1^{+-}0}^{(0)} = E_{\Sigma_u^-}$ and $V_{1^{+-}\pm1}^{(0)} = E_{\Pi_u}$.
For what concerns hybrid potentials of higher order in $1/m_h$, there are no available lattice determinations.
In the absence of them, these potentials have been modeled either using the short-distance multipole expansion~\cite{Brambilla:2018pyn},
or both the short-distance multipole expansion and the long-range effective string theory~\cite{Oncala:2017hop,Soto:2017one}.

\begin{table}[ht]
\begin{center}
\begin{tabular}{|c|c|c|c|c|}
\hline
Multiplet & $\,\,\,T\,\,\,$ & $J^{PC}(S=0)$ & $J^{PC}(S=1)$ & $E_\Gamma$\\
\hline
$H_1$& $1$ & $1^{--}$ & $(0,1,2)^{-+}$ &$E_{\Sigma^-_u}$, $E_{\Pi_u}$\\
$H_2$& $1$ & $1^{++}$ & $(0,1,2)^{+-}$ &$E_{\Pi_u}$\\
$H_3$& $0$ & $0^{++}$ & $1^{+-}$ &$E_{\Sigma^-_u}$\\
$H_4$& $2$ & $2^{++}$ & $(1,2,3)^{+-}$ &$E_{\Sigma^-_u}$, $E_{\Pi_u}$\\
\hline
\end{tabular}
\caption{Lowest-lying quarkonium hybrid multiplets.
 The number labeling $H$ reflects the order in which the state appears in the spectrum from lower to higher masses.
 $S$ is the total spin of the quark-antiquark pair, and $T$ is the sum of the orbital angular momentum of the quark-antiquark pair
 and the gluonic angular momentum.
 Note that the $T=0$ state is not the lowest mass state~\cite{Berwein:2015vca}.}
\label{tab:spin_multiplet}
\end{center}
\end{table}

The term $\hat{r}^{i\dagger}_{\lambda} ({\bm{\nabla}^2_r}/{m_h})\hat{r}^i_{\lambda'}$ in the Lagrangian~\eqref{lag:hyb}
can be split into a kinetic operator acting on the heavy quark-antiquark field and a nonadiabatic coupling:
$\hat{r}^{i\dagger}_{\lambda} ({\bm{\nabla}^2_r}/{m_h}) \hat{r}^i_{\lambda'} = \delta_{\lambda\lambda^\prime} \, {\bm{\nabla}^2_r}/{m_h}+ C^{\rm nad}_{1^{+-}\lambda\lambda^{\prime}}$,
with $C^{\rm nad}_{1^{+-}\lambda\lambda^{\prime}} = \hat{r}^{i\dagger}_{\lambda} [{\bm{\nabla}^2_r}/{m_h}, \hat{r}^i_{\lambda'}]$ being the nonadiabatic coupling.
Concerning the size of the different terms appearing in the Lagrangian~\eqref{lag:hyb},
the temporal derivative, the kinetic term and the leading-order (static) potential (up to a constant shift) are of order $m_h v^2$.
As discussed in Ref.~\cite{Brambilla:2017uyf}, this is also the size of the nonadiabatic coupling.
For diatomic molecules the counting is different, essentially because the size of the electron cloud and the distance of the two atoms are in the molecule of the same order,
while for hybrids the distance of the heavy quark-antiquark pair is of order $1/(m_h v)$, i.e., smaller than the size of the hadron, which is of order $\Lambda_{\text{QCD}}$.
As a consequence the adiabatic coupling may be treated as a perturbation in diatomic molecules, while it contributes at leading order to heavy quarkonium hybrids.
Another consequence is that, since in diatomic molecules there is no special hierarchy between these two lengths,
there is neither a special symmetry at short distances nor a corresponding degeneracy pattern, which are instead typical, as we have seen, of heavy quarkonium hybrids.

\begin{figure}[ht]
\centerline{\includegraphics[width=0.95\textwidth]{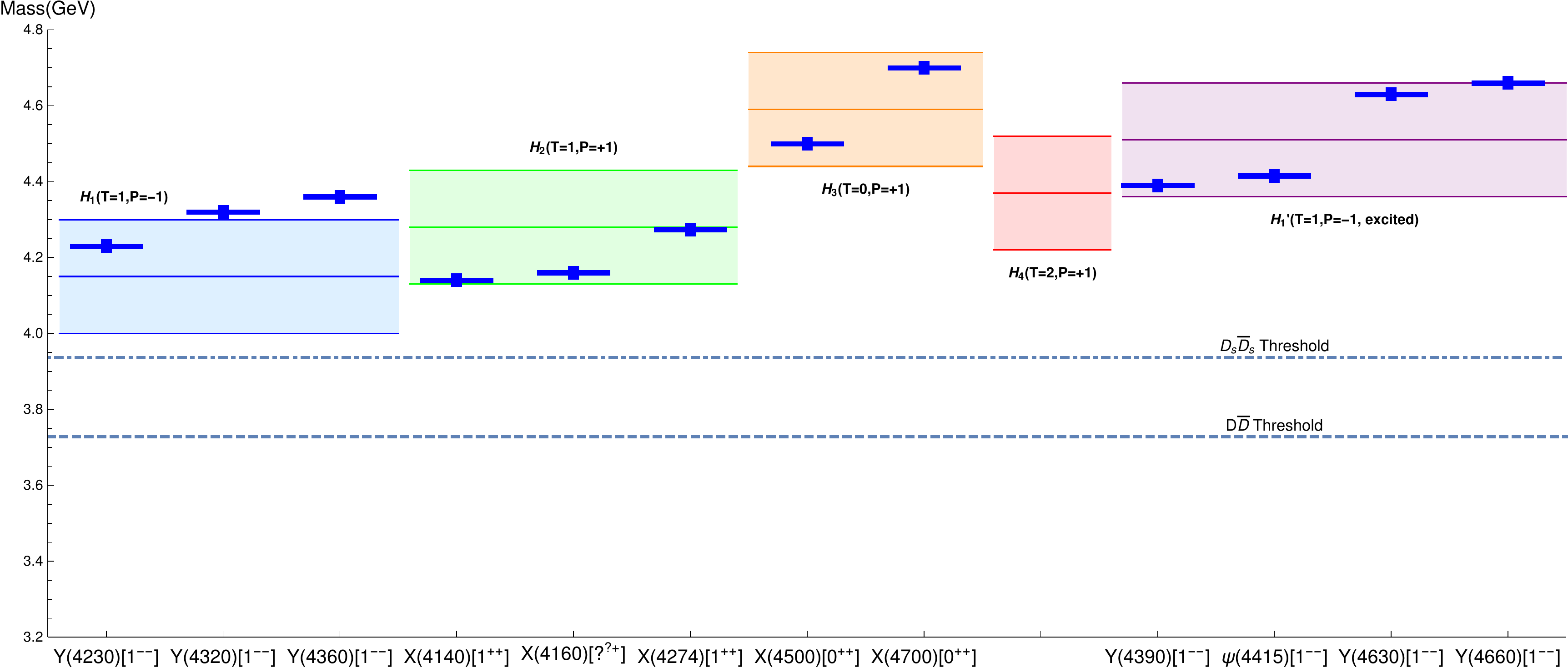}}
\caption{Mass spectrum of neutral exotic charmonium states obtained by solving Eqs.~\eqref{eq:coupledhadron} according to Ref.~\cite{Berwein:2015vca}.
 The experimental states of Sec.~\ref{Sect:3.1} that have matching quantum numbers are plotted in solid blue lines.
 The spin-symmetry multiplets are labeled $H_1$ ($T=1$, negative parity), $H_2$ ($T=1$, positive parity), $H_3$ ($T=0$, positive parity),
 $H_4$ ($T=2$, positive parity), and $H^{\prime}_1$ (first radially excited state with $T=1$ and negative parity).
 The multiplets have been plotted with error bands corresponding to a gluelump mass uncertainty of $0.15$~GeV. Figure updated from Refs.~\cite{Berwein:2015vca,Brambilla:2017uyf}.
\label{fig:exp101016}}
\end{figure}

The leading-order equations of motion for the fields $\Psi^{\dagger}_{1^{+-}\lambda}$ that follow from the Lagrangian~\eqref{lag:hyb} are a set of coupled Schr\"odinger equations~\cite{Berwein:2015vca}
\begin{equation}
 i\partial_0 \Psi_{1^{+-}\lambda} =
 \left[\left(-\frac{\bm{\nabla}^2_r}{m_h} + V_{1^{+-}\lambda}^{(0)}\right)\delta_{\lambda\lambda^{\prime}} -\sum_{\lambda^{\prime}}C^{\rm nad}_{1^{+-}\lambda\lambda^{\prime}}\right]\Psi_{\kappa\lambda^{\prime}}\,,
\label{eq:coupledhadron}
\end{equation}
which generalize Eq.~\eqref{pNRQCDSchr-equation}.
By solving them we obtain the eigenvalues $\mathcal{E}_N$ that give the masses $M_N$ of the states as $M_N = 2m_h + \mathcal{E}_N$.
Keeping in the Eqs.~\eqref{eq:coupledhadron} only the heavy quark-antiquark
kinetic energy and the hybrid static energies, $E_{\Sigma_u^-}$ and $E_{\Pi_u}$, amounts to the Born--Oppenheimer approximation.
Keeping only the diagonal terms amounts to the adiabatic approximation.
As we argued above, the exact leading-order equations include both diagonal and off-diagonal terms of the so-called non-adiabatic coupling,
as both diagonal and off-diagonal terms contribute at the same order to the energy levels of heavy quarkonium hybrids.

The coupled Schr\"odinger equations~\eqref{eq:coupledhadron} mix states with the same parity.
A physical consequence of this mixing is the so-called $\Lambda$-doubling,
i.e., the lifting of degeneracy between states with opposite parity.
The effect is also present in molecular physics, however, there
$\Lambda$-doubling is a subleading effect,
while it is a leading-order effect in the spectrum of quarkonium hybrids.
The eigenstates of the Eqs.~\eqref{eq:coupledhadron} are organized in the
multiplets shown in Table~\ref{tab:spin_multiplet}.
In the adiabatic approximation the multiplets $H_1$ and $H_2$ would be degenerate~\cite{Braaten:2013boa,Braaten:2014ita,Braaten:2014qka}.
The solution of the full set of equations~\eqref{eq:coupledhadron} is the spectrum shown in Fig.~\ref{fig:exp101016}.
The spectrum has been computed using for the charm quark the mass in the
renormalon subtraction (RS) scheme $m_{c\,\text{RS}}= 1.477(40)$~GeV~\cite{Pineda:2001zq,Bali:2003jq}
(for more details about masses in renormalon subtraction schemes see also Sec.~\ref{Sect:4.2.1}).
The gluelump masses, which enter in the normalization of the hybrid potentials,
have been computed in the same scheme and assigned an uncertainty of $\pm 0.15$~GeV,
which is the largest source of uncertainty in the hybrid masses~\cite{Bali:2003jq}.
Figure~\ref{fig:exp101016} clearly shows $\Lambda$-doubling, as the degeneracy between the $H_1$ and $H_2$ multiplets is lifted.
The experimental states plotted in Fig.~\ref{fig:exp101016} are neutral isoscalar states in the charmonium sector taken from Sec.~\ref{Sect:3.1} with matching $J^{PC}$ quantum numbers.
The figure does not imply that all these states should be identified with hybrids.
A critical analysis of their properties and different possible interpretations have been extensively presented in Sec.~\ref{Sect:3.1},
and in Secs.~\ref{Sect:4.1.3}, \ref{Sect:4.1.4} and~\ref{Sect:4.1.5}.
Promising candidates for charmonium hybrids or for states with a large hybrid component
are the $Y(4230)$ and $Y(4390)$ because of their significant width into $\pi^+\pi^-h_c$.
In the hybrid picture this decay does not need spin flipping of the heavy quark-antiquark pair, which is in a spin zero state.
Spin-flipping terms are suppressed in the heavy quark limit.
Nevertheless, it has been pointed out in Ref.~\cite{Oncala:2017hop} that mixing with spin one quarkonium states happens already at order $1/m_h$ (see below).
This possibly large mixing may allow for significant widths also into final states with spin one quarkonia, in particular $\pi^+\pi^-J/\psi$.
In Ref.~\cite{Berwein:2015vca} a similar study as the one summarized in Fig.~\ref{fig:exp101016}
has been done for hybrids in the bottomonium [$(b\bar{c})$] sector:
$H_1$ multiplets get a mass of $(10.79 \pm 0.15)$~GeV [$(7.48 \pm 0.15)$~GeV],
$H_2$ multiplets a mass of $(10.84 \pm 0.15)$~GeV [$(7.58 \pm 0.15)$~GeV],
$H_3$ multiplets a mass of $(11.06 \pm 0.15)$~GeV [$(7.85 \pm 0.15)$~GeV],
$H_4$ multiplets a mass of $(10.90 \pm 0.15)$~GeV [$(7.65 \pm 0.15)$~GeV],
and $H_1'$ multiplets a mass of $(10.98 \pm 0.15)$~GeV [$(7.76 \pm 0.15)$~GeV].
The RS bottom mass has been fixed at $m_{b\,\text{RS}}= 4.863(55)$~GeV.
From the experimental side, candidate states of bottomonium hybrids in the $H_1$ or $H_1'$ multiplets
are the $\Upsilon(10860)~[1^{--}]$, with a mass of $M_{\Upsilon(10860)} = (10891.1\pm3.2^{+0.6}_{-1.7})$~MeV
and the $\Upsilon(11020)~[1^{--}]$, with a mass of $M_{\Upsilon(11020)}=(10987.5^{+6.4}_{-2.5}\,^{+9.0}_{-2.1})$~MeV~\cite{Santel:2015qga}, see Sec.~\ref{Sect:3.2.3}.
To these we can add the recently observed signal by Belle with a mass of $M_{\Upsilon(10750)}=(10752.7\pm5.9\,^{+0.7}_{-1.1})$~MeV,
which may also qualify as an $H_1$ multiplet bottomonium hybrid candidate~\cite{Abdesselam:2019gth}.

The charmonium hybrid spectrum has been recently computed by the Hadron
Spectrum Collaboration using dynamical lattice QCD simulations:
in Ref.~\cite{Liu:2012ze} using anisotropic lattices with $2+1$ flavors at a pion mass of $396$~MeV,
and in Ref.~\cite{Cheung:2016bym} in an improved setting at a pion mass of $236$~MeV, see Fig.~\ref{fig:lattice:excitedcharmoniumspectrum}.
Lattice data are consistent with the computations reported in Fig.~\ref{fig:exp101016},
although the quoted lattice errors are somewhat smaller but these are only the statistical uncertainties
--- see Sec.~\ref{Sect:4.3:charmonia} for a discussion of these
lattice results and potential systematic uncertainties.
In the case of the $H_1$ multiplet the smaller errors make the data sensitive to the spin splittings of the states inside the multiplet.
The spin splittings provide relevant information on the heavy quarkonium hybrid spin interaction, as we will discuss at the end of this section.
The bottomonium hybrid spectrum has been studied, so far, only on quenched lattices:
in Ref.~\cite{Juge:1999ie} using anisotropic lattices and treating the bottom-antibottom pair non relativistically in the framework of lattice NRQCD,
and in Ref.~\cite{Liao:2001yh} using anisotropic lattices and treating the bottom-antibottom pair fully relativistically.
At the present state of our knowledge the results of Ref.~\cite{Juge:1999ie} appear consistent with the mass values of the bottomonium hybrid multiplets reported above from Ref.~\cite{Berwein:2015vca},
while those of Ref.~\cite{Liao:2001yh} show some major discrepancies. We refer to Ref.~\cite{Berwein:2015vca} for a discussion.

\begin{figure}[ht]
\begin{center}
\includegraphics[width=0.37\textwidth]{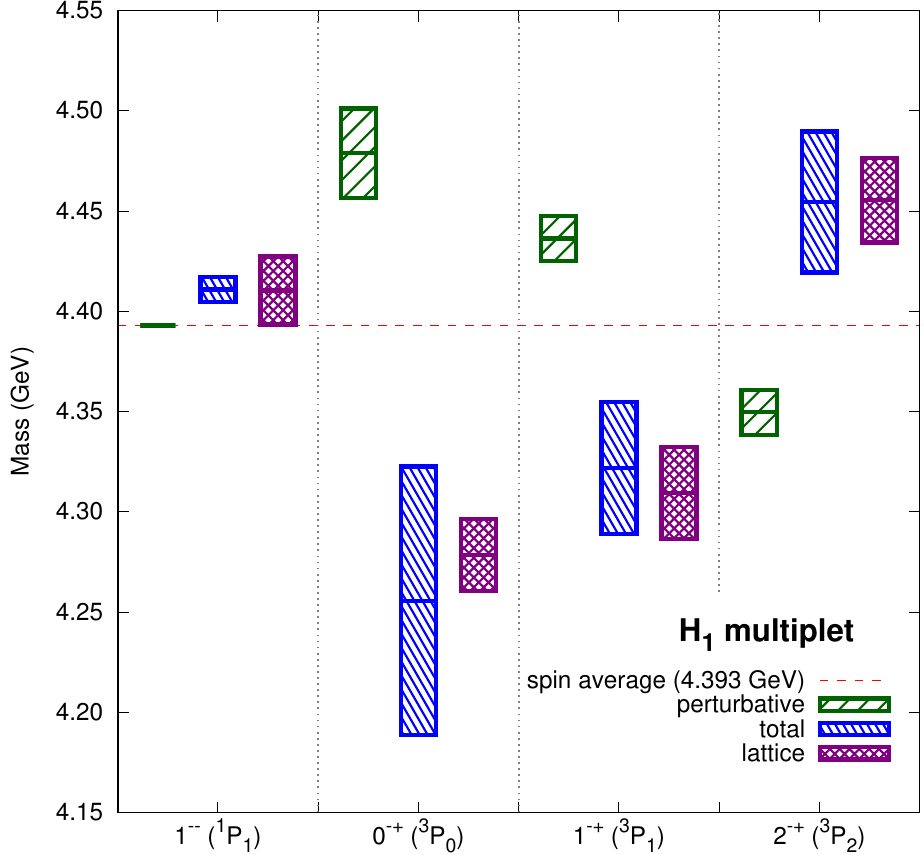}
\hspace*{0.50cm}
\includegraphics[width=0.37\textwidth]{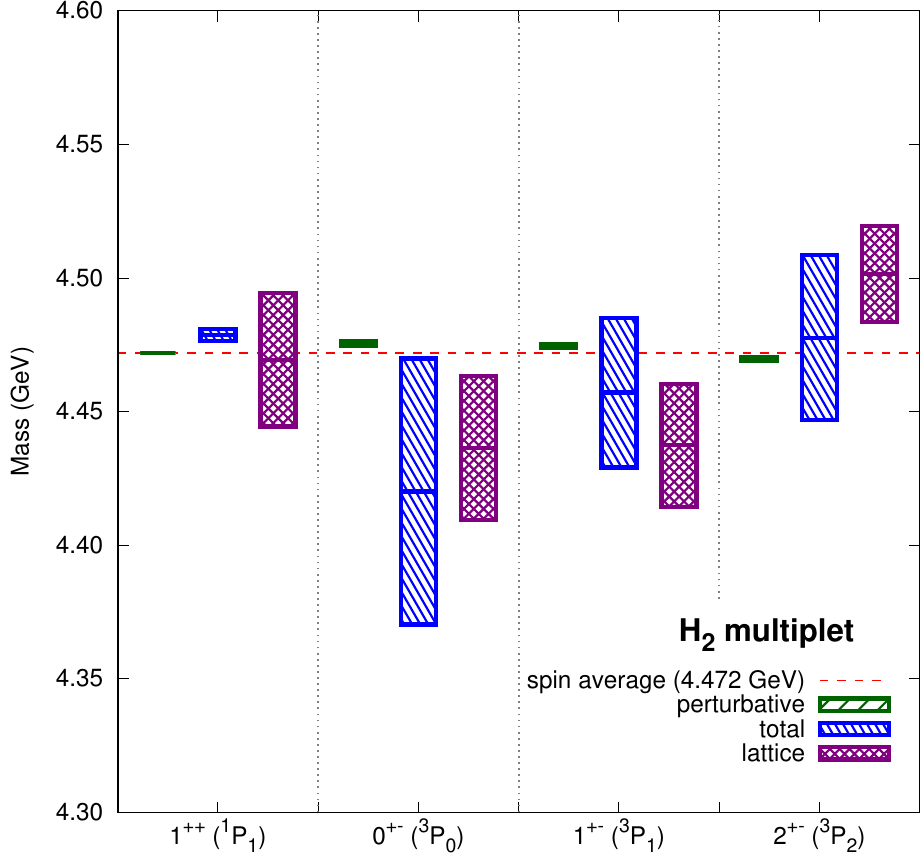}
\\[2ex]
\includegraphics[width=0.37\textwidth]{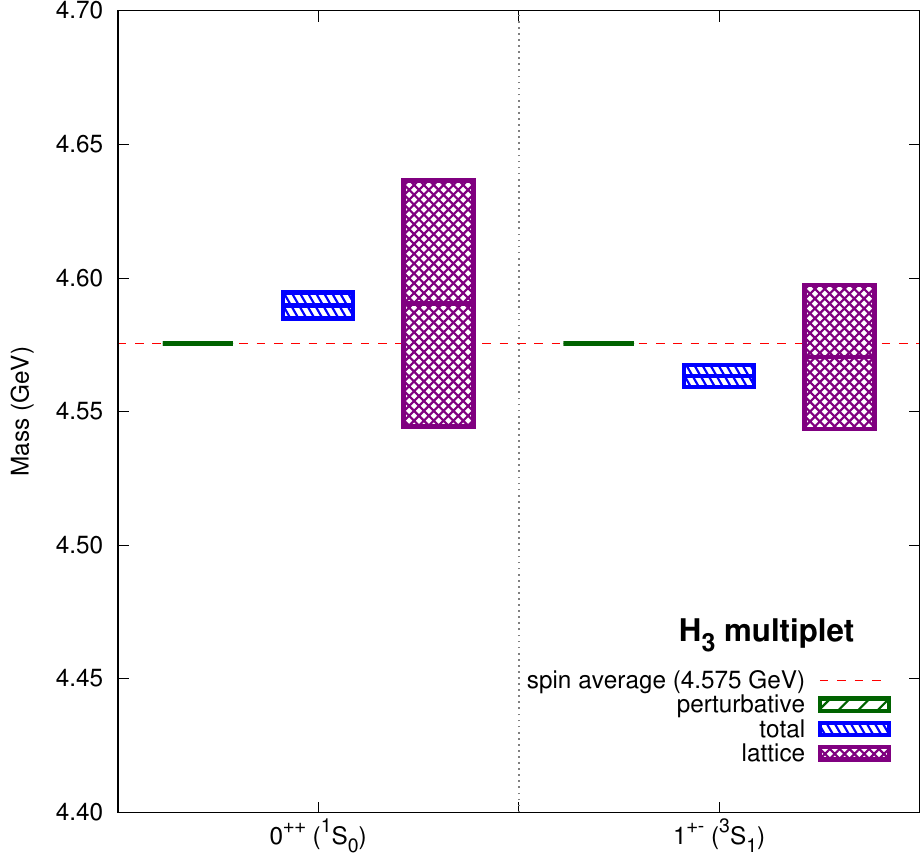}
\hspace*{0.50cm}
\includegraphics[width=0.37\textwidth]{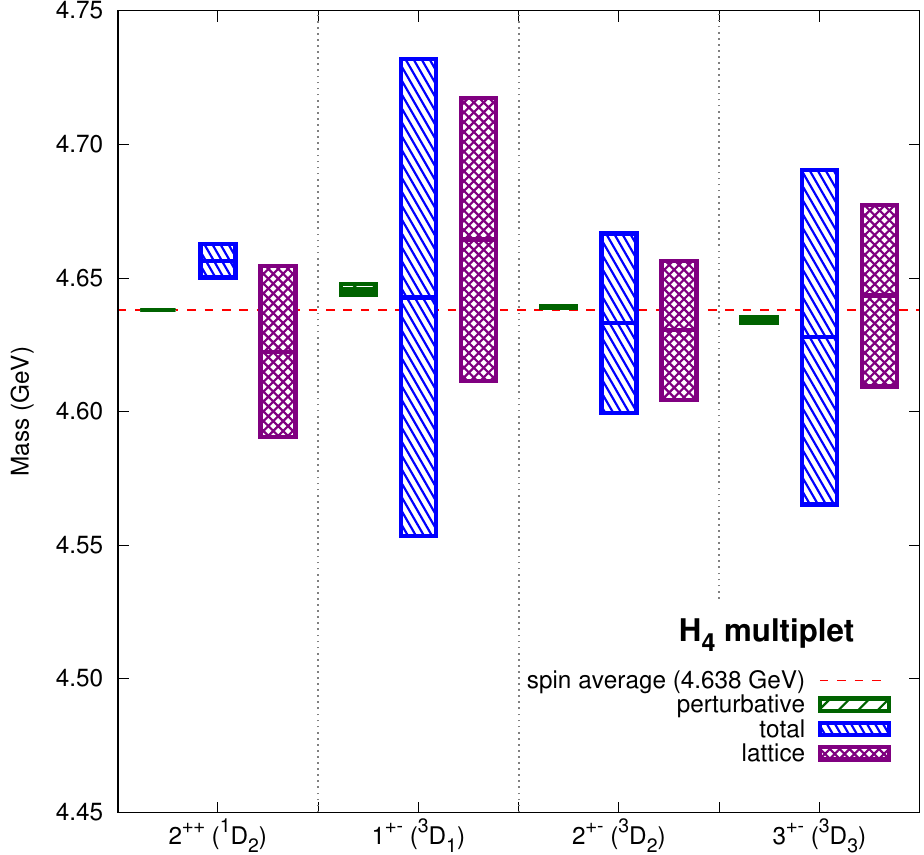}
\caption{Spectrum of the four lowest-lying charmonium hybrid multiplets.
The lattice results from Ref.~\cite{Cheung:2016bym} with $m_{\pi}\approx 240$~MeV are the most right (purple) boxes for each quantum number.
The perturbative contributions to the spin-dependent operators added to the spin average of the lattice results (red dashed lines) are the most left (green) boxes.
The central (blue) boxes for each quantum number are the full results from the spin-dependent operators including perturbative and non-perturbative contributions.
The unknown non-perturbative matrix elements are fitted to reproduce the lattice data.
Figure from Ref.~\cite{Brambilla:2018pyn}.
}
\label{fg:ccg_Cheung}
\end{center}
\end{figure}

The bottomonium and charmonium hybrid spectrum has been studied in the framework of QCD sum rules in Ref.~\cite{Chen:2013zia}
and the $b\bar{c}$ hybrid spectrum in Ref.~\cite{Chen:2013eha}.
Using hybrid operators and computing correlation functions and spectral functions up to dimension-six condensates, the sum rules stabilize and provide mass predictions.
The masses for the $H_1$ hybrids are between $3.4$~GeV and $4$~GeV in the charmonium sector, between $9.7$~GeV and $9.9$~GeV
in the bottomonium sector, and between $6.8$~GeV and $7.2$~GeV for the $c\bar{b}$ hybrids.
Hence, for this multiplet they are somewhat smaller than in the analysis of Ref.~\cite{Berwein:2015vca}.
The other multiplets are consistent with Ref.~\cite{Berwein:2015vca},
although the $1^{++}$ state of the charmonium $H_2$ multiplet and the $H_3$ multiplet tend to be heavier.
Similar observations also hold for the $1^{+}$ state of the $b\bar{c}$ $H_2$ multiplet and the $H_3$ multiplet,
while the bottomonium $0^{+-}$ hybrid state of the $H_2$ multiplet tends to be lighter.

\begin{figure}[ht]
\begin{center}
\includegraphics[width=0.37\textwidth]{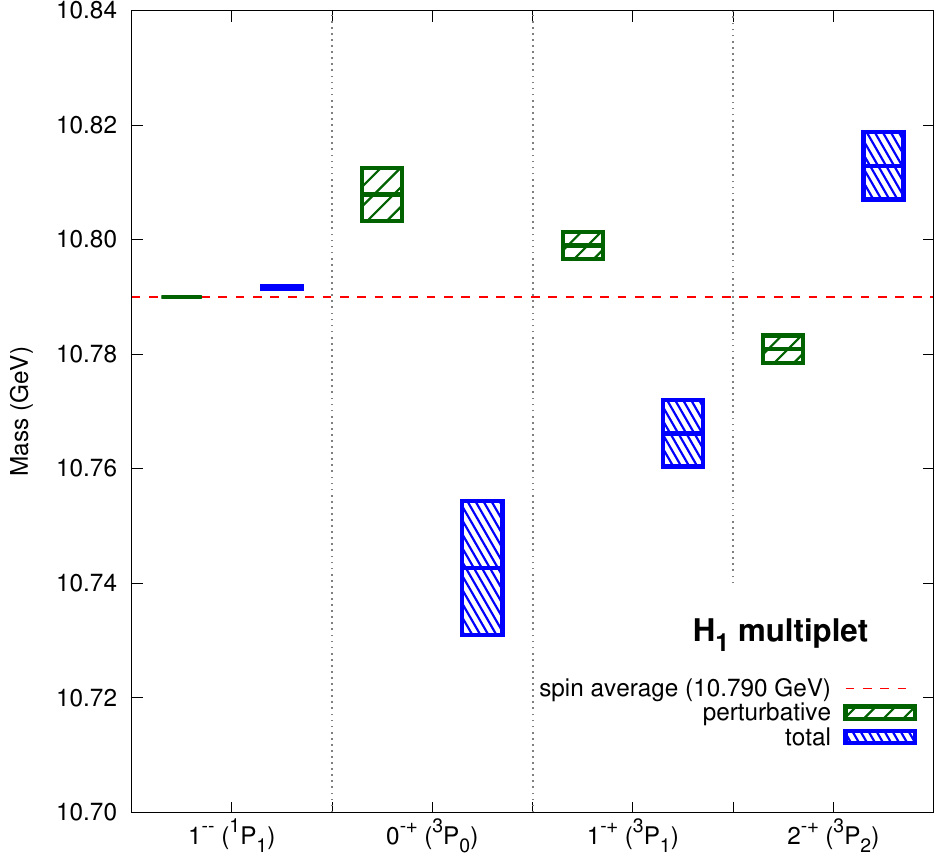}
\hspace*{0.50cm}
\includegraphics[width=0.37\textwidth]{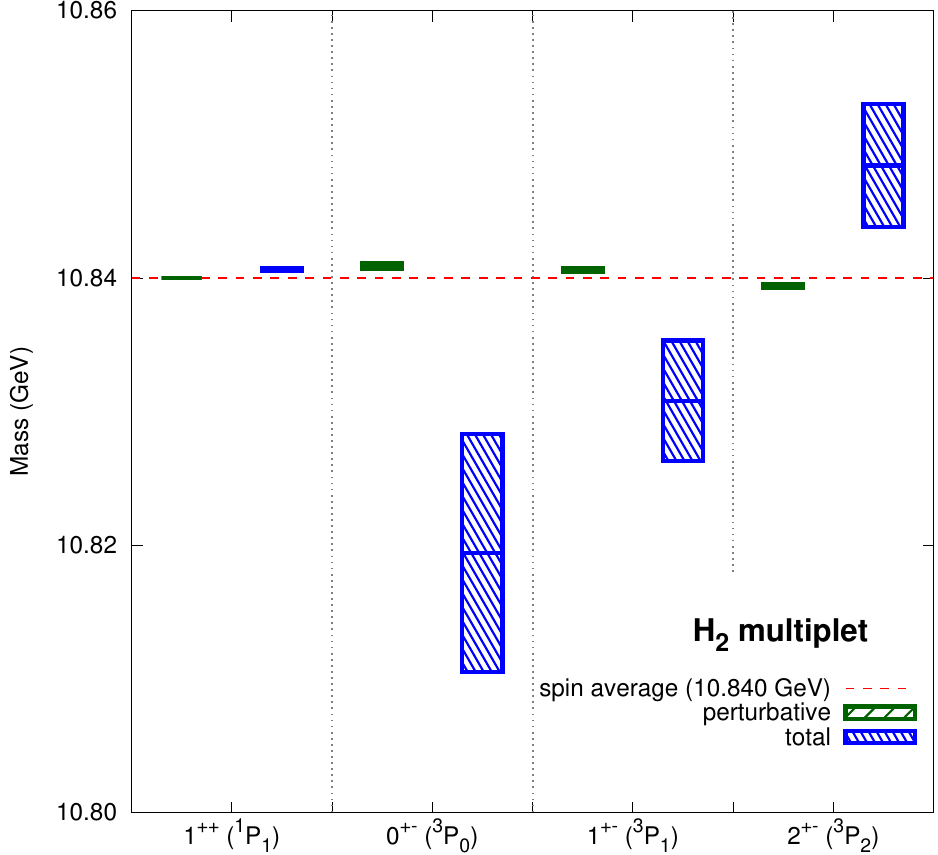}
\\[2ex]
\includegraphics[width=0.37\textwidth]{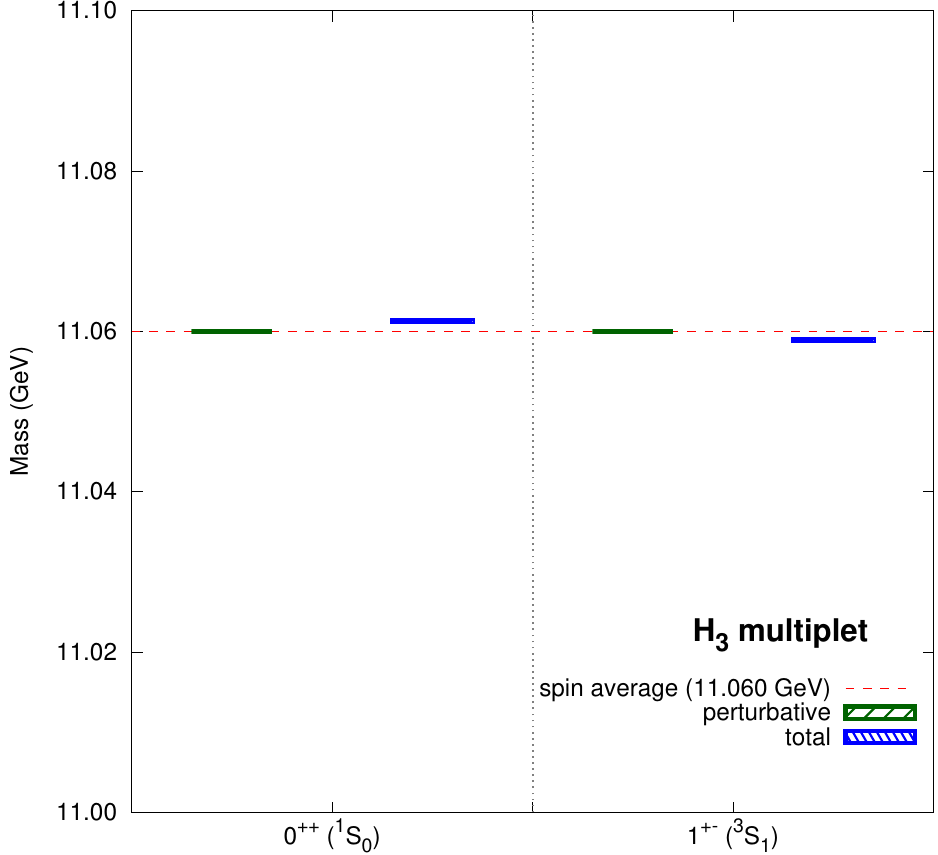}
\hspace*{0.50cm}
\includegraphics[width=0.37\textwidth]{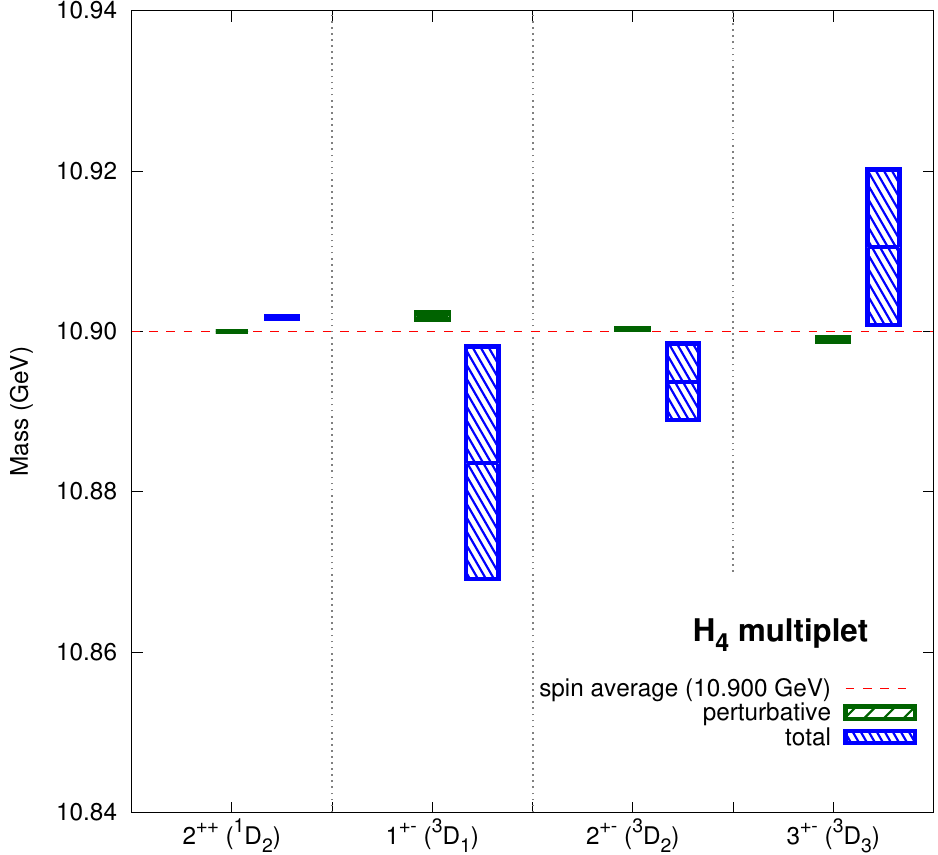}
\caption{Spectrum of the four lowest-lying bottomonium hybrid multiplets computed by adding the spin-dependent contributions to the spectrum obtained in Ref.~\cite{Berwein:2015vca}.
 The non-perturbative contributions are determined from the fit of the charmonium hybrid spectrum of Ref.~\cite{Cheung:2016bym}, see Fig.~\ref{fg:ccg_Cheung}.
 The average mass for each multiplet is displayed as a red dashed line.
 The results with only the perturbative contributions included and the full results with perturbative and non-perturbative contributions included
 are shown for each quantum number by the left (green) and the right (blue) boxes, respectively.
 Figure from Ref.~\cite{Brambilla:2018pyn}.
}
\label{fg:bbg_Cheung}
\end{center}
\end{figure}

As remarked above, the lattice data of Ref.~\cite{Cheung:2016bym} are sensitive to the charmonium hybrid spin splittings (but it is important to remember that these data suffer from systematic uncertainties).
In Ref.~\cite{Brambilla:2018pyn}, this information has been used to constrain the heavy quarkonium hybrid spin-dependent potential.
The heavy quarkonium hybrid spin-dependent potential may be organized, like in the quarkonium case, as an expansion in $1/m_h$, see Eqs.~\eqref{sec4.2.3.Vexpansion} and~\eqref{potentialSD2}.
Differently from the quarkonium case, the hybrid potential gets a first contribution already at order $1/m_h$~\cite{Oncala:2017hop}.
For the implications of this term to the spin splittings see Ref.~\cite{Soto:2017one};
heavy quarkonium hybrid spin splittings in the Born--Oppenheimer framework with NRQCD have been also analyzed in Ref.~\cite{Lebed:2017xih}.
The $1/m_h$ potential is the leading contribution to the spin-dependent potential.
It stems from the expectation value of the chromomagnetic operator, $c_F\psi^\dagger \bm{\sigma}\cdot \bm{B}/(2m_h)\psi + \text{charge conjugation}$, on the hybrid state
(the operator appears in the NRQCD Lagrangian~\eqref{eq:sect4:NRQCD}, and, in the framework of pNRQCD, among the operators listed in Ref.~\cite{Brambilla:2003nt}).
It is of order $\Lambda^2_{\text{QCD}}/m_h$.
The same operator does not contribute at leading order to matrix elements of quarkonium states as its projection on quark-antiquark color singlet states vanishes.
Hence, spin splittings are remarkably less suppressed in heavy quarkonium hybrids than in heavy quarkonia.
For the lowest-lying hybrid excitations of a gluonic operator with quantum numbers~$1^{+-}$, the $1/m_h$ spin-dependent potential takes the form
\begin{equation}
V_{1^{+-}\lambda\lambda'\,{\rm SD}}^{(1)}(\bm{r}) = V_{SK}(r)\left(\hat{r}^{i\dagger}_{\lambda}\bm{K}^{ij}\hat{r}^j_{\lambda'}\right)\cdot\bm{S}
+ V_{SK\,b}(r)\left[\left(\bm{r}\cdot \hat{\bm{r}}^{\dagger}_{\lambda} \right)\left(r^i\bm{K}^{ij} \hat{r}^j_{\lambda'} \right)\cdot\bm{S}
+ \left(r^i\bm{K}^{ij} \hat{r}^{j\dagger}_{\lambda}\right)\cdot\bm{S} \left(\bm{r}\cdot \hat{\bm{r}}_{\lambda'} \right)\right], \label{sdm2}
\end{equation}
where $\left({K}^{ij}\right)^k=i\epsilon^{ikj}$ is the angular momentum operator of the spin one gluons.
Also the $1/m_h^2$ spin-dependent potential presents some new structures with respect to the corresponding quarkonium potential~\eqref{potentialSD2}:
\begin{align}
 V_{1^{+-}\lambda\lambda'\,{\rm SD}}^{(2)}(\bm{r}) &=
 V^{(2)}_{LS\,a}(r)\left(\hat{r}^{i\dagger}_{\lambda}\bm{L}\,\hat{r}^i_{\lambda'} \right)\cdot\bm{S}
 + V^{(2)}_{LS\,b}(r) \hat{r}^{i\dagger}_{\lambda} \left(L^iS^j+S^iL^j\right) \hat{r}^j_{\lambda'}\nonumber\\
 & + V^{(2)}_{S^2}(r)\bm{S}^2\delta_{\lambda\lambda'}
 + V^{(2)}_{S_{12}\,a}(r) S_{12} \delta_{\lambda\lambda'}
 + V^{(2)}_{S_{12}\,b}(r) \hat{r}^{i\dagger}_{\lambda} \hat{r}^j_{\lambda'} \left(S^i_1S^j_2+S^i_2S^j_1\right),\quad\label{sdm3}
\end{align}
where $\bm{L}$ is the orbital angular momentum of the heavy-quark-antiquark pair, and the spin operators are defined as after Eq.~\eqref{potentialSD2}.
The terms proportional to $V^{(2)}_{LS\,a}(r)$, $V^{(2)}_{S^2}(r)$, and $V^{(2)}_{S_{12}\,a}(r)$ are present also in the quarkonium case.
The leading-order perturbative expressions of $V^{(2)}_{LS\,a}(r)$, $V^{(2)}_{S^2}(r)$, and $V^{(2)}_{S_{12}\,a}(r)$ follow from the perturbative expressions of the corresponding
quarkonium potentials in Eq.~\eqref{V2S12pNRQCD} after replacing an overall factor $-4/3$ by $1/6$, which is the same change that relates the Coulomb potential in the
SU(3) fundamental representation with the Coulomb potential in the SU(3) adjoint representation.
The functions $V^{(2)}_{LS\,a}(r)$, $V^{(2)}_{S^2}(r)$, and $V^{(2)}_{S_{12}\,a}(r)$ also get non-perturbative contributions.
Non-perturbative contributions are particularly important for $V_{SK}(r)$, $V_{SK\,b}(r)$, $V^{(2)}_{LS\,b}(r)$, and $V^{(2)}_{S_{12}\,b}(r)$,
since these terms vanish at leading order in perturbation theory.

In Fig.~\ref{fg:ccg_Cheung}, for each multiplet, the charmonium hybrid spin splittings computed in lattice QCD (purple, right boxes) are compared
with leading-order perturbation theory (green, left boxes) and with a fit that allows for non-perturbative contributions to the potentials listed in Eqs.~\eqref{sdm2} and~\eqref{sdm3}
(blue, central boxes). We see that the pattern of splittings induced by the perturbative contributions alone is opposite to the data.
It is also opposite to the quarkonium case, as the signs of all spin-dependent potentials is reversed at leading order.
Hence, non-perturbative contributions are crucial to bring the splittings in agreement with lattice data, consistently with the fact that the non-perturbative contribution
proportional to $V_{SK}(r)$ is the dominant spin-dependent potential for hybrids.
In Ref.~\cite{Brambilla:2018pyn}, the fitted contributions turn out to scale, in powers of $\Lambda_{\text{QCD}}$, as expected by the power counting of the EFT.
If the non-perturbative potentials may be organized according to the multipole expansion,
then information gained from the charmonium hybrid spin splittings can be used to predict, for each multiplet, the bottomonium hybrid spin splittings.
The predictions are shown in Fig.~\ref{fg:bbg_Cheung}. We see again that the characteristic pattern of spin splittings induced by the perturbative
contributions (green, left boxes) is reversed by the full result that includes the non-perturbative contributions (blue, right boxes).

\begin{table}[ht]
	\centering
	\begin{tabular}{|c|c|c|c|c|c|}
		\hline
			$nL_{T} \rightarrow n'L'$ & $\Delta E$ (MeV) & {$\Gamma$ (MeV) } \\ \hline					
 $c\bar{c}$ sector & &\\
 $1P_0\rightarrow 2S$ & {808} & {7.5(7.4)} \\
		$2(S/D)_1\rightarrow 1P$ & 861 & 22(19) \\
 $4(S/D)_1\rightarrow 1P$ & 1224 & 23(15) \\
		\hline
		\hline
 $b\bar{b}$ sector & &\\
 $1P_0\rightarrow 1S$ & {1569} & {44(23)} \\
		$1P_0\rightarrow 2S$ & {1002} & {15(9)} \\
		$2P_0\rightarrow 2S$ & {1290} &{2.9(1.3)} \\
		$2P_0\rightarrow 3S$ & {943} & {15(12)} \\
 $4P_0\rightarrow 1S$ & {2337} & {53(25)} \\
 $4P_0\rightarrow 2S$ & {1770} & {18(7)} \\
 $4P_0\rightarrow 3S$ & {1423} & {7.4(4.1)} \\
		{$2(S/D)_1\rightarrow 1P$} & {977} & {17(8)} \\
 {$3(S/D)_1\rightarrow 1P$} & {1176} & {29(14)} \\
 {$3(S/D)_1\rightarrow 2P$} & {818} & {5(3)} \\
 {$4(S/D)_1\rightarrow 2P$} & {891} & {33(25)} \\
 {$5(S/D)_1\rightarrow 1P$} & {1376} & {18(7)} \\
 {$5(S/D)_1\rightarrow 2P$} & {1018} & {14(8)} \\
		\hline
	\end{tabular}
	\caption{Decay widths for charmonium (above) and bottomonium (below) hybrids to lower-lying charmonia and bottomonia, respectively.
 $n$ and $n'$ are principal quantum numbers, $L(L+1)$ and $L'(L'+1)$ are eigenvalues of $\bm{L}^2$, where $\bm{L}$ is
 the orbital angular momentum of the heavy quarks, and $T(T+1)$ are eigenvalues of $(\bm{L+K})^2$, where $\bm{K}$ is the angular momentum of the gluons.
 $\Delta E$ is the energy difference between the hybrid and the quarkonium state. Mixing with quarkonia has been neglected.
 The hybrid states $P_0$ belong to the $H_3$ multiplet of Table~\ref{tab:spin_multiplet} and the states $(S/D)_1$ to the $H_1$ multiplet.
	 Table taken from Ref.~\cite{Oncala:2017hop}.}
	\label{tab:hybdecay}
\end{table}

A full theoretical description of hybrids, which may eventually lead to a certain identification of some of the $X$ and $Y$ quarkonium states, requires also the study of their decay properties.
An early study is Ref.~\cite{Kou:2005gt} that explains the apparent suppression of the decay $Y(4260) \to D^{(*)}\bar{D}^{(*)}$ with respect to the decay
$Y(4260) \to \pi^+\pi^- J/\psi$ by interpreting the $Y(4260)$ (now $Y(4230)$) as a charmonium hybrid state with a magnetic constituent gluon,
see Sec.~\ref{Sect:3.1.2}.
A relation of this model with the Born--Oppenheimer picture was first suggested in Refs.~\cite{Vairo:2006pc,Vairo:2009tn}.
Later it was worked out in Ref.~\cite{Berwein:2015vca}, see Fig.~\ref{fig:exp101016} and the possible $H_1$ assignment of the $Y(4230)$.
More recently, heavy quarkonium hybrid inclusive transitions into quarkonia have been studied in the BOEFT framework in Ref.~\cite{Oncala:2017hop}.
The outcome of that study is summarized in Table~\ref{tab:hybdecay}. The transition widths range from a few MeV to about 50~MeV.
Moreover, in Ref.~\cite{Oncala:2017hop} it was emphasized that the same NRQCD operator responsible
for the appearance of the $1/m_h$ spin-dependent potential \eqref{sdm2}
is also responsible for the appearance of a mixing potential between hybrids and ordinary quarkonia at order $1/m_h$.
The mixing potential mixes spin $0$ ($1$) hybrids with spin $1$ ($0$) quarkonia and may explain, in dependence of the strength of the mixing,
which is of order $\Lambda^2_{\text{QCD}}/m_h$ and non-perturbative, why some hybrid candidates appear to decay both into $\pi^+\pi^-J/\psi$ and $\pi^+\pi^-h_c$.
This could be the case of the $Y(4230)$, which, according to Ref.~\cite{Oncala:2017hop}, has a potentially large mixing with the ordinary charmonium $2D$, $1^{--}$, state.
Finally, selection rules for hadronic transitions of $X$, $Y$ and $Z$ mesons have been derived in Ref.~\cite{Braaten:2014ita}

\vspace{1.2cm}\noindent
$\bullet$ {\it Tetraquarks}
\vspace{0.3cm}

The BOEFT Lagrangian density in the isospin $I=1$ sector, relevant for tetraquarks,
may be constructed following the same line of reasoning that leads to the BOEFT for hybrids
and to the heavy-hadron chiral effective Lagrangian~\cite{Wise:1992hn}.
Its form is very similar to the BOEFT Lagrangian for hybrids given in Eq.~\eqref{lag:hyb}, but the isovector fields are $2\times 2$ matrices:
\begin{align}
Z_\kappa = Z_\kappa^i\sigma^i=\left(
\begin{array}{cc}
Z_\kappa^0 & \sqrt{2} Z_\kappa^+\\
\sqrt{2} Z_\kappa^- & -Z_\kappa^0 \\
\end{array}
\right)\,,
\end{align}
where $\sigma^i$ are the isospin Pauli matrices.
The BOEFT Lagrangian density in the $I=1$ sector has the form~\cite{TarrusCastella:2019rit}:
\begin{align}
{\mathcal L}_{\text{BOEFT for } I=1} &= \int d^3r \; \text{Tr} \left\{ Z^{\dag}_{0^{+-}}\left(iD_0-V^{\rm tetra}_{\Sigma^+_g}(r)+\frac{\bm{\nabla}^2_r}{m_h}\right)Z_{0^{+-}}\right\} \nonumber\\
&+ \int d^3r \; \sum_{\lambda\lambda'} \text{Tr}\left\{ Z^{\dag}_{1^{+-}\lambda} \left(iD_0 - V^{\rm tetra}_{1^{+-}\lambda\lambda'}(r)
+\hat{r}^{i\dagger}_{\lambda}\frac{\bm{\nabla}^2_r}{m_h}\hat{r}^i_{\lambda'}\right)Z_{1^{+-}\lambda'}\right\} \nonumber\\
&+ \int d^3r \; \sum_{\lambda\lambda'} \text{Tr}\left\{ Z^{\dag}_{1^{--}\lambda} \left(iD_0 - V^{\rm tetra}_{1^{--}\lambda\lambda'}(r)
+\hat{r}^{i\dagger}_{\lambda}\frac{\bm{\nabla}^2_r}{m_h}\hat{r}^i_{\lambda'}\right)Z_{1^{--}\lambda'}\right\} +...\,,
\label{lag:tetra1pm}
\end{align}
where the dots stand for terms involing higher orbital momentum and possibly mixing between states.
The fields $Z_{1^{+-}\lambda}$, $Z_{1^{--}\lambda}$, and $Z_{0^{+-}}$ depend on time, the c.m. coordinate $\bm{R}$ and the relative coordinate $\bm{r}$.
The trace is also over the isospin indices.
The covariant derivative for the $I=1$ fields reads $D_{\mu} Z_\kappa=\partial_{\mu}+[\Gamma_{\mu},\,Z_\kappa]$
with $\Gamma_{\mu}=\left(u^{\dagger}\partial_{\mu}u + u\partial_{\mu}u^{\dagger}\right)/2$
and $u=\text{exp}[i\pi^i\sigma^i/(2f_\pi)]$. The pion fields, $\pi^i$,
depend on $t$ and $\bm{R}$.
The direct use of the effective Lagrangian~\eqref{lag:tetra1pm} is limited by the fact that the potentials have not, even in their static limit, been measured on the lattice.
For recent studies in this direction we refer to Sec.~\ref{Sect:4.3:static}.
Hence, the situation is different from the hybrid case, where static hybrid energies are known since long time.
In particular, an analysis of the type performed for heavy quarkonium hybrids in the previous paragraphs is at the moment not possible for quarkonium tetraquarks.
In Refs.~\cite{Braaten:2013boa,Braaten:2014qka}, to circumvent this difficulty, it was assumed that the tetraquark static energies have the same shape as the hybrid ones.
In this way, it became possible for the authors to provide preliminary mass estimates and to make predictions on the ordering of the tetraquark spectrum.
Clearly, the lattice computation of the tetraquark static energies would provide a major input for studies based on the Born--Oppenheimer picture.

\subsubsection{Effective field theories at the hadron level}
\label{Sect:4.2.4}

\begin{figure}[t!]
\begin{center}
 \includegraphics[width=0.6\linewidth]{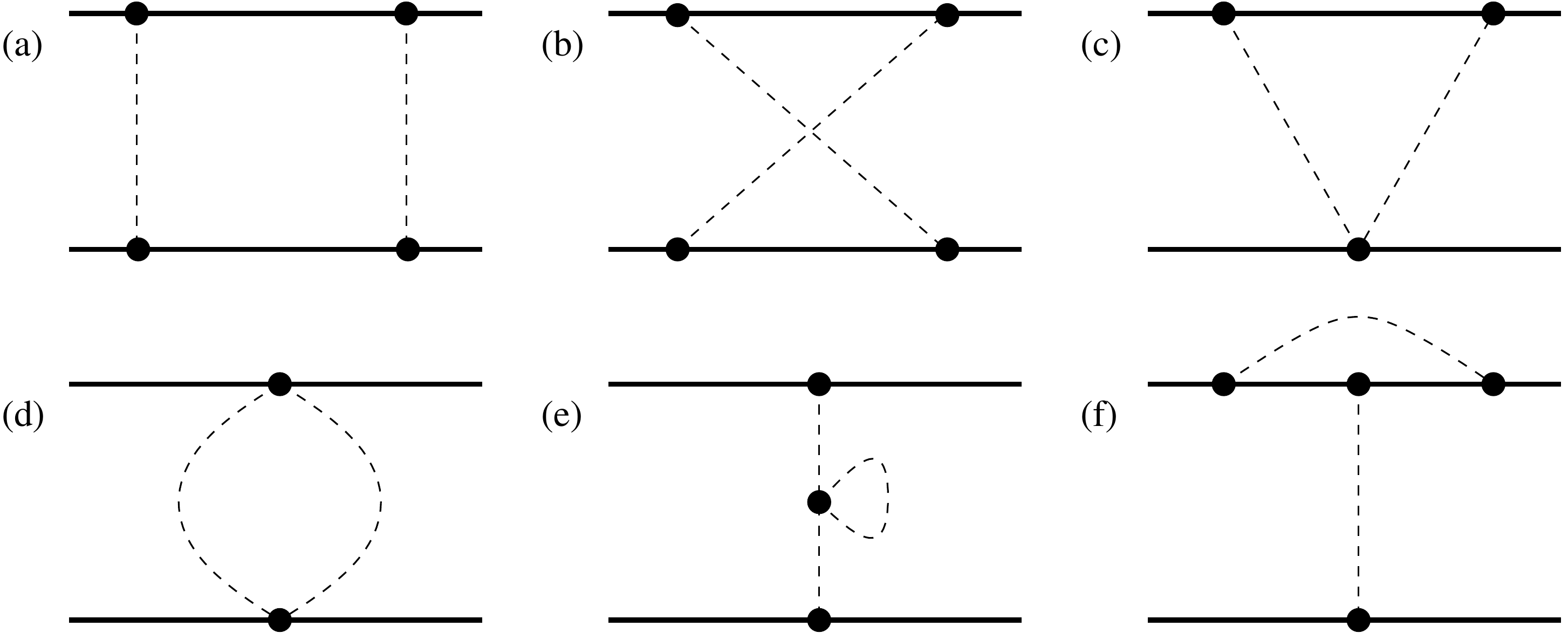}
\caption{Topologies contributing at the leading one-loop level to the scattering of two heavy particles:
the ladder diagram (a), the crossed box (b), the triangle (c), the football (d), the pion tadpole (e) and
the vertex correction (f).}
\label{fig:laddervsbox}
\end{center}
\end{figure}

Effective field theories for few-hadron systems can be constructed along the lines of the analogous effective field
theories for few-nucleon systems --- see Ref.~~\cite{Epelbaum:2008ga} for a comprehensive review of the subject.
Here we will focus on the treatment of few-hadron bound systems.
On the one hand, the formalism must be formulated non-perturbatively.
But, on the other hand, in an effective field theory one needs to identify a proper expansion parameter
in such a way that the results can be improved systematically by going to higher orders in this expansion.
It was Weinberg in 1990 who argued that those seemingly contradictory requirements can be brought together,
if one performs an expansion not of the full amplitude but only of the scattering potential which then should
be resummed by some wave equations~\cite{Weinberg:1990rz}. The argument is based on the following observation: Assume the scattering of
two heavy particles interacting via the exchange of a light meson. As
illustrated in Fig.~\ref{fig:laddervsbox}, at the leading one-loop level the scattering potential
contains six contributions: the ladder, the crossed box, the triangle, the football, the pion tadpole and the vertex correction.
 Clearly, the diagrams (a), (b) and (f) scale equally with respect
to the coupling constants, however, a detailed analysis reveals that the ladder diagram scales with an additional factor $\pi(2\mu/p)$ compared
to the other two, where $\mu$ denotes the reduced mass of the two heavy particles and $p$ is the typical momentum of the two-hadron state.
This kind of kinematic enhancement is also absent in diagrams (c), (d) and (e).
The enhancement factor comes from the two-hadron cut, which is present only in diagrams that have a time slice
that cuts only two heavy hadron lines (in the language of few-body physics these kinds of diagrams are called two--body reducible).
Thus, for momenta small compared to the masses of the two hadrons,
the ladder diagram is enhanced compared to all the other one-loop diagrams. This remains true at higher orders
in perturbation theory and thus justifies a resummation of at least part of the potential constructed to a given order
in perturbation theory, where ``potential'' means all contributions that are two-hadron irreducible.
In contrast to Weinberg's scheme, it was argued in Refs.~\cite{Kaplan:1998tg,Kaplan:1998we} that instead of the
potential the full amplitude should be analyzed under the assumption that
there is a pole present very near threshold.
Within this scheme only a momentum-independent counterterm needs to be resummed while pion exchange is perturbative.
However, in Ref.~\cite{Fleming:1999ee} it was shown that, as soon as the tensor force
of the one-pion exchange contributes, the expansion that results from the scheme of Refs.~\cite{Kaplan:1998tg,Kaplan:1998we} does not converge at least for the two-nucleon system in the deuteron channel.

\begin{figure}[t!]
\begin{center}
 \includegraphics[width=0.8\linewidth]{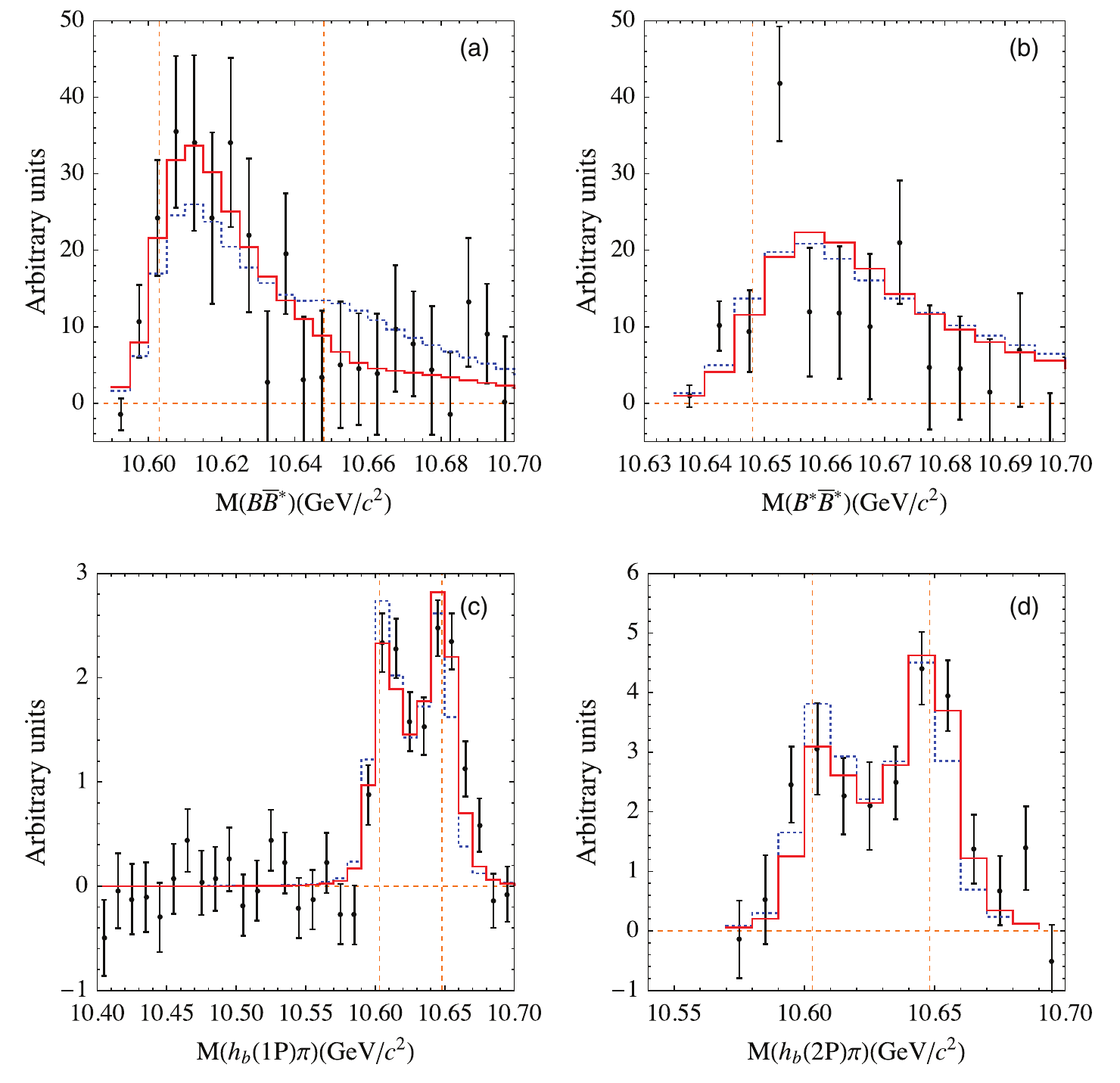}
\caption{Results for the line shapes of the $Z_b$ states reported in Ref.~\cite{Wang:2018jlv}: The red solid
line (blue dotted line) is based on a calculation that includes one-pion exchange as well as contact interactions (only contact
interactions).}
\label{fig:Zblineshapes}
\end{center}
\end{figure}

As in case of studies of the few-nucleon system, also for the interactions of heavy mesons both schools co-exist:
While some authors propose that only contact terms need to be resummed and one-pion exchange should be treated
perturbatively~\cite{AlFiky:2005jd,Fleming:2007rp,Valderrama:2012jv,Nieves:2012tt,HidalgoDuque:2012pq,Guo:2013sya},
others call for a resummation of both contact terms and pion exchange~\cite{Baru:2011rs,Baru:2013rta,Baru:2016iwj,Baru:2017gwo,Wang:2018jlv}.
For example, in Ref.~\cite{Fleming:2007rp} it is argued that the tensor force of the pion should be reduced significantly here compared
to the two-nucleon system as a result of a smaller effective pion mass and a smaller coupling that invalidate
the argument of Ref.~\cite{Fleming:1999ee}. However, in Ref.~\cite{Wang:2018jlv} it is shown that the fit to the
line shapes of $Z_b(10610)$ and $Z_b(10650)$, especially in the lower elastic channel, improves once
the tensor force of the pion is included, pointing at the need for a non-perturbative inclusion of the pion exchange --- see Fig.~\ref{fig:Zblineshapes} ---
and thus calling for an application of Weinberg's counting also in heavy meson systems. This work will be discussed in a little more detail below.
From now on we only investigate two--hadron molecular states. The possible interplay of scales in meson containing three hadrons
is discussed in Ref.~\cite{Hagen:2010wh} on the example for the $Y(4660)$.

The leading-order piece of the HQET Lagrangian shown in Eq.~(\ref{eq:sect4:HQET}) is independent of the heavy-quark flavor.
This is at the origin of the heavy-quark flavor symmetry, which is routinely employed to relate
features from the $c$-quark sector to the $b$-quark sector, as long as there is only a single heavy quark involved.
However, when it comes to doubly-heavy systems the flavor symmetry is no more a symmetry of the leading-order Lagrangian,
since both the potential and the kinetic energy of the heavy particle are of the same order.
See the discussion at the end of Sec.~\ref{Sect:4.2.2} for NRQCD, but the same reasoning applies to pNRQCD.
It should be remarked, however, that, differently from (p)NRQCD, where the potential seen by the two heavy quarks is known at short distances
and may be computed in perturbation theory, in the hadronic EFT this is not the case:
The short distance interaction always contains counter terms at leading order.
Hence, if one wants to construct an EFT that connects the $c$-quark with the $b$-quark sector,
the scaling of these counter terms with the hadron mass, $M$, must be either derived or imposed.
Clearly, different assumptions lead to different results:
In Refs.~\cite{Nieves:2011vw,Guo:2013sya} the binding energy of the heavier systems is found larger than that of the lighter systems;
the opposite is the case in Ref.~\cite{AlFiky:2005jd}, reflecting the fact that the former work assumes that the scattering potential of the two heavy-meson
system is independent of the heavy meson-mass $M$, whereas the latter derives a $1/M$ scaling based on a renormalisation group argument.
The role of the heavy flavor symmetry in the hadronic EFT was first studied systematically in Ref.~\cite{Baru:2018qkb}, with the conclusion that no common effective field
theory can be defined for bottomed and charmed molecular systems simultaneously, since the heavy meson mass cannot be removed consistently from the formalism.
As we have remarked, this is the same situation that we face in (p)NRQCD for bottomonia and charmonia.
On the other hand, as in case of (p)NRQCD, the heavy-particle spin symmetry holds at leading order also in any EFT for two--hadron systems.

The power counting to treat the heavy two-hadron systems is based on a velocity counting in a non relativistic
scheme. The typical velocity for some intermediate state is estimated via
\begin{equation}
v \sim \sqrt{|M-M_1-M_2|/\mu},
\label{eq:vestimate}
\end{equation}
where $M$ denotes the total energy of the intermediate state that couples to the
two mesons with masses $M_1$ and $M_2$; their reduced mass is denoted by $\mu$.
It may either be formulated based on a four-dimensional formalism or a three-dimensional formalism
related to time-ordered perturbation theory (for a comparison of the two schemes we refer to Appendix E
of Ref.~\cite{Hanhart:2003pg}). For example, in the former scheme the integral measure
counts as $v^5$, a heavy meson propagator as $1/v^2$ and a pion exchange (as long as one studies
the scattering of two ground-state open-flavor mesons off each other\footnote{Here the
$1/v^2$ enhancement from the propagator gets balanced by the factor $v^2$ in the numerator since pions
couple to heavy mesons in the $P$-wave.}) or a momentum-independent counter term as $v^L$, where $L$ denotes the partial wave.
In the latter case, the integral measure counts as $v^3$ and a time slice that cuts only heavy mesons counts
as $1/v^2$. Pion exchange and contact terms are counted as in the four-dimensional formalism the same way. Accordingly, a meson bubble
scales as $v^{(2L+1)}$ in both schemes. This shows that loops are most important for $S$-waves ($L=0$), however,
there is no parametric enhancement of the loops calling for their resummation (This statement is not in conflict with the
observation made at the beginning of this section, namely that two--hadron reducible diagrams are parametrically enhanced
compared to the irreducible ones). This is completely analogous to
the situation in the two--nucleon system and we refer to Ref.~\cite{Epelbaum:2008ga} for a detailed discussion.
What is generally done to arrive at a formalism where the heavy meson loops are resummed to all orders
is to argue that the coupling constants that multiply each loop are sufficiently large to call for this or to
 simply impose it, since a resummation is necessary to generate bound states (see, e.g., Refs.~\cite{Kaplan:1998tg,Kaplan:1998we,Fleming:2007rp}),

\begin{figure}[t!]
\begin{center}
 \includegraphics[width=1.\linewidth]{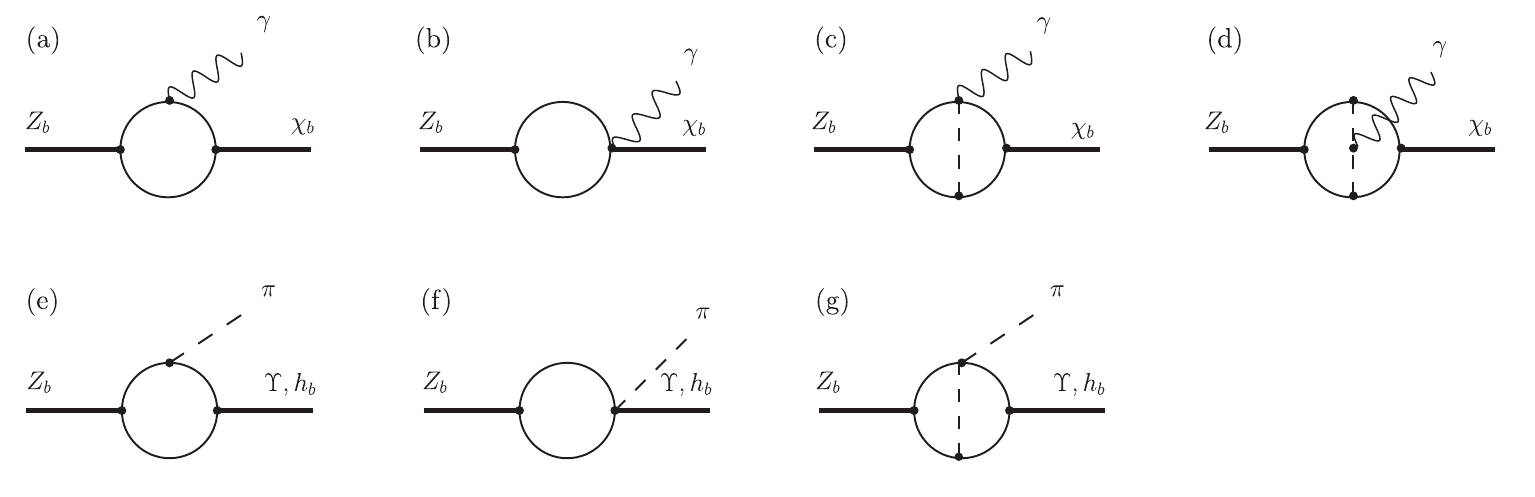}
\caption{Diagrams contributing to various $Z_b$ and $Z_b'$ decays at leading (a,e) and subleading order (b,c,d,f,g).
Thick (thin) solid lines denote the propagation of doubly (singly) heavy states, dashed lines show pions and wavy
lines photons.}
\label{fig:Zbdecays}
\end{center}
\end{figure}

The same construction can also be applied to decays. Since for a molecular state its coupling to
the molecule-forming channel is large, two-hadron loops always appear at leading order. However,
whether a given decay is sensitive to the long-range part of the wave function, which contains the
information on the molecular component, as explained in Sec.~\ref{Sect:4.1.5}, depends on the quantum numbers involved.
As an example, we present here the reasoning of Ref.~\cite{Cleven:2013sq} where the decays of
the $Z_b$ states were investigated from an effective field theory point of view based on the
non relativistic effective field theory introduced in Refs.~\cite{Guo:2009wr,Guo:2010ak}.
The relevant diagrams that appear at leading order as well as some subleading diagrams
are displayed in Fig.~\ref{fig:Zbdecays}.

We start with the radiative decays. Since the
$Z_b$ states carry the quantum numbers $1^{+-}$ and the photon is a vector with $J^{PC}=1^{--}$,
the quarkonium in the final state needs to have $C=+$ and $P=(-1)^{L+1}$, where $L$ denotes the angular momentum
between the outgoing quarkonium and the photon. To reach an allowed quark model bottomonium
in the final state, the decay must happen in a $P$-wave ($L=1$) and the final state should contain a $\chi_{bJ}$ state.
The latter couples to a pair of $B$-mesons in an $S$-wave and, accordingly, all vertices inside the loop are $S$-wave ones.
Then all decays must scale with the photon momentum $q$.\footnote{In Ref.~\cite{Cleven:2013sq}
the photon energy was used, which is equivalent since the photon is on shell.}
Adapting the rules outlined in the previous paragraph (we now use the four-dimensional counting only,
although the two schemes naturally continue to give the same scaling) we find for diagram (a):
\begin{equation}
v^5 \, (1/v^2)^3 q = q/v.
\label{eq:triangleest}
\end{equation}
It is shown in Ref.~\cite{Guo:2012tg} (see also Ref.~\cite{Guo:2017jvc})
that the proper scale used for $v$ in
the triangle topology
away from the Landau singularity (discussed in more detail at the end of this section)
is the average of the two velocities appearing. Since the velocity of the $B$-meson pair
right in front of the $\chi_{bJ}(mP)$-state [from Eq.~(\ref{eq:vestimate}) we find $v_\chi=0.12, 0.26, 0.37$ for $m=3,2,1$,
respectively] is much larger than the one right after the $Z_b$ decay ($v_z\sim 0.02$)
due to the proximity of the $Z_b$ states to the threshold, one may safely choose $v\sim v_\chi$ in Eq.~(\ref{eq:triangleest}).
Note that as soon as one takes $v_\chi$ as the hard scale of the problem, a different power counting
arises and the same factorisation theorem used for positronium decays may also be applied here ---
the two choices ($v_\chi$ as hard or soft scale) are contrasted on the example of pionic decays
of $X(3872)$ in Ref.~\cite{Mehen:2015efa}. See also the discussion in Sec.~\ref{Sect:4.1.5}.

For diagram (b) one finds
\begin{equation}
v_z^5 \, (1/v_z^2)^2 \, q = q v_z.
\label{eq:subleadingest1}
\end{equation}
Thus diagram (a) is enhanced compared to diagram (b) by about one order of magnitude. Gauge invariance demands diagrams
(c) and (d) to appear at the same order. For this diagram we find (the rules applied here are a simplified
version of what is presented in Ref.~\cite{Cleven:2013sq} that, however, does not change the conclusions)
\begin{equation}
v_z^5 \, (1/v_z^2)^2 \, qv_\chi/v_\chi^2 \, v_\chi^5\, (1/v_\chi^2)^2 = q v_z,
\label{eq:subleadingest2}
\end{equation}
where we used that the pion momentum is controlled by the larger one of the two-loop momenta.
Thus diagrams (c) and (d) are equally suppressed as diagram (b). Accordingly, diagram (a) allows
one to estimate the decay rate with a 10\% accuracy at the amplitude level,
which translates to a 20\% accuracy for the branching fraction.

We now turn to the pionic decays of the $Z_b$ states shown in the second line of Fig.~\ref{fig:Zbdecays}.
Parity conservation demands that the decay to the $h_b$ states, which couple to the $B$-pairs in
an $S$-wave, happens in a $P$-wave, while the decay to the $\Upsilon$ states, which couple in a $P$-wave
to the $B$-meson pairs, happens in an $S$-wave. Accordingly, we estimate that
\begin{equation}
v^5 \, (1/v^2)^3 q = q/v.
\label{eq:triangleest2}
\end{equation}
for the $h_b$ decays, where $v\sim v_h\sim v_\chi$ in line with the
discussion above, and
\begin{equation}
v^5 \, (1/v^2)^3 v = 1.
\label{eq:triangleest3}
\end{equation}
for the $\Upsilon$ decays, where $q$ denotes the outgoing pion momentum.
Moreover, the latter loop is divergent while the former is convergent.
Thus, there
must be a counter term at leading order to render the calculation of the $\Upsilon$-decays well defined,
while there is no leading order counterterm for the $h_b$ decays.

To estimate the diagram in Fig.~\ref{fig:Zbdecays}(g) we need to use that the pion rescattering
vertex scales with the pion energy ($E_\pi$) --- an insight that may be adapted from effective field theory
studies of the reaction $NN\to NN\pi$~\cite{Lensky:2005jc} (for a recent review see Ref.~\cite{Baru:2013zpa}).
Thus we find for this diagram with an $h_b$ in the final state
\begin{equation}
v_z^5 \, (1/v_z^2)^2 \, (E_\pi/\Lambda_\chi) q/v_h^2 \, v_h^5 \, (1/v_h^2)^2 = q^2/(\Lambda_\chi) v_z/v_h,
\label{eq:subleadingest3}
\end{equation}
where we introduced the typical hadronic scale $\Lambda_\chi\sim 1$ GeV to render the units equal
and identified the pion energy with the pion momentum, since both are of the same order of magnitude.
For the decay to the ground state $h_b$ the pion momentum can be of the order of $\Lambda_\chi$,
however, the remaining factors are sufficient to provide a suppression of diagram (g) by more than one order of magnitude.
For the $\Upsilon$ in the final state the analogous estimate gives
\begin{equation}
v_z^5 \, (1/v_z^2)^2 \, q v_h/v_h^2 \, v_h^5 \, (1/v_h^2)^2 v_h = q v_z v_h.
\label{eq:subleadingest33}
\end{equation}
Thus also for the $\Upsilon$ final states the higher loop diagrams are suppressed.
In Ref.~\cite{Cleven:2013sq} even higher loop diagrams are shown to be suppressed. Thus we may conclude that the effective field
theory described here allows one to calculate the decays of the $Z_b$ states to $\gamma\chi_{bJ}$
and $\pi h_{bJ}$ in a controlled way (with a 10\% uncertainty in the amplitude); these
transitions are dominated by the light degrees of freedom and are, therefore, sensitive to the molecular
component of the $Z_b$ states.
The same reasoning allows one to show that under certain conditions heavy meson loops can provide
very prominent contributions in the transitions of regular heavy quarkonia~\cite{Guo:2010ak,Guo:2016yxl,Guo:2011dv}.
In contrast to this, the
decays to the $\pi\Upsilon$ final states acquire short-range contributions already at leading order.
In this case, it is not possible to make quantitative predictions for the transitions within the molecular picture.
An analogous reasoning allows one to conclude that the transitions
$X(3872)\to \gamma J/\psi$ and $X(3872)\to \gamma \psi(2S)$ are sensitive to the short-range
part of the $X$ wave function and may not be used to draw conclusions on the molecular
nature of the $X(3872)$~\cite{Guo:2014taa} --- contrary to the claims of Refs.~\cite{Swanson:2004pp}
where a particular model was employed to estimate the rates.

\begin{figure}[t!]
 \begin{center}
 \includegraphics[width=0.7\linewidth]{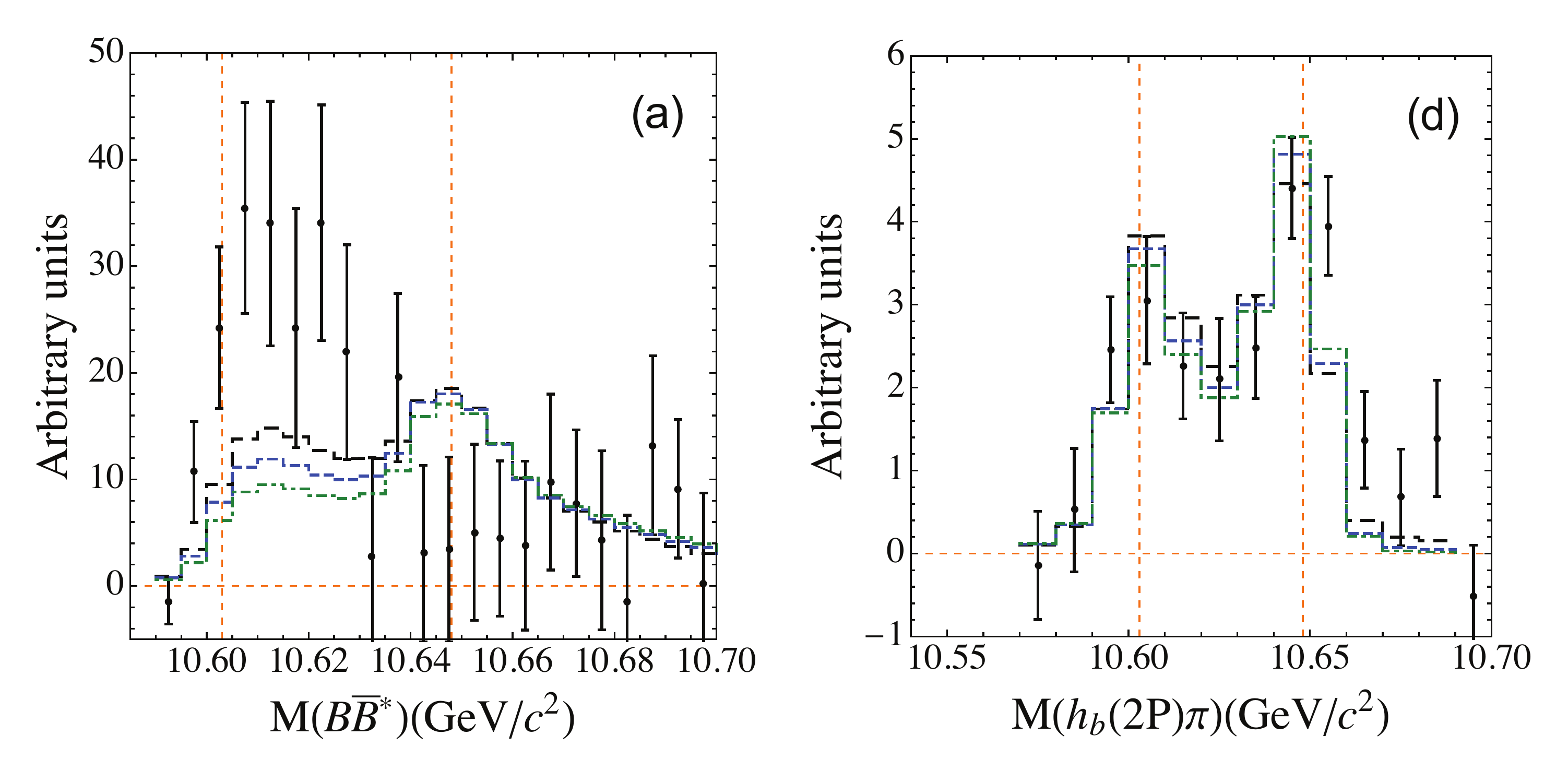}
\hspace{0.6cm}
\includegraphics[width=0.7\linewidth]{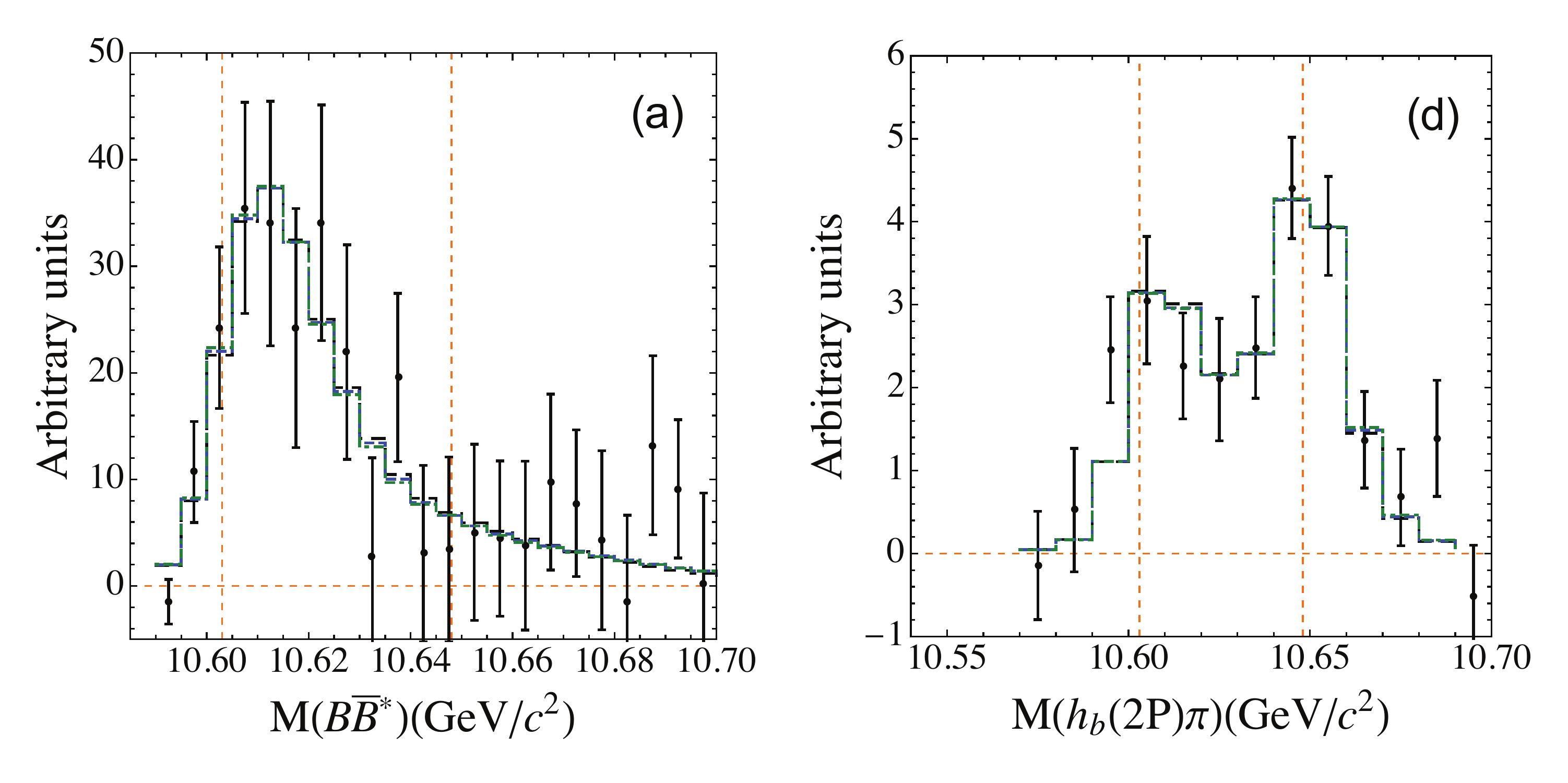}
\caption{ \label{fig:CutF}
The fitted line shapes for a calculation employing only the formal leading order potential (upper line) and
when the $S$-$D$ counter term is added (lower line) in the elastic $B\bar{B}^*$ and inelastic $\pi h_b(2P)$ channels with sharp cut-offs 800~MeV (black long-dashed),
1000~MeV (blue dashed), and 1200~MeV (green dot-dashed), respectively.
The experimental data are from Refs.~\cite{Belle:2011aa,Garmash:2015rfd}.}
 \end{center}
\end{figure}

In Ref.~\cite{Wang:2018jlv} the line shapes of the $Z_b$ states are studied within the effective
field theory formalism sketched above. At the first stage, only the naive leading order diagrams were considered:
Momentum independent counter terms as well as the one pion exchange. Note that, since the
pion exchange with its sizeable tensor force can connect $S$- to $D$-waves, the basis needs
to be enlarged compared to the inclusion of only $S$-waves commonly kept
when studying heavy meson molecules. The typical momentum in the $B^*\bar B$ channel for the
energies near the $B^*\bar B^*$ threshold is $p_{\rm typ}\sim\sqrt{M_B\delta}\sim 500$ MeV, where $\delta=M_{B^*}-M_B\approx 45$~MeV.
This implies, since $p_{\rm typ}/M_\pi> 1$, that there is no suppression of higher partial waves.
Moreover, the coupling to $D$-waves induces a sizeable regularisation scheme dependence --- cf. first
row of Fig.~\ref{fig:CutF}, where the scheme dependence is illustrated by the variation of
a sharp cut-off employed to render the loop diagrams finite; for each value of the cut-off a refit
to the data was performed, which is nothing but a numerical implementation of the renormalisation
program. In a properly renormalised EFT the contact terms should be capable of absorbing the
cut-off dependence. For the system at hand this appears possible only once a counter term that connects $S$- to $D$-waves is promoted
to leading order as can be seen from the second row in Fig.~\ref{fig:CutF}.\footnote{For the nucleon--nucleon system a formal argument
for the need to promote certain counter terms to lower order in the presence of the tensor force is provided in Ref.~\cite{Birse:2007sx}.}
It should be stressed, however, that not all pion effects were absorbed into the promoted
counter term: Not only does the inclusion of the one pion exchange improve the fit quality (compare red solid
and blue dotted lines in Fig.~\ref{fig:Zblineshapes}), the convergence of the approach is improved
as well: If momentum-dependent counter terms that connect $S$-to-$S$-waves are included in addition,
fitting without one-pion exchange calls for those formally subleading terms to be larger than the leading ones.
On the contrary, as soon as one pion exchange is included in the calculation, these additional terms provide a perturbative correction only.

\begin{table}[t!]
\begin{tabular}{lllcccccc}
$J^{PC}$& State & Threshold &Im~$p_{\rm in}$ & Im~$p_{B\bar B}$ & Im~$p_{B\bar{B^*}}$ & Im~$p_{B^*\bar{B}^*}$ &$E_{\rm pole}$ w.r.t. threshold (MeV) & Residue at $E_{\rm pole}$ \\
\hline
$1^{+-}$& $Z_b$ & $B\bar{B}^*$ &$+$ & & $-$ & $+$ &$(-0.9\pm 0.4)+i(1.0\pm 0.3)$ & $ (-1.4\pm 0.2)+i(0.5\pm 0.1)$\\
$1^{+-}$& $Z_b'$ & $B^*\bar{B}^*$ &$+$ & & $+$ & $-$ &$(-0.8\pm 0.5)+i(1.3\pm 0.4)$ & $(-1.4\pm 0.3)+i(0.7\pm0.1)$\\
$0^{++}$& $W_{b0}$ & $B\bar{B}$ &$+$ & $-$ & & $+$ &$(-1.0\pm 0.6)+i(1.0\pm 0.3)$ & $(-1.4\pm 0.3)+i(0.5\pm 0.1)$\\
$0^{++}$& $W_{b0}'$& $B^*\bar{B}^*$ &$+$ & $+$ & & $-$ &$(-1.2\pm 0.6)+i(0.9\pm 0.3)$ & $(-1.4\pm 0.3)+i(0.4\pm 0.1)$\\
$1^{++}$& $W_{b1}$ & $B\bar{B}^*$ &$+$ & & $-$ & &$(-0.3\pm 0.6)+i(1.6\pm 0.8)$ & $(-1.3\pm 0.4)+i(0.9\pm 0.1)$\\
$2^{++}$& $W_{b2}$ & $B^*\bar{B}^*$ &$+$ & & &$-$ &$(0.4\pm 0.6)+i(1.9\pm 0.9 )$ & $(-1.2\pm 0.4)+i(1.3\pm 0.2 )$
\end{tabular}
\caption{The pole positions and the residues $g^2$, normalized according to Eq.~(\ref{eq:gnorm}), in various $S$-wave $B^{(*)}\bar B^{(*)}$ channels for a
potential with contact terms only.
The energy $E_{\rm pole}$ is given relative to the nearest open-bottom threshold quoted in the third column.
The Riemann Sheet (RS) is defined by the signs of the imaginary parts of the corresponding momenta (quoted in the columns 4-7);
a missing sign indicates that this channel is uncoupled. Uncertainties correspond to a $1\sigma$
deviation in the parameters allowed by the fit to the data in the channels with $J^{PC}=1^{+-}$ where the $Z_b^{(\prime)}$ states reside \cite{Wang:2018jlv}.
 The poles are
calculated for the cutoff $\Lambda=1$ GeV.
}\label{tab:Contact}
\end{table}

\begin{table}[t]
\begin{tabular}{lllcccccc}
$J^{PC}$& State & Threshold &Im~$p_{\rm in}$ & Im~$p_{B\bar B}$ & Im~$p_{B\bar{B^*}}$ & Im~$p_{B^*\bar{B}^*}$ &$E_{\rm pole}$ w.r.t. threshold [MeV] & Residue at $E_{\rm pole}$ \\
\hline
$1^{+-}$& $Z_b$ & $B\bar{B}^*$ &$-$ & & $-$ & $+$ &$(-1.3\pm 0.2)-i(0.6\pm 0.1)$ & $(-0.6\pm 0.1)-i(0.1\pm 0.1) $ \\
$1^{+-}$& $Z_b'$ & $B^*\bar{B}^*$ &$-$ & & $-$ & $-$ &$(2.1\pm 2.2)-i(12.9\pm 2.4)$ & $(0.8\pm 0.1)-i(0.4 \pm 0.2)$ \\
$0^{++}$& $W_{b0}$ & $B\bar{B}$ &$+$ & $-$ & & $+$ & $(-8.5 \pm 2.8) + i (1.5 \pm 0.2)$ & $(-2.0 \pm 0.7) - i (0.1 \pm 0.3)$ \\
$0^{++}$& $W_{b0}'$& $B^*\bar{B}^*$ &$-$ & $-$ & & $-$ & $(-1.2\pm 0.1) - i (0.7 \pm 0.3)$ & $(-0.4 \pm 0.1) - i (0.2 \pm 0.1)$ \\
$1^{++}$& $W_{b1}$ & $B\bar{B}^*$ &$-$ & & $-$ & $+$ & $(25.0 \pm 2.6) - i (20.5 \pm 3.3)$ &$(0.9 \pm 0.1) - i (0.4 \pm 0.2)$\\
$2^{++}$& $W_{b2}$ & $B^*\bar{B}^*$ &$-$ & $-$ & $-$ &$-$ & $(4.0 \pm 2.1) - i (10.4 \pm 1.5)$ & $(0.4 \pm 0.1) - i (0.2 \pm 0.1)$
\end{tabular}
\caption{The same as in Table~\ref{tab:Contact} but with pion exchange included.}\label{tab:pf1}
\end{table}

The pole locations and residues that emerge from the described study of the $Z_b$ states,
as well as their spin partners $W_{bJ}$, are presented in Tab.~\ref{tab:Contact} for a potential with
contact terms only and in Tab.~\ref{tab:pf1} for the fully renormalised potential including
one pion exchange~\cite{Baru:2019xnh} --- note that for the $W_{bJ}$ states partial
waves up to $G$-waves need to be included in order to render the potential consistent with
the heavy quark spin symmetry~\cite{Baru:2016iwj}. In all cases, to reduce the complexity
of the problem when extracting the poles and residues, the inelastic channels are lumped into a single effective channel.
The sign convention of the energies is such that negative (positive) energies refer to below (above)
threshold poles. Moreover, a negative (positive) imaginary part of a pole-momentum for a given channel
refers to the unphysical (physical) sheet. Thus, in all cases the poles extracted from the analysis appear
on the unphysical sheet very close to the threshold and the
data are consistent with only a single near-threshold pole in line with a molecular interpretation for the states.
Since all poles are on the unphysical sheets, the Weinberg criterion to measure the molecular admixture
cannot be applied to the residues provided in the tables. However, we may still use Eq.~(\ref{eq:geffdef})
to estimate the order of magnitude of the residues one would expect for molecular states in the $B\bar B$ system bound by, say, 1 MeV only. The residues in the tables were
extracted from the diagonal elements of the
theoretical $B^{(*)}\bar B^{(*)}$ scattering
amplitudes using the definition
\begin{equation}
T_{\alpha\alpha}(M)=\frac{g_\alpha^2}{2M(M-M_R)} \ ,
\label{eq:gnorm}
\end{equation}
in combination with a non relativistic normalisation of the fields. Thus $g_{\rm eff}^2$ defined in Eq.~(\ref{eq:geffdef}) needs to
be divided by a factor $(2M_1)(2M_2)$ to match this normalisation, and we get
$$
\left.\frac{g_{\rm eff}^2}{4M_1M_2}\right|_{E_b=1 \ \mbox{\small MeV}}\sim 0.7 \ .
$$
In light of this number we should regard all residues listed in Tab.~\ref{tab:Contact} and Tab.~\ref{tab:pf1}
as large and in this sense providing further evidence for a molecular interpretation of the $Z_b$ states.

One feature, however, is special: When the pion exchange is included some states
appear as above-threshold poles with a small imaginary part. This was commonly believed to be a signature for
compact states, however, the proximity of the left--hand singularity from the pion exchange,
which introduces an additional small scale, seems to be responsible for this finding.
Here more research appears to be necessary --- especially since all poles but that of the $W_{b2}$ move
below threshold in the formalism without one pion exchange and the latter pole moves below
threshold as well as soon as the inelastic channels are switched off.

The effective field theory outlined in the chapter is based on the most general Lagrangian compatible with
the QCD symmetries. As such, a priori it does not involve any assumption about the nature of the states
that emerge once the dynamical equations are solved. However, an assumption enters as soon as the
power counting is employed. As argued in the beginning of this section, the scheme is built on the assumption
that the two-hadron relative momentum is the smallest scale of the problem. Then ladder diagrams
are enhanced and a potential picture is justified. According to Weinberg's criterion we call states
that emerge from the presence of the two-hadron cut hadronic molecules (see the discussion in Sec.~\ref{Sect:4.1.5}).
If there were an $s$-channel pole in the system (that could be generated from a nearby tetraquark state, for example),
it would appear as a pole of the kind $1/(s-s_0)$. Employing
a momentum/energy expansion of the counter terms means that those pole terms are effectively expanded
in $s/s_0$. Thus, the effective field theory would signal a close by bare
pole by a nonconvergence of the expansion of the contact terms. The analysis of, e.g., the
$Z_b$ states presented in Ref.~\cite{Wang:2018jlv} shows that the currently existing data
are consistent with the assumed power counting as soon as a properly renormalised one pion exchange is included
in the formalism. It is therefore fair to say that the data are consistent with
a molecular nature of the states. However, once the one pion exchange is omitted from the study,
the expansion of contact terms seems to converge badly. Better data are necessary in order to allow
one to draw more sound conclusions from the $Z_b$ line shapes. In particular, data for the line shapes of
the spin partner states would be extremely valuable. Theoretical predictions for these observables within the hadronic effective
field theory outlined in this section are provided in Ref.~\cite{Baru:2019xnh}.

It is important to note that under special kinematic circumstances the triangle diagrams
shown in the first column of Fig.~\ref{fig:Zbdecays} may get enhanced significantly
by the so-called triangle, or Landau, singularity~\cite{Landau:1959fi}. For a detailed discussion
we refer to Refs.~\cite{Wang:2016dtb,Guo:2017jvc,Szczepaniak:2015eza,Bayar:2016ftu}.
These logarithmic singularities can play an important role, whenever all intermediate
states in the triangle can go (near) on-shell simultaneously and additional kinematic conditions are met.
It should be stressed, however, that in many cases
the inverse of the average velocity (see above) is a good approximation to the triangle diagram~\cite{Guo:2017jvc}.
Triangle singularities might play an important role in different decays of vector
mesons~\cite{Wang:2013hga,Szczepaniak:2015eza,Bondar:2016pox},
and might explain the signal of the $Z_c(4430)$~\cite{Pakhlov:2011xj,Pakhlov:2014qva,Uglov:2016nql} or the narrow pentaquark signal~\cite{Guo:2015umn,Guo:2016bkl}.
In this context it is important to observe that the triangle singularities appear only under very special kinematic conditions. In particular,
if a state is seen, e.g., from very different initial states or energies, one can assume that the triangle singularities are not operative
for all of them. Moreover, when studying a three--body final state containing a particle pair that emerged from an elastic rescattering
form a particle pair in the triangle, no logarithmic singularity can be present according to Schmid's theorem~\cite{Schmid:1967ojm,Anisovich:1995ab} ---
in other words: Even when a triangle singularity is responsible for a structure that was interpreted as a resonance in
some inelastic channels, in the corresponding elastic channel there should not be any signal. On the other hand: If there were a pole
present it should naturally show up in all channels to which it can couple according to the selection rules. This feature can clearly be
tested experimentally.

In Ref.~\cite{Guo:2019qcn} it was proposed to exploit the interplay of the triangle singularity in $D^*\bar D^*\to \gamma X(3872)$ (with a very low
energy initial state as could be studied at LHCb) and the $X(3872)$ pole for a very high accuracy mass determination of the $X(3872)$.
In this reference, it is shown that the resulting line shape is very sensitive to the relative location of the logarithmic Landau singularity and the pole.

\subsection{Lattice QCD}
\label{Sect:4.3}
Lattice QCD is a method which enables first-principles
systematically-improvable calculations in QCD. The theory is discretised onto
a four-dimensional spacetime lattice, a grid of points, of finite volume and
quantities of interest are extracted from Euclidean (imaginary-time)
correlation functions which are computed numerically using Monte-Carlo methods.
For spectroscopy, the energy eigenstates of QCD in a finite volume follow from computations of
two-point correlation functions, $\langle \mathcal{O}^{\vphantom{\dagger}}_i(t) \, \mathcal{O}^\dagger_j(0) \rangle$,
where $\mathcal{O}_i(t)$ is an interpolating operator at discrete time $t$
with appropriate quantum numbers. The effect of working on a discrete lattice
can be accounted for by performing calculations with a number of different
lattice spacings and extrapolating to the continuum limit and, similarly, the
effect of working in a finite volume can be taken into account by performing
calculations with a number of different lattice volumes. Historically,
calculations with light (up and down) quarks having their physical mass were
too computationally expensive -- unphysically-heavy light quarks were used,
giving an unphysically-large $\pi$ mass, $M_\pi$, and a `chiral' extrapolation
to physical $M_\pi$ was performed. More recently, it has become feasible to
calculate some quantities using physical-mass, or very close to physical-mass, light quarks.

Calculations of the masses of the lower-lying hadrons in each flavour sector which are stable within
QCD, i.e., below the relevant thresholds for strong-decay, are now very mature
and give precise results with the various systematic uncertainties quantified.
See Refs.~\cite{Gray:2005ur,Burch:2009az,Meinel:2010pv,Namekawa:2011wt,Daldrop:2011aa,Mohler:2011ke,Dowdall:2011wh,McNeile:2012qf,Dowdall:2012ab,Donald:2012ga,Aoki:2012xaa,Dowdall:2013jqa,Kalinowski:2015bwa,Cichy:2016bci,DeTar:2018uko} for some recent work in the charmonium and bottomonium sectors.
Computations have begun to include the effects of isospin breaking
(different masses for the up and down quarks) and QED,
for now mainly in the light-hadron sector, e.g.,
Refs.~\cite{Duncan:1996xy,Blum:2007cy,Blum:2010ym,Aoki:2012st,Horsley:2012fw,deDivitiis:2013xla,Borsanyi:2013lga,Borsanyi:2014jba,Horsley:2015eaa,Horsley:2015vla,Giusti:2017dmp,Basak:2018yzz}.

However, most hadrons are not stable in QCD but instead are resonances
that decay to final states containing two or more lighter hadrons.
In particular, the various charmonium and bottomonium-like $X,Y,Z$ states
that are the subjects of this report are resonances or have masses very close to threshold.
These phenomena are more challenging to study in lattice QCD because it is
not possible to compute scattering amplitudes or the properties of resonances directly from
finite-volume Euclidean correlation functions -- asymptotic states can not be
defined in finite volume and, in contrast to an infinite volume, the spectrum
above threshold is discrete with allowed momenta quantised by the boundary conditions.

Nevertheless, infinite-volume scattering amplitudes can be determined
indirectly from a relation between them and the discrete spectrum of
finite-volume energy eigenstates. For elastic scattering, each energy level,
$E_\textrm{cm}$, gives the scattering phase shift at that energy,
$\delta(E_\textrm{cm})$. When more than one scattering channel or partial wave
is relevant, each energy provides a constraint on the scattering amplitudes at
that energy. Once the scattering amplitudes have been determined, analytically
continuing to complex $E_\textrm{cm}$, their singularity content can be
investigated and so the resonance and bound-state content inferred (including
the mass and width of any resonance and its coupling to various hadron-hadron
channels). This method was originally introduced by L\"{u}scher,
Refs.~\cite{Luscher:1985dn,Luscher:1986pf,Luscher:1990ux,Luscher:1991cf},
and has subsequently been generalised by many others -- it is now applicable to
hadron-hadron scattering with arbitrary spin, arbitrary overall momentum with
respect to the lattice, and with any number of coupled two-hadron scattering
channels, Refs.~\cite{Rummukainen:1995vs,Li:2003jn,Bedaque:2004kc,Feng:2004ua,Christ:2005gi,Kim:2005gf,He:2005ey,Lage:2009zv,Bernard:2010fp,Doring:2011vk,Doring:2011nd,Fu:2011xz,Leskovec:2012gb,Gockeler:2012yj,Doring:2012eu,Hansen:2012tf,Briceno:2012yi,Guo:2012hv,Li:2012bi,Agadjanov:2013kja,Briceno:2013hya,Briceno:2014oea,Li:2014wga,Lee:2017igf}.
Work is advancing to extend the approach to channels with three or
more hadrons, see Refs.~\cite{Briceno:2012rv,Polejaeva:2012ut,Kreuzer:2012sr,Hansen:2014eka,Meissner:2014dea,Hansen:2015zga,Hansen:2015zta,Hansen:2016fzj,Hansen:2016ync,Briceno:2017tce,Hammer:2017uqm,Hammer:2017kms,Mai:2017bge,Briceno:2018aml,Briceno:2018mlh,Mai:2018djl,Doring:2018xxx,Hansen:2019nir,Blanton:2019igq,Pang:2019dfe}.
A related approach for connecting finite-volume lattice QCD calculations to infinite-volume scattering amplitudes,
based on the same underlying ideas,
is to consider an effective Hamiltonian describing hadron-hadron interactions in a finite volume, such as unitarised chiral perturbation theory,
see e.g.\ Refs.~\cite{Hanhart:2008mx,Rios:2008zr,Doring:2011nd,Doring:2011vk,Doring:2011ip,Nebreda:2011di,Doring:2012eu,Chen:2012rp,Albaladejo:2013aka,Garzon:2013uwa,Wu:2014vma}.
Other approaches include using an optical potential, Ref.~\cite{Agadjanov:2016mao}, the histogram method in which
a probability distribution connected to the cross section is computed, Refs.~\cite{Bernard:2008ax,Giudice:2012tg},
and a method, advocated by HAL QCD, where a non-relativistic hadron-hadron potential is computed
from the Bethe--Salpeter wavefunction,
Refs.~\cite{Ishii:2006ec,Aoki:2009ji,Aoki:2011gt,HALQCD:2012aa,Aoki:2012tk,Aoki:2012bb,Aoki:2013cra,Aoki:2013tba,Namekawa:2017sxs,Yamazaki:2017gjl,Aoki:2017yru,Yamazaki:2018qut,Iritani:2018zbt}.
Ref.~\cite{Briceno:2017max} provides a more detailed review of the ``L\"{u}scher method''
and applications, and discusses some of the other approaches.

To robustly determine scattering amplitudes using the L\"{u}scher method,
many finite-volume energies must be extracted across the energy range being
considered. The reduced symmetry of the spatial lattice volume,
usually a cubic or rectangular box, compared to the infinite-volume continuum,
means that partial waves of different $J$ can `mix' and appear in the same spectrum --
sufficient energies must be extracted across a number of different quantum-number
channels in order to disentangle them.
Achieving the required constraint generally necessitates the computation of many
energies in a number of lattice volumes, for systems moving with
respect to the lattice, and/or using ``twisted boundary conditions'' on the
quark fields -- these all modify the quantisation of momentum and so provide additional $E_\textrm{cm}$, which give extra constraints.
Furthermore, this approach neglects corrections that are exponentially
suppressed in the spatial extent $L$, $\sim e^{-M_\pi L}$, and so requires a
large enough volume, $L M_\pi \gg 1$, for these to be negligible.
In recent years, computational power and algorithms have advanced sufficiently
to allow applications of this method to meaningfull computations of
hadron-hadron scattering. However, it must be kept in mind that, unlike the
lattice QCD studies of stable hadrons mentioned above, these investigations do
not generally have quantitative control over all the systematic uncertainties
(for example, those arising from working at a finite lattice spacing or with
unphysically-heavy light quarks).

As discussed in Ref.~\cite{Briceno:2017max}, a number of light-hadron
scattering channels have been studied in detail: many energy levels are
extracted, the scattering amplitudes are robustly determined and the resonant
content is inferred. For example, the $\rho$ resonance in isospin-1 $\pi\pi$
scattering has been studied by a number of groups for various light-quark masses and lattice spacings,
Refs.~\cite{Aoki:2007rd,Feng:2010es,Aoki:2011yj,Lang:2011mn,Dudek:2012xn,Pelissier:2012pi,Wilson:2015dqa,Bali:2015gji,Bolton:2015psa,Bulava:2016mks,Guo:2016zos,Hu:2016shf,Fu:2016itp,Hu:2017wli,Alexandrou:2017mpi,Andersen:2018mau}.

Lattice QCD calculations in the charmonium, bottomonium and
related exotic-flavor sectors are, however, less advanced. We will discuss
calculations with three levels of sophistication. The first class of
calculations neglect the fact that hadrons above threshold are unstable and
determine the energy eigenstates using solely fermion-bilinear operators (with
a structure resembling a single meson) -- the full finite-volume spectrum,
which includes multihadron levels, is not extracted. As discussed in
Refs.~\cite{Dudek:2010wm,Liu:2012ze,Briceno:2017max}, these results should
only be considered a guide to the pattern of narrow resonances, and the precise
resonance mass and other properties can not be obtained. It is also unclear
whether these calculations are sensitive to broad resonances and resonances
where coupled-channel effects are important.

The second class of calculations include operators with relevant
meson-meson-like structures and extract the finite-volume spectrum in the
energy region considered (though they may neglect some potentially relevant
meson-meson channels). However, these do not have enough constraints to
robustly determine scattering amplitudes using the L\"uscher method and are
often limited to looking for the presence of extra levels in the spectrum
compared to the spectrum if there were no meson-meson interactions.
This is likely to give a reasonable guide to the presence of a narrow resonance in
elastic scattering, but it is less clear for broader resonances and
coupled-channel scattering, see Ref.~\cite{Briceno:2017max}. The third class of
calculations extract the complete finite-volume spectrum in the energy region
considered and have enough constraints to determine the scattering amplitudes
and hence the resonance content. Only these should be considered rigorous
but the other less sophisticated calculations may provide useful qualitative guides.

We now discuss lattice QCD calculations of charmonia and related
exotic-flavor hadrons before moving to bottomonia and then investigations
using the potential between two infinitely-heavy (static) quarks. Unless
otherwise stated, all the computations are performed for a single lattice
spacing and have dynamical up, down, and strange quarks, i.e.\ they include the
effects of these quarks in the sea, and the two light (up and down) quarks
are degenerate. Some of the calculations have dynamical light quarks but
quenched (not dynamical) strange quarks -- this leads to a non-unitary theory
and an unknown systematic uncertainty. The use of quenched charm (or bottom)
quarks is not expected to have a significant effect far below $cccc$
(or $bbbb$) threshold.

\subsubsection{Charmonia and related flavour-exotic channels}
\label{Sect:4.3:charmonia}
\begin{figure}[tb]
\begin{center}
\includegraphics[width=0.65\linewidth]{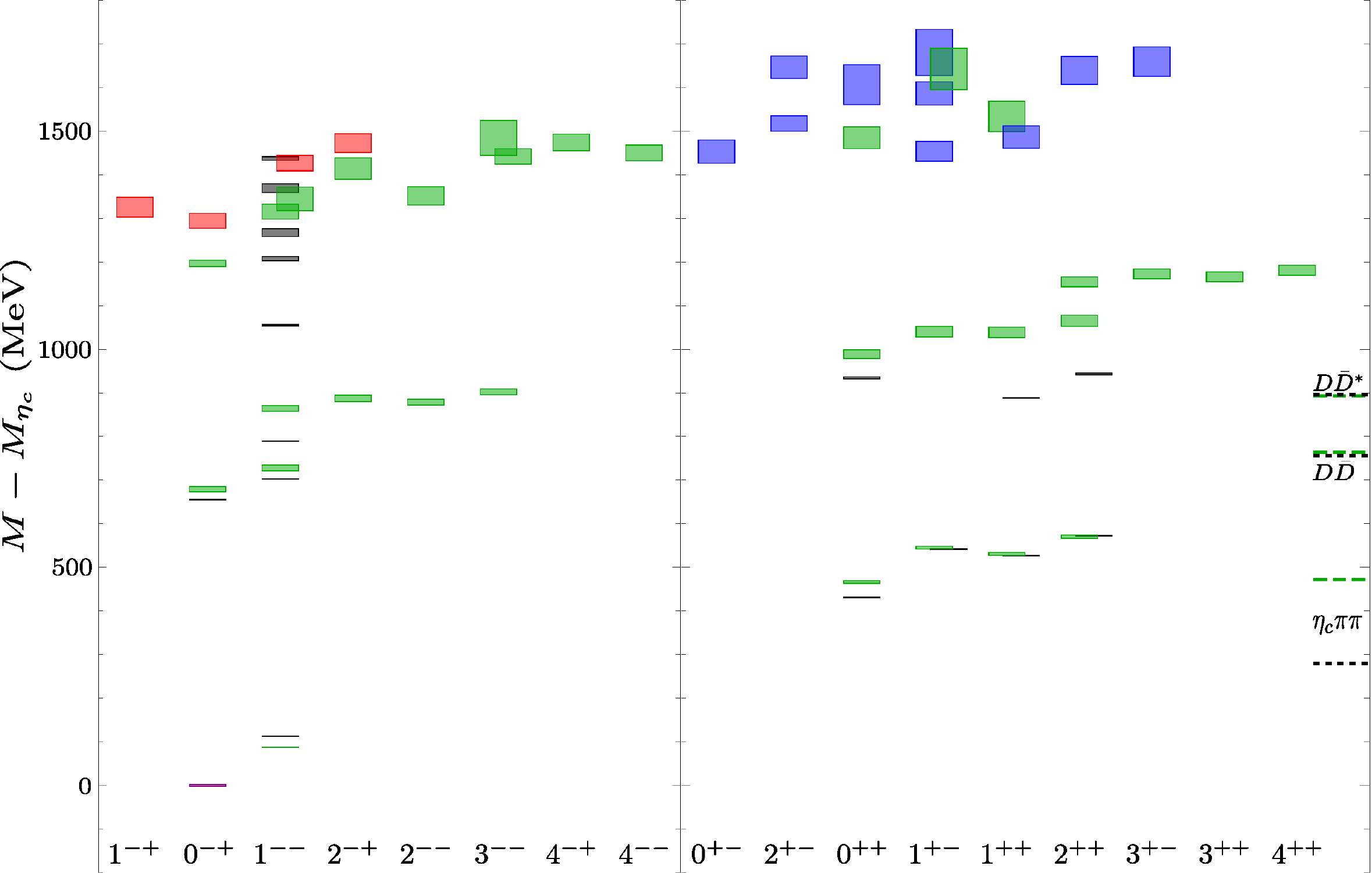}
\caption{From Ref.~\cite{Cheung:2016bym}. Charmonium spectrum labeled
by $J^{PC}$ from a lattice calculation with $M_\pi \approx 240$ MeV
where the unstable nature of states above threshold has been
neglected. Green, red and blue boxes represent the computed masses
and one-sigma statistical uncertainties, with red and blue boxes
highlighting states identified as hybrid mesons; black boxes are
experimental values. Some of the lower thresholds are shown as dashed
lines on the right of the figure (green are using computed masses and
black are using experimental masses).}
\label{fig:lattice:excitedcharmoniumspectrum}
\end{center}
\end{figure}

Neglecting the fact that they are unstable, higher-lying charmonia have been
studied in Refs.~\cite{Bali:2011rd,Mohler:2012na,Liu:2012ze,Cheung:2016bym}.
The results presented in Refs.~\cite{Liu:2012ze,Cheung:2016bym}, with
$M_\pi \approx 390$ and 240 MeV, one lattice spacing and a few volumes,
include hybrid mesons with exotic and non-exotic quantum numbers --
see Fig.~\ref{fig:lattice:excitedcharmoniumspectrum} for the results with $M_\pi \approx 240$ MeV.
These suggest an interesting hybrid-meson phenomenology with the lightest hybrid multiplet being consistent with a
quark-antiquark pair in an $S$ wave coupled to a $J^{PC} = 1^{+-}$ gluonic excitation, and an energy scale of $\sim$ 1.2 - 1.3 GeV.
The results have been discussed in the context of Born--Oppenheimer effective field theory in Sec.~\ref{Sect:4.2.3bis}.
However, as mentioned above, these spectra should only be considered a guide to the pattern of narrow resonances.

One challenge that lattice QCD investigations of charmonium resonances face
is the large number of channels to which such a resonance can potentially couple --
all these channels need to be considered to reach robust conclusions.
Currently, calculations are mostly limited to extracting finite-volume spectra
and looking for the presence of extra levels, and they have generally been exploratory or inconclusive.
More detailed computations are needed that extract a larger number of finite-volume energy levels and
allow scattering amplitudes to be determined and hence the state content to be investigated robustly.
All the following calculations neglect contributions where a charm
quark and a charm antiquark annihilate -- these are OZI suppressed and so
expected to be small (see also Ref.~\cite{Levkova:2010ft}).
In the NRQCD language of Sec.~\ref{Sect:4.2.2}, quark-antiquark annihilation is described
by four-fermion operators in the Lagrangian density, Eq.~\eqref{eq:sect4:NRQCD}.
These operators are of dimension 6 or higher, and they are therefore suppressed
by powers of the quark mass in the power counting of the effective field theory.

In Refs.~\cite{Prelovsek:2013cra,Padmanath:2015era}
$I=0$ $J^{PC} = 1^{++}$ $D \bar{D}^*$, $\omega J/\psi$, $\sigma \eta_c$, $\sigma \chi_{c1}$ scattering is
investigated, neglecting potential coupling to some other channels.
Interpolating operators with diquark-antidiquark structure are used in addition
to those with fermion-bilinear and meson-meson structures, though their
inclusion is not found to have a significant impact on the results.
Computations are performed with $M_\pi \approx 266$ MeV and dynamical light
quarks (quenched strange quarks). One rather small volume,
$M_\pi L \approx 2.7$, is employed with the system overall at rest leading to only
a few energy levels being determined. A bound state is found just below
$D \bar{D}^*$ threshold, a candidate for the $X(3872)$.
Refs.~\cite{Baru:2013rta,Baru:2015tfa,Jansen:2015lha} discuss how the binding energy of
the $X(3872)$ is expected to depend on the light quark mass.
A study of $\bar{c} c \bar{s} s$ ($D_s \bar{D}_s^*$, $\phi J/\psi$) in Ref.~\cite{Padmanath:2015era},
neglecting coupling to the other $I=0$ channels, finds no sign of a bound state or resonance,
so no candidate for the $X(4140)$.

\begin{figure}[tb]
\begin{center}
\includegraphics[width=0.55\linewidth]{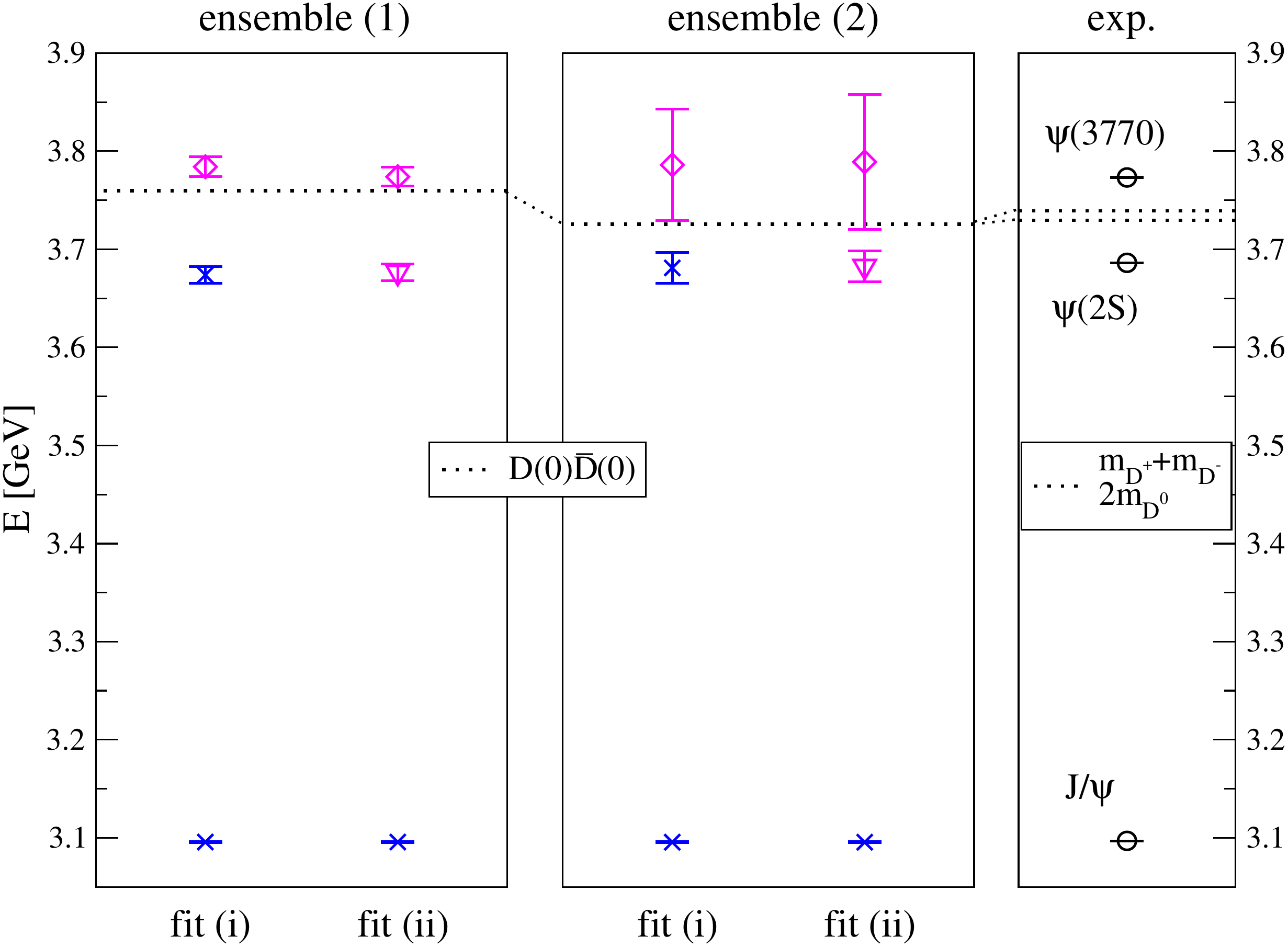}
\caption{From Ref.~\cite{Lang:2015sba}. Charmonium spectrum in the
$J^{PC} = 1^{--}$ channel from lattice calculations with $M_\pi
\approx 266$ MeV (ensemble 1) and $M_\pi \approx 156$ MeV (ensemble
2), compared to experimental results. Two different fit procedures
have been used (fit i and fit ii) as described in
Ref.~\cite{Lang:2015sba}.}
\label{fig:lattice:lang2015_vectors}
\end{center}
\end{figure}

Ref.~\cite{Lang:2015sba} studied $P$-wave $D\bar{D}$ ($1^{--}$) scattering
and $S$-wave $\omega J/\psi$, $D \bar{D}$ ($0^{++}$) scattering, neglecting
potential couplings to other channels. Two small lattices are used: one
with $M_\pi \approx 266$ MeV, $M_\pi L \approx 2.7$ and dynamical light quarks,
and the other with $M_\pi \approx 156$ MeV, $M_\pi L \approx 2.3$ and dynamical
light and strange quarks, and a small number of energy levels extracted.
A $1^{--}$ resonance is found and possibly a $\psi(3770)$ candidate as shown
in Fig.~\ref{fig:lattice:lang2015_vectors}.
Assuming elastic $D \bar{D}$ scattering there is a suggestion of a narrow $0^{++}$
resonance slightly below 4 GeV -- this does not appear to be in
particularly good agreement with the experimental data in Ref.~\cite{Chilikin:2017evr}.

Low-energy $\phi(\bar{s}s)$ $J/\psi$ scattering in $S$-wave and $P$-wave is
investigated in Ref.~\cite{Ozaki:2012ce} using a small volume,
$M_\pi L \approx 2$, with $M_\pi \approx 156$ MeV. Other open channels are
neglected as are contibutions where any quark and antiquark annihilate,
leading to unquantified systematic uncertainties. Twisted boundary conditions
are employed to extract a number of energy levels and determine the phase shifts close to threshold.
However, the twisted boundary conditions are only applied to valence quarks, not sea quarks,
leading to a non-unitary theory and another unknown systematic uncertainty.
The extracted $S-$ and $P$-wave phase shifts are consistent with weak
attraction and show no sign of a resonance such as a candidate for the $X(4140)$.


Moving to exotic-flavor channels, Ref.~\cite{Cheung:2017tnt} investigates
the doubly-charmed $I=0$ channel (quark content $cc\bar{q}\bar{q}$ where $q$
is an up or down quark) with $J^{P} = 0^+, 1^+, 2^+$ and the $I=1/2$ channel
(quark content $cc\bar{q}\bar{s}$) with $J^{P} = 0^+, 1^+$, and the hidden-charm
$I=1$ channel (quark content $c\bar{c} q\bar{q}$) with $I^{G} J^{P} = 1^+ 1^{+}, 1^- 0^{+}, 1^- 1^{+}$.
Calculations are performed with $M_\pi \approx 390$ MeV
on a single volume where $M_\pi L \approx 3.8$. Operators with both meson-meson and diquark-antidiquark structures are used,
though the latter are found not to be important in determining the spectrum.
In each channel studied, a number of energy levels are extracted and there is no sign of a bound state or narrow resonance.

In Ref.~\cite{Prelovsek:2014swa} (an extension of Ref.~\cite{Prelovsek:2013xba}), the hidden-charm $I^G = 1^+$, $J^{P}=1^{+}$
channel is studied considering $D \bar{D}^*, \pi J/\psi , \rho \eta_c, D^* \bar{D}^*, \pi \psi(2S) , \pi \psi(1D) $ scattering.
A small lattice volume is used, $M_\pi L \approx 2.7$, with $M_\pi \approx
266$ MeV and dynamical light quarks (quenched strange quarks).
Interpolating operators with meson-meson structure are employed along
with a few having diquark-antidiquark structure. A number of energy
levels are extracted -- these only give evidence for at most weak
interactions and there is no sign of a bound state or narrow resonance
[e.g.\ the $Z_c(3900)^+$] up to $\sim 4.2$ GeV. Ref.~\cite{Albaladejo:2016jsg} considers different possible scenarios for the $Z_c(3900)^+$ (resonance or virtual state) in comparison to these lattice results.
A similar study of the $I^G=1^-$ $J^{P} = 1^{+}$ channel
($D \bar{D}^*$, $\rho J/\psi$, $\pi \chi_{c0}$, $\pi \chi_{c1}$)
is presented in Refs.~\cite{Prelovsek:2013cra,Padmanath:2015era}.
A few energy levels are determined for the system overall at rest
and no evidence for a bound state or resonance is seen.

Ref.~\cite{Chen:2014afa} investigates low-energy $S$-wave $D
\bar{D}^*$ scattering with $I^G = 1^+$, $J^P = 1^+$, neglecting
possibly couplings to other open channels. Computations are performed
on three lattices with $M_\pi \approx$ 300, 420, 485 MeV and $M_\pi L \approx$ 3.3, 4.6, 5.3, respectively -- these have dynamical light
quarks (quenched strange quarks). Twisted boundary conditions are
used to extract a number of energy levels but the twisting is only
applied to valence quarks, not sea quarks, leading to a non-unitary
theory and an unquantified systematic uncertainty. The phase shifts
are found to correspond to weakly repulsive scattering and the
scattering length and effective range are determined. There is no
sign of a bound state or narrow resonance such as a $Z_c(3900)^+$ candidate.

The same setup, with the same limitations, is used in Ref.~\cite{Chen:2015jwa} to study low-energy $S$-wave $D^* \bar{D}^*$
scattering with $I^G = 1^+$, $J^P = 1^+$. Again, the interaction is
found to be weakly repulsive, the scattering length and effective
range are extracted, and there is no evidence for a bound state or
narrow resonance such as the $Z_c(4020)^+$. A similar study of
low-energy $D^* \bar{D}_1$ scattering in $S$-wave ($J^P = 0^-$) and
$P$-wave ($J^P = 1^+$) with $I^G = 1^+$ is presented in Ref.~\cite{Chen:2016lkl},
relevant for, inter-alia, the $Z_c(4430)^+$.
In this case, the results suggest an attractive interaction but a
definite conclusion on whether or not there is a bound state can not be reached.


The HAL QCD collaboration has pursued an alternative to the
L\"{u}scher approach in which lattice QCD is used to compute
correlation functions involving spatially-displaced interpolating
operators. These are related to Nambu--Bethe--Salpeter wave functions
and in turn to non-relativistic potentials which are used in the
Schr\"{o}dinger equation, see Refs.~\cite{Ishii:2006ec,Aoki:2009ji,HALQCD:2012aa,Aoki:2012tk,Aoki:2011gt,Aoki:2012bb,Aoki:2013cra,Aoki:2013tba,Namekawa:2017sxs,Yamazaki:2017gjl,Aoki:2017yru,Yamazaki:2018qut,Iritani:2018zbt}.
There has been a lot of discussion as to the reliability of the HAL
QCD approach and how it is employed. There is yet to be a successful
demonstration of its application to the $\rho$ resonance in $I=1$
$\pi\pi$ scattering. For a number of baryon-baryon channels, there
appears to be a discrepancy between results obtained using this method
and results obtain using the L\"{u}scher approach, though
investigations using the latter method have their own limitations --
see Refs.~\cite{Briceno:2017max,Yamazaki:2017gjl,Aoki:2017yru,Yamazaki:2018qut,Iritani:2018vfn} and references therein.
In Ref.~\cite{Haidenbauer:2019utu} the HAL QCD potentials for baryon-baryon scattering are critically examined
phenomenologically. It is found that especially in the $\Omega N$ and $\Omega\Omega$ channels, where
one pion change is not forbidden in the scattering potential, the range of the potential reported by the HAL QCD
studies is at odds with phenomenological expectations.

Ref.~\cite{Ikeda:2013vwa} investigates the flavor-exotic $cc\bar{u}\bar{d}$
and $cs\bar{u}\bar{d}$ channels. Reasonably large lattice volumes
($M_\pi L \gtrsim 6$) with $M_\pi \approx$ 410 MeV, 570 MeV, 700 MeV are used.
The HAL QCD method is used to extract $S$-wave potentials for $D D$, $\bar{K} D$, $D D^*$, and $\bar{K} D^*$,
these potentials are fit and the corresponding Schr\"{o}dinger equations are solved to compute the phase shifts.
The $I=0$ channels are found to be attractive, particularly $I=0$ $J^P = 1^+$ $D D^*$,
i.e.\ $cc\bar{u}\bar{d}$, and become more attractive as $M_\pi$ decreases.
However, there is no evidence for a bound state or resonance.
The $I=1$ channels are all found to be repulsive.

$S$-wave $\pi J/\psi$, $\rho \eta_c$, $\bar{D} D^*$ coupled-channel scattering
with $I=1$, $J^P = 1^+$ is studied in Ref.~\cite{Ikeda:2016zwx,Ikeda:2017mee}
using the same lattices as above. The HAL QCD method is employed and the
diagonal potentials are found to be weak, but the off-diagonal
$\pi J/\psi$ -- $\bar{D} D^*$ and $\rho \eta_c$ -- $\bar{D} D^*$ potentials are
strong. The $S$-matrix is computed from these potentials and poles are
searched for: a pole is found on an unphysical sheet far below $D \bar{D}^*$
threshold with a large imaginary part and it is suggested that the
$Z_c(3900)^+$ is a threshold cusp. These results are then used as an input to
a phenomenological study of the invariant mass spectra in $Y(4230) \to \pi\pi J/\psi$ and $Y(4230) \to \pi D \bar{D}^*$.


Going beyond spectroscopy and studying transitions between mesons can provide a more
stringent probe of their structure. Ref.~\cite{Dudek:2009kk} performed a quenched
calculation of radiative transition ampltidues involving some excited, high spin and
exotic charmonia, neglecting the unstable nature of states above threshold.
One highlight was the radiative decay of an exotic $J^{PC} = 1^{-+}$ hybrid meson, $\eta_{c1}$, which was
found to have a significant partial width, $\Gamma(\eta_{c1} \to J/\psi \gamma) \sim 100$~keV,
supporting suggestions from models that the photocouplings between $1^{-+}$ hybrids and
conventional mesons should be large. A large decay amplitude for the decay of a
$1^{--}$ hybrid to $\eta_c$ was found, again in line with phenomenological expectations.
In addition, it was shown how the pattern of different multipole amplitudes in the radiative
decay of $2^{++}$ charmonia to $J/\psi$ can distinguish $P$ wave and $F$ wave tensor mesons.
As well as studying the $\chi_{c2} \to J/\psi \gamma$ transition, Ref.~\cite{Yang:2012mya}
performed a quenched calculation of the $\eta_{c2}(2^-+) \to J/\psi \gamma$ transition
amplitudes, finding a small partial width and determining the three multipole amplitudes.
These investigations demonstrate the feasibility of using lattice QCD to study transitions
involving higher-lying charmonia and that such calculations can provide
phenomenologically-interesting results -- calculations with dynamical quarks allowing for the unstable nature of resonances are warranted.

\hfill

In summary, there have been a number of lattice QCD studies of charmonium-like
resonances and related scattering channels, but these all suffer from
significant limitations. Candidates for conventional resonances and the $X(3872)$
have been seen. However, there is currently no clear evidence for a
flavor-exotic hidden-charm or doubly-charmed state from lattice QCD
calculations, in contrast to the various experimental signals for charged
charmonium-like structures --- see Sec.~\ref{Sect:3.2}.

\subsubsection{Bottomonia and related flavour-exotic channels}
\label{Sect:4.3:bottomonia}
\begin{figure}[tb]
\begin{center}
\includegraphics[width=0.6\linewidth]{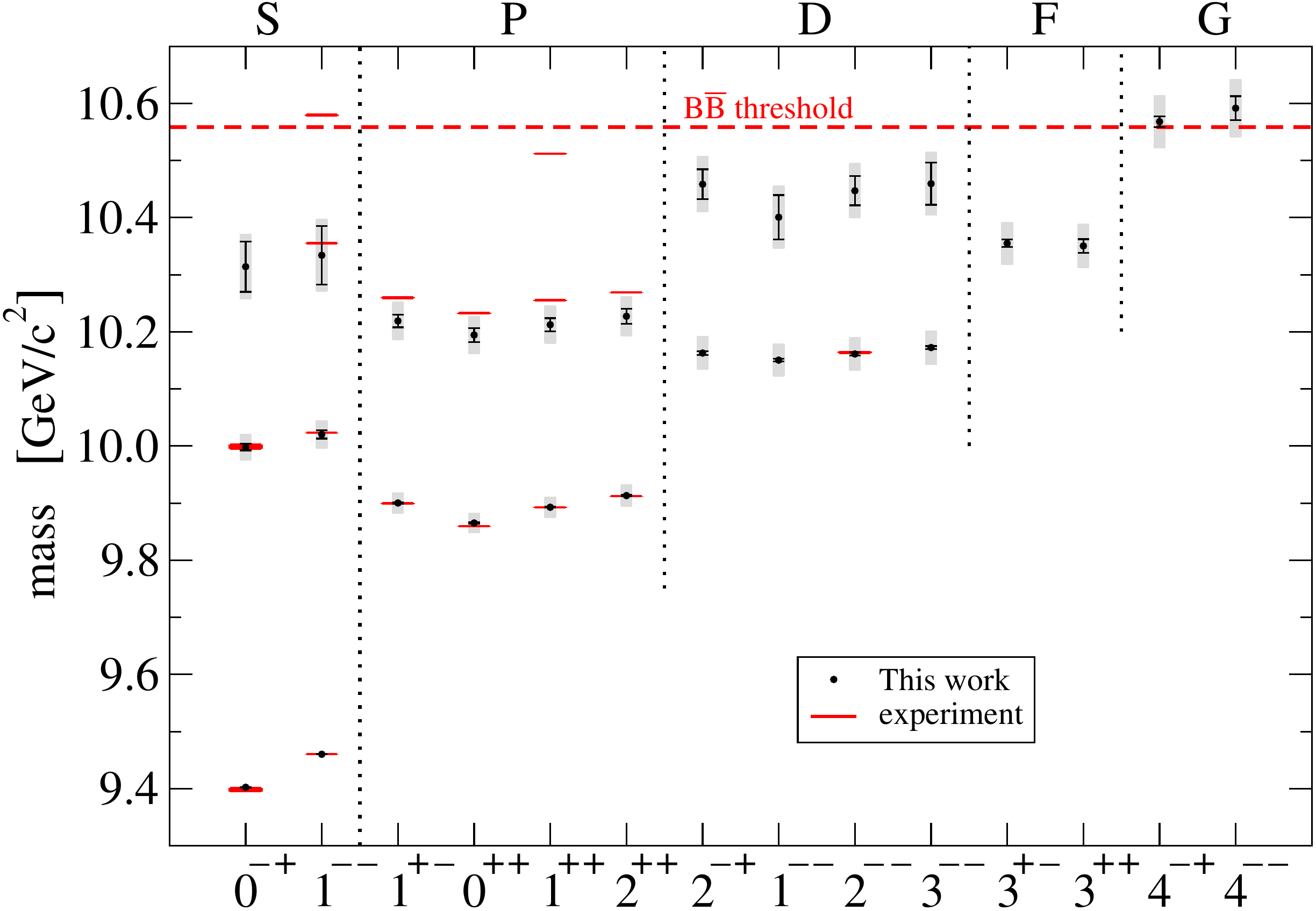}
\caption{From Ref.~\cite{Wurtz:2015mqa}. Bottomonium spectrum labeled
by $J^{PC}$ from a lattice calculation with $M_\pi \approx 156$ MeV.
Black points with errorbars are the lattice results with statistical
uncertainties, and grey bands give an estimate of the combined
systematic and statistical uncertainties. Red bands are experimental
masses.}
\label{fig:lattice:excitedbottomoniumspectrum}
\end{center}
\end{figure}

Some higher-lying $S$-, $P$-, $D$-, $F$- and $G$-wave bottomonia up to
$B\bar{B}$ threshold have been studied in Refs.~\cite{Lewis:2012ir,Wurtz:2015mqa} (along with $B$, $B_s$ and $B_c$ mesons).
A single lattice volume was used with $M_\pi \approx 156$ MeV and NRQCD $b$ quarks -- the results were found to be
in reasonable agreement with experiment -- see Fig.~\ref{fig:lattice:excitedbottomoniumspectrum}.

Ref.~\cite{Francis:2016hui} investigates the possibility of $J^P = 1^+$
flavor-exotic $I=0$ $u d \bar{b}\bar{b}$ and $I=1/2$
$q s \bar{b}\bar{b}$ systems. Three different light-quark masses are
considered corresponding to $M_\pi \approx$ 164, 299 and 415 MeV, each on one
volume ($M_\pi L \approx$ 6.1, 4.4, and 2.4, respectively), and NRQCD $b$
quarks are employed. A small number of energy levels are computed using
local meson-meson and diquark-antidiquark interpolating operators.
These give some evidence for bound states for all three light-quark masses: $(189 \pm 10 \pm 3)$ MeV below the $B B^*$ threshold for $u d \bar{b}\bar{b}$
and $(98 \pm 7 \pm 3)$ MeV below the $B_s B^*$ threshold for $q s \bar{b}\bar{b}$, though the robustness of the signal is not clear in all cases.

That investigation has recently been extended in Ref.~\cite{Francis:2018jyb},
which studies $J^P = 1^+$ $u d \bar{b}\bar{b}'$ ($I=0$), $u d \bar{b}'\bar{b}'$ ($I=0$), $q s \bar{b}\bar{b}'$ and $q s \bar{b}'\bar{b}'$ for $M_\pi \approx 299$ MeV,
where the mass of the $b'$ quark ranges from 0.59 to 6.3 $m_b$.
Evidence for binding is found in all cases and the dependence of the binding energy on
$m_{b'}$ is fit to phenomenologically-motivated forms. Extrapolating to
$m_{b'} = m_c$ appears to suggest that only $u d \bar{c} \bar{b}$ has a
possibility of being bound. The $J^P=1^+$ $I=0$ $u d \bar{c} \bar{b}$ is then
studied for all three $M_\pi$ -- the ground state is found to be consistent
with threshold for $M_\pi =$ 415 MeV and bound for $M_\pi =$ 163 MeV and
299 MeV, with the binding energy estimated to be between 15 and 61 MeV.
However, as the authors point out, unaccounted-for finite-volume effects may be significant.

\begin{figure}[tb]
\begin{center}
\includegraphics[width=0.48\linewidth, trim=0 0 510 0, clip]{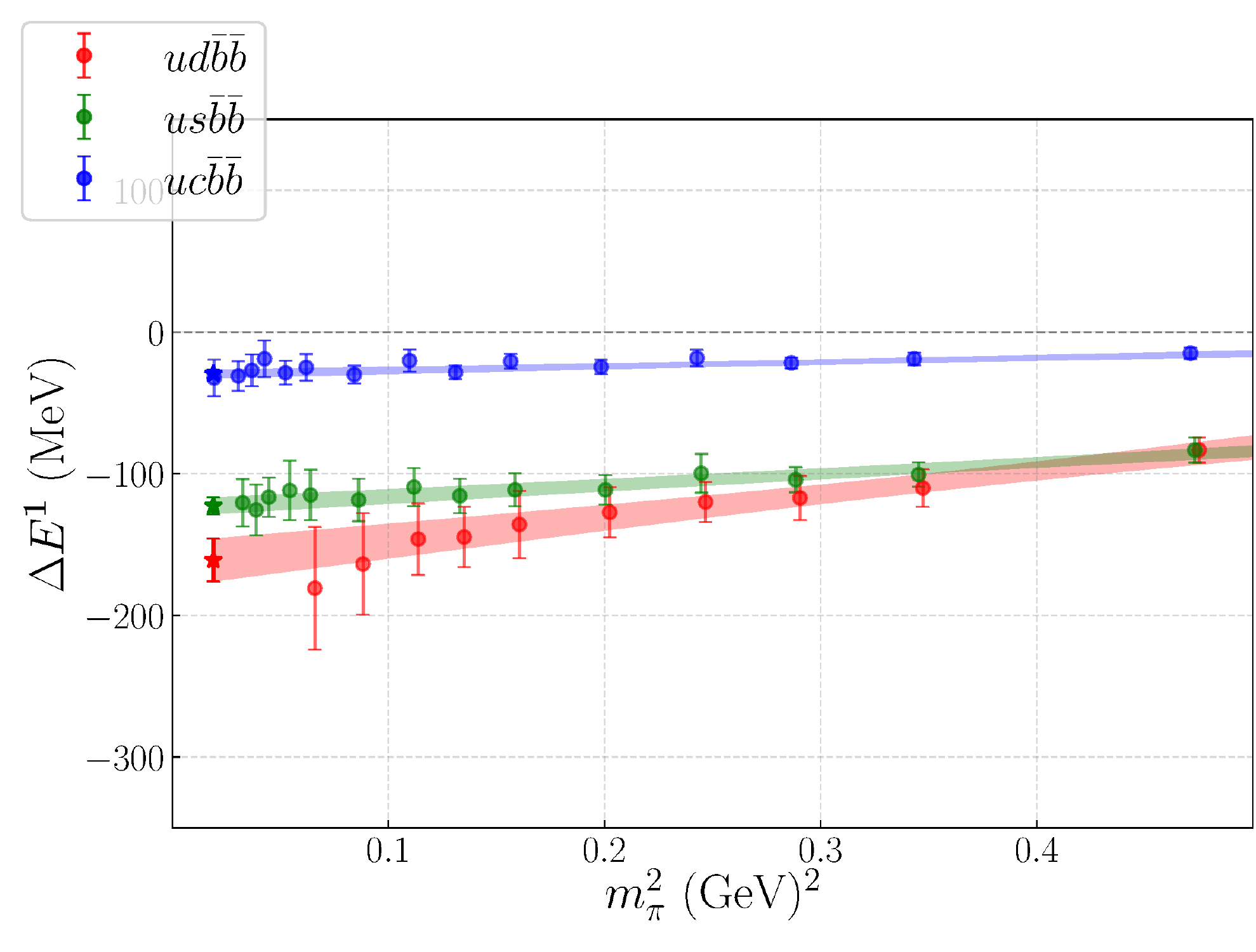}
\includegraphics[width=0.48\linewidth]{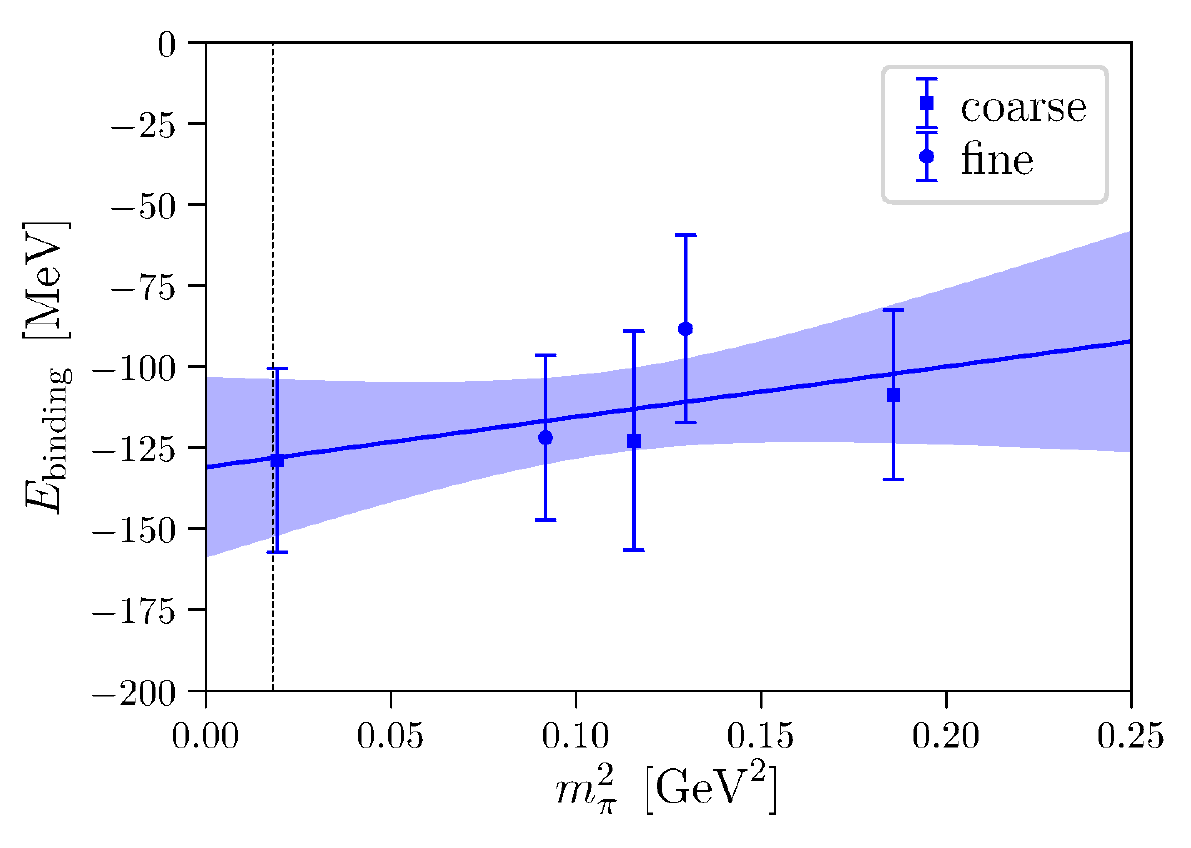}
\caption{Left: Dependence of the binding energies in the $J^P = 1^+$ $u d \bar{b} \bar{b}$ ($I=0$), $u s \bar{b} \bar{b}$ and $u c \bar{b} \bar{b}$ channels on the valence $M_\pi$ from Ref.~\cite{Junnarkar:2018twb} (the light sea-quark mass is fixed).
Right: Dependence of the binding energy in the $J^P = 1^+$ $u d \bar{b}\bar{b}$ ($I=0$) channel on $M_\pi$ from Ref.~\cite{Leskovec:2019ioa}.}
\label{fig:lattice:tetraquark_mpi}
\end{center}
\end{figure}

In Ref.~\cite{Junnarkar:2018twb} a number of $J=0,1$ double-bottom channels
are studied with three different lattice spacings ($a \approx$ 0.12, 0.09,
0.06 fm), each on a single volume, and with dynamical light, strange and charm quarks.
The sea light-quark mass corresponds to $M_\pi \sim 300$ MeV, but a different
action is used for the light valence quarks with a range of masses
corresponding to $M_\pi \approx$ 153 to 689 MeV -- this leads to a non-unitary
theory and an unquantified systematic uncertainty. Local meson-meson and
diquark-antidiquark operators are employed and a small number of levels
extracted, though in some cases the quality of the signal is not that good.
With $J^P = 1^+$, bound states are claimed in $u d \bar{b} \bar{b}$ ($I=0$) and $u s \bar{b} \bar{b}$
with respective masses $143 \pm 34$~MeV and $87 \pm 32$~MeV below $B B^*$ threshold
 -- see left panel of Fig.~\ref{fig:lattice:tetraquark_mpi}.
For $u c \bar{b} \bar{b}$, $s c \bar{b} \bar{b}$, $u d \bar{c} \bar{c}$ ($I=0$)
and $u s \bar{c} \bar{c}$, there is some suggestion of a
close-to-threshold state but finite-volume effects may be important. The
binding energy shows a general trend to increase as the light-quark masses decrease.
There is no evidence for bound states with $J^P = 0^+$ for $u u \bar{b} \bar{b}$ ($I=1$), $s s \bar{b} \bar{b}$, $c c \bar{b} \bar{b}$, $u u \bar{c} \bar{c}$ ($I=1$) and $s s \bar{c} \bar{c}$.

Very recently, Ref.~\cite{Leskovec:2019ioa} has reported on a more detailed
investigation of the $I=0$ $J^P = 1^+$ $u d \bar{b}\bar{b}$ channel.\footnote{Some preliminary results from an early stage of this study appeared in Ref.~\cite{Peters:2016isf} where, in addtion, $b$ quarks with masses ranging from one to five times the physical $b$-quark mass were considered.}
Five different lattices are used with dynamical light and strange quarks,
a range of lattice spacings ($a \approx$ 0.083 -- 0.114 fm), volumes ($M_\pi L \approx$ 3.9 to 5.8)
and light-quark masses ($M_\pi \approx$ 139 to 431 MeV); NRQCD $b$ quarks are employed.
Operators with a meson-meson-like structure ($B B^*$ and $B^* B^*$ ) are used in addition
to those with a local diquark-antidiquark structure, and a couple of energy levels
are extracted on each lattice.
The L\"{uscher} method was used to determine the infinite-volume binding energy through a
fit to an effective range expansion (though this was found to be consistent with the finite-volume binding energy),
giving a bound state $128 \pm 24 \pm 10$ MeV below $BB^*$ threshold, where the second error is an estimate
of systematic uncertainties -- see right panel of Fig.~\ref{fig:lattice:tetraquark_mpi}.

In an investigation of the $bb\bar{b}\bar{b}$ system, Ref.~\cite{Hughes:2017xie} uses four lattices with various light-quark
masses (including one lattice with physical-mass light quarks) and lattice spacings ($a =$ 0.06 -- 0.12 fm), and NRQCD $b$ quarks.
A number of meson-meson and diquark-antidiquark operators are employed and no sign of a
bound state is found with $J^{PC} = 0^{++}$, $1^{+-}$, $2^{++}$.


Turning briefly away from bottomonium-like systems, Ref.\ \cite{Lang:2016jpk} investigates the exotic-flavor bottom-strange $I=1$ $J^P = 0^+$ $\bar{b}s\bar{d}u$ channel relevant for the $X(5568)$ (see Sec.~\ref{Sect:3.2.1}). A number of $B_s \pi$ and $B\bar{K}$ operators are employed in a calculation on small volume, $M_\pi L \approx 2.3$, using dynamical light and strange quarks with $M_\pi \approx 156$ MeV. A few energy levels are extracted and the results are compared with the levels expected if a $X(5568)$ were present. No signs of strong $B_s \pi$ interactions are found and there is no evidence for a $X(5568)$ candidate. Ref.\ \cite{Lu:2016kxm} comments on some implications for the $X(5568)$ in light of these results.

\hfill

To summarise, there have been few lattice QCD investigations of excited bottomonia.
However, in contrast to the charm sector, there appears to be
evidence for a double-bottom $J^P = 1^+$ bound state in
$I=0$ $u d \bar{b}\bar{b}$ and $I=1/2$ $q s \bar{b}\bar{b}$, with the binding
energy increasing if the light-quark mass decreases or $m_b$ increases.
However, confirmation of some of these results is required, and
further work is needed to investigate a number of channels in more detail and
assess some of the systematics. In the next section we discuss lattice QCD calculations
of the potential between two static quarks -- these support the presence of
a $J^P = 1^+$ $I=0$ $u d \bar{b}\bar{b}$ bound state for large $m_b$.

\subsubsection{Static quark potentials}
\label{Sect:4.3:static}
Refs.~\cite{Bicudo:2012qt,Bicudo:2015vta} use lattice QCD to compute the
potential between two infinitely-heavy static antiquarks $\bar{Q}$ in the
presence of light quarks, fit to a phenomenologically-inspired form and then
investigate whether a bound state is present by solving the Schr\"{o}dinger
equation. These calculations have dynamical light quarks (quenched strange
quarks)\footnote{Because the twisted mass action is used for the quarks,
isospin is broken and this has to be taken into account in the calculations.}
and two small lattice volumes are used: one with $M_\pi \approx 340$ MeV, $a \approx 0.079$ fm and $M_\pi L \approx 3.3$ for systems involving $qq$ and
another with $M_\pi \approx$ 352 MeV, $a \approx 0.042$ fm and $M_\pi L \approx 2.4$ for systems involving $ss$ or $cc$.
They find evidence for a $J^P = 1^+$ bound state in $I=0$ $qq\bar{Q}\bar{Q}$, and evidence against a bound state
in $J^P = 0^+, 1^+$ or $2^+$ for $I=1$ $qq\bar{Q}\bar{Q}$, $ss\bar{Q}\bar{Q}$ and $cc\bar{Q}\bar{Q}$.

In Ref.~\cite{Bicudo:2015kna}, this work is extended by performing
calculations with multiple light-quark masses, corresponding to
$M_\pi \approx 340$ MeV, 480 MeV, and 650 MeV ($M_\pi L \approx$ 3.3, 4.6, 6.3),
and extrapolating to physical $M_\pi$. Again, evidence is found for a
$J^P = 1^+$ $I=0$ $qq\bar{Q}\bar{Q}$ bound state, with a mass $90^{+43}_{-36}$ MeV
below threshold, and against a $J^P = 0^+, 1^+$ or $2^+$ $I=1$ $qq\bar{Q}\bar{Q}$ bound state.
Ref.~\cite{Bicudo:2016ooe} makes an attempt to incorporate spin effects, computing static potentials for
various light-quark spin combinations, interpreting in terms of $BB$, $BB^*$
and $B^*B^*$ meson pairs, then using in a coupled-channel Schr\"{o}dinger equation.
Following this procedure, evidence remains for a $I=0$ $J^P=1^+$
$qq\bar{Q}\bar{Q}$ bound state but the spin effects reduce the binding, giving a mass $59^{+30}_{-38}$ MeV below threshold.
An extension in Ref.~\cite{Bicudo:2017szl} investigates the possibility of
resonances using the emergent wave method, predicting a $J^P = 1^-$ $I=0$
$qq\bar{Q}\bar{Q}$ resonance but no $I=1$ bound states or resonances.

An independent investigation presented in Ref.~\cite{Brown:2012tm} employs
a similar approach to calculate the static potential in $I=0,1$
$qq\bar{Q}\bar{Q}$ for a range of spin and parity channels.
One lattice volume is used with $M_\pi \approx 380$ MeV and $M_\pi L \approx 6$.
Results are consistent with those of Refs.\ \cite{Bicudo:2012qt,Bicudo:2015vta}:
there is evidence for a $I=0$ $J^P=1^+$ bound state with a mass $50.0 \pm 5.1$ MeV below threshold.


As well as molecular mesons and tetraquarks, another proposed exotic multiquark structure
is a hadroquarkonium state which has a compact $Q\bar{Q}$ core along with a
more diffuse distribution of light quarks --- see Sec.~\ref{Sect:4.1.3}.
Ref.~\cite{Alberti:2016dru} computes the static $Q\bar{Q}$ potential in the
presence of a variety of light mesons and baryons, and compares with the static potential in vacuum.
One lattice volume is used with $M_\pi \approx 223$ MeV and $M_\pi L \approx 4.6$.
A suggestion of weak attraction, a shift in the potential $\lesssim$ a few MeV, is found for $Q\bar{Q}$ with $\pi$, $K$, $\rho$, $K^*$,
$\phi$, $N$($1/2^+$), $\Xi$($1/2^+$), $\Delta$($3/2^+$), $\Xi^*$($3/2^+$), $N$($1/2^-$), $\Xi$($1/2^-$) or $\Xi^*$($3/2^+$).
Applying the results to a phenomenological study of charmonia (where the static approximation is not
necessarely a good approximation), the energy shifts in the presence of a light hadron are
found to be $\sim$ -0.9 to -2.6 MeV for $1S$ charmonia, -1 to -6.5 MeV for
$2S$ charmonia and -1 to -4.9 MeV for $1P$ charmonia. Because the shifts are
very small, the effects of having unphysical quark masses and working in a
finite volume may be significant -- these need to be checked before definite conclusions can be drawn.


Lattice calculations of hybrid static potentials, i.e.\ the potential
between a static quark and antiquark in the presence of a gluonic
excitation, in pure SU(3) gauge theory (no dynamical quarks) have a long history,
Refs.~\cite{Griffiths:1983ah,Campbell:1984fe,Campbell:1987nv,Michael:1990az,Perantonis:1990dy,Foster:1998wu,Juge:1999ie,Bali:2000vr,Juge:2002br,Bali:2003jq,Capitani:2018rox}
-- the results of these calculations and their phenomenological
implication were discussed above in Section~\ref{Sect:4.2.3bis}.
A related recent calculation of the color field density profiles of excited flux tubes between a static quark and antiquark is presented in Ref.~\cite{Bicudo:2018jbb}.

\hfill

In summary, lattice QCD calculations of the potential between static quarks
can provide an interesting window on the physics of the strong interaction.
They appear to support the existence of a bound state with $J^P = 1^+$ $I=0$ $qq\bar{Q}\bar{Q}$
in line with the phenomenological findings outlined in Sec.~\ref{Sect:4.1.4}.
However, because the binding is sensitive to the light- and heavy-quark masses and spin effects,
calculations with physical-mass quarks and addressing the various other systematics are necessary to draw stronger conclusions.


To conclude this section, Table~\ref{table:lattice_summary} gives a summary of lattice studies where candidates for the $X(3872)$ and flavour exotics have been seen. In Sec.~\ref{Sect.6.3} we give an outlook on prospects for future lattice calculations.

\begin{table}[ht]
\begin{center}
\begin{tabular}{|c|c|c|c|c|}
\hline
State & Reference & Approx.\ $M_\pi$ / MeV & $M_\pi L$ & Binding Energy / MeV \\
\hline
\multirow{2}{*}{$X(3872)$ $I=0$ $J^{PC} = 1^{++}$} & \cite{Prelovsek:2013cra} & 266 & 2.7 & $11 \pm 7$ \\
 & \cite{Padmanath:2015era} & 266 & 2.7 & $8 \pm 15$ or $9 \pm 8$ \\
\hline
$ud\bar{c}\bar{b}$ $I=0$ $J^P=1^+$ & \cite{Francis:2018jyb} & extrap.\ of 164, 299, 415 MeV & 2.4, 4.4, 6.1 & $\sim$ 15 -- 61 \\
\hline
\multirow{3}{*}{$ud\bar{b}\bar{b}$ $I=0$ $J^P=1^+$} & \cite{Francis:2016hui} & extrap.\ of 164, 299, 415 MeV & 2.4, 4.4, 6.1 & $189 \pm 10 \pm 3$ \\
 & \cite{Junnarkar:2018twb} & extrap.\ of various valence quark masses & various & $143 \pm 34$ \\
 & \cite{Leskovec:2019ioa} & extrap.\ of various (139 to 431 MeV) & various & $128 \pm 24 \pm 10$ \\
\hline
\multirow{2}{*}{$qs\bar{b}\bar{b}$ $I=1/2$ $J^P=1^+$} & \cite{Francis:2016hui} & extrap.\ of 164, 299, 415 MeV & 2.4, 4.4, 6.1 & $98 \pm 7 \pm 3$ \\
 & \cite{Junnarkar:2018twb} & extrap.\ of various valence quark masses & various & $87 \pm 32$ \\
\hline
\multirow{5}{*}{\begin{tabular}{c}$ud\bar{Q}\bar{Q}$ $I=0$ $J^P=1^+$ \\ (static quark potentials) \end{tabular}}
 & \cite{Brown:2012tm} & 380 & 6 & $50 \pm 5$ \\
 & \cite{Bicudo:2012qt} & 340 & 3.3 & $\sim$ 30 -- 57 \\
 & \cite{Bicudo:2015vta} & 340 & 3.3 & $90^{+46}_{-42}$ or $93^{+47}_{-43}$ \\
 & \cite{Bicudo:2015kna} & extrap.\ of 340, 480, 650 MeV & 3.3, 4.6, 6.3 & $90^{+43}_{-36}$ \\
 & \cite{Bicudo:2016ooe} & extrap.\ of 340, 480, 650 MeV & 3.3, 4.6, 6.3 & $59^{+30}_{-38}$ \\
\hline
\end{tabular}
\end{center}
\caption{Summary of lattice studies where candidates for the $X(3872)$ and flavour exotics have been seen, and the binding energy below the relevant threshold. See the text for more details.}
\label{table:lattice_summary}
\end{table}

\section{Summary and future prospects}
\label{sect:5}
With more and more experimental data available and the improvement of experimental techniques, a large number of $XYZ$ states have been reported in the
charmonium and bottomonium sectors which have properties at odds
with the quark model description in terms of charm--anti-charm and
bottom--anti-bottom systems, respectively.
This suggests the existence of new kinds of hadrons and, in this
report, we referred to any state that does not appear to fit with the
expectations for an ordinary $\bar{Q}Q$ meson or $qqq$ baryon as ``exotic'' or as an ``exotic candidate''.

For many states, their unambiguous identification as exotics, and even more their classification in a
specific class of exotics, is still a matter of intense investigation both experimentally and theoretically.
On one hand, experimental investigations aim to complete the phenomenological description of these states. On the other hand, theoretical studies aim to reduce the uncertainties in the computation of observables for ordinary quarkonia and exotics. A particular challenge for theoretical studies is to properly account for the mixing of exotics with ordinary quarkonia, which may play an important role for some states.

We have reported extensively on the latest experimental progress which in recent years has
come mostly from the Belle, BESIII, and LHCb experiments. This report
contains a description of all exotic candidates that have been observed
up to the current date. On the theory side, only lattice QCD has the potential to perform full self-consistent calculations from first principles. Currently, however, lattice QCD calculations of exotics adopt approximations and have unquantified systematic uncertainties that reduce their reliability.
Moreover, not all observables are currently accessible in lattice QCD calculations. For these reasons, also other methods are employed to describe exotics. Effective field theories allow to perform calculations of exotic states with a clear connection to QCD. They are limited by their specific ranges of applicability, and need to be supplemented by data or lattice QCD results. Nevertheless, they are systematically improvable and progress is steady. In addition to these systematic approaches there are various phenomenological models investigating different aspects of exotic states. While uncertainty estimates for their results are difficult if not impossible, phenomenological models can still provide valuable insights, in particular where other methods have not been applied yet. In the report, we have presented in some detail these different theoretical approaches.

To understand the existing $XYZ$ states and finally understand the
exotic hadrons, more efforts in both experiment and theory are needed. In experiment, we need to obtain more accurate information on the $XYZ$s, including lineshapes and the resonance parameters for various the $J^{PC}$ quantum numbers, the production and decay modes and so on. Whereas, on the
theoretical side, we need improved calculations to more cleanly discriminate exotic hadrons from conventional hadronic states. We close this report with a discussion of the experimental prospects before giving a theoretical outlook.

\subsection{Ongoing and future experiments}
\label{sect:6.1}
It is expected that data samples accumulated at the $B$ factory
experiments (BaBar and Belle), $\tau$-charm experiment (BESIII),
and hadron machine experiments (CDF, D0, LHCb, ATLAS, CMS) will
continue to be used for the study of the $XYZ$ states. At the same
time, the ongoing experiments will continue to accumulate more
experimental data: LHCb just started exploration of its full RUN1
and RUN2 data and is upgrading for more luminosity; BESIII will
continue its program with extended c.m.\ energy coverage and
improved data taking strategy; the GlueX experiment started data
taking in 2017 with an expectation to reach a higher sensitivity
for the search for pentaquarks and other exotic mesons in
photoproduction processes with more beam time available; Belle II
has started its pioneering run with the full detector in April
2019 and is expecting its target integrated luminosity of 50 ab$^{-1}$
in a few years.
Furthermore, some new experiments are expected to join the effort
of investigating the exotic hadrons in the future: the PANDA
experiment at Darmstadt, Germany, the super $\tau$-charm factories
STCF (HIEPA) in Hefei, China, and SCTF in Novosibirsk, Russia.
In this section, we give a brief description of future prospects
at the ongoing and planned experiments.

\vspace{0.3cm}\noindent
 {\it $\bullet$ The GlueX experiment}
\vspace{0.3cm}

The GlueX experiment, studying light meson spectroscopy with
an emphasis on the search for light hybrid states, is the flagship
experiment of the newly constructed Hall D at the Thomas Jefferson
National Accelerator Facility (JLab)~\cite{Ghoul:2015ifw}. The
experiment exploits photoproduction from a fixed hydrogen target,
using both linearly polarized and unpolarized photons.

\begin{figure*}[htbp]
\begin{center}
\includegraphics[height=6cm]{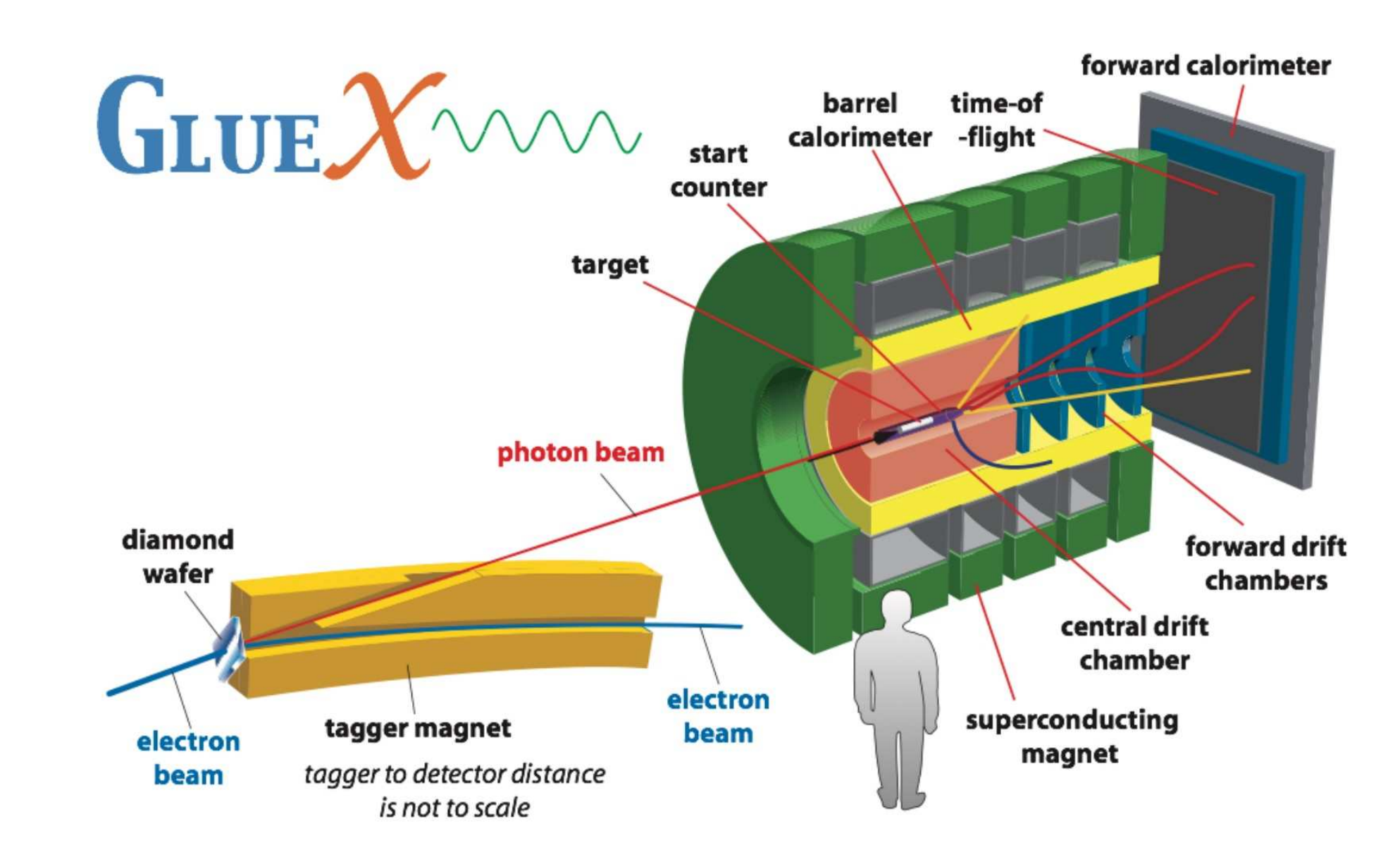}
\end{center}
\caption{Schematic of the GlueX detector and the photon beam tagger~\cite{Ghoul:2015ifw}.}
\label{gluex-detector}
\end{figure*}

The GlueX detector is illustrated in Fig.~\ref{gluex-detector}. An
electron beam of up to 12~GeV impinges on a thin diamond radiator
which produces a coherent bremsstrahlung photon beam with a
substantial linear polarization in a narrow energy range peaking
at 9~GeV. With a collimator suppressing the incoherent
bremsstrahlung spectrum, a linear polarization of 40\% is achieved
in the coherent peak at 9~GeV. The scattered electrons are used to
tag the energy of the photon beam. The main detector, shown in
Fig.~\ref{gluex-detector}, consists of a 2~T solenoid magnet with
the central and forward drift chamber for charged particles, a
lead/scintillating fiber barrel calorimeter and lead-glass forward
calorimeter for neutral particles and electron detection, and a
start counter and forward TOF wall for precision timing
measurements. The angular coverage of the detector is complete
down to opening angles from the beam of about two degrees. Pions
are reconstructed for momenta down to 100~MeV and protons to
300~MeV with a resolution from 1\% to 3\%.

The GlueX detector has been successfully commissioned and started
its physics data taking in spring 2017. Based on the accumulated
data, the analysis of a variety of physics topics is already
ongoing. For example, the cross section of $\gamma p \to J/\psi p$
has been measured at $8.2<E_\gamma<11.8$~GeV using
about $25\%$ of the total data accumulated by the GlueX experiment
from 2016 to date~\cite{Ali:2019lzf}. Although clear $J/\psi$
signals were observed, no statistically significant evidence was
found for the pentaquark observed by LHCb.

The observation of many well-known meson resonances as well as the
first successful reconstruction of the $J/\psi$ demonstrates good
prospects for a broad physics program with initial GlueX data. The
mapping of the entire light meson spectrum will be possible, with
a precise measurement of the properties of known resonances and,
ultimately, the candidates for exotic mesons. The completion of
the initial GlueX running is expected in 2019, after which several
upgrades are planned. A DIRC detector will be added in the forward
region to improve pion and kaon separation~\cite{Stevens:2016cia}.
This and a factor of five or more higher luminosity will allow to
reach a higher sensitivity for the search for pentaquarks and
other exotic mesons.

\vspace{0.3cm}\noindent
 {\it $\bullet$ The BESIII experiment}
\vspace{0.3cm}

Since its commissioning in 2008, the BESIII detector has been
operating successfully for more than 10 years. The end cap TOF
system has been upgraded into a MRPC-based detector with improved
particle identification capability. A cylindrical gas electron
multiplier (CGEM) based inner detector CGEM-IT is being
constructed to replace the inner MDC chamber with similar
performance as the previous chamber.

\begin{figure*}[htbp]
\begin{center}
\includegraphics[height=6cm]{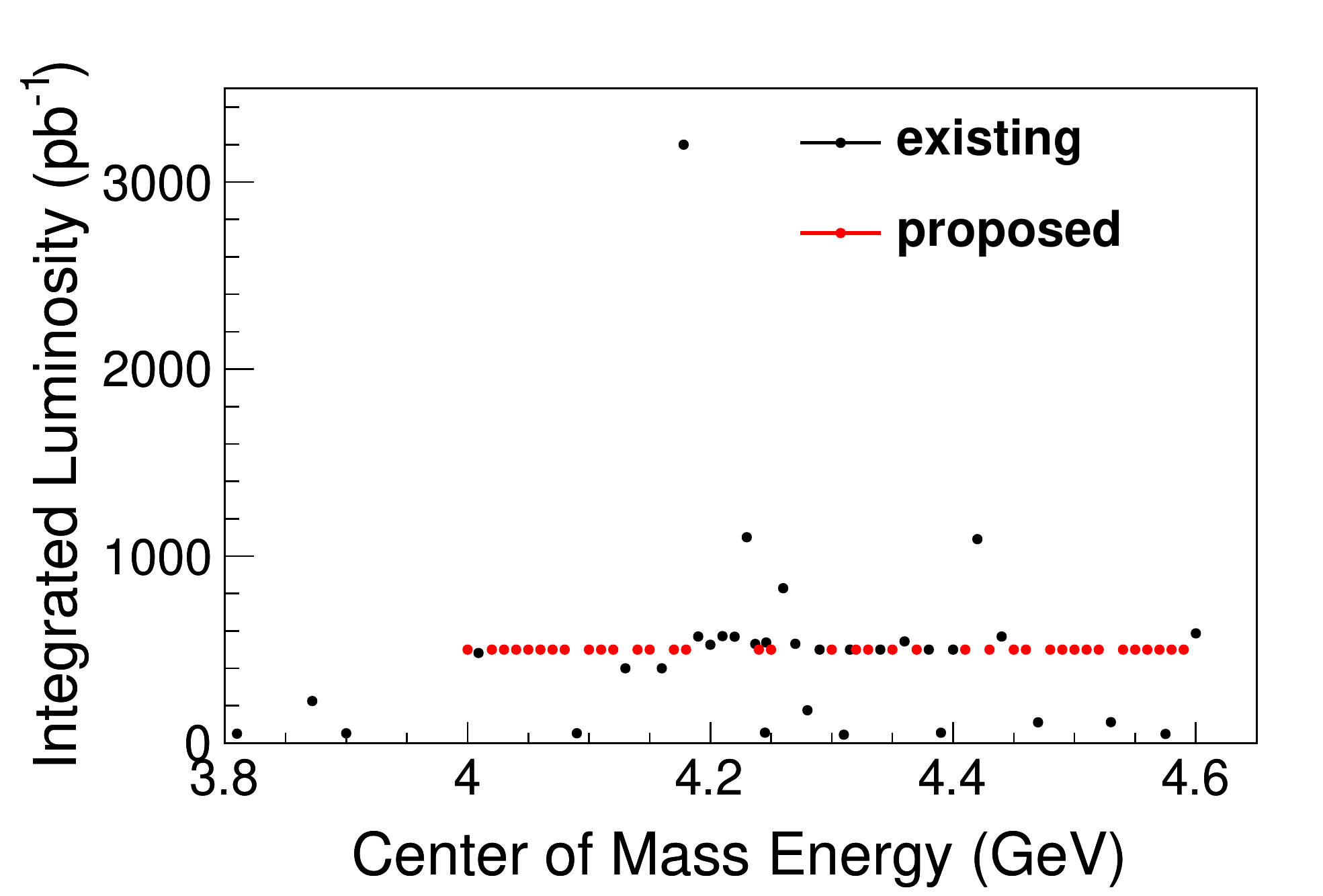}
\end{center}
\caption{The data sets accumulated by BESIII mainly for the study
of $\xyz$ states, where the data sets collected are
shown in black, and those considered in the near future for potential future measurements
are shown in red.}\label{bes3-data}
\end{figure*}

The BEPCII has delivered more than 20~fb$^{-1}$ integrated
luminosity at different energy points in the last 10 years and the
peak luminosity keeps on improving every year. BESIII is planning
to take more data above 4~GeV, 500~pb$^{-1}$ per point with a
10~MeV interval. The intervals were chosen to cover the
possibility for studying narrow $\xyz$ states. For example, the
$e^+ e^- \to \pi^+ \pi^- J/\psi$ cross section has shown rapid
changes between 4.20 and 4.23~GeV. The cross sections for other
reactions are likely to have such rapidly changing features, which
require fine energy scans. Figure~\ref{bes3-data} shows the data
sets accumulated by BESIII, and possible future plan.

To meet the requirement of the BESIII physics research, two
aspects of the BEPCII upgrade have been considered: one is to
increase the maximum c.m. energy from 4.6 to 4.9 GeV, and the other is
the top-up injection in collision operation. The former will allow
to fully cover the $Y(4630)$ and $Y(4660)$ resonances and a wide
range of $\EE\to K^+K^-\jpsi$ production, which may be a key mode
of searching for the $Z_{cs}$ state and might also provide deeper
insights for the $Y(4660)$. The latter will
increase the integrated luminosity by about 30\% in the same
period of data taking time. These upgrades are in progress and are
expected to be completed in 2019.

In addition, an even more ambitious upgrading plan of increasing
the peak luminosity by a factor of 10 has been
proposed~\cite{Bogomyagkov:2016pld}. In this scheme, the Crab
Waist collision with large Piwinski angle would be adopted
together with significant modification of the BEPCII parameters.
At the same time, BEPCII is also investigating a possibility of
moderate improvement of the peak luminosity by increasing the
number of bunches and the beam current, as well as installing more
radio frequency cavities. This latter scenario may need much less
machine shutdown time to accumulate similar amount of integrated
luminosity in 5 to 10 years than the realisation of the Crab Waist scheme.

\vspace{0.3cm}\noindent
 {\it $\bullet$ The Belle II experiment}
\vspace{0.3cm}

The Belle II detector is the sole experiment at the SuperKEKB
$\EE$ collider in Tsukuba, Japan. The goal of the Belle II
experiment, as a next generation flavor factory, is to search for
new physics in the flavor sector at the intensity frontier, and to
improve the precision of measurements of Standard Model
parameters. It started to record data in April 2018 and is in a unique
position regarding bottomonium physics, and an
experiment with full capability to explore QCD phenomenology and
study conventional/exotic states.

There are many experimental reasons that make Belle II perfectly
suited to study exotic states: (1)~Running at the $\Upsilon(4S)$
resonance produces a very clean sample of $B\bar{B}$ pairs. The
low background environment allows for reconstruction of final
states containing photons. (2)~Due to low track multiplicities and
detector occupancy, the reconstruction efficiency of
$B-$/$D-$mesons is high and the trigger bias is very low.
(3)~Since the initial state is known, ``missing mass'' analyses
can be performed to infer the existence of new particles via
energy/momentum conservation rather than demanding full
reconstruction of their final states. This gives the opportunity
to measure the absolute branching fractions of the $XYZ$ states.
(4)~Overcoming the statistical limitations of previous
experiments, Belle II will be in a unique position to perform, for
example, searches for resonant states via ISR processes, studies
of quarkonium-like spectroscopy via bottomonium decays and
bottomonium-like spectroscopy via hadronic and radiative
transitions among bottomonia, and studies of QCD bound states like
the deuteron and di-baryons.

The SuperKEKB accelerator is an upgrade of the KEKB $B$-factory
running in the region of the $\Upsilon$ resonances with the beam
energies of 7~GeV for the high energy ring and 4~GeV for the low
energy ring. The target luminosity of SuperKEKB is a factor of 40
larger than that of KEKB, i.e., $8 \times
10^{35}$~cm$^{-2}s^{-1}$. To achieve such a high luminosity, a
nanobeam collision scheme is taken. The maximum c.m.\ energy in
SuperKEKB is 11.24~GeV, just at the $\Lambda_b\bar{\Lambda}_b$
threshold which makes the possibility of studying the $\Lambda_b$
decays very uncertain.

The Belle II detector is a hermetic magnetic spectrometer
including several substantially upgraded or new subsystems. The
new vertex detector (VXD) consists of two subdetectors: a Pixel
Vertex Detector and a double-sided Silicon strip Vertex Detector.
An improvement by a factor of 2 on the vertex resolution compared
with the Belle vertex detector is obtained with this strategy. The
central tracking system is a large volume CDC surrounding the VXD.
To be able to operate at high event rates, the CDC has been modified
with smaller cells. The particle identification system includes
the Time-Of-Propagation system in the barrel region and the
Aerogel Ring Image Cherenkov detector in the forward region to
measure the time of propagation and the impact position of
Cherenkov photons, the EMC based on CsI(Tl) crystals to detect
photons and identify electrons, the K-Long and Muon detector. A
full discussion can be found in the Technical Design
Report~\cite{Abe:2010gxa}.

\begin{figure*}[htbp]
\begin{center}
\includegraphics[height=5cm]{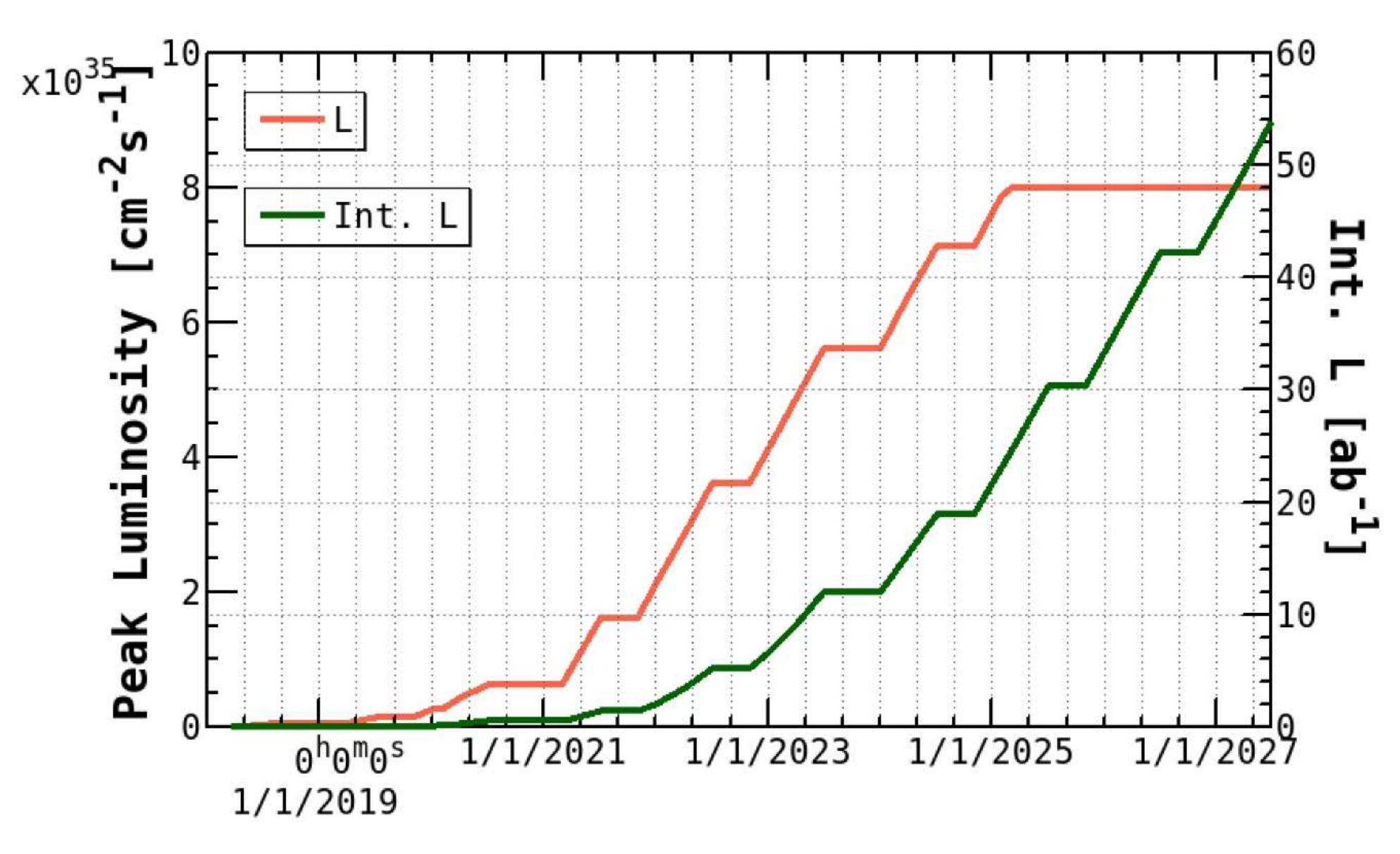}
\end{center}
\caption{Plan of instantaneous and integrated luminosity at SuperKEKB.}\label{belle2-lum}
\end{figure*}

The Belle II experiment started physics running with its full detector in
March 2019. It is foreseen to run for 9 months/year
with a three-month summer shutdown. It is planned to reach a peak
luminosity of $8\times 10^{35}$~cm$^2s^{-1}$ by 2025 and continue
running until 2027 to integrate more than 50~ab$^{-1}$. The
planned integrated luminosity is shown in Fig.~\ref{belle2-lum}.

\begin{figure}[htbp]
\begin{center}
\includegraphics[width=10cm]{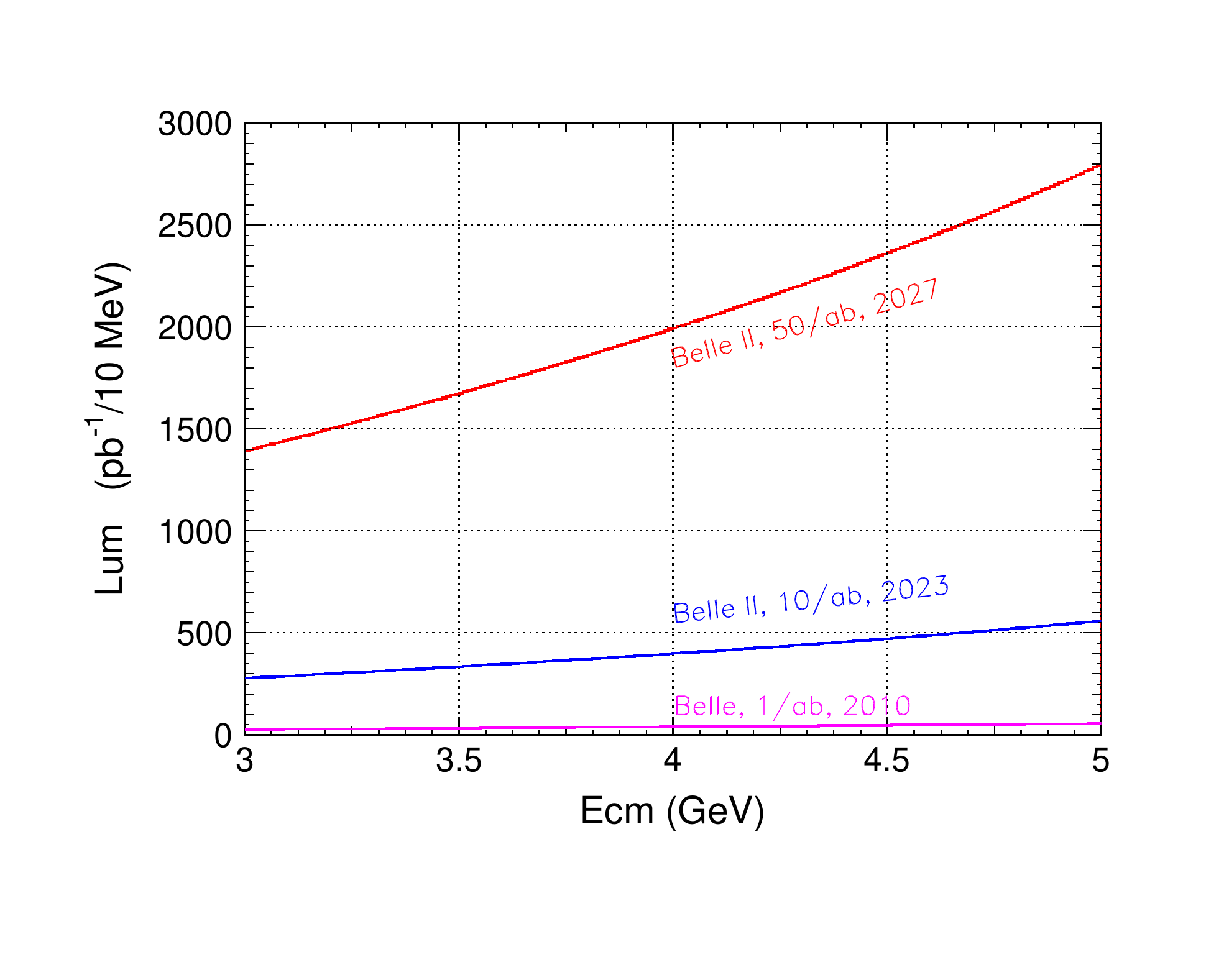}
\caption{Effective luminosity at low energy in the Belle and Belle
II $\Upsilon(4S)$ data samples.} \label{lum_belle2}
\end{center}
\end{figure}

In the near future, the study of $XYZ$ particles will benefit a
lot from the large data samples of the Belle II experiment in many
different ways~\cite{Bevan:2014iga}, among which ISR can produce
events in the same energy range covered by BESIII.
Figure~\ref{lum_belle2} shows the effective luminosity
distribution in the Belle and Belle II data samples. We can see
that, 50~ab$^{-1}$ of Belle II data correspond to
2,000--2,800~pb$^{-1}$ data for every 10~MeV from 4--5~GeV,
similar statistics will be reached for modes like $\EE\to \ppjpsi$
at Belle II and BESIII taking into account the fact that Belle II
has lower efficiency. Belle II has the advantage that data at
different energies will be accumulated at the same time, making
the analysis much simpler than at BESIII where data are
accumulated at many data points. Many ISR processes, like $e^+e^-
\to \pi^+ \pi^- J/\psi$, $\pi^+ \pi^- \psi(2S)$, $K^+ K^- J/\psi$,
$K^+ K^- \psi(2S)$, $\gamma X(3872)$, $\pi^+\pi^-\psi_2(1D)$,
$\pi^+ \pi^- h_c$, $\pi^+ \pi^- h_c(2P)$, $\omega \chi_{cJ}$,
$\phi \chi_{cJ}$, $\eta J/\psi$, $\eta' J/\psi$, $\eta \psi(2S)$,
$\eta h_c$, $(D\bar{D}^{\ast})^{\pm}\pi^{\mp}$, $(D^{\ast}
\bar{D}^{\ast})^{\pm}\pi^{\mp}$, $\Lambda_c^+ \bar{\Lambda}_c^-$,
$\Lambda_c^+ \bar{\Sigma}_c^-$, can be done at Belle II to search
for more production and decay modes of known $XYZ$ states or more
new charmonium-like states.

\vspace{0.3cm}\noindent
 {\it $\bullet$ The LHCb experiment}
\vspace{0.3cm}

\begin{figure*}[htbp]
\begin{center}
\includegraphics[height=3cm]{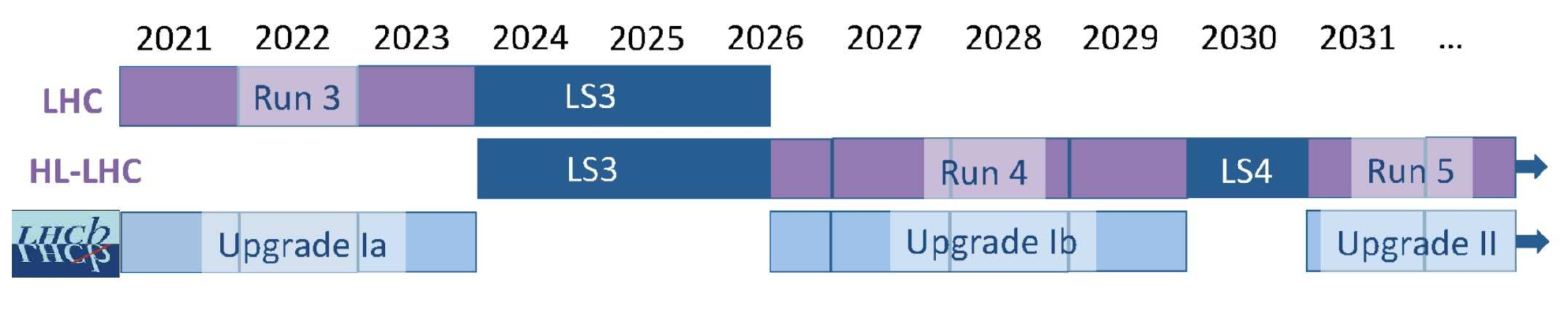}
\end{center}
\caption{Timeline of accelerator and experiment operation of the
LHC, HL-LHC, and LHCb over the decade 2021 to
2031~\cite{Bediaga:2018lhg}.}\label{lhcb-upgrade}
\end{figure*}

The LHCb experiment has demonstrated itself as an ideal laboratory
for flavor physics studies. The LHCb upgrade will fully exploit
the flavor-physics opportunities of the high-luminosity large
hadron collider (HL-LHC) to search for physics beyond the Standard
Model in an approach complementary to the energy frontier. The
timelines of operation and major shutdowns of the LHC, HL-LHC, and
LHCb are illustrated in Fig.~\ref{lhcb-upgrade}.

The LHCb Upgrade~I is currently in progress and data taking will
start in 2021 after the LHC Long Shutdown~2 (LS2). Consolidation
of the LHCb Upgrade~I detector is required during LS3. The
preparatory work for the LHCb Upgrade II will also be carried out
at this time. These changes are referred to as Upgrade~Ib. During
the period of LHCb Upgrade~I, the detector will be operated with
an instantaneous luminosity of $2\times 10^{33}$~cm$^{-2}s^{-1}$.
A total integrated luminosity of around 23~fb$^{-1}$ is
anticipated by the end of Run 3 and 50~fb$^{-1}$ by the end of Run
4 of the LHC. LHCb Upgrade~II will be installed during LS4, which
is expected to start data taking in 2031. At that time, the LHCb
will operate with an instantaneous luminosity of up to $2\times
10^{34}$~cm$^{-2}s^{-1}$, an order of magnitude above Upgrade~I.
LHCb will accumulate a data sample corresponding to a minimum of
300~fb$^{-1}$. To meet such a high instantaneous luminosity,
modest improvements of the Upgrade I and II detector will be
performed during LS3 and LS4. New upgrades in some subdetector
components include a pixel detector, a vertex detector, the RICH
system, and so on. They will improve the intrinsic performance of
the experiment in certain key areas. For more detail, we refer to
Ref.~\cite{Bediaga:2018lhg}.

The LHCb experiment has already strongly contributed to our
understanding of the $XYZ$ particles, e.g. via the determination of
the quantum numbers of the
$X(3872)$~\cite{Aaij:2013zoa,Aaij:2015eva}, and the observation of
pentaquark candidates $P_c(4380)^+$ and
$P_c(4450)^+$~\cite{Aaij:2015tga} [recently LHCb found that the
$P_c(4450)^+$ is composed of two narrow overlapping structures
$P_c(4440)^+$ and $P_c(4457)^+$~\cite{Aaij:2019vzc}]. The large
data set collected in the Upgrade II era will boost the
sensitivity of searches for pentaquark multiplets and heavy exotic
states with small production cross sections and low efficiency.
For examples: (1) the observation of radiative decays involving
exotic baryons will provide a new insight into the structure of
the pentaquark candidates by performing an amplitude analysis of
the $\Lambda_b \to J/\psi p \gamma K$ decay. The proposed improved
EMC is crucial for the feasibility of such a measurement in a
high-luminosity environment. (2) The neutral pentaquark candidates
can be searched for in the $\Lambda_c^+D^-$ system in the process
$\Lambda_b^0 \to \Lambda_c^+D^-\bar{K}^{*0}$. Such process suffers
from the low detection efficiency and the small branching
fractions of $\Lambda_c^+$ and $D^-$ decays. Such channel needs
very large data sets to compensate the low efficiency and small
product branching fraction. (3) The pentaquark states with
strangeness can be searched for in the $\Xi_b^- \to J/\psi \Lambda
K^-$ process, which has been seen at LHCb using Run~1
data~\cite{Aaij:2017bef}. About 300 signal events were observed,
thus no intermediate states were investigated. An increase of the
integrated luminosity by a factor of 100 would allow detailed
amplitude analyses with a similar sensitivity as in the pentaquark
discovery channel.

The enhancements of the Upgrade~I and~II
detector~\cite{Bediaga:2018lhg} together with the enormous sample
sizes of Run 5 will ensure that LHCb maintains its position in
exotic hadron studies, and provide a unique access to studies
involving the production of $B_c^+$, $\Lambda_b^0$, and $\Xi_b$
hadrons.

\vspace{0.3cm}\noindent
 {\it $\bullet$ The PANDA experiment}
\vspace{0.3cm}

The PANDA (antiProton ANnihilation in DArmstadt) experiment under
construction will be located at the FAIR (Facility for Antiproton
and Ion Research) complex in Darmstadt, Germany. It is dedicated
to study hadron physics. The FAIR accelerator complex will provide
particle beams for four main experimental pillars, one of which is
the PANDA experiment. At FAIR, a new proton LINAC will
preaccelerate protons to 70~MeV, and ultimately accelerate them to
about 30~GeV. The proton beam will hit a copper target acting as
the antiproton production target. Magnetic horns are then used to
filter the antiprotons of 3.7~GeV, which are collected and cooled
in the Collector Ring (CR), then injected in the High Energy
Storage Ring (HESR). Here, the antiprotons can be deaccelerated or
further accelerated to a range of 1.5~GeV and 15~GeV, which
correspond to c.m. energies in the range of 2.2~GeV and 5.5~GeV.
The full setup is designed to provide an instantaneous peaking
luminosity of $2\times 10^{32}$~cm$^{-2}s^{-1}$. The accumulated
integrated luminosity can reach 2~fb$^{-1}$ in about five months.

\begin{figure*}[htbp]
\begin{center}
\includegraphics[height=8cm]{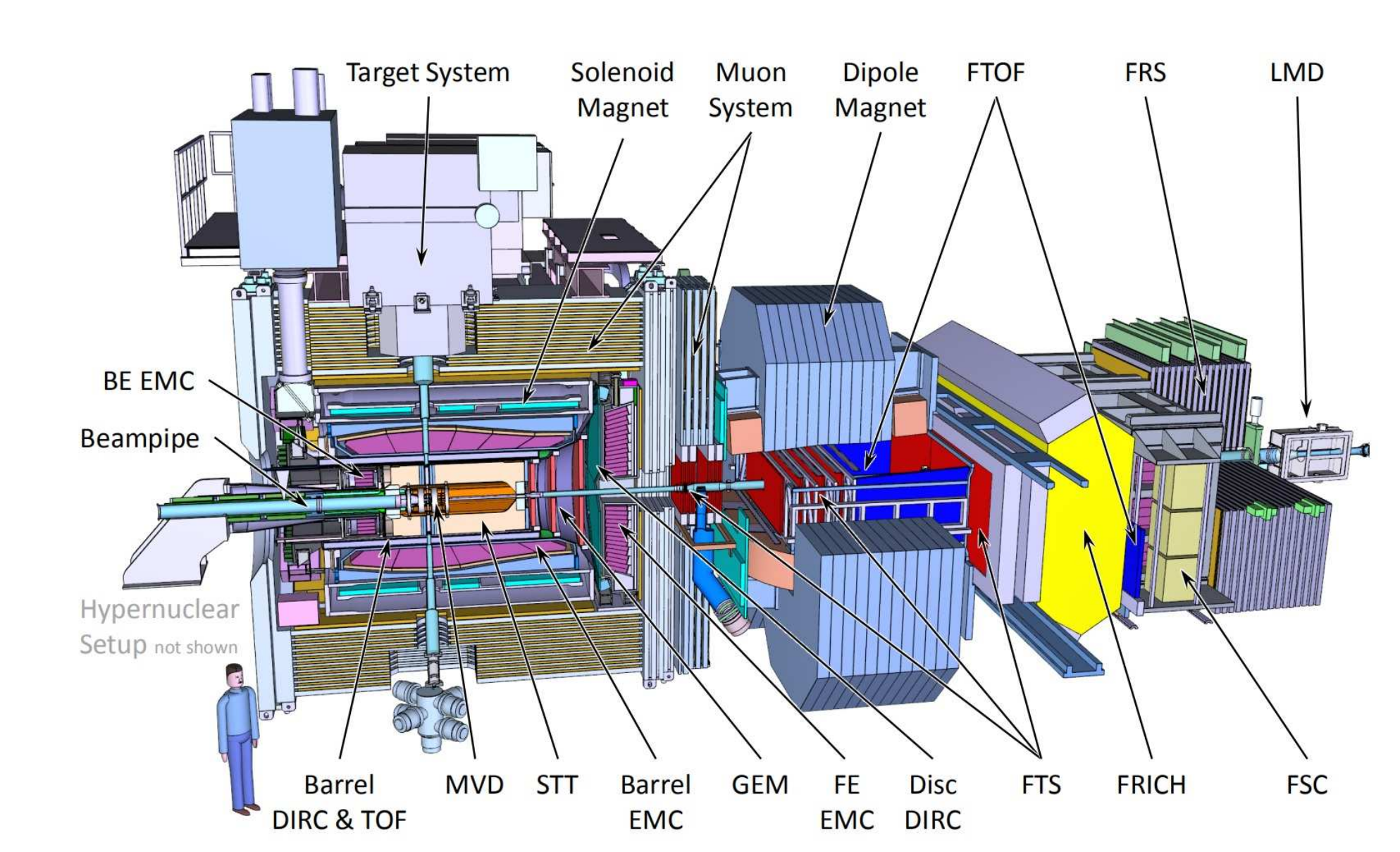}
\end{center}
\caption{The proposed PANDA experimental setup~\cite{PANDA:2018zjt}.}\label{panda}
\end{figure*}

The proposed PANDA detector with a 4$\pi$ geometrical acceptance
is shown in Fig.~\ref{panda}. It consists of the target
spectrometer surrounding the target area and the forward
spectrometer for the detection of particles produced in the
forward direction. The Micro Vertex Detector (MVD), surrounding
the target region, will provide precise vertex position
measurements with resolution of about 50~$\mu m$ perpendicular to
and 100~$\mu m$ along the beam axis. It consists of silicon pixel
and strip sensors. Tracking with a transverse momentum resolution
better than 1\% will be provided by Gas Electro Multiplier (GEM)
planes and a Straw Tube Tracker (STT) combined with the MVD and
the field of the 2 T solenoid magnet. Information from two
Detection of Internally Reflected Cherenkov Light (DIRC) detectors
and a Time-Of-Flight detector system (TOF) will be utilized to
perform particle identification (PID). The Electromagnetic
Calorimeter (EMC) will be used to detect photons. Muon PID will be
provided by the muon detector system surrounding the solenoid magnet.

The forward spectrometer, which consists of a forward tracking
system (FTS) of three pairs of straw tube planes, covers polar
angles below 10 and 5 degrees in the horizontal and vertical
planes, respectively. An Aerogel Ring Imaging Cherenkov Counter
(ARICH) and a Forward TOF system (FTOF) will be used for PID and
the Forward Spectrometer Calorimeter (FSC) provides photon
detection and electron/pion separation. A Forward Range System
(FRS) and the Luminosity Detector (LMD) complete the forward spectrometer.

Three different scenarios for the different phases of the
accelerator completion are expected and summarized in
Table~\ref{panda-data}, where the momentum spreads $dp/p$,
beam-energy resolutions d$E_{\rm cms}$, and integrated
luminosities $\cal{L}$ (at $\sqrt{s}=3.872$~GeV) are given for
three different HESR operation modes: The High Luminosity (HL),
High Resolution (HR), and initial ``Phase-1'' (P1)~\cite{Lehrach:2005ji}.

\begin{table*}[htbp]
 \centering
 \begin{tabular}{cccc}
 \hline
 HESR mode & $dp/p$ & d$E_{\rm cms}$ (keV) & $\cal{L}$ [1/(day $\times$ nb)]
 \\
 \hline
 HL & $1\times10^{-4}$ & 167.8 & 13680\\
 HR & $2\times10^{-5}$ & 33.6 & 1368 \\
 P1 & $5\times10^{-5}$ & 83.9 & 1170 \\
 \hline
 \end{tabular}
\caption{ \label{panda-data} Momentum spreads $dp/p$, beam-energy
resolutions d$E_{\rm cms}$, and integrated luminosities $\cal{L}$
(at $\sqrt{s}=3.872$~GeV) of the three different HESR operation
modes~\cite{Lehrach:2005ji}.}
\end{table*}

For the PANDA experiment, a significant contribution to the study
of the $XYZ$ states will be the precise measurement of the
energy-dependent cross section of a specific process over a
certain range of c.m. energies by adjusting the beam momentum at
high precision. This allows for the precise determination of some
parameters for some $XYZ$ states, for example the natural width
and line-shape measurements of very narrow resonances. Recently
PANDA used the $X(3872)$ as an example to study the achievable
sensitivities of measuring its width, since an absolute width
measurement is the key to understand the nature of the $X(3872)$
and distinguish between the various theoretical models. For such a
narrow state with $J^{PC}=1^{++}$, the PANDA experiment has a good ability
to perform precise energy scans of the resonance regions. Based on a comprehensive MC simulation study, the
achievable sensitivities of measuring $X(3872)$ width are achieved
by assuming 40 energy scan points and data-taking time period of
two days per point~\cite{PANDA:2018zjt}. The outcome of the
sensitivity study is very promising. For example, with an
assumption of the BW shape of the $X(3872)$ and an input signal
cross section of $\sigma(p \bar{p}\to X(3872))=50$~nb, a 3$\sigma$
precision on the measured $X(3872)$ width ($\Gamma_{\rm mesa}$),
$\Delta \Gamma_{\rm mesa}/\Gamma_{\rm mesa}$ better than 33\%, is
achieved for an assumed natural decay width larger than about
$\Gamma_0= 40$, 80, and 110~keV for HR, HL, and P1 running modes, respectively.

\vspace{0.3cm}\noindent
 {\it $\bullet$ The Super-Charm-Tau and Super-Tau-Charm Factories}
\vspace{0.3cm}

For a decade, the Budker Institute of Nuclear Physics in
Novosibirsk has been developing plans for a Super-Charm-Tau Factory (SCTF).
SCTF is an $e^+e^-$ collider complex for high-precision
measurements between 2 and 6~GeV with instantaneous luminosity up
to $10^{35}~{\rm cm}^{-2}s^{-1}$ and longitudinal polarization of
the initial $e^-$ beam. The feasibility of constructing a collider
with such instantaneous luminosity is based on the Crab-Waist
scheme of collisions with large Piwinski angle experimentally
tested at the $\phi$-factory DAFNE in Frascati. The chosen
collision scheme and machine lattice provide luminosity from $0.7$
to $2\times 10^{35}~{\rm cm}^{-2}s^{-1}$ at beam energies from 1
to 3~GeV. Such a facility should be a successor of the CLEO-c and
BESIII experiments integrating their best features and increasing
the useful yield of BESIII by two orders of magnitude. Integrated
luminosity of $\sim 10$~ab$^{-1}$ could be collected in 10 years.
The energy range covers rich physics from light quark mesons to
the $\tau^+\tau^-$ threshold, charm mesons, baryons, and exotic hadrons.

While SCTF cannot compete generally with Belle II and LHCb in
the size of the data samples, the advantages of SCTF are:
(1)~Threshold production of pairs of $\tau$ leptons and charmed
hadrons; (2)~Longitudinal polarization of initial electrons
facilitating searches for CP violation in decays of charm baryons
and $\tau$ leptons; (3)~Coherent production of $D^0\bar{D}^0$
mesons (measurement of phases); (4)~Double tagging (measurement of
absolute branching fractions).

More than 10 charmonium-(like) states discovered since 2003 remain
unclassified; in other words, their origin is still unclear.
Detailed studies of their properties that could lead to their
final classification demand huge data samples. With such collected
statistics coupled-channel analyses become possible, which result
in a consistent set of resonance parameters. Various production
mechanisms of such states exist:
\begin{enumerate}
 \item
All $\psi (Y)$ states with $J^{PC}=1^{--}$ will be directly
produced at $\sqrt{s}=M_Y$: $\psi(4260/4230)$, $\psi(4360)$,
$\psi(4660)$
 \item
Charged $Z_c$ states can be produced by scanning the $\sqrt{s}$
range and studying the $J/\psi\pi\pi$, $h_c\pi\pi$,
$D^{(*)}\bar{D}^{(*)}$ final states
 \item
Neutral $c\bar{c}$ states with other quantum numbers can be
studied in the recoil to $\pi\pi$, $\pi^0$, $\eta$, $\omega$ final
states
 \item
$C=+1$ states can be also produced in $\gamma\gamma$ collisions
 \item
For the SCTF with the maximal energy higher than 6~GeV, between 6
and 7~GeV double $c\bar{c}$ production becomes possible
\end{enumerate}

To accomplish the physical program of experiments at the SCTF a
universal magnetic detector should be designed with the following
features: (1)~Digitizing electronics and a data acquisition system
have to be capable of reading out events with a data rate of
300-400~kHz and average event size of 30~kB; (2)~High efficiency
for soft tracks; (3)~Excellent momentum resolution for charged
particles and good energy resolution for photons; (4)~High-quality
particle identification; (5)~Trigger is capable of selecting
events and rejecting background at high detector occupancy.

\begin{figure*}
\centering
\includegraphics*[width=0.7\textwidth]{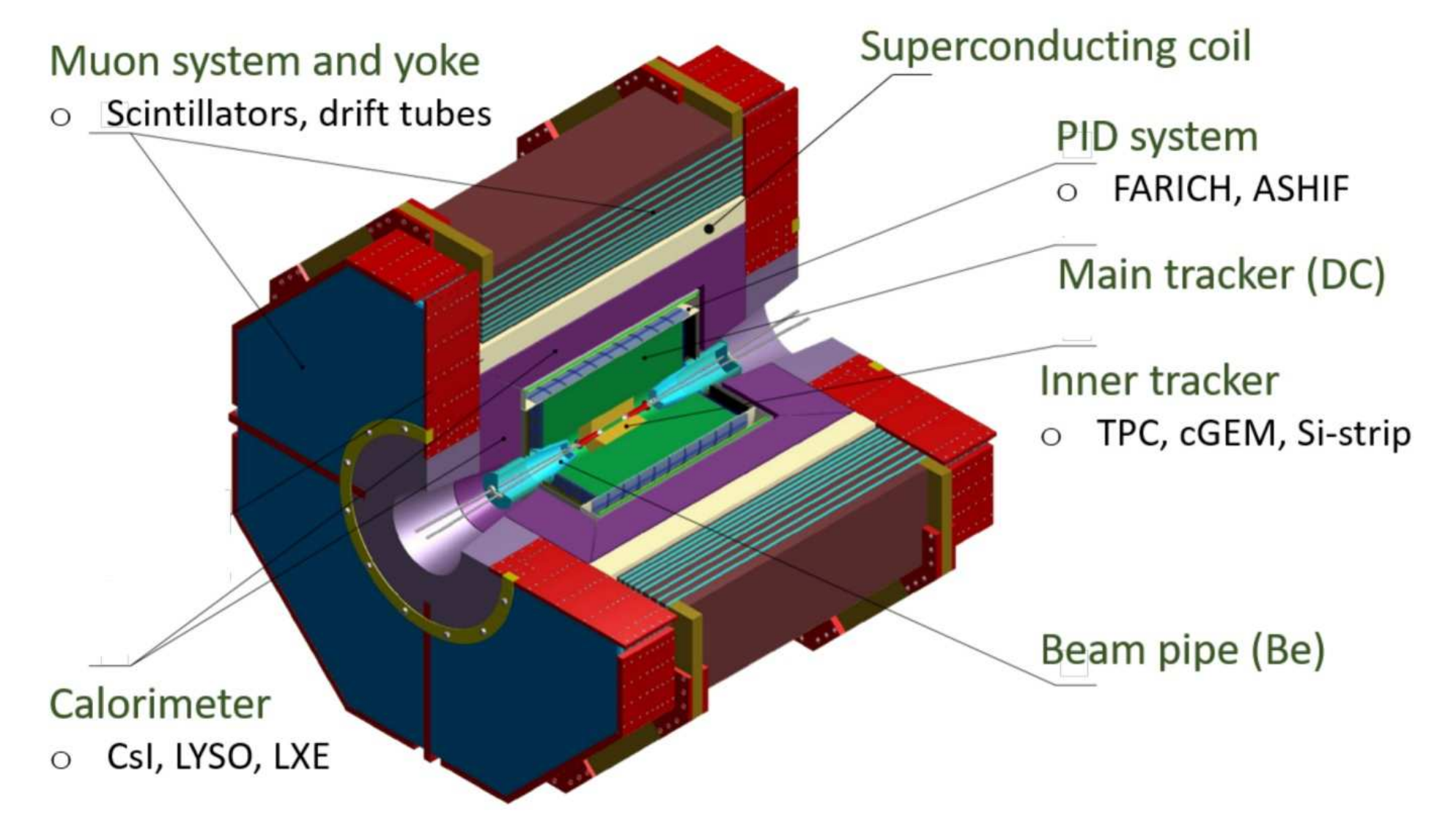}
\caption{Detector for Novosibirsk SCTF}
\label{sctf2}
\end{figure*}

The detector includes a standard set of subsystems shown in
Fig.~\ref{sctf2}: Beryllium beam pipe, inner tracker (time
projection chamber with micropattern gaseous detectors, four-layer
cylindrical GEM detector, stack of silicon strip layers), main
tracker (drift chamber with traditional hexagonal cells, low-mass
drift chamber with cluster counting), particle identification
system (FARICH - Focusing aerogel-based ring imaging Cherenkov
detector, ASHIPH - aerogel shifter photomultiplier system), EMC
(CsI crystal counters, LYSO, Liquid Xenon), thin superconducting
solenoid (thickness 0.1$X_0$), iron yoke with a built-in muon
system (scintillators, drift tubes).

It is believed that BEPCII will finish its mission in the next decade.
After that, a super $\tau$-charm factory, called High Intensity Electron
Positron Accelerator (HIEPA), is being proposed in China.
It will be a next generation electron-positron collider operating in the range of c.m.\ energies from
2 to 7 GeV with polarized electron beam in collision.
The design peak luminosity is $(0.5 \sim 1)\times 10^{35}~{\rm
cm}^{-2}s^{-1}$ at $\sqrt{s}=4$~GeV.
The HIEPA detector is designed to consist of a small-cell MDC with 48 layers,
an EMC, a cylindrical RICH for particle identification,
and a muon detector using the muon telescope detector method.
The SCTF and HIEPA share most of the common physics goals and interests -- the approval of either one of them (or even better both) will make a dedicated high precision study of the physics in tau-charm energy region possible for another one to two decades.


\subsection{Issues and opportunities in experiments}
\label{Sect.6.2}
With the large data samples accumulated at the $B$ factories,
$\tau$-charm facilities, and hadron machines, we have achieved a lot in
the study of the exotic states recently: new charged
charmonium-like states $\zc$ and $\zcp$ were discovered; the
spin-parity quantum numbers of the $\xx$, $\zc$, and $Z_c(4430)$
were determined; the $\y$ structure was found to be dominated by
the $Y(4230)$ with lower mass, narrower width, and more decay
modes; and so on. However, an understanding of these $XYZ$ states is
still primarily at the level of conjecture. In view of this
situation, there is still a lot to learn with the existing and
coming data samples to understand these states better. We stress
that the topics contained hereinafter are not meant to be
comprehensive, but are offered as examples in the hope that
progress will be spurred in various directions.

\begin{itemize}
\item In the $X$ sector:
 \begin{itemize}
 \item
Search for more decay modes [including confirmation of $X(3872)
\to \gamma \psp$], measure precisely its parameters and the
absolute branching fractions for the $X(3872)$.
 \item
Measure the production cross section of $\EE\to \gamma X(3872)$
and $\pp\psi(1\,^3D_2)$ [possibly also $\pp\psi(1\,^3D_3)$],
determine whether they are from resonance decays or continuum
production.
 \item
Study the other $X$ states, such as $X(3915)$ and $X(4140)$, in
$\EE$ annihilation and $B$ decays.
 \item
Confirm some marginal states, such as $X(3940)$, $X(4160)$, and
$X(4350)$.
 \item
Perform PWA analyses as much as possible with improved
parameterizations of the data to determine $J^P$ values and
measure resonant parameters.
 \item
Search for flavor analog exotic states like the $X_b$ [the bottomonium
analog of the $X(3872)$].
 \item
Search for a charmonium/bottomonium-like state with exotic $J^{PC}$.

 \end{itemize}
\item In the $Y$ sector:
 \begin{itemize}
 \item
Measure more precisely the line shapes of more final states in
$\EE$ annihilation, including open-charm and charmonium final
states.
 \item
Try coupled-channel analysis with more information.
 \item
Search for the $Y$ states in $B$ or other particle decays.
 \item
Search for the $Y$ states in more processes, such as $Y \to D_s
D_{s1}(2536)$, $D_s D_{s2}(2573)$, $D_s^* D_{s0}(2317)$, etc.
 \item
Search for the $Y$ states in a higher mass region, such as above
4.7~GeV.
 \item
Search for quantum number partners of the $Y$ states.

 \end{itemize}
\item In the $Z$ sector:
 \begin{itemize}
 \item
Measure the $Z_c$ production cross section in $\EE$ annihilation,
determine whether they are produced from resonance decays or
continuum production.
 \item
Search for $Z_c$ production in $B$ or other particle decays.
 \item
Determine the quantum numbers, measure the Argand plot of the
resonant amplitude, and search for more decay modes.
 \item
Search for $Z_{cs}$ states decaying into $K^\pm\jpsi$,
$D_s^-D^{*0}+c.c.$, or $D_s^{*-}D^0+c.c.$
 \item
Search for more $Z_c$ states and possible partners containing strangeness.
\item Study in more detail the $Z_b$ production from $\Upsilon(11020)$
which might highlight the role of the $B_1\bar B$ channel in this transition.
 \end{itemize}
\item In the $P_c$ sector:
 \begin{itemize}
 \item
Measure additional reactions to investigate the $P_c$, such as the
$J/\psi$ photoproduction off a nucleon using the $\gamma p \to
J/\psi p$ reaction.
 \item
Confirm the $P_c$ via other final states, for examples
$\Lambda_b^0 \to \psi(2S) p K^-$, $J/\psi \pi^+ \pi^- p K^-$,
$\chi_{c1} p K^-$, and $\chi_{c2} p K^-$.
 \item
Search for the $P_c$ states in $B$ decays, baryon decays,
quarkonium decays, and $\EE$ annihilation.
 \end{itemize}
\end{itemize}

\subsection{Issues and opportunities in theory}
\label{Sect.6.3}
In this section, we briefly outline possible developments in the theoretical approaches discussed in the main body of the review,
and in particular in Sec.~\ref{sect:4}, that may improve our understanding of the exotic $XYZ$ states.
In general, we regard it as crucial that, within each existing approach, as many observables as possible are evaluated,
since different features show up most prominently in different observables.
In particular, different assumptions about the underlying structure of the states lead to quite different
predictions for the masses of states other than the ones used to fix the parameters of a given scheme. 

We start by considering phenomenological approaches.
As emphasized in Sec.~\ref{Sect:4.1.1}, the main virtue of the quark model
is that it allows a first classification of states
and the computation of some observables by means of a simple theoretical apparatus.
The quark model has turned out to be quite useful in identifying possible exotic states
close to or above open heavy-flavor thresholds, where more rigorous alternatives
are often absent.
The very notion ``exotic'' is based on the observation of whether or not the studied
resonance fits into the quark-model scheme.
It is natural to expect that the quark model will continue to play a role in
the identification of ordinary quarkonium states and, climbing up in the
spectra, of the $XYZ$ states.
For the identification of states, such as the $XYZ$, close to or
above open-flavor thresholds, its most important evolution
should consist of incorporating more and more
realistically the effects due to the coupling of the heavy quark-antiquark pair with hadron pairs.
The inclusion of these effects goes under the generic name of unquenching the quark model.
Unquenching the quark model is not without difficulties.
The reason is that the results are extremely sensitive to the details of
the particular model framework and to the values of the parameters used therein.
Indeed, even small changes in the parameters may result in a sizeable shift
in the nodes of the wave functions of the mesons involved and, therefore, their overlap.
This, in turn, may result in sizeable changes of the mass shifts and decay widths of the hadrons under study.
Besides this, as was stressed in Sec.~\ref{Sect:4.1}, a proper inclusion of
strong thresholds does not reduce to just a modification of the
quark-model potential, but requires also an extension of the Fock space to
multiparticle components and coupled channels.
Multiparticle components include hybrids, hadroquarkonia, tetraquarks, hadronic molecules, and others.

Hybrids may be described in the quark model by extending it to include
constituent gluons, see Sec.~\ref{Sect:4.1.2}.
The connection of the model parameters with QCD is however very feeble.
This is more so for the constituent gluon than it is in general for the quark model,
as gluons are exactly massless in the QCD Lagrangian due to the gauge symmetry.
Hence, although it is reasonable to expect that the quark model with
constituent gluons will continue to serve as a reference in identifying hybrid state candidates,
it will not, by its own definition, be improvable in a significant way.
Even for what concerns the classification of states, we have seen that the
degeneracy patterns of higher multiplets may be different in other approaches.

Hadroquarkonium and related research have been presented in Sec.~\ref{Sect:4.1.3}.
To deepen our understanding of the hadroquarkonium dynamics
it would be important to clarify under which circumstances the QCD van
der Waals forces generate binding and when not.
For example, in certain cases spin symmetry can be employed to predict additional, as yet unobserved
states. Those predictions allow for a non-trivial test or falsifications of the approach as detailed in Sec.~\ref{Sect:4.1.3}.

If there exist compact tetraquarks with compact diquarks in the color antitriplet configuration as building blocks, as discussed in Sec.~\ref{Sect:4.1.4}, the most striking problem is the abundance of predicted states that,
at least at present, vastly exceeds the number of observations.
Although there are attempts to explain why compact tetraquarks with masses near
thresholds are produced more abundantly, further work is required.
It would be very valuable if within the compact tetraquark approach predictions for line shapes were also provided.

In analogy to the nucleon-nucleon system, the two most popular approaches to
hadronic molecules are phenomenological as well as EFT approaches.
Phenomenological approaches have been presented in Sec.~\ref{Sect:4.1.5}.
The form of the phenomenological potentials is typically deduced from various meson exchanges.
Although such models are rather attractive, since the potentials allow for a
straightforward interpretation, they have serious shortcomings.
To begin with, the potentials constructed and used in these models need
regularisation at short distances, which calls for additional modelling.
This results in significant differences from work to work, thus giving rise to
different, sometimes controversial, predictions.
Also, quantum-mechanical calculations employing a static potential inevitably
ignore multiparticle unitarity,
which might play an important, sometimes crucial, role in some systems.

Despite some advantages, phenomenological approaches, and the quark model
in particular, have some obvious limitations.
The most important one is that phenomenological models are not derived from QCD.
Hence they may miss relevant degrees of freedom and their parameters do not
have a field-theoretical definition.
As a result, even in cases where phenomenological models catch the main
physical features of the system under study,
they are not improvable in a systematic fashion.
This is why crucial progress may be expected if phenomenological
approaches get embedded in or substituted
by suitable effective field theories of QCD. The ones relevant for the
$XYZ$ states have been extensively presented in Sec.~\ref{Sect:4.2}.
Effective field theories provide proper definitions of the potentials,
couplings, low-energy constants and matrix elements describing the systems
of interest.
They are supposed to include in their description all relevant degrees
of freedom,
and are equipped by a power counting that allows, on one hand, to assess
the accuracy of each prediction
and, on the other one, to systematically improve it.

States below threshold are well described by pNRQCD (see Sec.~\ref{Sect:4.2.3}).
Potentials can be computed at short distances in perturbative QCD, and we
have documented the progress made in this direction in the last years
allowing computations with (almost) N$^3$LL accuracy.
At large distances the effective string theory provides valuable analytical tools.
Nevertheless, the most straightforward way to compute the potentials from
QCD is numerically via lattice QCD.
Despite many decades of work, still several improvements and even first-time
calculations are possible in this direction.
The precision of the already computed potentials (spin- and momentum-dependent
ones) should be improved, as some of them do not match properly with short-distance expectations.
Further, they should be computed (ideally) in 2+1+1 flavor lattice simulations.
For the time being only the static potential is available on dynamical lattices.
Finally, some potentials should be computed for the first time, in particular
the $1/m_h^2$ spin- and momentum-independent ones, where $m_h$ is the heavy-quark mass.
These are crucial to have eventually wave functions that include consistently
all leading relativistic corrections.

An effective field theory for hybrids and, still to be fully developed, for tetraquarks is
the BOEFT reviewed in Sec.~\ref{Sect:4.2.3bis}.
To be fully predictive, this EFT requires input from lattice QCD. Hybrid
potentials have been computed, also recently, but in the pure gauge theory.
Full QCD lattice calculations are still missing.
Even more urgent than the calculation of the dynamical hybrid potentials is
the calculation of the leading spin-dependent potential and quarkonium-hybrid mixing potential.
Both depend on the same $1/m_h$ suppressed operator.
Their determination would allow a consistent
determination of the fine structure of the hybrid spectrum and of mixing
with quarkonium states of different spin.
The mixing of spin $0$ ($1$) hybrids with spin $1$ ($0$) quarkonium states
may be crucial to identify those hybrid candidate states that
have been detected in both $\pi^+\pi^-J/\psi$ and $\pi^+\pi^-h_c$ decay channels.
Potentials in the isospin $1$ (tetraquark) sector are even less known than in the hybrid sector.
However, several groups are working in this direction and first results have appeared recently.
These studies will allow, for the first time, to put on a solid ground the
dynamics of systems made of two heavy and two light quarks,
also by highlighting possible analogies and symmetry relations with hybrid systems.

As we mentioned above, hadronic molecules have been studied not only within
phenomenological approaches but also, more recently, with EFTs at the hadron level.
In both cases, a Schr{\"o}dinger or Schr{\"o}dinger-like equation is solved
for a system of two hadrons connected by a potential.
The differences are, however, in the choice of the potential and in the
regularisation of the equations.
A detailed discussion of the EFT approach for hadronic molecules and a
comprehensive list of relevant references can be found in Sec.~\ref{Sect:4.2.4}.
Further theoretical developments of the EFT approach for molecules should
include a systematic improvement in the
theoretical accuracy by considering higher orders in the EFT
expansion, extension to the SU(3) flavor group for light quarks
and tests of the accuracy of the various approximate symmetries employed to construct the EFT potential.
For example, systematic studies of the breaking patterns of the heavy-quark
spin symmetry appear to be very important especially for the charm-quark sector.
Moreover, since within the EFT the light-quark mass dependence is under control (see, e.g., Ref.~\cite{Baru:2013rta}),
it should be possible to provide more predictions for the pole trajectories
as a function of the light-quark masses.
These contain important information on the nature of the
states~\cite{Guo:2012tg} and can be compared to lattice QCD evaluations.

Other rigorous approaches to the physics of $XYZ$ states are sum rules,
briefly reviewed in Sec.~\ref{Sect:4.sumrules} and direct determinations by
lattice QCD, discussed in Sec.~\ref{Sect:4.3}. As we have seen, lattice
QCD plays an important role in providing the crucial non-perturbative input to EFTs.
The combination of suitable EFTs with lattice QCD has the potential to access all relevant observables.
Nevertheless, lattice QCD alone may also provide direct determinations
of many observables.

In recent years there have been significant advances in first-principles lattice
QCD calculations of excited hadrons, near-threshold states and resonances.
However, as summarised in Sec.~\ref{Sect:4.3}, progress in studying resonances
and other phenomena relevant for the $XYZ$s in the charmonium, bottomonium
and related exotic flavor sectors has been generally limited to extracting finite-volume
energy levels rather than robustly determining infinite-volume scattering amplitudes.

No clear sign of any charged charmonium- or bottomonium-like state has been seen in
lattice calculations, in contrast to the various experimental signals for such structures.
A candidate for the $X(3872)$ has been found, Refs.~\cite{Prelovsek:2013cra,Padmanath:2015era},
but more detailed calculations are necessary. This is a case where precision
lattice calculations will be particularly difficult: the $X(3872)$ is very close to threshold
and isospin breaking effects are significant, so the quark masses will need to
be very precisely tuned and QED effects included --- such effects have begun to be
included in some calculations as mentioned in Sec.~\ref{Sect:4.3}. Furthermore,
because the $X(3872)$ is above three-hadron thresholds, three-hadron
scattering channels will need to be incorporated into the analysis.

As discussed in Sec.~\ref{Sect:4.3:bottomonia}, there appears to be accumulating evidence
for a double-bottom $J^P = 1^+$ bound state in $I=0$ $u d \bar{b}\bar{b}$ and
$I=1/2$ $q s \bar{b}\bar{b}$, with the binding getting stronger as the light-quark
mass decreases or the bottom-quark mass increases. This is supported by calculations
involving static quarks, Sec.~\ref{Sect:4.3:static}, and phenomenological
studies, Sec.~\ref{Sect:4.1.4}. Investigating similar exotic $\bar{b}b$ states
is more challenging, but this is another area where we can expect progress.

One useful probe that is available to lattice QCD calculations, and not possible in experiment,
is to observe how phenomena evolve as the quark masses are varied --- this can provide insight
into the structure of states and provide a window on the underlying physics. For example, a conventional
charmonium or bottomonium resonance would be expected to vary little in mass as the light-quark
mass changes, though the relative position of thresholds will change. On the other hand, a state
where compact tetraquark or meson-meson configurations are important would behave in a very different way.
As a concrete example of this in the light-meson sector, compare the behaviour of the
$\sigma$ and $\kappa$ resonances with the $\rho$ resonance in unitarised chiral perturbation theory,
Refs.~\cite{Hanhart:2008mx,Nebreda:2010wv}. The fact that states predicted in a simple model may become broad when analyzed more closely, and thus build the continuum, while others survive as new hadrons, was pointed out already in Ref.~\cite{Jaffe:1978bu}.

One reason that progress has been slower here than in the light-meson sector
is that many hadron-hadron channels become kinematically
open relatively close together in energy, including those involving hadrons
with non-zero spin --- this leads to numerous finite-volume energy
levels that must all be extracted and a number of coupled scattering
amplitudes that must be constrained simultaneously.
Techniques to extract coupled-channel scattering amplitudes, which have been demonstrated for light mesons,
Refs.~\cite{Dudek:2014qha,Wilson:2014cna,Wilson:2015dqa,Moir:2016srx,Dudek:2016cru,Briceno:2017qmb,Woss:2018irj},
suggest that there are good prospects for progress in the near future.
Another issue is the presence of three-hadron thresholds relatively low in energy.
To date, no scattering amplitudes involving more than two hadrons have been
extracted from lattice QCD calculations. Recent computations of
finite-volume energy levels involving three hadrons and developments in the formalism
for relating these to infinite-volume scattering amplitudes suggest that
we can expect the first lattice determinations of these in the not-too-distant future.
However, it is likely to take longer to see realistic applications in the charmonium
and bottomonium sectors. These are areas where the challenges overlap with those
encountered when analysing experimental data, the techniques developed there
may be useful and it may be fruitful to work together with the amplitude analysis community.

Studying transitions enables more stringent probes of the structure of the $XYZ$s,
both theoretically and experimentally, than is possible solely from spectroscopy.
Following the first calculations of radiative transition amplitudes involving a resonance,
$\rho \to \pi \gamma$, Refs.~\cite{Briceno:2015dca,Briceno:2016kkp,Alexandrou:2018jbt},
this is another area where lattice QCD calculations can be expected to make progress.

\hfill

With the ongoing and
planned experiments, and theoretical approaches being developed, there
are good prospects for increasing our understanding of the $XYZ$ states
and exotic hadrons in the near future.

\section*{Acknowledgements}

We would like to express our gratitude to Jaume Tarr\'us Castell\`a and Ulf-G. Mei\ss ner for useful comments
on the manuscript.
This work is supported in part by National Natural Science
Foundation of China (NSFC) under contract Nos. 11575017, 11521505, 11661141008, 11761141009, 11835012, and 11975076;
the Ministry of Science and Technology of China
under Contract Nos. 2015CB856701 and 2018YFA0403902;
Key Research Program of Frontier
Sciences, CAS, Grant No. QYZDJ-SSW-SLH011; the CAS Center for
Excellence in Particle Physics (CCEPP);
the Russian Science Foundation (A.N. was supported by Grant No. 18-12-00226) and the Ministry of Science and Education of
Russian Federation (S.E. was supported by grant 14.W03.31.0026);
the NSFC and Deutsche Forschungsgemeinschaft (DFG) through
funds provided to the Sino--German Collaborative Research Center ``Symmetries and the Emergence of Structure in QCD'' (NSFC Grant No.~11621131001,
DFG Grant No.~TRR110); the DFG
cluster of excellence ``Origins'';
the U.K. Science and Technology Facilities Council (STFC)
[grant number ST/P000681/1].

\bibliographystyle{elsarticle-num}
\bibliography{refs}

\end{document}